\documentclass[rmp,aps,twocolumn,showpacs,preprintnumbers,amsmath,amssymb]{revtex4}
\usepackage{graphicx}
\usepackage{bm}

\def\inbar{\,\vrule height1.5ex width.4pt depth0pt}
\def\IR{\relax{\rm I\kern-.18em R}}
\def\IC{\relax\hbox{$\inbar\kern-.3em{\rm C}$}}

\begin{document}
\title{Dilute ferromagnetic semiconductors: Physics and spintronic structures}

\author{Tomasz Dietl}
\email{dietl@ifpan.edu.pl}
\affiliation{Institute of Physics, Polish Academy of Sciences, al. Lotnik\'ow 32/46, PL-02 668 Warszawa, Poland}
\affiliation{Institute of Theoretical Physics, Faculty of Physics, University of Warsaw, ul. Ho\.za 69, PL-00 681 Warszawa, Poland}
\affiliation{WPI-Advanced Institute for Materials Research (WPI-AIMR), Tohoku University, 2-1-1 Katahira, Aoba-ku, Sendai 980-8577, Japan}

\author{Hideo Ohno}
\email{ohno@riec.tohoku.ac.jp}
\affiliation{Laboratory for Nanoelectronics and Spintronics, Research Institute of Electrical Communication, Tohoku University, Katahira 2-1-1, Aoba-ku, Sendai, Miyagi 980-8577, Japan}
\affiliation{WPI-Advanced Institute for Materials Research (WPI-AIMR), Tohoku University, 2-1-1 Katahira, Aoba-ku, Sendai 980-8577, Japan}
\affiliation{Center for Spintronics Integrated System, Tohoku University, 2-1-1 Katahira, Aoba-ku, Sendai 980-8577, Japan}
\date{\today}

\begin{abstract}
This review compiles results of experimental and theoretical studies on thin films and quantum structures of semiconductors with randomly distributed Mn ions, which exhibit spintronic functionalities associated with collective ferromagnetic spin ordering. Properties of  $p$-type Mn-containing III-V as well as II-VI, IV-VI, V$_2$-VI$_3$,  I-II-V, and elemental group IV semiconductors are described paying particular attention to the most thoroughly investigated system (Ga,Mn)As that supports the hole-mediated ferromagnetic order up to 190~K for the net concentration of Mn spins below 10\%. Multilayer structures showing efficient spin injection and spin-related magnetotransport properties as well as enabling magnetization manipulation by strain, light, electric fields, and spin currents are presented together with their impact on metal spintronics. The challenging interplay between magnetic and electronic properties in topologically trivial and non-trivial systems is described, emphasizing the entangled roles of disorder and correlation at the carrier localization boundary. Finally, the case of dilute magnetic insulators is considered, such as (Ga,Mn)N, where low temperature spin ordering is driven by short-ranged superexchange that is ferromagnetic for certain charge states of magnetic impurities.
\end{abstract}


\maketitle
\newpage
\tableofcontents

\section{Introduction}
\label{sec:introduction}
It has been appreciated for a long time that materials systems combining the tunability of semiconductors with the spin contrast specific to ferromagnets offer a rich spectrum of outstanding properties which are attractive {\em per se} as well as open prospects for an entirely new set of functionalities. From the materials physics perspective there appear three main roads bridging semiconductors and ferromagnets. The first of them is to use hybrid structures consisting of ferromagnetic metals and semiconductors or oxide semiconductors \cite{Zutic:2004_RMP}. Here, as an example of commercially relevant development, one can quote the trilayer structure FeCoB/MgO/FeCoB. In this magnetic tunnel junction, owing to specific symmetry mismatch of wave functions at high quality interfaces, the tunneling resistance increases more than sevenfold at room temperature when magnetization of two ferromagnetic layers becomes antiparallel \cite{Ikeda:2008_APL}. The second possible strategy is to turn ferrimagnetic oxides, such as (Zn,Ni)Fe$_2$O$_4$ showing spontaneous magnetization up 800~K, into good semiconductors by mastering carrier doping and interfacing of these compounds with main stream semiconductors and metals. The third road is to develop semiconductors supporting spontaneous spin polarization, preferably, up to above the room temperature.

This review focuses on epitaxially grown Mn-containing III-V but also II-VI, IV-VI, V$_2$-VI$_3$, I-II-V, and elemental group IV semiconductors, in which {\em randomly} distributed spins of Mn ions show collective ferromagnetic ordering. While the list of the dilute ferromagnetic semiconductors studied so-far is long, certainly the most extensively investigated compound is (Ga,Mn)As \cite{Ohno:1996_APL} and, accordingly, a large part of this review is devoted to this system. It has been demonstrated over the last decade or so that owing to previously unavailable combination of quantum structures and ferromagnetism in semiconductors, the engineered structures of these systems show a variety of new physical phenomena and functionalities. In fact, a series of accomplishments in this field accounts, to a large extent, for spreading of spintronic research over virtually all materials families.

At the same time, however, over the course of several years the studies of semiconductors showing ferromagnetic features have emerged as one of the most controversial fields of today's condensed matter physics and materials science. It becomes increasingly clear that there are four principal reasons for this state of the matter:
\begin{itemize}
\item Challenging the natural assumption [fulfilled in, {\em e.\,g.}, Mn-based II-VI dilute magnetic semiconductors (DMSs)] that the transition metal (TM) impurities substitute randomly distributed cation sites, it appears that depending on epitaxy conditions, co-doping with shallow dopants, and post-growth processing, the magnetic ions can assume interstitial positions and/or aggregate, also with other defect centers, which affects crucially magnetic properties.
\item  A number of growth and processing methods exposes the studied samples to contamination by magnetic nanoparticles and whenever they reside (in the film or substrate volume, at the surface or interface) can determine the magnetic response of the system, particularly in the thin film form.
\item These materials, including (Ga,Mn)As, show simultaneously intricate properties of mismatch semiconductors alloys [such as Ga(As,N)] and of doped semiconductors on the localization verge (such as GaAs:Si). The controversies here echo much dispute, and by some regarded as still unsettled questions, of whether in the relevant range of concentrations the impurity levels (derived from N or Si in the above two examples) are dissolved in the band continuum or form resonant or band gap states, respectively.
\item Considerable effort has been devoted to describe DMSs from first principles (\emph{ab initio}), employing various implementations of the density-functional theory (DFT), particularly involving the local spin-density approximation (LSDA) and its variants. It becomes increasingly clear that inaccuracies of this approach, such as the placement of $d$ levels too high in energy and the underestimation of the band gap, have twisted the field, for instance, by indicating  that the double exchange dominates in (In,Mn)As and that ferromagnetism exists in intrinsic (Zn,Co)O.
\end{itemize}

As emphasized in this review, the present understanding of the field and, in particular, the progress in resolving the above controversial issues as well as a successful modeling of spintronic functionalities are built on two experimental and two conceptual pillars:
\begin{itemize}
\item advanced nanoscale characterization allowing to asses the location and distribution of magnetic ions, dopants, defects, and carriers;

\item comprehensive spectroscopic data providing information on the position of levels introduced by TM ions, their spin and charge states, as well as the coupling to band and/or local states;

\item careful consideration of host band structure, taking into account thoroughly interband and spin-orbit couplings, confinement effects, as well as the presence of surface and edge states in topologically non-trivial cases;

\item  realization that the realm of quantum (Anderson-Mott) localization underlines transport and optical phenomena in carrier-controlled ferromagnetic DMSs.

\end{itemize}

According to the accumulated insight, most of magnetically doped semiconductors and semiconductor oxides, which exhibit ferromagnetic features can be grouped into two main classes:
\begin{enumerate}
\item Uniform DMSs, in which ferromagnetic behavior originates from {\em randomly} distributed TM cations. In most cases [the flagship example being (Ga,Mn)As] the spin-spin interactions are mediated by a high density $p$ of delocalized or weakly localized holes. The confirmed magnitude of the Curie temperature $T_{\text{C}}$ approaches 190~K in Ga$_{1-x}$Mn$_x$As \cite{Olejnik:2008_PRB,Wang:2008_APL} and Ge$_{1-x}$Mn$_x$Te \cite{Fukuma:2008_APLb,Hassan:2011_JCG} with saturation magnetization (in moderate fields, $\mu_0H \lesssim 5$~T) corresponding to less than 10\% of Mn cations. In the absence of itinerant carriers other coupling mechanisms, such as ferromagnetic superexchange in (Ga,Mn)N \cite{Bonanni:2011_PRB,Sawicki:2012_PRB},
  can account for ferromagnetic spin ordering. Since with no carriers the coupling is short ranged, the $T_{\text{C}}$ values reach only about 13~K at $x \approx 10$\% in Ga$_{1-x}$Mn$_x$N \cite{Stefanowicz:2013_PRB}.

\item Heterogenous DMSs, specified by a highly non-random distribution of magnetic elements. Here, ferromagnetic-like properties persisting typically to above room temperature are determined by nanoregions with high concentrations of magnetic cations, brought about by chemical or crystallographic phase separation \cite{Bonanni:2010_CSR}. To this family belong also numerous materials systems, in which ferromagnetic-like properties -- persisting up to high temperatures -- appear related rather to defects than to the presence of TM-rich regions \cite{Coey:2008_JPD}.

\end{enumerate}

The studies of the compounds belonging to the first class is undoubtedly the most mature. On the one hand, significant advances in epitaxy and post-growth processing allowed one to develop a class of ferromagnetic semiconductors, primarily (Ga,Mn)As, showing textbook thermodynamic and  micromagnetic characteristics, despite inherent alloy disorder and a relatively small concentration of the magnetic constituent.  More importantly, the progress in controlling and understanding of these materials has provided a basis for demonstrating novel methods enabling magnetization manipulation and switching as well as spin injection, sensing of the magnetic field, and controlling of the electric current by magnetization direction, the accomplishments having now a considerable impact on the metal spintronics \cite{Ohno:2010_NM}. At the same time, over the course of the years, ferromagnetic DMSs, particularly their magnetic phase diagrams $T_{\text{C}}(x,p)$ and micromagnetic properties, have become a test bench for various theoretical and computational methods of materials science.

In contrast, the control, understanding, and functionalization of the second class of materials systems is in its infancy. However, one may expect a number of developments in the years to come as the availability of materials systems with modulated semiconductor and metallic ferromagnetic properties at the nano-scale, which persist up to above the room temperature, opens new horizons for basic and applied research.

Our aim here is to survey various properties of {\em uniform Mn-based ferromagnetic DMSs}, which we refer to as dilute ferromagnetic semiconductors (DFSs). As seen in the Table of Contents, the main body of the present review consists of three major parts.

First we discuss epitaxial growth and nanocharacterization of DFSs (Sec.~II). We put a particular emphasis on the question of the position and spatial distribution of magnetic ions, which is essential in understanding pertinent properties of any DMSs. We also touch upon the issue of a non-uniform carrier distribution.

In the second part (Secs.~III-VI), we present various outstanding spintronic capabilities of DFSs and their quantum structures with nonmagnetic semiconductors. In particular, we describe how hole-mediated ferromagnetism allows for magnetization manipulation and switching not only by doping or co-doping but also by strain, electric field, and light (Sec.~III). Next the suitability of these systems for spin injection to non-magnetic semiconductors is discussed (Sec.~IV). We also show that in addition to properties specific to semiconductor quantum structures, these materials exhibit functionalities presently or previously discovered in magnetic multilayers, including magnetization switching by an electric current and various magnetotransport phenomena (Sec.~V) as well as inter-layer coupling, exchange bias, and ferromagnetic proximity effect (Sec.~VI).

Finally, in the third part (Secs.~VII-X), we present results on quantitative theoretical studies of thermodynamic, micromagnetic, and spintronic properties of DFSs. We start this part by describing the present understanding of the electronic structure of these systems and exchange coupling between localized spins and itinerant carriers (Sec.~VII).  Equipped with this information, we  present  theoretical models of superexchange (Sec.~VIII) and carrier-mediated ferromagnetism in DFSs (Sec.~IX). Exploiting detailed information on the band structure effects, spin-orbit coupling, and $p$-$d$ hybridization provided by extensive spectroscopic studies on relevant DMSs, these models allow for a computationally efficient interpretation of experimental findings with no adjustable parameters (Sec.~X). Along with emphasizing success of this experimentally constrained approach to the understanding of basic properties and spintronic capabilities of DFSs, we indicate unsettled issues awaiting  further experimental and theoretical investigations.

We conclude our review by discussing possible future directions in basic and applied studies of magnetically doped semiconductors (Sec.~XI).

In this review, we purposively refrain from describing a historical perspective, intermediate or disproved/unconfirmed developments, and a variety of qualitative considerations that have been put forward but not yet shaped into the form allowing for a quantitative verification {\em vis-\`a-vis} experimental results with no adjustable parameters. We refer readers interested in a survey of various models proposed over the course of the years to explain the nature of electronic states and ferromagnetism in these systems to review articles on the theory of DFSs from the perspective of  model Hamiltonians \cite{Jungwirth:2006_RMP} and {\em ab initio} approaches \cite{Sato:2010_RMP,Zunger:2010_P}.  A short paper presenting the topic in a condensed and tutorial way as well as explaining origins of various exchange mechanisms is also available \cite{Bonanni:2010_CSR}. Earlier book chapters review thoroughly the pioneering works on II-VI \cite{Furdyna:1988_B,Dietl:1994_B} and III-V \cite{Matsukura:2002_B} DMSs. Two other surveys present successes and limitations of Drude-Boltzmann type models in describing abundant experimental results on transport \cite{Jungwirth:2008_B} and optical \cite{Burch:2008_JMMM} phenomena in (Ga,Mn)As and related systems. Accordingly, we only briefly discuss these phenomena here, also realizing  that there are not yet theoretical frameworks allowing for the quantitative description of absolute values of dc or ac conductivity tensor components in the regime of quantum localization, even in the absence of $p$-$d$ coupling \cite{Lee:1985_RMP,Belitz:1994_RMP}.

\section{Growth and characterization}
\subsection{Growth methods and diagrams}
\label{sec:growth}

Some of DFSs can be grown by the thermal equilibrium Bridgman method, a primal example being IV-VI alloys, particularly p-Pb$_{1-x-y}$Sn$_y$Mn$_x$Te \cite{Story:1986_PRL,Eggenkamp:1995_PRB}, in which cation vacancies supplied a large concentration of holes mediating ferromagnetic coupling between Mn spins. The same growth technique delivered ferromagnetic Zn$_{1-x}$Mn$_x$Te:P \cite{Kepa:2003_PRL}, in which P acceptors provided holes after appropriate annealing. Interestingly, the Bridgman method was successfully used to obtain rhombohedral Bi$_{2-x}$Mn$_x$Te$_3$, a ferromagnetic topological insulator, in which Mn ions that introduced both spins and holes, were found to be randomly distributed up to at least $x = 0.09$ \cite{Hor:2010_PRB}. At the same time, solid state reaction was employed to synthesize polycrystalline p-Ge$_{1-x}$Mn$_x$Te \cite{Cochrane:1974_PRB} up to $x =0.5$ and p-Li(Zn$_{1-x}$Mn$_x$)As up to $x = 0.15$ \cite{Deng:2011_NC}, in which holes originated presumably from cation vacancies and Li substituting Zn, respectively.

However, rapid progress in the search for ferromagnetic DMSs stems, to a large extent, from the development of methods enabling material synthesis far from thermal equilibrium, primarily by molecular beam epitaxy (MBE) \cite{Ohno:1998_S}, but also by pulsed-laser deposition (PLD) \cite{Fukumura:2005_SST}, metalorganic vapor phase epitaxy (MOVPE) \cite{Bonanni:2007_SST}, atomic layer deposition (ALD) \cite{Lukasiewicz:2012_SST}, sputtering \cite{Fukumura:2005_SST}, ion implantation \cite{Pearton:2003_JAP}, and pulsed-laser melting of implanted layers \cite{Scarpulla:2008_JAP,Zhou:2012_APEX}. These methods have a potential to provide high-quality DMS films with a concentration of the magnetic constituent beyond the solubility limits at thermal equilibrium. Moreover, the use of these methods offers unprecedented opportunity for considering physical phenomena and device concepts for previously unavailable combination of quantum structures and ferromagnetism in semiconductors.

\begin{figure}
\includegraphics[width=3.2in]{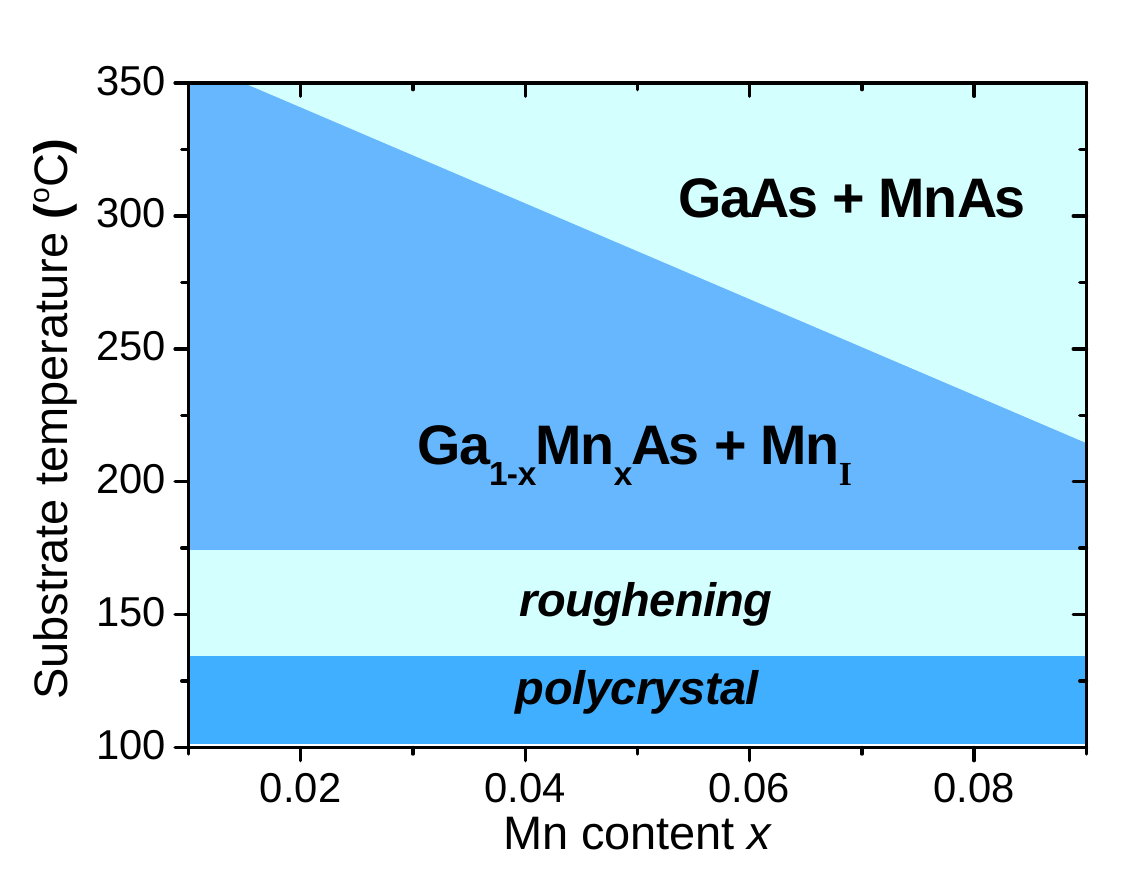}\vspace{-0mm}
\caption[]{(Color online) Schematic diagram of the temperature window for growth of dilute magnetic semiconductor Ga$_{1-x}$Mn$_x$As by low temperature  molecular beam epitaxy. With the increase of the Mn content $x$ the window shrinks and the concentration of Mn interstitials Mn$_{\text{I}}$ increases. The magnitude of biaxial strain is determined by the substrate lattice constant even beyond the critical thickness for the formation of misfit dislocations. Adapted from \onlinecite{Matsukura:2002_B}.}
\label{fig:growth}
\end{figure}

Figure \ref{fig:growth} outlines the growth phase diagram of (Ga,Mn)As \cite{Esch:1997_PRB,Ohno:1998_S,Matsukura:2002_B}, which appears to be generic to a wide class of DMSs. Because of low solubility of TM impurities, typically a fraction of a percent, and the associated tendency of TM cations to aggregate, the growth at high temperatures results in a nanocomposite system consisting of a TM-rich compound in a form of nanocrystals embedded in a TM-poor semiconductor matrix. This crystallographic phase separation can deteriorate crystal quality and surface morphology, as evidenced for the case of (Ga,Mn)Sb \cite{Abe:2000_PE}. Furthermore, it was shown that MnAs nanocrystals on the surface of GaAs serve as seeds nucleating growth of GaAs nanowires \cite{Sadowski:2007_NL}.

The lowering of the substrate temperature to the 200--300$^o$C range \cite{Munekata:1989_PRL,Ohno:1992_PRL,De_Boeck:1996_APL,Ohno:1996_APL,Esch:1997_PRB} makes it possible to surpass the thermal equilibrium solubility limit and, at the same time, to maintain the two-dimensional coherent growth, as witnessed by the smoothness of the surface and the persistence of electron diffraction stripes over the entire process of the film deposition.  The use of a cracker effusion cell for the anion source \cite{Campion:2003_JCG} as well as a careful adjustment of the ratio between cation and anion fluxes \cite{Myers:2006_PRB} allows one to minimize the concentration of points defects, such as As-antisite donors, which tend to form during low temperature MBE.

Importantly, owing to low deposition temperatures, strain associated with lattice mismatch to the substrate remains unrelaxed even for film thicknesses exceeding critical values for the formation of misfit dislocations under thermal equilibrium conditions.  A uniformly strained (Ga,Mn)As film with the thickness of $6.8 \mu$m was obtained employing (001) GaAs substrate for which lattice mismatch was $\Delta a/a \approx 0.4$\%  \cite{Welp:2004_APL}. The use of substrates with various lattice parameters and crystallographic orientations allows one to fabricate DFS films with tailored magnetic anisotropy characteristics (see, Sec.~\ref{sec:anisotropy}).

Additionally, the low-temperature (LT) epitaxy process makes it possible to increase substantially the electrical activity of shallow impurities. For instance, by assisting MBE growth with nitrogen plasma, it was possible to introduce a sizable concentration of holes indispensable to mediate ferromagnetic coupling between Mn spins in (Zn,Mn)Te \cite{Ferrand:2000_JCG,Ferrand:2001_PRB} and (Be,Mn)Te \cite{Hansen:2001_APL,Sawicki:2002_PSSB}. Another relevant approach is to employ the concept of modulation doping, successfully applied in (Cd,Mn)Te/(Cd,Mg,Zn)Te:N \cite{Haury:1997_PRL,Boukari:2002_PRL}, and also examined in the case of (Ga,Mn)As/(Al,Ga)As:Be \cite{Wojtowicz:2003_APL}.

In addition to III-V and II-VI DFSs, the MBE method has been employed for deposition of (Ge,Mn) \cite{Park:2002_S} and (Ge,Mn)Te \cite{Fukuma:2008_APLb,Hassan:2011_JCG,Knoff:2009_APPA,Lim:2011_JAP}.

\subsection{Importance of nanocharacterization}

As already mentioned in Introduction, the rich materials physics of ferromagnetic DMSs stems to a large extent from  non-anticipated forms of distributions and lattice positions assumed by magnetic ions, defects, and carriers in these systems as well as form their sensitivity to contamination by ferromagnetic nanoparticles \cite{Grace:2009_AM}. Importantly, rather than being specific to a given DMS, these striking properties depend sensitively on the employed substrate, growth conditions, co-doping, and post-growth processing. Four issues, relevant to DFSs, can be called into attention here.

\begin{enumerate}
\item Attractive interactions between magnetic impurities and their limited solubility can result in the highly non-random distribution of TM atoms over cation sites ({\em chemical phase separation}) or in TM precipitation in the form of compounds or elemental inclusions ({\em crystallographic phase separation}). Typically, the TM-rich nanocrystals formed in these ways dominate the magnetic response of the system. They are either randomly distributed over the film volume or tend to accumulate near the surface or interface. Atom diffusion on the growth surface is typically faster than in the bulk, which facilitates aggregation of magnetic cations to the form of TM-rich nanocrystals during the epitaxy.

\item  Even if nanocrystals are not assembled, the attractive force between TM cations can enhance the concentration of {\em nearest neighbor cation-substitutional TM dimers}. Moreover, since on the surface (comparing to bulk) certain crystal directions are not equivalent, the dimers--if stable during the entire growth process--can assume a directional distribution that lowers alloy symmetry and, hence, modify magnetic anisotropy.

\item The upper limit of achievable carrier density in a given host is usually determined by the mechanism of {\em self-compensation}. In the case of hole doping the effect consists of the appearance of compensating donor-like point defects once the Fermi level reaches an appropriately low energy in the valence band. These defects not only remove carriers from the Fermi level but can form with TM ions {\em defect complexes} characterized by non-standard magnetic properties. In fact, TM ions can form complexes also with other defects or impurities.

\item Even for a perfectly random distribution of magnetic ions and carrier dopants, due to the relevance of quantum localization effects, there appear significant nano-scale spatial fluctuations in the hole density. Because of the relationship between carriers and magnetism, the value of magnetization ceases to be spatially uniform. Another source of inhomogeneity are space charge layers often forming at the surface or interface.
\end{enumerate}

These outstanding properties of DMSs can be addressed by ever improving nanocharacterization tools involving synchrotron, electron microscopy, ion beam, and scanning probe methods. Some of experimental techniques relevant to DMSs have recently been reviewed \cite{Bonanni:2011_SST}. This collection contains also useful information about the methodology of magnetic measurements on thin DMS films.

Next the above issues are described in some details paying particular attention to the data obtained for (Ga,Mn)As. Enlisted are also methods allowing to determine the concentration of holes and Mn ions if their distribution is, at least approximately, random.

\subsection{Solubility limits and Mn distribution}

It is well known that the phase diagrams of a number of alloys exhibit a solubility gap in a certain concentration range.  Particularly low is the solubility of TM impurities in semiconductors, so that low-temperature epitaxy or ion implantation have to be employed to introduce a sizable amount of the magnetic constituent. An exception here is a large solubility of Mn in II-VI compounds, where Mn atoms remain distributed randomly over the substitutional cation sites up to concentrations often exceeding 50\% \cite{Pajaczkowska:1978_PCGC,Furdyna:1988_B}, even if the alloy is grown close to thermal equilibrium, as in the case of, {\em e.g.}, the Bridgman method.

The large solubility of Mn in II-VI compounds can be associated to the truly divalent character of Mn whose $d$ states little perturb the $sp^3$ tetrahedral bonds as both the lower $d^5$ (donor) and the upper $d^6$ (acceptor) Hubbard levels are respectively well below and above the band edges \cite{Dietl:1981_B,Dietl:2002_SST,Zunger:1986_B}. This qualitative picture is supported by first principles computations, showing a virtual absence of an energy change associated with bringing two Zn-substitutional Mn atoms to the nearest neighbor cation sites in (Zn,Mn)Te, $E_{\text{d}} = 21$~meV \cite{Kuroda:2007_NM}.

According to the pioneering {\em ab initio} work \cite{Schilfgaarde:2001_PRB} and to the subsequent developments \cite{Sato:2005_JJAP,Ye:2006_PRB,Kuroda:2007_NM,Da_Silva:2008_NJP}, a strong tendency to form non-random alloys occurs in the case of DMSs in which TM-induced states are close to the Fermi energy and thus contribute significantly, \emph{via} the $p$-$d$ hybridization, to the bonding as well as can supply or trap carriers. For instance, the pairing energy of two Ga-substitutional Mn atoms is computed to be $E_{\text{d}} = -120$~meV in GaAs and $-300$~meV in GaN \cite{Schilfgaarde:2001_PRB}.

\begin{figure}
\includegraphics[width=3.1in]{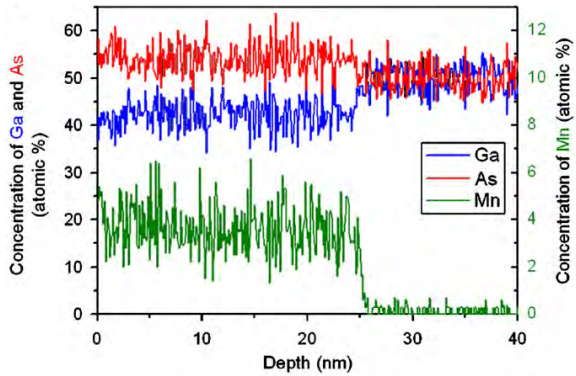}\vspace{-0mm}
\caption[]{(Color online) Composition profile along the growth direction obtained with a 0.1~nm step for a $10\times 10$~nm$^2$ slice of (Ga,Mn)As grown by LT-MBE on a GaAs substrate. A uniform Mn distribution (lowest curve, right scale) with no accumulation at the interface is documented for the (Ga,Mn)As layer. The data suggest an As surplus (uppermost curve). From \onlinecite{Kodzuka:2009_U}.}
\label{fig:GaMnAs_Kodzuka_2}
\end{figure}

\begin{figure}
\includegraphics[width=2.5in]{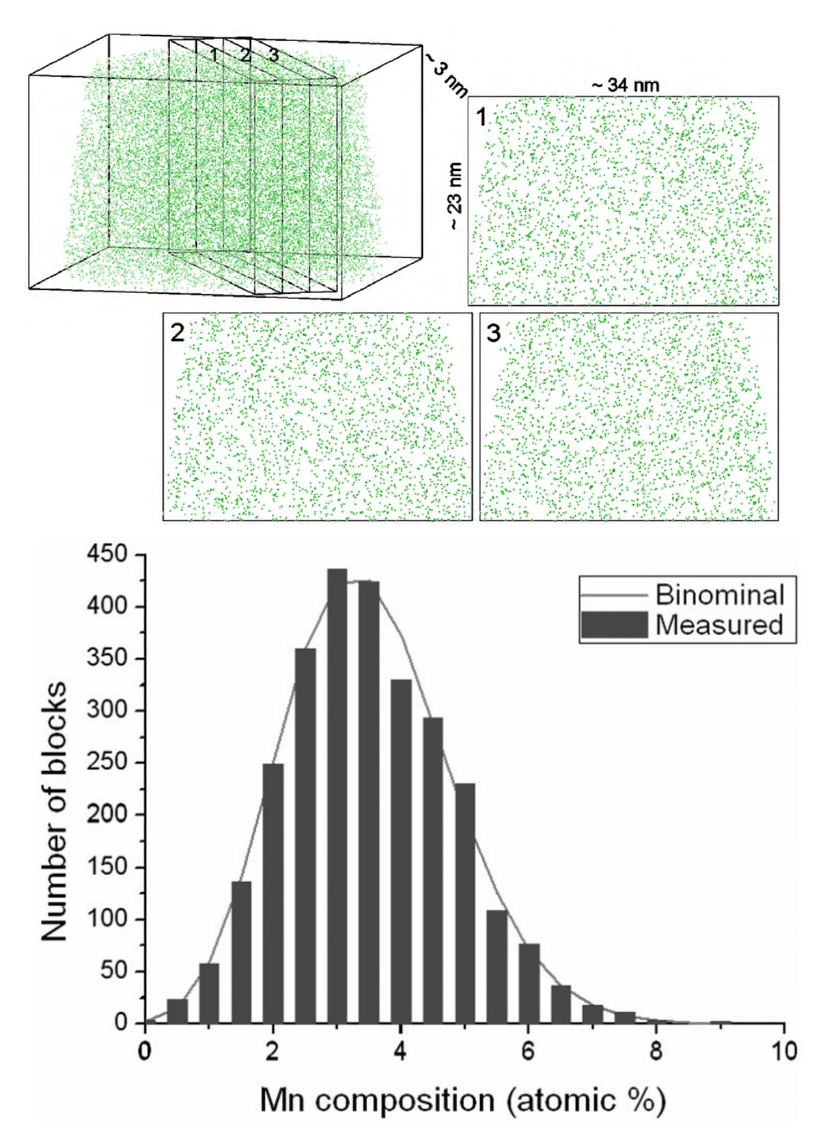}\vspace{-0mm}
\caption[]{(Color online) Mn distribution in 3~nm thick slices of Ga$_{1-x}$Mn$_x$As, grown by low temperature epitaxy, determined by the three dimensional atom probe technique (3DAP)  (upper panel). Frequency distribution of the Mn compositions in 200 slices, compared to the binomial distribution expected for a random alloy with $x = 7.2$\% shown by the solid line. The Mn composition determined from the lattice constant (XRD) is $x = 3.7$\% for this film. From \onlinecite{Kodzuka:2009_U}.}
\label{fig:GaMnAs_Kodzuka}
\end{figure}

However, as already mentioned in Sec.~\ref{sec:growth}, a sufficiently low magnitude of substrate temperature prevents the formation of hexagonal MnAs or zinc-blende Mn-rich (Mn,Ga)As nanocrystals in (Ga,Mn)As grown by MBE. Indeed, according to the newly developed three dimensional atom probe technique (3DAP) that allows one to obtain 3D maps of elements' distribution with a 1~nm resolution, the Mn distribution is uniformed along the growth direction and in-plane, without any evidence for Mn aggregation in the sample volume or Mn segregation at the interface, as shown in Figs.~\ref{fig:GaMnAs_Kodzuka_2} and \ref{fig:GaMnAs_Kodzuka}. However, within the attained resolution, the presence of short range correlations, that is a formation of dimers of trimmers, cannot be confirmed or ruled out. It is unclear at present to what extent this new method provides accurate information on the absolute values of the particular element concentration. The present data, as they stand, suggest a surplus of As and Mn in the studied slices.

The absence of Mn aggregation in (Ga,Mn)As obtained by low temperature MBE was confirmed by cross-sectional scanning tunneling tomography \cite{Richardella:2010_S}.

In the case of wurtzite (wz) (Ga,Mn)N grown by MOVPE \cite{Bonanni:2011_PRB} and MBE \cite{Kunert:2012_APL} a range of nanocharacterization methods indicate the absence of Mn aggregation in films grown under carefully adjusted conditions. A remarkable difference between (Ga,Mn)N and (Ga,Fe)N [in which the same methods reveal the formation of Fe-rich nanocrystals \cite{Bonanni:2008_PRL,Navarro:2011_PRB}] was explained by LSDA {\em ab initio} studies in terms of the repulsive ($E_{\text{d}} = 170$~meV) and attractive ($E_{\text{d}} = -120$~meV) interactions between the nearest neighbor cations pairs of Mn and Fe, respectively, on the growth surface (0001) of wz-GaN \cite{Gonzalez:2011_PRB}.

\subsection{Formation of Mn dimers}
\label{sec:dimers}

\begin{figure}
\includegraphics[width=3.1in]{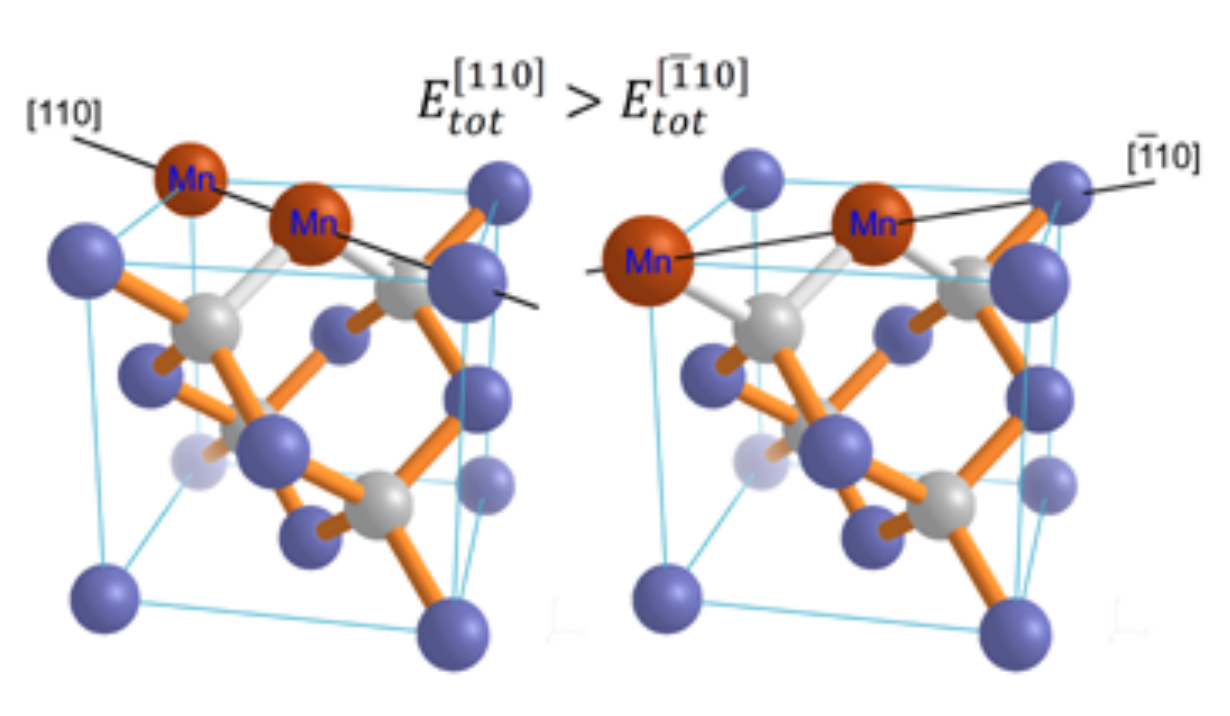}\vspace{-0mm}
\caption[]{(Color online) Mn dimers on the (001) GaAs surface, if residing along the $[110]$  direction, are not bridged by an As atom (a). Such a bonding exists for $[\bar{1}10]$ dimers (b), resulting in the lower energy.}
\label{fig:dimers}
\end{figure}

Figure~\ref{fig:dimers} presents the nearest neighbor Mn dimers residing on GaAs (001) surface along two crystallographic directions.  As seen, in the $[\bar{1}10]$ case the two Mn ions are connected by the same As atom, whereas there is no such As atom for the dimer along the [110] axis, implying that these two directions are not equivalent on the surface, in contrast to bulk dimers, for which there is an As bridge for these two cases -- one below, one above the dimer plane. Furthermore, since the Mn-Mn interaction discussed in the previous subsection is brought about by $p$-$d$ hybridization, one may expect much stronger attractive force for the $[\bar{1}10]$ pair comparing to the [110] case. Indeed, {\em ab initio} computations show that the corresponding difference in $E_{\text{d}}$ is as large as 1.0~eV \cite{Birowska:2012_PRL}, much higher than growth temperature. Thus, if barriers for Mn diffusion along the surface are sufficiently small, a non-volatile asymmetry in the pair distribution will set in in the whole film during the epitaxy. A group theory analysis demonstrated that the corresponding lowering of symmetry leads to the appearance of two additional terms in the $kp$ Hamiltonian, which are of the form of effective biaxial and shear strains, $\epsilon_{xx}$ and  $\epsilon_{xy}$, respectively \cite{Birowska:2012_PRL}.

The asymmetry in the dimer distribution invoked by the above model has not yet been directly confirmed by any nanocharacterization method. However, it was suggested \cite{Birowska:2012_PRL} that strain associated with the formation of dimers along the $[\bar{1}10]$ direction can trigger stacking faults propagating in the $(111)$ and $(11\bar{1})$ planes, as observed in (Ga,Mn)As by high resolution electron transmission microscopy \cite{Kong:2005_JAP} and synchrotron x-ray diffraction \cite{Kopecky:2011_PRB}.

\subsection{Self-compensation -- Mn interstitials and Mn complexes}
\label{sec:self-compensation}
In many cases TM impurities rather than residing in the substitution sites, prefer to occupy interstitial positions, the case of TM-doped Si \cite{Zunger:1986_B}, but also of Mn in GaAs, as suggested theoretically \cite{Masek:2001_APP} and found experimentally \cite{Yu:2002_PRB}. According to combined Rutherford backscattering (RBS) and particle-induced x-ray emission (PIXE) measurements  \cite{Yu:2002_PRB}, as-grown (Ga,Mn)As contains a significant portion of Mn occupying  interstitial positions, Mn$_{\text{I}}$, the defect found also in (In,Mn)Sb \cite{Wojtowicz:2003_APL} and (Al,Ga,Mn)As \cite{Rushforth:2008_PRB}. The presence of Mn$_{\text{I}}$ in (Ga,Mn)As was confirmed by extended x-ray absorption fine structure (EXAFS) spectroscopy \cite{Bacewicz:2005_JPCS,DAcapito:2006_PRB} and transmission electron microscopy (TEM) \cite{Glas:2004_PRL}. The interstitials appear to enlarge the (Ga,Mn)As lattice constant, according to x-ray diffraction \cite{Potashnik:2001_APL,Sadowski:2004_PRB,Mack:2008_APL} and theoretical studies \cite{Masek:2003_PRB}.

Whilst the Mn impurity in the cation-substitutional site acts as a single acceptors in III-V compounds, it becomes a double donor in the interstitial position of the GaAs lattice \cite{Masek:2001_APP,Yu:2002_PRB}. Since the formation of holes in the valence band, {\em i.~e.,} in the bonding states increases the system energy, the formation of hole compensating defects is energetically favored. In line with this self-compensation scenario, the relative concentration of Mn interstitial, $x_{\text{I}}/x$ increases with the total Mn content $x$, leading to a corresponding decrease of the hole concentration, $p= (x - 3x_{\text{I}})N_0$, where $N_0$ is the total cation concentration \cite{Yu:2002_PRB,Wang:2004_JAP}. Other compensating donor impurities or defects, such as As antisites often present in GaAs and related systems deposited at low temperatures (Sec.~\ref{sec:growth}), lowers the value of $p$ further on,
\begin{equation}
p = N_0(x - 3x_{\text{I}}) - zN_{\text{D}},
\label{eq:p}
 \end{equation}
where $z = 1$ and $z = 2$ for the single and double donors, respectively, of the concentration $N_{\text{D}}$. This formula indicates that it may not be possible to determine the hole concentration knowing only Mn concentrations ($x$ and $x_{\text{I}}$).

\begin{figure*}
\includegraphics[width=5.8in]{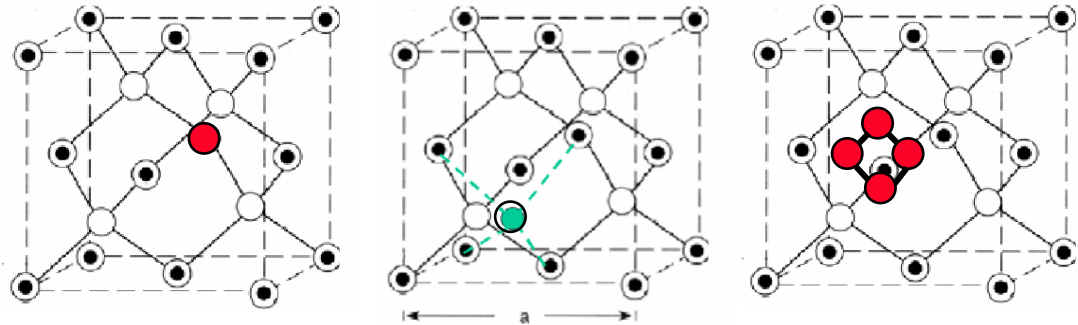}\vspace{-0mm}
\caption[]{(Color online) Location of Mn (full circles) in GaAs films (Ga - open circles with central dots; As - open circles) grown by low temperature molecular beam epitaxy as seen by particle-induced x-ray emission (PIXE): Ga-substitutional, interstitial, and Mn-rich small clusters incommensurate with the GaAs lattice. A tetrahedral interstitial position with cations as the nearest neighbors is shown but the experiment does not exclude that Mn occupies a tetrahedral position close to anions or a hexagonal interstitial site. From  \onlinecite{Furdyna:2008_B}.}
\label{fig:PIXE}
\end{figure*}

It appears natural to assume that mobile positively charged interstitials will occupy a void position next to the negatively charged Mn$_{\text{Ga}}$ acceptors, as shown in Fig.~\ref{fig:PIXE}. However, this conclusion appears in variance with the TEM studies indicating that Mn$_{\text{I}}$ occupies preferably a tetrahedral position with As as the nearest neighbors \cite{Glas:2004_PRL}. Also EXAFS spectroscopy \cite{DAcapito:2006_PRB}  has not yet provided evidences for the formation of Mn$_{\text{I}}$--Mn$_{\text{Ga}}$ dimers. On the other hand, the observation by x-ray magnetic circular dichroism (XMCD) of some non-ferromagnetic Mn inside (Ga,Mn)As films has been assigned to such dimers \cite{Kronast:2006_PRB}. This issue, as well as the strength of exchange couplings between band holes and Mn$_{\text{I}}$ in various positions, have not yet been settled theoretically \cite{Blinowski:2003_PRB,Masek:2003_PRB} and experimentally. At the same time, a strong antiferromagnetic interaction is expected for Mn$_{\text{I}}$-Mn$_{\text{Ga}}$ dimers \cite{Blinowski:2003_PRB,Masek:2003_PRB}. The corresponding formation of spin singlets, not coupled  to holes by an exchange interaction, could explain a reduction in the concentration of Mn spins,
\begin{equation}
x_{\text{eff}} = x - 2x_{\text{I}},
\label{eq:xeff}
\end{equation}
contributing to ferromagnetic order in as-grown (Ga,Mn)As \cite{Potashnik:2002_PRB,Wang:2004_JAP,Chiba:2008_JAP,Stefanowicz:2010_PRBb,Edmonds:2005_PRB}.
We note that, as discussed in Secs.~\ref{sec:doping} and \ref{sec:antiferro}, antiferromagnetic superexchange between {\em substitutional} Mn ions can reduce $x_{\text{eff}}$ and $T_{\text{C}}$ further on.

\subsection{Non-uniform carrier distribution}
\label{sec:carrier_distribution}

A spontaneous formation of a spatially non-uniform (modulated) carrier and magnetization distribution has been persistently suggested theoretically in the context of magnetic semiconductors \cite{Nagaev:1993_B} and considered also for DFSs \cite{Timm:2005_PRL}.

If magnetic ordering is mediated by carriers, spatially inhomogeneous magnetization can result from a non-uniform distribution of carrier density. One origin of such inhomogeneity is the formation of space charge layers at the interfaces or surfaces of DFSs, the effect examined quantitatively in gated metal-insulator-semiconductor structures of (Ga,Mn)As \cite{Sawicki:2010_NP,Nishitani:2010_PRB}. It was also argued \cite{Proselkov:2012_APL} that Coulomb repulsion between surface and interstitial donors produces a gradient in the concentration of Mn$_{\text{I}}$ and, thus, in the hole density and $T_{\text{C}}$, as seen in neutron \cite{Kirby:2006_PRB} and magnetization \cite{Proselkov:2012_APL} studies.

Furthermore, according to the physics of disorder-driven quantum localization in doped semiconductors \cite{Lee:1985_RMP,Altshuler:1985_B}, density of electronic states at the Fermi level $\rho_{\text{F}}$ does not exhibit any critical behavior in the vicinity of the Anderson-Mott transition. In contrast to $\rho_{\text{F}}$, the local density of states (LDOS) shows critical fluctuations in the vicinity of the transition, as recently visualized by scanning tunneling spectroscopy in (Ga,Mn)As \cite{Richardella:2010_S}. These fluctuations lead to a nanoscale electronic phase separation into regions with differing hole concentrations, the effect explaining \cite{Dietl:2008_JPSJ} a surprising appearance of Coulomb blockade peaks in the conductance of gated nanoconstrictions \cite{Wunderlich:2006_PRL,Schlapps:2009_PRB}.

\begin{figure}
\includegraphics[width=3.2in]{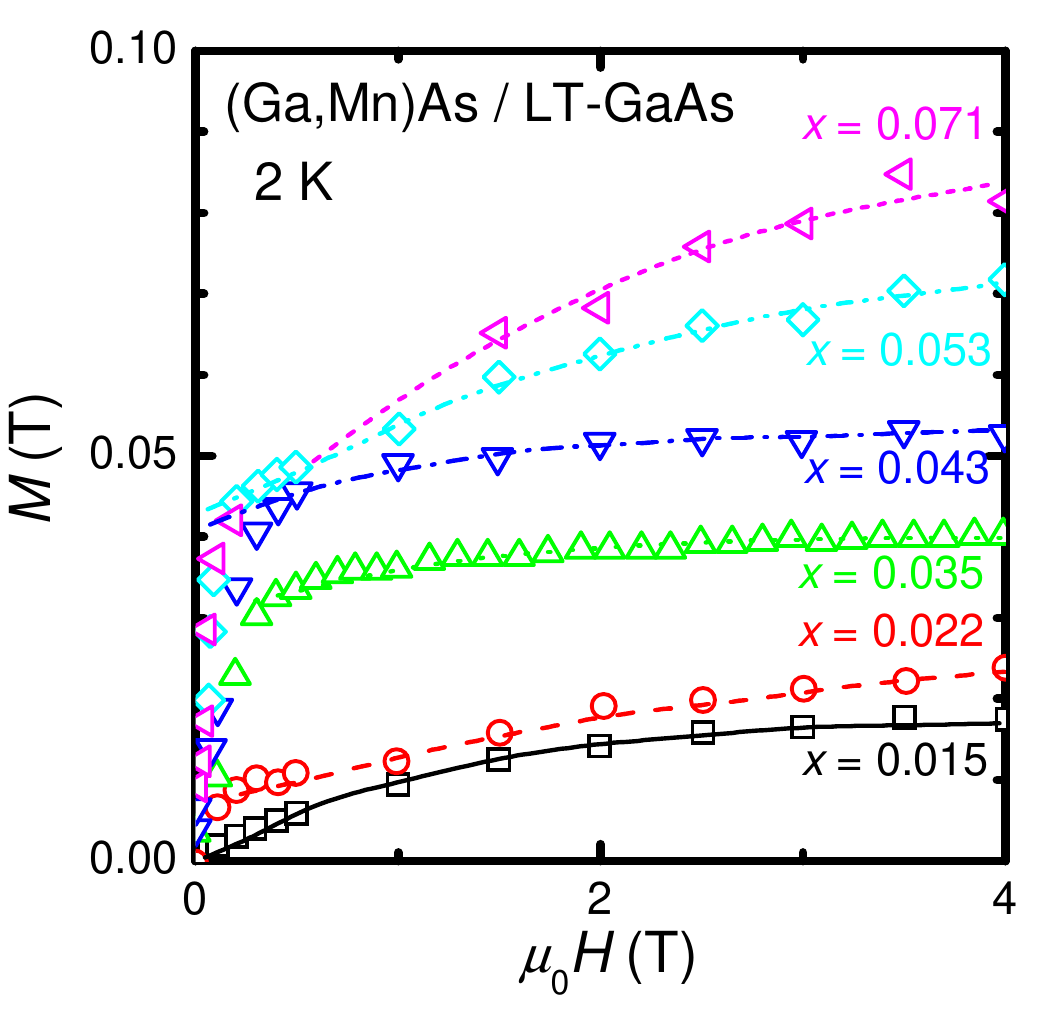}\vspace{-0mm}
\caption[]{(Color online) Magnetic field dependence of magnetization at 2~K for non-annealed films of Ga$_{1-x}$Mn$_x$As with the nominal value of $x$ ranging from 0.015 to 0.071. The magnetic field is applied perpendicular to the sample plane. The dashed lines are fits of the high field magnetization changes to the paramagnetic Brillouin function with the adjusted effective temperature $T_{\text{eff}} = 4$~K. The fitting implies that the concentration of paramagnetic ions is between 20\% ($x = 0.35$ and $0.43$) and 50\% ($x = 0.071$) of the total Mn composition.  From \onlinecite{Oiwa:1998_PB}.}
\label{fig:GaMnAs_Oiwa}
\end{figure}

The electronic phase separation is expected to be enhanced by competing long-range ferromagnetic and short-range antiferromagnetic interactions \cite{Dagotto:2001_PR}, particularly in the instances when carrier density is relatively low, as in II-VI and compensated III-V DFSs. These phenomena give rise to the coexistence of ferromagnetic with paramagnetic or superparamagnetic regions, even if the distribution of magnetic ions is perfectly uniform. Such a co-existence was seen in XMCD studies \cite{Takeda:2008_PRL} as well as in direct magnetic measurements \cite{Ferrand:2001_PRB,Oiwa:1998_PB,Sawicki:2010_NP}, as shown in Fig.~\ref{fig:GaMnAs_Oiwa}. Since a characteristic relaxation time of superparamagnetic particles is slower than 0.1~$\mu$s, they would generate a ferromagnetic-like response in muon rotation experiments \cite{Dunsiger:2010_NM}.

\subsection{Determination of carrier concentration}
\label{sec:carrier_concentration}
The determination of carrier density in DFS films from Hall effect measurements is highly challenging for three reasons.

First, the anomalous Hall effect often dominates, so that the evaluation of the normal Hall effect is only possible in high magnetic fields and at low temperatures, where spins of magnetic ions are saturated \cite{Omiya:2001_PE}. Under these conditions, taking into account the valence band structure of DFSs, the Hall resistance is expected to provide the hole concentration within an accuracy of about 20\% \cite{Jungwirth:2005_PRBa}. However, in this regime, a direct influence of the magnetic field on the hole magnetic moments reduces the hole spin polarization \cite{Dietl:2001_PRB,Sliwa:2006_PRB,Sliwa:2013_arXiv} and, hence, the anomalous Hall effect, linearly in the magnetic field. This results in an overestimation of the hole concentration, particularly in the high hole concentration range, where the ordinary Hall resistance is relatively small. In an extreme case of (In,Mn)Sb, where the magnetic moment of holes is large, a sign reversal the Hall resistance in the magnetic field was observed \cite{Mihaly:2008_PRL}.

Second, additional corrections to the Hall resistance come from quantum localization phenomena \cite{Altshuler:1985_B,Lee:1985_RMP}, which eventually lead to the divergence of the Hall coefficient in the vicinity of the metal-insulator transition \cite{Dietl:2008_JPSJ}, the effect persisting up to temperatures of the order of the acceptor binding energy \cite{Fritzsche:1960_PR}, about 1000~K in (Ga,Mn)As. Accordingly, the determined magnitude of carrier density directly from the Hall resistance, even at room temperature, is typically significantly underestimated at the localization boundary in DFSs such as (Ga,Mn)As \cite{Sheu:2007_PRL,Satoh:2001_PE}.

Third, carriers accumulated at interfaces or substrate often contribute to total conductance. Under these conditions, in order to determine the relevant Hall resistivity, magnetotransport measurements should be carried out over a wide field range and interpreted in terms of multichannel formulae \cite{Bonanni:2007_PRB}. The presence of an electron layer at the (In,Mn)As/GaSb interface is thought to lead to an underestimated value of hole density in (In,Mn)As \cite{Liu:2004_PE}.

In view of the above difficulties other methods were successfully employed to determine hole density in (Ga,Mn)As: (i) electrochemical capacitance-voltage profiling \cite{Yu:2002_PRB}; (ii) Raman-scattering intensity analysis of the coupled plasmon--LO-phonon mode \cite{Seong:2002_PRB}, and (iii) infrared spectroscopy providing the hole concentration from dynamic conductivity integrated over the frequency \cite{Chapler:2013_PRB}.

\subsection{Determination of alloy composition}
A routine, non-destructive, and accurate determination of an average alloy composition $x$ is by no means straightforward in the case of DFS thin films. The intensity of TM flux during the growth and the character of reflection high-energy electron diffraction (RHEED) or \emph{ex-situ} secondary ion mass spectroscopy (SIMS) serve to evaluate the nominal TM concentration $x$. For the purpose of calibration the electron probe microanalysis (EPMA)--requiring usually films thicker than 1~$\mu$m--or the relation between the flux and thickness of the end compound, say, MnAs \cite{Ohya:2007_APL} have been employed. The calibration can also be used to establish the composition dependence of the lattice constant $a(x)$, which can readily be determined by x-ray diffraction (XDR) measurements. Here the sensitivity of $a(x)$ to the carrier and defect density \cite{Potashnik:2001_APL,Masek:2003_PRB,Sadowski:2004_PRB,Mack:2008_APL} has to be considered. Channeling Rutherford backscattering (c-RBS) \cite{Yu:2002_PRB,Kunert:2012_APL} and particle induced x-ray emission (c-PIXE) \cite{Yu:2002_PRB} experiments also allow to determine Mn content.

Recently, a three dimensional atom probe technique (3DAP) is being developed \cite{Kodzuka:2009_U}, which together with already frequently used electron energy loss spectroscopy (EELS) \cite{Jamet:2006_NM} and energy dispersive x-ray spectroscopy (EDS) \cite{Kuroda:2007_NM},  have the potential to provide TM composition, also in the case of thin films.

In the case of DMSs the composition can also be assessed from, interesting by its own, magnetic measurements, the method requiring a modeling of magnetism. However, it is now appreciated that because epitaxial films are thin and the concentration of magnetic impurities is typically low, magnetic response of DMS layers can be significantly perturbed by spurious magnetic moments and a limited resolution of typical magnetometers \cite{Sawicki:2011_SST}. Accordingly, prior to deposition of DMS films, magnetic properties of the substrate have to be carefully assessed. Furthermore, results of magnetic measurements on DMSs samples should be systematically compared to data obtained for films nominally undoped with magnetic ions but otherwise grown, co-doped, and processed in the identical way as the DMS samples in question.

\begin{figure}
\includegraphics[width=3.2in]{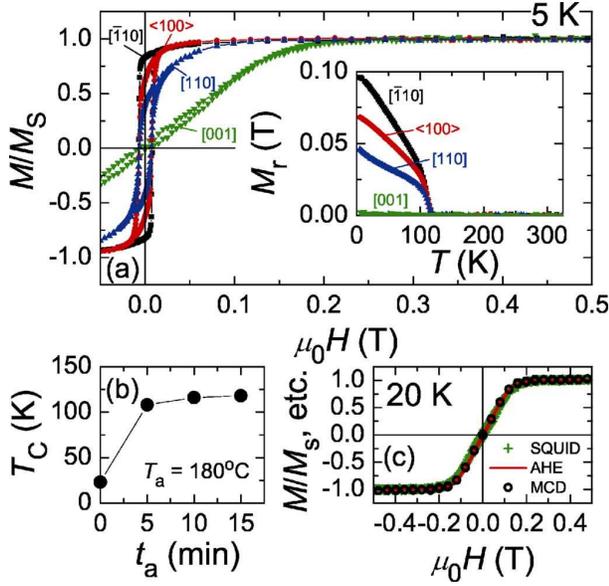}\vspace{-0mm}
\caption[]{(Color online) Properties of a ferromagnetic 5~nm-thick (001) Ga$_{1-x}$Mn$_x$As film of nominal composition $x =0.20$ grown by MBE at $170^o$C. (a) SQUID magnetometer measurements of magnetization loops for four different orientations of the magnetic field in respect to crystallographic axes: $[100]$, $[1\bar{1}0]$, $[110]$, and $[001]$ at 5~K after annealing at $180^o$C for 15~min. The data are normalized by the value of saturation magnetization $M_{\text{Sat}} =0.113$~T which points to the effective Mn concentration $x_{\text{eff}} = 0.1$. The inset shows temperature dependence of remanent magnetization $M_{\text{r}}$ for the same orientations, which imply that the easy axis is along $[1\bar{1}0]$.  (b) The increase of the Curie temperature with the annealing time. (c)  Normalized magnitudes of magnetization, Hall resistance (the anomalous Hall effect - AHE), and magnetic circular dichroism (MCD) at 1.82~eV, \emph{i.~e.,} in the region of interband optical transitions. All data were obtained at 20~K for the magnetic field along $[001]$. From \onlinecite{Chiba:2007_APL}.}
\label{fig:GaMnAs_Chiba07}
\end{figure}

An example of the application of this technique for a ferromagnetic (Ga,Mn)As is illustrated in Fig.~\ref{fig:GaMnAs_Chiba07}, where the findings revealed a rather high value of saturation magnetization, $M_{\text{Sat}} = 90 \pm 5$ emu/cm$^{3}$ \cite{Chiba:2007_APL}.  However, assuming the magnetic moment of $5\mu_{\text{B}}$ per Mn ion [\emph{i.e.} neglecting a small hole contribution \cite{Sliwa:2006_PRB}], the magnitude of $M_{\text{Sat}}$ leads to the effective concentration of Mn spins $x_{\text{eff}}$ more than 2 times smaller than the nominal value $x = 0.20$ obtained for this film from a linear extrapolation of the Mn flux calibration for $x < 0.1$. This discrepancy, noted also by other groups \cite{Mack:2008_APL,Wang:2008_APL}, is discussed further on in the subsection that follows (Sec.~\ref{sec:doping}).

The destructive influence of compensating donor defects, such as Mn$_{\text{I}}$ on the hole and effective Mn concentrations lowers the magnitude of $T_{\text{C}}$ significantly. However, as discussed in Sec.~\ref{sec:doping}, the concentration of interstitials can be considerably reduced by low temperature annealing.

According to RBS-PIXE studies of (Ga,Mn)As mentioned above, in addition to Ga-substitutional and interstitial positions, Mn atoms assume locations incommensurate with the GaAs lattice, refereed to as "random", which can involve a half of the total number of Mn ions \cite{Chiba:2008_JAP}. It has been suggested that the "random" incorporation  corresponds, at least partly, to Mn gathered on the surface as a result of out diffusion of interstitial Mn occurring during the growth or annealing of thin layers \cite{Yu:2005_APL,Chiba:2008_JAP}. Such a scenario is supported by the study combining synchrotron XRD and a technique of x-ray standing-wave fluorescence at grazing incidence \cite{Holy:2006_PRB},  which shows that (Ga,Mn)As consists of a uniform single-crystal film covered by a thin surface Mn-rich layer containing Mn atoms at random non-lattice sites. After annealing, the concentration of interstitial Mn and the corresponding lattice expansion of the epilayer are reduced, the effect being accompanied by an increase in the density of randomly distributed Mn atoms in the disordered surface layer \cite{Rader:2009_PSSB}, where Mn ions are oxidized \cite{Edmonds:2004_PRL,Edmonds:2004_APL,Yu:2005_APL,Olejnik:2008_PRB,Schmid:2008_PRB}.

Another kind of a self-compensation mechanism was found in (Ga,Mn)N. In this material, the Mn acceptor level resides in the mid-gap region (Sec.~\ref{sec:doping}), so that a co-doping by shallow acceptors, such as Mg, is necessary to produce holes in the valence band. It turned out, however, that Mg-Mn complexes are formed in MOVPE grown GaN:Mn:Mg, hampering hole doping of the valence band \cite{Devillers:2012_SR}.

\section{Control of ferromagnetism}
\label{sec:control}
In this section we discuss the most prominent feature of DFSs, that is the possibility of manipulating their magnetic properties, including Curie temperature, saturation magnetization, and magnetic anisotropy,  by growth conditions, doping, strain, electric field, light, and electric current. We also present devices in which electric current is controlled by magnetization direction. Theoretical modeling of pertinent ferromagnetic effects in DFSs is presented in Sec.~\ref{sec:theory}, whereas Sec.~\ref{sec:comparison} contains a comparison of experimental findings and theoretical predictions.

\subsection{Changing of hole density by doping, co-doping, and post-growth processing}
\label{sec:doping}
The existence of a strong interaction between subsystems of localized spins and effective mass carriers is the signature of DMSs \cite{Furdyna:1988_B,Dietl:1994_B}. This interaction accounts for giant Zeeman splitting of bands, spin-disorder scattering, the formation of magnetic polarons, and the mediation by itinerant carriers of ferromagnetic coupling between localized Mn spins. As  predicted theoretically \cite{Dietl:1997_PRB}, and observed experimentally for (Zn,Mn)O:Al \cite{Andrearczyk:2001_ICPS}, this coupling is relatively weak in the case of electrons in DMSs. In contrast, ferromagnetic interactions between diluted spins are rather strong when mediated by delocalized or weakly localized holes \cite{Story:1986_PRL,Ohno:1992_PRL,Ohno:1996_APL,Haury:1997_PRL,Ferrand:2000_JCG,Sheu:2007_PRL,Jungwirth:2010_PRL}. In fact, they can overcome competing short-range antiferromagnetic superexchange occurring between Mn$^{2+}$ ions in DMSs.  Thus, along with the dependence on the magnetic ion density $x$, ferromagnetic properties of DMSs can be controlled by changing the net acceptor concentration as well as by gating (Sec.~\ref{sec:electric_field}) or illumination (Sec.~\ref{sec:light}). Conversely, experimentally observing that $T_{\text{C}}$ does not vary with $x$ usually means that magnetic impurities are not randomly distributed ({\em i.~e.}, their local concentration  does not depend on the average value $x$). Similarly, the lack of dependence on carriers' density indicates that carriers may not account for ferromagnetic order.  Several issues, discussed in more details further on, has to be taken into account in this context:

\begin{enumerate}
\item Similarly to other doped semiconductors, holes in DFSs undergo Anderson-Mott localization if their concentration is smaller than a critical value $p_{\text{c}}$.

\item Due to a contribution of $p$-$d$ coupling to the hole binding energy, the value of $p_{\text{c}}$  is shifted to higher hole densities in DFSs, as compared to corresponding non-magnetic counterparts.

\item The carrier-mediated ferromagnetism appears already in the weakly localized regime $p < p_{\text{c}}$) but no long-range and, thus, efficient ferromagnetic coupling takes place in the strongly localized regime, $p\ll p_{\text{c}}$, where holes are tightly bound to the parent acceptors.

\item Deeply in the metallic phase, $p\gg p_{\text{c}}$, ferromagnetic features show typically textbook thermodynamic and micromagnetic properties, despite disorder inherent to doped semiconductor alloys.

\item Because of the self-compensation mechanism (Sec.~\ref{sec:self-compensation}), the introduction of a sizable acceptor concentration may not result in a correspondingly large hole concentration.

\end{enumerate}

There is no quantitative theory for $p_{\text{c}}$ but empirically its magnitude is typically within the range $p_{\text{c}}^{1/3}a_{\text{B}} = 0.26 \pm 0.05$, if the effective Bohr radius  $a_{\text{B}}$ is  evaluated from the binding energy $E_{\text{I}}$ of the relevant acceptor in the limit $p=0$ according to one of the prescriptions \cite{Edwards:1978_PRB}: $a_{\text{B}} = \hbar/(2m^*E_{\text{I}})^{1/2}$ (quantum defect theory) or $a_{\text{B}} = e^2/(8\pi\epsilon_0\epsilon_{\text{r}} E_{\text{I}})$, where $\epsilon_{\text{r}}$ is the static dielectric constant. Employing the latter for GaAs:Be and GaAs:Mn, for which $E_{\text{I}}$ = 28.6~meV \cite{Fiorentini:1995_PRB} and 112.4~meV \cite{Linnarsson:1997_PRB}, one obtains $p_{\text{c}} =(2.3 \pm 1.6)\times 10^{18}$ and $1.4 \pm 1 \times 10^{20}$~cm$^{-3}$, respectively. It is worth noting, however, that in DFSs $p_{\text{c}}$ also depends on the magnitude of magnetization and even on its orientation in respect to crystallographic axes \cite{Pappert:2006_PRL}.

\begin{figure}[ht]
\begin{center}
\includegraphics[width=3.1in]{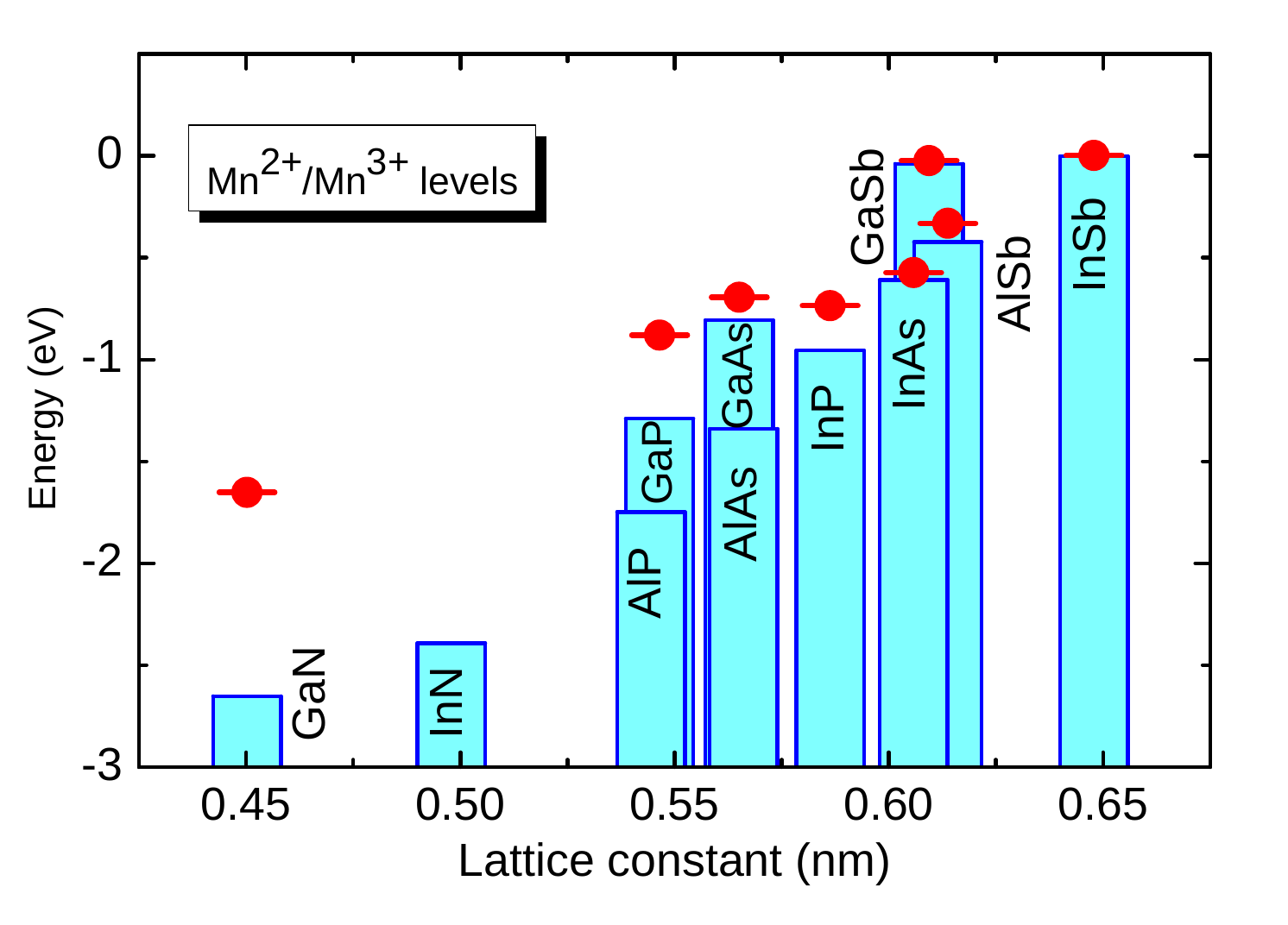}
\caption[]{(Color online) Compilation of experimental energies of Mn levels in the gap of III-V compounds with respect to the valence-band edges. Adapted from \onlinecite{Dietl:2002_PRB}.}
 \label{fig:III_V_EI}
\end{center}
\end{figure}

As shown in Fig.~\ref{fig:III_V_EI}, $E_{\text{I}}$ increases rather dramatically on going from antimonides to nitrides through arsenides and phosphides in Mn-doped III-V compounds. The values of $E_{\text{I}}$ were primarily determined from optical data but also from transport studies in the strongly localized regime \cite{Wolos:2009_PSSC}, where--at temperatures above the hopping regime--the activation energy of conductivity $\epsilon_1 = E_{\text{I}}$. A deviation of $E_{\text{I}}$ from values expected for effective mass acceptors, known as a central cell correction or a chemical shift, was interpreted \cite{Dietl:2002_PRB,Dietl:2008_PRB,Mahadevan:2004_APL,Sato:2010_RMP} in terms of the hybridization-induced repulsion between $t_2$ states originating from Mn $d$ levels and $p$-like  valence band states from which the acceptor state is built. The role of this mechanism increases with the decreasing cation-anion bond length and energy distance between the valence band top and Mn level, eventually resulting in a transition to the strong coupling limit, where the hole binding energy is dominated rather by the $p$-$d$ interaction than by the acceptor Coulomb potential. The emergence of an impurity band in the energy gap with increasing $p$-$d$ coupling was also captured by the dynamic mean-field approximation \cite{Chattopadhyay:2001_PRL}.

Because of these differences in magnitudes of $E_{\text{I}}$, critical densities for the metal-insulator transition (MIT) vary significantly within the Mn-based III-V DMSs family \cite{Dietl:2008_PRB,Wolos:2009_PSSC}. For instance, a comparison of uncompensated Ga$_{1-x}$Mn$_x$As, Ga$_{1-x}$Mn$_x$P, and Ga$_{1-x}$Mn$_x$N with similar Mn content $x \approx 6\%$, shows that holes are respectively delocalized \cite{Jungwirth:2007_PRB}, at the localization boundary \cite{Scarpulla:2005_PRL}, and in the strongly localized regime where no carrier-mediated mechanism of spin-spin coupling operates \cite{Sarigiannidou:2006_PRB,Stefanowicz:2013_PRB}.

\begin{figure}[ht]
\begin{center}
\includegraphics[width=3.4in]{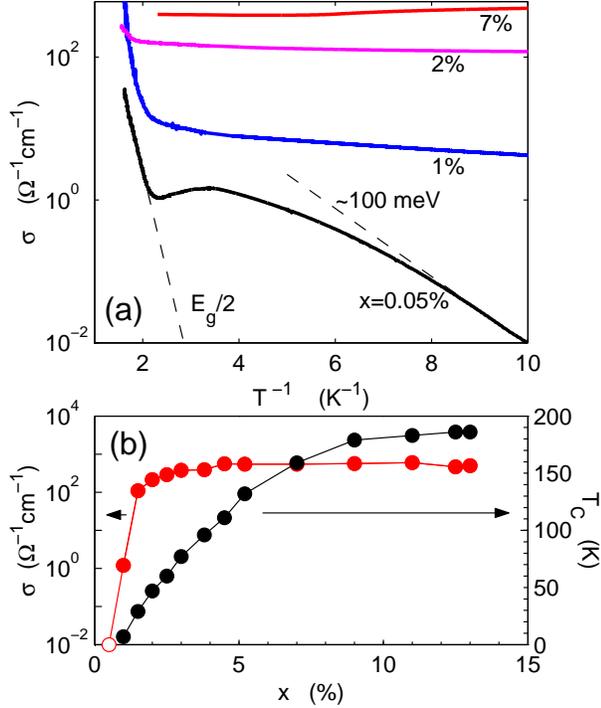}
\caption[]{(Color online) (a) Conductivity {\em vs.} inverse temperature in annealed samples Ga$_{1-x}$Mn$_x$As showing a transition from the insulator to metallic behavior on increasing $x$. (b) Correlation between conductivity at 4.2~K (left scale) and Curie temperature (right scale) in such samples. An abrupt increase of the conductivity at $x \gtrsim 1.5$\% witnesses an insulator-to-metal transition on Mn doping. According to the magnitude of saturation magnetization, the effective (net) Mn concentration attains 8\% for the highest nominal values of $x$. From \onlinecite{Jungwirth:2010_PRL}, supplementary material.}
 \label{fig:GaMnAs_Jungwirth_2010}
\end{center}
\end{figure}

Except for (Ga,Mn)N, considerable hole conductivities $\sigma$ are characteristic to DFSs, where actually a correlation between the magnitudes of $\sigma$ and $T_{\text{C}}$ is seen, as shown in Fig.~\ref{fig:GaMnAs_Jungwirth_2010} for (Ga,Mn)As. In most situations, $\sigma(T)$ remains non-zero at $T \rightarrow 0$, implying metallic conductance. However, in some important cases, {\em e.~g.}, (Ga,Mn)As with $x \lesssim 2$\% \cite{Jungwirth:2007_PRB,Sheu:2007_PRL}, (Ga,Mn)P \cite{Winkler:2011_APL}, and (Zn,Mn)Te:N \cite{Ferrand:2001_PRB}, $\sigma(T)$ vanishes at $T \rightarrow 0$ but $T_{\text{C}}$ remains non-zero. Altogether, these data indicate that the ferromagnetism occurs not only on the metal side of the MIT but in a non critical way penetrates into the weakly localized regime, where high-temperature activation energy of conductivity  $\epsilon_2 < E_{\text{I}}$ provides information on the distance between the mobility edge and the Fermi level \cite{Fritzsche:1960_PR}. However, on moving deeply into the insulator phase $T_{\text{C}}$ vanishes or at least becomes smaller than the explored temperature range down to 2~K in (Ga,Mn)As \cite{Sheu:2007_PRL}.

\begin{figure}
\includegraphics[width=3.2in]{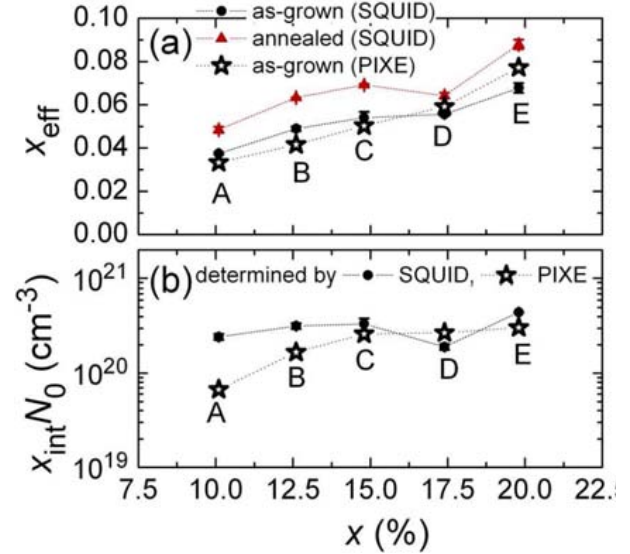}\vspace{-0mm}
\caption[]{(Color online) Determination of Mn composition and location in (Ga,Mn)As. (a) Effective Mn composition $x_{\text{eff}}$ determined by magnetization measurements (closed symbols, SQUID) before (circles) and after (triangles) annealing and RBS-PIXE measurements (open stars, PIXE) as function of the nominal Mn composition $x$ in 4 and 5 nm thick (Ga,Mn)As layers. (b) The interstitial Mn concentration $x_{\text{I}}N_0$ determined as the difference in $x_{\text{eff}}$ after and before annealing from the magnetization measurements (closed circles, SQUID) and from RBS-PIXE measurements before annealing (open stars, PIXE). From \onlinecite{Chiba:2008_JAP}.}
\label{fig:GaMnAs_Chiba08}
\end{figure}

The above mentioned detrimental effect of interstitials on the ferromagnetism of (Ga,Mn)As can be partly reduced by an annealing process \cite{Hayashi:2001_APL,Potashnik:2001_APL,Edmonds:2002_APL,Ku:2003_APL,Chiba:2003_S,Sorensen:2003_APL} that promotes the diffusion of the Mn$_{\text{I}}$ ions to the surface, where they partake in the formation of an antiferromagnetic MnO thin film \cite{Edmonds:2004_APL,Yu:2005_APL,Olejnik:2008_PRB,Schmid:2008_PRB} or an  MnAs monolayer, if the surface is covered by As \cite{Adell:2007_PRB}. This post-growth thermal treatment leads to a substantial increase in the magnitudes of conductivity, $T_{\text{C}}$, and spontaneous magnetization, to the values shown in Figs.~\ref{fig:GaMnAs_Jungwirth_2010}, \ref{fig:GaMnAs_Chiba07}, and \ref{fig:GaMnAs_Chiba08}. A similar effect of low-temperature annealing is observed in (In,Mn)As \cite{Hashimoto:2002_JCG}.

The efficiency of annealing appeared enhanced in nanodots \cite{Eid:2005_APL} or nanowires \cite{Chen:2011_NL}. In contrast, the process of out diffusion was self-limiting if the annealing was performed in an oxygen-free atmosphere or the surface was covered by a cap \cite{Chiba:2003_S}. The diffusion of Mn$_{\text{I}}$ towards the surface can already occur during the growth, the process being particularly efficient in thin samples \cite{Yu:2005_APL,Chiba:2008_JAP}. Accordingly, the concentration of Mn$_{\text{I}}$ in such samples is relatively low, below 2\%, as shown in Fig.~\ref{fig:GaMnAs_Chiba08}.

\begin{figure}
\includegraphics[width=3.2in]{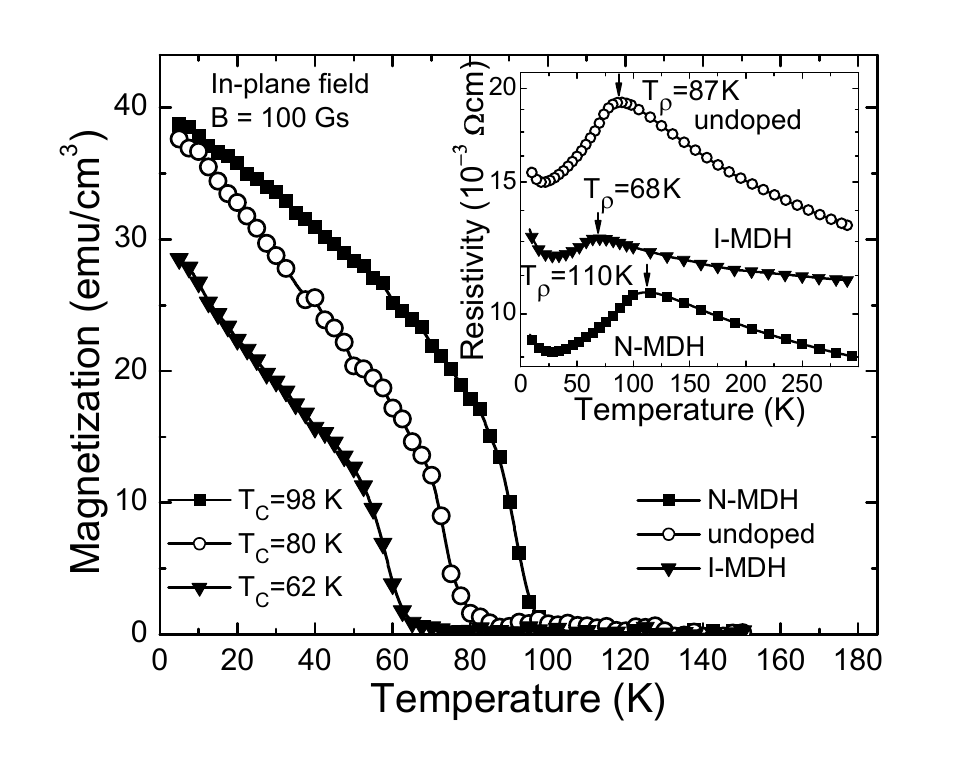}\vspace{-0mm}
\caption[]{ Temperature dependence of remanent magnetization and resistivity (inset) for three Ga$_{0.76}$Al$_{0.24}$As/Ga$_{1-x}$Mn$_x$As/Ga$_{0.76}$Al$_{0.24}$As quantum well (QW) structures. The width of the QW is 5.6~nm, $x = 0.06$. Beryllium acceptors were introduced either into the first barrier (grown before the ferromagnetic QW), or into the second barrier; or the sample was undoped, as marked. From \onlinecite{Wojtowicz:2003_APLb}.}
\label{fig:GaMnAs_Be}
\end{figure}

A natural question arises whether co-doping of (Ga,Mn)As by non-magnetic acceptors, say Be, could enlarge $T_{\text{C}}$ over the values displayed in Fig.~\ref{fig:GaMnAs_Jungwirth_2010}. It could be expected that an antiferromagnetic character of carrier-mediated interaction, showing up when carrier density becomes greater than the magnetic impurity concentration, can drive the system towards a spin-glass phase, as observed in Pb$_{1-x-y}$Sn$_y$Mn$_x$Te \cite{Eggenkamp:1995_PRB}. It turned out, however,  that the presence of additional holes during the growth of (Ga,Mn)As layer increases the concentration of Mn interstitial donors (Mn$_{\text{I}}$) by the self-compensation mechanism. This diminishes net hole {\em and} Mn densities, so that $T_{\text{C}}$ gets actually reduced \cite{Wojtowicz:2003_APLb,Yu:2004_APL}. However, $T_{\text{C}}$ is increased if additional holes are transferred to (Ga,Mn)As layer {\em after} its epitaxy has been completed. Such engineering of ferromagnetism in Ga$_{1-y}$Al$_y$As/(Ga,Mn)As/Ga$_{1-y}$Al$_y$As quantum structures is presented in Fig.~\ref{fig:GaMnAs_Be}. As seen, modulation doping by Be in the back barrier diminishes $T_{\text{C}}$, as then the Fermi level assumes a high position during the growth of (Ga,Mn)As layer which results in the Mn$_{\text{I}}$ formation. In contrast, when Be is introduced in the front barrier, \emph{i.e.}, after the growth of (Ga,Mn)As, the concentration of Mn$_{\text{I}}$ is small and $T_{\text{C}}$ becomes high.

\begin{figure}
\includegraphics[width=3.2in]{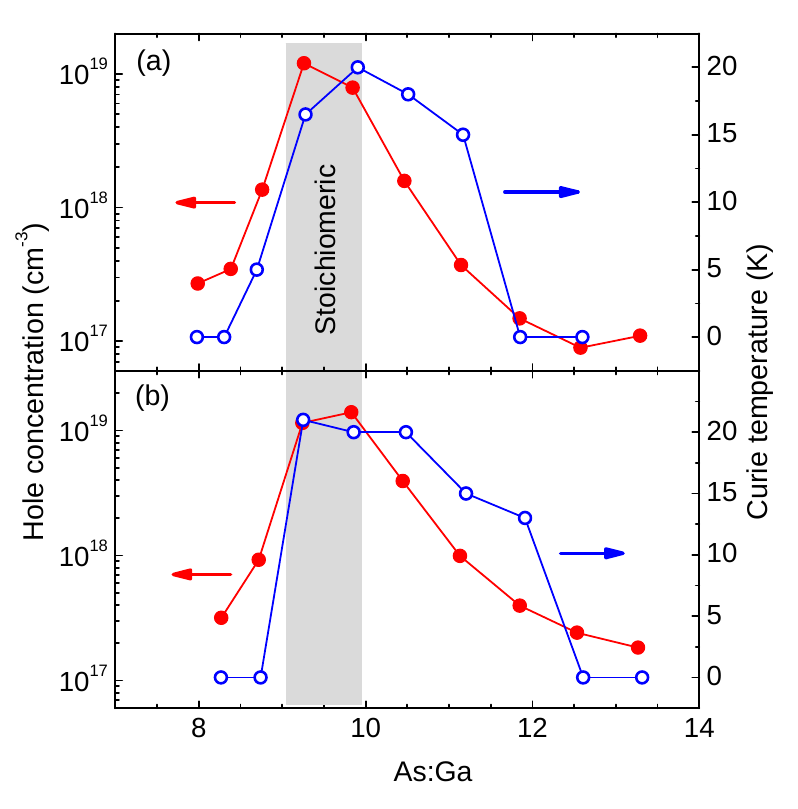}\vspace{-0mm}
\caption[]{(Color online) Effect of growth conditions on the Curie temperature in (Ga,Mn)As. The hole density $p$ from the Hall effect at 300~K and the Curie temperature $T_{\text{C}}$ are plotted as a function of As:Ga flux ratio for two wafers with different Mn concentrations, (a) 1.25\%, and (b) 1.50\%. Lines guide the eye. The stoichiometric region is shaded grey, where effects of disorder-induced hole localization (caused by roughness on the Ga-rich side and by As antisites on the As-rich side) are minimized. Adapted from \onlinecite{Myers:2006_PRB}.}
\label{fig:GaMnAs_Myers}
\end{figure}

Not surprisingly, the magnitude of  $T_{\text{C}}$ in DFSs can be lowered by incorporating compensating donor impurities or donor defects. In the case of (Ga,Mn)As the effect was observed by co-doping with Sn \cite{Satoh:2001_PE}, Si \cite{Wang:2008_PE}, Te and  S \cite{Scarpulla:2008_JAP}, As vacancies \cite{Mayer:2010_PRB}, and As antisite defects \cite{Myers:2006_PRB,Sheu:2007_PRL}, as shown in Fig.~\ref{fig:GaMnAs_Myers}. On the other hand, co-doping with donors during the epitaxy can facilitate the incorporation of Mn in the substitutional positions \cite{Wang:2008_PE}, the effect particulary appealing if the donors could then be removed by post-growth processing \cite{Bergqvist:2011_PRB}.

\begin{figure}
\includegraphics[width=3.2in]{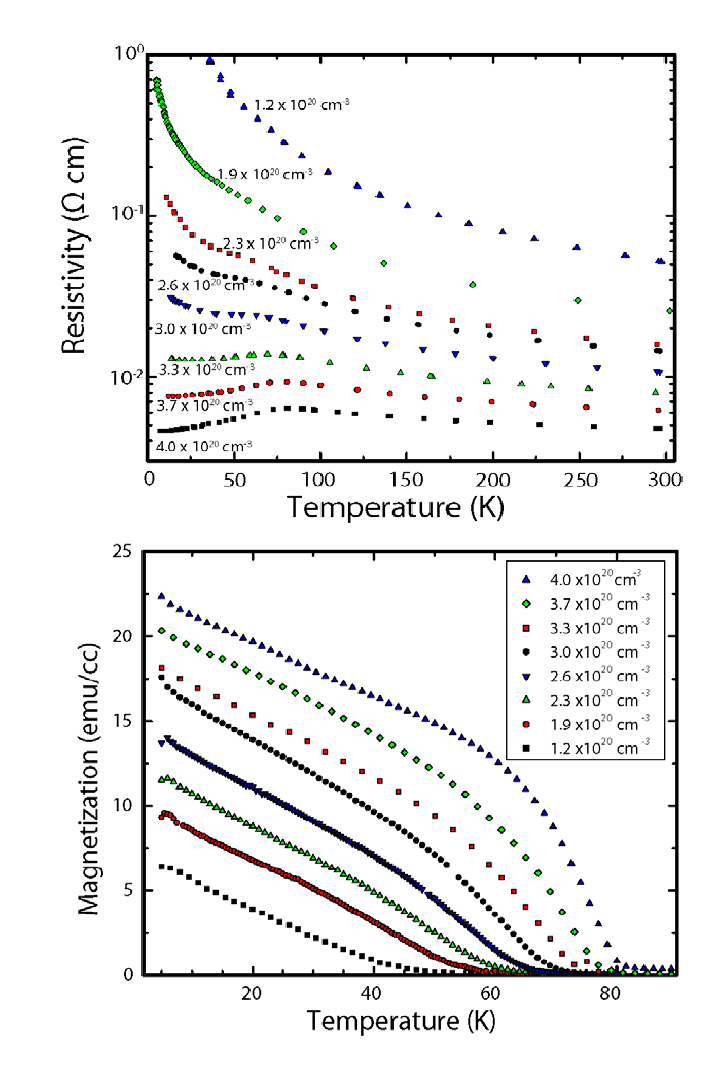}
\caption[]{(Color online) Temperature dependence of resistivity (upper panel) and magnetization (lower panel) at various hole densities changed by irradiation of a Ga$_{0.955}$Mn$_{0.045}$As film by Ne$^{+}$ ions. From \onlinecite{Mayer:2010_PRB}.}
\label{fig:GaMnAs_Mayer}
\end{figure}

Figure \ref{fig:GaMnAs_Mayer} presents temperature dependence of resistance and magnetization in a series of Ga$_{0.955}$Mn$_{0.045}$As samples differing by irradiation doses of Ne$^+$ ions that generate hole compensating defects. The resistance extrapolated to zero temperature is finite in the weakly irradiated samples but it diverges in the strongly compensated case, witnessing the presence of the irradiation-induced MIT. Interestingly, a ferromagnetism is observed on its either side but the magnitudes of both $T_{\text{C}}$ and saturation spontaneous magnetization $M_{\text{Sat}} = M(T\rightarrow 0, H\rightarrow 0)$ decreases gradually when the degree of compensation increases. This suggests that on reducing the net hole concentration, at a given value of Mn density, not only $T_{\text{C}}$ but also the concentration of Mn spins participating in the long range ferromagnetic order at $T = 0$, $x_{\text{eff}}$, gets smaller. Similar results were obtained for hydrogenated samples \cite{Thevenard:2007_PRB} and a (Ga,Mn)P film irradiated with Ar$^+$ ions \cite{Winkler:2011_APL}. These data support the scenario of the electronic and, hence, magnetic phase separation in the vicinity of the MIT in DFSs, the effect enlarged by competing ferromagnetic and antiferromagnetic interactions (see, Sec.~\ref{sec:carrier_distribution}).

\begin{figure}
\includegraphics[width=3.2in]{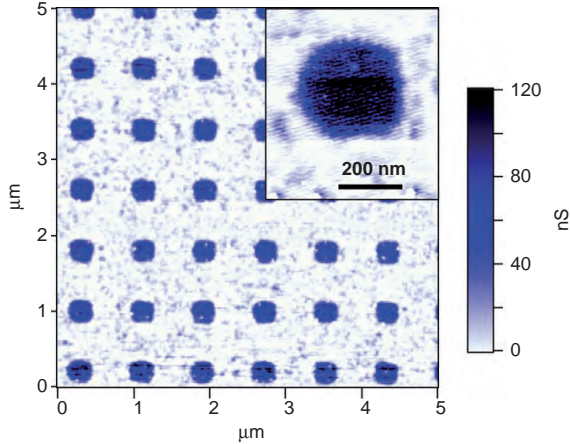}\vspace{-0mm}
\caption[]{(Color online) Conductance atomic force microscopy image of sub-micron (Ga,Mn)As features produced with selective hydrogenation. Inset shows a scan on a single feature. From \onlinecite{Farshchi:2007_PB}.}
\label{fig:GaMnAs_Farshchi}
\end{figure}

Also post-growth hydrogenation reduces the hole density and turns (Ga,Mn)As into a paramagnet \cite{Goennenwein:2004_PRL,Thevenard:2007_PRB,Farshchi:2007_PB}. This process is entirely reversible by annealing below $200^o$C \cite{Thevenard:2007_PRB}. By employing local reactivation using confined low-power pulsed-laser annealing or by hydrogenation through a mask it was possible to pattern ferromagnetic structures with features size below 100~nm \cite{Farshchi:2007_PB}, as shown in Fig.~\ref{fig:GaMnAs_Farshchi}.

The ferromagnetism can be weakened even if the net acceptor concentration is kept constant but additional disorder enhances hole localization. For example, partial anion substitution (As $\rightarrow$ P or As $\rightarrow$  N) \cite{Stone:2008_PRL} or structure roughness introduced by excess Ga \cite{Myers:2006_PRB,Stone:2008_PRL}, the case depicted in Fig.~\ref{fig:GaMnAs_Myers}, lowered significantly $T_{\text{C}}$ values of (Ga,Mn)As. This strong sensitivity of DFS properties to electronic and structural disorder, together with uncertainties associated with the experimental determination of $x_{\text{eff}}$ and $p$,  account presumably for a dispersion in reported $T_{\text{C}}$ values at given Mn and hole concentrations, and impede theoretical interpretation of $T_{\text{C}}$ \cite{Samarth:2012_NM,Wang:2013_PRB}.



\begin{figure}
\includegraphics[width=3.4in]{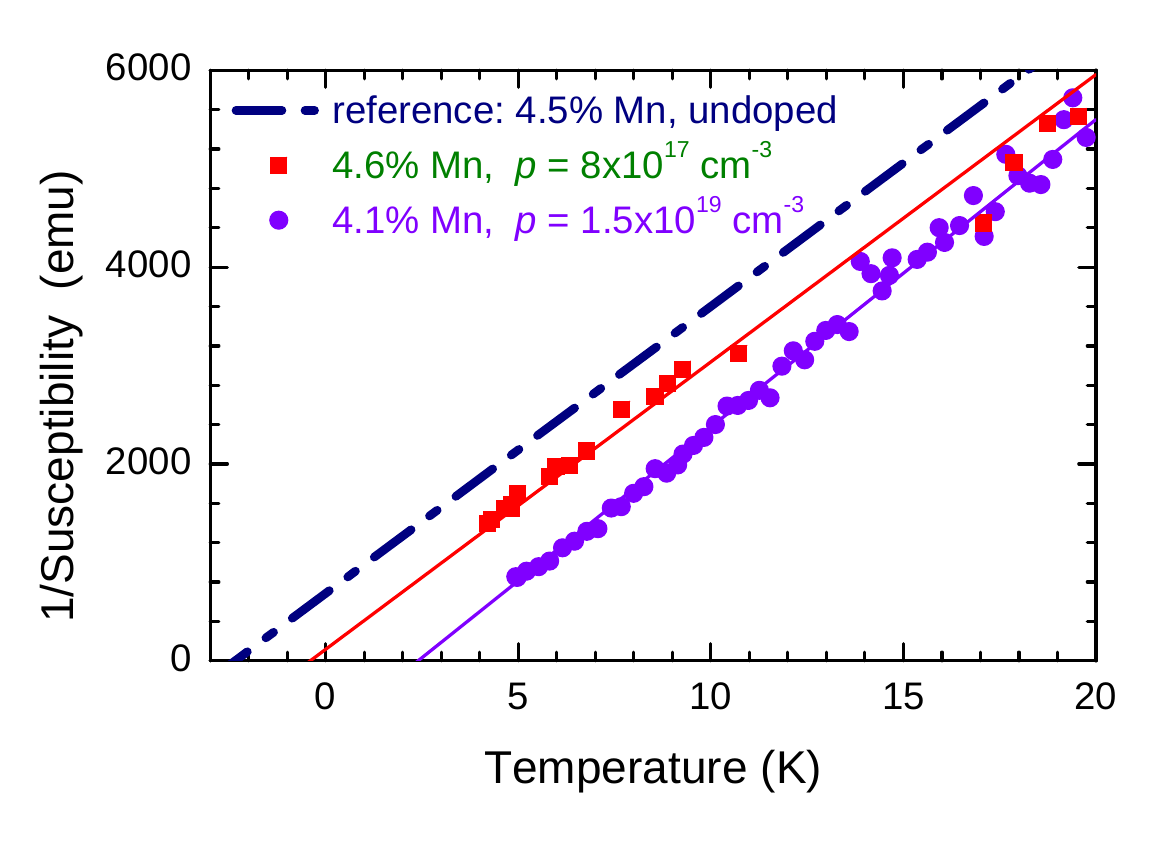}\vspace{-0mm}
\caption[]{(Color online) Inverse magnetic susceptibility from SQUID measurements (squares) for two p-Zn$_{1-x}$Mn$_x$Te samples with similar Mn composition $x \approx 0.045$ but different hole concentrations. Solid lines show linear fit. The dotted line presents the dependence expected for an undoped sample with a similar Mn content. From \onlinecite{Ferrand:2001_PRB}.}
\label{fig:ZnMnTe_Ferrand}
\end{figure}

Figure~\ref{fig:ZnMnTe_Ferrand} illustrates how co-doping of (Zn,Mn)Te with shallow N acceptors triggers the ferromagnetism, the effect demonstrated also for (Cd,Mn)Te:N \cite{Haury:1997_PRL}, (Zn,Mn)Te:N \cite{Ferrand:2000_JCG}, (Be,Mn)Te:N \cite{Sawicki:2002_PSSB},  and (Zn,Mn)Te:P \cite{Kepa:2003_PRL}. Similarly, ferromagnetism of  Pb$_{1-x -y}$Sn$_y$Mn$_x$Te is brought about by holes originating from native defects--primarily cation vacancies--generated by post-growth annealing \cite{Story:1986_PRL}.

Also in the case of II-VI DMSs, such as (Zn,Mn)Te:N, the effect of self-compensation challenges the progress in raising the Curie temperature, where the magnitude of achievable hole density by nitrogen doping decreases with the Mn concentration \cite{Ferrand:2001_PRB}. Furthermore, the MIT is shifted to higher hole concentrations, as the acceptor binding energy is enhanced by magnetic polaron effects \cite{Jaroszynski:1985_SSC,Ferrand:2001_PRB}. Moreover, in the strong coupling limit, the Mn ion can act as a hole trap, which hampers the possibility of obtaining holes in the valence band \cite{Dietl:2008_PRB}. This situation takes presumably place in (Zn,Mn)O, as witnessed by the presence of a relatively large subbandgap absorption corresponding to the photoionization process: Mn$^{2+} + \nu \rightarrow $Mn$^{2+}$ + h + e, where the hole is bound to Mn$^{2+}$ and the electron transferred to the conduction band \cite{Godlewski:2010_OM}.

\subsection{Controlling magnetic anisotropy by hole density and strain}
\label{sec:anisotropy}

Together with magnitudes of $T_{\text{C}}$ and $M_{\text{Sat}}$, the character and strength of magnetic anisotropy determine possible functionalities of any ferromagnet. In the subsections below experimentally demonstrated manipulations with orientation of magnetization are discussed taking into account strain engineering by lattice mismatch to substrates and strain relaxation in nanostructures as well as by piezoelectric and elastic actuators.  Microscopic theory of these phenomena is outlined in Sec.~\ref{sec:magnetization-theory}, whereas its comparison to experimental findings is presented in Sec.~\ref{sec:anisotropy-comparison}.

\subsubsection{Magnetic anisotropy in films and nanostructures}
Extensive magnetic \cite{Sawicki:2006_JMMM}, ferromagnetic resonance (FMR) \cite{Liu:2006_JPCM,Cubukcu:2010_PRB}, magnetotransport \cite{Tang:2003_PRL,Gould:2008_NJP,Glunk:2009_PRB} and magnetooptical \cite{Hrabovsky:2002_APL,Welp:2003_PRL} studies of (Ga,Mn)As and (Ga,Mn)(As,P) films deposited coherently on (001) substrates [typically GaAs substrate or relaxed (Ga,In)As buffer layer] allowed to establish how the system energy depends on the direction of magnetization $\vec{M}$ at a given external magnetic field $\vec{H}$ and biaxial strain imposed by lattice mismatch, quantified by a relative difference between the lattice parameter of the substrate and the free standing layer,
\begin{eqnarray}
\epsilon_{xx} = \epsilon_{yy} = \Delta a/a; \nonumber \\
\epsilon_{zz} =-2\epsilon_{xx}c_{12}/c_{11},
\label{eq:epsilon}
\end{eqnarray}
where the ratio of elastic moduli $c_{12}/c_{11} = 0.453$ in GaAs.

\begin{figure}
\includegraphics[width=2.5in]{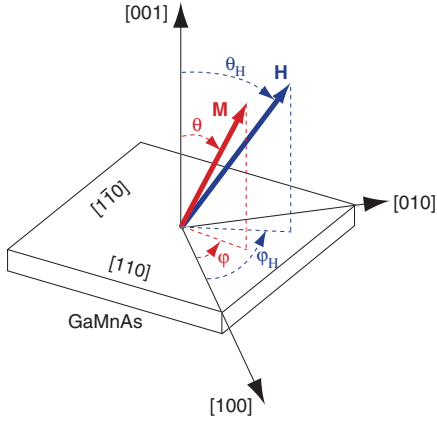}\vspace{-0mm}
\caption[]{(Color online) In the coordinate system employed in this paper $x$, $y$, and $z$ axes are along [001], [010], and [001] crystallographic axes, respectively. From \onlinecite{Liu:2006_JPCM}.}
\label{fig:coordinates}
\end{figure}

In a single domain state, according to the Stoner�Wohlfarth formalism, the functional of free energy density contains contributions from the Zeeman energy, shape (demagnetization) and  crystalline magnetic anisotropies, $\cal{F} = \cal{F}_{\text{Z}} + \cal{F}_{\text{d}} + \cal{F}_{\text{cr}}$. To determine spatial orientation of $\vec{M}$,  $\cal{F}$ is minimized with respect to $\theta$ and $\phi$ defined in Fig.~\ref{fig:coordinates}.  It has been established that in order to describe experimental data, $\cal{F}_{\text{cr}}$ has to contain {\em at least} three contributions (taken in the lowest order): cubic as well as in-plane and out-of-plane uniaxial anisotropy terms. Their relative magnitudes were found dependent on magnetization, hole density, and strain leading to a range of spectacular phenomena, such as spin reorientation transitions on varying temperature \cite{Welp:2003_PRL,Wang:2005_PRL,Sawicki:2004_PRB,Kamara:2012_JNN,Thevenard:2006_PRB}, hole density \cite{Sawicki:2004_PRB,Sawicki:2005_PRB,Khazen:2008_PRB,Thevenard:2005_APL,Thevenard:2006_PRB} or strain imposed by piezoelectric stressors \cite{Bihler:2008_PRB,Rushforth:2008_PRB,Overby:2008_APL,Casiraghi:2012_APL}. This strong sensitivity to strain means also that for anisotropy-related studies, samples should be mounted in a way minimizing thermal stress. Importantly, magnetization orientation can also be manipulated by gate voltage, electric current, and light, as described in Secs.\,III.C-G.

Taking ${\cal{F}}(\vec{M} \parallel [100])$ as a reference energy the particular contributions to $\cal{F}$ in terms of assume then the form,
\begin{eqnarray}
{\cal{F}}_{\text{Z}} = -\mu_0\vec{M}\vec{H} = \\
\nonumber
-\mu_0MH[\cos\theta\cos\theta_H + \sin\theta\sin\theta_H\cos(\phi -\phi_H)],
\end{eqnarray}
\begin{equation}
{\cal{F}}_{\text{d}} = \frac{1}{2}\mu_0M^2m_z^2,
\end{equation}
and
\begin{eqnarray}
{\cal{F}}_{\text{cr}} =
K_{\text{C}}(m_x^2m_y^2  + m_x^2m_z^2 + m_y^2m_z^2)+\\
\nonumber
 + K_{xy}m_xm_y + K_{zz}m_z^2,
\label{eq:F_cr}
\end{eqnarray}
where we have introduced magnetization directional cosines $m_x = \sin\theta\cos\phi$, $m_y =\sin\theta\sin\phi$,  and $m_z = \cos\theta$;  $(\theta,\phi)$ and $(\theta_H,\phi_H)$ are azimuthal and polar angles of $\vec{M}$ and $\vec{H}$, respectively (see, Fig.~\ref{fig:coordinates}) and $K_i$ are sample and temperature dependent fitting parameters (crystalline anisotropy energies)  to experimental dependence $\vec{M}(\vec{H})$.\footnote{Differing conventions of parameterizing $\cal{F}_{\text{cr}}$ exist in the literature. For instance, the cubic term is often decomposed into in-plane and perpendicular-to-plane components, which increases the number of fitting parameters but is {\em a priori} justified by symmetry in the presence of a biaxial strain.} These energies are related to the anisotropy magnetic fields, $\mu_0H_i = 2K_i/M$, describing the strength of the applied field allowing aligning of magnetization along the hard axes. As required by time reversal symmetry, $\cal{F}_{\text{d}}$ and $\cal{F}_{\text{cr}}$ are even functions of $M$.

\begin{figure}
\includegraphics[width=3.1in]{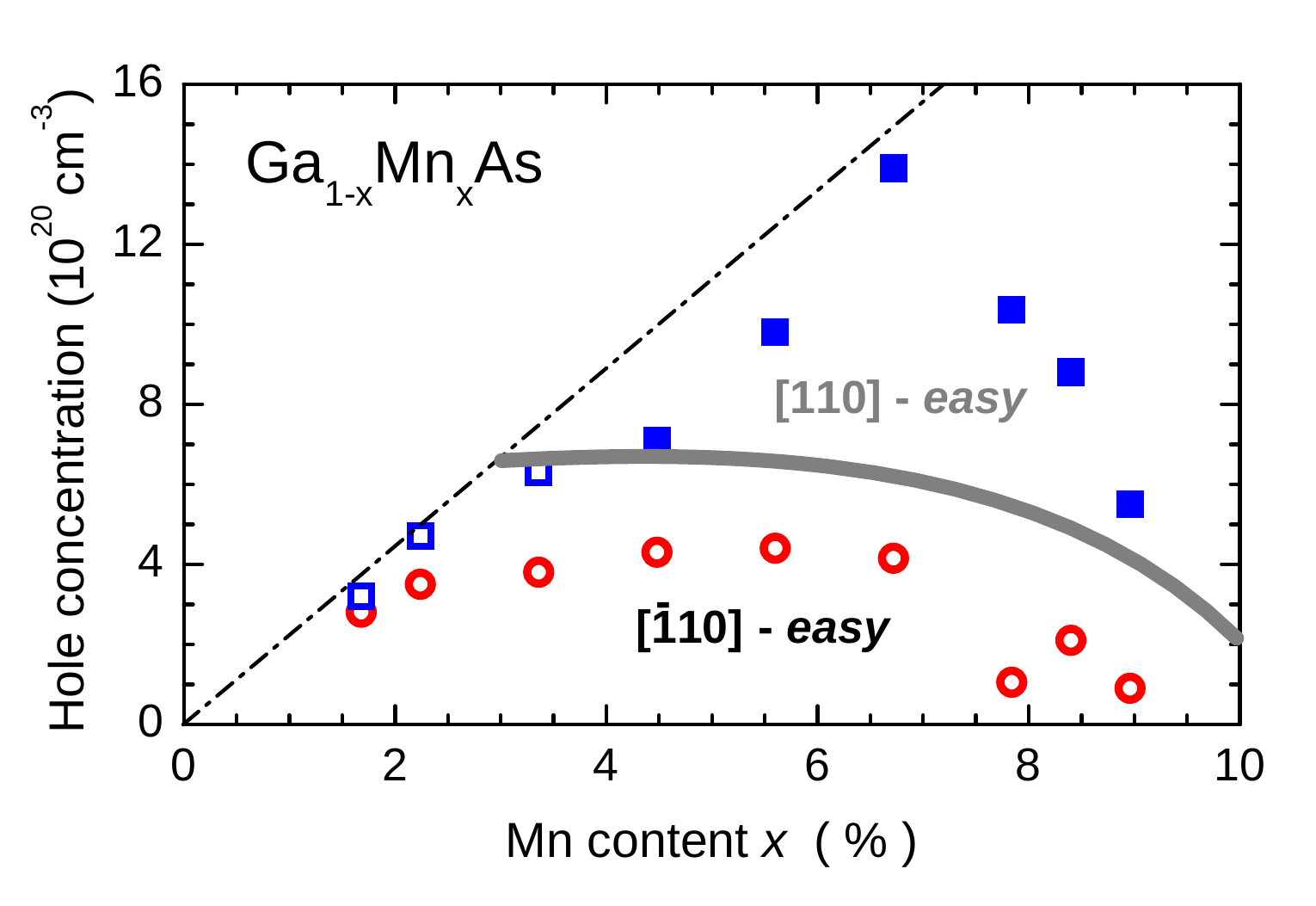}\vspace{-0mm}
\caption[]{(Color online) Crystallographic orientation of the uniaxial easy axis depending on the hole and Mn concentration for as-grown (circles) and annealed (squares) 50-nm thick Ga$_{1-x}$Mn$_x$As films. Open symbols mark samples with the uniaxial easy axis oriented along the $[\bar{1}10]$ direction, full symbols denote samples exhibiting the easy axis along [110]. Half-filled squares mark the two samples exhibiting easy axis rotation to [110] on increasing temperature. The dashed line marks the compensation free p-type Mn doping level in (Ga,Mn)As. The thick gray line separates the two regions of hole densities where, independently of being annealed or not, at elevated temperatures the layers consistently show the same crystallographic alignment of the uniaxial easy axis. From \onlinecite{Sawicki:2005_PRB}.}
\label{fig:GaMnAs_Sawicki_2005}
\end{figure}

\begin{figure}
\includegraphics[width=2.9in]{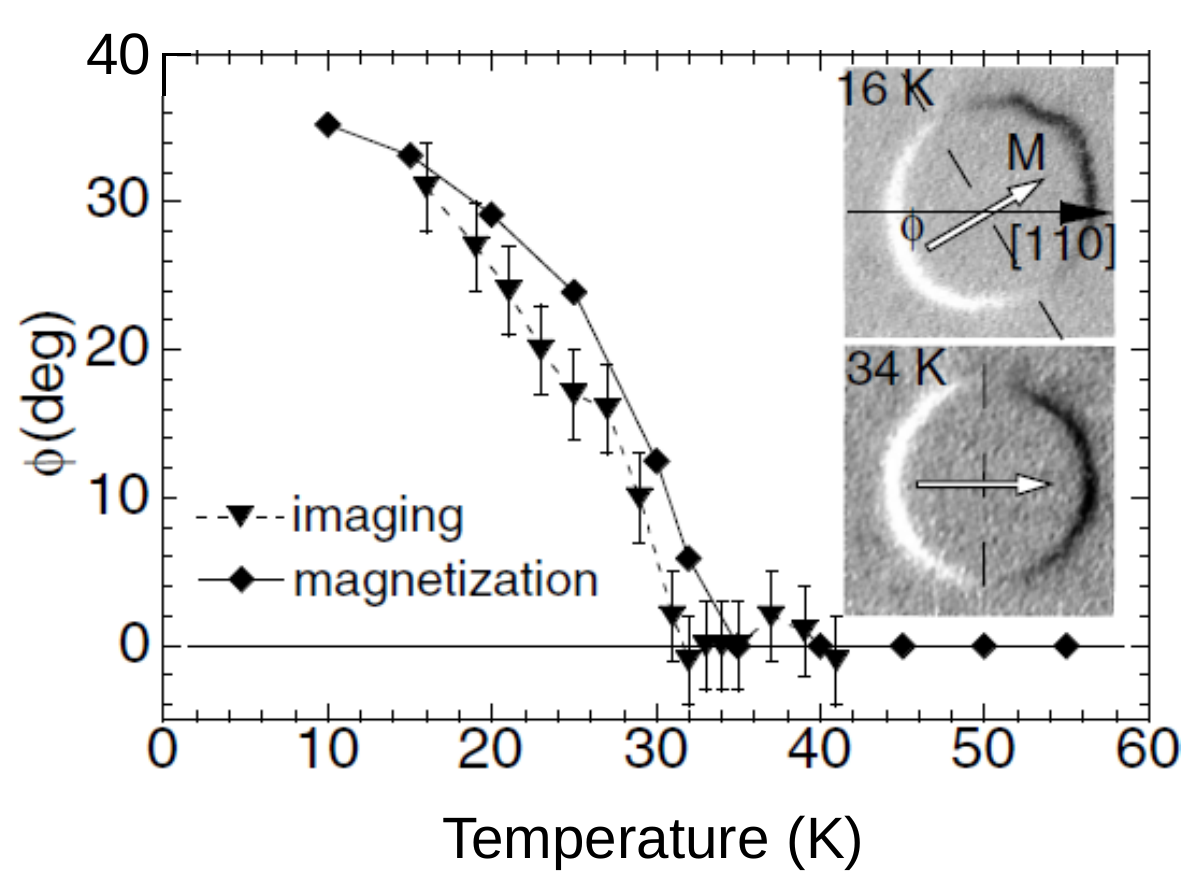}\vspace{-0mm}
\caption[]{ Temperature dependence of the angle of the easy axis with respect to the [110] direction. The inset shows the magneto-optical contrast around an 80-$\mu$m hole at 16 and 34 K, which identifies the moment orientation at a zero magnetic field. From \onlinecite{Welp:2003_PRL}.}
\label{fig:GaMnAs_Welp}
\end{figure}

Because of a relatively low magnitude of spontaneous magnetization (typically $\mu_0M \lesssim 0.1$~T), the strength of the shape anisotropy field, $\mu_0H_{\text{d}} = \mu_0M$, is substantially smaller in DFSs than in ferromagnetic metals. In contrast, the magnitude of crystalline anisotropy is rather sizable. According to experimental studies referred to above, each of the three contributions to crystalline magnetic anisotropy, displayed in Eq.~\ref{eq:F_cr},  shows a specific pattern:

\ \\
{\em Cubic anisotropy} -- Independently of epitaxial strain and hole density, the value of $K_{\text{C}}$ was found positive in (Ga,Mn)As (showing that the cubic easy axis is along $\langle 100 \rangle$) and corresponds to $\mu_0H_{\text{C}}$ of the order of 0.1~T at  $T \ll T_{\text{C}}$.  It decays rather fast with temperature, $K_{\text{C}}\sim M^4(T)$, consistently with the expected isotropy of linear response functions in cubic systems requiring that $\partial^2K_{\text{C}}/\partial M^2 \rightarrow 0$ for $M \rightarrow 0$. In contrast, a negative value of $K_{\text{C}}$ (corresponding to a $\langle 110 \rangle$ cubic easy direction) was reported for (In,Mn)As \cite{Liu:2005_APL} and (Ga,Mn)P \cite{Bihler:2007_PRB}.

\ \\
{\em In-plane uniaxial anisotropy} -- No such anisotropy, first observed in magnetotransport experiments on (Ga,Mn)As \cite{Katsumoto:1998_pssb}, is expected for the $D_{\text{2d}}$ symmetry group corresponding to biaxially strained (001) zinc-blende crystals. It was demonstrated that the corresponding anisotropy field was independent of the film thickness \cite{Welp:2004_APL}, pointing to the bulk, not surface or interface, origin of this anisotropy, the conclusion consistent with no effect of film thickening by etching on its presence \cite{Sawicki:2005_PRB}. However, as noted in Sec.~\ref{sec:dimers}, according to theory \cite{Birowska:2012_PRL},  a surplus of the Mn dimer concentration along the $[\bar{1}10]$ direction comparing to the [110] case is expected. This  lowers the symmetry to  $C_{\text{2v}}$ (even in the absence of any strain), for which distinct in-plane and out-of-plane uniaxial anisotropies are allowed. Experimentally, the value of an effective shear-like component $K_{xy}$ is usually positive \cite{Zemen:2009_PRB} ({\em i.~e.},  the corresponding easy axis points along $[\bar{1}10]$ direction), and $\mu_0H_{xy}$ is typically of the order of 0.02~T at  $T \ll T_{\text{C}}$, so that it is smaller than $\mu_0H_{\text{C}}$.  It was found \cite{Sawicki:2005_PRB} that at appropriately high hole concentrations the uniaxial easy axis flips to the [110] direction, as shown in Fig.~\ref{fig:GaMnAs_Sawicki_2005},  the effect often occurring only at sufficiently high temperatures, $T_{\text{C}}/2 \lesssim T \leqslant T_{\text{C}}$ \cite{Sawicki:2005_PRB,Proselkov:2012_APL,Kopecky:2011_PRB}.  Furthermore, since comparing to $K_{\text{C}}$, $K_{xy}$ decays slower with temperature, $K_{xy}\sim M^2(T)$, a spin reorientation transition  $\langle 100 \rangle \rightarrow [\bar{1}10]$ is observed on increasing temperature in the range $T \lesssim T_{\text{C}}/2$ \cite{Welp:2003_PRL,Wang:2005_PRL,Kamara:2012_JNN}, the effect illustrated in Fig.~\ref{fig:GaMnAs_Welp}. A competition between cubic and in-plane uniaxial magnetic anisotropies were also found for (In,Mn)As \cite{Liu:2005_APL} and (Ga,Mn)P \cite{Bihler:2007_PRB}.

\ \\
{\em Out-of-plane uniaxial anisotropy} -- According to experimental studies of (Ga,Mn)As on In$_y$Ga$_{1-y}$As \cite{Glunk:2009_PRB} and of (Ga,Mn)As$_{1-y}$P$_y$ on GaAs \cite{Cubukcu:2010_PRB} as a function of $y$ and, thus, epitaxial (biaxial) strain $\epsilon_{zz}$, the anisotropy energy $K_{zz}$  can be decomposed into two contributions.

\begin{figure}
\includegraphics[width=3in]{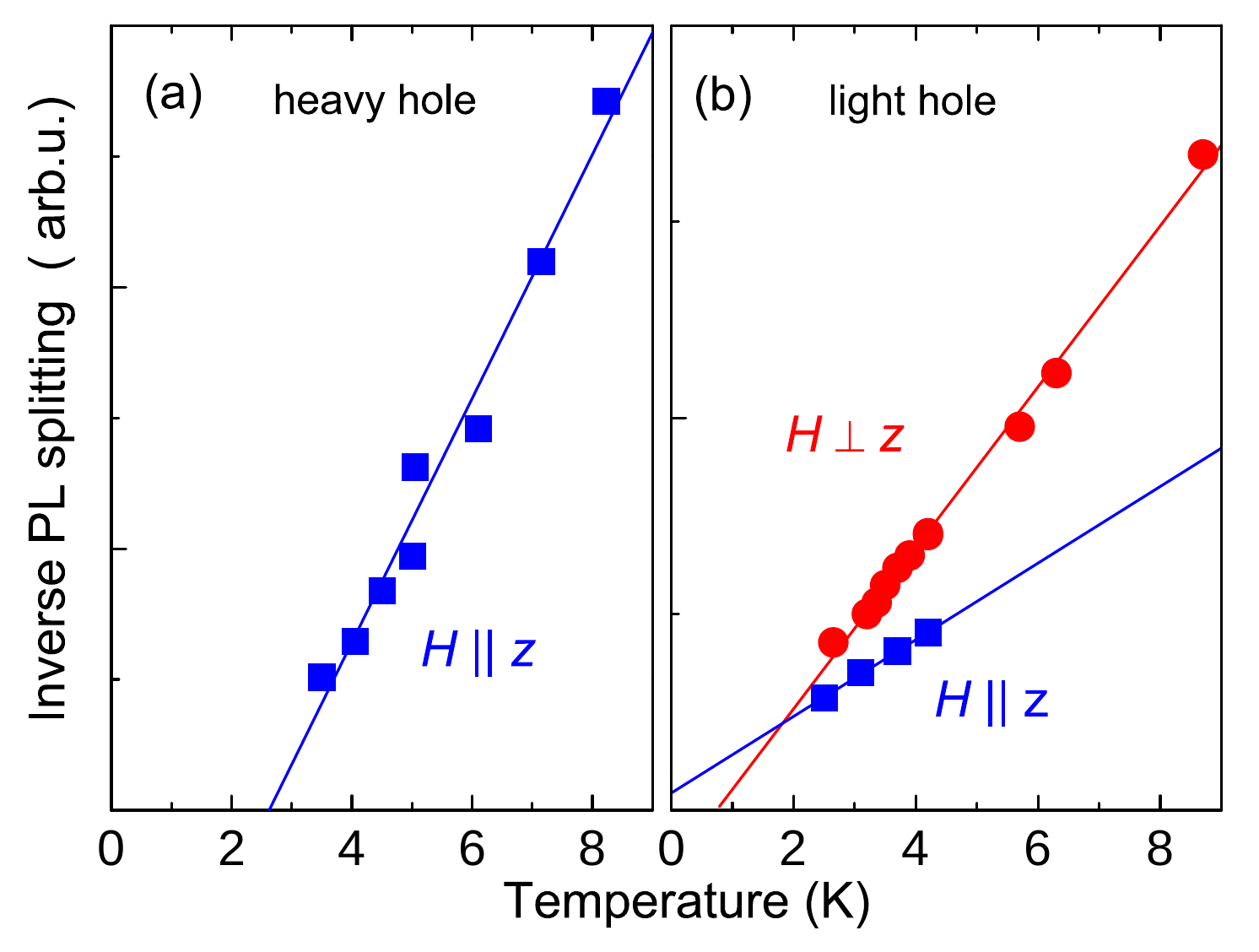}\vspace{-0mm}
\caption[]{(Color online) Determination of easy directions in p-(Cd,Mn)Te modulation doped quantum wells with only the ground state hole subband occupied under compressive (a) and tensile epitaxial strain (b). Curie-Weiss behavior above Curie temperature was obtained from photoluminescence measurements in the magnetic field parallel and perpendicular to the growth axis $z$ (points). The line splitting, proportional to weak-field Mn magnetization, is presented in relative units. The straight lines are drawn through experimental points. Adapted from \onlinecite{Kossacki:2004_PE}.}
\label{fig:CdMnTe_Kossacki}
\end{figure}

One is linear in $\epsilon_{zz}$, $\mu_0H_{zz} = A\epsilon_{zz}$, corresponding to the in-plane and perpendicular-to-plane crystalline magnetic anisotropy for compressive and tensile strain, respectively. According to studies up to $|\epsilon_{zz}| \approx 0.4$\%, $A$ can reach a magnitude of the order of +1~T/\% \cite{Glunk:2009_PRB,Cubukcu:2010_PRB} but $|A|$ decreases, or even changes sign when diminishing hole density at a fixed compressive  \cite{Khazen:2008_PRB,Sawicki:2004_PRB,Thevenard:2005_APL} or tensile strain \cite{Thevenard:2006_PRB}. For hole concentrations $p \approx 10^{20}$~cm$^{-3}$ corresponding to the vicinity of the spin reorientation transitions  $(001) \rightleftarrows [001]$, the transition can occur on changing temperature \cite{Sawicki:2004_PRB,Thevenard:2006_PRB}. Similarly to (Ga,Mn)As with high hole concentrations, also (In,Mn)As shows perpendicular-to-plane orientation of the easy axis for a tensile strain, imposed by either (Ga,Al)Sb \cite{Munekata:1993_APL,Liu:2004_PE,Ohno:2000_N} or InAs substrate \cite{Zhou:2012_APEX}, whereas the easy axis is in-plane under a compressive strain [produced by an (In,Al)As substrate] \cite{Liu:2005_APL}. Interestingly, an opposite relation between the strain character and easy axis direction, consistent with the findings for (Ga,Mn)As with low carrier density, was observed for p-(Cd,Mn)Te:N \cite{Kossacki:2004_PE}, as shown in Fig.~\ref{fig:CdMnTe_Kossacki} as well as for (Al,Ga,Mn)As \cite{Takamura:2002_APL} and (Ga,Mn)P \cite{Bihler:2007_PRB}, presumably because of low net hole concentrations in all these cases.

Another contribution to $K_{zz}$ found in (Ga,Mn)As films is strain independent term leading to a non-zero out-of-plane uniaxial anisotropy term even for $\epsilon_{zz} = 0$ \cite{Glunk:2009_PRB,Cubukcu:2010_PRB}. The corresponding value is of the order of $\mu_0H_{zz} \approx 0.1$~T. Its positive sign means that this anisotropy, along with the demagnetization term, enlarges a tendency to the in-plane orientation of the easy axis.  As already mentioned, this contribution, unexpected within the group theory for a zinc-blende alloys having a random distribution of constituents, is assigned to a surplus of $[\bar{1}10]$ Ga-substitutional Mn dimers \cite{Birowska:2012_PRL}.

\ \\

Appropriately modified forms of $\cal{F}_{\text{cr}}$ were found to describe $\vec{M}(\vec{H})$ for (Ga,Mn)As grown on a (113)A GaAs substrate \cite{Wang:2005_PRBb,Limmer:2006_PRB,Dreher:2010_PRB,Stefanowicz:2010_PRBb}. In this case, however, four (not two) contributions to ${\cal{F}}_{cr}$ are allowed by symmetry and, in fact, describe magnetic and FMR data \cite{Stefanowicz:2010_PRBb}. They correspond to cubic $K_{\text{C}}$, biaxial $K_{zz}$, and two shear-like,  $K_{xy}$ and $K_{xz} = K_{yz}$ anisotropy energies (the axes of the coordinate system are taken along main crystallographic directions). Similarly to the case of (001) substrates discussed above, the spin reorientation transition from the biaxial $\langle 100 \rangle$ anisotropy at low temperatures to uniaxial anisotropy with the easy axis along the $[\bar{1}10]$ direction at high temperatures is observed (around 25 K).  As evidenced by investigations of the polar magnetooptical Kerr effect, a declined orientation of the easy axes with respect to the film plane and the film normal allows the perpendicular-to-plane component of magnetization to be reversed by an in-plane magnetic field \cite{Stefanowicz:2010_PRBb}.

A specific strain distribution in (Ga,Mn)As nanostructures, either in the form of nanobars patterned lithographically \cite{Humpfner:2007_APL,King:2011_PRB} or shells deposited onto GaAs nanowires \cite{Rudolph:2009_NL}, was found to result in the easy axis orientation along the nanostructure long axis. The magnitude of the observed anisotropy field was much larger than expected for the corresponding shape anisotropy, pointing to the importance of crystalline and strain effects.  In the case of rectangular nanobars the epitaxial in-plane strain is retained along the bar long axis but it is partly relaxed in the transverse direction, as confirmed by finite element calculations \cite{Wenisch:2007_PRL,King:2011_PRB} and observed by x-ray reciprocal space mapping \cite{Wenisch:2007_PRL,King:2011_PRB}. It was possible to rotate the easy axis by $90^o$ by nanopatterning \cite{King:2011_PRB}.

Magnitudes of possible surface or interface magnetic anisotropies have not yet been assessed for DFSs.

\subsubsection{Piezoelectric and elastic actuators}

\begin{figure}
\includegraphics[width=3.4in]{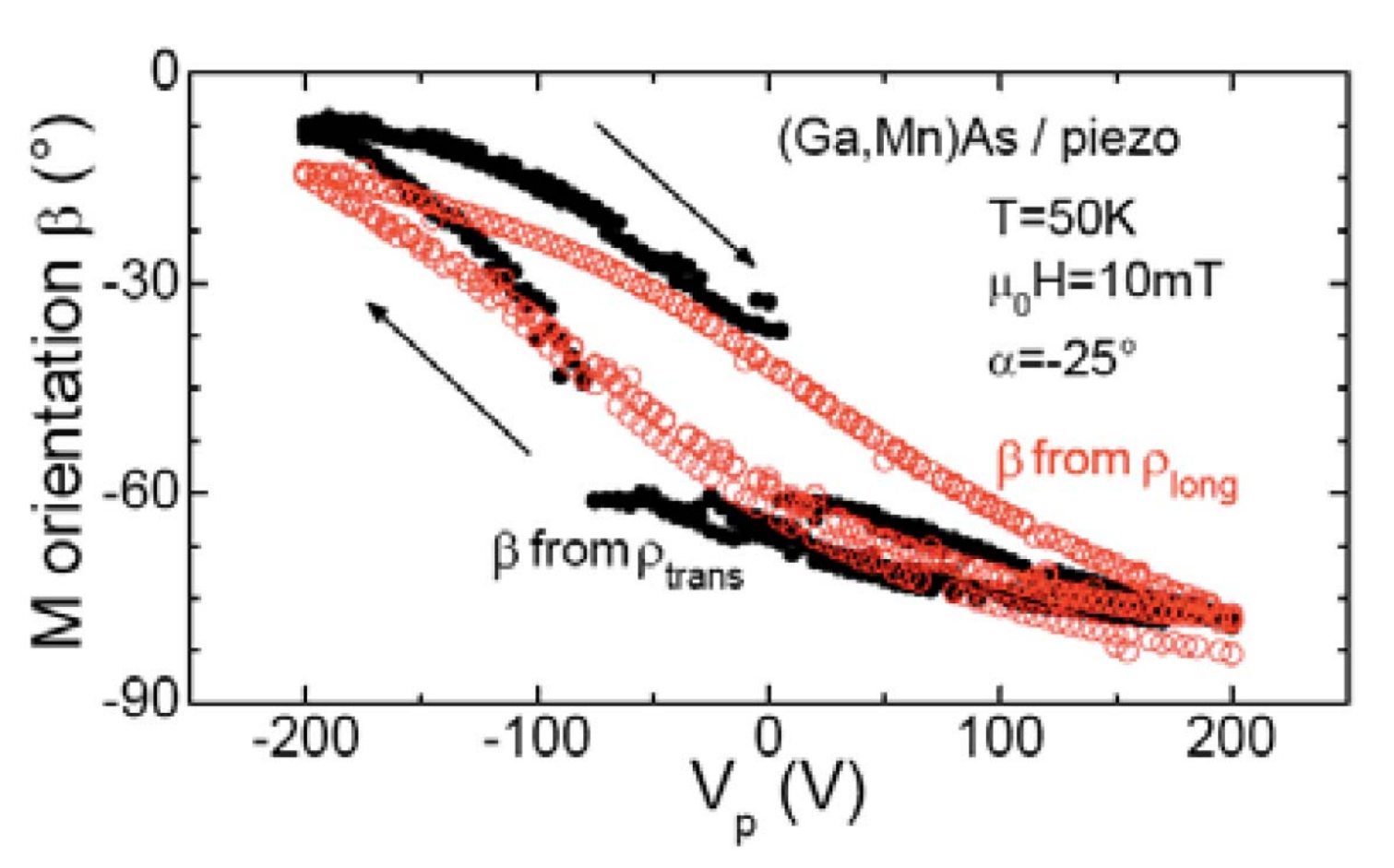}
\caption[]{(Color online) Voltage-induced in-plane magnetization rotation at 50 K and in 10~mT, determined from longitudinal  and transverse (Hall) resistances (open and full circles, respectively), in a Ga$_{0.955}$Mn$_{0.045}$As film cemented to a piezoelectric actuator, demonstrating a reversible change of magnetization direction by about $70^o$  by the application of a voltage. Magnetic field and magnetization angles ($\alpha$ and $\beta$, respectively) are measured in respect to the $[110]$ direction of current and main expansion of the actuator. The hysteretic behavior is caused by the actuator. From \onlinecite{Goennenwein:2008_pss}.}
\label{fig:GaMnAs_Goennenwein}
\end{figure}

A strong sensitivity of magnetic anisotropy to strain makes it possible to manipulate magnetization directions by an electric field in hybrid structures consisting of a DFS film cemented to a piezoelectric actuator. This appealing method was successfully demonstrated for (Ga,Mn)As by applying a voltage-controlled strain along either $\langle 110 \rangle$
\cite{Goennenwein:2008_pss,Rushforth:2008_PRB,Casiraghi:2012_APL}
or $\langle 100 \rangle$ \cite{Overby:2008_APL,Bihler:2008_PRB} directions of (Ga,Mn)As. In this way, rotation of the easy axis form either $[\bar{1}10]$  or $\langle 100 \rangle$ directions by about $70^o$ was possible at appropriately selected temperature and magnetic field values, as shown in Fig.~\ref{fig:GaMnAs_Goennenwein}. Importantly, an elaborated sequence of applied magnetic fields and voltages was found to switch magnetization in an irreversible fashion, showing a road for developing a novel voltage-controlled memory cell \cite{Bihler:2008_PRB}.

\begin{figure}
\includegraphics[width=3.2in]{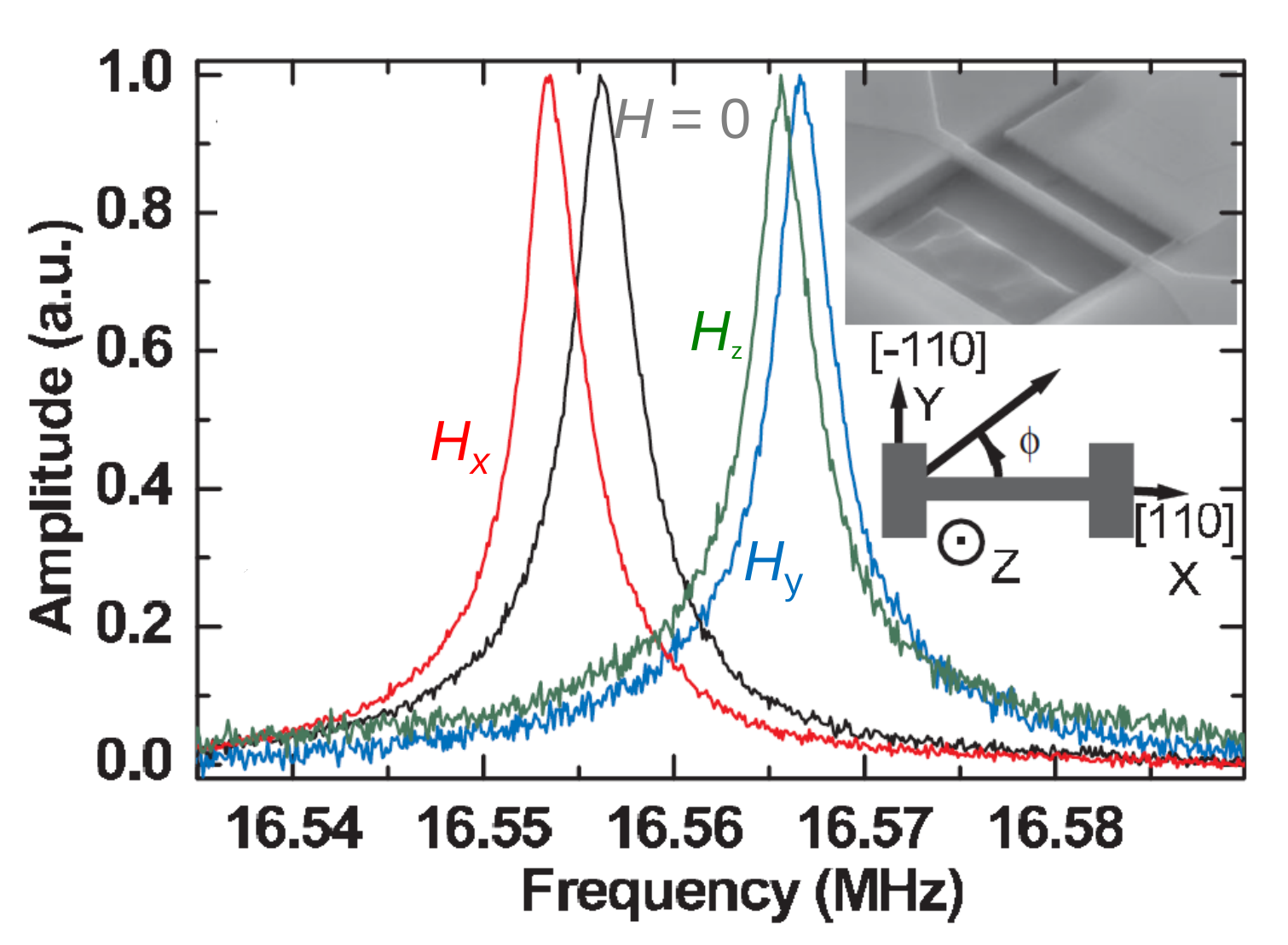}
\caption[]{(Color online) Frequency response for various $\mu_0H = 1$~T field directions ($z$ is out of the plane) of a suspended beam containing as-grown 50-nm thick Ga$_{0.948}$Mn$_{0.052}$As. Amplitude is normalized for clarity. Inset: axis directions and the scanning electron microscope image of the beam of the length 6~$\mu$m and thickness 0.18~$\mu$m  with an Au side gate. From \onlinecite{Masmanidis:2005_PRL}.}
\label{fig:GaMnAs_Masmanides}
\end{figure}

Magneto-elastic properties of (Ga,Mn)As were also studied by determining eigenfrequencies of  nanoelectromechanical resonator as a function of temperature and magnetic field orientation, as shown in Fig.~\ref{fig:GaMnAs_Masmanides} \cite{Masmanidis:2005_PRL}. This experiment was described by considering a contribution ${\cal{F}}_{\text{me}}$ to the system free energy  associated with additional strain $\epsilon_{ij}$ imposed by vibrations. By combining ${\cal{F}}_{\text{me}}$ with an elastic energy of the host lattice, it was possible to determine how stress $\sigma_{ij}$ and, thus, the frequency of the (Ga,Mn)As nanoelectromechanical resonator should vary with the magnetic field orientation. It was found that the low temperature results can be described by first order magnetostriction coefficients, $\lambda_{100}$ and $\lambda_{111}$, but above 20~K a second order magnetostriction, characterized by a parameter $h_3$, had to be taken into account in order to describe the data.

\subsection{Manipulation by an electric field}
\label{sec:electric_field}
\begin{figure}
\includegraphics[scale=1.05]{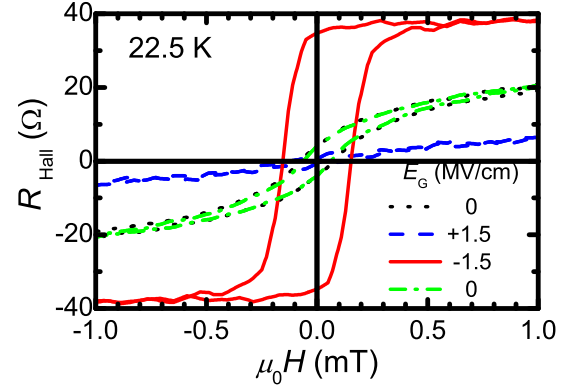}
\caption[]{(Color online) Magnetization hysteresis loops determined by measurements of anomalous Hall effect at constant temperature of 22.5 K for various gate voltages in field-effect transistor with (In,Mn)As channel. The data in a wider field range are shown in the inset. Adapted from \onlinecite{Ohno:2000_N}.}
\label{fig:FET}
\end{figure}

\begin{figure}
\includegraphics[scale=0.80]{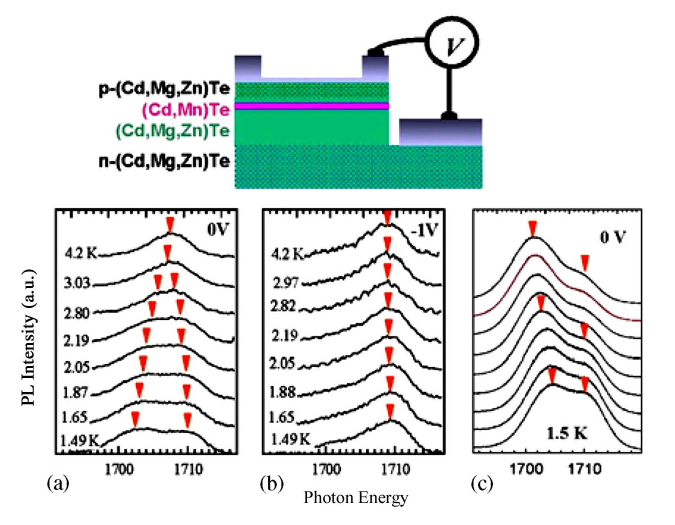}
\caption[]{(Color online) Effect of temperature (a), bias voltage (b), and illumination (c) on photoluminescence of structure consisting of modulation doped p-(Cd,Mn)Te quantum well and n-type barrier. Zero-field line splitting (marked by arrows) witnesses the appearance of a ferromagnetic ordering (a) which does not show up if the quantum well is depleted from the holes by reverse bias of p-i-n diode (b). Low-temperature splitting is enhanced by additional illumination by white light (c), which increases hole concentration in the quantum well. Adapted from \onlinecite{Boukari:2002_PRL}.}
\label{fig:LED}
\end{figure}

Since ferromagnetism in DFSs is hole-mediated, one can turn on and off the magnetic phase by controlling the number of holes in the system without changing the temperature, which can be done electrostatically by applying an electric field $E_{\text{G}}$ to the ferromagnetic semiconductor layer of interest. This was demonstrated using a thin (In,Mn)As (~5~nm) as a channel layer of a metal-insulator-semiconductor field-effect transistor (MISFET) with a polyimide insulator \cite{Ohno:2000_N} and a modulation doped p-(Cd,Mn)Te 8~nm thick quantum well placed in the intrinsic region of a \emph{p-i-n} diode \cite{Boukari:2002_PRL}. An appropriate strain engineering resulted in the easy axis perpendicular to the layer plane and allowed to probe magnetization through the anomalous Hall effect \cite{Ohno:2000_N} and splitting of the luminescence line \cite{Haury:1997_PRL}, as shown in Figs.~\ref{fig:FET} and \ref{fig:LED}, respectively.

Another effect of gating is the change of the coercive force $H_{\text{c}}$ at which magnetization reverses its direction; greater (smaller) $H_{\text{c}}$ for negative (positive) $E_{\text{G}}$. By using this phenomenon, a new scheme of magnetization reversal, an electric-field assisted magnetization reversal, was demonstrated for (In,Mn)As \cite{Chiba:2003_S}. Under a certain applied field $H$, $H_{\text{c}}$ is electrically modified from its original $|H_{\text{c}}| > |H|$ to $|H_{\text{c}}'| < |H|$, thereby electrically triggering the magnetization reversal. Once $T_{\text{C}}$ becomes high enough, this scheme can be useful for future magnetic field recording technology, where the magnetic anisotropy for retaining information becomes so large that it is almost impossible to change the magnetization direction by field alone.

Although there have only been limited success on (Ga,Mn)As \cite{Nazmul:2004_JJAP}, recent progress in low-temperature deposition of high-quality gate oxides by atomic layer deposition made it possible to observe electrical modulation of ferromagnetism in (Ga,Mn)As \cite{Chiba:2006_APL,Sawicki:2010_NP,Chiba:2010_PRL}, (Ga,Mn)Sb \cite{Chang:2013_APL},  and (Ge,Mn) \cite{Xiu:2010_NM}. For such oxides, the gate-induced changes in the areal carrier density reach $3\times 10^{13}$~cm$^{-2}$, which for a typical value of the Thomas-Fermi screening length would result in the amplitude of the hole variation about  $3\times 10^{20}$~cm$^{-3}$, and the correspondingly high modulation of $T_{\text{C}}$. However, the Fermi level is often pinned by gap surface states, which limit the $T_{\text{C}}$ changes to about 20~K in (Ga,Mn)As \cite{Chiba:2006_APL,Sawicki:2010_NP,Nishitani:2010_PRB,Chiba:2010_PRL}, even if a polymer
electrolyte is employed \cite{Endo:2010_APLb}. A theoretical  description of these and related data \cite{Stolichnov:2011_PRB} is discussed in Sec.~\ref{sec:Curie}.

As elaborated in the previous section, also magnetic anisotropy, which determines the magnetization direction, depends on the hole concentration in DFSs. By applying an electric field and by using anisotropic magnetoresistance \cite{Chiba:2008_N} as well as direct magnetization measurements \cite{Chiba:2008_N,Sawicki:2010_NP}, the effect of the electric field on in-plane magnetization orientation was evidenced in MISFET of (Ga,Mn)As and high-$k$ oxide as a gate insulator. As demonstrated by Hall effect measurements, a fourfold change in the value of the out-of-plane uniaxial anisotropy field was achieved by gating an ultrathin ferromagnetic (Ga,Mn)As/(Ga,Mn)(As,P) bilayer \cite{Niazi:2013_APL}.

An important variant of gating is the application of ferro{\em electric} overlayers allowing for a non-volatile and subnanosecond change in interfacial hole density, the method successfully employed to demonstrate the manipulation of $T_{\text{C}}$ by an electric field in (Ga,Mn)As \cite{Riester:2009_APL,Stolichnov:2008_NM,Stolichnov:2011_PRB}.

Manipulation of magnetism by gating was also demonstrated for (Ge,Mn) films \cite{Park:2002_S} and quantum dots \cite{Xiu:2010_NM}. In the former case, an enhancement and a reduction of $T_{\text{C}}$ was demonstrated at 50~K by applying $\mp 5$~V through an SiN$_x$ gate insulator to a 60-nm thick Ge$_{0.977}$Mn$_{0.023}$ film on Ge(001). Self-assembled Ge$_{0.95}$Mn$_{0.05}$ dots were deposited on a p-Si substrate and covered by a 40-nm thick Al$_2$O$_3$ gate insulator. In was shown by SQUID measurements that positive gate voltage up to 40~V reduced a saturation value of magnetization tenfold at 50 K and by 30\% at 100~K.

Another interesting case constitutes magnetically doped topological insulator (Bi,Mn)$_2$(Te,Se)$_3$ showing in the bulk form hole-mediated ferromagnetism with $T_{\text{C}}$ of 12~K \cite{Checkelsky:2012_NP}. In contrast, no conductivity and ferromagnetism were observed in few nm-thick flakes put on a SiO$_2$/doped-Si wafer, presumably because of cleavage-induced hole compensating defects. However, the application of a strong negative electric field across SiO$_2$ allowed to restore hole conductivity and ferromagnetism characterized by $T_{\text{C}}$ up to 12~K.

\subsection{Current-induced magnetization switching}
\label{sec:current}
\begin{figure}
\includegraphics[scale=0.8]{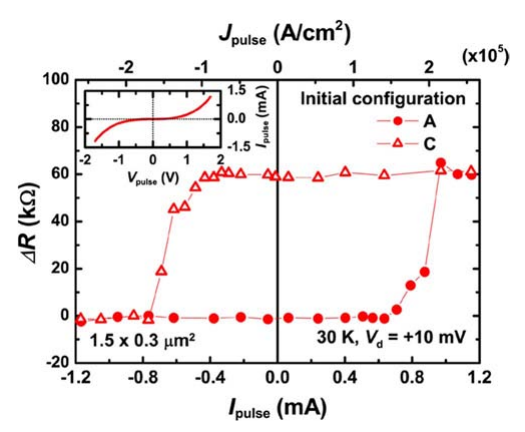}
\caption[]{(Color online) Resistance changes $\Delta R$ after current pulse injection for $1.5\times 0.3$~$\mu$m$^2$ device at 30~K. Closed circles show $\Delta R$ for initially parallel magnetizations (configuration A) and open triangles for initial configuration C (antiparallel magnetizations). The inset shows current-voltage characteristic of the device. From \onlinecite{Chiba:2004_PRL}.}
\label{fig:CIMS}
\end{figure}

When current flows through a ferromagnetic layer, the current becomes spin-polarized. In magnetic tunnel junctions (MTJs), the flow of spins from one ferromagnetic electrode to the other across the tunnel barrier exerts a torque between the two electrodes, and its direction depends on the flow of spin, {\em i.e.}, on the current direction.  In sufficiently small MTJs, the torque can reach a threshold value above which magnetization reversal takes place. This is so-called current-induced magnetization switching (CIMS) that was observed in submicron (Ga,Mn)As MTJs  \cite{Chiba:2004_PRL,Elsen:2006_PRB}, as shown in Fig.~\ref{fig:CIMS}. The critical current density $j_{\text{c}}$ for switching is of the order of 10$^4$--10$^5$~A/cm$^2$, and can be qualitatively understood by Slonczewski's spin-transfer torque model \cite{Chiba:2004_PRL}, although for the assumed value of the Gilbert damping constant $\alpha_{\text{G}}$ the model resulted in an order of magnitude greater $j_{\text{c}}$ than the observed one.

Another appealing method, particularly in the context of DFSs, is magnetization manipulation by an effective magnetic field produced by an electric current through spin-orbit coupling, as opposed to the Oersted effect. It is well known that in confined 2D systems there appear terms linear in $k$ coupled to the electron spin. In particular, the corresponding Rashba field was shown to generate current-dependent shift of electron spin-resonance in an asymmetrical quantum well of n-Si \cite{Wilamowski:2007_PRL}. It was suggested theoretically \cite{Bernevig:2005_PRB} and demonstrated experimentally that sufficiently strong current, {\em via} spin-orbit coupling in strained (Ga,Mn)As (see, Eq.~\ref{eq:k_linear}), can serve to rotate magnetization by 90$^o$ \cite{Chernyshov:2009_NP} or even by 180$^o$ \cite{Endo:2010_APL} in (Ga,Mn)As films.

\subsection{Current-induced domain wall motion}
\label{sec:current_domains}
Magnetic domains are formed in a ferromagnet as a result of competition between exchange energy (which tries to align all the spins) and magnetostatic energy (which tries to align magnets antiparallel). The transition region between the domains, in which localized spins (Mn spins of magnetization $M$ in our case) gradually changes its direction, is called a domain wall (DW). Spin polarized currents interact with DW and once a threshold $j_{\text{c}}$ is passed can displace the DW, resulting in magnetization reversal of a region swept by the DW. Although such a current-induced DW motion has been of interest for many years in the context of metallic ferromagnets, the DW switching without assistance of a magnetic field was first demonstrated for (Ga,Mn)As films  \cite{Yamanouchi:2004_N}.

\begin{figure}
\includegraphics[width=3.4in]{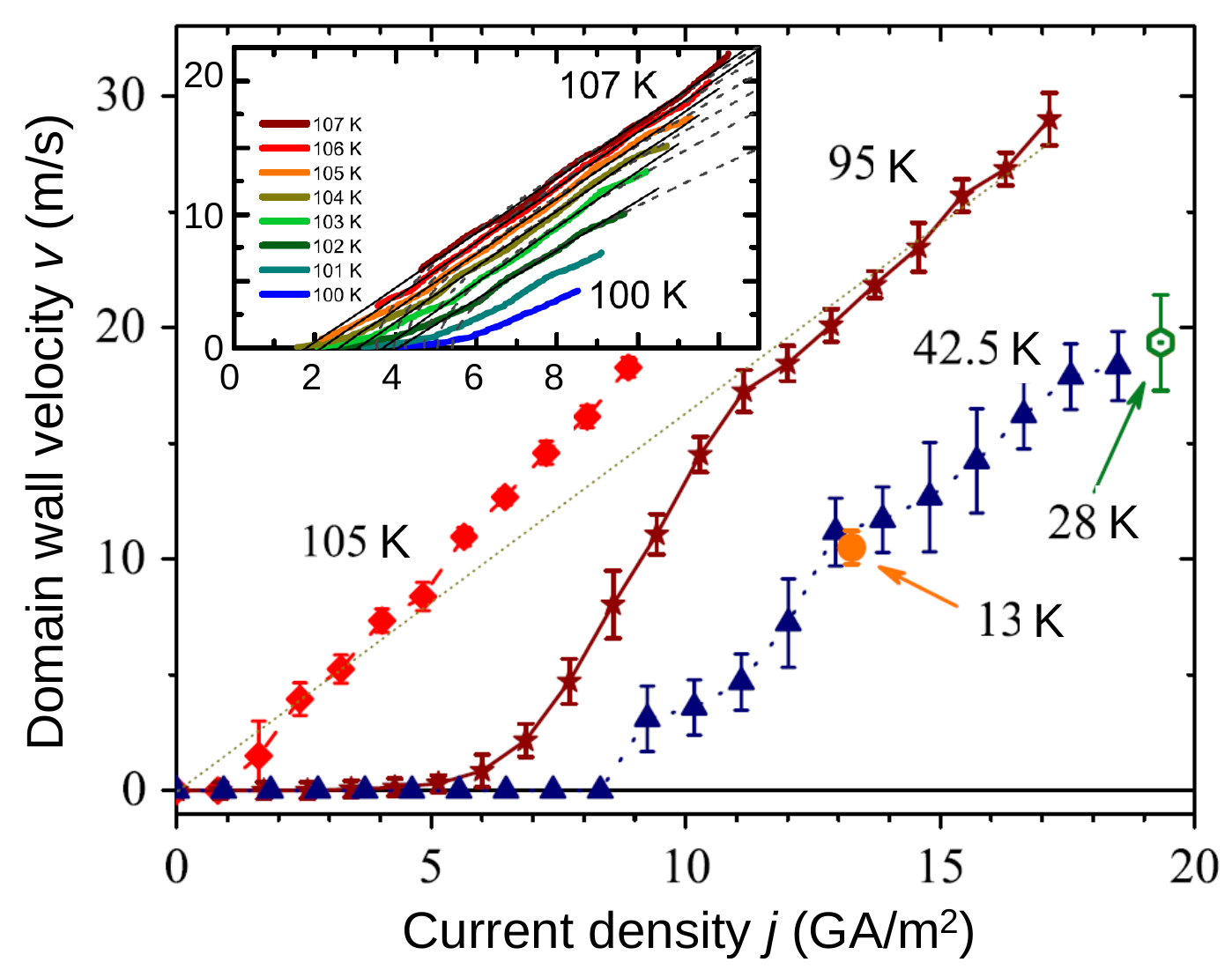}
\caption[]{(Color online) Velocity of magnetic domain wall as a function of current density at various temperatures (corrected for current-induced heating) below $T_{\text{C}}$ for (Ga$_{0.9}$Mn$_{0.1}$)(As$_{0.89}$P$_{0.11}$) [adapted from  \onlinecite{Curiale:2012_PRL}] and Ga$_{0.955}$Mn$_{0.045}$As [inset, adapted from \onlinecite{Yamanouchi:2006_PRL}]. The thin dashed limes show the expected theoretical dependence.}
\label{fig:DW}
\end{figure}

In microtracks of (Ga,Mn)As \cite{Yamanouchi:2004_N,Yamanouchi:2006_PRL,Yamanouchi:2007_S,Adam:2009_PRB} and (Ga,Mn)(As,P) \cite{Curiale:2012_PRL,Wang:2010_APL} with the easy axis  perpendicular to the film plane,  DW displacement under current pulses was monitored by magneto-optical Kerr microscopy. As shown in Fig.~\ref{fig:DW}, a similar dependence of DW velocity $v$ on the current density $j$ were found for these two material systems \cite{Yamanouchi:2006_PRL,Curiale:2012_PRL}.

These findings were interpreted by the generalized Landau-Lifshitz-Gilbert equation containing, as displayed in Appendix, current-induced adiabatic and non-adiabatic spin torques, accounting for transfer of spin momenta from current carriers to Mn ions. In particular,  \onlinecite{Yamanouchi:2006_PRL,Wang:2010_APL} presumed the dominance of the adiabatic spin torque, {\em i.~e.,} $\beta_{\text{w}}/\alpha_{\text{G}} \ll 1 $, the assumption leading to \cite{Tatara:2004_PRL} $v = A(j^2-j^2_{\text{c}})^{1/2}$ for $j >j_{\text{c}}$. Here, in terms of spin current polarization $P_{\text{c}}$ and DW width $\delta_{\text{W}}$ (discussed theoretically in Secs.~\ref{sec:magnetic_stiffeess} and \ref{sec:structures_theory}, respectively), $A = g\mu_{\text{B}}P_{\text{c}}/2eM$ and $j_{\text{c}} = 2eK\delta_{\text{W}}/\pi\hbar P_{\text{c}}$, where in the perpendicular case the relevant magnetic anisotropy energy of DW spins is $K = \mu_0M^2/(2 + 4\delta_{\text{W}}/\pi t)$, where $t$ is the film thickness. The experimental values of both $A$ and $j_{\text{c}}$, implied by the data in Fig.~\ref{fig:DW}, are in quantitative agreement with this theory. Moreover, this approach explains why no current-induced DW displacement was observed for the in-plane easy axis \cite{Tang:2006_PRB}, as $j_{\text{c}}$ is then determined by $K_{\text{cr}} \gg K$. Within this model, DW motion at $j > j_{\text{c}}$ is accompanied by in-plane Mn spin  precession.

On the other hand, \onlinecite{Curiale:2012_PRL} (see also \onlinecite{Adam:2009_PRB}) interpreted their data assigning non-zero values of $j_{\text{c}}$ to extrinsic DW pinning and allowing for a large magnitude of the non-adiabatic spin torque (present due to spin-orbit interactions), $\beta_{\text{w}}/\alpha_{\text{G}} \gtrsim 1$ \cite{Zhang:2004_PRL,Garate:2009_PRBc,Hals:2009_PRL}.  In this case, $v = Aj$ at $j \gg j_{\text{c}}$, the regime corresponding within this model to steady state DW flow without Mn spin precession.

In the subthreshold regime, $j < j_{\text{c}}$, the DW velocity was found to decay exponentially when reducing $j$, indicating that DW displacement proceeded through current-induced creep \cite{Yamanouchi:2006_PRL,Yamanouchi:2007_S,Curiale:2012_PRL}. Over the exploited range of $j$ and $T$, $v(j)$ assumed a scaling form \cite{Yamanouchi:2006_PRL,Yamanouchi:2007_S} $\ln v  = a - bj^{-\mu}$, where $a \propto (T_{\text{C}} - T)^{\rho}$, $b \propto (T_{\text{C}} - T)^{\sigma}$. Empirically \cite{Yamanouchi:2006_PRL,Yamanouchi:2007_S}, $\mu = 0.4 \pm 0.1$ and $\sigma = 2\pm 0.2$ (the value of $\rho$ is uncertain but probably around 1). Interestingly, for DW creep generated by a magnetic field $H$, the scaling equation assumed a similar form (with $j$ replaced by $H$). However, in agreement with theoretical considerations, the values of the exponents were different,  emphasizing the unalike role of these two DW driving  mechanisms \cite{Yamanouchi:2007_S}.

Domain wall dynamics in the magnetic field, particularly mobility, pining, and flexing were examined in (Ga,Mn)As and (Ga,Mn)(As,P) with perpendicular easy axis by spatially resolved magnetooptical and Hall effects \cite{Balk:2011_PRL,Dourlat:2008_PRB,Thevenard:2011_PRB}.

An issue obviously related to current-induced DW motion is DW resistance that, in general, consists of extrinsic $R^{\text{ext}}$ and intrinsic $R^{\text{int}}$ components. The former is brought about by a non-uniform current distribution associated with differences in magnitudes of conductivity tensor components $\sigma_{ij}(\vec{M})$ on the two sides of the DW. It was demonstrated \cite{Chiba:2006_PRL,Roberts:2007_PRB,Wang:2010_JMMM,Xiang:2007_PRB}, by solving the current continuity equation $\mbox{div}[\hat{\sigma}(x,y)\mbox{grad}V(x,y)] =  0$, that the extrinsic term dominates in (Ga,Mn)As with both perpendicular \cite{Chiba:2006_PRL,Wang:2010_JMMM} and in-plane easy axis \cite{Tang:2004_N} as well as explains the corresponding magnetoresistance \cite{Xiang:2007_PRB}.

Nevertheless,  if DW cross sections $A$ were sufficiently small, $R^{\text{int}}$ could be revealed, as shown for 25~nm thick Ga$_{0.95}$Mn$_{0.05}$As bars of the width from 150 down to 4~$\mu$m  and with the strain-induced perpendicular orientation of the easy axis \cite{Chiba:2006_PRL,Wang:2010_JMMM}. For samples containing etched steps that pinned DWs, $R^{\text{int}}A \simeq 0.5$~$\Omega \mu$m$^2$ \cite{Chiba:2006_PRL} and 0.15~$\Omega \mu$m$^2$ \cite{Wang:2010_JMMM} for films with $T_{\text{C}} = 80$ and 122~K, respectively. These values  are much larger  than $R^{\text{int}}A$ evaluated from the measured magnitude of anisotropic magnetoresistance (AMR) for the Bloch DW in strained films in question.  However, if DWs were pinned by linear defects, the value below experimental resolution, $R^{\text{int}}A = 0.01 \pm 0.02$~$\Omega \mu$m$^2$, was found (after subtracting the AMR contribution) for the sample with $T_{\text{C}} = 122$~K \cite{Wang:2010_JMMM}. Theoretically predicted DW resistances in (Ga,Mn)As (see, Sec.~\ref{sec:structures_theory}) are within this range.

\subsection{Magnetization manipulation by light}
\label{sec:light}
It was demonstrated that light irradiation affects magnetic properties and, in particular, changes the magnitude of the coercive field in (In,Mn)As/GaSb heterostructures \cite{Koshihara:1997_PRB,Oiwa:2001_APL}. The effect was attributed to persistent photoconductivity, that is with the light-induced increase of hole density in (In,Mn)As associated with trapping of photoelectrons by deep levels, which was not reversible at a given temperature -- in order to return to the original state, the sample had to be heated.

In the case of Mn-based II-VI DMS reversible tuning of magnetism by light was demonstrated in the case of modulation doped p-(Cd,Mn)Te/(Cd,Mg,Zn)Te:N heterostructures  \cite{Haury:1997_PRL,Boukari:2002_PRL}, as depicted in Fig.~\ref{fig:LED} and discussed theoretically in Sec.~\ref{sec:T_c_II_VI}.  Interestingly,  illumination with photons of energies above the barrier band gap destroys ferromagnetic order if the magnetic quantum well (QW) resides in an undoped (intrinsic) region of a p-i-p structure. Here, the holes are effectively transferred from the QW to acceptors in the (Cd,Mg,Zn)Te barrier \cite{Haury:1997_PRL,Boukari:2002_PRL}. In contrast, illumination enhances the magnitude of spontaneous magnetization in the case of a p-i-n diode in which photoholes accumulate in the (Cd,Mn)Te QW \cite{Boukari:2002_PRL}, as shown in Fig.~\ref{fig:LED}.

Reversible changes of magnetization by {\em circularly} polarized light were witnessed by Hall effect measurements for (Ga,Mn)As and Mn $\delta$-doped GaAs \cite{Oiwa:2002_PRL,Nazmul:2004_JJAP}. The magnitude of the Hall voltage (and hence, presumably, the magnitude of magnetization along the growth direction) either increased or decreased depending on the helicity of impeding light.

All-optical  switching of magnetization between two non-equivalent cubic in-plane directions was demonstrated in (Ga,Mn)As microbar employing a scanning laser magnetooptical microscope \cite{Aoyama:2010_JAP}. Lithography-induced strain relaxation contributed significantly to the magnitude of uniaxial anisotropy. External magnetic field served to magnetize the sample along the harder cubic direction but was not applied during the switching. Light served primarily to elevate temperature to $T>T_{\text{C}}/2$ at which cubic and uniaxial anisotropy energies became nearly equal (Sec.~\ref{sec:anisotropy}.

Another interesting case is (Ge,Mn)Te, which deposited at low temperature is amorphous and paramagnetic, presumably because dangling bonds associated with lattice point defects (vacancies) are reconstructed in the amorphous network and do not provide holes. A laser or electron beam triggers a local lattice recrystallization, allowing to pattern ferromagnetic nanostructures \cite{Knoff:2011_PSSB}.

Particularly informative and relevant for fast magnetization manipulation is subpicosecond magneto-optical two-color Kerr spectroscopy and related magnetization sensitive time-resolved methods. Here, we discuss experiments in which illumination generated incoherent magnetization dynamics; the data pointing to coherent magnetization precession are described in the subsequent subsection.

In the case of a Ga$_{0.98}$Mn$_{0.02}$As film with in-plane magnetization, {\em circularly} polarized 0.1~ps pulses with the fluence of 10 $\mu$J/cm$^2$ resulted in a transient Kerr effect \cite{Kimel:2004_PRL}. The determined spectral dependence of the Kerr effect was similar to that observed in a static magnetic field of 1~mT along the growth direction \cite{Kimel:2004_PRL}.

Extensive time-resolved studies with {\em linearly} polarized pumping pulses were carried out for (Ga,Mn)As \cite{Kojima:2003_PRB,Wang:2007_PRL} and (In,Mn)As \cite{Wang:2005_PRLb}, and revealed the presence of fast ($<1$~ps) and slow (~100 ps) processes. The fast component rapidly grew with pump power, saturated at high fluences ($>10$~mJ/cm$^2$), and indicated a quenching of ferromagnetism on a subpicosecond timescale, also when the holes were excited {\em via} intra valence band transitions \cite{Wang:2008_PRB}. A detailed quantitative theoretical study \cite{Cywinski:2007_PRB} demonstrated that the inverse Overhauser effect, that is dynamic demagnetization of Mn spins by $sp-d$ spin exchange with  photocarriers, accounted for the fast process, whose timescale was determined by carriers' energy relaxation.  In contrast, the slow component at low fluences ($\sim$10~$\mu$J/cm$^2$) corresponded to a recovery of ferromagnetic order or even enhancement of $T_{\text{C}}$ by an enlarged carrier density, the effect appearing on the timescale of spin-lattice relaxation and persisting up to photohole lifetime \cite{Wang:2007_PRL}. However, a substantial rise of lattice temperature dominated at high fluences leading to a complete destruction of ferromagnetism \cite{Wang:2005_PRLb,Cywinski:2007_PRB}.

\subsection{Coherent control of magnetization precession}

In a series of experiments on (Ga,Mn)As trains of  subpicosecond pulses of light with photon energies near the band gap \cite{Hashimoto:2008_PRL,Qi:2009_PRB,Nemec:2012_NP} or picosecond strain pulses \cite{Bombeck:2013_PRB,Scherbakov:2010_PRL} triggered oscillations of Kerr rotation as a function of time. These findings were assigned to a tilt of the magnetization vector $\vec{M}$, followed by coherent precession of $\vec{M}$ around its equilibrium orientation. This tilt was brought about by illumination-induced modification of the magnetic anisotropy field $\vec{H}_{\text{eff}}$ generated by a transient change of temperature \cite{Qi:2009_PRB} or strain \cite{Bombeck:2013_PRB,Scherbakov:2010_PRL}. Also evidences were found for the presence of non-thermal effects generated by light pulses, such as a transient torque produced by a burst of spin polarized photoelectrons \cite{Nemec:2012_NP} or an influence of photoholes on magnetic anisotropy \cite{Hashimoto:2008_PRL}. Altogether, studies of time-resolved Kerr rotation as well as of magnetization precession driven by an a.~c. magnetic field (FMR, Sec.~\ref{sec:anisotropy}) or electric current \cite{Fang:2011_NN} have demonstrated that the Landau-Lifshitz-Gilbert equation (recalled in Appendix) describes adequately magnetization dynamics in DFSs. Actually, an explicit solution of this equation was derived providing a frequency and damping of magnetization precession in a given magnetic field in terms of the Gilbert damping constant $\alpha_{\text{G}}$ and anisotropy fields $H_i$ specific to DFSs \cite{Qi:2009_PRB,Nemec:2013_NC}.

In another study \cite{Luo:2010_SCh} magnetization precession in (Ga,Mn)As was found to be overdamped but polarization dependent transient out-of-plane component of the magnetization was visible. A transient out-of-plane magnetization was also detected for a linearly-polarized pump, if the sample was exposed to an in-plane magnetic field prior to optical measurements.

It was shown experimentally \cite{Wang:2009_APL} and discussed theoretically \cite{Kapetanakis:2009_PRL} that excitations with near ultraviolet photons lead to coherent magnetization rotation in (Ga,Mn)As driven by photocarrier coherences and nonthermal populations excited in the $\langle 111\rangle$ equivalent directions of the Brillouin zone.   A subsequent theoretical work proposed a protocol for all-optical switching between four metastable magnetic states in DFSs \cite{Kapetanakis:2011_APL}.

\section{Spin injection}
\label{sec:injection}

Ferromagnetic semiconductors can be used as an epitaxially integrated spin-polarized carrier emitter into nonmagnetic structures working without or in a weak external magnetic field. Electrical spin injection from (Ga,Mn)As to nonmagnetic GaAs has been shown to be possible in a device structure integrated with a GaAs-based nonmagnetic light-emitting diode (LED) as a detector of spin-polarized holes \cite{Ohno:1999_N,Young:2002_APL} or electrons in Esaki diodes \cite{Kohda:2001_JJAP,Johnston-Halperin:2002_PRB,Dorpe:2004_APL,Dorpe:2005_PRB,Kohda:2006_APL}. By measuring circular polarization of electroluminescence, one can determine the spin polarization of injected carriers from (Ga,Mn)As. Because of carrier confinement in the LED emission region, the heavy hole subband is usually relevant in the radiative recombination process. Hence, according to corresponding selection rules, this method allows to detect carriers with spins polarized along the growth direction, which give rise to circularly polarized vertical (surface) emission \cite{Oestreich:1999_N,Jonker:2000_PRB,Fiederling:2003_APL}. Since in the structures studied so-far (Ga,Mn)As easy axis was in-plane, an out-of-plane magnetic field was applied to either orient Mn magnetization along the growth direction  \cite{Kohda:2001_JJAP,Johnston-Halperin:2002_PRB,Young:2002_APL,Kohda:2006_APL} or---in an oblique magnetic field configuration \cite{Dorpe:2004_APL,Dorpe:2005_PRB}---to generate additionally a spin component along the growth direction by spin precession. In these experiments emission in $\sigma^+$ polarization prevails demonstrating antiferromagnetic coupling between holes and Mn spins in (Ga,Mn)As.

By employing an Esaki diode as spin-injector in the magnetic field tilted 45$^o$ out-of-plane, electroluminescence circular polarization $P_{\text{EL}}$ reached the saturation magnitude of 21\% for Ga$_{0.92}$Mn$_{0.08}$As with $T_{\text{C}} = 120$~K \cite{Dorpe:2004_APL}. For this magnetizing field direction {\em and} the selection rules specified above, the determine value of $P_{\text{EL}}$ leads to spin current polarization injected from (Ga,Mn)As, $\Pi_{\text{inj}} = 40$\% at 4.6~K, where the experimentally determined depolarization factor $T_{\text{s}}/\tau =0.74$ \cite{Dorpe:2004_APL} is taken into account.\footnote{The selection rules assumed here imply $\Pi_{\text{inj}}$ twice smaller than that quoted originally.} A 6\% anisotropy in $P_{\text{EL}}$ was observed by rotating magnetization projection between $[110]$ and $[\bar{1}10]$ \cite{Dorpe:2005_PRB}.

\begin{figure}
\includegraphics[width =3.2in]{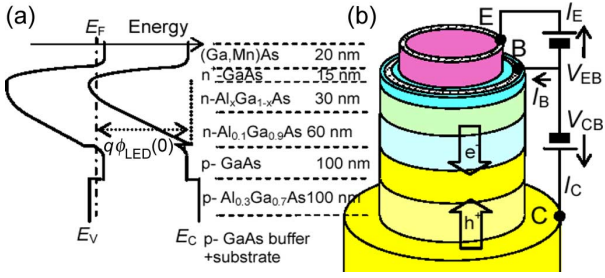}
\caption[]{(Color online) Spin injection from (Ga,Mn)As. (a) Schematic band diagram (a) of the three terminal device (b) allowing to bias independently the Esaki and light emitting diode ($V_{\text{EB}}$ and $V_{\text{CB}}$, respectively);(c) circular polarization $P_{\text{EL}}$ of light emitted along the growth and magnetization direction {\em vs.} $V_{\text{EB}}$ at various $V_{\text{CB}}$ (adapted from \onlinecite{Kohda:2006_APL}). (d) Spin current polarization $\Pi$ for spin injection to n-GaAs ($V < 0$) or spin extraction from n-GaAs ($V > 0$) $via$ contacts of the Esaki diode in non-local magnetotransport measurements (adapted from \onlinecite{Ciorga:2009_PRB}).}
\label{fig:spin_injection}
\end{figure}

By the use of a three terminal device structures to control bias voltages of an Esaki diode and a LED (spin-detector) independently, as shown in Fig.~\ref{fig:spin_injection}, the efficiency of the electron spin injection {\em via} band-to-band Zener tunneling from p-type Ga$_{0.943}$Mn$_{0.057}$As to n-type GaAs and then to LED was measured as a function of bias voltage. Emission in $\sigma^+$ polarization prevailed, confirming antiferromagnetic coupling between holes and Mn spins in (Ga,Mn)As. The values of $P_{\text{EL}}$ up to 32.4\% were attained for a (Ga,Mn)As emitter with $T_{\text{C}} = 70$~K \cite{Kohda:2006_APL}. Since in this case the magnetic field is along the growth direction and $T_{\text{s}}/\tau =0.64$ \cite{Kohda:2006_APL}, one obtains $\Pi_{\text{inj}} = 47\pm 1$\% at 10~K, where a 1\% correction for a non-zero $P_{\text{EL}}$ without emitter current is taken into account.

Electrical injection and detection of spin-polarized electrons were demonstrated in a single wafer all semiconductor lateral structure, incorporating  Ga$_{0.95}$Mn$_{0.05}$As/n$^+$-GaAs Esaki diodes acting as both spin injecting (or extracting) and spin detecting contacts to n-GaAs \cite{Ciorga:2009_PRB}. Prior to processing, $T_{\text{C}}$ of (Ga,Mn)As was 65~K. Spin precession and the spin-valve effect were observed in the nonlocal signal. Figure~\ref{fig:spin_injection}(d) shows $\Pi_{\text{inj}}$ and $\Pi_{\text{ext}}$ for the reverse and forward bias $V_{\text{EB}}$, respectively, determined under the assumption that $\Pi_{\text{inj(ext)}}$ is equal to the spin detection efficiency, which is strictly valid at $V_{\text{EB}}\rightarrow 0$. As seen, $\Pi_{\text{inj(ext)}} = 51\pm 2$\% at 4.2~K in this case.

Further evidences for spin injection from (Ga,Mn)As were provided by studies of Andreev
reflection \cite{Braden:2003_PRL,Panguluri:2005_PRB,Piano:2011_PRB}, the spin-Seebeck effect \cite{Jaworski:2010_NM}, and  spin pumping under FMR conditions \cite{Chen:2013_NC}. Andreev reflection was also detected in the case of (In,Mn)As \cite{Panguluri:2004_APL,Geresdi:2008_PRB}.

Theoretically expected values of spin current polarization are presented in Sec.~\ref{sec:structures_theory}.

\section{Spintronic magnetoresistance structures}
\label{sec:magnetoresistance_structures}
\subsection{Anisotropic magnetoresistance and Hall effects}
Owing to the strong spin-orbit interaction and typically lower carrier densities comparing to ferromagnetic metals, DFSs show sizable magnitudes of anisotropic magnetoresistance (AMR), planar and anomalous Hall effects as well as of related thermomagnetic \cite{Pu:2006_PRL,Pu:2008_PRL,Jaworski:2011_PRL} and magnetooptical phenomena in the subbandgap spectral region \cite{Acbas:2009_PRL}. Magnetization-dependent transport effects have been playing a crucial role in determining magnetization magnitude and orientation in variety of DFSs, as discussed in Secs.~\ref{sec:anisotropy} and \ref{sec:electric_field}.  Quantitative theory aiming at evaluation of conductivity tensor components $\sigma_{i,j}(\vec{M})$ in (Ga,Mn)As-type DFSs, developed in the lowest order in disorder for films of (Ga,Mn)As and related systems, was already reviewed {\em vis-\`a vis} results of extensive experimental studies \cite{Jungwirth:2006_RMP,Jungwirth:2008_B,Nagaosa:2010_RMP}. An open and interesting question is how quantum localization and confinement will affect magnitudes of magnetization-dependent charge and heat transport in these ferromagnets. A breakdown of the proportionality between magnetization and the anomalous Hall effect found in thin and high quality (Ga,Mn)As films at low temperatures \cite{Chiba:2010_PRL} is just one example showing that considerable further effort will be devoted towards understanding of transport phenomena in DFSs.

\subsection{Colossal magnetoresistance}
\label{sec:colossal}
A direct manifestation of interplay between magnetism and localization in magnetic semiconductors, as well as in DMSs and DFSs \cite{Dietl:2008_JPSJ}, are colossal magnetoresistance (CMR) phenomena and a related effect of critical scattering \cite{Novak:2008_PRL}. A peculiarity of CMR in DFSs is its strong dependence on the orientation of the magnetic field in respect to crystallographic axes \cite{Gareev:2010_APL,Katsumoto:1998_pssb}. In general terms, magnetization rotation $\delta \vec{M}$ results in a shift of the Fermi level, related to a change of anisotropy energy according to $ \Delta\epsilon_{\text{F}}(\delta\vec{M}) = \text{d}{\Delta\cal{F}}_{\text{cr}}(\delta\vec{M})/\text{d}p$. As $\epsilon_{\text{F}}$ controls the critical hole concentration $p_{\text{c}}$ corresponding to the metal-insulator transition (see, Secs.~\ref{sec:doping} and \ref{sec:MIT}),  colossal effects are seen in transport \cite{Gareev:2010_APL,Katsumoto:1998_pssb} and tunneling \cite{Pappert:2006_PRL} for samples with hole densities close to $p_{\text{c}}$. Importantly,  the influence of quantum localization persists well beyond the immediate vicinity of $p_{\text{c}}$.

\subsection{Coulomb blockade}
\label{sec:SET}

One of signs indicating that quantum-localization effects persist up to $p \gg p_{\text{c}}$ are signatures of the Coulomb blockade found in nanoconstrictions of (Ga,Mn)As \cite{Wunderlich:2006_PRL,Schlapps:2009_PRB}, pointing to substantial nano-scale fluctuations in the hole density. Interestingly, in these experiments conductance oscillations were not only generated by sweeping the gate voltage but also by changing the direction of magnetization. The latter results from the dependence of the Fermi energy $\epsilon_{\text{F}}$ on $\delta\vec{M}$, as discussed in the previous subsection. This dependence leads to: (i) a charge redistribution within the nanoconstriction, as $\Delta\epsilon_{\text{F}}(\delta\vec{M})$ depends on $p$ that show spatial fluctuations \cite{Wunderlich:2006_PRL}; (ii) changes in the localization and fluctuation landscape, as $\epsilon_{\text{F}}$ controls $p_{\text{c}}$. At sufficiently small values of $p-p_{\text{c}}(\vec{M})$, astonishingly large magnitudes of AMR were found in various nanostructures of (Ga,Mn)As at low temperatures \cite{Giddings:2005_PRL}.

\subsection{Giant and tunneling magnetoresistance devices}
\label{sec:TMR}
Like in their metal counterpart, in trilayer structures of (Ga,Mn)As the magnitude of vertical resistance increases if magnetization in the two ferromagnmetic electrodes assumes an anti-parallel alignment, either if non-magnetic central layer is of conductive p-GaAs \cite{Chung:2010_PRB} or forms a tunneling barrier, the case of AlAs \cite{Tanaka:2001_PRL,Chun:2002_PRB}, GaAs \cite{Chiba:2004_PE}, ZnSe \cite{Saito:2005_PRL} or paramagnetic (Al,Mn)As \cite{Ohya:2009_APL}. In the latter case, the magnitude of tunneling magnetoresistance (TMR), $(R_{\uparrow\downarrow} -R_{\uparrow\uparrow})/R_{\uparrow\uparrow}$, attained  175\% at 2.6 K for the barrier thickness $d = 4$~nm.

\begin{figure}
\includegraphics[width=3.1in]{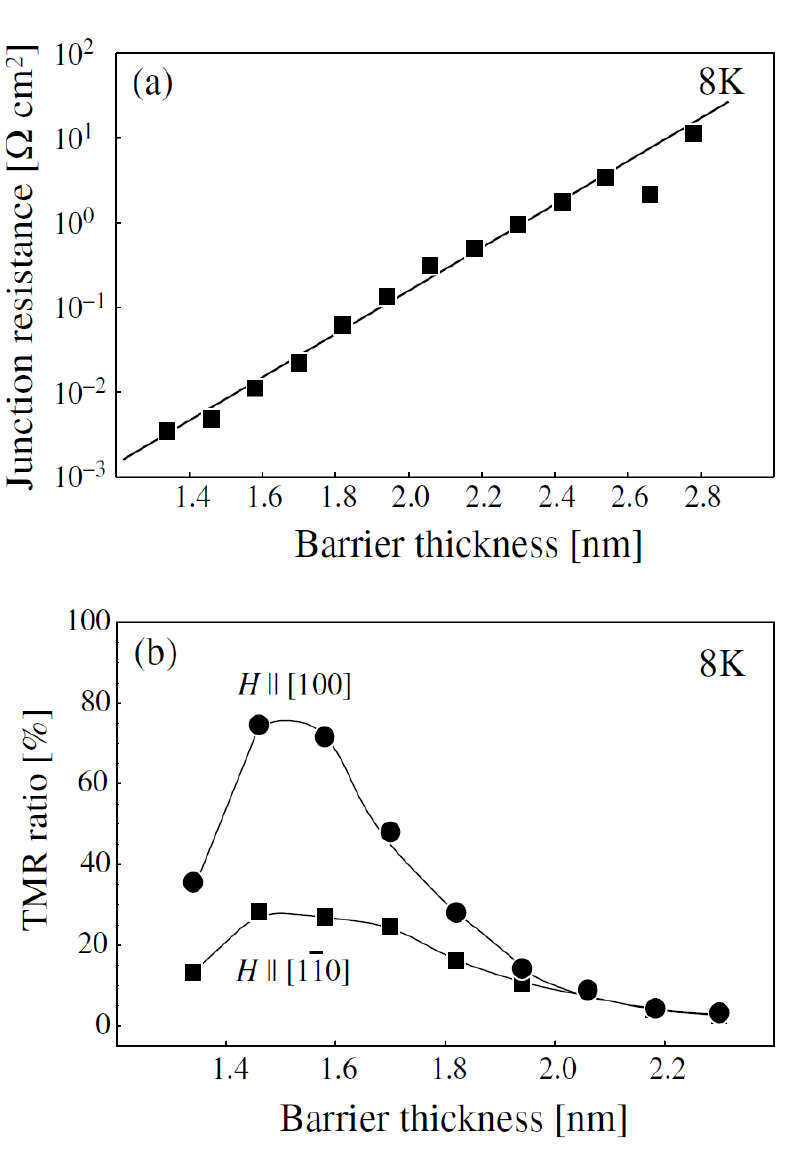}
\caption[]{Junction resistance (a) and the magnitude of TMR measured for two in-plane directions of the magnetic field (b) at 8~K as a function of the barrier width for the MTJ structure show in the inset (b). Adapted from \onlinecite{Tanaka:2001_PRL}.}
\label{fig:GaMnAs_TMR_Tanaka}
\end{figure}

Do magnetic tunnel junctions (MTJs) of DFSs exhibit specific features?  As shown in Fig.~\ref{fig:GaMnAs_TMR_Tanaka}, the magnitude of TMR,  {\em grows} (up to 76\% at 8~K) when the width of the AlAs barrier diminishes down to 1.5~nm \cite{Tanaka:2001_PRL}. At the same time, the values of TMR are seen to depend on the {\em direction} (in respect to crystallographic axes) of the magnetic field  employed to reverse sequentially magnetization at a coercive force of particular ferromagnetic electrodes. In this family of phenomena, known as tunneling anisotropic magnetoresistance (TAMR), particularly spectacular is the case of the junction with one nonmagnetic electrode, {\em e.g.,} (Ga,Mn)As/AlO$_x$/Au \cite{Gould:2004_PRL}, in which the vertical resistance decreases (by 3\%) when rotating magnetization from an easy to a hard in-plane direction. Actually, a linear dependence between the decrease of the MTJ resistance (up to 10\%) and the energy of magnetic anisotropy was found for various magnetization orientations in (Ga,Mn)As/ZnSe/(Ga,Mn)As \cite{Saito:2005_PRL}. Another characteristic feature is a fast decay of TMR magnitude with the bias voltage $V$ -- in (Ga,Mn)As/ GaAs/ (Ga,Mn)As at 4.7~K, the TMR magnitudes drops from 100\% to 20\% when $V$ increases to 0.1~V \cite{Chiba:2004_PE}. These findings are compared to theoretical expectations in Sec.~\ref{sec:structures_theory}.

There are a range of MTJ properties escaping up to now from a straightforward quantitative modeling. In particular, $I(V)$ characteristics of MnAs/AlAs/(Ga,Mn)As \cite{Chun:2002_PRB} and (Ga,Mn)As/GasAs/(Ga,Mn)As MTJs \cite{Pappert:2006_PRL,Ruster:2005_PRL} showed a Coulomb gap, a prominent manifestation of how important are correlation effects in quantum localization, as discussed in Secs.~\ref{sec:colossal} and \ref{sec:MIT}. Presumably, these effects, together with temperature dependent magnetization of Mn spins residing in a depleted interfacial layer, accounted for a threefold increase,  up to 290\%, of TMR between 4.7 and 0.3 K in an (Ga,Mn)As/GasAs/(Ga,Mn)As MTJ \cite{Chiba:2004_PE}. A question also arises to what extent the Coulomb gap affected the magnitude and magnetic anisotropy of tunneling thermopower in a GaAs:Si/GaAs/(Ga,Mn)As MTJ \cite{Naydenova:2011_PRL}.

\subsection{Double barrier structures}
\label{sec:RTD}
\begin{figure}[bt]
\begin{center}
\includegraphics[width=3.3in]{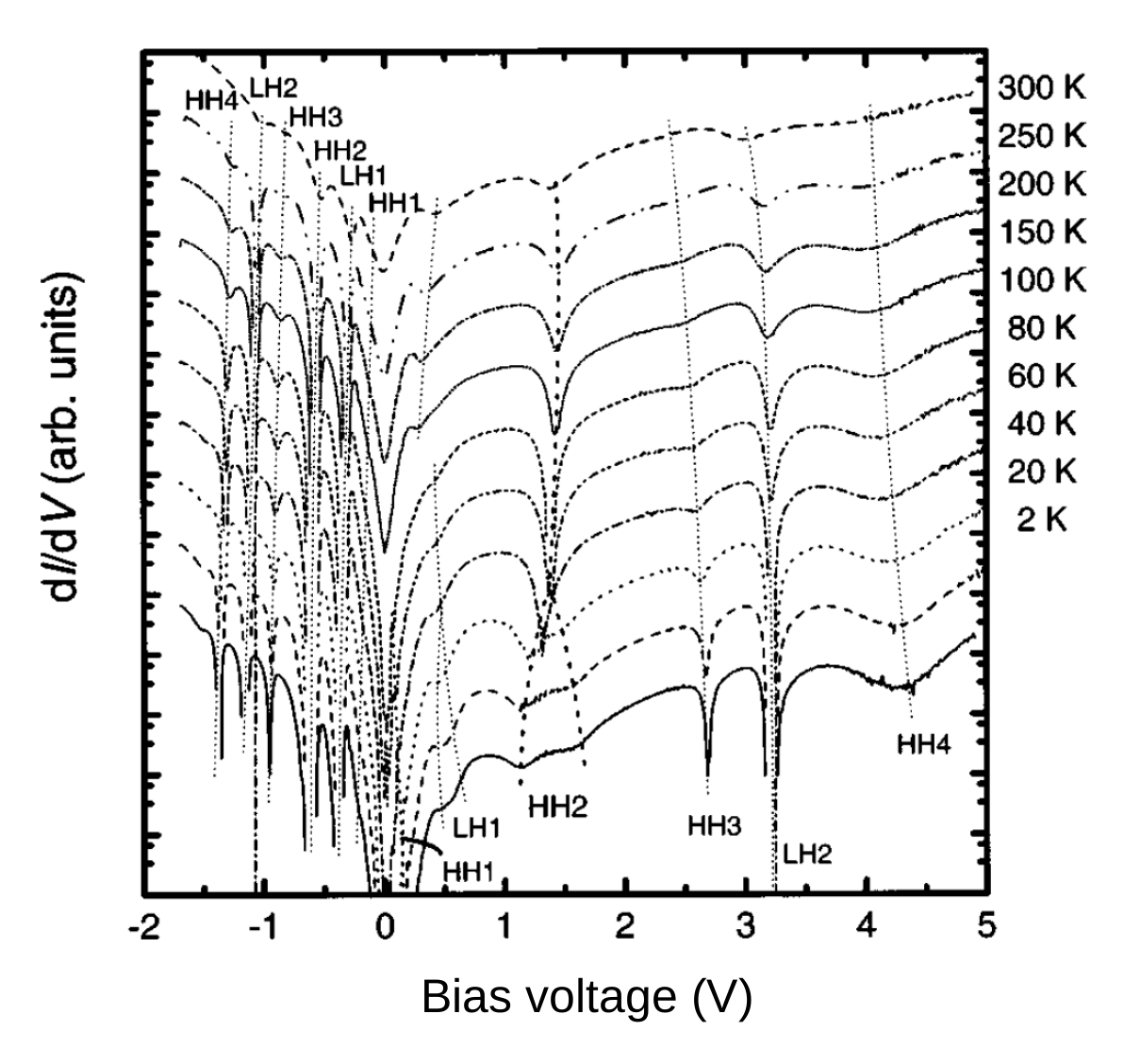}
\caption[]{Dynamic conductance d$I$/d$V$ at various temperatures for a resonant tunneling diode containing a GaAs quantum well and a (Ga,Mn)As electrode. From \onlinecite{Ohno:1998_APL}.}
 \label{fig:GaMnAs_RTD_Ohno}
\end{center}
\end{figure}

A series of works \cite{Ohno:1998_APL,Elsen:2007_PRL,Tran:2009_APL,Muneta:2012_APL} were devoted to tunneling phenomena in double barrier MBE-grown (Ga,Mn)As/AlAs/GaAs/AlAs/p-GaAs:Be structures deposited onto p$^+$-GaAs:Be substrates. To minimize effects of inter-diffusion, few nm-thick GaAs separators were additionally inserted between AlAs barriers and the p-type electrodes. Except for the top (Ga,Mn)As electrode, all layers were deposited at high temperatures ($T_{\text{g}} \gtrsim 600^o$C).  As shown in Fig.~\ref{fig:GaMnAs_RTD_Ohno}, sharp peaks in the dynamic conductance d$I$/d$V$ as a function of bias voltage $V$ were observed \cite{Ohno:1998_APL}. These and related results \cite{Elsen:2007_PRL,Tran:2009_APL,Muneta:2012_APL} led to a number of conclusions. In particular,  the appearance of resonances for both polarities of bias voltage and the magnitudes of their relative distances demonstrated the presence of resonant tunneling $via$ quantized hole subbands in the GaAs quantum well  \cite{Ohno:1998_APL,Elsen:2007_PRL}. Second, the absence at positive bias of resonances corresponding to the hole ground state subbands in GaAs quantum well (HH1 and LH1) indicated, in accord with the TMR results discussed in the previous subsection (Sec.~\ref{sec:TMR}), that  the Fermi level of (Ga,Mn)As resides about 0.1~eV above the valence band top of GaAs \cite{Elsen:2007_PRL}. Finally, up to 5 times larger distances between resonances  if holes were injected from (Ga,Mn)As (positive bias) than in the case when (Ga,Mn)As was a collecting electrode (negative bias) pointed to an asymmetry in the structure layout \cite{Ohno:1998_APL,Elsen:2007_PRL,Muneta:2012_APL}. This asymmetry was linked to a much lower value of the hole concentration ($p \simeq 10^{18}$~cm$^{-3}$) and, thus, longer depletion length in GaAs:Be  comparing to (Ga,Mn)As, leading to a rather different effective barrier width on the collector side for the two polarities  \cite{Elsen:2007_PRL}, as shown in Fig.~\ref{fig:GaMnAs_RTD_Ohno}.

In these devices three magnetic signatures were observed. First, a spontaneous and temperature dependent splitting of two peaks was revealed when holes were injected from (Ga,Mn)As, the effect visible in Fig.~\ref{fig:GaMnAs_RTD_Ohno} \cite{Ohno:1998_APL}. The temperature dependence of the splitting showed a Brillouin-type behavior with $T_{\text{C}} \simeq 70$~K. Second, for the same bias, the magnitudes of resonance peaks were found dependent on magnetization orientation -- they were reduced by about 10\% when magnetization was turned from the easy axis [100] to the hard [001] direction \cite{Elsen:2007_PRL}. The effect was examined quantitatively \cite{Elsen:2007_PRL} within the $p$-$d$ Zener model exposed in Sec.~\ref{sec:theory} and the outcome is shown in Sec.~\ref{sec:structures_theory}. Third, it was demonstrated that aforementioned  magnetization rotation resulted in a shift of resonance positions for negative bias, the effect assigned to a decrease of the work function when magnetization was moved away from the easy direction \cite{Tran:2009_APL}. Magnetic effects, were also detected by in structures containing on its top an additional AlAs barrier and a (Ga,Mn)As layer \cite{Muneta:2012_APL}. Here, a change of series resistance associated with parallel and antiparallel arrangement of (Ga,Mn)As magnetizations led to a shift of resonance positions. These and related experiments demonstrate, therefore, how to gate electric current by manipulating with magnetization.

Double barrier structures deposited at low temperatures ($T_{\text{g}} \simeq 230^o$C), in which the bottom GaAs:Be layer was replaced by (Ga,Mn)As, showed significantly  different properties \cite{Mattana:2003_PRL}. In particular, the absence of resonances and the magnitude of TMR (about 40\%) demonstrated that rather than resonant, sequential tunneling accounted for hole transport, the effect pointing out to a shortening of the phase coherence time below the dwell time in GaAs wells grown by LT-MBE \cite{Mattana:2003_PRL}. Even lower values of TMR (below 2\%) were found in similar structures containing (In,Ga)As wells \cite{Ohya:2005_APL}, indicating that spin relaxation time became shorter than the dwell time in this case.

In another type of investigated structures, the GaAs quantum well in the original design (Fig.~\ref{fig:GaMnAs_RTD_Ohno}) was replaced by a (Ga,Mn)As layer of various thicknesses up to 20~nm grown by LT-MBE \cite{Ohya:2007_PRB,Ohya:2010_PRL}. In these devices, a resonance in d$I$/d$V$ was observed at $V \simeq -0.1$~V, accompanied by one or two satellite features visible in d$^2I$/d$V^2$ at $V < 0$. A slight shift of the resultant oscillatory pattern was resolved on going from parallel to antiparallel magnetization orientation of the two (Ga,Mn)As layers, leading to a TMR-like behavior with a relative change of current for parallel and antiparallel magnetization orientations reaching 40\%. It was demonstrated that the position and the magnitude of these oscillations in d$I$/d$V$ and TMR {\em vs.} $V$ can be efficiently controlled by a third electrode biasing the (Ga,Mn)As quantum well \cite{Ohya:2010_APL}, a valuable step towards development of a spin transistor. A similar oscillatory behavior of d$^2I$/d$V^2$ at $V < 0$ was also detected in simpler structures, in which the top barrier and the (Ga,Mn)As electrode were replaced by an Au film deposited directly onto the lower (Ga,Mn)As layer, resulting in the layout Au/(Ga,Mn)As/AlAs/p-GaAs:Be, where the (Ga,Mn)As layer exhibited $T_{\text{C}}$ up to 154~K \cite{Ohya:2011_NP}. Corresponding results were also obtained for structures in which (Ga,Mn)As was replaced by (Ga,In,Mn)As ($T_{\text{C}}$ up to 135~K) or (In,Mn)As ($T_{\text{C}} \simeq 47$~K) and p-GaAs:Be by p-(Ga,In)As:Be, employing p-InP substrates in this case \cite{Ohya:2012_PRB}.

\begin{figure}[bt]
\begin{center}
\includegraphics[width=3.5in]{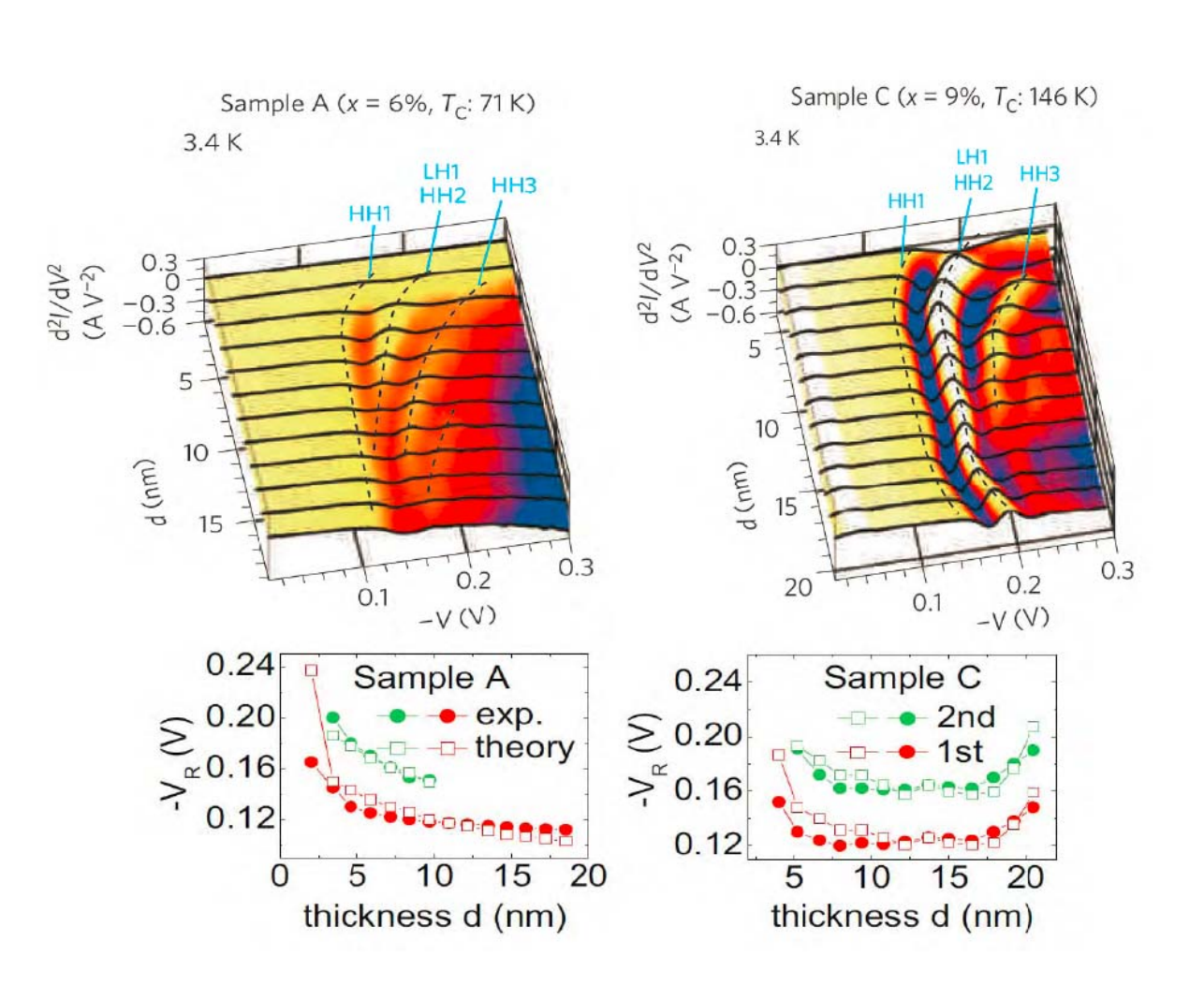}
\caption[]{Current-voltage characteristics in tunnel structures containing Ga$_{1-x}$Mn$_x$ wells. Upper panel: experimental data on d$^2I$/d$V^2$ for two tunneling structures containing Ga$_{1-x}$Mn$_x$ wells of different thickness $d$ obtained by consecutive etching, and for which values of the area resistance product $AR(d)$ were available. Dashed lines present positions $V_{\text{R}}$ of the features in d$^2I$/d$V^2$, calculated with four adjustable parameters, assuming resonant tunneling $via$ quantized hole states in (Ga,Mn)As wells (from \onlinecite{Ohya:2011_NP,Ohya:2011_arXiv}). Lower panel: calculated positions of the features from $V_{\text{R}}(d) = V_{\text{R}}(d) =V_{\text{R}}(d_{\text{m}})RA(d)/RA(d_{\text{m}})$, where $d_{\text{m}}$ is an intermediate thickness (from \onlinecite{Dietl:2011_arXiv}).}
 \label{fig:GaMnAs_RTD_Tanaka}
\end{center}
\end{figure}

These comprehensive investigations showed consistently that the oscillatory pattern in d$^2I$/d$V^2$: (i) appeared for $V < 0$; (ii) tended to spread and move towards higher voltages $|V|$ when the thickness of the bottom (Ga,Mn)As layer decreased, as shown in Fig.~\ref{fig:GaMnAs_RTD_Tanaka};  (iii) did not reveal any temperature-dependent splitting of quantized hole states.

In order to interpret these findings it was suggested in these works that: (i) the oscillations in d$^2I$/d$V^2$ witness resonant tunneling $via$ quantized subbands in the relevant (Ga,In,Mn)As quantum well embedded by AlAs-rich barriers (or AlAs and the Schottky barrier underneath the Au film); (ii) particular features correspond to subsequent hole subbands starting from the ground state HH1 level, the assumption allowing to describe (with four adjustable parameters) the position and evolution of the features with the layer thickness; (iii) since a negative voltage has to be applied to reach the ground state subband in (Ga,In,Mn)As, the hole Fermi level is pinned by an impurity band located about 50~meV above the valence band top, implying that up to $10^{21}$~cm$^{-3}$ holes reside in a narrow band separated from the valence band in both (Ga,Mn)As and (In,Mn)As; (iv) the valence band states are entirely imminent to the presence of  Mn ions, which results in the lack of exchange splitting and high coherency of quantized states even for a 20~nm thick (Ga,Mn)As quantum well.

However, the above model was found questionable  \cite{Dietl:2011_arXiv}, particularly taking into account previous results on tunneling in (Ga,Mn)As \cite{Richardella:2010_S} as well as in double well structures involving (Ga,Mn)As layers \cite{Ohno:1998_APL,Elsen:2007_PRL,Tran:2009_APL,Mattana:2003_PRL}. It was suggested  \cite{Dietl:2011_arXiv} that the findings can be interpreted by assigning the features in d$^2I$/d$V^2$ at $T \ll T_{\text{C}}$ to sequential hole tunneling transitions from quantized hole subbands in the accumulation layer of GaAs:Be to continuum of states determined by quenched disorder in (Ga,Mn)As, followed by transitions to the top electrode. The features originating from quantum states in GaAs:Be can be resolved in this case since competing resonances associated with the quantum states in the well are washout by disorder in (Ga,Mn)As. They appear at $V_{\text{R}} <0$, where $|eV_{\text{R}}|$ scales with  the Mn concentration dependent valence band offset between (Ga,Mn)As and GaAs.  This new interpretation, contradicting the presence of an impurity band, is consistent with: (i) the failure to observe the genuine impurity band directly by tunneling spectroscopy; (ii) the absence of resonant tunneling when the well consisted of disordered GaAs grown by LT-MBE \cite{Mattana:2003_PRL}; (iii) the presence of the features only for $V<0$, in contrast to the case of GaAs well grown by HT-MBE, where resonances appeared for both bias polarities, as expected for resonant tunneling \cite{Ohno:1998_APL,Elsen:2007_PRL}; (iv) the non-occurrence of corresponding features when GaAs:Be was replaced by (Ga,Mn)As \cite{Mattana:2003_PRL}; (v) the evolution of the feature positions with the device resistance (showing correlation with the thickness of (Ga,Mn)As layers, see Fig.~\ref{fig:GaMnAs_RTD_Tanaka}); (vi) the presence of TMR-like behavior but at the same time the lack of spin splitting of electronic states giving rise to the tunneling features (the splitting of the features is at least two orders of magnitude smaller than exchange splitting of states in any known uniform magnetic semiconductors with a corresponding magnitude of magnetization); (vii) the similarity of the results for wells of ferromagnetic (Ga,Mn)As grown on AlAs/GaAs:Be, and (In$_{0.53}$Ga$_{0.47}$,Mn)As and (In,Mn)As grown on AlAs/In$_{0.53}$Ga$_{0.47}$As:Be.

In related structures (Ga,Mn)As/ AlAs/(Ga,Mn)As/ (Al,Ga)As/ GaAs:Be, where the AlAs and (Al,Ga)As barriers were 1.5 and 100~nm thick, respectively, negative dynamic resistivity features with various degree of sharpness were seen in the $I(V)$ dependence in a number of tested devices \cite{Likovich:2009_PRB}. These features underwent a shift to higher values of bias $V$ for antiparallel magnetization orientations, which resulted, in the most prominent case, in a TMR-like signal as large as 30\%.

\subsection{Read-write devices}
\begin{figure*}[tb]
\includegraphics[width=6.1in]{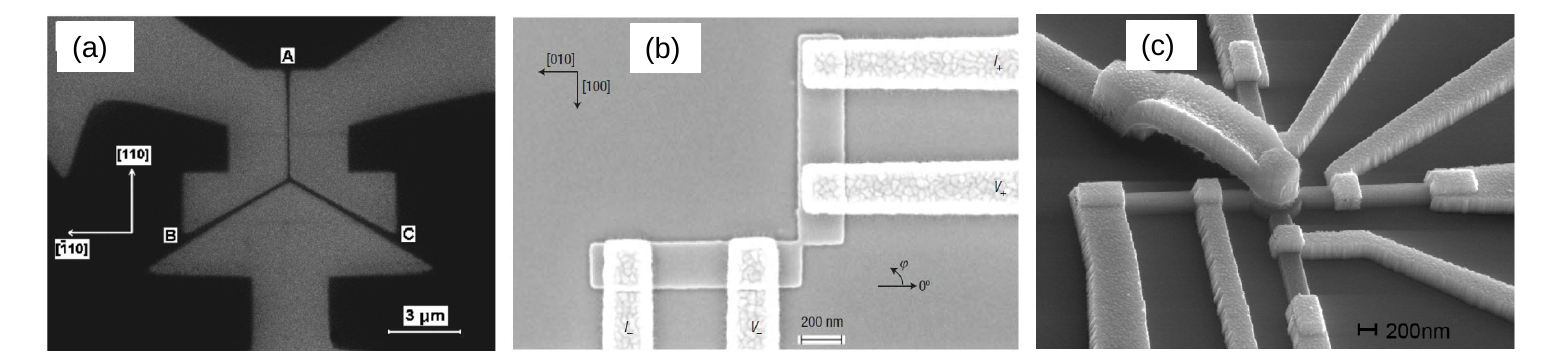}\vspace{-0mm}
\caption[]{Memory cells working at 4.2~K involving three (a), two (b), and four (c) 200~nm wide and few $\mu$m long (Ga,Mn)As nanobars forming a junction and contacted to current leads (from \onlinecite{Figielski:2007_APL}, \onlinecite{Pappert:2007_NP}, and \onlinecite{Mark:2011_PRL}, respectively). Due to strain relaxation magnetization is aligned along the nanobar long axes but its orientation in particular bars can be preselected by applying and removing an external magnetic field of an appropriate in-plane direction [bit writing in (a,b) and memory cell initiation in (c)]. Owing to the AMR effect associated with a domain wall in the junction [enhanced by carrier depletion in (b)], the value of two terminal resistance can tell relative magnetization directions in particular bars [bit reading in (a,b)]. In (c), magnetization direction along one of two cubic axes in the central disk (of diameter 650~nm) can be preselected by a current pulse along the pair of wires that are spin-polarized in the required direction [bit writing in (c)]. Tunneling resistance of the Au/AlO$_x$/(Ga,Mn)As MTJ deposited over the central disk depends on magnetization orientation in the disk [bit reading in (c)].}
\label{fig:GaMnAs_Figielski}
\end{figure*}

Figure \ref{fig:GaMnAs_Figielski} highlights layout and operation principle of (Ga,Mn)As-based magnetic memory cells, in which AMR-related phenomena allowed for bit reading, whereas either an external magnetic field or spin-polarized electric current served to bit writing.

\section{Interlayer coupling, ferromagnetic proximity effect, and exchange bias}
\label{sec:interlayer}

\subsection{Interlayer coupling}
Low-temperature magnetotransport studies of (Ga,Mn)As/(Al$_y$Ga$_{1-y}$As/ (Ga,Mn)As trilayer structures reveled ferromagnetic coupling between (Ga,Mn)As layers, whose strength decayed with temperature and Al content, $0.14 \leq y \leq 1$, in the 2.8~nm thick  Al$_y$Ga$_{1-y}$As spacer \cite{Chiba:2000_APL}. A ferromagnetic interlayer interaction was also found by neutron investigations of (Ga,Mn)As/GaAs superlattices \cite{Kepa:2001_PRB,Sadowski:2002_TSF,Chung:2010_PRB} for the whole explored range of GaAs thicknesses, $0.7 \leq d \leq 7$~nm \cite{Kepa:2001_PRB,Sadowski:2002_TSF,Chung:2010_PRB}. However, for GaAs:Be spacers with hole density of $1.2\times 10^{20}$~cm$^{-3}$, the coupling was still ferromagnetic for $d = 1.2$ and 2.3~nm but became antiferromagnetic when increasing $d$ to 3.5 and 7.1~nm \cite{Chung:2010_PRB}. Since ferromagnetism is spatially inhomogeneous in (Ga,Mn)As, a long range dipole-dipole coupling can account for this observation \cite{Kakazei:2005_JAP}.

In the case of MnAs/p-GaAs/(Ga,Mn)As, a ferromagnetic coupling was found, whose strength monotonically decayed with the thickness of the p-GaAs layer in the studied range $1 \leq d \leq 5$~nm \cite{Wilson:2010_PRB}.

Theoretical modeling of interlayer coupling is discussed in Sec.~\ref{sec:inter_layer_theory}.

\subsection{Ferromagnetic proximity effect}

The Fe and Mn L$_{2;3}$ XMCD spectra recorded at room temperature for Fe/(Ga,Mn)As heterostructures demonstrated the presence of Mn spin ordering antiparallel to Fe spins extending 2~nm \cite{Maccherozzi:2008_PRL} or 0.7~nm \cite{Olejnik:2010_PRB} into (Ga,Mn)As. The uncovered character of the ferromagnetic proximity effect was reproduced by DFT computations \cite{Maccherozzi:2008_PRL}. Interestingly, if the thickness of (Ga,Mn)As  was reduced down to 5~nm, the ferromagnetic proximity effect allowed to shift up by 35~K the temperature range in which both spontaneous magnetization and spin injection to n-GaAs through a Fe/(Ga,Mn)As/n-GaAs Esaki diode could be detected \cite{Song:2011_PRL}. These robust spin selective contacts made it possible to probe electrically the spin Hall effect in n-GaAs \cite{Ehlert:2012_PRB}.

\subsection{Exchange bias}
As mentioned above, the coupling of Mn ions in (Ga,Mn)As to an Fe overlayer is antiferromagnetic. Accordingly, {\em below} $T_{\text{C}}$ of (Ga,Mn)As,  its magnetic properties can be described in terms of exchange bias, leading to enlarged coercivity and a history dependent shift of the hysteresis loop center away from the zero magnetic field. Such phenomena were noted for MnAs/(Ga,Mn)As \cite{Zhu:2007_APL,Wilson:2010_PRB} and Fe/(Ga,Mn)As \cite{Olejnik:2010_PRB}, and interpreted by the exchange spring model \cite{Wilson:2010_PRB}. Magnetization processes related to exchange bias were found and examined for MnO/(Ga,Mn)As heterostructures, in which N\'eel and Curie temperatures were comparable \cite{Eid:2004_APL,Ge:2007_PRB}. Similarly, MnO and MnTe exchange-biased (Ge,Mn)Te \cite{Lim:2012_JAP}. Furthermore, nanocrystalline precipitates of ferromagnetic MnAs in (Ga,Mn)As \cite{Wang:2006_APL} and of antiferromagnetic MnTe in (Ge,Mn)Te \cite{Lechner:2010_APL} resulted in an enhancement of the coercivity field.

\section{Electronic states}
\label{sec:electronic_states}
\subsection{Vonsovsky's model and Mott-Hubbard localization}
Experimental results discussed in the subsequent subsections (Secs.\,VII.B.1-2) indicate that magnetic moments of Mn in DFSs are localized, not itinerant. According to the Vonsovsky model \cite{Vonsovsky:1946_ZETF}, the relevant electron states can then be divided into two categories  \cite{Dietl:1981_B}: (i) localized magnetic $d$-like levels described by the Anderson impurity model \cite{Anderson:1961_PR} or its derivatives \cite{Parmenter:1973_PRB}; (ii) effective-mass band states that can be treated within tight binding or $kp$ methods \cite{Kohn:1955_PR}, employed commonly for quantitative simulations of functionalities specific to semiconductors, their alloys and quantum structures. Importantly, pertinent properties of holes, for instance valence band lineups in semiconductor heterostructures, result from hybridization between anion $p$ orbitals and cation $d$ orbitals  \cite{Wei:1987_PRL}. In the case of open $d$ shells, this hybridization leads additionally to a strong $p$-$d$ exchange interaction \cite{Bhattacharjee:1983_PBC,Dietl:1981_B} accounting for outstanding spintronic properties of DFSs. There exists also an $s-d$ exchange interaction in DMSs but because of its relatively small magnitude, the corresponding $T_{\text{C}}$ values were found to be below 1~K \cite{Dietl:1997_PRB,Andrearczyk:2001_ICPS}.

In the next two sections (Secs.~\ref{sec:superexchange} and \ref{sec:Zener}), Vonsovskii's electronic structure is employed to describe spin-spin exchange interactions within the superexchange and $p$-$d$ Zener models. This is followed (Sec.~\ref{sec:comparison}) by discussing the applicability of this approach to a quantitative description of spintronic functionalities of DFSs. We note in passing that since implementations of density functional theories within local density approximations (LDA) cannot handle adequately the physics of the Vonsovsky model, particularly the Mott-Hubbard localization, other {\em ab initio} approaches are being developed for DFSs, for instance, incorporating into the LDA hybrid functionals \cite{Stroppa:2009_PRB} or the dynamic mean-field approximation \cite{DiMarco:2013_NC}.

\subsection{Mn localized magnetic moments}
\label{sec:magnetic moments}
\subsubsection{Magnetic resonances}
\label{sec:magnetic_resonance}
In the case of III-V and also III-VI compounds as well as group IV semiconductors, Mn ions introduce both spins and holes. According to electron paramagnetic studies in the impurity limit $ x \lesssim 10^{-3}$, the Land\'e factor of neutral Mn acceptors (Mn$^{3+}$) in GaAs:Mn is $g = 2.77$ \cite{Schneider:1987_PRL,Szczytko:1999_PRB}, the value consistent with a moderate binding energy  $E_{\text{I}} =110$~meV (see, Fig.~\ref{fig:III_V_EI}) and an antiferromagnetic character of $p$-$d$ exchange coupling between the hole spin $J = 3/2$ and the Mn$^{2+}$ center in a high spin $S = 5/2$ state \cite{Schneider:1987_PRL}. Spin resonance in GaP:Mn \cite{Kreissl:1996_PRB} as well as Mn-related optical spectra in GaN:Mn \cite{Wolos:2008_B,Bonanni:2011_PRB}, were successfully described in terms of the group theory for Ga-substitutional localized Mn$^{3+}$ centers corresponding to $S = 2$. Magnetization studies suggest that this spin state of Mn ions persists up to at least $x = 0.1$ \cite{Kunert:2012_APL}.

In contrast, in GaN:Mn samples containing compensating donor impurities, the character of hyperfine splitting and $g = 2.01\pm 0.05$ \cite{Bonanni:2011_PRB,Graf:2003_PRB,Wolos:2008_B} demonstrated the presence of Mn$^{2+}$ ions ($S=5/2, L =0$). Similar spectra were found on  increasing Mn concentrations in (Ga,Mn)As \cite{Szczytko:1999_PRB,Fedorych:2002_PRB} and (In,Mn)As \cite{Szczytko:2001_PRB}. In the case of arsenides, however, the presence of Mn$^{2+}$ spectra indicates  detaching of holes from individual negatively charged Mn$^{2+}$ acceptors at $x \gtrsim 0.001$ rather than compensation by donors.

For still higher Mn concentrations ($x\gtrsim 0.02$),  extensive ferromagnetic resonance studies, carried out for (In,Mn)As \cite{Liu:2005_APL}, (Ga,Mn)As \cite{Liu:2006_JPCM,Khazen:2008_PRB}, and (Ga,Mn)P \cite{Bihler:2007_PRB}, pointed to the Land\'e factor $g = 1.93 \pm 0.5$ at low temperature. A slight deviation from the value $g = 2.00$ expected for Mn$^{2+}$, suggests an admixture of orbital momentum, brought presumably about by spin polarized holes present in these DFSs below $T_{\text{C}}$. The value $S = 5/2$ was also evaluated from the neutron scattering length in (Ga,Mn)As \cite{Kepa:2001_PRB}. In contrast, according to extensive magnetization measurements \cite{Sawicki:2012_PRB,Kunert:2012_APL}, trivalent Mn$^{3+}$ configuration dominates up to at least $x=0.1$ in Ga$_{1-x}$Mn$_x$N. This finding is consistent with a large ionization energy of Mn acceptors in GaN (see, Fig.~\ref{fig:III_V_EI}), leading to strong localization of holes on individual Mn ions even at high Mn content $x$.

In the case of II-VI \cite{Furdyna:1988_B,Dietl:1994_B} and IV-VI DMSs \cite{Bauer:1992_SST}, Mn ions substitute divalent cations and assume Mn$^{2+}$ charge states characterized by a high spin and vanishing orbital momentum ($S=5/2, L =0$).
This spin state was confirmed by ferromagnetic resonance studies on ferroelectric and ferromagnetic (Ge,Mn)Te \cite{Dziawa:2008_B}.

\subsubsection{High energy spectroscopy}
It is worth recalling that ultraviolet and soft x-ray methods probe usually film regions adjacent to the surface, so that an adequate surface preparation is of paramount importance. With this reservation, we note that the picture presented above, namely that in all Mn-based DFSs but (Ga,Mn)N, Mn assumes single valent 2+, $S = 5/2$ configuration, was strongly supported by photoemission and x-ray spectroscopy. In particular, Mn ions in both (In,Mn)As \cite{Okabayashi:2002_PRB} and (Ga,Mn)As were found in a single valence state characterized by the Mn $d$ electron count $n_d = 5.3 \pm 0.1$ \cite{Okabayashi:1999_PRB}. A measurable enhancement over $n_d = 5$ can be interpreted as the presence of $p$-$d$ hybridization leading to a non-zero occupancy of the $d^6$ Mn level by quantum hopping from As valence states.

Similarly, XMCD studies at the  Mn $L$-edge corroborated the 2+ and $S=5/2$ configuration of Mn ions in ferromagnetic (In,Mn)As \cite{Chiu:2005_APL,Zhou:2012_APEX}, (Ga,Mn)As \cite{Wu:2005_PRB,Edmonds:2006_PRL}, and (Ga,Mn)P \cite{Stone:2006_APL}. Furthermore, a shift of orbital momentum from Mn to As with increasing $x$ was found in (Ga,Mn)As \cite{Wadley:2010_PRB}.

In the case of (Ga,Mn)N, XMCD data at the Mn $K$- \cite{Sarigiannidou:2006_PRB} and $L$-edge \cite{Freeman:2007_PRB} confirmed the 3+ and $S =2$ state of Mn in ferromagnetic (Ga,Mn)N. It was found, however, that surface donor defects turned adjacent Mn ions into divalent Mn$^{2+}$ states \cite{Freeman:2007_PRB}, visualized also by photoemission studies \cite{Hwang:2005_PRB}. Coexistence of Mn$^{2+}$ and Mn$^{3+}$ was also observed in x-ray absorption spectroscopy (XAS) \cite{Sonoda:2006_JPCM}. In contrast, in uncompensated samples of (Ga,Mn)N both x-ray absorption near-edge structure (XANES) \cite{Bonanni:2011_PRB} and x-ray emission spectroscopy (XES) \cite{Devillers:2012_SR} pointed to 3+ charge state of Mn in GaN.

\subsection{Anderson-Mott localization of carriers}
\label{sec:MIT}
As already mentioned in Sec.~\ref{sec:doping},  interplay between hole localization and hole-mediated ferromagnetism is arguably the most characteristic feature of DFSs. In particular, $p$-$d$ hybridization that accounts for exchange coupling between localized spins and itinerant holes,  shifts at the same time the metal-insulator transition (MIT) to higher hole concentrations.

It is worth noting that current {\em ab initio} methods designed to handle disorder, such as the coherent potential approximation, are {\em not} capturing the physics of the Anderson-Mott localization. There are preliminary quantum Monte Carlo approaches aiming at elaborating computational schemes that might provide quantitative information in the regime of quantum localization in many body interacting systems \cite{Fleury:2008_PRL}. Nevertheless, for the time being it is safe to argue that the current theory of the Anderson-Mott MIT does {\em not} offer quantitative predictions on: (i) the magnitude of the critical hole concentration $p_{\text{c}}$ corresponding to the MIT; (ii) the absolute value of conductivities $\sigma_{ij}$, and (iii) the nature of excitations at high energies $\omega \gtrsim 1/\tau$.  Empirically, some of these excitations exhibit single impurity characteristics, even on the metallic side of the MIT. This duality of behaviors is described phenomenologically within the so-called two-fluid model of electronic states \cite{Paalanen:1991_PB}, the approach exploited extensively to understand DMSs \cite{Dietl:2000_S,Dietl:2008_JPSJ}, and now acquiring some theoretical support \cite{Terletska:2011_PRL}.

However, the current theory does provide quantitative and experimentally testable information on the values of critical exponents as well as on the dependence of $\sigma_{ij}(p)$ on dimensionality, frequency, temperature, magnetic field, spin scattering, and spin splitting in the metallic regime $k_{\text{F}}\ell > 1$, where $\ell$ is the microscopic mean free path \cite{Altshuler:1985_B,Lee:1985_RMP,Belitz:1994_RMP,Dietl:2008_JPSJ}. The appearance of these specific dependences, known as quantum corrections to conductivity, heralds the failure of Drude-Boltzmann-like approaches in capturing the physics accounting for the magnitudes of $\sigma_{ij}(p)$. Importantly, a quantitative study of the quantum corrections can  provide information on the thermodynamic density of states (DOS) $\rho_{\text{F}} = \partial p/\partial \epsilon_{\text{F}}$, which does not show any critical behavior across the MIT and assumes a value specific to the relevant carrier band \cite{Altshuler:1985_B,Lee:1985_RMP,Belitz:1994_RMP}. According to the same theory, with an accuracy of typically better than 20\% (corresponding to a magnitude of the relevant Landau parameter of the Fermi liquid), the corresponding effective mass is equal to $m^*$ for low energy intraband charge excitations, provided by, {\em e.~.g.}, cyclotron resonance studies.

In the region $k_{\text{F}}\ell < 1$, corresponding usually to $p \lesssim p_{\text{c}}$, renormalization group equations \cite{Finkelstein:1990_SSR,Lee:1985_RMP,Belitz:1994_RMP} can serve to asses the evolution of relevant characteristics, such as localization radius, dielectric constant, and one-particle DOS, with $p_{\text{c}} - p$. This DOS shows a Coulomb gap at the Fermi level on the insulator side of the MIT, $p < p_{\text{c}}$, which evolves into a Coulomb anomaly at $p > p_{\text{c}}$. The theory shows that in the weakly localized regime, the localization length $\xi$ is much longer than the effective Bohr radius $a_{\text{B}}$ of a single acceptor, so that band characteristics are preserved at distances smaller than $\xi$. Furthermore, because of large screening by weakly localized carriers, there are few, if any, bound states associated with individual acceptors.

In the case of low dimensional systems (2D and 1D) a cross-over from the weakly to strongly localized regime, rather than the MIT, occurs at $k_{\text{F}}\ell \approx 1$. The absence of metallic regime means that, in principle, localization phenomena are relevant at any hole density.

\subsection{Where do holes reside in DFSs?}
As discussed in Sec.~\ref{sec:doping}, the Fermi level is pinned by the deep Mn acceptor impurity band in (Ga,Mn)N, where charge transport might proceed only {\em via} phonon-assisted hopping. A broadly disputed question \cite{Samarth:2012_NM,Wang:2013_PRB} then arises, where do reside delocalized or weakly localized holes mediating ferromagnetic coupling in (Ga,Mn)As and related systems?

\subsubsection{Photoemission}
\begin{figure}[bt]
\begin{center}
\includegraphics[width=3.7in]{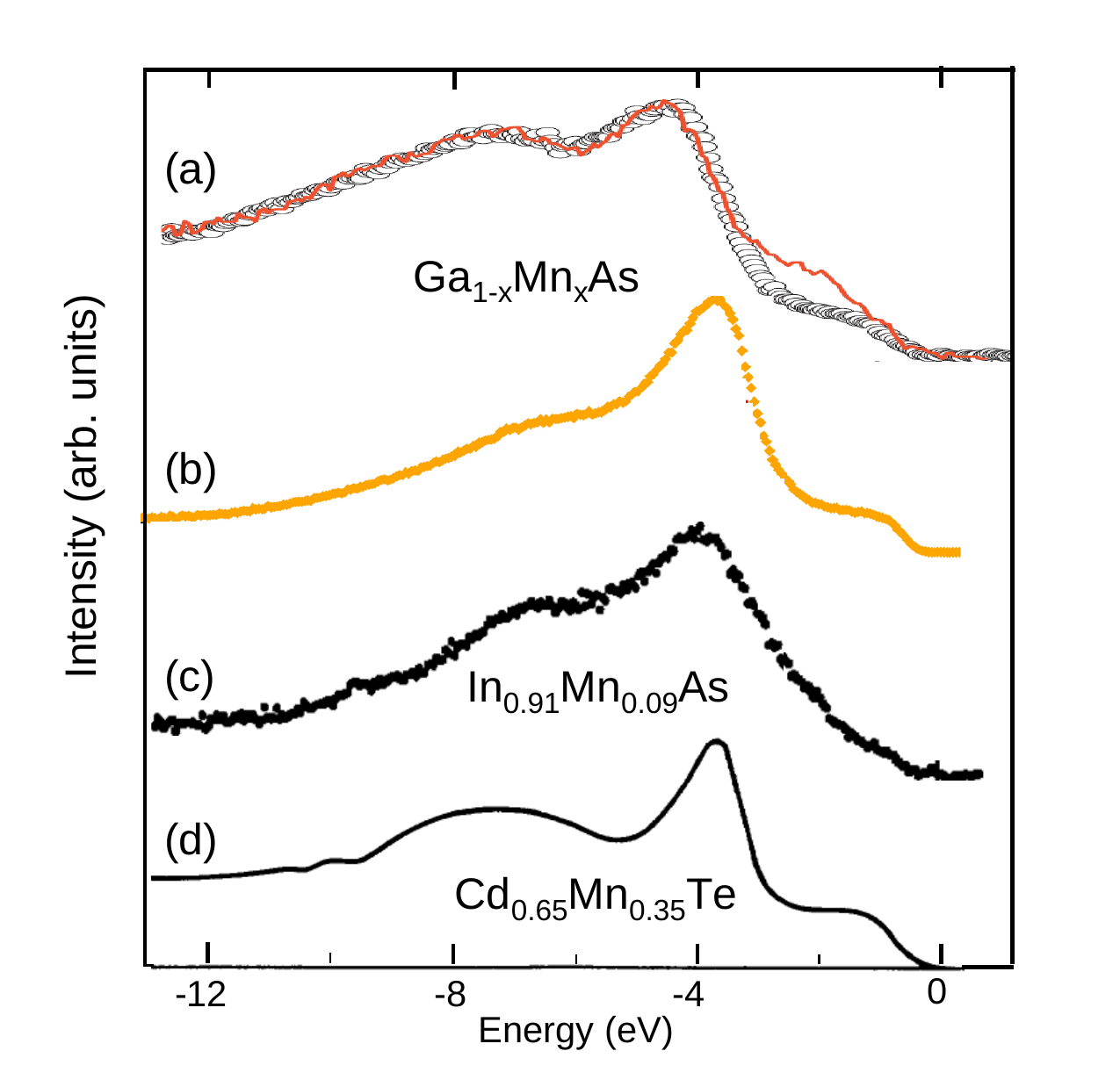}
\caption[]{(Color online) Partial DOS below the Fermi energy (taken as a reference) brought about by Mn 3$d$ states, as obtained from photoemission studies at room temperature on various DMSs and employing differing experimental methods: (a) as-grown Ga$_{0.931}$Mn$_{0.069}$As (thin line, \onlinecite{Okabayashi:1999_PRB}) and annealed Ga$_{1-x}$Mn$_{x}$As with $T_{\text{C}} = 160$~K (points, \onlinecite{Rader:2009_PSSB}); (b) Ga$_{0.94}$Mn$_{0.06}$As \cite{DiMarco:2013_NC}; (c)  In$_{0.91}$Mn$_{0.09}$As \cite{Okabayashi:2002_PRB}; (d) Cd$_{0.65}$Mn$_{0.35}$Te \cite{Ley:1987_PRB}. The data in (a,b,d) were obtained as a difference between on-resonant [(a) and (d) $h\nu =
50$~eV (M-line); (b) $h\nu = 641$~eV (L-line)] and off-resonant spectra; in (c) a difference between spectra for In$_{0.91}$Mn$_{0.09}$As and InAs at 70~eV is shown.}
 \label{fig:photoemission}
\end{center}
\end{figure}

In addition to quantitative information on the $d$ electron count (presented above) and $p$-$d$ hybridization (discussed in Sec.~\ref{sec:p-d}), photoemission studies allow examining a shift of the Fermi level and modifications to the band structure introduced by Mn ions. In Fig.~\ref{fig:photoemission} the contribution of Mn $3d$ states to valence band DOS, obtained by angle integrated photoemission at various resonant excitation energies, are presented for (Ga,Mn)As containing above 5\% of Mn. Corresponding findings are also shown for (In,Mn)As and (Cd,Mn)Te.

Several important conclusions emerge from these as well as from more recent spectra \cite{Kobayashi:2013_arXiv}.  First, a general agreement between the (Ga,Mn)As data obtained for different energies of exciting photons (and, thus, absorption length) indicates that these results are not substantially affected by surface effects. Second, there is a considerable similarity between findings for (Ga,Mn)As, (In,Mn)As, and (Cd,Mn)Te: a center of gravity of the $3d$-state contribution is at ~4~eV below the Fermi level, and an additional local maximum appears deeper in the valence band. This indicates that the physics of $p$-$d$ hybridization is similar in these systems. Third, the DOS magnitude decays to zero through a knee or a weak maximum at 0.2-0.4~eV on approaching the Fermi energy with no trace of an impurity band above the top of the valence band. Since the Mn concentration is low (~2-10\%) in DFSs and the main weight of Mn $d$ states is well below $\epsilon_{\text{F}}$, the $d$ states do not accommodate holes, and their contribution to the hole wave function is below 1\%. \footnote{In one photoemission work a dispersion-less impurity-like band was detected above the valence band top in (Ga,Mn)As \cite{Okabayashi:2001_PRB}.  This observation was not confirmed by subsequent studies by the same group \cite{Rader:2004_PRB,Kobayashi:2013_arXiv} and others \cite{Gray:2012_NM,DiMarco:2013_NC}.}

Another important finding of photoemission works was the demonstration that the {\em total} DOS tends to zero at $\epsilon_{\text{F}}$ in (Ga,Mn)As \cite{Okabayashi:1999_PRB} despite high values of hole concentrations. This reconfirmed the presence of a Coulomb anomaly at the Fermi level in this system, of the half-width 0.1-0.2~eV, as discussed in the subsections above and below (Secs.~\ref{sec:MIT} and \ref{sec:hole_mass}, respectively).  This conclusion was supported by angle-resolved photoemission studies on (In,Mn)As \cite{Okabayashi:2002_PRB} and (Ga,Mn)As \cite{Gray:2012_NM,Kobayashi:2013_arXiv}, which revealed a characteristic depression in DOS on approaching the $\Gamma$ point.

\subsubsection{Hole effective mass in III-V DMSs}
\label{sec:hole_mass}
Cyclotron resonance measurements on ferromagnetic (In,Mn)Sb and (In,Mn)As ($x = 0.02$) in high magnetic field ($B > 100$~T) was explained by Landau level positions calculated from the eight-band $kp$ model for InSb and InAs, respectively, indicating that the itinerant holes reside in the valence band of the host semiconductor \cite{Matsuda:2011_JPCS}.

\begin{figure}[bt]
\begin{center}
\includegraphics[width=3.4in]{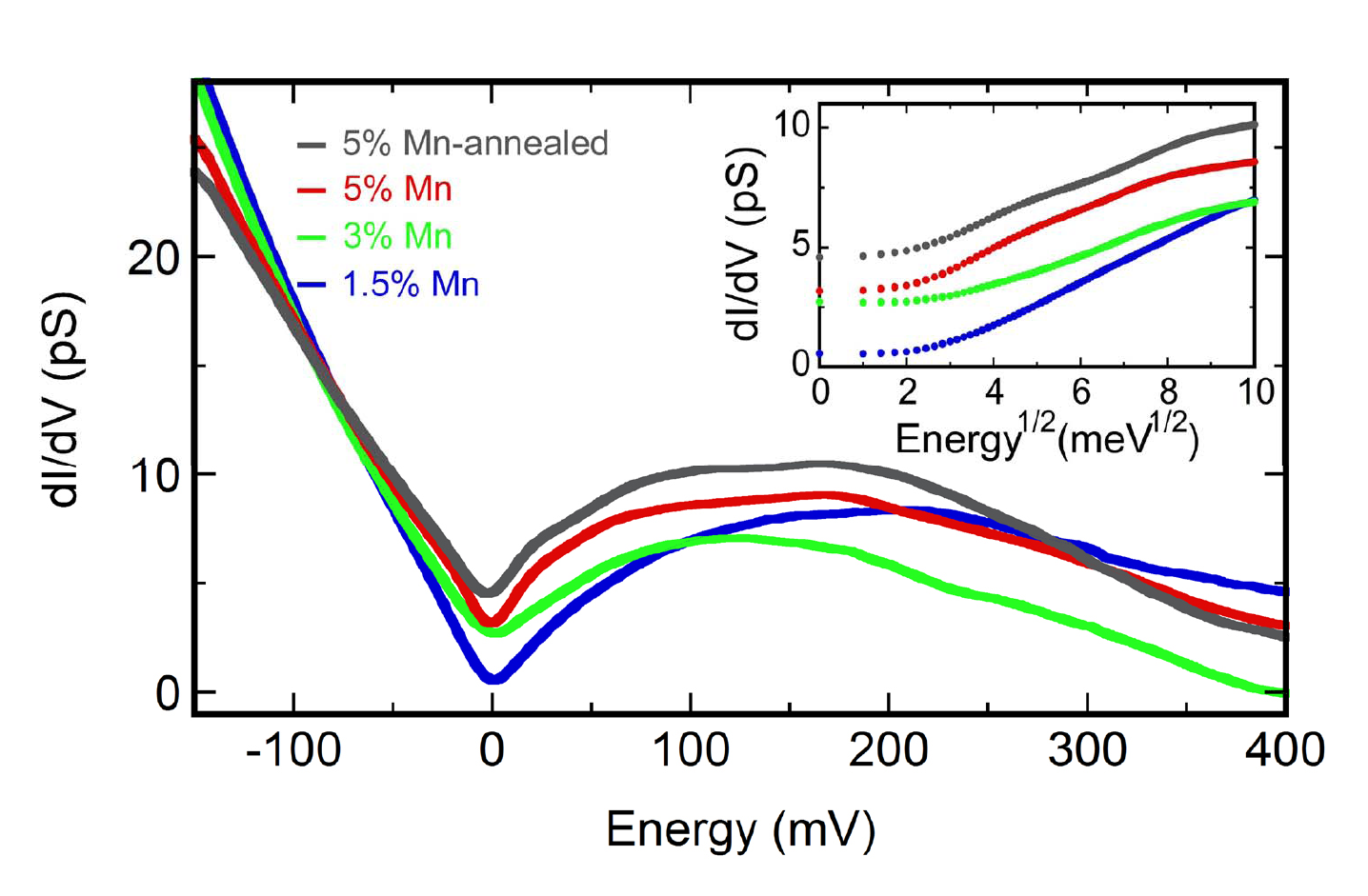}
\caption[]{(Color online) The spatially averaged differential conductance for Ga$_{1-x}$Mn$_x$As with various Mn concentrations $x$ obtained by scanning tunneling microscopy. The inset shows the same data as the main panel, with the square root of the voltage on the horizontal axis, the dependence expected theoretically. From \onlinecite{Richardella:2010_S}.}
 \label{fig:GaMnAs_Richardella}
\end{center}
\end{figure}

No detection of cyclotron resonance has been reported for (Ga,Mn)As, where holes are at the localization boundary, so that Landau level broadening $\hbar/\tau$ precludes the observation of cyclotron resonance. The proximity to the Anderson-Mott type of the MIT was well documented in (Ga,Mn)As by the appearance of a zero-bias anomaly in tunneling $I(V)$ characteristics \cite{Chun:2002_PRB,Pappert:2006_PRL,Richardella:2010_S}.  As shown in Fig.~\ref{fig:GaMnAs_Richardella}, the DOS minimum at $\epsilon_{\text{F}}$, at least in the region adjacent to the surface,  fills up rather slowly on enlarging the Mn concentration and, thus, the hole density  beyond $p_{\text{c}}$ of (Ga,Mn)As.

This finding can be explained by noting that according to multiband $kp$ \cite{Dietl:2001_PRB,Sliwa:2011_PRB} and multiorbital tight binding computations for (Ga,Mn)As \cite{Werpachowska:2010_PRBa}, the effective mass of holes at $\epsilon_{\text{F}}$ increases by a factor of 2 when the hole concentration $p$ changes from $10^{19}$ to $10^{21}$~cm$^{-3}$. A correspondingly slow growth of $k_{\text{F}}l \propto 1/m^{*2}$ with hole density makes that the region dominated by localization effects extends far beyond the immediate vicinity to the MIT in this compound.

Scanning probe tunneling spectroscopy provided maps of local DOS (LDOS) in (Ga,Mn)As with various Mn concentrations $x$ \cite{Richardella:2010_S}. A log-normal distribution of LDOS was found even in the limit of weak localization \cite{Richardella:2010_S}, corroborating that quantum interference, rather than trapping by individual impurities, accounts for hole localization in (Ga,Mn)As in the Mn concentration range relevant for ferromagnetism \cite{Dietl:2000_S,Dietl:2008_JPSJ}.

\begin{figure}[bt]
\begin{center}
\includegraphics[width=3.6in]{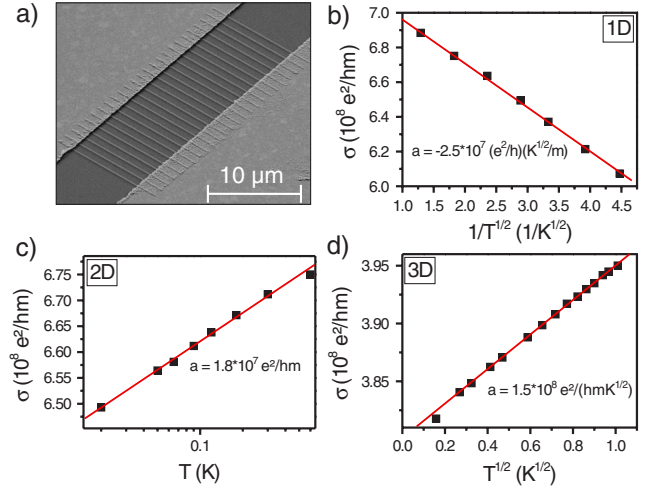}
\caption[]{(Color online) Low-temperature conductivity in (Ga,Mn)As samples of various dimensionality: (a,b) collection of {\em quasi}-1D wires connected in parallel; (c,d) {\em quasi}-2D and -3D Hall bars, respectively. The character (solid lines) and magnitudes (slopes $a$) of the observed temperature dependence are expected theoretically for disorder-modified hole-hole interactions in (Ga,Mn)As valence band. From \onlinecite{Neumaier:2009_PRL}.}
 \label{fig:GaMnAs_Neumaier}
\end{center}
\end{figure}

Another manifestation of Anderson-Mott localization in (Ga,Mn)As is a large magnitude of quantum corrections to conductivity that can be diminished by a magnetic field, temperature, and frequency \cite{Matsukura:2004_PE}. Figure~\ref{fig:GaMnAs_Neumaier} shows $\sigma(T)$ below 1~K in ferromagnetic (Ga,Mn)As samples of various dimensionality \cite{Neumaier:2009_PRL}. Taking into account effects of disorder-modified hole-hole interactions, the magnitudes of slopes $a$  in the dependence $\sigma(T)$ provide information on the thermodynamic DOS  $\rho(\epsilon_{\text{F}})$ in the 1D and 3D cases \cite{Altshuler:1985_B,Lee:1985_RMP}. According   to the quantitative analysis \cite{Dietl:2008_JPSJ,Neumaier:2009_PRL,Sliwa:2011_PRB}, the values of $a$ indicate that the effective mass of holes at the Fermi level in (Ga,Mn)As differs by less than by a factor of 2 comparing to that of the disorder-free GaAs valence band.

Since the external magnetic field has no effect on $\sigma(T)$ \cite{Neumaier:2008_PRB}, the single-particle Anderson localization term, destroyed presumably by a demagnetizing field, does {\em not} contribute to the observed temperature dependence of conductivity in this ferromagnetic semiconductor.\footnote{Some authors \cite{Honolka:2007_PRB,Mitra:2010_PRB} analyzing $\sigma(T)$ up to 4~K, found that $\sigma(T) = \sigma_0 +  AT^{\alpha}$, where $\alpha = 1/3$. This dependence was interpreted in terms of a renormalization group equation \cite{Altshuler:1985_B} applicable close to the MIT, where $ \sigma_0 <  AT^{\alpha}$ and then  $1/3 \lesssim \alpha \lesssim 1/2$ in the 3D case \cite{Lee:1985_RMP,Belitz:1994_RMP}. Furthermore, the apparent value of $\alpha$ can be reduced above 1~K by a cross-over to the regime, where the effect of scattering by magnetic excitations onto quantum corrections to conductivity becomes significant \cite{Dietl:2008_JPSJ}.}

The existence of sizable quantum localization effects indicates that the real part of intraband optical conductivity, $\sigma_1(\omega)$, in addition to a dispersion expected within the Drude theory for a GaAs-type complex valence band \cite{Sinova:2002_PRB,Hankiewicz:2004_PRB}, should show a significant drop with decreasing $\omega$ down to $\hbar\omega \approx k_{\text{B}}T$. The low energy gaps in DOS and conductivity not only share the same physical origin but involve the same energy scale and dispersion at low energies. Quantitative formulae describing the disappearance of quantum localization contributions to $\sigma$ with $\omega$ are theoretically known for a metallic case and low-energy excitations in a simple band, {\em i.~e.,} for $ \omega \lesssim 1/\tau \lesssim \epsilon_F$  \cite{Altshuler:1985_B,Lee:1985_RMP}, where $\hbar/\tau$ is of the order of 0.1~eV in ferromagnetic Ga$_{1-x}$Mn$_x$As.  Actually, interplay between Anderson-Mott effects and a Drude-like decay of intra-band conductance at high frequencies leads to a maximum in $\sigma(\omega)$, found at $\omega_{\text{m}} \sim 0.2$~eV  in (Ga,Mn)As \cite{Burch:2008_JMMM}. Unfortunately, any detailed interpretation \cite{Burch:2008_JMMM,Jungwirth:2007_PRB,Kojima:2007_PRB} of $\omega_{\text{m}}$ and its shift with $x$ or $T$ is rather inconclusive as {\em no} theory for $\sigma(\omega)$ is available in this cross-over regime, particularly for the complex valence band and in the presence of spin-disorder scattering. However, a comparison of $\sigma(\omega)$ across the MIT in (Ga,Be)As and (Ga,Mn)As \cite{Chapler:2011_PRB} provides a strong confirmation of the aforementioned persistence of localization effects even for $p \gg p_{\text{c}}$ in (Ga,Mn)As as well as demonstrates a similarity in the DOS and conductivity gaps in this DFS.

\begin{figure}[bt]
\begin{center}
\includegraphics[width=3.5in]{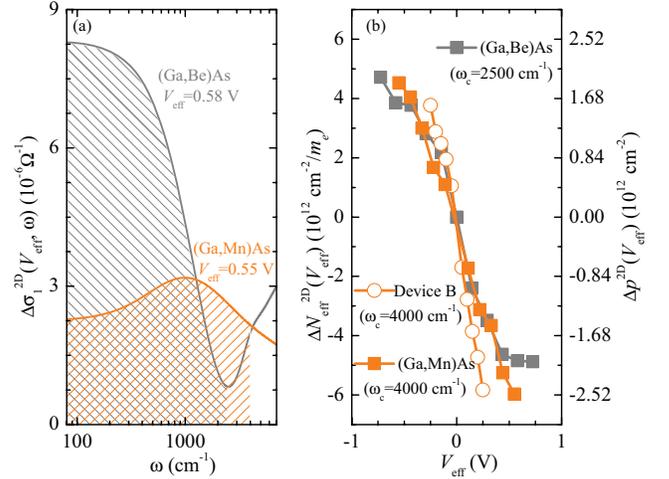}
\caption[] {(Color online) Panel (a) shows a gate-induced change in optical conductivity for a (Ga,Mn)As-based device and a (Ga,Be)As-based device at two values of the gate voltage $V_{\text{eff}}$. The orange and gray shaded regions indicate the area included in application of the optical sum rule for the (Ga,Mn)As and (Ga,Be)As accumulation layers, respectively. Panel (b) shows the change in the integrated spectral weight at all gate voltages for the (Ga,Be)As-, (Ga,Mn)As-, and second (Ga,Mn)As- (device B) based devices.  Using an appropriate calibration procedure, the right axis is converted into the two-dimensional change in hole concentration, $\Delta p^{\text{2D}}$. From \onlinecite{Chapler:2012_PRB}.}
 \label{fig:GaMnAs_Chapler}
\end{center}
\end{figure}

In contrast to $\sigma_1(\omega)$,  an integral of $\sigma_1(\omega)$ over $\omega$ depends uniquely on the ratio of the hole density $p$ and a combination of the heavy and light hole masses, $m_{\text{op}}$, for any strength of disorder according to the optical conductivity sum rule for intraband excitations \cite{Sinova:2002_PRB}. Figure~\ref{fig:GaMnAs_Chapler} presents a comparison  of $\sigma_1(\omega)$ and $\Delta N =\Delta p^{\text{2D}}/m_{\text{op}}$ for (Ga,Be)As and (Ga,Mn)As determined at room temperature as a function of the gate voltage $V_{\text{eff}}$ that changes the hole concentration. The values of $\Delta N$ were  obtained by the integration of $\sigma_1(\omega)$ up to a finite value $\omega_{\text{c}}$. A similar magnitudes of $\Delta N$ in both systems together with the evaluated value $m_{\text{op}} \lesssim 0.42m_0$ \cite{Chapler:2012_PRB} confirmed that the hole band of (Ga,Mn)As retains basic characteristics of the GaAs valence band for which $m_{\text{op}}/m_0 = 0.25 - 0.29$ was theoretically predicted \cite{Sinova:2002_PRB}.

\begin{figure}[bt]
\begin{center}
\includegraphics[width=3.4in]{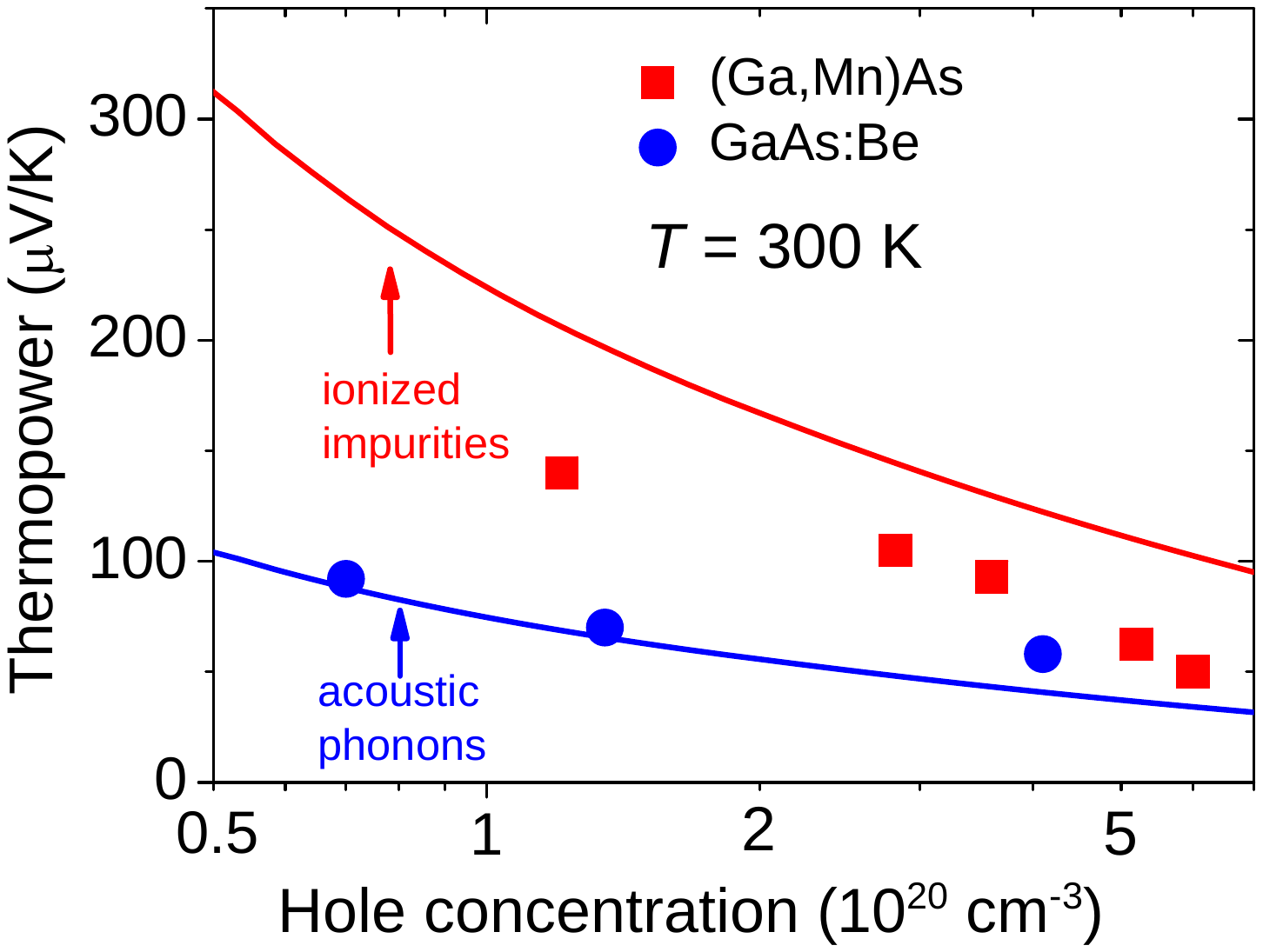}
\caption[]{(Color online) Room temperature thermoelectric power in (Ga,Mn)As and GaAs:Be as a function of hole density changed by ion irradiation (points). From \onlinecite{Mayer:2010_PRB}. Lines are calculated (neglecting phonon drag contribution) for the GaAs valence band [using the standard six band Luttinger model and parameters \cite{Dietl:2001_PRB}], assuming that ionized impurity and acoustic phonon scattering dominates (upper and lower curve, respectively); the actual value of thermopower $S$ should lie between lines obtained for heavy and light hole bands in each case. From \onlinecite{Sliwa:2011_PRB}.}
 \label{fig:GaMnAs_thermopower}
\end{center}
\end{figure}

This conclusion was corroborated by the magnitudes of room temperature thermoelectric power $S$ in (Ga,Be)As and (Ga,Mn)As examined in the same hole concentration range \cite{Mayer:2010_PRB}, as shown in Fig.~\ref{fig:GaMnAs_thermopower}. It is worth noting that neglecting a phonon drag contribution, $S$  is proportional to a logarithmic derivative of conductivity over energy and, thus, to a first approximation, to thermodynamic DOS per one carrier. A large magnitude of $S$ would therefore be expected if the Fermi level were pinned by a large DOS of an impurity band \cite{Heremans:2012_EES}.

In summary, according to the data discussed in this and previous subsection, holes in (Ga,Mn)As and related DFSs reside in a host-like valence band that is, however, strongly affected by the proximity to the MIT. This conclusion is further supported by the outcome of photoreflectance studies \cite{Yastrubchak:2011_PRB}.

\subsection{Experimental studies of $p$-$d$ exchange energy}
\label{sec:p-d}
The energy distribution of Mn $3d$ states shown in Fig.~\ref{fig:photoemission} can serve to evaluate parameters of the Anderson Hamiltonian characterizing hybridization between $p$-like valence bands in tetrahedrally coordinated semiconductors and $d$ states of Mn ions. This was carried out employing a configuration-interaction method to describe photoemission and x-ray absorbtion spectra in II-VI and III-V DMSs \cite{Mizokawa:2002_PRB,Hwang:2005_PRB}. By using the Schrieffer-Wolf transformation one then obtains the magnitudes of energies $N_0\beta$ \cite{Kacman:2001_SST} and $N_0W$ \cite{Benoit:1992_PRB}, characterizing  respectively spin-dependent and spin-independent parts of the local potential introduced by individual Mn ions. These energies are proportional to square of hybridization matrix element $V_{pd}$, and inversely proportional to distances of the Fermi level to $d^5$ and $d^6$ states, $\epsilon_d$ and $U-\epsilon_d$.

When the Mn potential is too weak to bind a hole, and a possible Coulomb contribution (existing in III-V DMSs) is screened by carriers, the first order perturbation theory (virtual-crystal and molecular-field approximations) describes the valence band offset and spin splitting of valence band states leading to, d$E_{\text{v}}(x)$/d$x = N_0W$ and  ${\cal{H}}_{pd} = \beta\vec{s}\vec{M}/g\mu_{\text{B}}$, where $M$ is magnetization of Mn ions and $g = 2.0$ is their Land\'e factor. If, however, the perturbation introduced by a single Mn impurity is so strong that a bound state appears, the influence of the Mn ion ensemble on the band valence structure has to be treated in a non-perturbative way \cite{Dietl:2008_PRB}. Reversed signs of the band offset and of spin splitting can appear in this strong coupling case. In this regime, in addition to $W$ and $\beta$,  the spectrum at $k =0$, for a given hole mass value $m^*$, is determined by at least one more parameter, namely, the spatial extend $b$ of the perturbation introduced by Mn or alternatively by the ratio of the corresponding potential well depth to its minimum value giving rise to a bound state, $U/U_{\text{c}}$ \cite{Benoit:1992_PRB,Dietl:2008_PRB}.

Exciton magnetospectroscopy has been the primary source of information on exchange splittings of bands, and thus on $\beta$ in DMSs without carriers. Magnetooptical studies of hole-doped (Cd,Mn)Te quantum wells \cite{Haury:1997_PRL,Boukari:2002_PRL,Kossacki:2004_PRB} demonstrated that interband transitions are considerably affected by hole-hole interactions as well as by the Moss-Burstein shift that accounted for the sign inversion of MCD comparing to the case of undoped (Cd,Mn)Te \cite{Haury:1997_PRL}. Despite that modulation doping was employed, evidences for scattering broadening of DOS were also found \cite{Boukari:2002_PRL}. Polarization-resolved magnetoabsorption measurements in the band gap region of ferromagnetic (Ga,Mn)As \cite{Szczytko:1999_PRB}, were also interpreted taking into account the Moss-Burstein shift and scattering broadening in the heavy hole band \cite{Szczytko:2001_PRB}.

\begin{figure}[tb]
\centering
\includegraphics[width=9.5cm]{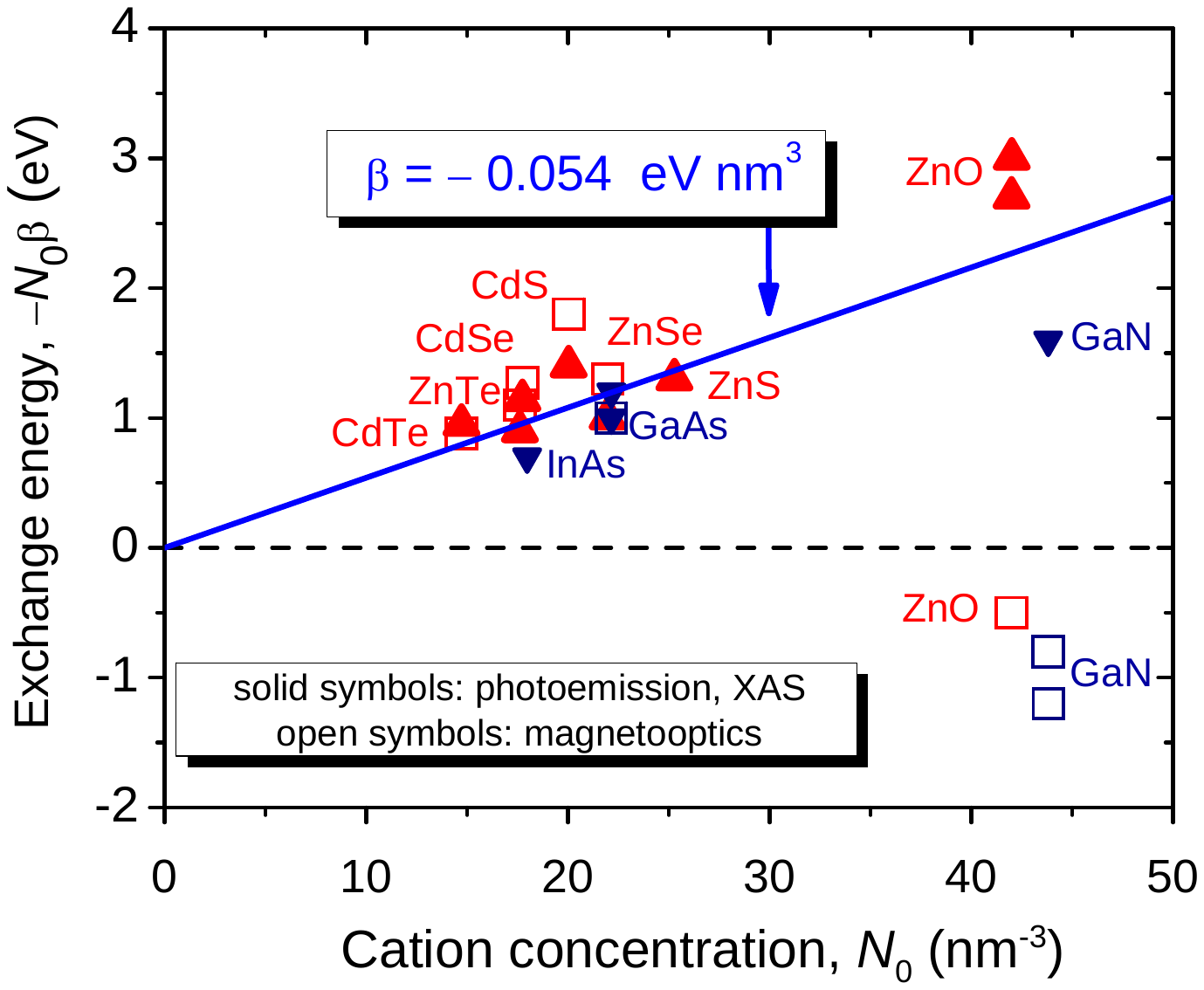}
\caption[]{(Color online) Compilation of experimentally determined energies of the $p$-$d$ exchange interaction $N_0\beta$  for various Mn-based DMSs as a function of the cation concentration $N_0$. Solid symbols denote the values evaluated from photoemission and x-ray absorption spectra for (Cd,Mn)VI \cite{Mizokawa:1993_PRB}, (Zn,Mn)VI \cite{Mizokawa:2002_PRB}, (Zn,Mn)O \cite{Okabayashi:2004_JAP}, (In,Mn)As \cite{Okabayashi:2002_PRB}, (Ga,Mn)As \cite{Okabayashi:1998_PRB,Okabayashi:1999_PRB}, (Ga,Mn)N \cite{Hwang:2005_PRB}. The values shown by open symbols were determined within the molecular field approximation from
excitonic splittings in the magnetic
field in (Cd,Mn)Te \cite{Gaj:1979_SSC}, (Zn,Mn)Te \cite{Twardowski:1984_SSCb}, (Zn,Mn)Se \cite{Twardowski:1984_SSCa}, (Cd,Mn)Se \cite{Arciszewska:1986_JPCS}, (Cd,Mn)S \cite{Benoit:1992_PRB}, (Zn,Mn)O \cite{Pacuski:2011_PRB}, (Ga,Mn)N \cite{Pacuski:2007_PRB,Suffczynski:2011_PRB}, and from band splittings in the magnetic
field in (Ga,Mn)As \cite{Szczytko:2001_PRB}.  Solid line corresponds to a constant value of $\beta$ across the DMS series.} \label{fig:beta}
\end{figure}

In Fig.~\ref{fig:beta} the magnitudes of $N_0\beta$ determined from high energy spectroscopy and interband magnetooptical studies are collected. The reversed signs of the apparent $N_0\beta$ values determined from excitonic magnetoreflectivity within the molecular field approximation for (Zn,Mn)O and (Ga,Mn)N show that these systems are in the strong coupling regime \cite{Dietl:2008_PRB}. To a first approximation $\beta = - 54$~meV\,nm$^3$ describes the available data for various tetrahedrally coordinated DMSs in the weak coupling regime (except for mercury chalcogenides \cite{Furdyna:1988_B}, where $|\beta|$ magnitudes appear somewhat smaller). It corresponds to $N_0\beta = -1.2$~eV for (Ga,Mn)As, the value consistent with the results of photoemission \cite{Okabayashi:1998_PRB,Okabayashi:1999_PRB}, magnetoabsorption \cite{Szczytko:2001_PRB}, and the energy difference between the states corresponding to a parallel and antiparallel spin arrangement of a bound hole bound and a Mn ion in the limit of low Mn concentrations, $\Delta\epsilon = 8 \pm 3$~meV \cite{Linnarsson:1997_PRB,Averkiev:1987_SPS}. The antiferromagnetic character of this coupling was also corroborated by a direction of current-induced spin torque in (Ga,Mn)As (Secs.~\ref{sec:current} and  \ref{sec:current_domains}) and a sign of circular polarization in spin-LEDs (Sec.~\ref{sec:injection}).

Since the pioneering studies of interband MCD in (Ga,Mn)As \cite{Ando:1998_JAP}, this techniques has been widely employed to asses effects of $p$-$d$ coupling onto the valence band of III-V DFSs.  In particular, theory of optical absorption and MCD involving six valence subbands and conduction band was developed for thin films of carrier-controlled DFSs \cite{Dietl:2001_PRB}, and showed to describe  puzzling MCD data for (Ga,Mn)As \cite{Beschoten:1999_PRL} with one fitting parameter --  the Mn-induced band gap offset. Its sign (corresponding to gap narrowing) and value (about 0.2~eV for $x =5$\%) are consistent with a net magnitude of many body effects \cite{Dietl:2001_PRB} and $p$-$d$ hybridization in the weak coupling limit \cite{Dietl:2008_PRB} and, moreover, with their independent experimental determinations \cite{Ohno:2002_PE,Elsen:2007_PRL,Thomas:2007_APL,Fujii:2013_PRL}.  A number of subsequent studies and interpretations of MCD and related optical phenomena in III-V DFSs were already reviewed \cite{Burch:2008_JMMM}. Unfortunately, the combined effects of strong disorder and correlation, relaxing selection rules and accounting for band gap narrowing, has so-far made interpretations of the findings somewhat ambiguous, the viewpoint put also forward in a recent attempt to describe the available MCD data in terms of a multi-orbital tight binding model \cite{Turek:2009_PRB}.

\section{Superexchange}
\label{sec:superexchange}

Owing to a relatively large distance between magnetic ions in  DMSs, no direct exchange coupling between $d$-like orbitals localized on Mn ions is expected. Thus, rather indirect coupling involving band states accounts for spin-spin interactions.  In this section we describe effects of short-range superexchange that dominates in the absence of carriers and competes with long-range carrier-mediated interactions if the concentration of band carriers is sufficiently high.

\subsection{Antiferromagnetic superexchange}
\label{sec:antiferro}

\subsubsection{II-VI DMSs}
A vast majority of nonmetallic TM compounds are antiferromagnets or ferrimagnets. In the absence of carriers, short-range antiferromagnetic coupling determines magnetic properties of DMSs containing Mn$^{2+}$ ions, the case of, for instance, Mn-based II-VI DMSs \cite{Shapira:2002_JAP}.  The relevant coupling between localized spins---the superexchange \cite{Anderson:1950_PR,Goodenough:1958_JPCS,Kanamori:1959_JPCS}---proceeds $via$ $p$-$d$ hybridization with bands of anions residing on the path between the TM spins in question. For this indirect interaction, to the lowest relevant order perturbation theory, the exchange energy $J_{ij}$  is proportional to $|V_{pd}|^4$ and decays fast with the distance $R_{ij}$ between magnetic ions.

In random antiferromagnets, such as intrinsic II-VI DMSs, frustration of interactions in spin triads and larger Mn clusters leads to spin-glass freezing. According to comprehensive studies, critical temperature $T_{\text{f}}$ is about 1~K at $x = 0.1$ and grows with $x$ as $T_{\text{f}} \propto x^m$, where  $m =2.3 \pm 0.1$ in Mn-doped cadmium and zinc chalcogenides \cite{Twardowski:1987_PRB}.  Scaling invariance \cite{Rammal:1982_B} implies then that the exchange energy $J_{ij}$ decays with the spin-spin distance as $R_{ij}^{-\lambda}$, where $\lambda = md = 6.8\pm 0.3$ for 3D systems.

In these systems, it is usually possible to parametrize experimental values of magnetization $M(T,H)$  by the paramagnetic Brillouin function for $S = 5/2$ \cite{Gaj:1979_SSC},
\begin{equation}
M(T,H) = g\mu_{\text{B}}N_0x_{\text{eff}}\text{B}_S\left[\frac{g\mu_{\text{B}}H}{k_{\text{B}}(T + T_{\text{AF}})}\right],
\end{equation}
where $x_{\text{eff}} < x$ and $T_{\text{AF}}>0$ describe a reduction of magnetization by antiferromagnetic interactions. The values of these parameters increases with temperature \cite{Spalek:1986_PRB},  $x_{\text{eff}} \rightarrow x$ and $T_{\text{AF}} \rightarrow -\Theta_0$, where
 \begin{equation}
\Theta_0 = \frac{1}{3}S(S + 1)\sum_jz_jJ_j.
\end{equation}
Here, the summation extends over the subsequent cation coordination spheres; $z_j$ is the number of cations in the
sphere $j$, and $J_j\equiv J_{ij} <0$ is the corresponding Mn-Mn exchange energy in the Hamiltonian ${\cal{H}}_{ij} = -J_{ij}\vec{S}_i\vec{S}_j$. The values of $J_{ij}$ were successfully modeled for II-VI Mn-based DMSs by combining {\em ab initio} and tight-binding-like approaches \cite{Larson:1988_PRB}.

As found for {\em ferromagnetic} p-(Cd,Mn)Te \cite{Haury:1997_PRL,Boukari:2002_PRL} and p-(Zn,Mn)Te \cite{Ferrand:2001_PRB} as for n-(Zn,Mn)O \cite{Andrearczyk:2001_ICPS}, the magnitude of  antiferromagnetic superexchange is larger for the nearest neighbor Mn pairs than ferromagnetic coupling at carrier densities achievable by now in these systems. This means that the magnitude of $T_{\text{C}}$ is weakened by non-zero values of $T_{\text{AF}}$ and $x -x_{\text{eff}}$. Furthermore, it was found by Monte Carlo simulations \cite{Lipinska:2009_PRB} that AF interactions account for a relatively fast spin dynamics in a ferromagnetic phase, provided that the holes visit only a part of the region occupied by TM spins. In such a situation, occurring for instance in (Cd,Mn)Te quantum wells, acceleration of spin dynamics and the associated decrease of coercivity are brought about by TM flip-flops at the boundary of the hole wave functions, where the molecular field produced by the hole spins vanishes.

\subsubsection{(Ga,Mn)As and related compounds}
In such compounds,  a strong antiferromagnetic interaction between Mn ions in an interstitial and a neighboring substitutional position reduces $x_{\text{eff}}$ substantially, particulary in non-annealed samples (see, Sec.~\ref{sec:self-compensation}). Relevant information on possible interactions between pairs of {\em substitutional} Mn spins was provided by studies of donor compensated samples, in which carrier-mediated ferromagnetic interactions are strongly reduced but superexchange is expected to be left intact. A sizable decrease of $x_{\text{eff}}$ values at low temperature under such conditions in (Ga,Mn)As (see, Sec.~\ref{sec:doping}) confirmed the presence of intrinsic antiferromagnetic coupling between Mn spins. This conclusion was substantiated by {\em ab initio}  \cite{Kudrnovsky:2004_PRB,Chang:2007_PRB} and tight binding \cite{Jungwirth:2005_PRBa} studies of DFSs, which demonstrated that the role played by antiferromagnetic superexchange can be rather significant -- the magnitude of the corresponding exchange energy was evaluated to be about 50\% of the ferromagnetic contribution for the nearest neighbor Ga-substitutional Mn pairs at $x =6$\%.

Altogether, the accumulated data indicate that in the case of {\em un}compensated III-V DFSs with Mn$^{2+}$ ions the value of $x_{\text{eff}}$ is controlled by {\em interstitial} Mn in the whole relevant temperature and magnetic field range. Furthermore, somewhat weaker but also short ranged antiferromagnetic interactions between {\em substitutional} Mn ions reduce the magnitude of net ferromagnetic spin-spin coupling and, thus, lower $T_{\text{C}}$ further on.

\subsubsection{Antiferromagnetic interactions in (Ga,Mn)N}
Optical \cite{Wolos:2008_B,Graf:2003_PSSB} and photoemission investigations \cite{Hwang:2005_PRB} showed that the acceptor Mn$^{3+}$/Mn$^{2+}$ level appear in the mid-gap region in GaN (see, Fig.~\ref{fig:III_V_EI}). Owing to correspondingly small effective Bohr radius, no indications of hole delocalization were found up to at least $x =0.1$. This means that holes reside in the Mn$^{3+}$/Mn$^{2+}$ impurity band, subject to Mott-Hubbard localization at weak or strong compensation, and to Anderson-Mott localization if the impurity band is partly occupied. It was found that if Mn$^{2+}$ ions prevailed (due to residual donor impurities or defects such as nitrogen vacancies \cite{Yang:2009_APL}),  antiferromagnetic interactions between Mn$^{2+}$ ions controlled magnetic properties \cite{Granville:2010_PRB,Zajac:2001_APL}, leading to spin-glass freezing at 4.5~K for $x \approx 0.1$ \cite{Dhar:2003_PRB}. The corresponding magnitudes of $x_{\text{eff}}$ and $T_{\text{AF}}$ \cite{Granville:2010_PRB,Zajac:2001_APL} were then similar to those of II-VI DMSs.

\subsection{Ferromagnetic superexchange}
\label{sec:ferro-superexchange}
\subsubsection{Double exchange {\em vs.} superexchange}

The case of chromium spinels and europium chalcogenides demonstrates that ferromagnetism is possible without band carriers. According to the time-honored nomenclature, one distinguishes two kinds of ferromagnetic coupling mechanisms operating in the absence {\em band} carriers:

\ \\
{\em Double exchange} -- This mechanism contributes when relevant magnetic ions are in two different charge states \cite{Zener:1951_PRb,Anderson:1955_PR} and if the system of TM electrons is on the metallic side or in the vicinity to the Anderson-Mott transition, the case of {\em e.~.g.} manganites \cite{Dagotto:2001_PR}. In this situation, collective ferromagnetic ordering is triggered by a lowering of electron kinetic energy (increase in the width of the $d$ band), appearing if all ions assume the same spin direction, the arrangement making quantum hopping efficient. The double exchange can be regarded as a strong coupling limit of the $p$-$d$ Zener model discussed in Sec.\,\ref{sec:Zener}, as sufficiently large $p$-$d$ hybridization leads to the appearance of an impurity band in the energy gap (see, Secs.~\ref{sec:doping} and \ref{sec:p-d}). Within this model $T_{\text{C}}$ attains a maximum if the impurity band is half-filled, so that--in the case relevant here--the concentrations of Mn$^{2+}$ and Mn$^{3+}$ ions are approximately equal.

\ \\
{\em Ferromagnetic superexchange} -- According to Anderson-Goodenough-Kanamori rules  \cite{Anderson:1950_PR,Goodenough:1958_JPCS,Kanamori:1959_JPCS} superexchange is ferromagnetic for certain charge states of TM ions and bond arrangements. This mechanism outperforms the double exchange if all magnetic ions are in the same charge state or if TM electrons are strongly localized, so that the impurity bandwidth is rather determined by disorder than by quantum hopping.
\ \\

It was found \cite{Blinowski:1996_PRB} employing a tight binding approximation that superexchange is ferromagnetic in the case of Cr$^{2+}$ ions in a tetrahedral environment,  $J_{ij} > 0$. Experimental studies of (Zn,Cr)Se \cite{Karczewski:2003_JSNM} and (Zn,Cr)Te \cite{Saito:2003_PRL} revealed indeed the presence of ferromagnetism in these systems, which however was largely determined by aggregation of Cr cations \cite{Karczewski:2003_JSNM,Kuroda:2007_NM}.

Ferromagnetic coupling mediated by bound holes in the strongly localized regime was also predicted within a tight binding approximation for a neutral or singly ionized pair of substitutional Mn acceptors in GaAs, neglecting on-site Coulomb repulsion $U$ and intrinsic antiferromagnetic interaction between Mn$^{2+}$ spins \cite{Strandberg:2010_PRB}.

\subsubsection{Ferromagnetic superexchange in (Ga,Mn)N}
\begin{figure}[tb]
\centering
\includegraphics[width=8.5cm]{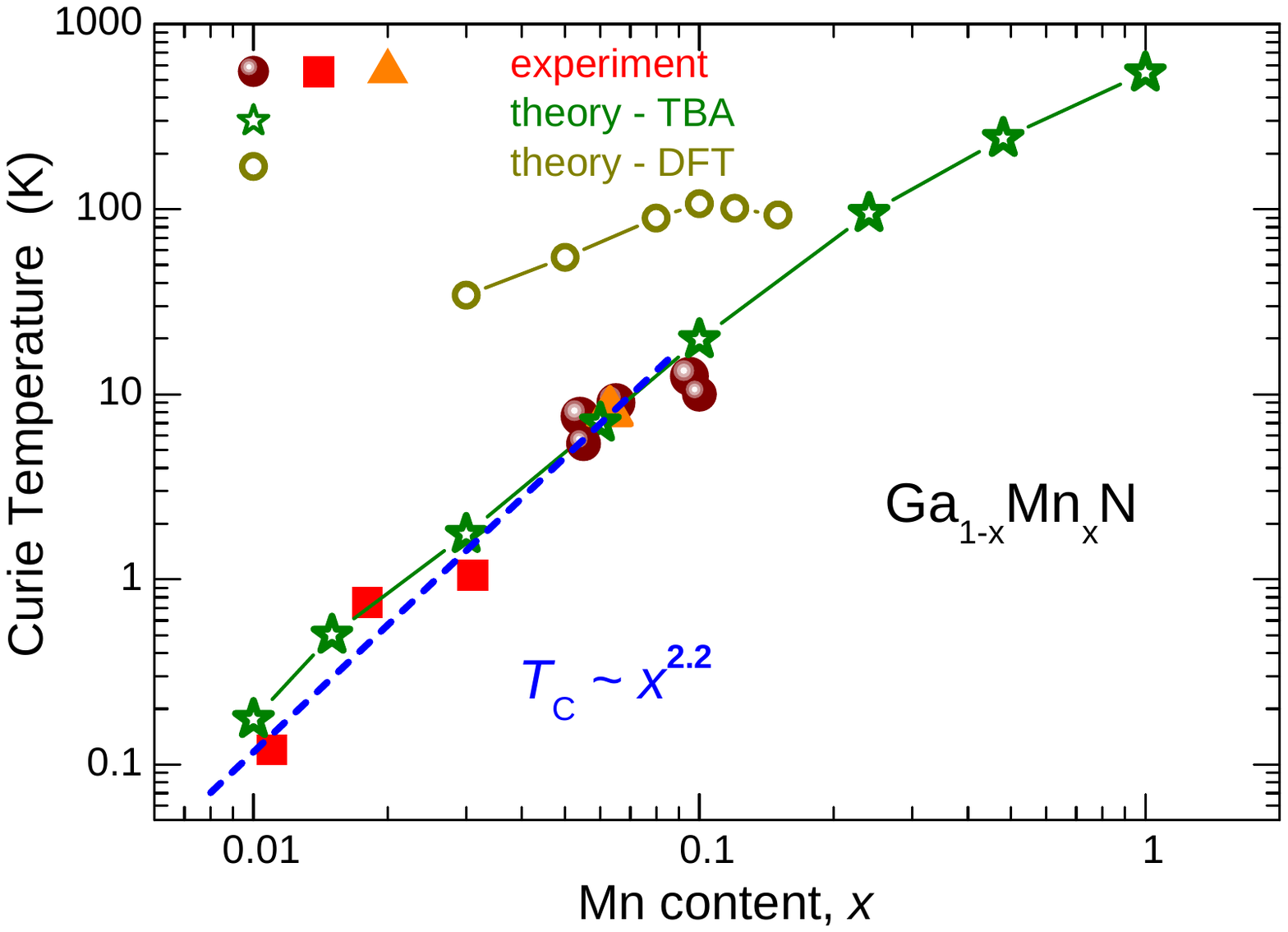}
  \caption[]{(Color online) Experimental Curie temperatures $T_{\text{C}}$ as a function of Mn content $x$ in Ga$_{1-x}$Mn$_x$N (solid squares -- MOVPE samples, \onlinecite{Sawicki:2012_PRB}; solid circles -- MBE samples, \onlinecite{Stefanowicz:2013_PRB}; solid triangle -- MBE sample, \onlinecite{Sarigiannidou:2006_PRB}) compared to theory within the tight binding approximation (that provided the magnitudes of exchange integrals) and Monte Carlo simulations, serving to determine $T_{\text{C}}$ (stars, \onlinecite{Stefanowicz:2013_PRB}). Dashed line shows the scaling dependence  $T_{\text{C}} \propto x^{m}$ with $m=2.2$.  {\em Ab initio} and Monte Carlo results are presented for a comparison (open circles -- DFT, \onlinecite{Sato:2010_RMP}). Adapted from \onlinecite{Stefanowicz:2013_PRB}.}
  \label{fig:GaMnN_Tc}
\end{figure}

Extensive nanocharacterization of wz-(Ga,Mn)N obtained by MOVPE \cite{Stefanowicz:2010_PRBa,Bonanni:2011_PRB} and MBE \cite{Sarigiannidou:2006_PRB,Kunert:2012_APL} demonstrated that under carefully adjusted growth conditions, within an experimental margin below 10\%, all Mn ions are distributed randomly over Ga-substitutional positions, and assume a 3+, $S = 2$ charge and spin state, characterized by a non-zero value of orbital momentum. Empirically, $M(T,H)$ in the paramagnetic region, $x \lesssim 1$\%, and for the magnetic field perpendicular to the wurtzite $c$-axis is well described by the Brillouin function for $S=2$ with the Mn Land\'e factor $g \simeq 2.5$. This electronic configuration is analogous to Cr$^{2+}$ and, indeed, ferromagnetic coupling between Mn spins was found in (Ga,Mn)N \cite{Bonanni:2011_PRB,Sarigiannidou:2006_PRB,Kunert:2012_APL,Stefanowicz:2013_PRB}. Actually, owing to the absence of competing antiferromagnetic interactions and the high magnitude of cation density $N_0 = 4.39\times 10^{22}$~cm$^{-3}$, the largest magnitude of magnetization ever reported for any DMS was observed for Ga$_{0.905}$Mn$_{0.095}$N, $\mu_0M \simeq 190$~mT at $\mu_0H = 6.5$~T \cite{Kunert:2012_APL}. Importantly, ferromagnetic ordering was found at low temperatures in these samples. According to the magnetic phase diagram displayed in Fig.~\ref{fig:GaMnN_Tc}, $T_{\text{C}} \simeq 13$~K at $x = 0.1$ and $T_{\text{C}}(x) \propto x^m$, where $m = 2.2\pm 0.2$, \cite{Sawicki:2012_PRB,Stefanowicz:2013_PRB}. Since such a value of $m$ was observed for spin-glass freezing in II-VI DMSs (see, Sec.\,VIII.A.1), it was concluded that the superexchange is the dominant spin-spin coupling mechanism.

The superexchange scenario was substantiated by the evaluation of the exchange integral $J_{ij}$ as a function of the Mn-Mn distance within the aforementioned tight binding theory \cite{Blinowski:1996_PRB} and then $T_{\text{C}}(x)$ by Monte Carlo simulations \cite{Sawicki:2012_PRB,Stefanowicz:2013_PRB}. Within this approach, Mn ions were described in terms of Parmenter's \cite{Parmenter:1973_PRB} rotationally invariant generalization of the Anderson Hamiltonian for the relevant electronic configuration of the TM taking into account the Jahn-Teller distortion \cite{Gosk:2005_PRB,Stroppa:2009_PRB}, whereas the host band structure was modeled by the $sp^3s^*$ tight binding approximation, employing the established parametrization for GaN in the cubic approximation. Other parameters of the model were taken from infrared and visible \cite{Graf:2003_PSSB} as well as photoemission and soft x-ray absorption spectroscopy \cite{Hwang:2005_PRB} of (Ga,Mn)N.

As shown in Fig.~\ref{fig:GaMnN_Tc}, the dependence of $T_{\text{C}}$ on $x$ in insulating (Ga,Mn)N with Mn$^{3+}$ ions is well reproduced by the impurity band theory in question (stars). The examination of the critical behavior around $T_{\text{C}}$ provided information on the variance $\Delta x$ of macroscopic inhomogeneities in the Mn distribution, evaluated to be $\Delta x \approx 0.2$\% \cite{Stefanowicz:2013_PRB}.

\section{Theory of carrier mediated ferromagnetism}
\label{sec:theory}

This section presents the $p$-$d$ Zener model \cite{Dietl:2000_S} and its limitations. This model of ferromagnetism in p-type DFSs is built exploiting information summarized in Sec.~\ref{sec:electronic_states} on the relevant electronic states and coupling between localized Mn spins and itinerant holes. The presence of competing antiferromagnetic interactions is considered making use of findings presented in Sec.~\ref{sec:superexchange}. The model is parametrized by a small set of independently determined material parameters, it is numerically efficient and univocal. As shown in Sec.~\ref{sec:comparison}, the $p$-$d$ Zener model explains qualitatively, and often quantitatively, a palette of comprehensive results concerning ferromagnetic characteristics and their control in films, heterostructures, and nanostructures of (Ga,Mn)As, p-(Cd,Mn)Te, and related compounds, as collected in Sec.~\ref{sec:control}. By combining the model with disorder-free Landauer-B\"uttiker or Drude-Boltzmann formalisms, a theoretical description of spin-transport and magnetooptical devices has been attempted, although a proximity to the charge localization verge renders such a modeling not always applicable.

\subsection{The mean-field {$p$-$d$} Zener model}
\label{sec:Zener}
\subsubsection{The model}

Zener noted in the 1950s the role of band carriers in promoting ferromagnetic ordering between localized spins in magnetic metals. This ordering can be viewed as driven by the lowering of the carriers energy associated with their redistribution between spin subbands, split by the $sp-d$ exchange coupling to the localized spins. A more detail quantum treatment indicates, however, that the sign of the resulting interaction between localized spins oscillates with the spin-spin distance according to the celebrated Ruderman-Kittel-Kasuya-Yosida (RKKY) formula. However, the Zener and RKKY models were found equivalent within the continuous medium and  mean-field approximation \cite{Dietl:1997_PRB}. These approximations are valid as long as the period of RKKY oscillations, $R = \pi/k_{\text{F}}$ is large compared to an average distance between localized spins. Hence, the technically simpler mean-field Zener approach is meaningful in the regime usually relevant to DFSs, $p \lesssim xN_0$. Owing to higher DOS and larger exchange coupling to Mn spins, holes are considerably more efficient in mediating  spin-dependent interactions between localized spins in DFS. This hole-mediated Zener/RKKY ferromagnetism is enhanced by exchange interactions within the carrier liquid \cite{Dietl:1997_PRB,Jungwirth:1999_PRB}. Such interactions account for ferromagnetism of metals (the Stoner mechanism) and contribute to the magnitude of the Curie temperature $T_{\text{C}}$ in DFSs.

It is convenient to apply the Zener model of carrier-mediated ferromagnetism by introducing the functional of free energy density, ${\cal{F}}[\vec{M}(\vec{r})]$. The choice of the local magnetization $\vec{M}(\vec{r})$ as an order parameter means that the spins are treated as classical vectors, and that spatial disorder inherent to magnetic alloys is neglected. In the case of magnetic semiconductors ${\cal{F}}[\vec{M}(\vec{r})]$ consists of two terms,
\begin{equation}
{\cal{F}}[\vec{M}(\vec{r})] = {\cal{F}}_S[\vec{M}(\vec{r})] + {\cal{F}}_{\text{c}}[\vec{M}(\vec{r})],
\end{equation}
which describe, for a given magnetization  profile $\vec{M}(\vec{r})$, the free energy densities of the Mn spins in the absence of any carriers and of the carriers in the presence of the Mn spins, respectively. A visible asymmetry in the treatment of the carries and of the spins corresponds to an adiabatic approximation: the dynamics of the spins in the absence of the carriers is assumed to be much slower than that of the carriers. Furthermore, in the spirit of the virtual-crystal and molecular-field approximations, the classical continuous field $\vec{M}(\vec{r})$ controls the effect of the spins upon the carriers. Now, the thermodynamics of the system is described by the partition function $Z$, which can be obtained by a functional integration of the Boltzmann factor over all magnetization profiles $\vec{M}(\vec{r})$,
 \begin{equation}
 Z \sim \int{\text{D}}\vec{M}(\vec{r}) \exp\{-\int d\vec{r}{\cal{F}}[\vec{M}(\vec{r})]/k_{\text{B}}T\},
 \end{equation}
the approach developed in the context of DMSs for bound magnetic polarons \cite{Dietl:1983_PRB,Dietl:1983_JMMM}, and directly applicable for spin physics in quantum dots as well.

In the mean-field approximation, which should be valid for spatially extended systems and long-range spin-spin interactions, a term corresponding to the minimum of ${\cal{F}}[\vec{M}(\vec{r})]$ is assumed to determine $Z$ with a sufficient accuracy, the conclusion supported by Monte Carlo simulations discussed in Sec.~\ref{sec:limitations}.

If the energetics is dominated by spatially uniform magnetization  $\vec{M}$, the spin part of the free energy density in the magnetic field $\vec{H}$ can be written in the form
\begin{equation}
{\cal{F}}_{\text{S}}[\vec{M}] = \int_0^{\vec{M}} {\text{d}} \vec{M}_o\cdot\vec{h}(\vec{M}_o) - \vec{M}\cdot\vec{H}.
\end{equation}
Here, $\vec{h}(\vec{M}_o)$ denotes the inverse function to $\vec{M}_o(\vec{h})$, where $\vec{M}_o$ is the available experimentally macroscopic magnetization of the spins in the absence of carriers in the field $\vec{h}$ and temperature $T$, whose anisotropy is typically weak for Mn$^{2+}$ ions in the orbital singlet state. As discussed in Sec.~\ref{sec:antiferro}, it is usually possible to parametrize $M_o(h)$ by the Brillouin function B$_S(T,H)$ that takes the presence of intrinsic short-range antiferromagnetic interactions into account. Near $T_{\text{C}}$ and for $H = 0$, $M$ is sufficiently small to take $M_o(T,h) = \chi(T)h$, where $\chi(T)$ is the magnetic susceptibility of localized spins in the absence of carriers. Under these conditions,
\begin{equation}
{\cal{F}}_S[M] = M^2/2\chi(T),
\end{equation}
which shows that the increase of ${\cal{F}}_S$ with $M$ slows down with lowering temperature, where $\chi(T)$ grows. Turning to ${\cal{F}}_{\text{c}}[M]$ we note that owing to the giant Zeeman splitting of the bands proportional to $M$, the energy of the carriers, and thus ${\cal{F}}_{\text{c}}[M]$, decreases with $|M|$, ${\cal{F}}_{\text{c}}[M] -{\cal{F}}_{\text{c}}[0]\sim -M^2$. Accordingly, a minimum of ${\cal{F}}[M]$ at non-zero $M$ can develop in $H = 0$ at sufficiently low temperatures signalizing the appearance of a ferromagnetic order.

It had been postulated \cite{Dietl:2000_S}, and checked employing 40  orbitals' tight binding approximation (TBA) \cite{Werpachowska:2010_PRBb},  that the minimal Hamiltonian necessary to properly describe effects of the complex structure of the valence band in tetrahedrally coordinated semiconductors upon ${\cal{F}}_{\text{c}}[M]$ is the Luttinger  six bands' $kp$  model with the Bir-Pikus strain terms, supplemented by the $p$-$d$ exchange contribution taken in the virtual-crystal and molecular-field approximations,
\begin{equation}
{\cal{H}}_{pd} = \beta \vec{s}\cdot\vec{M}/g\mu_{\text{B}}.
\end{equation}
This term leads to spin splittings of the valence subbands, whose magnitudes---owing to the crucial role of the spin-orbit coupling---depend on the magnitude and direction of the hole wave vectors $\vec{k}$ in a complex way even for spatially uniform magnetization  $\vec{M}$. Within this formalism, the spin-orbit interaction results from the $p$-type symmetry of the periodic parts of the Bloch wave functions and the corresponding spin-orbit splitting of the valence band at the $\Gamma$ point of the Brillouin zone into $J=3/2$ and $J=1/2$ hole subbands, the $J =3/2$ subband exhibiting an additional splitting in the presence of confinement and/or strain. This effect is distinct from the $\vec{k}$-dependent Dresselhaus or Rashba spin-splitting appearing in the conduction band in the absence of inversion symmetry or if the cubic symmetry is perturbed, respectively. The incorporation of the spin-orbit interaction into the valence band model is essential \cite{Dietl:2000_S}, as it controls the magnitude of $T_{\text{C}}$ and accounts for magnetic anisotropy in DFSs with Mn in the high spin {2+} charge state for which single-ion magnetic anisotropy is small according to magnetic resonance studies \cite{Fedorych:2002_PRB,Qazzaz:1995_SSC}.

It would be technically difficult to incorporate such effects to the RKKY model, as the spin-orbit coupling leads to non-scalar terms in the spin-spin Hamiltonian. At the same time, the indirect exchange associated with the virtual spin excitations between the valence subbands, the Bloembergen-Rowland mechanism \cite{Dietl:1994_B}, is automatically included. The model allows for strain, confinement, and was developed for both zinc-blende and wurtzite materials \cite{Dietl:2001_PRB}. Furthermore, the direct influence of the magnetic field upon the hole spectrum was taken into account \cite{Dietl:2001_PRB,Jungwirth:2006_PRB,Sliwa:2006_PRB}. The aforementioned Stoner enhancement was described by introducing a Fermi-liquid-like parameter $A_{\text{F}}$ \cite{Dietl:1997_PRB,Haury:1997_PRL,Jungwirth:1999_PRB}, which enlarges the Pauli susceptibility of the hole liquid, typically by 20\% in the 3D case. No disorder effects were taken into account on the ground that their influence on {\em thermodynamic} properties is relatively weak except for the strongly localized regime. Obviously, a more elaborated parametrization of the valence band is necessary in many cases. For instance, the eight bands' model was employed to compute the infrared \cite{Hankiewicz:2004_PRB} and Hall conductivity \cite{Werpachowska:2010_PRBa} in (Ga,Mn)As, whereas the multi-orbitals' tight-binding approaches served to describe $T_{\text{C}}$ \cite{Vurgaftman:2001_PRB,Werpachowska:2010_PRBa,Jungwirth:2005_PRBa} and interlayer coupling \cite{Sankowski:2005_PRB} in  GaAs/(Ga,Mn)As superlattices.

Having the hole energies, the free energy density ${\cal{F}}_{\text{c}}[\vec{M}]$ was evaluated according to the procedure suitable for Fermi liquids of arbitrary degeneracy, {\em i.e.}, taking the carrier entropy into account. By minimizing  ${\cal{F}}[\vec{M}] = {\cal{F}}_S[\vec{M}] + {\cal{F}}_{\text{c}}[\vec{M}]$ with respect to $\vec{M}$ at a given $T$, $\vec{H}$, and hole concentration $p$, Mn spin magnetization  $\vec{M(T,H)}$ was obtained as a solution of the mean-field equation,
\begin{equation}
\vec{M}(T,H) = x_{\text{eff}}N_0g\mu_{\text{B}}S\mbox{B}_S\left[\frac{g\mu_{\text{B}}(-\partial {\cal{F}}_{\text{c}}[\vec{M}]/\partial \vec{M} + \vec{H})}{k_{\text{B}}(T+T_{AF})}\right],
\label{eq:MF}
\end{equation}
where the peculiarities of the valence band structure, such as the presence of various hole subbands, spin-orbit coupling, crystalline cubic and strain-induced anisotropies are hidden in $F_{\text{c}}[\vec{M}]$.

\subsubsection{Theory of the Curie temperature}
\label{sec:Tc_theory}
Near the Curie temperature $T_{\text{C}}$ and at $H = 0$, where $M$ is small and the free energy is an even function of $M$, one expects ${\cal{F}}_{\text{c}}[M] - {\cal{F}}_{\text{c}}[0] \sim -M^2$. It is convenient to parameterize this dependence by a generalized carrier spin susceptibility $\tilde{\chi}_{\text{c}}$, which is related to the magnetic susceptibility of the carrier liquid according to $\chi_{\text{c}} = A_{\text{F}}(g^*\mu_{\text{B}})^2\tilde{\chi}_{\text{c}}$. In terms of $\tilde{\chi}_{\text{c}}$,
\begin{equation}
{\cal{F}}_{\text{c}}[M] = {\cal{F}}_{\text{c}}[0] - A_{\text{F}} \tilde{\chi_{\text{c}}}\beta^2M^2/2(g\mu_{\text{B}})^2.
\end{equation}
By expanding B$_S(M)$ for small $M$ and introducing the spin susceptibility of the magnetic ions in the absence of carriers, $\tilde{\chi}_S = \chi/(g\mu_{\text{B}})^2$, one arrives at the mean-field formula making it possible to determine $T_{\text{C}}$,
 \begin{equation}
 A_{\text{F}}\beta^2\tilde{\chi}_S(T_{\text{C}},\vec{q})\tilde{\chi}_{\text{c}}(T_{\text{C}}, \vec{q}) = 1,
 \label{eq:MFA}
\end{equation}
 where $\beta$ should be replaced by the $s-d$ exchange integral $\alpha$ in the case of electrons and $\vec{q}$ denotes the Fourier component of the magnetization texture, for which $T_{\text{C}}$ attains the highest value. For the spatially uniform magnetization $q=0$ , in terms of $x_{\text{eff}}$ and $T_{\text{AF}}$,
 \begin{equation}
 T_{\text{C}} = T_{\text{F}} - T_{\text{AF}},
 \label{eq:T_C}
 \end{equation}
 where $T_{\text{F}}$ is given by
\begin{equation}
T_{\text{F}} = x_{\text{eff}}N_0S(S+1)A_{\text{F}}\tilde{\chi}_{\text{c}}(T_{\text{C}})\beta^2/3k_{\text{B}},
\label{eq:T_F}
\end{equation}
with the cation concentration $N_0 = 4/a_0^3$, $4/(\sqrt{3}a^2c)$, and $8/a_0^3$ for the zinc-blende, wurtzite, and elemental diamond-structure DFSs, respectively. As discussed in Sec.~\ref{sec:p-d}, for holes in tetrahedrally bound semiconductors the exchange integral $\beta = - 54$~meV\,nm$^3$, except perhaps for mercury chalcogenides, in which the value of $\beta$ appears somewhat smaller \cite{Furdyna:1988_B}. For other lattice structures, different combinations of hybridization matrix elements describe an appropriate exchange integral characterizing coupling between carriers and localized spins \cite{Dietl:1994_PRB}.

In the 3D case, typically, $A_{\text{F}} \approx 1.2$. For a strongly degenerate carrier liquid $|\epsilon_{\text{F}}|/k_{\text{B}}T \gg 1$, $\tilde{\chi}_{\text{c}} = \rho_{\text{s}}/4$, where $\rho_{\text{s}}$ is the total DOS for intra-band spin excitations at the Fermi level, typically reduced by spin-orbit interactions from DOS for charge excitations $\rho_{\text{F}}$. An analytic form of $\rho_{\text{s}}$ was derived for the four band Luttinger model  \cite{Ferrand:2001_PRB}. In the absence of spin-orbit interactions and in the 3D case it is given by $\rho_{\text{s}} = \rho_{\text{F}} = m^*_{\text{DOS}}k_{\text{F}}/\pi^2\hbar^2$.  In this case and for $A_{\text{F}} = 1$, $T_{\text{F}}$ assumes the well-known form, obtained already in the 1940s in the context of carrier-mediated nuclear ferromagnetism \cite{Frohlich:1940_PRSL} and in the 1970s in the context of DMSs \cite{Pashitskii:1979_SPSS}. In general, however,  $\tilde{\chi}_{\text{c}}$ has to be determined numerically by computing ${\cal{F}}_{\text{c}}[M]$ for a given band structure and degeneracy of the carrier liquid.

The above model predicts $T_{\text{F}}$ to be much higher for holes than for electrons for two reasons: (i) the density of states $\rho_{\text{s}}$ is typically lower in the conduction band (though effects of spin-orbit interactions are weaker); (ii)   the $s-d$ exchange integral $\alpha$ is typically over 4 times smaller than the $p$-$d$ integral $\beta$. Section~\ref{sec:comparison} presents a comparison of these predictions to experimental data.

As described earlier, $T_{\text{F}}$ can be computed by minimizing the free
energy, and without referring to the explicit form of the
Kohn-Luttinger amplitudes $u_{i{\vec{k}}}$. Since near $T_{\text{C}}$
the relevant magnetization $M$ is small, $\tilde{\chi}_{\text{c}}$ can also be determined from the linear response
theory. The corresponding $\rho_{\text{s}}$ assumes the form \cite{Dietl:2001_PRB},
\begin{equation}
\rho_{\text{s}} = \lim_{q \rightarrow 0} 8\sum_{ij{\vec{k}}}
\frac{|\langle u_{i,{\vec{k}}}|s_M|u_{j,{\vec{k}}+{\vec{q}}}\rangle|^2 f_i({\vec{k}})[1 -
f_j({\vec{k}}+{\vec{q}})]} {E_j({\vec{k}}+{\vec{q}})- E_i({\vec{k}})},
\label{eq:linh}
\end{equation}
where $s_M$ is the component of spin operator along the direction
of magnetization and $f_i({\vec{k}})$ is the Fermi-Dirac
distribution function for the $i$-th valence band subband. A quantitative analysis
 demonstrated that typically a 30\% contribution to $T_{\text{C}}$
originates from interband polarization (the Bloembergen-Rowland mechanism) involving light and heavy hole subbands \cite{Dietl:2001_PRB}.
This formalism was extended to {\em bulk} states in topological insulators, in which $u_{i{\vec{k}}}$ for both the valence and conduction
bands have a $p$-like symmetry, so that appreciable $T_{\text{F}}$ can be expected from interband polarization even if the Fermi energy resides in the band gap \cite{Yu:2010_S}.

\subsection{Theory of carrier-controlled Curie temperature in reduced dimensionality and topological insulator systems}

In thin films, heterostructures, and superlattices, owing to the formation of interfacial space charge layers, the hole density and corresponding Curie temperatures $T_{\text{C}}[p(z)]$  are non-uniform even for a uniform distribution of acceptors and donors. The role of non-uniformity in the carrier distribution grows on reducing the thickness $t$ of magnetic layers, and is particularly relevant in those structures in which $p(z)$ can be tuned electrostatically, for instance, by the gate voltage.  When $t$ is larger than the phase coherence length $L_{\phi}$, the region with the highest $T_{\text{C}}$ value determines $T_{\text{C}}$ of the whole structure. If, however, $t < L_{\phi}$ the value of local magnetization $M(z)$ and $T_{\text{C}}$ are determined by the distribution $p(z)$ across the {\em whole} channel thickness. In this regime two situations were considered:

If disorder is strong,  $\ell < t$, the scattering broadening makes dimensional quantization irrelevant although quantum mechanical non-locality remains important. Under these conditions the magnitude of layer's $T_{\text{F}}$ can be expressed as \cite{Sawicki:2010_NP,Nishitani:2010_PRB},
\begin{equation}
T_{\text{F}} = \int \mbox{d}z T_{\text{F}}[p(z)]\int \mbox{d}z p(z)^2/[\int \mbox{d}z p(z)]^2,
\label{eq:T_F_FET}
\end{equation}
where $T_{\text{F}}(p)$ is to be determined from the relevant 3D model and $p(z)$ is to be evaluated from the Poisson equation taking into account the pinning of the Fermi energy by surface states. It was predicted within the $p$-$d$ Zener model that the energy position $E_{\text{s}}$ of surface states should strongly affect the efficiency of $T_{\text{C}}$ tuning by the gate voltage $V_{\text{G}}$ (see, \onlinecite{Ohno:2013_JAP} and Sec.~\ref{sec:comparison}).

The opposite limit of a weak disorder $\ell \gg t$, relevant to modulation-doped II-VI heterostructures, was also considered \cite{Dietl:1997_PRB,Haury:1997_PRL}.  Owing to a typically large confinement-induced splitting between heavy and light hole subbands, only one ground state heavy hole subband is occupied, for which the $p$-$d$ exchange is of the Ising type, ${\cal{H}}_{pd} =-N_0\beta s_zS_z$, so that
\begin{equation}
T_{\text{F}} = N_0x_{\text{eff}}S(S+1)A_{\text{F}}\beta^2m^*/(12\pi\hbar^2k_{\text{B}}\tilde{L}_{\text{W}}),
\label{eq:T_F_2D}
\end{equation}
Here $m^*$ denotes the in-plane effective mass and $\tilde{L}_{\text{W}}$ is an effective width of the region occupied by carriers relevant to ferromagnetism given by \cite{Haury:1997_PRL,Dietl:1999_MSEB},
\begin{equation}
\tilde{L}_{\text{W}} = 1/\int \mbox{d}z|\phi(z)|^4.
\label{eq:L_W}
\end{equation}
where $\phi(z)$ is an envelope function of the relevant 2D subband. As seen, owing to a step-like form of DOS in the 2D case, $T_{\text{F}}$ does not depend on the hole density in this case. The expression for $T_{\text{F}}$ was generalized further on to the case of arbitrary degeneracy of the hole liquid and by including effects of disorder {\em via} scattering broadening of DOS \cite{Boukari:2002_PRL}.

The case of high carrier concentrations leading to the occupation of several hole subbands in (Ga,Mn)As based multilayer structures was also considered by incorporating an LSDA approach to two \cite{Jungwirth:1999_PRB,Giddings:2008_PRB} and four \cite{Fernandez-Rossier:2002_PRB} band $kp$ models, whereas the multi-orbitals' tight-binding approaches served to describe $T_{\text{C}}$ \cite{Vurgaftman:2001_PRB,Werpachowska:2010_PRBa} and interlayer coupling \cite{Sankowski:2005_PRB} in  GaAs/(Ga,Mn)As superlattices.

Theoretical approaches were developed allowing to evaluate Curie temperature for ferromagnetic ordering of magnetic impurities mediated by Dirac carriers at the surface of 3D topological insulators \cite{Liu:2009_PRL,Abanin:2011_PRL}. An Ising type of exchange was assumed, $H_{\text{ex}} =-N_0J_z s_zS_z$, leading to a gapped dispersion given by,
\begin{equation}
\epsilon(\vec{k}) = \pm [(J_zM/2g\mu_{\text{B}})^2 + (\hbar v_{\text{f}}k)^2]^{1/2},
\end{equation}
where $v_{\text{f}}$ is the Fermi velocity. For such a case (cf.~\onlinecite{Liu:2009_PRL}), in our notation,
\begin{equation}
T_{\text{F}} = N_0x_{\text{eff}}S(S+1)A_{\text{F}}rJ_z^2(E_{\text{c}} -|\epsilon_{\text{F}}|)/(24\pi\hbar^2k_{\text{B}}v_{\text{f}}^2\tilde{L}_{\text{W}}),
\end{equation}
where $N_0 = 12/(\sqrt{3}a^2c)$ and $4/a_0^3$ in hexagonal III-V ({\em e.~g.}, (Bi$_{1-x}$Mn$_x)_2$Se$_3$) and cubic II-VI or IV-VI ({\em e.~g.}, Sn$_{1-x}$Mn$_x$Te) compounds, respectively; $r$ is the number of Dirac cones at a given surface; $E_{\text{c}}$ is a cut-off energy associated with the termination of the Dirac surface band, and $\tilde{L}_{\text{W}}$ is the penetration depth of Dirac carriers related to their envelop function according to Eq.~\ref{eq:L_W}.

A formalism suitable to evaluate $T_{\text{F}}$ determined by bulk states of topological insulators is presented at the end of Sec.~\ref{sec:Tc_theory}.

The formation of spin-density waves is expected in the case of carrier-mediated ferromagnetism in 1D systems \cite{Dietl:1999_MSEB}.

\subsection{Theory of magnetization and hole polarization}
\label{sec:magnetization-theory}
The mean-field Eq.~\ref{eq:MF} allowed to determine Mn magnetization $M(T,H)$, particularly $M(T)$ at $T \le T_{\text{C}}$ \cite{Dietl:2001_PRB} and $M(H)$ at $T_{\text{C}}$ \cite{Sliwa:2011_PRB}.  The same formalism also provided quantitative information on the value of thermodynamic hole spin polarization \cite{Dietl:2001_PRB},
\begin{equation}
{\mathcal{P}} = \frac{2g\mu_B}{\beta p}\frac{\partial
{\cal{F}}_c(M)}{\partial M},
\label{eq:pol}
\end{equation}
which, despite the spin-orbit interaction, can approach 90\% in the relevant range of hole and Mn densities in (Ga,Mn)As but gets reduced down to about 50\% at high hole densities \cite{Dietl:2001_PRB}.

Furthermore, hole magnetization $M_{\text{c}}$, which determines the magnitude of spontaneous magnetization $M_{\text{s}}(T)= M(T) + M_{\text{c}}(M)$ was evaluated taking into account the effect of a magnetic field on the valence band \cite{Dietl:2001_PRB,Sliwa:2006_PRB,Sliwa:2013_arXiv}. It was found that holes reduce (Ga,Mn)As magnetization by about 10\%, so that the value of the magnetic moment per one Mn ion can be taken as $\mu\simeq 4.5\mu_{\text{B}}$ in ferromagnetic samples of (Ga,Mn)As weakly compensated by donors. The hole contribution is, therefore, about 2 times smaller than would be in the absence of spin-orbit coupling and for fully spin-polarized hole gas.

Theoretical studies of magnetic stiffness discussed in Sec.~\ref{sec:magnetic_stiffeess} made it possible to evaluate a reduction of $M(T)$ by spin wave excitations \cite{Konig:2001_PRB,Werpachowska:2010_PRBb}.
\subsection{Theory of magnetic anisotropy and magneto-elasticity}
\label{sec:anisotropy-theory}

Since Mn$^{2+}$ ions are in the orbital singlet $^6$A$_1$ state in DFSs, a single ion magnetic anisotropy is small \cite{Fedorych:2002_PRB,Edmonds:2006_PRL}, so that the dominant contribution comes from spin-orbit effects within hole band states \cite{Dietl:1997_PRB,Dietl:2000_S}. Owing to an interplay of the spin-orbit interaction with the crystal structure anisotropy, strain, and confinement the characteristic crystalline magnetic anisotropy fields $H_{\text{a}}$ are typically larger than the shape term $\mu_o H_{\text{d}} = \mu_oM_{\text{s}} \approx 0.13$~T for a (Ga,Mn)As thin film with $x_{\text{eff}} = 10$\% \cite{Dietl:2001_PRB}. Similarly, in the case of (Ga,Mn)As nanobars, crystalline magnetic anisotropy determined by strain distribution specific to free standing strained nanostructures dominates over the shape term \cite{Humpfner:2007_APL}.

Accordingly, the theoretically expected character and magnitude of crystalline magnetic anisotropy were obtained by considering how the carrier free energy density ${\cal{F}}_{\text{c}}[\vec{M}]$ depends on the direction of the magnetization vector $\vec{M}$ with respect to crystallographic axes at various values of epitaxial strain \cite{Dietl:2000_S}. Following subsequent detail studies for (Ga,Mn)As epilayers \cite{Dietl:2001_PRB,Abolfath:2001_PRB},  further theoretical analysis of anisotropy energy coefficients $K_i$ and anisotropy fields $H_i$ were carried out for the canonical (001) films \cite{Zemen:2009_PRB} as well as, additionally, for an arbitrary (11n) substrate orientation \cite{Stefanowicz:2010_PRBb}, the accomplishments discussed {\em vis \`a vis} experimental findings in Sec.~\ref{sec:anisotropy-comparison}. The formalism was developed for an arbitrary form of the strain tensor $\bm{\epsilon}$ and it is valid as long as non-linear strain effects are not significant. It was checked that terms linear in products of $k_i$ and $\epsilon_{ij}$ can be neglected \cite{Stefanowicz:2010_PRBb}.

A sizable strength of crystalline magnetic anisotropy and related magneto-elastic phenomena, comparable to ferromagnetic metals despite much smaller magnetic ion concentrations, comes from a large spin-orbit splitting of the valence band (about 0.3~eV for arsenides and 1~eV for tellurides), greater than the kinetic energy of holes \cite{Dietl:2001_PRB}.

\subsection{Theory of micromagnetic parameters and spin wave dispersion}
\label{sec:magnetic_stiffeess}
Similarly to other ferromagnets, a description of magnetization processes for various orientations of the external magnetic field $\vec{H}$ as well as the understanding of the domain structure require information not only on the magnetic anisotropy but also on the exchange stiffness.  These two micromagnetic characteristics correspond to energy penalties associated with (i) deviation of magnetization orientation from an easy direction, as described above and (ii) local twisting of magnetization from its global direction, respectively.

The exchange stiffness $A$ and the related spin wave dispersion $\omega(\vec{q})=Dq^2$, where $D = 2g\mu_{\text{B}}A/M$,  were theoretically determined for DFSs by examining the $q$ dependent part of the {\em hole} spin susceptibility $\tilde{\chi}_{\text{c}}(\vec{q})$ at a given average Mn magnetization $M$ \cite{Konig:2001_PRB,Brey:2003_PRB,Werpachowska:2010_PRBb}. In general, taking the presence of a spin-orbit interaction into account, $D$ is a tensor and, moreover, terms linear in $q$ can appear. Their magnitude and possible effects were analyzed within a multi-orbital tight binding model for thin films of (Ga,Mn)As \cite{Werpachowska:2010_PRBb}. A magnetic cycloid ground state was predicted for a few monolayer thick (Ga,Mn)As films.

These works  made it possible to evaluate the width of the Bloch domain wall,
\begin{equation}
\delta_{\text{W}} = \pi(A/K)^{1/2},
\end{equation}
which is the shortest length scale of the micromagnetic theory. It was found \cite{Dietl:2001_PRBb} that over the relevant range of material parameters $\delta_{\text{W}} \gtrsim 15$~nm stays much longer than a mean distance between holes and Mn ions in ferromagnetic (Ga,Mn)As. This evaluation substantiated the validity of the continuous medium approximation, employed in the approach exposed in this chapter. Moreover, it pointed out that the time honored micromagnetic theory, presented for (Ga,Mn)As-type ferromagnets in Appendix, and corresponding software packages, are also applicable to these systems.

The Gilbert damping parameter $\alpha_{\text{G}}$ due to particle-hole excitations in the (Ga,Mn)As valence band was evaluated first neglecting \cite{Sinova:2004_PRB,Tserkovnyak:2004_APL}, and then taking into account quantitatively important vertex corrections within the four band model \cite{Garate:2009_PRBb}. A monotonic decrease of $\alpha_{\text{G}}$ with the hole scattering rate was found. Within a similar model, a magnitude of non-adiabatic spin torque $\beta_{\text{w}}$ was evaluated and found to be of the order of one \cite{Hals:2009_PRL}. It would be interesting to find out how a finite value of spin-orbit splitting between $\Gamma_8$ and $\Gamma_7$ bands as well as localization and correlation effects will affect these conclusions.

A verification of the present theory by examining the temperature dependence of magnetization and specific heat is presented in Sec.~\ref{sec:magnetization},  whereas Sec.~\ref{sec:spin_waves} contains a comparison of experimental results to theoretical predictions on the domain structure and spin wave excitations.  Section~\ref{sec:current_domains} contains information on experimental determination of $\beta_{\text{w}}$.

\subsection{Limitations of the mean-field {$p$-$d$} Zener model}
\label{sec:limitations}
\ \\
{\em Material parameters} -- The model is parametrized by the lattice constant $a_0$; spin-orbit splitting $\Delta_0$; Luttinger parameters ($\gamma_1$,  $\gamma_2$, and $\gamma_3$ in the zinc-blende case when the six band Luttinger Hamiltonian is employed); exchange integral $\beta$; Landau's Fermi liquid parameter $A_{\text{F}}$, and--for non-zero strain--by elastic moduli $c_{ij}$ and two deformation potentials of the valence band, $b$ and $d$. Two of these parameters, $\beta$ and $A_{\text{F}}$, are known by now with an accuracy not better than 10\%,  which leads to the accumulated error in calculated $T_{\text{C}}$ values of the order of 25\%. Additionally, a quantitative verification of any DFS theory is challenging because of difficulties in assessing real hole and Mn concentrations that, moreover, are often non-uniformly distributed over the film volume, as discussed in Sec.~\ref{sec:growth}.

\ \\
{\em Thermodynamic magnetization fluctuations} --
The question how various corrections to the mean-field and continuous medium approximation affect theoretical values of $T_{\text{C}}$  was addressed in some details \cite{Jungwirth:2002_PRB,Brey:2003_PRB,Timm:2005_PRB,Jungwirth:2005_PRBa,Popescu:2006_PRB,Yildirim:2007_PRL}. It was found that the mean-field $p$-$d$ Zener model remains quantitatively valid for (Ga,Mn)As and related systems, typical lowering of $T_{\text{C}}$ values by magnetization fluctuations being below 20\%, though a value of 30\% was found in the most recent study \cite{Yildirim:2007_PRL}, if a correction for the classical spin approximation adopted in that work is taken into account.  According to Monte Carlo simulations for the 2D case, the fluctuations of magnetization diminish $T_{\text{C}}$ by a factor of 2 in the absence of competing antiferromagnetic interactions, whereas  in the presence of these interactions, a net quantitative correction to the mean-field approximation is much reduced \cite{Lipinska:2009_PRB}.

\ \\
{\em Antiferromagnetic interactions} --
According to results presented in Sec.~\ref{sec:antiferro}, carrier-mediated interactions compete with short-range superexchange coupling between Mn ions in cation-substitutional and/or interstitial positions. As discussed above, the presence of these antiferromagnetic interactions can be incorporated into the $p$-$d$ Zener model by introducing two parameters, $x_{\text{eff}} < x$ and $T_{\text{AF}} > 0$. Additionally, the short-range antiferromagnetic interaction enhances the importance of the antiferromagnetic portion of the RKKY coupling leading, for hole densities comparable to the concentration of localized spins, to a further reduction of $T_{\text{C}}$ values comparing to those expected from Eq.~\ref{eq:T_C}  \cite{Ferrand:2001_PRB}. Actually, in this limit, $p \gtrsim N_0x$, randomness of the interaction type (ferro {\em vs.} antiferro) associated with RKKY oscillations can drive the system towards a spin-glass phase rather than towards a ferromagnetic ground state expected within the mean-field approximation \cite{Eggenkamp:1995_PRB}. The stability of the ferromagnetic phase is, however, much enhanced in III-V and II-VI DFSs by multiband structure and strong anisotropy of the valence band \cite{Timm:2005_PRB}.

\ \\
{\em Kondo effect} -- The theory is developed for Mn concentrations high enough that magnetic ordering temperature is higher than the Kondo temperature, evaluated for II-VI p-type DMSs to be of the order of 1~K \cite{Dietl:1997_PRB}.

\ \\
{\em Effects of disorder and localization} --
The understanding of the interplay between carrier-mediated ferromagnetism and carrier localization is an emerging field of research \cite{Sheu:2007_PRL,Dietl:2008_JPSJ,Sawicki:2010_NP,Richardella:2010_S}. The relevant questions here are how the presence of spins affect carrier localization and how carrier-mediated ferromagnetism is influenced by localization.  According to experimental investigations of (Ga,Mn)As \cite{Matsukura:1998_PRB} and (Zn,Mn)Te \cite{Ferrand:2001_PRB}, the magnitude of $T_{\text{C}}$, similarly to other thermodynamic properties, shows no critical behavior at the MIT.  A non-critical behavior of $T_{\text{C}}$ across the MIT stems from the scaling theory of the Anderson-Mott transition. This theory implies that an average hole localization length, which diverges at the MIT, remains much greater than the mean distance between acceptors for the experimentally important range of hole densities.  Thus, holes can be regarded as band-like at the length scale relevant to coupling between magnetic ions. Hence, the spin-spin exchange interactions are effectively mediated by the itinerant carriers, so that the $p$-$d$ Zener model can serve to evaluate $T_{\text{C}}$, also on the insulator side of the MIT as long as holes remain only weakly localized. This view was supported by results of inelastic neutron scattering of nearest neighbor Mn pairs in p-(Zn,Mn)Te \cite{Kepa:2003_PRL}. In this experiment, the hole-induced change in the pair interaction energy shows the value expected for the band carriers despite that the studied sample was on the insulator side of the MIT.

As already mentioned, disorder introduces a certain life-time broadening of DOS, the effect equivalent to the lowering of $T_{\text{C}}$ by a finite mean free path within the RKKY theory and being captured within, {\em e.~g.}, the coherent potential approximation \cite{Jungwirth:2005_PRBa}. The broadening can also be phenomenologically introduced to the $p$-$d$ Zener model \cite{Dietl:1997_PRB,Boukari:2002_PRL}, typically diminishing the magnitude of DOS at $\epsilon_{\text{F}}$ and, thus of $T_{\text{C}}$.   It was suggested \cite{Dietl:2001_PRB} that this effect would destroy theoretically predicted oscillations in the magnitude of the cubic anisotropy field as a function of the Fermi level (hole concentration), as their period is smaller than the expected broadening energy of relevant $k$ states. It is important to note that in contrast to the one-particle DOS which determines, for instance, photoemission spectra and tunneling current \cite{Altshuler:1985_B,Pappert:2006_PRL,Richardella:2010_S}, the DOS for intra-band excitations, relevant to $T_{\text{C}}$ does not exhibit any interaction and disorder-induced Coulomb anomaly at the Fermi energy as well as does not show a critical behavior across the MIT.

\begin{figure}
\centering
\includegraphics[scale=0.35]{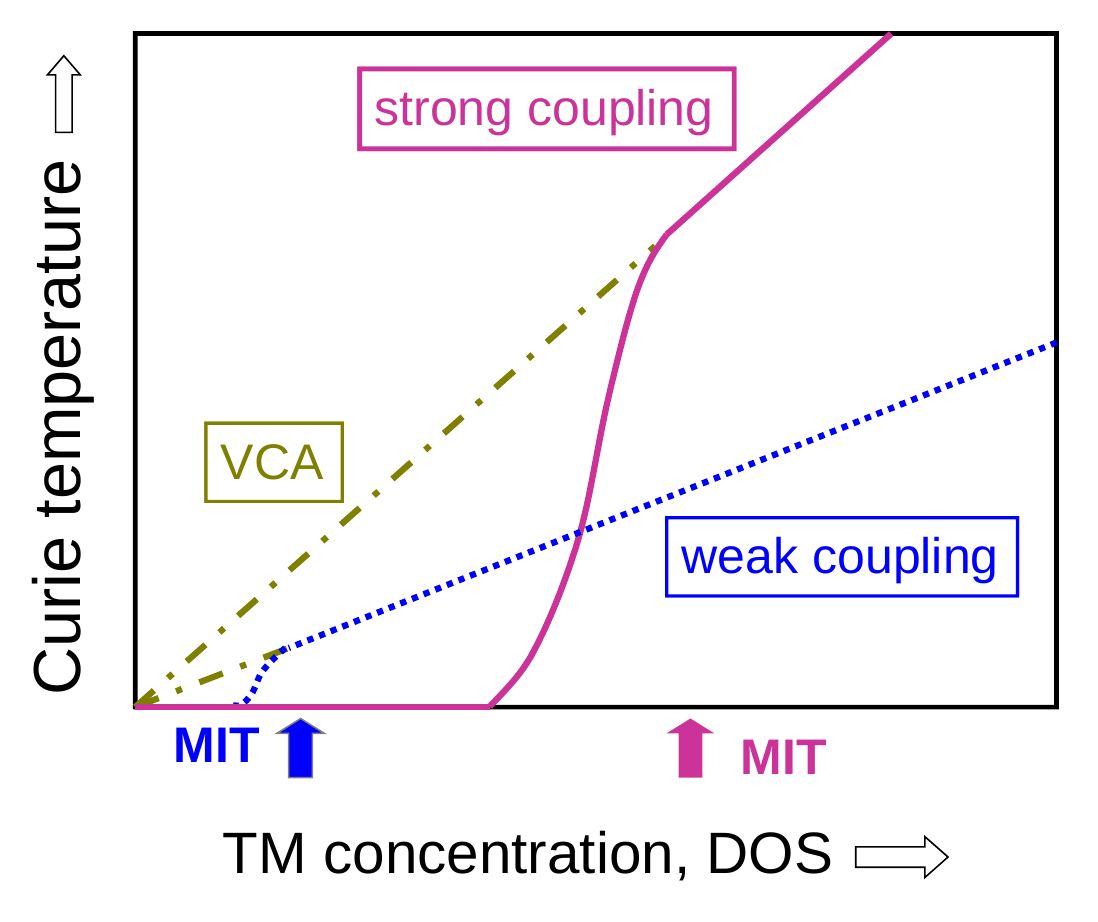}
\caption[]{(Color online) Schematic dependence of $T_{\text{C}}$ on the magnetic ion concentration and density of hole states at the Fermi level for a weak and a strong $p$-$d$ hybridization. Higher values of $T_{\text{C}}$ are predicted within  molecular field and virtual crystal approximations (VCA) for the strong coupling. However, the region where the holes are not localized and mediate the spin-spin interaction is wider in the weak coupling case. From \onlinecite{Dietl:2008_PRB}.}
\label{fig:MIT_VCA}
\end{figure}

Importantly, not only ferromagnetic correlations but also disorder effects, particularly, carrier localization depends on the strength of the $p$-$d$ interaction in DFSs. This dual effect of the $p$-$d$ coupling is sketched in Fig.~\ref{fig:MIT_VCA} \cite{Dietl:2008_PRB}. According to results presented in Secs.~\ref{sec:electronic_states} and \ref{sec:p-d}, on going from the weak to the strong coupling regime, {\em i.~e.,} to materials with a short bond length (nitrides, oxides), the magnitude of $p$-$d$ hybridization and, hence, the TM binding energy $E_{\text{I}}$ get progressively enhanced, which shifts the critical hole density $p_{\text{c}}$ for the MIT towards correspondingly higher values, narrowing the hole concentration range where the carrier-mediated ferromagnetism can appear \cite{Dietl:2008_PRB}. At the same time, however,  according to Eq.~\ref{eq:T_F}, the magnitude of the characteristic ferromagnetic temperature $T_{\text{F}}$ increases with $N_0$, that is  when the cation-anion distance diminishes \cite{Dietl:2000_S}. It is still an open question whether the MIT and, thus, the region of high $T_{\text{C}}$ values can be experimentally achieved in nitrides and oxides.

Another consequence of carrier localization is the presence of static nano-scale fluctuations in the local DOS, discussed in Sec.~\ref{sec:carrier_distribution}. These fluctuations, typically accompanied by competing ferromagnetic and antiferromagnetic interactions, lead to a static phase separation into regions differing in the magnitude of hole density and, therefore, in the strength of ferromagnetic correlations \cite{Dietl:2000_S,Dietl:2007_JPCM,Sawicki:2010_NP}. Such an electronic phase separation leads to the appearance of randomly oriented ferromagnetic bubbles that start to develop at $T^* >  T_{\text{C}}$ \cite{Mayr:2002_PRB,Geresdi:2008_PRB} and tend to order at $T \ll T_{\text{C}}$ \cite{Sawicki:2010_NP}. Within this picture, the localization-induced disappearance of carrier-mediated ferromagnetism proceeds {\em via} a growing participation of superparamagnetic regions, leading to melting away of the percolating ferromagnetic cluster. Eventually, when antiferromagnetic interactions will start to dominate, a spin-glass phase will set in at low temperatures  (see, Sec.~\ref{sec:antiferro}). While excellent micromagnetic properties are expected deeply in the metallic regime, where the $p$-$d$ Zener model should be quantitatively correct, the diminished volume of ferromagnetic regions at lower hole densities makes the model only qualitatively valid. Alternatively, in weakly compensated DFS samples, a ferromagnetic superexchange or double exchange are expected to appear in the strongly localized regime (see, Sec.~\ref{sec:ferro-superexchange}).

\section{Comparison to experimental results}
\label{sec:comparison}
In this section a detailed comparison of experimental and theoretical results is presented for III-V and II-VI DFSs, for which relevant material parameters have been already determined. It is expected that further works on other compounds will allow for a quantitative description of magnetism also in those systems.

\subsection{Curie temperature}
\subsubsection{Chemical trends in III-V DFSs}
\label{sec:Chemical_trends}

\begin{figure}[ht]
\begin{center}
\includegraphics[width=7.0cm]{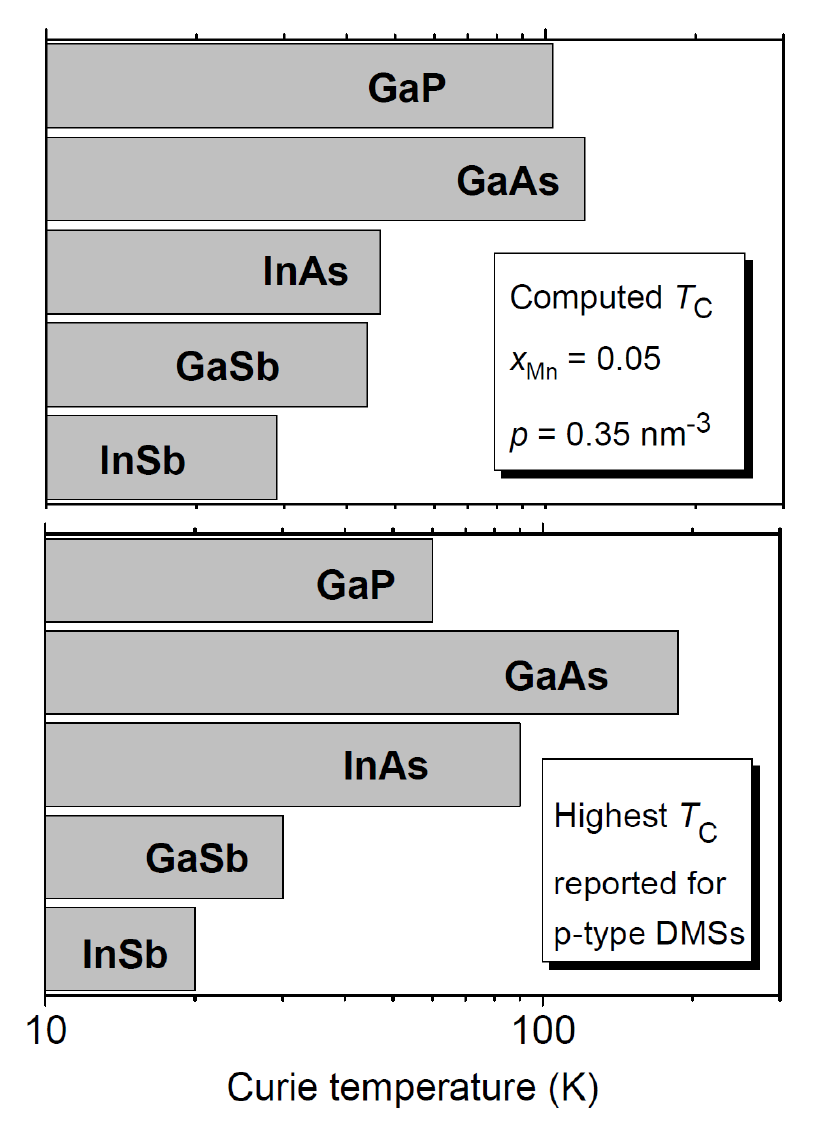}
\caption[]{Predictions of the $p$-$d$ Zener model compared to experimental data for p-type (III,Mn)V DMSs. Upper panel: computed values of the Curie temperature $T_{\text{C}}$ for various p-type semiconductors containing 5\% of Mn and $3.5\times 10^{20}$ holes per cm$^3$ \cite{Dietl:2000_S}; the value for (In,Mn)Sb is taken from \cite{Jungwirth:2002_PRB}. Lower panel: the highest reported values for (Ga,Mn)P \cite{Scarpulla:2005_PRL}; (Ga,Mn)As \cite{Olejnik:2008_PRB,Wang:2008_APL,Chen:2009_APL}; (In,Mn)As \cite{Schallenberg:2006_APL}); (Ga,Mn)Sb \cite{Abe:2000_PE}; (In,Mn)Sb \cite{Wojtowicz:2003_APL}. Adapted from \onlinecite{Dietl:2010_NM}.}
\end{center}
\label{fig:TC}
\end{figure}

In Fig.~\ref{fig:TC} the highest values of $T_{\text{C}}$ found to date in p-type Mn-based III-V DMSs are reported \cite{Scarpulla:2005_PRL,Olejnik:2008_PRB,Wang:2008_APL,Chen:2009_APL,Schallenberg:2006_APL,Abe:2000_PE,Wojtowicz:2003_APL}, and compared to the early predictions of the $p$-$d$ Zener model \cite{Dietl:2000_S,Jungwirth:2002_PRB} for fixed values of the Mn and hole concentrations. We see that the theory reproduces the chemical trends and describes semi-quantitatively the absolute values of $T_{\text{C}}$. The observed trend reflects a decrease of the $p$-$d$ exchange energy $N_0\beta$ for larger cation-anion distances as well as an enhanced role of the competing spin-orbit interaction in materials with heavier anions. However, a comparison of (In,Mn)As and (Ga,Sb)Mn or (Ga,Mn))As and (Ga,Mn)P in Fig.~\ref{fig:TC} indicates that the values of hole effective masses in particular compounds are relevant, too.

\subsubsection{Curie temperatures in (Ga,Mn)As and related systems}
\label{sec:Curie}

\begin{figure}
\includegraphics[width=8.2cm]{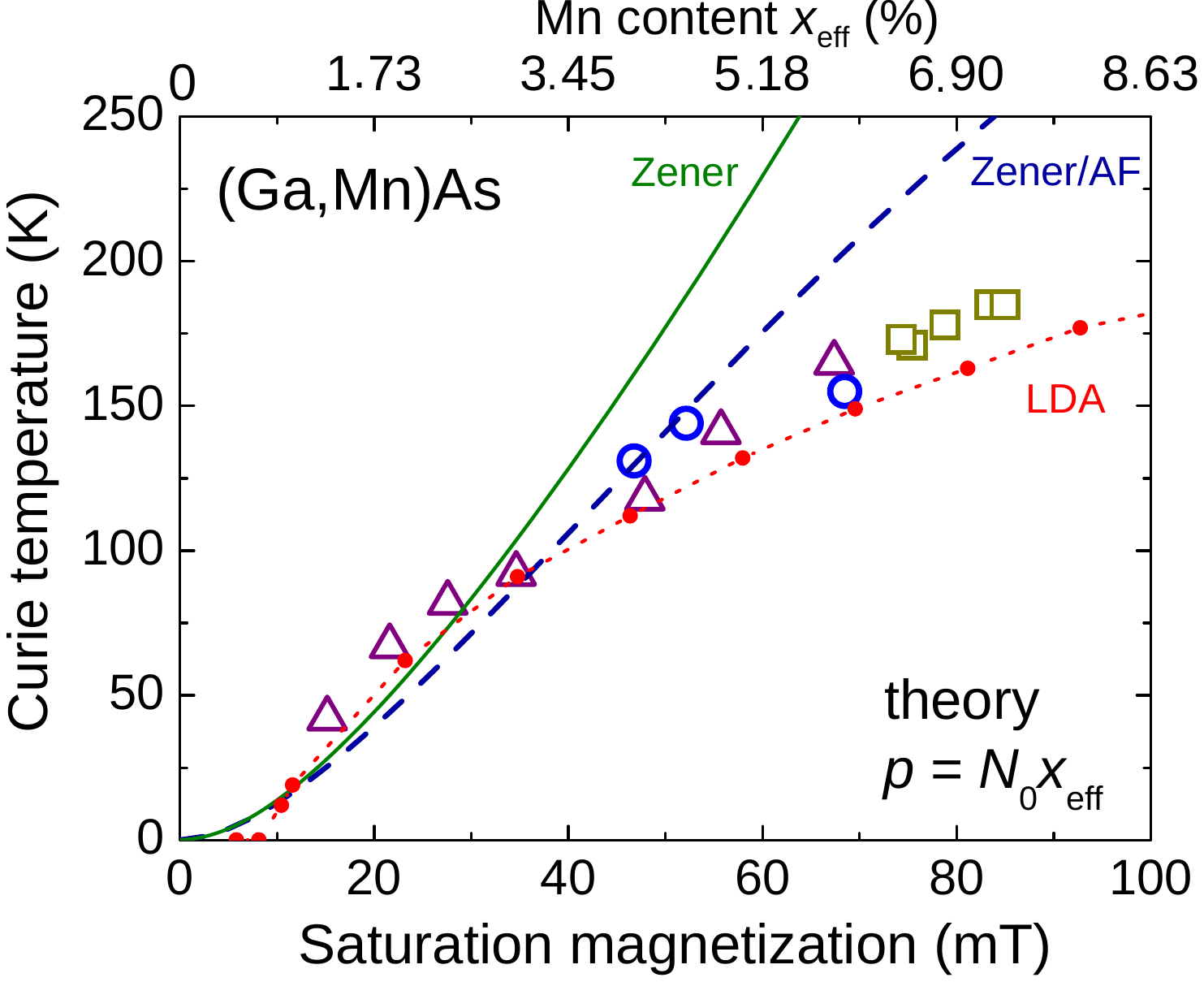}
\caption[]{(Color online) Curie temperature $T_{\text{C}}$ as a function of saturation magnetization $M_{\text{Sat}}$ for annealed (Ga,Mn)As films grown in various molecular beam epitaxy (MBE) systems \cite{Wang:2008_APL}. The solid line represents the $p$-$d$ Zener model \cite{Dietl:2000_S} assuming that the hole concentrations are equal to Mn concentrations $N_0x_{\text{eff}}$ contributing to saturation magnetization, which overestimates $T_{\text{C}}$ in the presence of Mn interstitials. Dashed line was obtained including phenomenologically antiferromagnetic interactions between Ga-substitutional Mn spins. Results of computations within one of {\em ab initio} approaches \cite{Sato:2010_RMP} are also shown for a comparison.}
\label{fig:GaMnAs_TC_Ms_theory}
\end{figure}

\begin{figure}
\includegraphics[width=8.6cm]{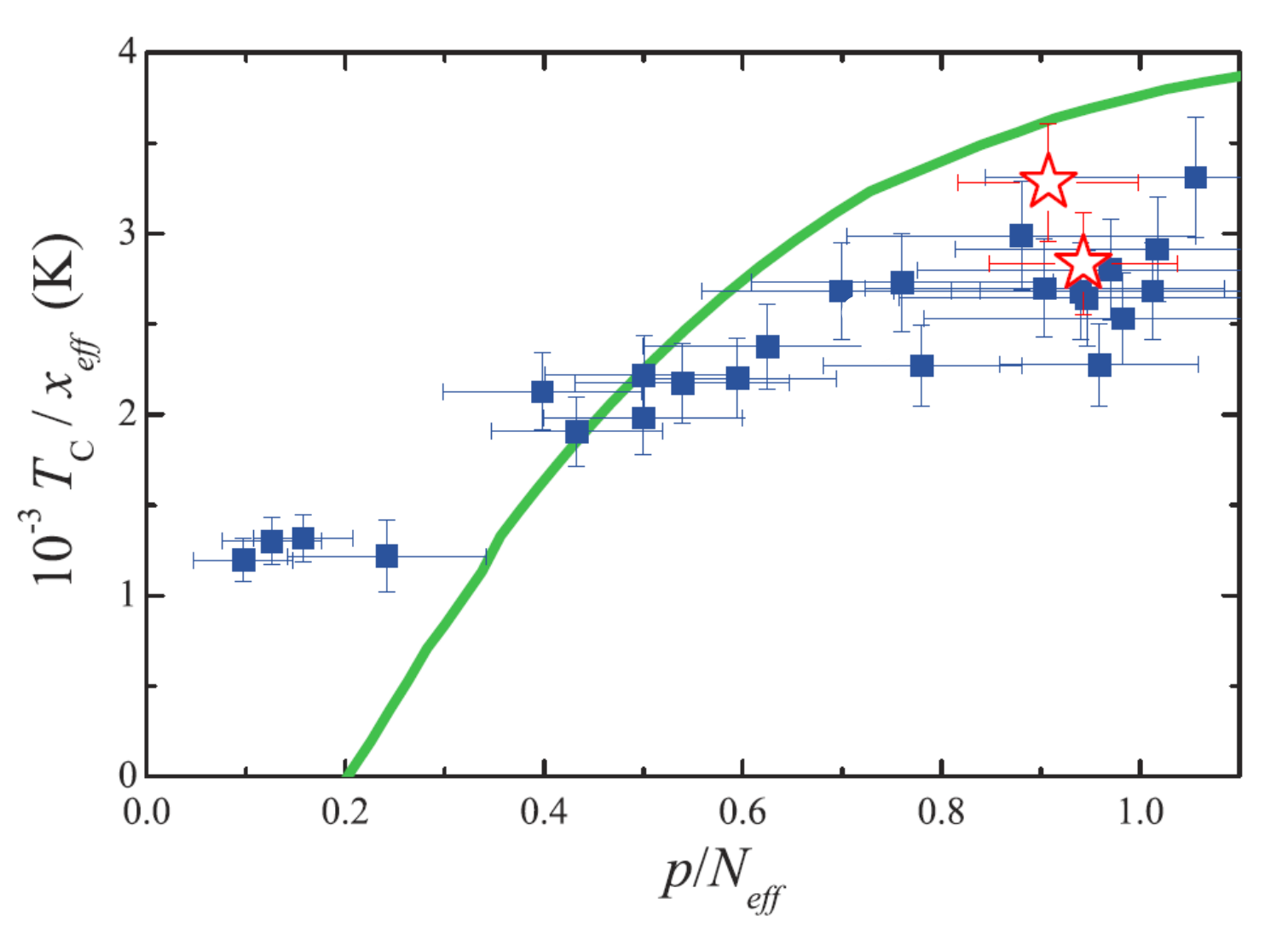}
\caption[]{(Color online) Curie temperature $T_{\text{C}}$ versus hole concentration $p$ normalized by the effective Mn density   ($T_{\text{C}}/x_{\text{eff}}$ and $p/N_0x_{\text{eff}}$, respectively) for annealed (Ga,Mn)As thin films, where squares and stars represent samples with hole density $p$ determined from high field Hall effect and ion channeling measurements \cite{Rushforth:2008_PRB}, respectively. The solid curve shows the tight-binding theory \cite{Jungwirth:2005_PRBa}. Adapted from \onlinecite{Wang:2013_PRB}.
\label{fig:GaMnAs_Tc_Jungwirth}}
\end{figure}

Figure~\ref{fig:GaMnAs_TC_Ms_theory} presents the experimentally established values of $T_{\text{C}}$ in a representative series of annealed (Ga,Mn)As thin films as a function of saturation magnetization $M_{\text{Sat}}$ determined at low temperatures, compared to the expectation of the mean-field $p$-$d$ Zener model. In order to generate the theoretical curve (solid line), the  calculation scheme and the set of standard material parameters proposed previously \cite{Dietl:2000_S,Dietl:2001_PRB} are employed.\footnote{This set is: lattice parameter $a_0 = 5.65$~{\AA}; spin-orbit splitting $\Delta_0 = 0.34$~eV; Luttinger parameters: $\gamma_1 = 6.85$,  $\gamma_2 = 2.1$; $\gamma_3 =  2.9$; exchange energy $N_0\beta = -1.2$~eV; Landau's Fermi liquid parameter $A_{\text{F}} = 1.2$.}  It is assumed that hole density and the effective Mn concentration are equal and related to $M_{\text{Sat}}$ according to $p = N_0x_{\text{eff}} = M_{\text{Sat}}/\mu$, where $\mu =4.5\mu_{text{B}}$ takes into account a contribution of holes to the total magnetization. A similar comparison is shown in Fig.~\ref{fig:GaMnAs_Tc_Jungwirth}, where the theoretical curve was obtained by tight-binding theory within coherent potential and mean-field approximations, adjusting parameters to reproduce the empirical band structure of GaAs and spin-splitting of (Ga,Mn)As \cite{Jungwirth:2005_PRBa}.

The comparison of experimental and theoretical results shown in Fig.~\ref{fig:GaMnAs_TC_Ms_theory} allows to draw several important conclusions concerning limitations of the $p$-$d$ Zener model, listed in Sec.~\ref{sec:limitations}. In particular, higher experimental than theoretical $T_{\text{C}}$ values at low $M_{\text{Sat}}$ stem, presumably, from nano-scale magnetization fluctuations at the localization boundary making that a portion of Mn spins does not participate in the ferromagnetic order. Under these conditions, an experimentally determined average value of $M_{\text{Sat}}$ is smaller than relevant $M_{\text{Sat}}$ corresponding to the ferromagnetic percolation cluster setting up at $T_{\text{C}}$.

In the high $M_{\text{Sat}}$ range, in turn, the magnitude of $T_{\text{C}}$ saturates faster with $M_{\text{Sat}}$ than expected theoretically. In addition to a correction for the effect of thermodynamic magnetization fluctuations, two other effects appear to come into play in this regime.

First, a relative importance of short range antiferromagnetic interactions between Ga-substitutional Mn ions increases with the Mn concentration. As discussed in Sec.~\ref{sec:antiferro}, these interactions do not affect $x_{\text{eff}}$ in (Ga,Mn)As but make $T_{\text{AF}} > 0$. Since a dependence $T_{\text{AF}}(x_{\text{eff}},T)$ is unknown for (Ga,Mn)As, guided by results for II-VI DMSs and (Ga,Mn)N (Sec.~\ref{sec:superexchange}), one can assume $T_{\text{AF}} = (M_{\text{Sat}}/A)^m$, where $A$ is a fitting parameter and $m = 2.3$. The dashed line in Fig.~\ref{fig:GaMnAs_TC_Ms_theory} has been obtained with $\mu_0A =10.6$~mT.

Second, it is rather probable that Mn interstitials are not entirely removed by annealing in this range of Mn content (see, Secs.~\ref{sec:self-compensation} and \ref{sec:doping}). If this is the case, the hole concentration is diminished according to Eqs.~\ref{eq:p} and \ref{eq:xeff}, $p = N_0(x_{\text{eff}} - x_{\text{I}} - zN_{\text{D}})$. In particular, for $M_{\text{Sat}} = 71$~mT, which corresponds to $x_{\text{eff}} = 6.1$\%, the $p$-$d$ Zener model with antiferromagnetic interactions reproduces $T_{\text{C}} = 170$~K for $x_{\text{I}} = 1.8$\% (if $N_{\text{D}} = 0)$, meaning that $x = 9.7$\%. A systematically observed saturation in values $x_{\text{eff}}$ and $T_{\text{C}}$ for $x > 10$\% \cite{Mack:2008_APL,Chiba:2007_APL,Ohya:2007_APL} suggests that substitutional incorporation of Mn is particularly difficult in such heavily Mn doped (Ga,Mn)As samples.

Finally, it should be recalled referring to Fig.~\ref{fig:GaMnAs_Tc_Jungwirth} that the agreement between the experimental and theoretical values could presumably be improved further on by noting that the hole concentration $p$ is underestimated and overestimated by the Hall effect measurements in the range of low and high concentrations of Mn acceptors, respectively, as discussed in Sec.~\ref{sec:carrier_concentration}.

\begin{figure}
\includegraphics[width=8.6cm]{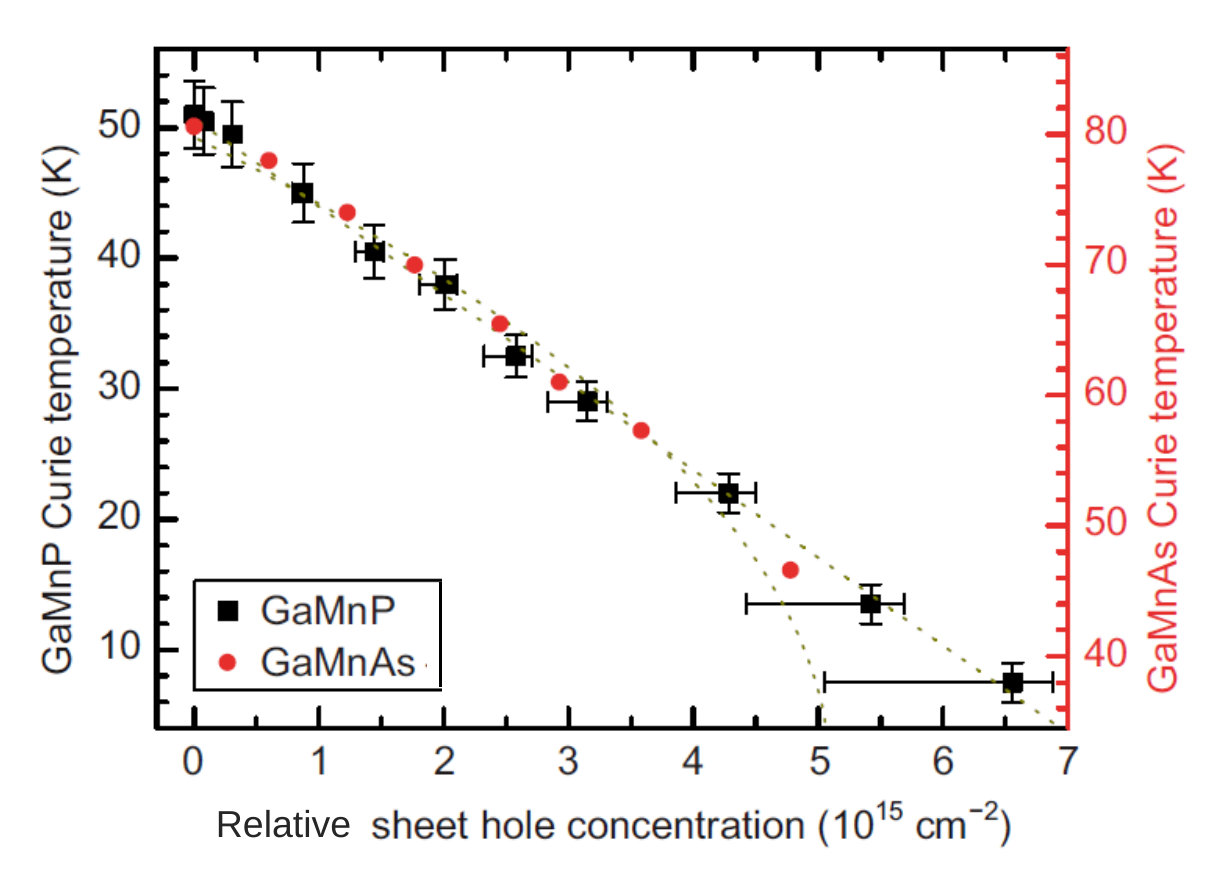}
\caption[]{(Color online) Curie temperature $T_{\text{C}}$ as a function of sheet hole density change by consecutive irradiations with
Ne$^+$ and Ar$^+$ ions in Ga$_{1-x}$Mn$_x$As, $x = 0.045$, grown by low-temperature MBE \cite{Mayer:2010_PRB} and in Ga$_{1-x}$Mn$_x$P, $x \approx 0.038$, obtained by ion implantation and pulsed-laser melting \cite{Winkler:2011_APL}, respectively.  Dotted lines show limiting trends in $T_{\text{C}} \propto p^{\gamma}$, where $\gamma$ =0.5 and 1, in accord with the theoretical anticipation, $\gamma = $ 0.6--0.8 \cite{Dietl:2001_PRB,Nishitani:2010_PRB}. For comparison, the {\em ab initio} approach providing the data shown in Fig.~\ref{fig:GaMnAs_TC_Ms_theory} predicted ferromagnetic ordering to vanish already entirely ($T_{\text{C}} =0$) for hole density reduced twofold by compensation \cite{Sato:2010_RMP}. After \onlinecite{Winkler:2011_APL}.}
\label{fig:GaMnP_Winkler}
\end{figure}

In view of the above discussion, particularly welcome are studies of $T_{\text{C}}$ as a function of hole density in a single sample, as such a dependence is virtually independent of poorly known values of $T_{\text{AF}}$ and background concentrations of compensating donors, $N_0x_{\text{I}}$ and $N_{\text{D}}$. According to numerical results for the $p$-$d$ Zener model \cite{Dietl:2001_PRB},  $\gamma = \mbox{d}\ln T_{\text{C}}/\mbox{d} \ln p$ = 0.6--0.8 in the relevant region of hole densities. This prediction was confirmed experimentally by tracing the dependence $T_{\text{C}}(p)$ in (Ga,Mn)As \cite{Mayer:2010_PRB} and (Ga,Mn)P \cite{Winkler:2011_APL} films irradiated by ions that produce hole compensating donor defects, the original data for (Ga,Mn)As shown already in Fig.~\ref{fig:GaMnAs_Mayer}. As depicted in  Fig.~\ref{fig:GaMnP_Winkler}, the $T_{\text{C}}(p)$ results point to $\gamma =$0.5�-1.0, in agreement with the theory.

\begin{figure}
\includegraphics[width=9cm]{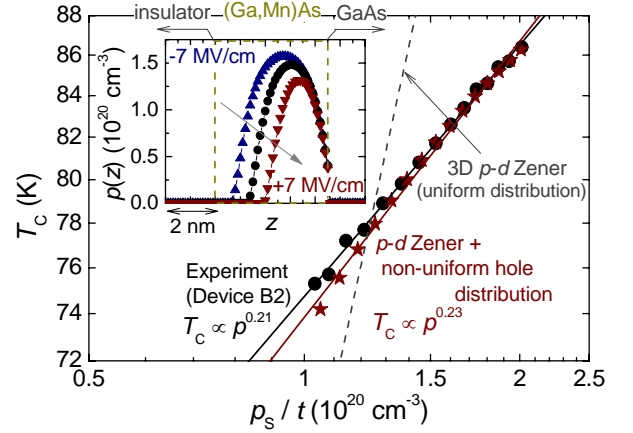}
\caption[]{(Color online) Curie temperature $T_{\text{C}}$ in a metal-oxide-semiconductor structure, in which the gate electric field $E_{\text{G}}$ changes the hole distribution (inset) and density (the areal hole concentration $p_{\text{s}}$ normalized to the  thickness $t$ of the (Ga,Mn)As channel). The solid line represents the generalized $p$-$d$ Zener model for thin layers (Eq.~\ref{eq:T_F_FET}), whereas the dotted line shows the dependence predicted by the $p$-$d$ Zener model for the 3D case, corroborated by the data in Fig.~\ref{fig:GaMnP_Winkler}. From \onlinecite{Nishitani:2010_PRB}.}
\label{fig:GaMnAs_TC_Nishitani}
\end{figure}

However, detailed studies of changes in $T_{\text{C}}$ induced by the gate voltage $V_{\text{g}}$ in metal-insulator-semiconductor (MIS) structures of (Ga,Mn)As \cite{Sawicki:2010_NP,Nishitani:2010_PRB} led to an entirely different value, $\gamma =  \mbox{d}\ln T_{\text{C}}/\mbox{d} \ln(-V_{\text{g}}) = 0.19\pm 0.02$ \cite{Nishitani:2010_PRB}. As shown in Fig.~\ref{fig:GaMnAs_TC_Nishitani}, this finding was elucidated by the $p$-$d$ Zener model generalized to the case of a non-uniform hole distribution obtained by solving the Poisson equation in thin (Ga,Mn)As layers (Eq.~\ref{eq:T_F_FET}), in which the Fermi level at the surface is pinned in the gap region by surface states \cite{Sawicki:2010_NP,Nishitani:2010_PRB}.

\begin{figure}
\includegraphics[width=8cm]{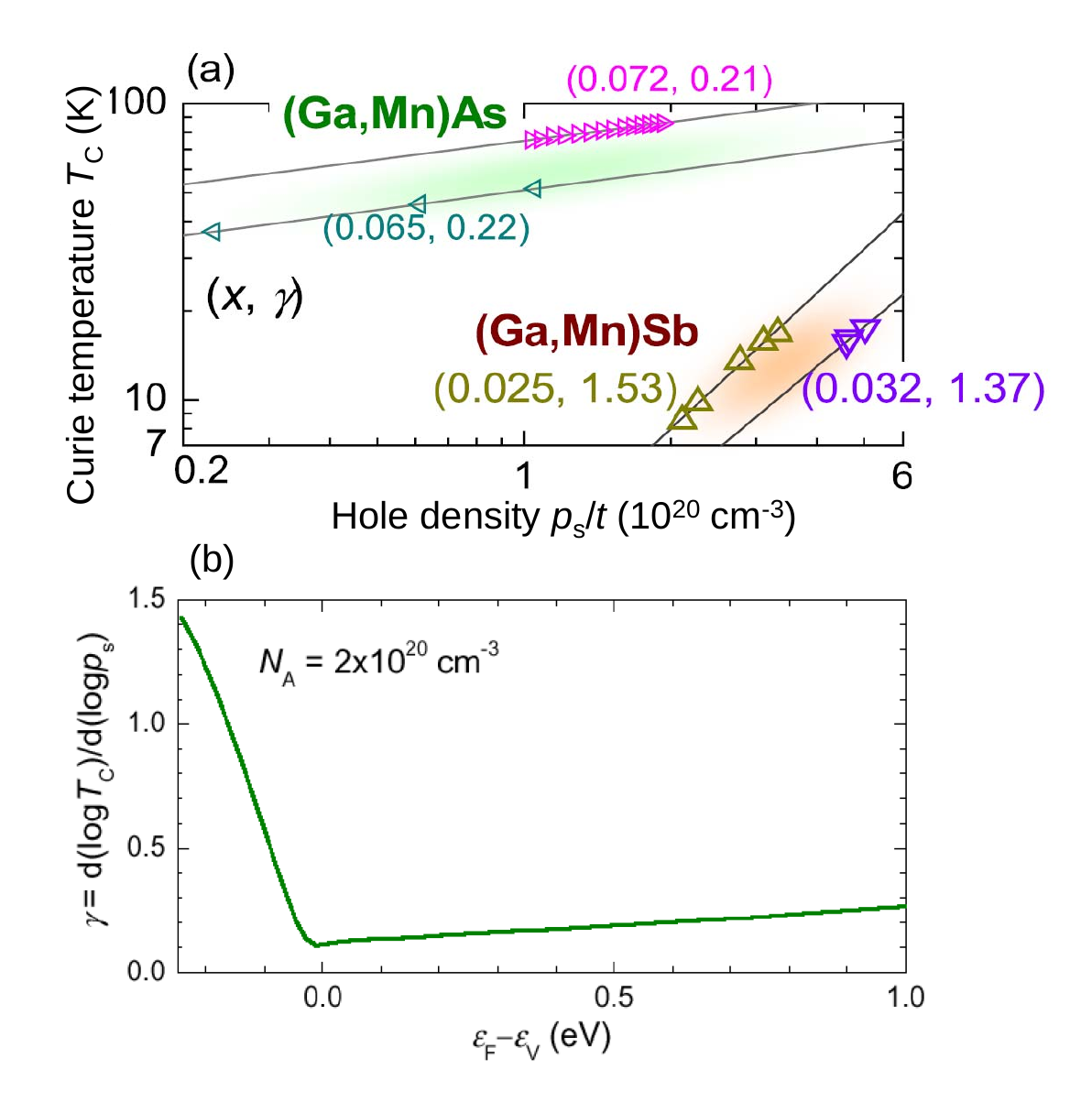}
\caption[]{(Color online) (a) Curie temperatures $T_{\text{C}}$ for various surface densities of holes $p_{\text{s}}$ changed by gate voltage in Ga$_{1-x}$Mn$_{x}$)As and Ga$_{1-x}$Mn$_{x}$Sb channels of the thickness $t=5$~nm. The values of $\gamma =  \mbox{d}\ln T_{\text{C}}/\mbox{d} \ln n_{\text{s}}$ are also shown (from \onlinecite{Chang:2013_APL}). (b) Computed values of $\gamma$  as a function of the position of the Fermi level at the surface in respect to the valence band top for Mn acceptor concentration $2\times 10^{20}$~cm$^{-3}$ and band structure parameters of GaAs;  the Fermi energy is known to be pinned by surface states in the band gap of GaAs and in the valence band of GaSb (from \onlinecite{Ohno:2013_JAP}).}
\label{fig:pinning}
\end{figure}

However, a much higher value, $\gamma \gtrsim 1$,  was observed for MIS structures of (Ga,Mn)Sb, as shown in Fig.~\ref{fig:pinning}(a) \cite{Chang:2013_APL}. According to theoretical evaluations from Eq.~\ref{eq:T_F_FET}, the efficiency of $T_{\text{C}}$ tuning by $V_{\text{G}}$ and, hence, the magnitude of $\gamma$ depends strongly on the energy position of surface states in respect to the valence band top. Since in (Ga,Mn)Sb, in contrast to (Ga,Mn)As, the Fermi level is pinned in the valence band by surface states, a large value of $\gamma$ is theoretically expected, as depicted in Fig.~\ref{fig:pinning}(b).

A relation between $T_{\text{C}}$ and film resistance $R$, changed by ferroelectric overlayers, was determined to be  $\delta =  - \mbox{d}\ln T_{\text{C}}/\mbox{d} \ln R =0.35 \pm 0.05$ for a number of thin (Ga,Mn)As films \cite{Stolichnov:2011_PRB}. Information on changes in hole mobility is needed to compare $\delta$ and $\gamma$.

Finally, results of a study of $T_{\text{C}}$ as a function of hydrostatic pressure $P$ \cite{Gryglas:2010_PRB} were found consistent with the diagram in Fig.~\ref{fig:MIT_VCA}: $T_{\text{C}}$ increases with $P$ according to the $p$-$d$ Zener model at a high hole concentration but it decreases in a sample close to the localization boundary. An increase of $T_{\text{C}}$ with $P$, of the magnitude corroborating the $p$-$d$ Zener model, was also reported for (In,Mn)Sb \cite{Csontos:2005_NM}.

\subsubsection{Curie temperatures in II-VI DFSs}
\label{sec:T_c_II_VI}
\begin{figure}
\includegraphics[width=8.1cm]{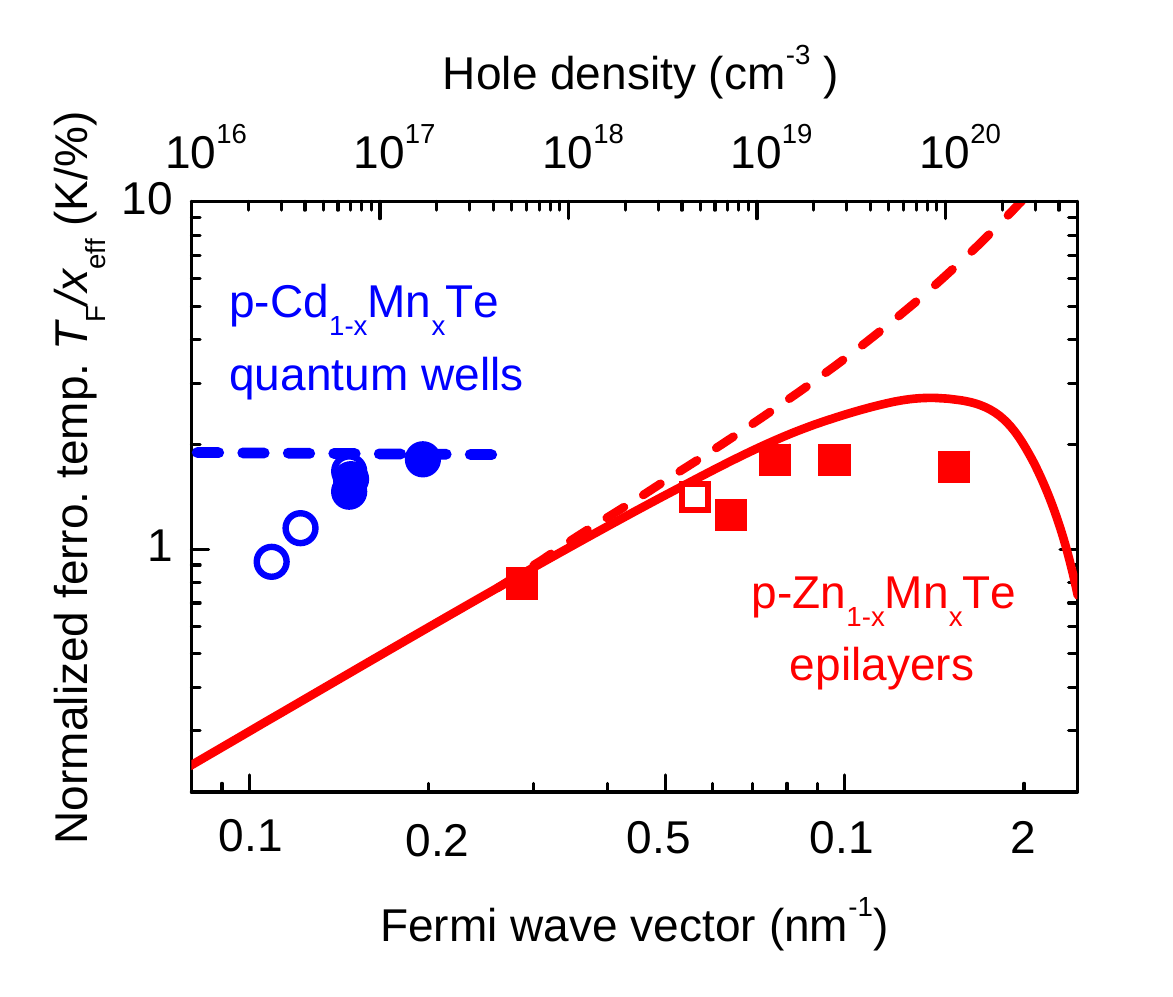}
\caption[]{(Color online) Ferromagnetic temperature $T_{\text{F}}= T_{\text{C}}+ T_{\text{AF}}$ normalized to the effective Mn concentration as a function of the Fermi wave vector (lower scale) and the hole density (upper scale) for epilayers of p-type  Zn$_{1-x}$Mn$_x$Te doped with N (full squares, \onlinecite{Ferrand:2001_PRB}) and  P (empty square, \onlinecite{Andrearczyk:2001_ICPS}) as well as for modulation doped p-(Cd,Mn)Te quantum wells (full circles, \onlinecite{Haury:1997_PRL}; empty circles, \onlinecite{Boukari:2002_PRL}). The dashed line represents the 2D \cite{Dietl:1999_MSEB} and 3D $p$-$d$ Zener model \cite{Dietl:2000_S}. Solid line is theoretical for the 3D case, taking into account the antiferromagnetic portion of the RKKY interaction  \cite{Ferrand:2001_PRB}. The dashed line represents the 3D $p$-$d$ Zener model \cite{Dietl:2000_S}.}
\label{fig:II_VI_TC}
\end{figure}

Since Mn in II-VI DMSs does not provide any carriers, it is possible to vary the Mn and hole concentration independently as well as to prepared modulation-doped quantum wells, in which the mean free path is longer than the well width. However, a relatively strong short-range antiferromagnetic (AF) coupling between Mn spins in II-VI DMSs, reduces $T_{\text{C}}$ significantly, as discussed in Sec.~\ref{sec:antiferro}. The corresponding values od $T_{\text{AF}}(x)$ and $x_{\text{eff}}(x)$ were determined from magnetization or spin-splitting studies for undoped DMSs, confirming that at low Mn concentrations $x \lesssim 5$\%,  $x_{\text{eff}} = x(1-x)^{12}$, as AF coupling between the nearest neighbor Mn pairs is there essential. As shown in Fig.~\ref{fig:II_VI_TC}, taking the presence of the AF interactions into account, the magnitudes of $T_{\text{F}}$ can be described quantitatively in p-(Zn,Mn)Te \cite{Ferrand:2001_PRB}.

In the case of 2D quantum wells of p-(Cd,Mn)Te, the data shown in Fig.~\ref{fig:II_VI_TC}, substantiated the validity of Eq.~\ref{eq:T_F_2D} and, in particular, the dimensional enhancement of DOS and thus of $T_{\text{F}}$ in the range of low carrier densities.  The parameter $A_{\text{F}} = 2.0$ was adjusted to explain the magnitude of $T_{\text{F}}$ \cite{Haury:1997_PRL,Boukari:2002_PRL}, a value consistent with its independent magnetooptical evaluation \cite{Kossacki:2004_PRB}. A more detailed analysis, combining solving of the Schroedinger equation for a given Mn distribution with Monte Carlo simulations for competing FM and AF interactions, confirmed the presence of scattering broadening of DOS, and the associated reduction of $T_{\text{F}}$ at low hole concentrations \cite{Lipinska:2009_PRB}. Furthermore, the simulations explained a specific shape of $M(H)$ in terms of fast Mn dynamics, even below $T_{\text{C}}$, caused by AF coupling to Mn spins residing beyond the region penetrated by holes.

A particularly relevant is the question whether the Zener theory can be extended to n-type DMSs. So far an indication of ferromagnetism was found by the observation of resistance hysteresis in n-Zn$_{1-x}$Mn$_x$O:Al  with $x = 3$\% and $n = 1.4\times 10^{20}$~cm$^{-3}$, which persisted up to 160~mK \cite{Andrearczyk:2001_ICPS}. Such a value of $T_{\text{C}}$, factor of 20 lower than in p-type Zn$_{1-x}$Mn$_x$Te with a similar Mn content, is--in fact--expected theoretically from Eq.~\ref{eq:MFA}:  for similar values of $A_{\text{F}}$ and $\rho_{\text{s}}$ in both systems, one order of magnitude difference between $\alpha^2$ and $\beta^2$ implies that the Mn spin susceptibility in the absence of carriers, $\chi_{S}(T_{\text{C}})$, has to be greater by a similar factor, which was realized by lowering temperature below 200~mK.

\subsection{Interlayer coupling}
\label{sec:inter_layer_theory}
Detailed theoretical studies of interlayer exchange energy were carried out for (Ga,Mn)As/GaAs superlattices within the multiorbital TBA \cite{Sankowski:2005_PRB}. The case of the (Al,Ga)As spacer was also considered. This theory predicted the regions of hole and Mn densities as well as spacer thicknesses and Al concentrations, in which an antiferromagnetic interaction between (Ga,Mn)As layers should appear. However, only ferromagnetic coupling has so-far been observed experimentally for undoped spacers, as discussed in Sec.~\ref{sec:interlayer}. It would be interesting to find out whether the parameter space where the antiferromagnetic interaction exists would become narrower if disorder and hole redistribution between particular LT-grown layers were incorporated into the theory in a self consistent manner. Furthermore, the role of dipole-dipole interactions is to be considered, too.

\subsection{Magnetization and specific heat}
\label{sec:magnetization}

It appears not easy to separate experimentally the hole contribution $M_{\text{c}}$ to the total magnetization $M_{\text{s}}$ and, in particular, to verify whether it reduces $M_{\text{s}}$ by 10\%, as predicted theoretically. However, measurements of XMCD at the As $K$-edge did provide the values of spin and orbital magnetic moments \cite{Freeman:2008_PRB,Wadley:2010_PRB} in agreement with the theory of hole magnetization \cite{Sliwa:2013_arXiv}. Furthermore, a recent study \cite{Ciccarelli:2012_APL}, using (Ga,Mn)As as a gate for the Coulomb blockade in an Al dot, allowed to determine the dependence of the Fermi level position on the magnetic field. By using thermodynamic relations $\epsilon_{\text{F}} = -\partial F_{\text{c}}/\partial p$ and $M_{\text{c}} = -\partial F_{\text{c}}/\partial H$, where $F_{\text{c}}$ is the carrier free energy we obtain $\partial\epsilon_{\text{F}}/\partial H = \partial M_{\text{c}}/\partial p$. For $x = 3$\% and assuming $x_{\text{I}} = 0$ or 0.5\% the theory \cite{Sliwa:2006_PRB,Sliwa:2013_arXiv} leads to $-\partial M_{\text{c}}/\partial p = 14$ or 15~$\mu$eV/T, respectively, in good agreement with the experimental value of $-\partial\epsilon_{\text{F}}/\partial \mu_0H = 18\pm 3$~$\mu$eV/T for Ga$_{0.097}$Mn$_{0.03}$As in the magnetic fields saturating Mn spins, $\mu_0H > 7.5~T$ \cite{Ciccarelli:2012_APL}.

It appears that there are three main ingredients underlying the temperature dependence of spontaneous magnetization in (Ga,Mn)As-type DFSs, which we discuss {\em vis \`a vis} theoretical expectations.

\begin{figure}
\includegraphics[width=8.5cm]{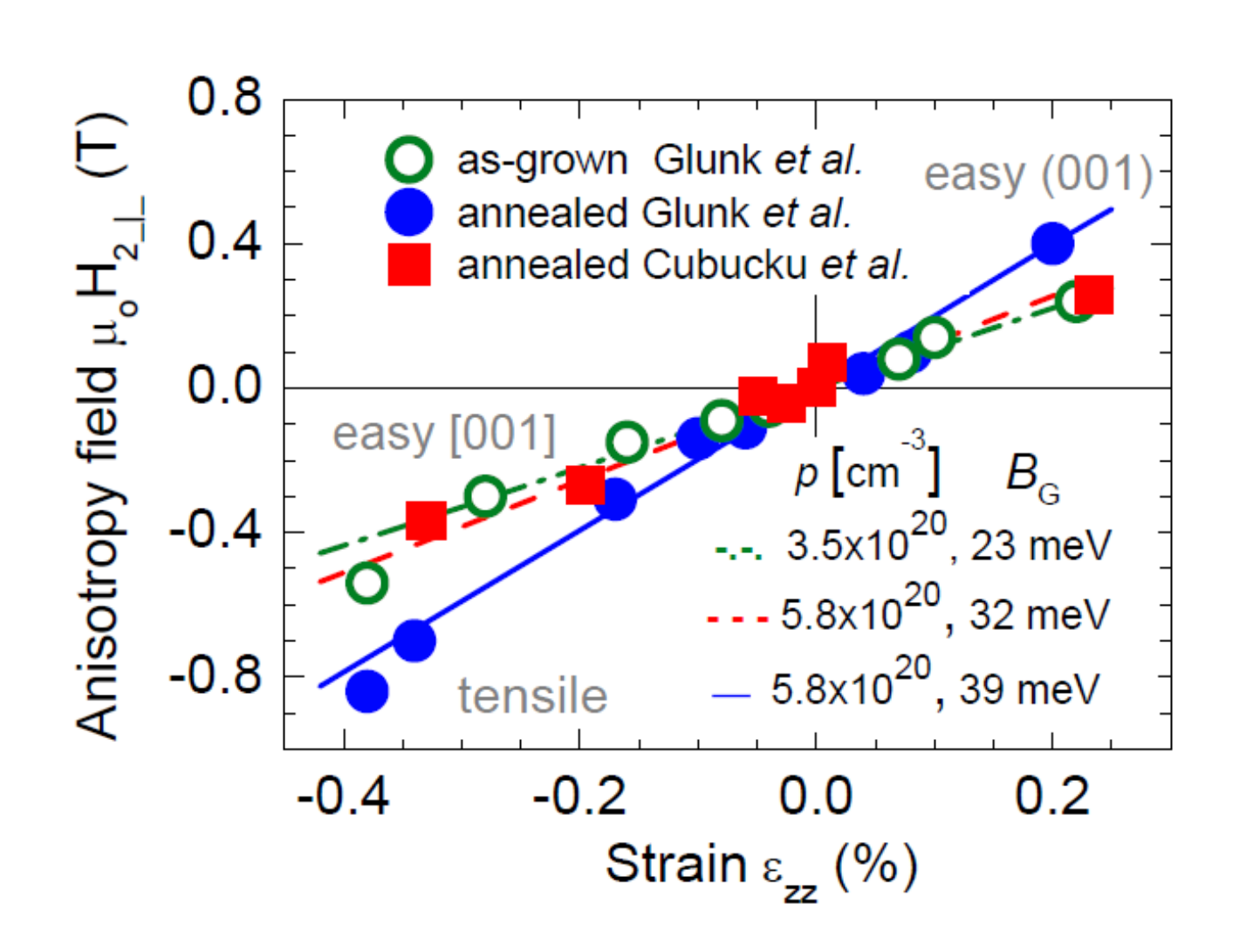}
\caption[]{(Color online) Temperature dependence of magnetization in (Ga,Mn)As \cite{Potashnik:2002_PRB,Gourdon:2007_PRB} compared to the Brillouin (dashed line) and Bloch theories neglecting and taking the spin gap into account (doted and solid lines, respectively). From \onlinecite{Werpachowska:2010_PRBb}.}
\label{fig:GaMnAs_M_T}
\end{figure}

\begin{itemize}
\item
The mean-field Eq.~\ref{eq:MF} allowed to determine TM magnetization $M(T,H)$. It was predicted that the dependence $M(T)$ should evolve from the Brillouin-like convex form at high hole densities towards a concave shape at the Fermi energy smaller than the low temperature spin splitting of the carrier band \cite{Dietl:2001_PRB}. Such a change in the magnetization behavior on reducing carrier density at a given Mn concentration was indeed observed in (Ga,Mn)As \cite{Mayer:2010_PRB}. However, the concave shape is also expected, and commonly observed \cite{Sheu:2007_PRL}, when the proximity of the Anderson-Mott localization results in the formation of superparamagnetic-like regions, whose magnetization grows relatively slowly on decreasing temperature \cite{Sawicki:2010_NP}.

\item
It was shown that away from the localization boundary, the dependence $M_{\text{s}}(T)$ obeys the Bloch $T^{3/2}$ law, demonstrating the importance of spin wave excitations \cite{Potashnik:2002_PRB}. However, the magnitude of the spin wave stiffness $D$ which was obtained in this way for a large series of samples, was by about factor of 2 greater than expected theoretically \cite{Werpachowska:2010_PRBb}. This discrepancy was resolved \cite{Werpachowska:2010_PRBb} by taking into account the presence of the spin gap brought by the anisotropy field (Eq.~\ref{eq:dispersion}). As shown, in Fig.~\ref{fig:GaMnAs_M_T}, the theory described  the experimental dependence $M_{\text{s}}(T)$ with no adjustable parameters.

\item
A complex temperature dependence of magnetization was revealed in samples for which a combination of strain, hole and Mn density values resulted in the temperature-induced spin reorientation transition \cite{Sawicki:2004_PRB,Sawicki:2005_PRB,Wang:2005_PRL}. A simple single-domain model was found to describe both $\vec{M}_s(T)$ in the whole temperature range and a critical-like behavior of a.c. magnetic susceptibility in the vicinity of the transition \cite{Wang:2005_PRL}.
\end{itemize}

\begin{figure}
\includegraphics[width=8.5cm]{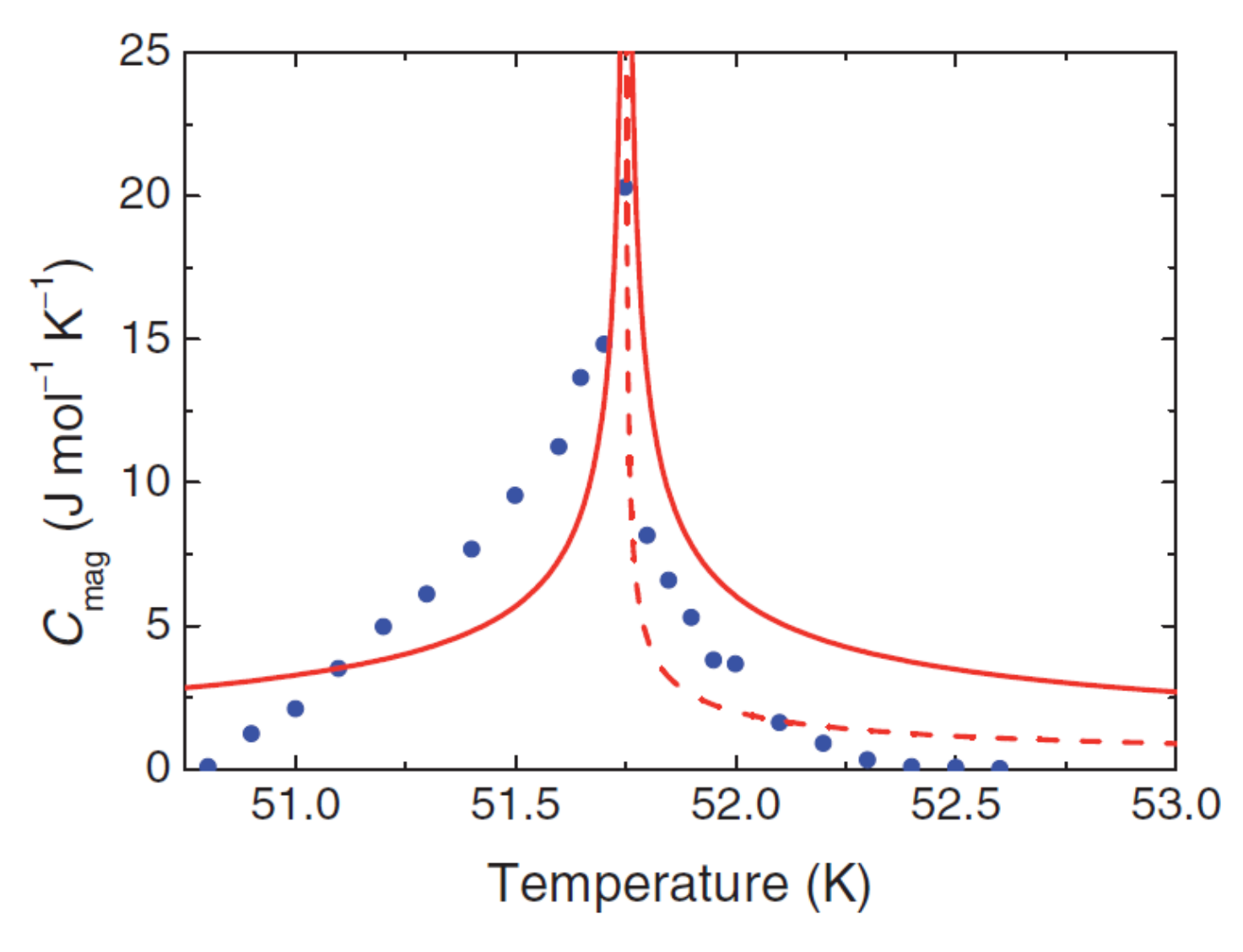}
\caption[]{(Color online) Experimentally determined specific heat anomaly at the Curie temperature $T_{\text{C}}$ in as-grown Ga$_{0.974}$Mn$_{0.026}$As, (points, \onlinecite{Yuldashev:2010_APE}). Theoretical temperature dependence of the magnetic specific heat calculated with no adjustable parameters for the Heisenberg and Ising models (solid and dashed lines, respectively, \onlinecite{Sliwa:2011_PRB}).}
\label{fig:GaMnAs_Sliwa_C}
\end{figure}

Within the Ginzburg-Landau approach, spin wave stiffness $D$ controls also the lambda-like anomaly of specific heat $C(T)$ near $T_{\text{C}}$. As shown in Fig.~\ref{fig:GaMnAs_Sliwa_C}, the theory \cite{Sliwa:2011_PRB} described the experimental data \cite{Yuldashev:2010_APE} reasonably well, particularly assuming that owing to uniaxial anisotropy, the Ising universality class applies to (Ga,Mn)As.

\subsection{Magnetic anisotropy and magneto-elastic phenomena}
\label{sec:anisotropy-comparison}
\begin{figure}
\includegraphics[width=8.5cm]{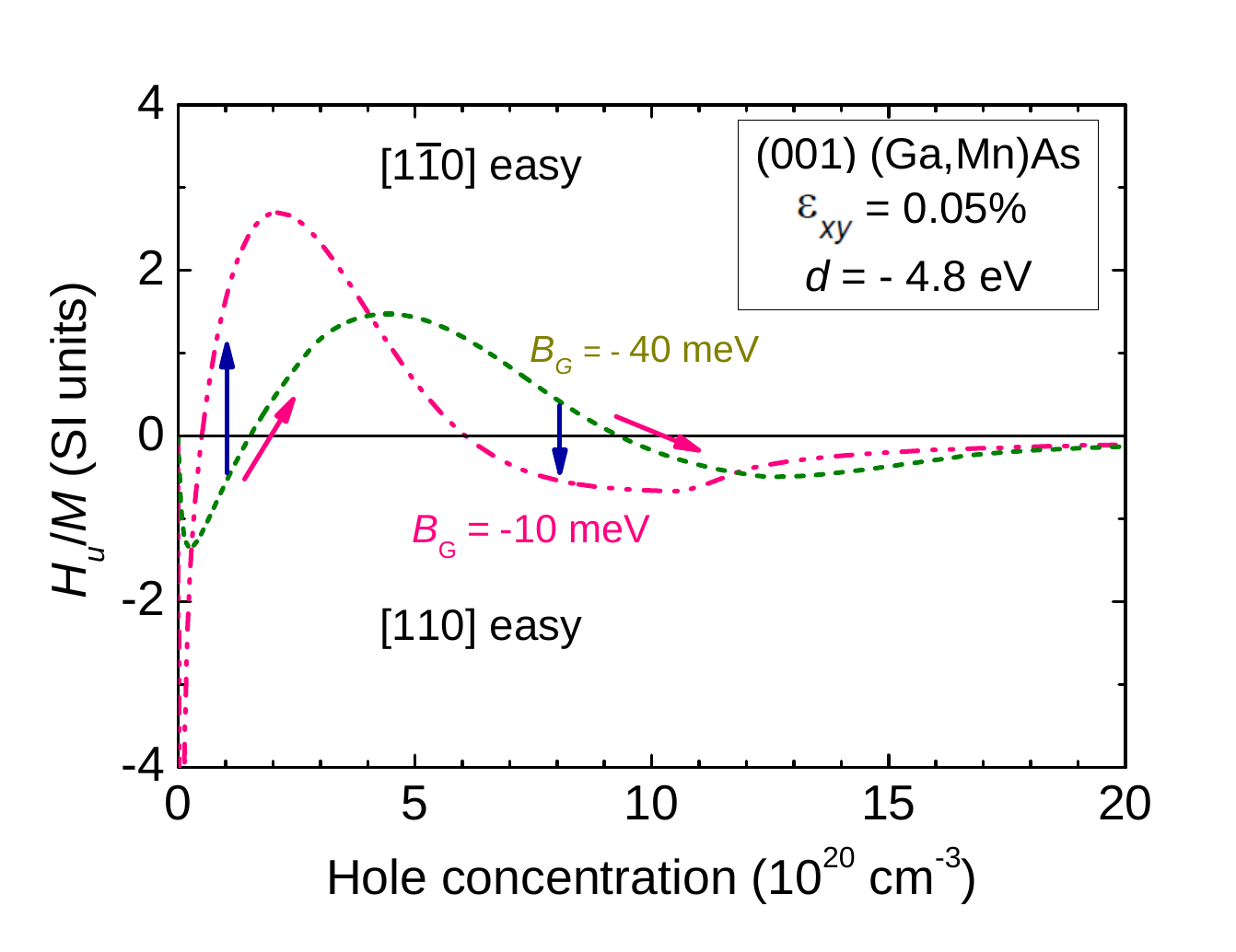}
\caption[]{(Color online) Theoretical in-plane uniaxial anisotropy field in (001) (Ga,Mn)As, normalized by magnetization, as a function of hole concentration at shear strain $\epsilon_{xy}=0.05$\% for two values of the valence-band spin splitting parameter, $B_{\text{G}} = A_{\text{F}}\beta M(T)/6g\mu_{\text{B}}$ (adapted from \onlinecite{Zemen:2009_PRB,Stefanowicz:2010_PRBb}). Arrows show expected theoretically and observed experimentally \cite{Chiba:2008_N,Sawicki:2005_PRB,Sawicki:2010_NP} spin reorientation transitions on increasing hole density or temperature.}
\label{fig:GaMnAs_uniaxial}
\end{figure}

A comparison of experimental data summarized in Sec.~\ref{sec:anisotropy}) to the theoretical model (Sec.~\ref{sec:magnetization-theory}), developed assuming literature values of deformation potentials and elastic moduli (see, e.~g., \onlinecite{Stefanowicz:2010_PRBb}), leads to a number of conclusions concerning particular contributions to bulk crystalline magnetic anisotropy in DFSs.

\ \\
{\em Cubic anisotropy}: As anticipated taking disorder into consideration \cite{Dietl:2001_PRB}, the experimental value of the cubic anisotropy field does not show any noticeable oscillatory behavior as a function of the hole concentration, expected within the disorder-free theory. At the same time, the observed order of magnitude  $\mu_0H_{\text{c}} \simeq 0.1$~T at $T \ll T_{\text{C}}$ and temperature dependence are consistent with the predicted amplitude of oscillations \cite{Dietl:2001_PRB,Abolfath:2001_PRB,Zemen:2009_PRB,Stefanowicz:2010_PRBb}. However, why the cubic easy axis assumes predominantly $\langle 100 \rangle$ orientations in (Ga,Mn)As, whereas the $\langle 110 \rangle$ directions are preferred in (In,Mn)As and (Ga,Mn)P has not yet been theoretically explained.

\ \\
{\em In-plane uniaxial anisotropy}: The sign and values of $\mu_0H_{xy}$ and $\mu_0H_{zz}$ found experimentally confirm the existence of a theoretically expected surplus of Mn dimers oriented along the $[\bar{1}10]$ direction for (001) (Ga,Mn)As \cite{Birowska:2012_PRL}, though the dimer formation can depend sensitively on the surface reconstruction, partial pressure of As, growth rate and temperature. The corresponding lowering of symmetry is described by effective strains $\epsilon_{xy}$ and $\epsilon_{zz}$ that can be incorporated into the $p$-$d$ Zener model. The in-plane uniaxial anisotropy field $\mu_0H_{xy}$ obtained is this way for (Ga,Mn)As \cite{Zemen:2009_PRB,Stefanowicz:2010_PRBb} is shown in Fig.~\ref{fig:GaMnAs_uniaxial} as a function of the hole concentration for two values of spontaneous magnetization $M(T)$. In this range, $\mu_0H_{xy}$ is linear in $\epsilon_{xy}$, so that the magnitudes of $p$ and $M$ corresponding to the spin reorientation transitions [{\em i.~e.}, $H_{xy}(p,\Delta) = 0]$ do not depend on the actual asymmetry in the Mn dimer distribution, which can vary from sample to sample, depending on epitaxy conditions. These theoretical results are in accord with directions of spin-reorientation transitions $[\bar{1}10]\rightleftarrows [110]$ observed as a function of either hole density or temperature in both high \cite{Sawicki:2005_PRB} and low hole concentration range, where gating was employed to vary hole concentrations \cite{Chiba:2008_N,Sawicki:2010_NP}.

\begin{figure}
\includegraphics[width=8.5cm]{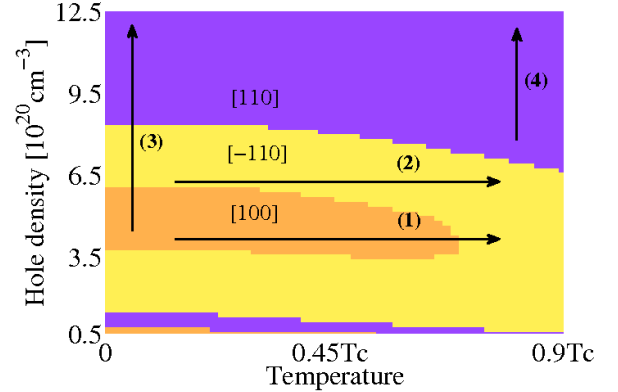}
\caption[]{(Color online) Theoretical hole density - temperature diagrams of crystal directions with the largest projection of the magnetic easy axis at $x_{\text{eff}} = 7$\%, $\epsilon_{xy} = 0.03$\%, $\epsilon_{zz}= 0.2$\%. Arrows mark spin reorientation transitions driven by change of temperature or hole density. From \onlinecite{Zemen:2009_PRB}.}
\label{fig:GaMnAs_Zemen}
\end{figure}

Furthermore, the theory confirms a weaker temperature dependence of $H_{xy}$ comparing to $H_{\text{c}}$, which according to theoretical results shown in Fig.~\ref{fig:GaMnAs_Zemen} for (Ga,Mn)As leads to the spin reorientation transition from $\langle 100 \rangle$ to $[\bar{1}10]$ on increasing temperature, in agreement with the experimental observations \cite{Welp:2003_PRL,Wang:2005_PRL,Kamara:2012_JNN}.

The same approach was successfully applied to explain the direction and the magnitude of easy axis rotation as a function of the voltage applied to a piezoelectric actuator containing a (Ga,Mn)As film cemented to its surface along one of $\langle 110 \rangle$ directions \cite{Rushforth:2008_PRB}. In this case a real and known strain $\epsilon_{xy}$ is imposed by the actuator.

\begin{figure}
\includegraphics[width=8.5cm]{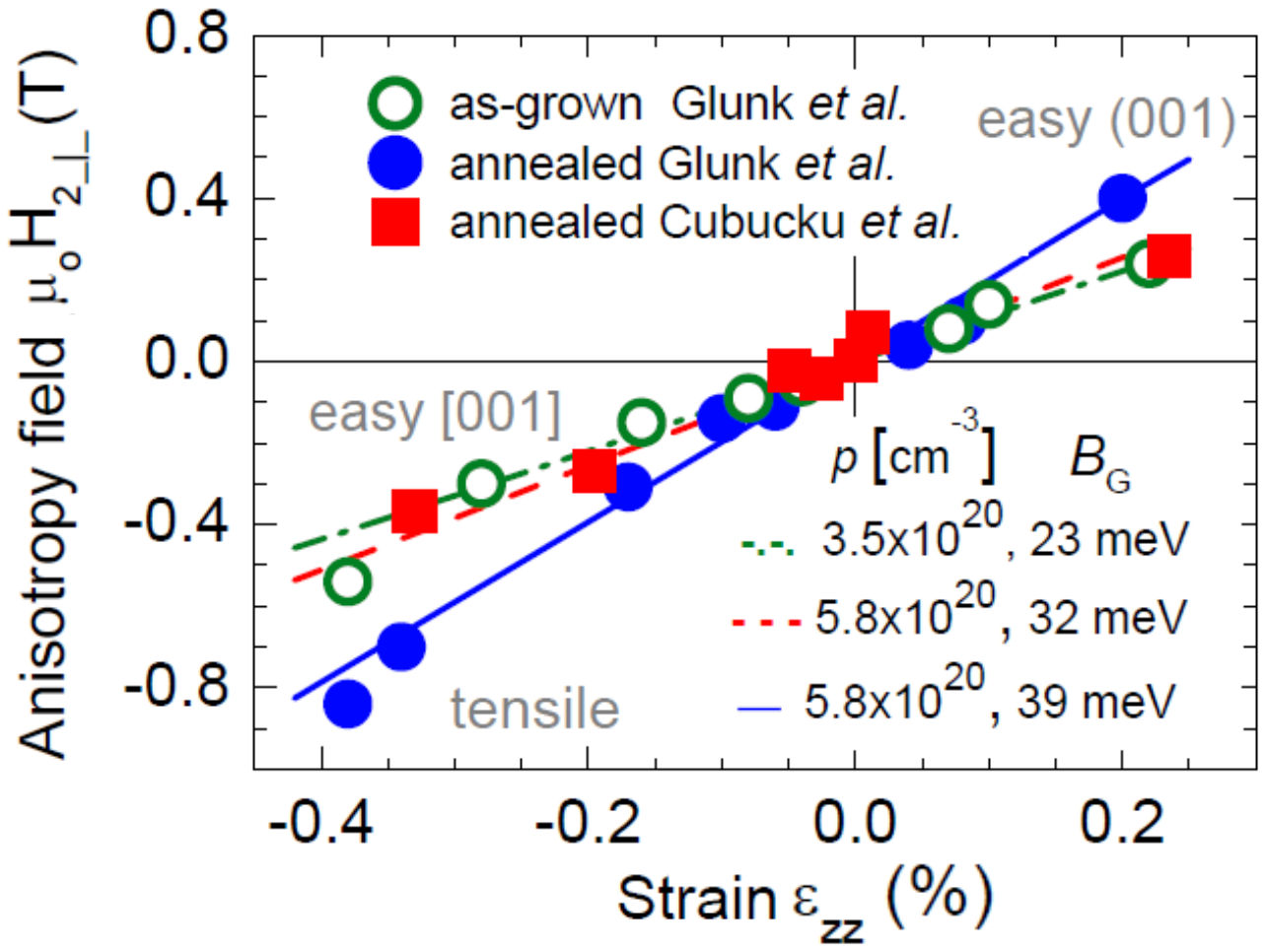}
\caption[]{(Color online) A uniaxial anisotropy field induced by epitaxial strain in (Ga,Mn)As/(Ga,In)As (circles, \onlinecite{Glunk:2009_PRB}) and in (Ga,Mn)(As,P)/GaAs (squares, \onlinecite{Cubukcu:2010_PRB}) compared to predictions of the $p$-$d$ Zener model for the selected values of the hole concentrations and the parameter $B_\text{G}$ characterizing the valence band spin-splitting \cite{Glunk:2009_PRB}. The value $B_\text{G} = 30$~meV corresponds to saturation magnetization at $x_{\text{eff}} \simeq 0.05$.}
\label{fig:GaMnAs_Ku_Limmer}
\end{figure}

Theoretical description of the magnitude of $K_{xz}$ within the $p$-$d$ Zener model for (113) (Ga,Mn)As \cite{Stefanowicz:2010_PRBb} pointed also to the presence of a symmetry lowering perturbation. The current theory of dimer-related magnetic anisotropy for (001) (Ga,Mn)As \cite{Birowska:2012_PRL} has not yet been extended to other orientations of the substrate, so that this finding awaits for a theoretical interpretation.

\ \\
{\em Out-of-plane uniaxial anisotropy}: As exemplified in Fig.~\ref{fig:GaMnAs_Ku_Limmer}, the theory \cite{Dietl:2001_PRB} describes quantitatively the magnitude anisotropy field $\mu_oH_{zz}$ in (001) (Ga,Mn)As at 4~K as well as its dependence on the epitaxial strain $\epsilon_{zz}$, as determined by various groups \cite{Glunk:2009_PRB,Cubukcu:2010_PRB} for samples with typical hole and Mn concentrations,  $3\times10^{20} \lesssim p \lesssim 6\times 10^{20}$~cm$^{3}$, $0.035 \lesssim x_{\text{eff}} \lesssim 0.065$. At given $p$, $H_{zz}$ varies linearly with spontaneous magnetization $M_{\text{s}}(T)$ (at not too high $M_{\text{s}}$), which accounts for the dependence of $H_{zz}$ on temperature. A good agreement between the experiment and theory for magnetic anisotropy generated by epitaxial strain was also found for (113) (Ga,Mn)As \cite{Stefanowicz:2010_PRBb}.

\begin{figure}
\includegraphics[width=8.5cm]{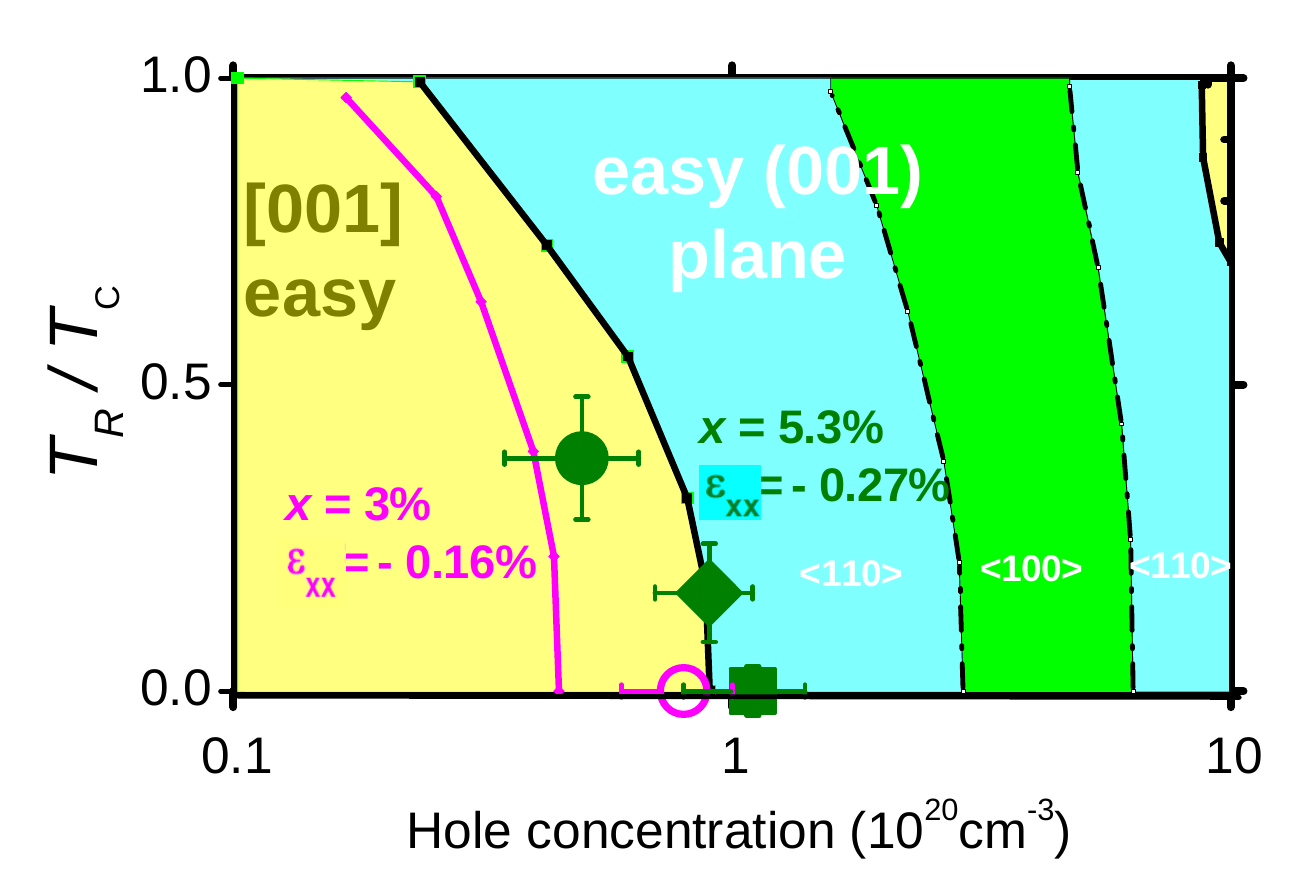}
\caption[]{(Color online) Experimental (solid points) and computed values (thick lines) of the ratio of the spin reorientation temperature to Curie temperature for a flip of the easy axis from the perpendicular to in-plane direction in Ga$_{1-x}$Mn$_{1-x}$As. The hole concentration in the $x = 5.3$\% sample is changed by annealing. No spin reorientation transition was found for the $x = 3$\% sample (empty circle), in agreement with the theory for this Mn concentration. Dashed lines mark expected temperatures for the reorientation of the easy axis between $\langle 100\rangle$ and $\langle 110\rangle$ in-plane directions. The $\langle 100\rangle$ orientation of the in-plane easy axis is observed for these samples. Adapted from \onlinecite{Sawicki:2004_PRB}.}
\label{fig:SRT}
\end{figure}

By decreasing hole density down to $p \approx 10^{20}$~cm$^{-3}$, a temperature-dependent spin reorientation transition takes place from in-plane to perpendicular easy axis orientations (see, Sec.~\ref{sec:anisotropy}),  in agreement with theoretical predictions within the six bands' $kp$ model \cite{Dietl:2001_PRB}, as shown in Fig.~\ref{fig:SRT}. The perpendicular alignment of the easy axis for compressive strain is actually expected within a $kp$ four band model of the valence band,  which is valid at low hole concentrations. This model implies the perpendicular and in-plane orientation of the total orbital momentum $\vec{J}$ of holes for compressive and tensile strain, respectively, explaining the corresponding alignment of Mn spins in (Al,Ga,Mn)As \cite{Takamura:2002_APL} and (Ga,Mn)P \cite{Bihler:2007_PRB} in the low hole concentration regime.

According to theoretical results displayed in Fig.~\ref{fig:SRT}, at high hole densities $p\gtrsim 10^{21}$~cm$^{-3}$, a subsequent spin reorientation transition between in-plane and perpendicular-to-plane magnetic anisotropy is expected theoretically for (Ga,Mn)As \cite{Sawicki:2004_PRB,Zemen:2009_PRB,Werpachowska:2010_PRBa}, which has not yet been found experimentally.

In the case of patterned nanobars, a starting point of a theoretical analysis was a strain distribution, as determined by finite element computations and x-ray diffraction for nanobars patterned along various crystallographic directions from two different (Ga,Mn)As wafers \cite{King:2011_PRB}. A linear dependence of the magnetic anisotropy energy on strain allowed to develop theory in terms a mean strain over the bar cross-section. As a whole, the theory confirmed that strain relaxation accounts for the alignment of the easy axis along the long edge of nanobars, as observed \cite{Humpfner:2007_APL,King:2011_PRB}. However, while the computed magnitude and temperature dependence of magnetic anisotropy characteristics were found in a satisfactory agreement with the data for nanobars patterned from one wafer, there were significant quantitative discrepancies in the case of another series of nanobars \cite{King:2011_PRB}. It might be that the single domain approximation breaks down in some samples with highly non-uniform strain distribution.

Another important effect of strain in zinc-blende crystals is the appearance in the hole dispersion of terms linear in $k$, coupled to the hole total orbital momentum $\vec{J}$, given within the four band model by \cite{Ivchenko:1995_B},
\begin{equation}
{\cal{H}}_{k\epsilon} = C_5\vec{\phi}\vec{J} + C_6\vec{\psi}\vec{J}.
\label{eq:k_linear}
\end{equation}
Here $C_i$ are relevant $kp$ parameters (deformation potentials); $\phi_x = k_y\epsilon_{xy} - k_z\epsilon_{xz}$
[and cyclic permutations (c.~p.)] and $\psi_x = k_x(\epsilon_{yy} - \epsilon_{zz})$ (and c.~p.), where $\epsilon_{ij}$ denote the sum of the deformation-induced and effective components of the strain tensor. Remarkably, the form of ${\cal{H}}_{k\epsilon}$ implies that electric current, by leading to a non-zero value of $<\vec{k}>$ in its direction, generates an effective magnetic field that can orient hole spins and, thus, serve to switch the direction of magnetization. Such an effect was demonstrated experimentally \cite{Chernyshov:2009_NP,Endo:2010_APL}, and interpreted within this $kp$ formalism.
\ \\

Finally, we note that the theory \cite{Dietl:1997_PRB,Kossacki:2004_PE} readily explains why in p-type (Cd,Mn)Te quantum wells under compressive strain the easy axis is along the growth direction $z$ (Fig.~\ref{fig:CdMnTe_Kossacki}), as for this strain configuration and confinement the heavy hole subband is occupied, for which $J_z =\pm 3/2$ and, thus, $s_z =\pm 1/2$.  In contrast, for tensile strain light hole subband is involved, so that $J_z =\pm 1/2)$, and hence the hole spins and, thus, the easy axis assume the in-plane orientation.


\subsection{Domain structure, exchange stiffness, and spin waves}
\label{sec:spin_waves}

Macroscopically large magnetic domains were observed in (Ga,Mn)As with in-plane magnetic anisotropy \cite{Welp:2003_PRL}, which confirmed excellent micromagnetic properties of this system. High-resolution electron holography technique provided direct images of domain wall magnetization profiles of such films \cite{Sugawara:2008_PRL}. The N\'eel type domain walls were found of the width ranging from approximately 40 to 120~nm, the values consistent with the magnitude of $(A/K_{\text{C}})^{1/2} \approx 40$~nm computed within the $p$-$d$ Zener model for the studied films \cite{Sugawara:2008_PRL}.

In the case of a 200~nm thick (Ga,Mn)As film under tensile strain imposed by an (In,Ga)As substrate, with the easy axis  perpendicular to the plane, periodic stripe domains  were revealed by a micro-Hall scanning probe \cite{Shono:2000_APL}. It was found \cite{Dietl:2001_PRBb} that the values of the energy $K_{u}$ of uniaxial magnetic anisotropy  and of the exchange stiffness $A$, both determined from the $p$-$d$ Zener model, lead to the low-temperature width of the stripes $W = 1.1$~$\mu$m, which compares favorably with the experimental value $W = 1.5$~$\mu$m. However, the data suggest that a decrease of $A$ with temperature is slower than expected theoretically.

\begin{figure}
\includegraphics[width=8.5cm]{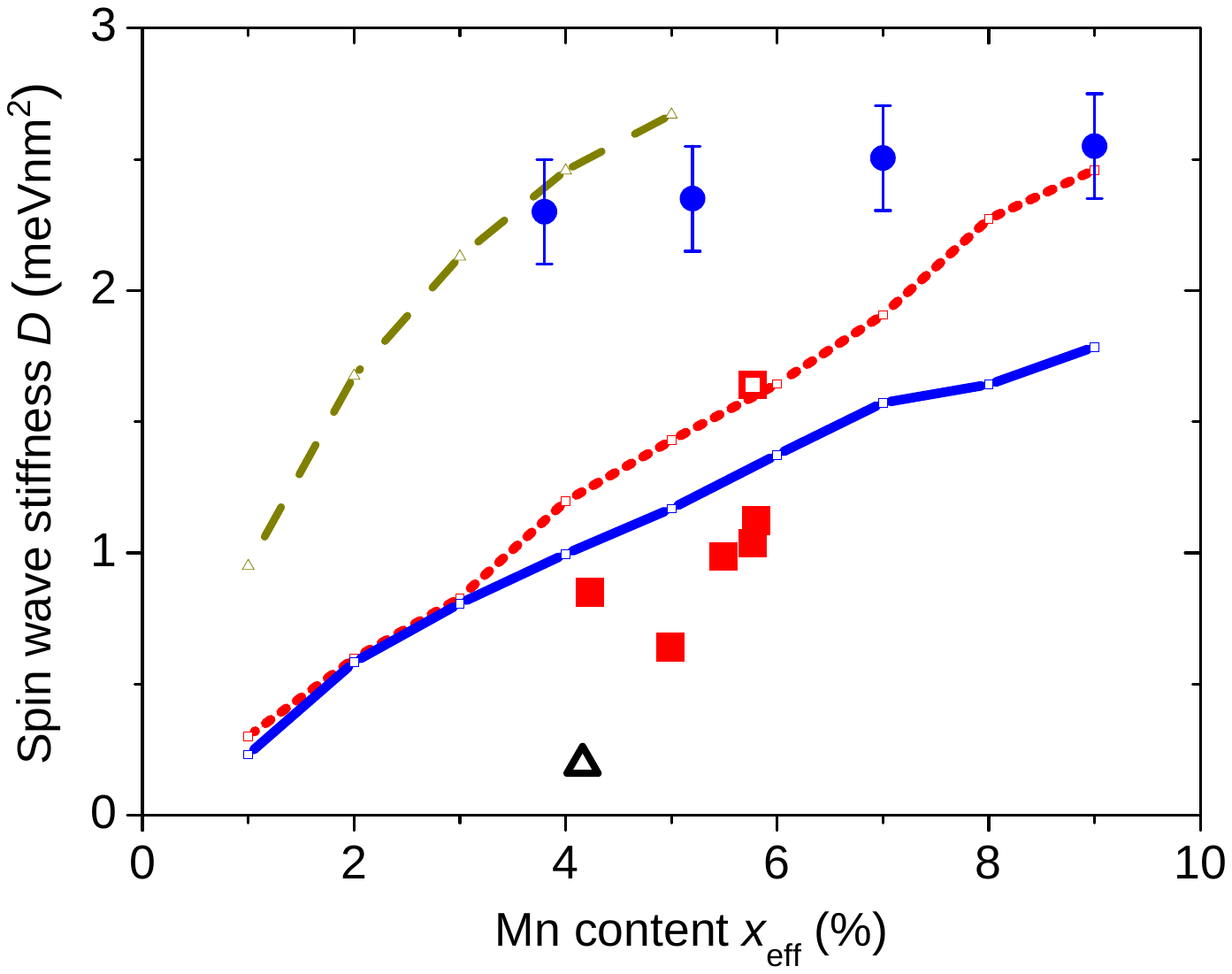}
\caption[]{(Color online) Spin-wave stiffness obtained from time-resolved magnetooptical studies of Ga$_{1-x}$Mn$_x$As at 15~K  (solid circles, \onlinecite{Nemec:2013_NC}) and from widths of domain stripes \cite{Haghgoo:2010_PRB} measured at 4~K (empty square) or determined by extrapolation to low temperatures the values obtained at $T=0.4T_{\text{C}}$ (solid squares) for (Ga,Mn)(As,P) on GaAs and (Ga,Mn)As on (In,Ga)As (solid triangle).  Solid and dotted lines represent zero-temperature $p$-$d$ Zener modeling with no adjustable parameter for $p= 0.3N_0x$ and $p = N_0x$, respectively, in (Ga,Mn)As \cite{Werpachowska:2010_PRBb}. Zero-temperature spin wave stiffness obtained by {\em ab initio} computations is shown for a comparison (dashed line, \onlinecite{Bouzerar:2007_EPLb}).}
\label{fig:GaMnAs_D_x}
\end{figure}

More recently, the values of $M_{\text{s}}$, $K_{\text{u}}$, and $A$ were determined for a series of (Ga,Mn)As and (Ga,Mn)(As,P) 50-nm films with perpendicular magnetic anisotropy by combining magnetometry and ferromagnetic resonance with Kerr microscopy that allowed to determine the period of stripe domains \cite{Haghgoo:2010_PRB}. As shown in Fig.~\ref{fig:GaMnAs_D_x}, the low temperature value of $D = 2g\mu_{\text{B}}A/M$ deduced from the data agrees with the expectations of the $p$-$d$ Zener model for $T \ll T_{\text{C}}$. On the other hand, a decrease of $A$ and, thus of $D$ with temperature for this sample is {\em faster} in this case \cite{Haghgoo:2010_PRB} than expected theoretically. Accordingly, the magnitude of $D$ obtained for a series of samples at $T=0.4T_{\text{C}}$, even enlarged by 20\% implied by temperature variation of $M_{\text{s}}$ between $T=0.4T_{\text{C}}$ and $T \ll T_{\text{C}}$, are lower than theoretically predicted (Fig.~\ref{fig:GaMnAs_D_x}).

Spin-wave signatures were clearly resolved in ferromagnetic resonance (FMR) \cite{Fedorych:2002_PRB,Liu:2007_PRB,Bihler:2009_PRBa} and pump-probe differential magnetic Kerr experiments \cite{Wang:2007_PRB} on (Ga,Mn)As films. However, a quantitative interpretation of resonance energies in terms of dimensional quantization of the spin wave spectrum was so-far possible only assuming the presence of long range inhomogeneities along the growth direction, taken in the form of a triangular \cite{Bihler:2009_PRBa} or a parabolic well \cite{Liu:2007_PRB}. Similarly, heavily debated are effects of pinning, magnetic anisotropy, and modes at surfaces and interfaces \cite{Bihler:2009_PRBa,Wang:2007_PRB,Liu:2007_PRB}. The experimental values of spin-wave stiffness $D$ obtained by various experiments show a rather large dispersion \cite{Werpachowska:2010_PRBb}.

Recently, spin-wave-like resonances were detected by time-resolved magnetooptics on a series of thin annealed (Ga,Mn)As samples \cite{Nemec:2013_NC}. Energy differences between particular resonances and scaling with sample thickness indicated that dimensional quantization of bulk spin waves in spatially uniformed ferromagnetic slabs was observed. The values of  $D$ obtained in this way are larger by a factor of about 2 than expected within the $p$-$d$ Zener model, as shown in Fig.~\ref{fig:GaMnAs_D_x}.

\subsection{Spintronic structures}
\label{sec:structures_theory}
Modeling of spin-injection efficiency, tunneling magnetoresistance, and domain wall resistance in various structures of DFSs is particularly appealing from the theoretical perspective as useful results can be obtained neglecting disorder entirely.  Below we present a quantitative outcome of the disorder-free Landauer-B\"uttiker coherent transport theory combined with either multiband $kp$ or multiorbital tight-binding approaches, the latter better handling the inversion symmetry breaking (Dresselhaus terms) and interfacial Rashba effects as well as the tunneling \emph{via} $\vec{k}$ points away from the zone center. Within the employed formalisms, the holes are assumed to reside in the GaAs-like valence band, subject to the $p$-$d$ exchange interaction treated within the virtual crystal and molecular field approximations with the standard values of the $sp-d$ exchange energies, $N_0\alpha =0.2$~eV and $N_0\beta = -1.2$~eV. As shown in this section, a number of prominent spintronic characteristics, including spin injection efficiency, the magnitude of TMR and TAMR effects as well as domain-wall resistance (reviewed in Secs.~\ref{sec:injection}, \ref{sec:TMR}, and \ref{sec:current_domains}, respectively), are captured by such modeling, though some aspects of experimental results, such as zero-bias anomaly in tunneling spectra \cite{Chun:2002_PRB,Pappert:2006_PRL,Richardella:2010_S}, a strong temperature dependence of TMR magnitude at $T \ll T_{\text{C}}$ \cite{Chiba:2004_PE} or very large values of TAMR in nanostructures \cite{Ruster:2005_PRL,Giddings:2005_PRL} or junctions \cite{Pappert:2006_PRL} point to the importance of correlation effects at the localization boundary, which have not been taken into account in theoretical models developed up to now.

\subsubsection{Spin current polarization}
\label{spin_current_comparison}
\begin{figure}
\includegraphics[width=8.5cm]{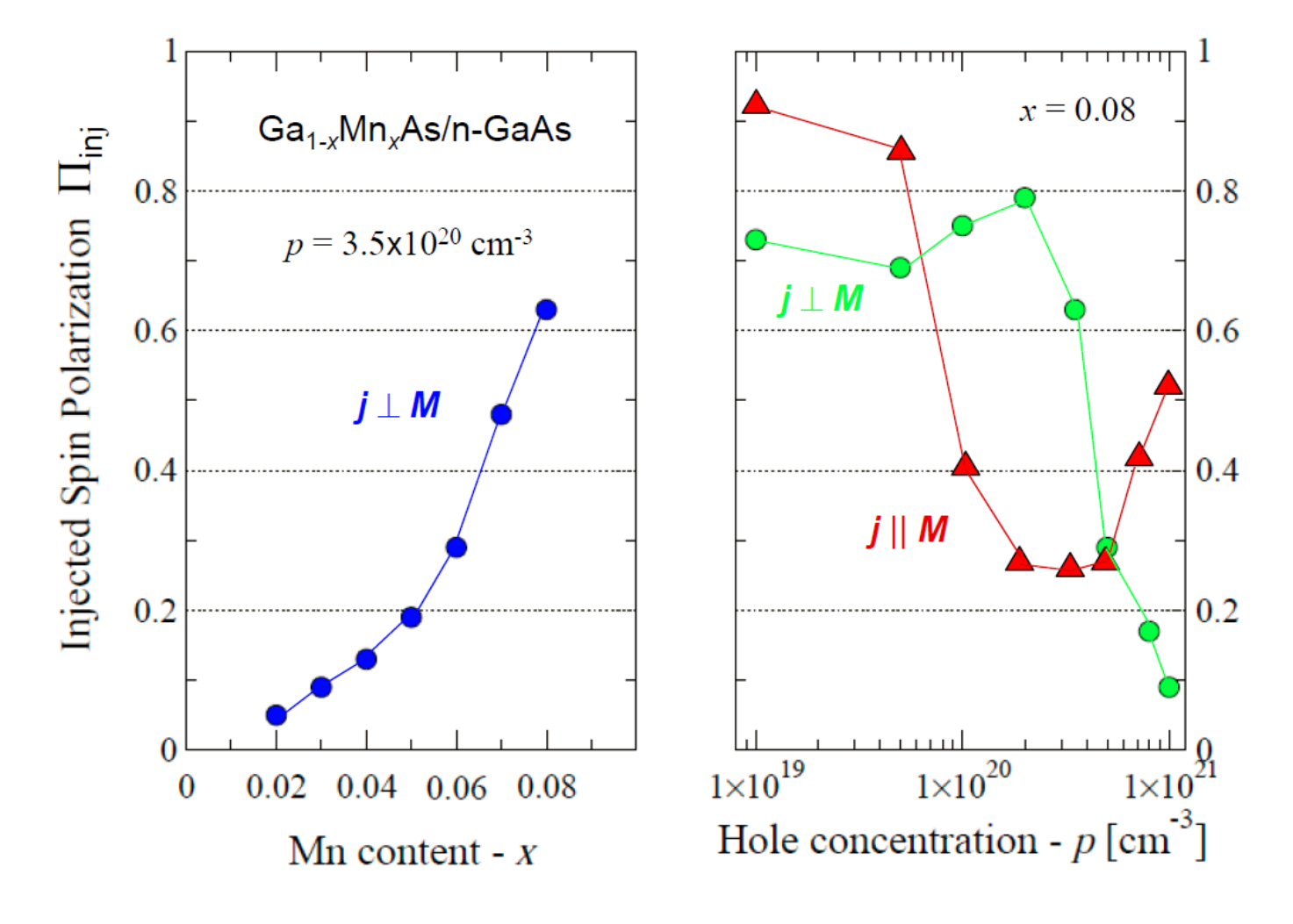}
\caption[]{(Color online) Computed spin polarization (points) of Zener tunneling current at zero temperature and low bias voltage 0.01~V in Esaki diodes of unstrained  Ga$_{1-x}$Mn$_x$As and n-GaAs  (inset) with the electron concentration $n = 10^{19}$~cm$^{-3}$ as a function of the effective Mn density (left panel) and hole concentration for two orientations of magnetization in respect to the current direction (right panel). Adapted from \onlinecite{Sankowski:2006_PE,Sankowski:2007_PRB}.}
\label{fig:GaMnAs_Sankowski_Pi}
\end{figure}

Spin current polarization $\Pi_{\text{inj}}$ in Esaki diodes was computed within the forty orbitals' $sp^3d^5s^*$ tight-binding model  \cite{Dorpe:2005_PRB,Sankowski:2006_PE,Sankowski:2007_PRB}. Figure~\ref{fig:GaMnAs_Sankowski_Pi} presents theoretical values of $\Pi_{\text{inj}}$ at low bias for various Mn and hole concentrations for a Ga$_{1-x}$Mn$_x$As/n-GaAs Esaki diode with the depletion region consisting of 4 double-monolayers and assuming $n = 10^{19}$~cm$^{-3}$ in GaAs.  As the magnitude and the dependence of hole spin splitting on the orientation of $\vec{k}$ with respect to $\vec{M}$ are different in particular valence band subbands, the predicted values of $\Pi_{\text{inj}}$ vary strongly not only with $x_{\text{eff}}$, $p$, and $n$ but also with the angle $\theta$ between $\vec{j}$ and $\vec{M}$.

As mentioned in Sec.~\ref{sec:injection}, three experiments \cite{Dorpe:2004_APL,Kohda:2006_APL,Ciorga:2009_PRB} carried out at $T \ll T_{\text{C}}$ and $\theta = 45^o, 0^o, 90^o$ for samples with nominal Mn content $x = 0.08, 0.057, 0.05$ and $T_{\text{C}} = 120, 70, 65$~K  led to $\Pi_{\text{inj}} =0.4, 0.47$, and 0.51, respectively. The values of $T_{\text{C}}$ for particular samples imply a certain degree of compensation by interstitial Mn and/or antisite defects, so that $x_{\text{eff}}  \leqslant x$ and $p < N_0x$. Taking this into account, we conclude that the theoretical results summarized in Fig.~\ref{fig:GaMnAs_Sankowski_Pi} are consistent with the experimental findings, though a quantitative comparison requires more accurate information on the magnitudes of $x_{\text{eff}}$ and $p$. Furthermore, the theory \cite{Dorpe:2005_PRB} explained a decay of $\Pi_{\text{inj}}$ to zero at the bias of the order of 0.2~V \cite{Dorpe:2004_APL,Kohda:2006_APL,Ciorga:2009_PRB} (see, Fig.~\ref{fig:spin_injection}). Since spin injection is a surface-sensitive phenomenon, the theory \cite{Dorpe:2005_PRB,Sankowski:2007_PRB} predicted a 6\% difference in $\Pi_{\text{inj}}$ for $M \parallel [110]$ in comparison to the case $M \parallel [1\bar{1}0]$ at $x_{\text{eff}} = 0.08$ and $p = 3.5\times 10^{20}$~cm$^{-3}$, the effect found experimentally \cite{Dorpe:2005_PRB}. This difference is about factor of 10 greater than that generated by effective shear strain (see, Sec.~\ref{sec:anisotropy-comparison}), $\epsilon_{xy} \lesssim 0.1$\% \cite{Sankowski:2007_PRB}.

It is worth noting that the magnitudes of $\Pi_{\text{inj}}$ depicted in Fig.~\ref{fig:GaMnAs_Sankowski_Pi} are only approximately equal to the values of spin current polarization $P_{\text{c}}$ provided by Andreev reflection or domain wall velocity. With this taken into account, the theoretical results on $\Pi_{\text{inj}}$ are consistent with available data on $P_{\text{c}}$  \cite{Curiale:2012_PRL}.

An effective and tunable by bias injection of spin polarized electrons was predicted theoretically, within an eight band $kp$ model, for interband resonant tunneling in double-barrier InAs/AlSb/Ga$_x$Mn$_{1-x}$Sb/AlSb/InAs heterostructures \cite{Petukhov:2003_PRB}.

\subsubsection{Magnetic tunnel junctions}

\begin{figure}
\includegraphics[width=8.2cm]{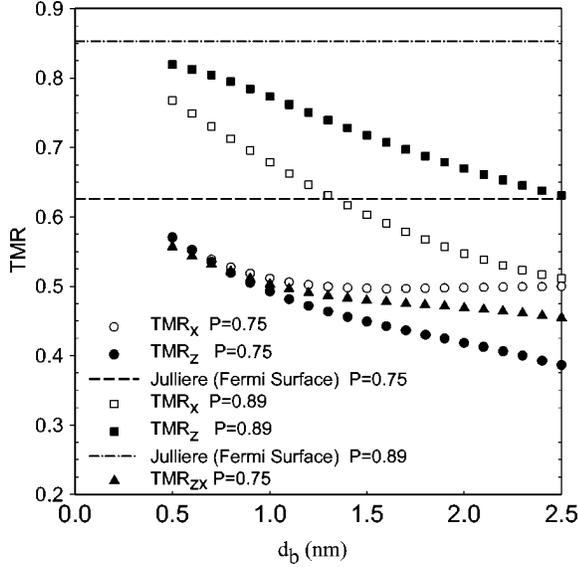}
\caption[]{Computed magnitudes of TMR$_{\text{p}} = (R_{\uparrow\downarrow} - R_{\uparrow\uparrow})/R_{\uparrow\downarrow}$ for the hole concentration $10^{20}$~cm$^{-3}$, current along the [001] direction, and magnetization along the [100], [001], and  [101] directions (TMR$_x$, TMR$_z$, and TMR$_{xz}$, respectively). The data are shown for two values of hole spin polarization $P = 0.75$ and 0.89 corresponding to, at $T\ll T_{\text{C}}$, $x_{\text{eff}} \simeq 0.042$ and 0.066, respectively. The values expected from Julliere's formula  TMR$_{\text{p}} = 2P^2/(1+P^2)$ are shown by horizontal lines. From \onlinecite{Brey:2004_APL}.}
\label{fig:GaMnAs_Brey}
\end{figure}

Figure \ref{fig:GaMnAs_Brey} shows the magnitudes of TMR$_{\text{p}}$ = TMR/(TMR + 1), where TMR is defined in Sec.~\ref{sec:TMR}, for a trilayer structure containing (Ga,Mn)As electrodes separated by a non-magnetic 0.3~eV-high hole barrier, computed within a six band $kp$ approach employing Luttinger parameters of GaAs and assuming realistic values of hole spin polarization $P$ and concentration $p$ in (Ga,Mn)As.  As seen, for $p = 10^{20}$~cm$^{-3}$ and $P = 0.75$ ($x_{\text{eff}} \simeq 0.042$; \onlinecite{Dietl:2001_PRB}), the theory satisfactorily describes the experimental TMR magnitude as well as its dependence on the barrier thickness and magnetization orientation, as observed for Ga$_{0.96}$Mn$_{0.04}$As /AlAs/ Ga$_{0.968}$Mn$_{0.032}$As (\onlinecite{Tanaka:2001_PRL}; see, Fig.~\ref{fig:GaMnAs_TMR_Tanaka}).

The tight binding model discussed in the previous subsection in the context of spin current polarization was also successfully employed \cite{Sankowski:2006_PE,Sankowski:2007_PRB} to describe low temperature magnitudes of TMR in trilayer structures with (Ga,Mn)As electrodes and AlAs \cite{Tanaka:2001_PRL} or GaAs \cite{Chiba:2004_PE} barriers.  Furthermore, the theory reproduced a fast decrease of TMR with the device bias as well as it indicated that the magnitude of TAMR should not exceed 10\% under usual strain conditions and for hole densities corresponding to the metal side of the metal-to-insulator transition \cite{Sankowski:2007_PRB}.

\begin{figure}
\includegraphics[width=9.0cm]{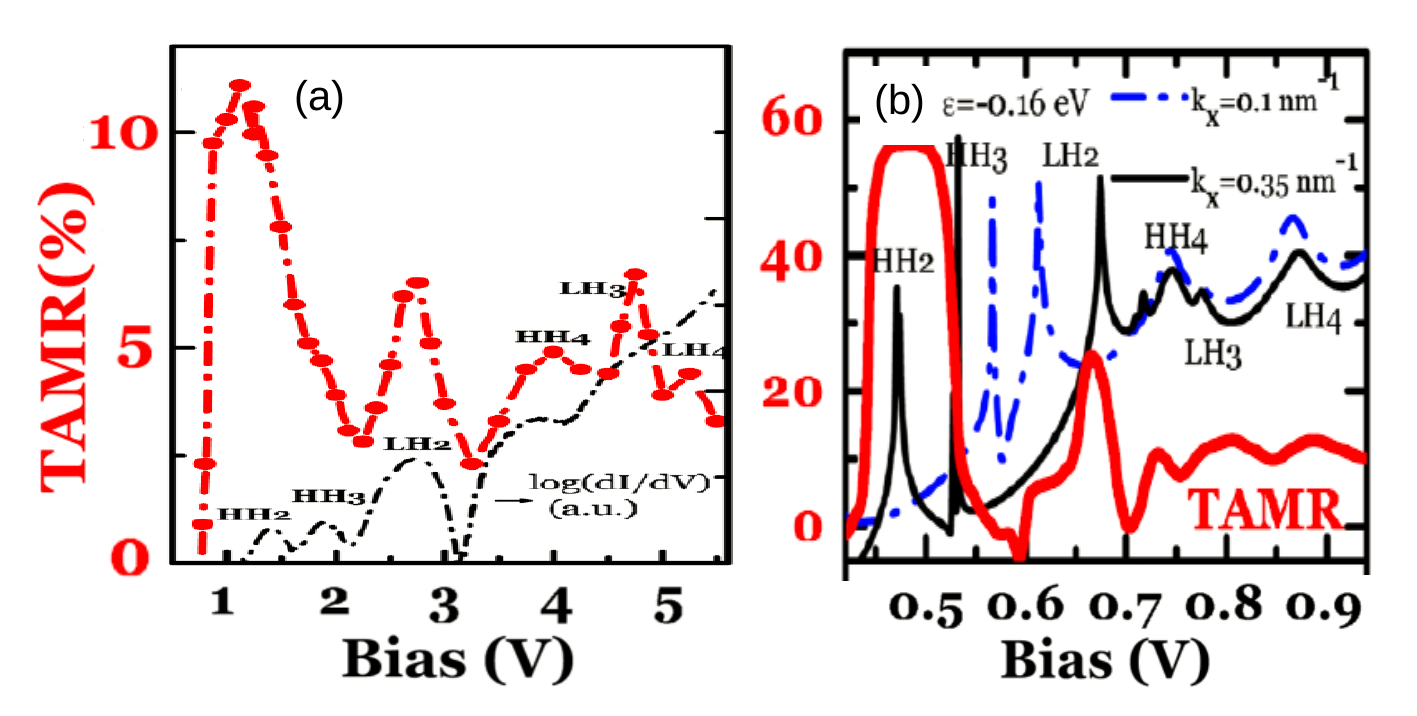}
\caption[]{(Color online) Comparison of experimental (a) and computed (b) values of differential conductance d$I$/d$V$ (thin lines) and TAMR =
$(I^{[001]} - I^{[100]})/I^{[100]}$ (thick lines), where the indices [001] and [100] describe magnetization orientation, as a function of bias $V$ in a resonant tunneling diode with a (Ga,Mn)As hole emitter, grown along the [001] direction. Experimental and theoretical data come into agreement if renormalization of $V$ and TAMR values by a factor of 5 implied by a series resistance is taken into account.  The computation was performed for indicated values of average hole energy $\epsilon$  and in plane momentum $k_x$. From \onlinecite{Elsen:2007_PRL}.}
\label{fig:GaMnAs_Elsen}
\end{figure}

Theoretical studies were put forward to examine TMR and TAMR in double barrier structures within a six band $kp$ theory.  It was demonstrated that spin-dependent resonant tunneling could dramatically enhance TMR in resonant tunneling diodes containing both emitter and collector of (Ga,Mn)As \cite{Petukhov:2002_PRL}. This prediction has not yet been confirmed experimentally presumably because it is not easy to ensure coherent tunneling in RTDs grown by LT MBE \cite{Mattana:2003_PRL}. In contrast, the six band $kp$ formalism explained a character of TAMR oscillations as a function of bias in a double barrier structure with {\em one} (Ga,Mn)As electrode, as depicted in Fig.~\ref{fig:GaMnAs_Elsen} \cite{Elsen:2007_PRL}.

\subsubsection{Domain wall resistance}

A spherical four band model \cite{Nguyen:2006_PRL} and a twenty orbitals' $sp^3s^*$ tight-binding approximation \cite{Oszwaldowski:2006_PRB} were employed to evaluate the intrinsic domain-wall resistivity $R^{\text{int}}A$ in unstrained (Ga,Mn)As within the disorder free Landauer-B\"uttiker formalism.  These studies demonstrated that owing to the spin-orbit interaction $R^{\text{int}}A > 0$ even if the domain-wall width is much longer than the de Broglie wavelength of holes at the Fermi level.  Various crystallographic orientation of (Ga,Mn)As nanowires containing either Bloch or N\'eel domain walls were considered \cite{Oszwaldowski:2006_PRB}. It was found that the computed values $10^{-2} \gtrsim R^{\text{int}}A  \gtrsim 10^{-3}$~$\Omega\mu$m$^2$ depending on hole density, are at least one order of magnitude smaller than the ones determined experimentally for domain walls pinned by etched steps (\onlinecite{Chiba:2006_PRL,Wang:2010_JMMM}; see, Sec.~\ref{sec:current_domains}) but consistent with the much lower value  $R^{\text{int}}A = 0.01 \pm 0.02 $~$\Omega\mu$m$^2$, found in the case of the wall pining by linear defects \cite{Wang:2010_JMMM}.

\section{Summary and outlook}
\label{sec:comments}

A series of accomplishments presented in this review has documented a prominent role of dilute ferromagnetic semiconductors, especially (Ga,Mn)As, in bridging science and technology of semiconductors and magnetic materials. Indeed, several spintronic functionalities revealed in DFSs are now extensively explored in ferromagnetic metals, examples include electrical spin injection to semiconductors, electric-field control of magnetism, single domain wall motion by spin-torque transfer in the absence of a magnetic field, and tunneling anisotropic magnetoresistance in sequential, resonant, and Coulomb blockade regimes. Conversely, the Stoner-Wohlfarth, Landau-Lifshitz-Gilbert, and  Berger-Slonczewski formalisms, developed initially for magnetic metals, have been successfully employed to describe spintronic characteristics of DFS-based ferromagnets.

An outstanding aspect of DFSs is that input parameters to these formalisms can be theoretically evaluated by incorporating exchange coupling between carriers and localized spins into the computationally efficient $kp$ or tight binding approaches, employed routinely to model semiconductor properties and devices.  As emphasized in this review, the $p$-$d$ Zener model describes semi-quantitatively, and often quantitatively, a number of thermodynamic and micromagnetic properties of tetrahedrally coordinated DFSs containing delocalized or weakly localized valence band holes, including the Curie temperature in various dimensionality systems, Mn spin and hole magnetization, anisotropy fields, and exchange stiffness as a function of hole and Mn ion concentrations. It remains to be seen whether progress in the experimental determination of these concentrations will bring experimental and theoretical results even closer.

Furthermore, a number of findings reviewed here documented substantial progress in assessing the role played by Anderson-Mott localization, the competition between ferromagnetic and antiferromagnetic interactions, solubility limit, self-compensation, and the transition to the strong coupling case with decreasing of the lattice parameter -- the challenges that had been identified \cite{Dietl:2000_S} as possible obstacles on the way to synthesize a DFS supporting ferromagnetic order up to above room temperature. In particular, as currently known, the influence of antiferromagnetic interactions and self-compensation limit $T_{\text{C}}$ to about 200~K so far in (Ga,Mn)As. At the same time, the importance of quantum localization in (Ga,Mn)As and related systems makes a quantitative description of static and dynamic conductivities difficult as there is no appropriate theory, even for non-magnetic semiconductors, in this regime. The strong coupling, in turn, shifts the insulator-to-metal transition to a non-achievable Mn and hole concentration range as of today in nitrides and oxides, so that rather than hole-mediated coupling, ferromagnetic superexchange accounts for $T_{\text{C}}$ up to 13~K in Ga$_{0.9}$Mn$_{0.1}$N with the Fermi level residing in the Mn acceptor impurity band.
A striking consequence of solubility limits is a self-organized assembling of magnetic nanocrystals inside a semiconductor host by chemical or crystallographic phase separation. These heterogeneous magnetic systems have apparent $T_{\text{C}}$ usually well above room temperature and, accordingly, a number of groups look for possible spintronic functionalities of such nanocomposites.

Is it then possible to obtain a high $T_{\text{C}}$ uniform DFS? The view that the $p$-$d$ Zener mechanism can result in the robust ferromagnetism is supported by the case of double perovskite compounds, such as Sr$_2$CrReO$_6$, where this mechanism leads to magnitudes of $T_{\text{C}}$ as high as 625~K, despite the fact that the distance between localized spins is as large as 0.6 -- 0.7~nm \cite{Serrate:2007_JPCM}, much greater than the separation of 0.5~nm between next nearest neighbor cations in GaN and ZnO. In light of this estimate, the search for a room temperature uniform DFS will continue to be an active field of research. Here, in addition to 3$d$ TM impurities in various hosts, other spin dopants will be considered, including $4d$ TMs and elements with open $f$ shells as well as spin carrying defects. However, independently of the progress in achieving a high $T_{\text{C}}$ system, (Ga,Mn)As and related compounds as well as a (Ga,Mn)N-type of ferromagnets will continue to constitute an important playground for exploring novel phenomena, functionalities, and concepts at the intersection of semiconductor physics and magnetism.

In addition to current interest in various magnetically doped semiconductors, oxides, and organic materials, a lot of attention will be devoted to four emerging families of compounds: (i) high Curie temperature ferrimagnetic spinel oxides and Heusler compounds such as Mn$_2$CoAl \cite{Ouardi:2013_PRL}, awaiting mastering of defect and carrier control; (ii) high N\'eel temperature semiconductors, {\em e.~g.} LiMnAs,  for antiferromagnetic spintronics \cite{Jungwirth:2011_PRB}; (iii) topological insulators, in which ferromagnetism might be mediated by Dirac electrons, {\em e.~g.}, (Bi,Mn)$_2$Te$_3$ \cite{Checkelsky:2012_NP}; (iv) derivatives of FeAs- and CuO-based superconductors, such as (K,Ba)(Zn,Mn,Fe)$_2$As$_2$ compounds \cite{Zhao:2013_NC} 
for studies of interplay between $p$-$d$ Zener ferromagnetism, antiferromagnetic superexchange, and superconductivity. Here, nanocharacterization protocols, elaborated over the recent years for DMSs \cite{Bonanni:2011_SST}, will play the essential role in the meaningful development of new materials.

One may anticipate that an increasing number of studies will be devoted to hybrid structures combining DFSs with other ferromagnets, antiferromagnets, superconductors, and topological insulators as well as to nanostructured systems, such as magnetically doped rings, nanowires, nanoconstrictions, quantum dots, and colloidal nanocrystals,  including possibly more complex structures, such as nano electro mechanical systems. Recent progress in the fabrication of (Zn,Mn)Te/(Zn,Mg)Te core/shell nanowires \cite{Wojnar:2012_NL} allows one to search for exotic ground states in modulation-doped 1D magnetic systems. Similarly, studies of magnetic quantum dots \cite{Abolfath:2007_NJP} and colloidal nanocrystals \cite{Beaulac:2008_AFM,White:2008_ChM} bridge on the one hand the physics of bound magnetic polaron and carrier-controlled ferromagnetism, and on the other, the electronic and nuclear magnetism.

In this review, we primarily focussed on properties and functionalities resulting from the {\emph{collective}} spin phenomena. Another ultimate limit of DMSs research constitutes works on manipulation of a \emph{single TM spin}. This field of {\em solotronics} \cite{Koenraad:2011_NM} began in DMSs by exploiting a single Mn in II-VI \cite{Cibert:2008_B,Goryca:2009_PRLb} and III-V \cite{Kudelski:2007_PRL} self-assembled quantum dots.  Another single-impurity phenomenon, not yet explored in the context of DMSs, is the Kondo effect expected in the case of a localized spin coupled by an antiferromagnetic exchange to a Fermi sea of carriers.

On the theoretical side, enduring progress in the reliability of {\em ab initio} methods is expected. From the DFS perspective, in addition to the issue of results' convergence as a function of the energy cut-off, density of $k$ points, and the number of atoms in the supercells, three experimentally relevant challenges, of differing numerical complexity, can be given: (i) the incorporation of the spin-orbit interaction that significantly affects the band structure of DFSs and accounts for magnetic anisotropy effects -- some progress in that direction has already been reported \cite{Mankovsky:2011_PRB}; (ii) the improved treatment of the exchange-correlation functional, so that more realistic values of the band-gap and $d$-level positions can be predicted as well as Mott-Hubbard localization of $d$ electrons handled adequately -- here also improved computational schemes are being implemented (see, {\em e.~g.},~\onlinecite{Stroppa:2009_PRB,DiMarco:2013_NC}); (iii) the development of methods that would be able to tackle Anderson-Mott localization phenomena, such as the appearance of the Coulomb gap in the density of states at the Fermi level.

In retrospect, striking properties and functionalities found in DFSs, not only influenced semiconductor and metal spintronics but, to a large extent, have accounted for a spread of spintronic research over many other materials families.  Search for novel magnetic semiconductors has led to the discovery of FeAs-based superconductors, nucleated the concept of magnetism without magnetic ions in oxides, and demonstrated a surprising influence of the Fermi energy upon the position and distribution of magnetic ions in semiconductors. Research on topological aspects of the anomalous Hall effect in bands coupled by spin-orbit interactions paved the way for uncovering spin Hall effects and topological matter. It might be, therefore, expected that studies of magnetically doped semiconductors, insulators, and organic materials will continue to bring unanticipated and inspiring discoveries in the years to come.

\section*{Acknowledgments}
We are grateful to our colleagues world-wide for their contributions, as
indicated in the reference list. T.~D. would like to thank Alberta
Bonanni, Maciej Sawicki, and Cezary \'Sliwa and H.~O. Fumihiro Matsukura
for many years of fruitful collaboration. T.~D. acknowledges support from the
FunDMS Advanced Grant (No.~227690) of the European Research Council within
the "Ideas" 7th Framework Programme of the European Commission and the National Center
of Science in Poland (Decision No.~2011/02/A/ST3/00125). The research of H.~O. has
been supported by the Grant-in-Aids from MEXT/JSPS, the GCOE Program at Tohoku
University, the Research and Development for Next-Generation Information
Technology Program from MEXT, and the FIRST program from JSPS. H.~O. and
T.~D. were supported by the Ohno Semiconductor Spintronics Project,
an ERATO project of JST.

\appendix
\section{Micromagnetic theory}
As surveyed in this review, the structure and motion of magnetic domains  as well as a number of FMR and time-resolved magnetooptical experiments on (Ga,Mn)As and related compounds have demonstrated that the time-honored micromagnetic theory of ferromagnets applies to these systems. The use of the continuous medium approximation inherent to this theory is justified by a sizable width of domain walls that is much larger than an average distance between Mn spins. The micromagnetic theory provides a spatial and temporary evolution of magnetization in given values of a magnetic field and spin current. The magnitudes of magnetic anisotropy fields $H_{i}$ and exchange stiffness $A$ constitute input parameters in the static case, whereas in the dynamic situation the Land\'e factor $g$ and the Gilbert constant $\alpha_{\text{G}}$ enter additionally into theory. For a specific case of current-induced domain wall motion, non-adiabatic spin torque $\beta_{\text{w}}$ is one more input parameter.  As discussed in Sec.~\ref{sec:theory}, the magnitudes of these material parameters can be theoretically evaluated in a rather straightforward way, which constitutes a unique aspect of DFSs.

The starting point of the micromagnetic theory is the Landau-Lifshitz-Gilbert (LLG) equation according to which the dynamics of the local spin direction $\vec{m} = \vec{M}/M$ is determined by a competition of a torque, due to an effective magnetic field $\vec{H}_{\text{eff}}$ and/or an electric current $\vec{j}$, with a damping term, characterized by the Gilbert constant $\alpha_{\text{G}}$,
\begin{equation}
\dot{\vec{m}} =
-\gamma\vec{m}\times\vec{H}_{\text{eff}} - (1 -\beta_{\text{w}}\vec{m}\times) (\vec{\tau}_j\cdot\nabla)\vec{m} +  \alpha_{\text{G}}\vec{m}\times\dot{\vec{m}}.
\label{eq:LLG}
\end{equation}
Here $\gamma = g\mu_{\text{B}}/\hbar$; $\vec{H}_{\text{eff}}$ is given by a variational derivative,
$\vec{H}_{\text{eff}}(\vec{r})= -\delta{\cal{F}}[\vec{M}(\vec{r} )]/\delta \vec{M}(\vec{r})$; $\beta_{\text{w}}$ describes the magnitude of a non-adiabatic out-of-plane current torque brought about by the spin-orbit interaction, and $\vec{\tau}_j= g\mu_{\text{B}}P_{\text{c}}\vec{j}/(2eM)$, where $P_{\text{c}}$ is the current spin polarization, and $M$ is localized spin magnetization.
Interestingly, by the Onsager reciprocity relations, not only the spin current in the presence of a spin texture $\vec{m} =\vec{m}(\vec{r})$ induces magnetization dynamics $ \vec{m} = \vec{m}(t)$, but also $\vec{m}(\vec{r},t)$  can generate an electric current \cite{Hals:2009_PRL}. While the LLG equation (together with initial conditions) allows to determine $\vec{m}(\vec{r},t)$, its static limit $\dot{\vec{m}} =0$ is also interesting, as it makes it possible to find expected spin textures $\vec{m}(\vec{r})$ under stationary conditions, {\em i.~e.}, the domain structure.

It is seen from the form of the LLG equation that only contributions to $\vec{H}_{\text{eff}}$, which can have components perpendicular to $\vec{m}$, are relevant. These include external and the shape dependent dipolar (demagnetization) fields coming from surrounding magnetic moments, $\vec{H} + \vec{H}_\text{d}[\vec{M}(\vec{r})]$, as well as the crystalline anisotropy field $\vec{H}_\text{a}$ and the magnetization stiffness,
\begin{equation}
\vec{H}_{\text{ex}} = - \frac{2A}{M}\vec{\nabla}^2\vec{m}.
\end{equation}
This term originates from the free energy change $\delta {\cal{F}}$ associated with a twisting of $\vec{M}(\vec{r})$. To the lowest order in the gradient of the magnetization components in an isotropic medium, it has a general form,
\begin{equation}
\delta {\cal{F}} =  A|\vec{\nabla}\vec{m}|^2,
\end{equation}
where  $A$ is called the exchange stiffness, as its magnitude scales with the strength of the exchange coupling between the spins. In the absence of an electric current and damping, these equations lead to the precession modes (spin waves) of the form,
\begin{equation}
\hbar\omega(\vec{q}) = g\mu_{\text{B}}H_{\text{o}} + Dq^2,
\label{eq:dispersion}
\end{equation}
where the magnetic field $\vec{H}_{\text{o}}$ determining the spin gap is an appropriate sum of $\vec{H}$, $\vec{H}_{\text{d}}$, and $\vec{H}_{\text{a}}$, and $D = 2g\mu_{\text{B}}A/M$, as reviewed elsewhere \cite{Liu:2006_JPCM}. Boundary conditions as well as additional modes and spin-wave pinning at surfaces and interfaces in thin films and other nanostructures will modify this dispersion relation. Furthermore, since the precession involves the system of coupled Mn and hole spins, the apparent Mn Land\'e factor $g$ will be reduced or enhanced for antiparallel and parallel orientation of corresponding magnetizations, respectively \cite{Liu:2006_JPCM,Sliwa:2006_PRB}, as mentioned in Sec.~\ref{sec:magnetic_resonance}.

In the static and single domain approximation, for which $\vec{m}$ is time and position independent, its direction points along $\vec{H}_{\text{eff}}$ and is determined by minimization of ${\cal{F}}$ in respect to spherical angles $\theta$ and $\phi$, as discussed in Sec.~\ref{sec:anisotropy} for the specific case of (001) and (113) substrates.

%


\begin{thebibliography}{498}%
\makeatletter
\providecommand \@ifxundefined [1]{%
 \ifx #1\undefined \expandafter \@firstoftwo
 \else \expandafter \@secondoftwo
\fi
}%
\providecommand \@ifnum [1]{%
 \ifnum #1\expandafter \@firstoftwo
 \else \expandafter \@secondoftwo
\fi
}%
\providecommand \natexlab [1]{#1}%
\providecommand \enquote [1]{``#1''}%
\providecommand \bibnamefont  [1]{#1}%
\providecommand \bibfnamefont [1]{#1}%
\providecommand \citenamefont [1]{#1}%
\providecommand\href[0]{\@sanitize\@href}%
\providecommand\@href[1]{\endgroup\@@startlink{#1}\endgroup\@@href}%
\providecommand\@@href[1]{#1\@@endlink}%
\providecommand \@sanitize [0]{\begingroup\catcode`\&12\catcode`\#12\relax}%
\@ifxundefined \pdfoutput {\@firstoftwo}{%
 \@ifnum{\z@=\pdfoutput}{\@firstoftwo}{\@secondoftwo}%
}{%
 \providecommand\@@startlink[1]{\leavevmode}%
 \providecommand\@@endlink[0]{}%
}{%
 \providecommand\@@startlink[1]{%
  \leavevmode
  \pdfstartlink
   attr{/Border[0 0 1 ]/H/I/C[0 1 1]}%
   user{/Subtype/Link/A<</Type/Action/S/URI/URI(#1)>>}%
  \relax
 }%
 \providecommand\@@endlink[0]{\pdfendlink}%
}%
\providecommand \url  [0]{\begingroup\@sanitize \@url }%
\providecommand \@url [1]{\endgroup\@href {#1}{\urlprefix}}%
\providecommand \urlprefix [0]{URL }%
\providecommand \Eprint[0]{\href }%
\@ifxundefined \urlstyle {%
  \providecommand \doi [1]{doi:\discretionary{}{}{}#1}%
}{%
  \providecommand \doi [0]{doi:\discretionary{}{}{}\begingroup
  \urlstyle{rm}\Url }%
}%
\providecommand \doibase [0]{http://dx.doi.org/}%
\providecommand \Doi[1]{\href{\doibase#1}}%
\providecommand \bibAnnote [3]{%
  \BibitemShut{#1}%
  \begin{quotation}\noindent
    \textsc{Key:}\ #2\\\textsc{Annotation:}\ #3%
  \end{quotation}%
}%
\providecommand \bibAnnoteFile [2]{%
  \IfFileExists{#2}{\bibAnnote {#1} {#2} {\input{#2}}}{}%
}%
\providecommand \typeout [0]{\immediate \write \m@ne }%
\providecommand \selectlanguage [0]{\@gobble}%
\providecommand \bibinfo [0]{\@secondoftwo}%
\providecommand \bibfield [0]{\@secondoftwo}%
\providecommand \translation [1]{[#1]}%
\providecommand \BibitemOpen[0]{}%
\providecommand \bibitemStop [0]{}%
\providecommand \bibitemNoStop [0]{.\EOS\space}%
\providecommand \EOS [0]{\spacefactor3000\relax}%
\providecommand \BibitemShut [1]{\csname bibitem#1\endcsname}%
\bibitem[{\citenamefont{Abanin}\ and\
  \citenamefont{Pesin}(2011)}]{Abanin:2011_PRL}%
  \BibitemOpen
  \bibfield{author}{%
  \bibinfo {author} {\bibnamefont{Abanin}, \bibfnamefont{D.~A.}},\ and\
  \bibinfo {author} {\bibfnamefont{D.~A.}\ \bibnamefont{Pesin}}}%
  , \bibinfo {year} {2011},\ \bibfield{title}{%
  \enquote{\bibinfo {title} {Ordering of magnetic impurities and tunable
  electronic properties of topological insulators},}\ }%
  \bibfield{journal}{%
  \bibinfo {journal} {Phys. Rev. Lett.}\ }%
  \textbf{\bibinfo {volume} {106}},\ \bibinfo {pages} {136802}%
  \bibAnnoteFile{NoStop}{Abanin:2011_PRL}%
\bibitem[{\citenamefont{Abe}\ \emph{et~al.}(2000)\citenamefont{Abe},
  \citenamefont{Matsukura}, \citenamefont{Yasuda}, \citenamefont{Ohno},\ and\
  \citenamefont{Ohno}}]{Abe:2000_PE}%
  \BibitemOpen
  \bibfield{author}{%
  \bibinfo {author} {\bibnamefont{Abe}, \bibfnamefont{E.}}, \bibinfo {author}
  {\bibfnamefont{F.}~\bibnamefont{Matsukura}}, \bibinfo {author}
  {\bibfnamefont{H.}~\bibnamefont{Yasuda}}, \bibinfo {author}
  {\bibfnamefont{Y.}~\bibnamefont{Ohno}},\ and\ \bibinfo {author}
  {\bibfnamefont{H.}~\bibnamefont{Ohno}}}%
  , \bibinfo {year} {2000},\ \bibfield{title}{%
  \enquote{\bibinfo {title} {Molecular beam epitaxy of {III-V} diluted magnetic
  semiconductor {(Ga,Mn)Sb}},}\ }%
  \bibfield{journal}{%
  \bibinfo {journal} {Physica}\ }%
  \textbf{\bibinfo {volume} {E 7}},\ \bibinfo {pages} {981}%
  \bibAnnoteFile{NoStop}{Abe:2000_PE}%
\bibitem[{\citenamefont{Abolfath}\ \emph{et~al.}(2001)\citenamefont{Abolfath},
  \citenamefont{Jungwirth}, \citenamefont{Brum},\ and\
  \citenamefont{MacDonald}}]{Abolfath:2001_PRB}%
  \BibitemOpen
  \bibfield{author}{%
  \bibinfo {author} {\bibnamefont{Abolfath}, \bibfnamefont{M.}}, \bibinfo
  {author} {\bibfnamefont{T.}~\bibnamefont{Jungwirth}}, \bibinfo {author}
  {\bibfnamefont{J.}~\bibnamefont{Brum}},\ and\ \bibinfo {author}
  {\bibfnamefont{A.H.}\ \bibnamefont{MacDonald}}}%
  , \bibinfo {year} {2001},\ \bibfield{title}{%
  \enquote{\bibinfo {title} {Theory of magnetic anisotropy in
  {III$_{1-x}$Mn$_{x}$V} ferromagnets},}\ }%
  \bibfield{journal}{%
  \bibinfo {journal} {Phys. Rev. B}\ }%
  \textbf{\bibinfo {volume} {63}},\ \bibinfo {pages} {054418}%
  \bibAnnoteFile{NoStop}{Abolfath:2001_PRB}%
\bibitem[{\citenamefont{Abolfath}\ \emph{et~al.}(2007)\citenamefont{Abolfath},
  \citenamefont{Hawrylak},\ and\ \citenamefont{Zutic}}]{Abolfath:2007_NJP}%
  \BibitemOpen
  \bibfield{author}{%
  \bibinfo {author} {\bibnamefont{Abolfath}, \bibfnamefont{R.~M.}}, \bibinfo
  {author} {\bibfnamefont{P.}~\bibnamefont{Hawrylak}},\ and\ \bibinfo {author}
  {\bibfnamefont{I.}~\bibnamefont{Zutic}}}%
  , \bibinfo {year} {2007},\ \bibfield{title}{%
  \enquote{\bibinfo {title} {Electronic states of magnetic quantum dots},}\ }%
  \bibfield{journal}{%
  \bibinfo {journal} {New J. Phys.}\ }%
  \textbf{\bibinfo {volume} {9}},\ \bibinfo {pages} {353}%
  \bibAnnoteFile{NoStop}{Abolfath:2007_NJP}%
\bibitem[{\citenamefont{Acbas}\ \emph{et~al.}(2009)\citenamefont{Acbas},
  \citenamefont{Kim}, \citenamefont{Cukr}, \citenamefont{Nov\'ak},
  \citenamefont{Scarpulla}, \citenamefont{Dubon}, \citenamefont{Jungwirth},
  \citenamefont{Sinova},\ and\ \citenamefont{Cerne}}]{Acbas:2009_PRL}%
  \BibitemOpen
  \bibfield{author}{%
  \bibinfo {author} {\bibnamefont{Acbas}, \bibfnamefont{G.}}, \bibinfo {author}
  {\bibfnamefont{M.-H.}\ \bibnamefont{Kim}}, \bibinfo {author}
  {\bibfnamefont{M.}~\bibnamefont{Cukr}}, \bibinfo {author}
  {\bibfnamefont{V.}~\bibnamefont{Nov\'ak}}, \bibinfo {author}
  {\bibfnamefont{M.~A.}\ \bibnamefont{Scarpulla}}, \bibinfo {author}
  {\bibfnamefont{O.~D.}\ \bibnamefont{Dubon}}, \bibinfo {author}
  {\bibfnamefont{T.}~\bibnamefont{Jungwirth}}, \bibinfo {author}
  {\bibfnamefont{Jairo}\ \bibnamefont{Sinova}},\ and\ \bibinfo {author}
  {\bibfnamefont{J.}~\bibnamefont{Cerne}}}%
  , \bibinfo {year} {2009},\ \bibfield{title}{%
  \enquote{\bibinfo {title} {Electronic structure of ferromagnetic
  semiconductor {Ga$_{1-x}$Mn$_x$As} probed by subgap magneto-optical
  spectroscopy},}\ }%
  \bibfield{journal}{%
  \bibinfo {journal} {Phys. Rev. Lett.}\ }%
  \textbf{\bibinfo {volume} {103}},\ \bibinfo {pages} {137201}%
  \bibAnnoteFile{NoStop}{Acbas:2009_PRL}%
\bibitem[{\citenamefont{Adam}\ \emph{et~al.}(2009)\citenamefont{Adam},
  \citenamefont{Vernier}, \citenamefont{Ferr\'e}, \citenamefont{Thiaville},
  \citenamefont{Jeudy}, \citenamefont{L{ema\^itre}}, \citenamefont{Thevenard},\
  and\ \citenamefont{Faini}}]{Adam:2009_PRB}%
  \BibitemOpen
  \bibfield{author}{%
  \bibinfo {author} {\bibnamefont{Adam}, \bibfnamefont{J.-P.}}, \bibinfo
  {author} {\bibfnamefont{N.}~\bibnamefont{Vernier}}, \bibinfo {author}
  {\bibfnamefont{J.}~\bibnamefont{Ferr\'e}}, \bibinfo {author}
  {\bibfnamefont{A.}~\bibnamefont{Thiaville}}, \bibinfo {author}
  {\bibfnamefont{V.}~\bibnamefont{Jeudy}}, \bibinfo {author}
  {\bibfnamefont{A.}~\bibnamefont{L{ema\^itre}}}, \bibinfo {author}
  {\bibfnamefont{L.}~\bibnamefont{Thevenard}},\ and\ \bibinfo {author}
  {\bibfnamefont{G.}~\bibnamefont{Faini}}}%
  , \bibinfo {year} {2009},\ \bibfield{title}{%
  \enquote{\bibinfo {title} {Nonadiabatic spin-transfer torque in {(Ga,Mn)As}
  with perpendicular anisotropy},}\ }%
  \bibfield{journal}{%
  \bibinfo {journal} {Phys. Rev. B}\ }%
  \textbf{\bibinfo {volume} {80}},\ \bibinfo {pages} {193204}%
  \bibAnnoteFile{NoStop}{Adam:2009_PRB}%
\bibitem[{\citenamefont{Adell}\ \emph{et~al.}(2007)\citenamefont{Adell},
  \citenamefont{Adell}, \citenamefont{Ilver}, \citenamefont{Ka{\'n}ski},
  \citenamefont{Sadowski},\ and\ \citenamefont{Domaga{\l}a}}]{Adell:2007_PRB}%
  \BibitemOpen
  \bibfield{author}{%
  \bibinfo {author} {\bibnamefont{Adell}, \bibfnamefont{M.}}, \bibinfo {author}
  {\bibfnamefont{J.}~\bibnamefont{Adell}}, \bibinfo {author}
  {\bibfnamefont{L.}~\bibnamefont{Ilver}}, \bibinfo {author}
  {\bibfnamefont{J.}~\bibnamefont{Ka{\'n}ski}}, \bibinfo {author}
  {\bibfnamefont{J.}~\bibnamefont{Sadowski}},\ and\ \bibinfo {author}
  {\bibfnamefont{J.~Z.}\ \bibnamefont{Domaga{\l}a}}}%
  , \bibinfo {year} {2007},\ \bibfield{title}{%
  \enquote{\bibinfo {title} {Mn enriched surface of annealed {(Ga,Mn)As} layers
  annealed under arsenic capping},}\ }%
  \bibfield{journal}{%
  \bibinfo {journal} {Phys. Rev. B}\ }%
  \textbf{\bibinfo {volume} {75}},\ \bibinfo {pages} {054415}%
  \bibAnnoteFile{NoStop}{Adell:2007_PRB}%
\bibitem[{\citenamefont{Altshuler}\ and\
  \citenamefont{Aronov}(1985)}]{Altshuler:1985_B}%
  \BibitemOpen
  \bibfield{author}{%
  \bibinfo {author} {\bibnamefont{Altshuler}, \bibfnamefont{B.~L.}},\ and\
  \bibinfo {author} {\bibfnamefont{A.~G.}\ \bibnamefont{Aronov}}}%
  , \bibinfo {year} {1985},\ in\ \emph{\bibinfo {booktitle} {Electron-Electron
  Interactions in Disordered Systems}},\ \bibinfo {editor} {edited by\ \bibinfo
  {editor} {\bibfnamefont{A.~L.}\ \bibnamefont{Efros}}\ and\ \bibinfo {editor}
  {\bibfnamefont{M.}~\bibnamefont{Pollak}}}\ (\bibinfo {publisher} {North
  Holland, Amsterdam})\ p.~\bibinfo {pages} {1}%
  \bibAnnoteFile{NoStop}{Altshuler:1985_B}%
\bibitem[{\citenamefont{Anderson}(1950)}]{Anderson:1950_PR}%
  \BibitemOpen
  \bibfield{author}{%
  \bibinfo {author} {\bibnamefont{Anderson}, \bibfnamefont{P.~W.}}}%
  , \bibinfo {year} {1950},\ \bibfield{title}{%
  \enquote{\bibinfo {title} {Antiferromagnetism. theory of superexchange
  interaction},}\ }%
  \bibfield{journal}{%
  \bibinfo {journal} {Phys. Rev.}\ }%
  \textbf{\bibinfo {volume} {79}},\ \bibinfo {pages} {350}%
  \bibAnnoteFile{NoStop}{Anderson:1950_PR}%
\bibitem[{\citenamefont{Anderson}(1961)}]{Anderson:1961_PR}%
  \BibitemOpen
  \bibfield{author}{%
  \bibinfo {author} {\bibnamefont{Anderson}, \bibfnamefont{P.~W.}}}%
  , \bibinfo {year} {1961},\ \bibfield{title}{%
  \enquote{\bibinfo {title} {Localized magnetic states in metals},}\ }%
  \bibfield{journal}{%
  \bibinfo {journal} {Phys. Rev.}\ }%
  \textbf{\bibinfo {volume} {124}},\ \bibinfo {pages} {41}%
  \bibAnnoteFile{NoStop}{Anderson:1961_PR}%
\bibitem[{\citenamefont{Anderson}\ and\
  \citenamefont{Hasegawa}(1955)}]{Anderson:1955_PR}%
  \BibitemOpen
  \bibfield{author}{%
  \bibinfo {author} {\bibnamefont{Anderson}, \bibfnamefont{P.~W.}},\ and\
  \bibinfo {author} {\bibfnamefont{H.}~\bibnamefont{Hasegawa}}}%
  , \bibinfo {year} {1955},\ \bibfield{title}{%
  \enquote{\bibinfo {title} {Considerations on double exchange},}\ }%
  \bibfield{journal}{%
  \bibinfo {journal} {Phys. Rev.}\ }%
  \textbf{\bibinfo {volume} {100}},\ \bibinfo {pages} {675}%
  \bibAnnoteFile{NoStop}{Anderson:1955_PR}%
\bibitem[{\citenamefont{Ando}\ \emph{et~al.}(1998)\citenamefont{Ando},
  \citenamefont{Hayashi}, \citenamefont{Tanaka},\ and\
  \citenamefont{Twardowski}}]{Ando:1998_JAP}%
  \BibitemOpen
  \bibfield{author}{%
  \bibinfo {author} {\bibnamefont{Ando}, \bibfnamefont{K.}}, \bibinfo {author}
  {\bibfnamefont{T.}~\bibnamefont{Hayashi}}, \bibinfo {author}
  {\bibfnamefont{M.}~\bibnamefont{Tanaka}},\ and\ \bibinfo {author}
  {\bibfnamefont{A.}~\bibnamefont{Twardowski}}}%
  , \bibinfo {year} {1998},\ \bibfield{title}{%
  \enquote{\bibinfo {title} {Magneto-optic effect of the ferromagnetic diluted
  magnetic semiconductor {Ga$_{1-x}$Mn$_{x}$As}},}\ }%
  \bibfield{journal}{%
  \bibinfo {journal} {J. Appl. Phys.}\ }%
  \textbf{\bibinfo {volume} {83}},\ \bibinfo {pages} {6548}%
  \bibAnnoteFile{NoStop}{Ando:1998_JAP}%
\bibitem[{\citenamefont{Andrearczyk}\
  \emph{et~al.}(2001)\citenamefont{Andrearczyk},
  \citenamefont{Jaroszy{\'n}ski}, \citenamefont{Sawicki}, \citenamefont{{Le Van
  Khoi}}, \citenamefont{Dietl}, \citenamefont{Ferrand},
  \citenamefont{Bourgognon}, \citenamefont{Cibert}, \citenamefont{Tatarenko},
  \citenamefont{Fukumura}, \citenamefont{Jin}, \citenamefont{Koinuma},\ and\
  \citenamefont{Kawasaki}}]{Andrearczyk:2001_ICPS}%
  \BibitemOpen
  \bibfield{author}{%
  \bibinfo {author} {\bibnamefont{Andrearczyk}, \bibfnamefont{T.}}, \bibinfo
  {author} {\bibfnamefont{J.}~\bibnamefont{Jaroszy{\'n}ski}}, \bibinfo {author}
  {\bibfnamefont{M.}~\bibnamefont{Sawicki}}, \bibinfo {author}
  {\bibnamefont{{Le Van Khoi}}}, \bibinfo {author}
  {\bibfnamefont{T.}~\bibnamefont{Dietl}}, \bibinfo {author}
  {\bibfnamefont{D.}~\bibnamefont{Ferrand}}, \bibinfo {author}
  {\bibfnamefont{C.}~\bibnamefont{Bourgognon}}, \bibinfo {author}
  {\bibfnamefont{J.}~\bibnamefont{Cibert}}, \bibinfo {author}
  {\bibfnamefont{S.}~\bibnamefont{Tatarenko}}, \bibinfo {author}
  {\bibfnamefont{T.}~\bibnamefont{Fukumura}}, \bibinfo {author}
  {\bibfnamefont{Zhengwu}\ \bibnamefont{Jin}}, \bibinfo {author}
  {\bibfnamefont{H.}~\bibnamefont{Koinuma}},\ and\ \bibinfo {author}
  {\bibfnamefont{M.}~\bibnamefont{Kawasaki}}}%
  , \bibinfo {year} {2001},\ \enquote{\bibinfo {title} {Ferromagnetic
  interactions in p- and n-type {II-VI} diluted magnetic semiconductors},}\ in\
  \emph{\bibinfo {booktitle} {Proceedings 25th International Conference on
  Physics of Semiconductors, Osaka, Japan, 2000}},\ \bibinfo {editor} {edited
  by\ \bibinfo {editor} {\bibfnamefont{N.}~\bibnamefont{Miura}}\ and\ \bibinfo
  {editor} {\bibfnamefont{T.}~\bibnamefont{Ando}}}\ (\bibinfo {publisher}
  {Springer, Berlin})\ p.\ \bibinfo {pages} {235}%
  \bibAnnoteFile{NoStop}{Andrearczyk:2001_ICPS}%
\bibitem[{\citenamefont{Aoyama}\ \emph{et~al.}(2010)\citenamefont{Aoyama},
  \citenamefont{Kobayashi},\ and\ \citenamefont{Munekata}}]{Aoyama:2010_JAP}%
  \BibitemOpen
  \bibfield{author}{%
  \bibinfo {author} {\bibnamefont{Aoyama}, \bibfnamefont{J.}}, \bibinfo
  {author} {\bibfnamefont{S.}~\bibnamefont{Kobayashi}},\ and\ \bibinfo {author}
  {\bibfnamefont{H.}~\bibnamefont{Munekata}}}%
  , \bibinfo {year} {2010},\ \bibfield{title}{%
  \enquote{\bibinfo {title} {All-optical 90-degree switching of magnetization
  in a ferromagnetic {Ga$_{0.98}$Mn$_{0.02}$As} microbar},}\ }%
  \bibfield{journal}{%
  \bibinfo {journal} {J. Appl. Phys.}\ }%
  \textbf{\bibinfo {volume} {107}},\ \bibinfo {pages} {09C301}%
  \bibAnnoteFile{NoStop}{Aoyama:2010_JAP}%
\bibitem[{\citenamefont{Arciszewska}\ and\
  \citenamefont{Nawrocki}(1986)}]{Arciszewska:1986_JPCS}%
  \BibitemOpen
  \bibfield{author}{%
  \bibinfo {author} {\bibnamefont{Arciszewska}, \bibfnamefont{M.}},\ and\
  \bibinfo {author} {\bibfnamefont{M.}~\bibnamefont{Nawrocki}}}%
  , \bibinfo {year} {1986},\ \bibfield{title}{%
  \enquote{\bibinfo {title} {Determination of the band structure parameters of
  {Cd$_{0.95}$Mn$_{0.05}$Se} from magnetoabsorption measurements},}\ }%
  \bibfield{journal}{%
  \bibinfo {journal} {J. Phys. Chem. Solids}\ }%
  \textbf{\bibinfo {volume} {47}},\ \bibinfo {pages} {309}%
  \bibAnnoteFile{NoStop}{Arciszewska:1986_JPCS}%
\bibitem[{\citenamefont{Averkiev}\ \emph{et~al.}(1987)\citenamefont{Averkiev},
  \citenamefont{Gutkin}, \citenamefont{Osipov},\ and\
  \citenamefont{Reshchikov}}]{Averkiev:1987_SPS}%
  \BibitemOpen
  \bibfield{author}{%
  \bibinfo {author} {\bibnamefont{Averkiev}, \bibfnamefont{N.~S.}}, \bibinfo
  {author} {\bibfnamefont{A.~A.}\ \bibnamefont{Gutkin}}, \bibinfo {author}
  {\bibfnamefont{E.~B.}\ \bibnamefont{Osipov}},\ and\ \bibinfo {author}
  {\bibfnamefont{M.~A.}\ \bibnamefont{Reshchikov}}}%
  , \bibinfo {year} {1987},\ \bibfield{title}{%
  \enquote{\bibinfo {title} {Role of the exchange interaction in
  piezospectroscopic effects associated with {Mn} centers in {GaAs}},}\ }%
  \bibfield{journal}{%
  \bibinfo {journal} {Sov. Phys. Semicond.}\ }%
  \textbf{\bibinfo {volume} {21}},\ \bibinfo {pages} {1119}%
  \bibAnnoteFile{NoStop}{Averkiev:1987_SPS}%
\bibitem[{\citenamefont{Bacewicz}\ \emph{et~al.}(2005)\citenamefont{Bacewicz},
  \citenamefont{Twar{\'o}g}, \citenamefont{Malinowski},
  \citenamefont{Wojtowicz}, \citenamefont{Liu},\ and\
  \citenamefont{Furdyna}}]{Bacewicz:2005_JPCS}%
  \BibitemOpen
  \bibfield{author}{%
  \bibinfo {author} {\bibnamefont{Bacewicz}, \bibfnamefont{R.}}, \bibinfo
  {author} {\bibfnamefont{A.}~\bibnamefont{Twar{\'o}g}}, \bibinfo {author}
  {\bibfnamefont{A.}~\bibnamefont{Malinowski}}, \bibinfo {author}
  {\bibfnamefont{T.}~\bibnamefont{Wojtowicz}}, \bibinfo {author}
  {\bibfnamefont{X.}~\bibnamefont{Liu}},\ and\ \bibinfo {author}
  {\bibfnamefont{J.}~\bibnamefont{Furdyna}}}%
  , \bibinfo {year} {2005},\ \bibfield{title}{%
  \enquote{\bibinfo {title} {Local structure of {Mn} in {(Ga,Mn)As} probed by
  x-ray absorption spectroscopy},}\ }%
  \bibfield{journal}{%
  \bibinfo {journal} {J. Phys. Chem. Solids}\ }%
  \textbf{\bibinfo {volume} {66}},\ \bibinfo {pages} {235}%
  \bibAnnoteFile{NoStop}{Bacewicz:2005_JPCS}%
\bibitem[{\citenamefont{Balk}\ \emph{et~al.}(2011)\citenamefont{Balk},
  \citenamefont{Nowakowski}, \citenamefont{Wilson}, \citenamefont{Rench},
  \citenamefont{Schiffer}, \citenamefont{Awschalom},\ and\
  \citenamefont{Samarth}}]{Balk:2011_PRL}%
  \BibitemOpen
  \bibfield{author}{%
  \bibinfo {author} {\bibnamefont{Balk}, \bibfnamefont{A.~L.}}, \bibinfo
  {author} {\bibfnamefont{M.~E.}\ \bibnamefont{Nowakowski}}, \bibinfo {author}
  {\bibfnamefont{M.~J.}\ \bibnamefont{Wilson}}, \bibinfo {author}
  {\bibfnamefont{D.~W.}\ \bibnamefont{Rench}}, \bibinfo {author}
  {\bibfnamefont{P.}~\bibnamefont{Schiffer}}, \bibinfo {author}
  {\bibfnamefont{D.~D.}\ \bibnamefont{Awschalom}},\ and\ \bibinfo {author}
  {\bibfnamefont{N.}~\bibnamefont{Samarth}}}%
  , \bibinfo {year} {2011},\ \bibfield{title}{%
  \enquote{\bibinfo {title} {Measurements of nanoscale domain wall flexing in a
  ferromagnetic thin film},}\ }%
  \bibfield{journal}{%
  \bibinfo {journal} {Phys. Rev. Lett.}\ }%
  \textbf{\bibinfo {volume} {107}},\ \bibinfo {pages} {077205}%
  \bibAnnoteFile{NoStop}{Balk:2011_PRL}%
\bibitem[{\citenamefont{Bauer}\ \emph{et~al.}(1992)\citenamefont{Bauer},
  \citenamefont{Pascher},\ and\ \citenamefont{Zawadzki}}]{Bauer:1992_SST}%
  \BibitemOpen
  \bibfield{author}{%
  \bibinfo {author} {\bibnamefont{Bauer}, \bibfnamefont{G.}}, \bibinfo {author}
  {\bibfnamefont{H.}~\bibnamefont{Pascher}},\ and\ \bibinfo {author}
  {\bibfnamefont{W.}~\bibnamefont{Zawadzki}}}%
  , \bibinfo {year} {1992},\ \bibfield{title}{%
  \enquote{\bibinfo {title} {Magneto-optical properties of semimagnetic lead
  chalcogenides},}\ }%
  \bibfield{journal}{%
  \bibinfo {journal} {Semicond. Sci. Technol.}\ }%
  \textbf{\bibinfo {volume} {7}},\ \bibinfo {pages} {703}%
  \bibAnnoteFile{NoStop}{Bauer:1992_SST}%
\bibitem[{\citenamefont{Beaulac}\ \emph{et~al.}(2008)\citenamefont{Beaulac},
  \citenamefont{Archer}, \citenamefont{Ochsenbein},\ and\
  \citenamefont{Gamelin}}]{Beaulac:2008_AFM}%
  \BibitemOpen
  \bibfield{author}{%
  \bibinfo {author} {\bibnamefont{Beaulac}, \bibfnamefont{R.}}, \bibinfo
  {author} {\bibfnamefont{P.~I.}\ \bibnamefont{Archer}}, \bibinfo {author}
  {\bibfnamefont{S.~T.}\ \bibnamefont{Ochsenbein}},\ and\ \bibinfo {author}
  {\bibfnamefont{D.~R.}\ \bibnamefont{Gamelin}}}%
  , \bibinfo {year} {2008},\ \bibfield{title}{%
  \enquote{\bibinfo {title} {{Mn$^{2+}$}-doped {CdSe} quantum dots: New
  inorganic materials for spin-electronics and spin-photonics},}\ }%
  \bibfield{journal}{%
  \bibinfo {journal} {Adv. Funct. Mater.}\ }%
  \textbf{\bibinfo {volume} {18}},\ \bibinfo {pages} {3873}%
  \bibAnnoteFile{NoStop}{Beaulac:2008_AFM}%
\bibitem[{\citenamefont{Belitz}\ and\
  \citenamefont{Kirkpatrick}(1994)}]{Belitz:1994_RMP}%
  \BibitemOpen
  \bibfield{author}{%
  \bibinfo {author} {\bibnamefont{Belitz}, \bibfnamefont{D.}},\ and\ \bibinfo
  {author} {\bibfnamefont{T.~R.}\ \bibnamefont{Kirkpatrick}}}%
  , \bibinfo {year} {1994},\ \bibfield{title}{%
  \enquote{\bibinfo {title} {The {Anderson-Mott} transition},}\ }%
  \bibfield{journal}{%
  \bibinfo {journal} {Rev. Mod. Phys.}\ }%
  \textbf{\bibinfo {volume} {66}},\ \bibinfo {pages} {261}%
  \bibAnnoteFile{NoStop}{Belitz:1994_RMP}%
\bibitem[{\citenamefont{{Benoit \`a la Guillaume}}\
  \emph{et~al.}(1992)\citenamefont{{Benoit \`a la Guillaume}},
  \citenamefont{Scalbert},\ and\ \citenamefont{Dietl}}]{Benoit:1992_PRB}%
  \BibitemOpen
  \bibfield{author}{%
  \bibinfo {author} {\bibnamefont{{Benoit \`a la Guillaume}},
  \bibfnamefont{C.}}, \bibinfo {author}
  {\bibfnamefont{D.}~\bibnamefont{Scalbert}},\ and\ \bibinfo {author}
  {\bibfnamefont{T.}~\bibnamefont{Dietl}}}%
  , \bibinfo {year} {1992},\ \bibfield{title}{%
  \enquote{\bibinfo {title} {Wigner-{Seitz} approach to spin splitting in
  diluted magnetic semiconductors},}\ }%
  \bibfield{journal}{%
  \bibinfo {journal} {Phys. Rev. B}\ }%
  \textbf{\bibinfo {volume} {46}},\ \bibinfo {pages} {9853}%
  \bibAnnoteFile{NoStop}{Benoit:1992_PRB}%
\bibitem[{\citenamefont{Bergqvist}\
  \emph{et~al.}(2011)\citenamefont{Bergqvist}, \citenamefont{Sato},
  \citenamefont{Katayama-Yoshida},\ and\
  \citenamefont{Dederichs}}]{Bergqvist:2011_PRB}%
  \BibitemOpen
  \bibfield{author}{%
  \bibinfo {author} {\bibnamefont{Bergqvist}, \bibfnamefont{L.}}, \bibinfo
  {author} {\bibfnamefont{K.}~\bibnamefont{Sato}}, \bibinfo {author}
  {\bibfnamefont{H.}~\bibnamefont{Katayama-Yoshida}},\ and\ \bibinfo {author}
  {\bibfnamefont{P.~H.}\ \bibnamefont{Dederichs}}}%
  , \bibinfo {year} {2011},\ \bibfield{title}{%
  \enquote{\bibinfo {title} {Computational materials design for
  high-{${T}_{c}$} {(Ga,Mn)As} with {Li} codoping},}\ }%
  \bibfield{journal}{%
  \bibinfo {journal} {Phys. Rev. B}\ }%
  \textbf{\bibinfo {volume} {83}},\ \bibinfo {pages} {165201}%
  \bibAnnoteFile{NoStop}{Bergqvist:2011_PRB}%
\bibitem[{\citenamefont{Bernevig}\ and\
  \citenamefont{Vafek}(2005)}]{Bernevig:2005_PRB}%
  \BibitemOpen
  \bibfield{author}{%
  \bibinfo {author} {\bibnamefont{Bernevig}, \bibfnamefont{B.~A.}},\ and\
  \bibinfo {author} {\bibfnamefont{O.}~\bibnamefont{Vafek}}}%
  , \bibinfo {year} {2005},\ \bibfield{title}{%
  \enquote{\bibinfo {title} {Piezo-magnetoelectric effects in $p$-doped
  semiconductors},}\ }%
  \bibfield{journal}{%
  \bibinfo {journal} {Phys. Rev. B}\ }%
  \textbf{\bibinfo {volume} {72}},\ \bibinfo {pages} {033203}%
  \bibAnnoteFile{NoStop}{Bernevig:2005_PRB}%
\bibitem[{\citenamefont{Beschoten}\
  \emph{et~al.}(1999)\citenamefont{Beschoten}, \citenamefont{Crowell},
  \citenamefont{Malajovich}, \citenamefont{Awschalom},
  \citenamefont{Matsukura}, \citenamefont{Shen},\ and\
  \citenamefont{Ohno}}]{Beschoten:1999_PRL}%
  \BibitemOpen
  \bibfield{author}{%
  \bibinfo {author} {\bibnamefont{Beschoten}, \bibfnamefont{B.}}, \bibinfo
  {author} {\bibfnamefont{P.~A.}\ \bibnamefont{Crowell}}, \bibinfo {author}
  {\bibfnamefont{I.}~\bibnamefont{Malajovich}}, \bibinfo {author}
  {\bibfnamefont{D.~D.}\ \bibnamefont{Awschalom}}, \bibinfo {author}
  {\bibfnamefont{F.}~\bibnamefont{Matsukura}}, \bibinfo {author}
  {\bibfnamefont{A.}~\bibnamefont{Shen}},\ and\ \bibinfo {author}
  {\bibfnamefont{H.}~\bibnamefont{Ohno}}}%
  , \bibinfo {year} {1999},\ \bibfield{title}{%
  \enquote{\bibinfo {title} {Magnetic circular dichroism studies of
  carrier-induced ferromagnetism in {(Ga$_{1-x}$Mn$_{x}$)As}},}\ }%
  \bibfield{journal}{%
  \bibinfo {journal} {Phys. Rev. Lett.}\ }%
  \textbf{\bibinfo {volume} {83}},\ \bibinfo {pages} {3073}%
  \bibAnnoteFile{NoStop}{Beschoten:1999_PRL}%
\bibitem[{\citenamefont{Bhattacharjee}\
  \emph{et~al.}(1983)\citenamefont{Bhattacharjee}, \citenamefont{Fishman},\
  and\ \citenamefont{Coqblin}}]{Bhattacharjee:1983_PBC}%
  \BibitemOpen
  \bibfield{author}{%
  \bibinfo {author} {\bibnamefont{Bhattacharjee}, \bibfnamefont{A.~K.}},
  \bibinfo {author} {\bibfnamefont{G.}~\bibnamefont{Fishman}},\ and\ \bibinfo
  {author} {\bibfnamefont{B.}~\bibnamefont{Coqblin}}}%
  , \bibinfo {year} {1983},\ \bibfield{title}{%
  \enquote{\bibinfo {title} {Virtual bound state model for the exchange
  interaction in semimagnetic semiconductors such as {Cd$_{1-x}$Mn$_{x}$Te}},}\
  }%
  \bibfield{journal}{%
  \bibinfo {journal} {Physica}\ }%
  \textbf{\bibinfo {volume} {B+C 117-118}},\ \bibinfo {pages} {449}%
  \bibAnnoteFile{NoStop}{Bhattacharjee:1983_PBC}%
\bibitem[{\citenamefont{Bihler}\ \emph{et~al.}(2008)\citenamefont{Bihler},
  \citenamefont{Althammer}, \citenamefont{Brandlmaier},
  \citenamefont{Gepr\"ags}, \citenamefont{Weiler}, \citenamefont{Opel},
  \citenamefont{Schoch}, \citenamefont{Limmer}, \citenamefont{Gross},
  \citenamefont{Brandt},\ and\ \citenamefont{Goennenwein}}]{Bihler:2008_PRB}%
  \BibitemOpen
  \bibfield{author}{%
  \bibinfo {author} {\bibnamefont{Bihler}, \bibfnamefont{C.}}, \bibinfo
  {author} {\bibfnamefont{M.}~\bibnamefont{Althammer}}, \bibinfo {author}
  {\bibfnamefont{A.}~\bibnamefont{Brandlmaier}}, \bibinfo {author}
  {\bibfnamefont{S.}~\bibnamefont{Gepr\"ags}}, \bibinfo {author}
  {\bibfnamefont{M.}~\bibnamefont{Weiler}}, \bibinfo {author}
  {\bibfnamefont{M.}~\bibnamefont{Opel}}, \bibinfo {author}
  {\bibfnamefont{W.}~\bibnamefont{Schoch}}, \bibinfo {author}
  {\bibfnamefont{W.}~\bibnamefont{Limmer}}, \bibinfo {author}
  {\bibfnamefont{R.}~\bibnamefont{Gross}}, \bibinfo {author}
  {\bibfnamefont{M.~S.}\ \bibnamefont{Brandt}},\ and\ \bibinfo {author}
  {\bibfnamefont{S.~T.~B.}\ \bibnamefont{Goennenwein}}}%
  , \bibinfo {year} {2008},\ \bibfield{title}{%
  \enquote{\bibinfo {title} {{Ga$_{1-x}$Mn$_{x}$As$/$}piezoelectric actuator
  hybrids: {A} model system for magnetoelastic magnetization manipulation},}\
  }%
  \bibfield{journal}{%
  \bibinfo {journal} {Phys. Rev. B}\ }%
  \textbf{\bibinfo {volume} {78}},\ \bibinfo {pages} {045203}%
  \bibAnnoteFile{NoStop}{Bihler:2008_PRB}%
\bibitem[{\citenamefont{Bihler}\ \emph{et~al.}(2007)\citenamefont{Bihler},
  \citenamefont{Kraus}, \citenamefont{Huebl}, \citenamefont{Brandt},
  \citenamefont{Goennenwein}, \citenamefont{Opel}, \citenamefont{Scarpulla},
  \citenamefont{Stone}, \citenamefont{Farshchi},\ and\
  \citenamefont{Dubon}}]{Bihler:2007_PRB}%
  \BibitemOpen
  \bibfield{author}{%
  \bibinfo {author} {\bibnamefont{Bihler}, \bibfnamefont{C.}}, \bibinfo
  {author} {\bibfnamefont{M.}~\bibnamefont{Kraus}}, \bibinfo {author}
  {\bibfnamefont{H.}~\bibnamefont{Huebl}}, \bibinfo {author}
  {\bibfnamefont{M.~S.}\ \bibnamefont{Brandt}}, \bibinfo {author}
  {\bibfnamefont{S.~T.~B.}\ \bibnamefont{Goennenwein}}, \bibinfo {author}
  {\bibfnamefont{M.}~\bibnamefont{Opel}}, \bibinfo {author}
  {\bibfnamefont{M.~A.}\ \bibnamefont{Scarpulla}}, \bibinfo {author}
  {\bibfnamefont{P.~R.}\ \bibnamefont{Stone}}, \bibinfo {author}
  {\bibfnamefont{R.}~\bibnamefont{Farshchi}},\ and\ \bibinfo {author}
  {\bibfnamefont{O.~D.}\ \bibnamefont{Dubon}}}%
  , \bibinfo {year} {2007},\ \bibfield{title}{%
  \enquote{\bibinfo {title} {Magnetocrystalline anisotropy and magnetization
  reversal in {Ga$_{1-x}$Mn$_{x}$P} synthesized by ion implantation and
  pulsed-laser melting},}\ }%
  \bibfield{journal}{%
  \bibinfo {journal} {Phys. Rev. B}\ }%
  \textbf{\bibinfo {volume} {75}},\ \bibinfo {pages} {214419}%
  \bibAnnoteFile{NoStop}{Bihler:2007_PRB}%
\bibitem[{\citenamefont{Bihler}\ \emph{et~al.}(2009)\citenamefont{Bihler},
  \citenamefont{Schoch}, \citenamefont{Limmer}, \citenamefont{Goennenwein},\
  and\ \citenamefont{Brandt}}]{Bihler:2009_PRBa}%
  \BibitemOpen
  \bibfield{author}{%
  \bibinfo {author} {\bibnamefont{Bihler}, \bibfnamefont{C.}}, \bibinfo
  {author} {\bibfnamefont{W.}~\bibnamefont{Schoch}}, \bibinfo {author}
  {\bibfnamefont{W.}~\bibnamefont{Limmer}}, \bibinfo {author}
  {\bibfnamefont{S.~T.~B.}\ \bibnamefont{Goennenwein}},\ and\ \bibinfo {author}
  {\bibfnamefont{M.~S.}\ \bibnamefont{Brandt}}}%
  , \bibinfo {year} {2009},\ \bibfield{title}{%
  \enquote{\bibinfo {title} {Spin-wave resonances and surface spin pinning in
  {Ga$_{1-x}$Mn$_{x}$As} thin films},}\ }%
  \bibfield{journal}{%
  \bibinfo {journal} {Phys. Rev. B}\ }%
  \textbf{\bibinfo {volume} {79}},\ \bibinfo {pages} {045205}%
  \bibAnnoteFile{NoStop}{Bihler:2009_PRBa}%
\bibitem[{\citenamefont{Birowska}\ \emph{et~al.}(2012)\citenamefont{Birowska},
  \citenamefont{\'{S}liwa}, \citenamefont{Majewski},\ and\
  \citenamefont{Dietl}}]{Birowska:2012_PRL}%
  \BibitemOpen
  \bibfield{author}{%
  \bibinfo {author} {\bibnamefont{Birowska}, \bibfnamefont{M.}}, \bibinfo
  {author} {\bibfnamefont{C.}~\bibnamefont{\'{S}liwa}}, \bibinfo {author}
  {\bibfnamefont{J.~A.}\ \bibnamefont{Majewski}},\ and\ \bibinfo {author}
  {\bibfnamefont{T.}~\bibnamefont{Dietl}}}%
  , \bibinfo {year} {2012},\ \bibfield{title}{%
  \enquote{\bibinfo {title} {Origin of bulk uniaxial anisotropy in zinc-blende
  dilute magnetic semiconductors},}\ }%
  \bibfield{journal}{%
  \bibinfo {journal} {Phys. Rev. Lett.}\ }%
  \textbf{\bibinfo {volume} {108}},\ \bibinfo {pages} {237203}%
  \bibAnnoteFile{NoStop}{Birowska:2012_PRL}%
\bibitem[{\citenamefont{Blinowski}\ and\
  \citenamefont{Kacman}(2003)}]{Blinowski:2003_PRB}%
  \BibitemOpen
  \bibfield{author}{%
  \bibinfo {author} {\bibnamefont{Blinowski}, \bibfnamefont{J.}},\ and\
  \bibinfo {author} {\bibfnamefont{P.}~\bibnamefont{Kacman}}}%
  , \bibinfo {year} {2003},\ \bibfield{title}{%
  \enquote{\bibinfo {title} {Spin interactions of interstitial {Mn} ions in
  ferromagnetic {GaMnAs}},}\ }%
  \bibfield{journal}{%
  \bibinfo {journal} {Phys. Rev. B}\ }%
  \textbf{\bibinfo {volume} {67}},\ \bibinfo {pages} {121204}%
  \bibAnnoteFile{NoStop}{Blinowski:2003_PRB}%
\bibitem[{\citenamefont{Blinowski}\
  \emph{et~al.}(1996)\citenamefont{Blinowski}, \citenamefont{Kacman},\ and\
  \citenamefont{Majewski}}]{Blinowski:1996_PRB}%
  \BibitemOpen
  \bibfield{author}{%
  \bibinfo {author} {\bibnamefont{Blinowski}, \bibfnamefont{J.}}, \bibinfo
  {author} {\bibfnamefont{P.}~\bibnamefont{Kacman}},\ and\ \bibinfo {author}
  {\bibfnamefont{J.~A.}\ \bibnamefont{Majewski}}}%
  , \bibinfo {year} {1996},\ \bibfield{title}{%
  \enquote{\bibinfo {title} {Ferromagnetic superexchange in {Cr}-based diluted
  magnetic semiconductors},}\ }%
  \bibfield{journal}{%
  \bibinfo {journal} {Phys. Rev. B}\ }%
  \textbf{\bibinfo {volume} {53}},\ \bibinfo {pages} {9524}%
  \bibAnnoteFile{NoStop}{Blinowski:1996_PRB}%
\bibitem[{\citenamefont{Boeck}\ \emph{et~al.}(1996)\citenamefont{Boeck},
  \citenamefont{Oesterholt}, \citenamefont{Esch}, \citenamefont{Bender},
  \citenamefont{Bruynseraede}, \citenamefont{Hoof},\ and\
  \citenamefont{Borghs}}]{De_Boeck:1996_APL}%
  \BibitemOpen
  \bibfield{author}{%
  \bibinfo {author} {\bibnamefont{Boeck}, \bibfnamefont{J.~De}}, \bibinfo
  {author} {\bibfnamefont{R.}~\bibnamefont{Oesterholt}}, \bibinfo {author}
  {\bibfnamefont{A.~Van}\ \bibnamefont{Esch}}, \bibinfo {author}
  {\bibfnamefont{H.}~\bibnamefont{Bender}}, \bibinfo {author}
  {\bibfnamefont{C.}~\bibnamefont{Bruynseraede}}, \bibinfo {author}
  {\bibfnamefont{C.~Van}\ \bibnamefont{Hoof}},\ and\ \bibinfo {author}
  {\bibfnamefont{G.}~\bibnamefont{Borghs}}}%
  , \bibinfo {year} {1996},\ \bibfield{title}{%
  \enquote{\bibinfo {title} {Nanometer-scale magnetic {MnAs} particles in
  {GaAs} grown by molecular beam epitaxy},}\ }%
  \bibfield{journal}{%
  \bibinfo {journal} {Appl. Phys. Lett.}\ }%
  \textbf{\bibinfo {volume} {68}},\ \bibinfo {pages} {2744}%
  \bibAnnoteFile{NoStop}{De_Boeck:1996_APL}%
\bibitem[{\citenamefont{Bombeck}\ \emph{et~al.}(2013)\citenamefont{Bombeck},
  \citenamefont{J\"ager}, \citenamefont{Scherbakov}, \citenamefont{Linnik},
  \citenamefont{Yakovlev}, \citenamefont{Liu}, \citenamefont{Furdyna},
  \citenamefont{Akimov},\ and\ \citenamefont{Bayer}}]{Bombeck:2013_PRB}%
  \BibitemOpen
  \bibfield{author}{%
  \bibinfo {author} {\bibnamefont{Bombeck}, \bibfnamefont{M.}}, \bibinfo
  {author} {\bibfnamefont{J.~V.}\ \bibnamefont{J\"ager}}, \bibinfo {author}
  {\bibfnamefont{A.~V.}\ \bibnamefont{Scherbakov}}, \bibinfo {author}
  {\bibfnamefont{T.}~\bibnamefont{Linnik}}, \bibinfo {author}
  {\bibfnamefont{D.~R.}\ \bibnamefont{Yakovlev}}, \bibinfo {author}
  {\bibfnamefont{X.}~\bibnamefont{Liu}}, \bibinfo {author}
  {\bibfnamefont{J.~K.}\ \bibnamefont{Furdyna}}, \bibinfo {author}
  {\bibfnamefont{A.~V.}\ \bibnamefont{Akimov}},\ and\ \bibinfo {author}
  {\bibfnamefont{M.}~\bibnamefont{Bayer}}}%
  , \bibinfo {year} {2013},\ \bibfield{title}{%
  \enquote{\bibinfo {title} {Magnetization precession induced by
  quasitransverse picosecond strain pulses in (311) ferromagnetic
  {(Ga,Mn)As}},}\ }%
  \bibfield{journal}{%
  \bibinfo {journal} {Phys. Rev. B}\ }%
  \textbf{\bibinfo {volume} {87}},\ \bibinfo {pages} {060302}%
  \bibAnnoteFile{NoStop}{Bombeck:2013_PRB}%
\bibitem[{\citenamefont{Bonanni}(2007)}]{Bonanni:2007_SST}%
  \BibitemOpen
  \bibfield{author}{%
  \bibinfo {author} {\bibnamefont{Bonanni}, \bibfnamefont{A.}}}%
  , \bibinfo {year} {2007},\ \bibfield{title}{%
  \enquote{\bibinfo {title} {Ferromagnetic nitride-based semiconductors doped
  with transition metals and rare earths},}\ }%
  \bibfield{journal}{%
  \bibinfo {journal} {Semicond. Sci. Technol.}\ }%
  \textbf{\bibinfo {volume} {22}},\ \bibinfo {pages} {R41}%
  \bibAnnoteFile{NoStop}{Bonanni:2007_SST}%
\bibitem[{\citenamefont{Bonanni}(2011)}]{Bonanni:2011_SST}%
  \BibitemOpen
  \bibfield{author}{%
  \bibinfo {author} {\bibnamefont{Bonanni}, \bibfnamefont{A.}}}%
  , \bibinfo {year} {2011},\ \bibfield{title}{%
  \enquote{\bibinfo {title} {{(Nano)characterization} of semiconductor
  materials and structures},}\ }%
  \bibfield{journal}{%
  \bibinfo {journal} {Semicon. Sci. Technol.}\ }%
  \textbf{\bibinfo {volume} {26}},\ \bibinfo {pages} {060301}%
  \bibAnnoteFile{NoStop}{Bonanni:2011_SST}%
\bibitem[{\citenamefont{Bonanni}\ and\
  \citenamefont{Dietl}(2010)}]{Bonanni:2010_CSR}%
  \BibitemOpen
  \bibfield{author}{%
  \bibinfo {author} {\bibnamefont{Bonanni}, \bibfnamefont{A.}},\ and\ \bibinfo
  {author} {\bibfnamefont{T.}~\bibnamefont{Dietl}}}%
  , \bibinfo {year} {2010},\ \bibfield{title}{%
  \enquote{\bibinfo {title} {A story of high-temperature ferromagnetism in
  semiconductors},}\ }%
  \bibfield{journal}{%
  \bibinfo {journal} {Chem. Soc. Rev.}\ }%
  \textbf{\bibinfo {volume} {39}},\ \bibinfo {pages} {528}%
  \bibAnnoteFile{NoStop}{Bonanni:2010_CSR}%
\bibitem[{\citenamefont{Bonanni}\ \emph{et~al.}(2007)\citenamefont{Bonanni},
  \citenamefont{Kiecana}, \citenamefont{Simbrunner}, \citenamefont{Li},
  \citenamefont{Sawicki}, \citenamefont{Wegscheider}, \citenamefont{Quast},
  \citenamefont{Przybyli{\'n}ska}, \citenamefont{Navarro-Quezada},
  \citenamefont{Jakie{\l}a}, \citenamefont{Wo{\l}o{\'s}},
  \citenamefont{Jantsch},\ and\ \citenamefont{Dietl}}]{Bonanni:2007_PRB}%
  \BibitemOpen
  \bibfield{author}{%
  \bibinfo {author} {\bibnamefont{Bonanni}, \bibfnamefont{A.}}, \bibinfo
  {author} {\bibfnamefont{M.}~\bibnamefont{Kiecana}}, \bibinfo {author}
  {\bibfnamefont{C.}~\bibnamefont{Simbrunner}}, \bibinfo {author}
  {\bibfnamefont{T.}~\bibnamefont{Li}}, \bibinfo {author}
  {\bibfnamefont{M.}~\bibnamefont{Sawicki}}, \bibinfo {author}
  {\bibfnamefont{M.}~\bibnamefont{Wegscheider}}, \bibinfo {author}
  {\bibfnamefont{M.}~\bibnamefont{Quast}}, \bibinfo {author}
  {\bibfnamefont{H.}~\bibnamefont{Przybyli{\'n}ska}}, \bibinfo {author}
  {\bibfnamefont{A.}~\bibnamefont{Navarro-Quezada}}, \bibinfo {author}
  {\bibfnamefont{R.}~\bibnamefont{Jakie{\l}a}}, \bibinfo {author}
  {\bibfnamefont{A.}~\bibnamefont{Wo{\l}o{\'s}}}, \bibinfo {author}
  {\bibfnamefont{W.}~\bibnamefont{Jantsch}},\ and\ \bibinfo {author}
  {\bibfnamefont{T.}~\bibnamefont{Dietl}}}%
  , \bibinfo {year} {2007},\ \bibfield{title}{%
  \enquote{\bibinfo {title} {Paramagnetic {GaN:Fe} and ferromagnetic
  {(Ga,Fe)N}: {The} relationship between structural, electronic, and magnetic
  properties},}\ }%
  \bibfield{journal}{%
  \bibinfo {journal} {Phys. Rev. B}\ }%
  \textbf{\bibinfo {volume} {75}},\ \bibinfo {pages} {125210}%
  \bibAnnoteFile{NoStop}{Bonanni:2007_PRB}%
\bibitem[{\citenamefont{Bonanni}\ \emph{et~al.}(2008)\citenamefont{Bonanni},
  \citenamefont{Navarro-Quezada}, \citenamefont{Li},
  \citenamefont{Wegscheider}, \citenamefont{Mat\v{e}j},
  \citenamefont{Hol{\'y}}, \citenamefont{Lechner}, \citenamefont{Bauer},
  \citenamefont{Rovezzi}, \citenamefont{D'Acapito}, \citenamefont{Kiecana},
  \citenamefont{Sawicki},\ and\ \citenamefont{Dietl}}]{Bonanni:2008_PRL}%
  \BibitemOpen
  \bibfield{author}{%
  \bibinfo {author} {\bibnamefont{Bonanni}, \bibfnamefont{A.}}, \bibinfo
  {author} {\bibfnamefont{A.}~\bibnamefont{Navarro-Quezada}}, \bibinfo {author}
  {\bibfnamefont{Tian}\ \bibnamefont{Li}}, \bibinfo {author}
  {\bibfnamefont{M.}~\bibnamefont{Wegscheider}}, \bibinfo {author}
  {\bibfnamefont{Z.}~\bibnamefont{Mat\v{e}j}}, \bibinfo {author}
  {\bibfnamefont{V.}~\bibnamefont{Hol{\'y}}}, \bibinfo {author}
  {\bibfnamefont{R.~T.}\ \bibnamefont{Lechner}}, \bibinfo {author}
  {\bibfnamefont{G.}~\bibnamefont{Bauer}}, \bibinfo {author}
  {\bibfnamefont{M.}~\bibnamefont{Rovezzi}}, \bibinfo {author}
  {\bibfnamefont{F.}~\bibnamefont{D'Acapito}}, \bibinfo {author}
  {\bibfnamefont{M.}~\bibnamefont{Kiecana}}, \bibinfo {author}
  {\bibfnamefont{M.}~\bibnamefont{Sawicki}},\ and\ \bibinfo {author}
  {\bibfnamefont{T.}~\bibnamefont{Dietl}}}%
  , \bibinfo {year} {2008},\ \bibfield{title}{%
  \enquote{\bibinfo {title} {Controlled aggregation of magnetic ions in a
  semiconductor: An experimental demonstration},}\ }%
  \bibfield{journal}{%
  \bibinfo {journal} {Phys. Rev. Lett.}\ }%
  \textbf{\bibinfo {volume} {101}},\ \bibinfo {pages} {135502}%
  \bibAnnoteFile{NoStop}{Bonanni:2008_PRL}%
\bibitem[{\citenamefont{Bonanni}\ \emph{et~al.}(2011)\citenamefont{Bonanni},
  \citenamefont{Sawicki}, \citenamefont{Devillers}, \citenamefont{Stefanowicz},
  \citenamefont{Faina}, \citenamefont{Li}, \citenamefont{Winkler},
  \citenamefont{Sztenkiel}, \citenamefont{Navarro-Quezada},
  \citenamefont{Rovezzi}, \citenamefont{Jakie{\l}a}, \citenamefont{Grois},
  \citenamefont{Wegscheider}, \citenamefont{Jantsch},
  \citenamefont{Suffczy{\'n}ski}, \citenamefont{D'Acapito},
  \citenamefont{Meingast}, \citenamefont{Kothleitner},\ and\
  \citenamefont{Dietl}}]{Bonanni:2011_PRB}%
  \BibitemOpen
  \bibfield{author}{%
  \bibinfo {author} {\bibnamefont{Bonanni}, \bibfnamefont{A.}}, \bibinfo
  {author} {\bibfnamefont{M.}~\bibnamefont{Sawicki}}, \bibinfo {author}
  {\bibfnamefont{T.}~\bibnamefont{Devillers}}, \bibinfo {author}
  {\bibfnamefont{W.}~\bibnamefont{Stefanowicz}}, \bibinfo {author}
  {\bibfnamefont{B.}~\bibnamefont{Faina}}, \bibinfo {author}
  {\bibfnamefont{Tian}\ \bibnamefont{Li}}, \bibinfo {author}
  {\bibfnamefont{T.~E.}\ \bibnamefont{Winkler}}, \bibinfo {author}
  {\bibfnamefont{D.}~\bibnamefont{Sztenkiel}}, \bibinfo {author}
  {\bibfnamefont{A.}~\bibnamefont{Navarro-Quezada}}, \bibinfo {author}
  {\bibfnamefont{M.}~\bibnamefont{Rovezzi}}, \bibinfo {author}
  {\bibfnamefont{R.}~\bibnamefont{Jakie{\l}a}}, \bibinfo {author}
  {\bibfnamefont{A.}~\bibnamefont{Grois}}, \bibinfo {author}
  {\bibfnamefont{M.}~\bibnamefont{Wegscheider}}, \bibinfo {author}
  {\bibfnamefont{W.}~\bibnamefont{Jantsch}}, \bibinfo {author}
  {\bibfnamefont{J.}~\bibnamefont{Suffczy{\'n}ski}}, \bibinfo {author}
  {\bibfnamefont{F.}~\bibnamefont{D'Acapito}}, \bibinfo {author}
  {\bibfnamefont{A.}~\bibnamefont{Meingast}}, \bibinfo {author}
  {\bibfnamefont{G.}~\bibnamefont{Kothleitner}},\ and\ \bibinfo {author}
  {\bibfnamefont{T.}~\bibnamefont{Dietl}}}%
  , \bibinfo {year} {2011},\ \bibfield{title}{%
  \enquote{\bibinfo {title} {Experimental probing of exchange interactions
  between localized spins in the dilute magnetic insulator {(Ga,Mn)N}},}\ }%
  \bibfield{journal}{%
  \bibinfo {journal} {Phys. Rev. B}\ }%
  \textbf{\bibinfo {volume} {84}},\ \bibinfo {pages} {035206}%
  \bibAnnoteFile{NoStop}{Bonanni:2011_PRB}%
\bibitem[{\citenamefont{Boukari}\ \emph{et~al.}(2002)\citenamefont{Boukari},
  \citenamefont{Kossacki}, \citenamefont{Bertolini}, \citenamefont{Ferrand},
  \citenamefont{Cibert}, \citenamefont{Tatarenko}, \citenamefont{Wasiela},
  \citenamefont{Gaj},\ and\ \citenamefont{Dietl}}]{Boukari:2002_PRL}%
  \BibitemOpen
  \bibfield{author}{%
  \bibinfo {author} {\bibnamefont{Boukari}, \bibfnamefont{H.}}, \bibinfo
  {author} {\bibfnamefont{P.}~\bibnamefont{Kossacki}}, \bibinfo {author}
  {\bibfnamefont{M.}~\bibnamefont{Bertolini}}, \bibinfo {author}
  {\bibfnamefont{D.}~\bibnamefont{Ferrand}}, \bibinfo {author}
  {\bibfnamefont{J.}~\bibnamefont{Cibert}}, \bibinfo {author}
  {\bibfnamefont{S.}~\bibnamefont{Tatarenko}}, \bibinfo {author}
  {\bibfnamefont{A.}~\bibnamefont{Wasiela}}, \bibinfo {author}
  {\bibfnamefont{J.~A.}\ \bibnamefont{Gaj}},\ and\ \bibinfo {author}
  {\bibfnamefont{T.}~\bibnamefont{Dietl}}}%
  , \bibinfo {year} {2002},\ \bibfield{title}{%
  \enquote{\bibinfo {title} {Light and electric field control of ferromagnetism
  in magnetic quantum structures},}\ }%
  \bibfield{journal}{%
  \bibinfo {journal} {Phys. Rev. Lett.}\ }%
  \textbf{\bibinfo {volume} {88}},\ \bibinfo {pages} {207204}%
  \bibAnnoteFile{NoStop}{Boukari:2002_PRL}%
\bibitem[{\citenamefont{Bouzerar}(2007)}]{Bouzerar:2007_EPLb}%
  \BibitemOpen
  \bibfield{author}{%
  \bibinfo {author} {\bibnamefont{Bouzerar}, \bibfnamefont{G.}}}%
  , \bibinfo {year} {2007},\ \bibfield{title}{%
  \enquote{\bibinfo {title} {Magnetic spin excitations in diluted ferromagnetic
  systems: {The} case of {Ga$_{1-x}$Mn$_{x}$As}},}\ }%
  \bibfield{journal}{%
  \bibinfo {journal} {Europhys. Lett.}\ }%
  \textbf{\bibinfo {volume} {79}},\ \bibinfo {pages} {57007}%
  \bibAnnoteFile{NoStop}{Bouzerar:2007_EPLb}%
\bibitem[{\citenamefont{Braden}\ \emph{et~al.}(2003)\citenamefont{Braden},
  \citenamefont{Parker}, \citenamefont{Xiong}, \citenamefont{Chun},\ and\
  \citenamefont{Samarth}}]{Braden:2003_PRL}%
  \BibitemOpen
  \bibfield{author}{%
  \bibinfo {author} {\bibnamefont{Braden}, \bibfnamefont{J.~G.}}, \bibinfo
  {author} {\bibfnamefont{J.~S.}\ \bibnamefont{Parker}}, \bibinfo {author}
  {\bibfnamefont{P.}~\bibnamefont{Xiong}}, \bibinfo {author}
  {\bibfnamefont{S.~H.}\ \bibnamefont{Chun}},\ and\ \bibinfo {author}
  {\bibfnamefont{N.}~\bibnamefont{Samarth}}}%
  , \bibinfo {year} {2003},\ \bibfield{title}{%
  \enquote{\bibinfo {title} {Direct measurement of the spin polarization of the
  magnetic semiconductor {(Ga,Mn)As}},}\ }%
  \bibfield{journal}{%
  \bibinfo {journal} {Phys. Rev. Lett.}\ }%
  \textbf{\bibinfo {volume} {91}},\ \bibinfo {pages} {056602}%
  \bibAnnoteFile{NoStop}{Braden:2003_PRL}%
\bibitem[{\citenamefont{Brey}\ and\
  \citenamefont{{G{\'o}mez-Santos}}(2003)}]{Brey:2003_PRB}%
  \BibitemOpen
  \bibfield{author}{%
  \bibinfo {author} {\bibnamefont{Brey}, \bibfnamefont{L.}},\ and\ \bibinfo
  {author} {\bibfnamefont{G.}~\bibnamefont{{G{\'o}mez-Santos}}}}%
  , \bibinfo {year} {2003},\ \bibfield{title}{%
  \enquote{\bibinfo {title} {Magnetic properties of {GaMnAS} from an effective
  {Heisenberg Hamiltonian}},}\ }%
  \bibfield{journal}{%
  \bibinfo {journal} {Phys. Rev. B}\ }%
  \textbf{\bibinfo {volume} {68}},\ \bibinfo {pages} {115206}%
  \bibAnnoteFile{NoStop}{Brey:2003_PRB}%
\bibitem[{\citenamefont{Brey}\ \emph{et~al.}(2004)\citenamefont{Brey},
  \citenamefont{Tejedor},\ and\
  \citenamefont{{Fern{\'a}ndez-Rossier}}}]{Brey:2004_APL}%
  \BibitemOpen
  \bibfield{author}{%
  \bibinfo {author} {\bibnamefont{Brey}, \bibfnamefont{L.}}, \bibinfo {author}
  {\bibfnamefont{C.}~\bibnamefont{Tejedor}},\ and\ \bibinfo {author}
  {\bibfnamefont{J.}~\bibnamefont{{Fern{\'a}ndez-Rossier}}}}%
  , \bibinfo {year} {2004},\ \bibfield{title}{%
  \enquote{\bibinfo {title} {Tunnel magneto-resistance in {GaMnAs}: going
  beyond {Julli\`{e}re} formula},}\ }%
  \bibfield{journal}{%
  \bibinfo {journal} {Appl. Phys. Lett.}\ }%
  \textbf{\bibinfo {volume} {85}},\ \bibinfo {pages} {1996}%
  \bibAnnoteFile{NoStop}{Brey:2004_APL}%
\bibitem[{\citenamefont{Burch}\ \emph{et~al.}(2008)\citenamefont{Burch},
  \citenamefont{Awschalom},\ and\ \citenamefont{Basov}}]{Burch:2008_JMMM}%
  \BibitemOpen
  \bibfield{author}{%
  \bibinfo {author} {\bibnamefont{Burch}, \bibfnamefont{K.S.}}, \bibinfo
  {author} {\bibfnamefont{D.D.}\ \bibnamefont{Awschalom}},\ and\ \bibinfo
  {author} {\bibfnamefont{D.N.}\ \bibnamefont{Basov}}}%
  , \bibinfo {year} {2008},\ \bibfield{title}{%
  \enquote{\bibinfo {title} {Optical properties of {III--Mn--V} ferromagnetic
  semiconductors},}\ }%
  \bibfield{journal}{%
  \bibinfo {journal} {J. Magn. Magn. Mat.}\ }%
  \textbf{\bibinfo {volume} {320}},\ \bibinfo {pages} {3207}%
  \bibAnnoteFile{NoStop}{Burch:2008_JMMM}%
\bibitem[{\citenamefont{Campion}\ \emph{et~al.}(2003)\citenamefont{Campion},
  \citenamefont{Edmonds}, \citenamefont{Zhao}, \citenamefont{Wang},
  \citenamefont{Foxon}, \citenamefont{Gallagher},\ and\
  \citenamefont{Staddon}}]{Campion:2003_JCG}%
  \BibitemOpen
  \bibfield{author}{%
  \bibinfo {author} {\bibnamefont{Campion}, \bibfnamefont{R.~P.}}, \bibinfo
  {author} {\bibfnamefont{K.~W.}\ \bibnamefont{Edmonds}}, \bibinfo {author}
  {\bibfnamefont{L.~X.}\ \bibnamefont{Zhao}}, \bibinfo {author}
  {\bibfnamefont{K.~Y.}\ \bibnamefont{Wang}}, \bibinfo {author}
  {\bibfnamefont{C.~T.}\ \bibnamefont{Foxon}}, \bibinfo {author}
  {\bibfnamefont{B.~L.}\ \bibnamefont{Gallagher}},\ and\ \bibinfo {author}
  {\bibfnamefont{C.~R.}\ \bibnamefont{Staddon}}}%
  , \bibinfo {year} {2003},\ \bibfield{title}{%
  \enquote{\bibinfo {title} {High quality {GaMnAs} films grown with {As}
  dimers},}\ }%
  \bibfield{journal}{%
  \bibinfo {journal} {J. Cryst. Growth}\ }%
  \textbf{\bibinfo {volume} {247}},\ \bibinfo {pages} {42}%
  \bibAnnoteFile{NoStop}{Campion:2003_JCG}%
\bibitem[{\citenamefont{Casiraghi}\
  \emph{et~al.}(2012)\citenamefont{Casiraghi}, \citenamefont{Rushforth},
  \citenamefont{Zemen}, \citenamefont{Haigh}, \citenamefont{Wang},
  \citenamefont{Edmonds}, \citenamefont{Campion},\ and\
  \citenamefont{Gallagher}}]{Casiraghi:2012_APL}%
  \BibitemOpen
  \bibfield{author}{%
  \bibinfo {author} {\bibnamefont{Casiraghi}, \bibfnamefont{A.}}, \bibinfo
  {author} {\bibfnamefont{A.~W.}\ \bibnamefont{Rushforth}}, \bibinfo {author}
  {\bibfnamefont{J.}~\bibnamefont{Zemen}}, \bibinfo {author}
  {\bibfnamefont{J.~A.}\ \bibnamefont{Haigh}}, \bibinfo {author}
  {\bibfnamefont{M.}~\bibnamefont{Wang}}, \bibinfo {author}
  {\bibfnamefont{K.~W.}\ \bibnamefont{Edmonds}}, \bibinfo {author}
  {\bibfnamefont{R.~P.}\ \bibnamefont{Campion}},\ and\ \bibinfo {author}
  {\bibfnamefont{B.~L.}\ \bibnamefont{Gallagher}}}%
  , \bibinfo {year} {2012},\ \bibfield{title}{%
  \enquote{\bibinfo {title} {Piezoelectric strain induced variation of the
  magnetic anisotropy in a high {Curie} temperature {(Ga,Mn)As} sample},}\ }%
  \bibfield{journal}{%
  \bibinfo {journal} {Appl. Phys. Lett.}\ }%
  \textbf{\bibinfo {volume} {101}},\ \bibinfo {pages} {082406}%
  \bibAnnoteFile{NoStop}{Casiraghi:2012_APL}%
\bibitem[{\citenamefont{Chang}\ \emph{et~al.}(2007)\citenamefont{Chang},
  \citenamefont{Park}, \citenamefont{Sato},\ and\
  \citenamefont{Katayama-Yoshida}}]{Chang:2007_PRB}%
  \BibitemOpen
  \bibfield{author}{%
  \bibinfo {author} {\bibnamefont{Chang}, \bibfnamefont{Y.~H.}}, \bibinfo
  {author} {\bibfnamefont{C.~H.}\ \bibnamefont{Park}}, \bibinfo {author}
  {\bibfnamefont{K.}~\bibnamefont{Sato}},\ and\ \bibinfo {author}
  {\bibfnamefont{H.}~\bibnamefont{Katayama-Yoshida}}}%
  , \bibinfo {year} {2007},\ \bibfield{title}{%
  \enquote{\bibinfo {title} {First-principles study of the superexchange
  interaction in {(Ga,Mn)V (V = N, P, As, and Sb)}},}\ }%
  \bibfield{journal}{%
  \bibinfo {journal} {Phys. Rev. B}\ }%
  \textbf{\bibinfo {volume} {76}},\ \bibinfo {pages} {125211}%
  \bibAnnoteFile{NoStop}{Chang:2007_PRB}%
\bibitem[{\citenamefont{Chapler}\ \emph{et~al.}(2012)\citenamefont{Chapler},
  \citenamefont{Mack}, \citenamefont{Ju}, \citenamefont{Elson},
  \citenamefont{Boudouris}, \citenamefont{Namdas}, \citenamefont{Yuen},
  \citenamefont{Heeger}, \citenamefont{Samarth}, \citenamefont{Ventra},
  \citenamefont{Segalman}, \citenamefont{Awschalom}, \citenamefont{Wang}, ,\
  and\ \citenamefont{Basov}}]{Chapler:2012_PRB}%
  \BibitemOpen
  \bibfield{author}{%
  \bibinfo {author} {\bibnamefont{Chapler}, \bibfnamefont{B.~C.}}, \bibinfo
  {author} {\bibfnamefont{S.}~\bibnamefont{Mack}}, \bibinfo {author}
  {\bibfnamefont{L.}~\bibnamefont{Ju}}, \bibinfo {author}
  {\bibfnamefont{T.~W.}\ \bibnamefont{Elson}}, \bibinfo {author}
  {\bibfnamefont{B.~W.}\ \bibnamefont{Boudouris}}, \bibinfo {author}
  {\bibfnamefont{E.}~\bibnamefont{Namdas}}, \bibinfo {author}
  {\bibfnamefont{J.~D.}\ \bibnamefont{Yuen}}, \bibinfo {author}
  {\bibfnamefont{A.~J.}\ \bibnamefont{Heeger}}, \bibinfo {author}
  {\bibfnamefont{N.}~\bibnamefont{Samarth}}, \bibinfo {author}
  {\bibfnamefont{M.~Di}\ \bibnamefont{Ventra}}, \bibinfo {author}
  {\bibfnamefont{R.~A.}\ \bibnamefont{Segalman}}, \bibinfo {author}
  {\bibfnamefont{D.~D.}\ \bibnamefont{Awschalom}}, \bibinfo {author}
  {\bibfnamefont{F.}~\bibnamefont{Wang}}, ,\ and\ \bibinfo {author}
  {\bibfnamefont{D.~N.}\ \bibnamefont{Basov}}}%
  , \bibinfo {year} {2012},\ \bibfield{title}{%
  \enquote{\bibinfo {title} {Infrared conductivity of hole accumulation and
  depletion layers in {(Ga,Mn)As}- and {(Ga,Be)As}-based electric field-effect
  devices},}\ }%
  \bibfield{journal}{%
  \bibinfo {journal} {Phys. Rev. B}\ }%
  \textbf{\bibinfo {volume} {86}},\ \bibinfo {pages} {165302}%
  \bibAnnoteFile{NoStop}{Chapler:2012_PRB}%
\bibitem[{\citenamefont{Chapler}\ \emph{et~al.}(2013)\citenamefont{Chapler},
  \citenamefont{Mack}, \citenamefont{Myers}, \citenamefont{Frenzel},
  \citenamefont{Pursley}, \citenamefont{Burch}, \citenamefont{Dattelbaum},
  \citenamefont{Samarth}, \citenamefont{Awschalom},\ and\
  \citenamefont{Basov}}]{Chapler:2013_PRB}%
  \BibitemOpen
  \bibfield{author}{%
  \bibinfo {author} {\bibnamefont{Chapler}, \bibfnamefont{B.~C.}}, \bibinfo
  {author} {\bibfnamefont{S.}~\bibnamefont{Mack}}, \bibinfo {author}
  {\bibfnamefont{R.~C.}\ \bibnamefont{Myers}}, \bibinfo {author}
  {\bibfnamefont{A.}~\bibnamefont{Frenzel}}, \bibinfo {author}
  {\bibfnamefont{B.~C.}\ \bibnamefont{Pursley}}, \bibinfo {author}
  {\bibfnamefont{K.~S.}\ \bibnamefont{Burch}}, \bibinfo {author}
  {\bibfnamefont{A.~M.}\ \bibnamefont{Dattelbaum}}, \bibinfo {author}
  {\bibfnamefont{N.}~\bibnamefont{Samarth}}, \bibinfo {author}
  {\bibfnamefont{D.~D.}\ \bibnamefont{Awschalom}},\ and\ \bibinfo {author}
  {\bibfnamefont{D.~N.}\ \bibnamefont{Basov}}}%
  , \bibinfo {year} {2013},\ \bibfield{title}{%
  \enquote{\bibinfo {title} {Ferromagnetism and infrared electrodynamics of
  {Ga$_{1-x}$Mn$_{x}$As}},}\ }%
  \bibfield{journal}{%
  \bibinfo {journal} {Phys. Rev. B}\ }%
  \textbf{\bibinfo {volume} {87}},\ \bibinfo {pages} {205314}%
  \bibAnnoteFile{NoStop}{Chapler:2013_PRB}%
\bibitem[{\citenamefont{Chapler}\ \emph{et~al.}(2011)\citenamefont{Chapler},
  \citenamefont{Myers}, \citenamefont{Mack}, \citenamefont{Frenzel},
  \citenamefont{Pursley}, \citenamefont{Burch}, \citenamefont{Singley},
  \citenamefont{Dattelbaum}, \citenamefont{Samarth}, \citenamefont{Awschalom},\
  and\ \citenamefont{Basov}}]{Chapler:2011_PRB}%
  \BibitemOpen
  \bibfield{author}{%
  \bibinfo {author} {\bibnamefont{Chapler}, \bibfnamefont{B.~C.}}, \bibinfo
  {author} {\bibfnamefont{R.~C.}\ \bibnamefont{Myers}}, \bibinfo {author}
  {\bibfnamefont{S.}~\bibnamefont{Mack}}, \bibinfo {author}
  {\bibfnamefont{A.}~\bibnamefont{Frenzel}}, \bibinfo {author}
  {\bibfnamefont{B.~C.}\ \bibnamefont{Pursley}}, \bibinfo {author}
  {\bibfnamefont{K.~S.}\ \bibnamefont{Burch}}, \bibinfo {author}
  {\bibfnamefont{E.~J.}\ \bibnamefont{Singley}}, \bibinfo {author}
  {\bibfnamefont{A.~M.}\ \bibnamefont{Dattelbaum}}, \bibinfo {author}
  {\bibfnamefont{N.}~\bibnamefont{Samarth}}, \bibinfo {author}
  {\bibfnamefont{D.~D.}\ \bibnamefont{Awschalom}},\ and\ \bibinfo {author}
  {\bibfnamefont{D.~N.}\ \bibnamefont{Basov}}}%
  , \bibinfo {year} {2011},\ \bibfield{title}{%
  \enquote{\bibinfo {title} {Infrared probe of the insulator-to-metal
  transition in {Ga$_{1-x}$Mn$_x$As} and {Ga$_{1-x}$Be$_x$As}},}\ }%
  \bibfield{journal}{%
  \bibinfo {journal} {Phys. Rev. B}\ }%
  \textbf{\bibinfo {volume} {84}},\ \bibinfo {pages} {081203}%
  \bibAnnoteFile{NoStop}{Chapler:2011_PRB}%
\bibitem[{\citenamefont{Chattopadhyay}\
  \emph{et~al.}(2001)\citenamefont{Chattopadhyay}, \citenamefont{Sarma},\ and\
  \citenamefont{Millis}}]{Chattopadhyay:2001_PRL}%
  \BibitemOpen
  \bibfield{author}{%
  \bibinfo {author} {\bibnamefont{Chattopadhyay}, \bibfnamefont{A.}}, \bibinfo
  {author} {\bibfnamefont{S.~Das}\ \bibnamefont{Sarma}},\ and\ \bibinfo
  {author} {\bibfnamefont{A.~J.}\ \bibnamefont{Millis}}}%
  , \bibinfo {year} {2001},\ \bibfield{title}{%
  \enquote{\bibinfo {title} {Transition temperature of ferromagnetic
  semiconductors: a dynamical mean fieldstudy},}\ }%
  \bibfield{journal}{%
  \bibinfo {journal} {Phys. Rev. Lett.}\ }%
  \textbf{\bibinfo {volume} {87}},\ \bibinfo {pages} {227202}%
  \bibAnnoteFile{NoStop}{Chattopadhyay:2001_PRL}%
\bibitem[{\citenamefont{Checkelsky}\
  \emph{et~al.}(2012)\citenamefont{Checkelsky}, \citenamefont{Ye},
  \citenamefont{Y.~Onose},\ and\ \citenamefont{Tokura}}]{Checkelsky:2012_NP}%
  \BibitemOpen
  \bibfield{author}{%
  \bibinfo {author} {\bibnamefont{Checkelsky}, \bibfnamefont{J.~G.}}, \bibinfo
  {author} {\bibfnamefont{{Jianting}}\ \bibnamefont{Ye}}, \bibinfo {author}
  {\bibfnamefont{Y.~Iwasa}\ \bibnamefont{Y.~Onose}},\ and\ \bibinfo {author}
  {\bibfnamefont{Y.}~\bibnamefont{Tokura}}}%
  , \bibinfo {year} {2012},\ \bibfield{title}{%
  \enquote{\bibinfo {title} {Dirac-fermion-mediated ferromagnetism in a
  topological insulator},}\ }%
  \bibfield{journal}{%
  \bibinfo {journal} {Nat. Phys.}\ }%
  \textbf{\bibinfo {volume} {8}},\ \bibinfo {pages} {729}%
  \bibAnnoteFile{NoStop}{Checkelsky:2012_NP}%
\bibitem[{\citenamefont{Chen}\ \emph{et~al.}(2009)\citenamefont{Chen},
  \citenamefont{Yan}, \citenamefont{Xu}, \citenamefont{Lu},
  \citenamefont{Wang}, \citenamefont{Deng}, \citenamefont{Qian},
  \citenamefont{Ji},\ and\ \citenamefont{Zhao}}]{Chen:2009_APL}%
  \BibitemOpen
  \bibfield{author}{%
  \bibinfo {author} {\bibnamefont{Chen}, \bibfnamefont{L.}}, \bibinfo {author}
  {\bibfnamefont{S.}~\bibnamefont{Yan}}, \bibinfo {author}
  {\bibfnamefont{P.~F.}\ \bibnamefont{Xu}}, \bibinfo {author}
  {\bibfnamefont{J.}~\bibnamefont{Lu}}, \bibinfo {author}
  {\bibfnamefont{W.~Z.}\ \bibnamefont{Wang}}, \bibinfo {author}
  {\bibfnamefont{J.~J.}\ \bibnamefont{Deng}}, \bibinfo {author}
  {\bibfnamefont{X.}~\bibnamefont{Qian}}, \bibinfo {author}
  {\bibfnamefont{Y.}~\bibnamefont{Ji}},\ and\ \bibinfo {author}
  {\bibfnamefont{J.~H.}\ \bibnamefont{Zhao}}}%
  , \bibinfo {year} {2009},\ \bibfield{title}{%
  \enquote{\bibinfo {title} {Low-temperature magnetotransport behaviors of
  heavily {Mn}-doped {(Ga,Mn)As} films with high ferromagnetic transition
  temperature},}\ }%
  \bibfield{journal}{%
  \bibinfo {journal} {Appl. Phys. Lett.}\ }%
  \textbf{\bibinfo {volume} {95}},\ \bibinfo {pages} {182505}%
  \bibAnnoteFile{NoStop}{Chen:2009_APL}%
\bibitem[{\citenamefont{Chen}\ \emph{et~al.}(2013)\citenamefont{Chen},
  \citenamefont{Matsukura},\ and\ \citenamefont{Ohno}}]{Chen:2013_NC}%
  \BibitemOpen
  \bibfield{author}{%
  \bibinfo {author} {\bibnamefont{Chen}, \bibfnamefont{Lin}}, \bibinfo {author}
  {\bibfnamefont{F.}~\bibnamefont{Matsukura}},\ and\ \bibinfo {author}
  {\bibfnamefont{H.}~\bibnamefont{Ohno}}}%
  , \bibinfo {year} {2013},\ \bibfield{title}{%
  \enquote{\bibinfo {title} {Direct-current voltages in {(Ga,Mn)As} structures
  induced by ferromagnetic resonance},}\ }%
  \bibfield{journal}{%
  \bibinfo {journal} {Nat. Commun.}\ }%
  \textbf{\bibinfo {volume} {4}},\ \bibinfo {pages} {2055}%
  \bibAnnoteFile{NoStop}{Chen:2013_NC}%
\bibitem[{\citenamefont{Chen}\ \emph{et~al.}(2011)\citenamefont{Chen},
  \citenamefont{Yang}, \citenamefont{Yang}, \citenamefont{Zhao},
  \citenamefont{Misuraca}, \citenamefont{Xiong},\ and\ \citenamefont{von
  Molna\`{r}}}]{Chen:2011_NL}%
  \BibitemOpen
  \bibfield{author}{%
  \bibinfo {author} {\bibnamefont{Chen}, \bibfnamefont{Lin}}, \bibinfo {author}
  {\bibfnamefont{Xiang}\ \bibnamefont{Yang}}, \bibinfo {author}
  {\bibfnamefont{Fuhua}\ \bibnamefont{Yang}}, \bibinfo {author}
  {\bibfnamefont{Jianhua}\ \bibnamefont{Zhao}}, \bibinfo {author}
  {\bibfnamefont{J.}~\bibnamefont{Misuraca}}, \bibinfo {author}
  {\bibfnamefont{Peng}\ \bibnamefont{Xiong}},\ and\ \bibinfo {author}
  {\bibfnamefont{S.}~\bibnamefont{von Molna\`{r}}}}%
  , \bibinfo {year} {2011},\ \bibfield{title}{%
  \enquote{\bibinfo {title} {Enhancing the {Curie} temperature of ferromagnetic
  semiconductor {(Ga,Mn)As} to 200 {K} via nanostructure engineering},}\ }%
  \bibfield{journal}{%
  \bibinfo {journal} {Nano Lett.}\ }%
  \textbf{\bibinfo {volume} {11}},\ \bibinfo {pages} {2584}%
  \bibAnnoteFile{NoStop}{Chen:2011_NL}%
\bibitem[{\citenamefont{Chernyshov}\
  \emph{et~al.}(2009)\citenamefont{Chernyshov}, \citenamefont{Overby},
  \citenamefont{Liu}, \citenamefont{Furdyna}, \citenamefont{Lyanda-Geller},\
  and\ \citenamefont{Rokhinson}}]{Chernyshov:2009_NP}%
  \BibitemOpen
  \bibfield{author}{%
  \bibinfo {author} {\bibnamefont{Chernyshov}, \bibfnamefont{A.}}, \bibinfo
  {author} {\bibfnamefont{M.}~\bibnamefont{Overby}}, \bibinfo {author}
  {\bibfnamefont{Xinyu}\ \bibnamefont{Liu}}, \bibinfo {author}
  {\bibfnamefont{J.~K.}\ \bibnamefont{Furdyna}}, \bibinfo {author}
  {\bibfnamefont{Y.}~\bibnamefont{Lyanda-Geller}},\ and\ \bibinfo {author}
  {\bibfnamefont{L.~P.}\ \bibnamefont{Rokhinson}}}%
  , \bibinfo {year} {2009},\ \bibfield{title}{%
  \enquote{\bibinfo {title} {Evidence for reversible control of magnetization
  in a ferromagnetic material by means of spin-orbit magnetic field},}\ }%
  \bibfield{journal}{%
  \bibinfo {journal} {Nat. Phys.}\ }%
  \textbf{\bibinfo {volume} {5}},\ \bibinfo {pages} {656}%
  \bibAnnoteFile{NoStop}{Chernyshov:2009_NP}%
\bibitem[{\citenamefont{Chiba}\ \emph{et~al.}(2000)\citenamefont{Chiba},
  \citenamefont{Akiba}, \citenamefont{Matsukura}, \citenamefont{Ohno},\ and\
  \citenamefont{Ohno}}]{Chiba:2000_APL}%
  \BibitemOpen
  \bibfield{author}{%
  \bibinfo {author} {\bibnamefont{Chiba}, \bibfnamefont{D.}}, \bibinfo {author}
  {\bibfnamefont{N.}~\bibnamefont{Akiba}}, \bibinfo {author}
  {\bibfnamefont{F.}~\bibnamefont{Matsukura}}, \bibinfo {author}
  {\bibfnamefont{Y.}~\bibnamefont{Ohno}},\ and\ \bibinfo {author}
  {\bibfnamefont{H.}~\bibnamefont{Ohno}}}%
  , \bibinfo {year} {2000},\ \bibfield{title}{%
  \enquote{\bibinfo {title} {Magnetoresistance effect and interlayer coupling
  of {(Ga,Mn)As} trilayer structures},}\ }%
  \bibfield{journal}{%
  \bibinfo {journal} {Appl. Phys. Lett.}\ }%
  \textbf{\bibinfo {volume} {77}},\ \bibinfo {pages} {1873}%
  \bibAnnoteFile{NoStop}{Chiba:2000_APL}%
\bibitem[{\citenamefont{Chiba}\
  \emph{et~al.}(2004{\natexlab{a}})\citenamefont{Chiba},
  \citenamefont{Matsukura},\ and\ \citenamefont{Ohno}}]{Chiba:2004_PE}%
  \BibitemOpen
  \bibfield{author}{%
  \bibinfo {author} {\bibnamefont{Chiba}, \bibfnamefont{D.}}, \bibinfo {author}
  {\bibfnamefont{F.}~\bibnamefont{Matsukura}},\ and\ \bibinfo {author}
  {\bibfnamefont{H.}~\bibnamefont{Ohno}}}%
  , \bibinfo {year} {2004}{\natexlab{a}},\ \bibfield{title}{%
  \enquote{\bibinfo {title} {Tunneling magnetoresistance in {(Ga,Mn)As}-based
  heterostructures with a {GaAs} barrier},}\ }%
  \bibfield{journal}{%
  \bibinfo {journal} {Physica}\ }%
  \textbf{\bibinfo {volume} {E 21}},\ \bibinfo {pages} {966}%
  \bibAnnoteFile{NoStop}{Chiba:2004_PE}%
\bibitem[{\citenamefont{Chiba}\
  \emph{et~al.}(2006{\natexlab{a}})\citenamefont{Chiba},
  \citenamefont{Matsukura},\ and\ \citenamefont{Ohno}}]{Chiba:2006_APL}%
  \BibitemOpen
  \bibfield{author}{%
  \bibinfo {author} {\bibnamefont{Chiba}, \bibfnamefont{D.}}, \bibinfo {author}
  {\bibfnamefont{F.}~\bibnamefont{Matsukura}},\ and\ \bibinfo {author}
  {\bibfnamefont{H.}~\bibnamefont{Ohno}}}%
  , \bibinfo {year} {2006}{\natexlab{a}},\ \bibfield{title}{%
  \enquote{\bibinfo {title} {Electric-field control of ferromagnetism in
  {(Ga,Mn)As}},}\ }%
  \bibfield{journal}{%
  \bibinfo {journal} {Appl. Phys. Lett.}\ }%
  \textbf{\bibinfo {volume} {89}},\ \bibinfo {pages} {162505}%
  \bibAnnoteFile{NoStop}{Chiba:2006_APL}%
\bibitem[{\citenamefont{Chiba}\ \emph{et~al.}(2007)\citenamefont{Chiba},
  \citenamefont{Nishitani}, \citenamefont{Matsukura},\ and\
  \citenamefont{Ohno}}]{Chiba:2007_APL}%
  \BibitemOpen
  \bibfield{author}{%
  \bibinfo {author} {\bibnamefont{Chiba}, \bibfnamefont{D.}}, \bibinfo {author}
  {\bibfnamefont{Y.}~\bibnamefont{Nishitani}}, \bibinfo {author}
  {\bibfnamefont{F.}~\bibnamefont{Matsukura}},\ and\ \bibinfo {author}
  {\bibfnamefont{H.}~\bibnamefont{Ohno}}}%
  , \bibinfo {year} {2007},\ \bibfield{title}{%
  \enquote{\bibinfo {title} {Properties of {Ga$_{1-x}$Mn$_{x}$As} with high
  {Mn} composition $(x>0.1)$},}\ }%
  \bibfield{journal}{%
  \bibinfo {journal} {Appl. Phys. Lett.}\ }%
  \textbf{\bibinfo {volume} {90}},\ \bibinfo {pages} {122503}%
  \bibAnnoteFile{NoStop}{Chiba:2007_APL}%
\bibitem[{\citenamefont{Chiba}\
  \emph{et~al.}(2004{\natexlab{b}})\citenamefont{Chiba}, \citenamefont{Sato},
  \citenamefont{Kita}, \citenamefont{Matsukura},\ and\
  \citenamefont{Ohno}}]{Chiba:2004_PRL}%
  \BibitemOpen
  \bibfield{author}{%
  \bibinfo {author} {\bibnamefont{Chiba}, \bibfnamefont{D.}}, \bibinfo {author}
  {\bibfnamefont{Y.}~\bibnamefont{Sato}}, \bibinfo {author}
  {\bibfnamefont{T.}~\bibnamefont{Kita}}, \bibinfo {author}
  {\bibfnamefont{F.}~\bibnamefont{Matsukura}},\ and\ \bibinfo {author}
  {\bibfnamefont{H.}~\bibnamefont{Ohno}}}%
  , \bibinfo {year} {2004}{\natexlab{b}},\ \bibfield{title}{%
  \enquote{\bibinfo {title} {Current-driven magnetization reversal in a
  ferromagnetic semiconductor {(Ga,Mn)As/GaAs/(Ga,Mn)As} tunnel junction},}\ }%
  \bibfield{journal}{%
  \bibinfo {journal} {Phys. Rev. Lett.}\ }%
  \textbf{\bibinfo {volume} {93}},\ \bibinfo {pages} {216602}%
  \bibAnnoteFile{NoStop}{Chiba:2004_PRL}%
\bibitem[{\citenamefont{Chiba}\
  \emph{et~al.}(2008{\natexlab{a}})\citenamefont{Chiba},
  \citenamefont{Sawicki}, \citenamefont{Nishitani}, \citenamefont{Nakatani},
  \citenamefont{Matsukura},\ and\ \citenamefont{Ohno}}]{Chiba:2008_N}%
  \BibitemOpen
  \bibfield{author}{%
  \bibinfo {author} {\bibnamefont{Chiba}, \bibfnamefont{D.}}, \bibinfo {author}
  {\bibfnamefont{M.}~\bibnamefont{Sawicki}}, \bibinfo {author}
  {\bibfnamefont{Y.}~\bibnamefont{Nishitani}}, \bibinfo {author}
  {\bibfnamefont{Y.}~\bibnamefont{Nakatani}}, \bibinfo {author}
  {\bibfnamefont{F.}~\bibnamefont{Matsukura}},\ and\ \bibinfo {author}
  {\bibfnamefont{H.}~\bibnamefont{Ohno}}}%
  , \bibinfo {year} {2008}{\natexlab{a}},\ \bibfield{title}{%
  \enquote{\bibinfo {title} {Magnetization vector manipulation by electric
  fields},}\ }%
  \bibfield{journal}{%
  \bibinfo {journal} {Nature}\ }%
  \textbf{\bibinfo {volume} {455}},\ \bibinfo {pages} {515}%
  \bibAnnoteFile{NoStop}{Chiba:2008_N}%
\bibitem[{\citenamefont{Chiba}\ \emph{et~al.}(2010)\citenamefont{Chiba},
  \citenamefont{Werpachowska}, \citenamefont{Endo}, \citenamefont{Nishitani},
  \citenamefont{Matsukura}, \citenamefont{Dietl},\ and\
  \citenamefont{Ohno}}]{Chiba:2010_PRL}%
  \BibitemOpen
  \bibfield{author}{%
  \bibinfo {author} {\bibnamefont{Chiba}, \bibfnamefont{D.}}, \bibinfo {author}
  {\bibfnamefont{A.}~\bibnamefont{Werpachowska}}, \bibinfo {author}
  {\bibfnamefont{M.}~\bibnamefont{Endo}}, \bibinfo {author}
  {\bibfnamefont{Y.}~\bibnamefont{Nishitani}}, \bibinfo {author}
  {\bibfnamefont{F.}~\bibnamefont{Matsukura}}, \bibinfo {author}
  {\bibfnamefont{T.}~\bibnamefont{Dietl}},\ and\ \bibinfo {author}
  {\bibfnamefont{H.}~\bibnamefont{Ohno}}}%
  , \bibinfo {year} {2010},\ \bibfield{title}{%
  \enquote{\bibinfo {title} {Anomalous {Hall} effect in field-effect structures
  of {(Ga,Mn)As}},}\ }%
  \bibfield{journal}{%
  \bibinfo {journal} {Phys. Rev. Lett.}\ }%
  \textbf{\bibinfo {volume} {104}},\ \bibinfo {pages} {106601}%
  \bibAnnoteFile{NoStop}{Chiba:2010_PRL}%
\bibitem[{\citenamefont{Chiba}\
  \emph{et~al.}(2006{\natexlab{b}})\citenamefont{Chiba},
  \citenamefont{Yamanouchi}, \citenamefont{Matsukura}, \citenamefont{Dietl},\
  and\ \citenamefont{Ohno}}]{Chiba:2006_PRL}%
  \BibitemOpen
  \bibfield{author}{%
  \bibinfo {author} {\bibnamefont{Chiba}, \bibfnamefont{D.}}, \bibinfo {author}
  {\bibfnamefont{M.}~\bibnamefont{Yamanouchi}}, \bibinfo {author}
  {\bibfnamefont{F.}~\bibnamefont{Matsukura}}, \bibinfo {author}
  {\bibfnamefont{T.}~\bibnamefont{Dietl}},\ and\ \bibinfo {author}
  {\bibfnamefont{H.}~\bibnamefont{Ohno}}}%
  , \bibinfo {year} {2006}{\natexlab{b}},\ \bibfield{title}{%
  \enquote{\bibinfo {title} {Domain-wall resistance in ferromagnetic
  {(Ga,Mn)As}},}\ }%
  \bibfield{journal}{%
  \bibinfo {journal} {Phys. Rev. Lett.}\ }%
  \textbf{\bibinfo {volume} {96}},\ \bibinfo {pages} {096602}%
  \bibAnnoteFile{NoStop}{Chiba:2006_PRL}%
\bibitem[{\citenamefont{Chiba}\ \emph{et~al.}(2003)\citenamefont{Chiba},
  \citenamefont{Yamanouchi}, \citenamefont{Matsukura},\ and\
  \citenamefont{Ohno}}]{Chiba:2003_S}%
  \BibitemOpen
  \bibfield{author}{%
  \bibinfo {author} {\bibnamefont{Chiba}, \bibfnamefont{D.}}, \bibinfo {author}
  {\bibfnamefont{M.}~\bibnamefont{Yamanouchi}}, \bibinfo {author}
  {\bibfnamefont{F.}~\bibnamefont{Matsukura}},\ and\ \bibinfo {author}
  {\bibfnamefont{H.}~\bibnamefont{Ohno}}}%
  , \bibinfo {year} {2003},\ \bibfield{title}{%
  \enquote{\bibinfo {title} {Electrical manipulation of magnetization reversal
  in a ferromagnetic semiconductor},}\ }%
  \bibfield{journal}{%
  \bibinfo {journal} {Science}\ }%
  \textbf{\bibinfo {volume} {301}},\ \bibinfo {pages} {943}%
  \bibAnnoteFile{NoStop}{Chiba:2003_S}%
\bibitem[{\citenamefont{Chiba}\
  \emph{et~al.}(2008{\natexlab{b}})\citenamefont{Chiba}, \citenamefont{Yu},
  \citenamefont{Walukiewicz}, \citenamefont{Nishitani},
  \citenamefont{Matsukura},\ and\ \citenamefont{Ohno}}]{Chiba:2008_JAP}%
  \BibitemOpen
  \bibfield{author}{%
  \bibinfo {author} {\bibnamefont{Chiba}, \bibfnamefont{D.}}, \bibinfo {author}
  {\bibfnamefont{K.~M.}\ \bibnamefont{Yu}}, \bibinfo {author}
  {\bibfnamefont{W.}~\bibnamefont{Walukiewicz}}, \bibinfo {author}
  {\bibfnamefont{Y.}~\bibnamefont{Nishitani}}, \bibinfo {author}
  {\bibfnamefont{F.}~\bibnamefont{Matsukura}},\ and\ \bibinfo {author}
  {\bibfnamefont{H.}~\bibnamefont{Ohno}}}%
  , \bibinfo {year} {2008}{\natexlab{b}},\ \bibfield{title}{%
  \enquote{\bibinfo {title} {Properties of {Ga$_{1-x}$Mn$_{x}$As} with high
  {($x>0.1$)}},}\ }%
  \bibfield{journal}{%
  \bibinfo {journal} {J. Appl. Phys.}\ }%
  \textbf{\bibinfo {volume} {103}},\ \bibinfo {pages} {07D136}%
  \bibAnnoteFile{NoStop}{Chiba:2008_JAP}%
\bibitem[{\citenamefont{Chiu}\ \emph{et~al.}(2005)\citenamefont{Chiu},
  \citenamefont{Wessels}, \citenamefont{Keavney},\ and\
  \citenamefont{Freeland}}]{Chiu:2005_APL}%
  \BibitemOpen
  \bibfield{author}{%
  \bibinfo {author} {\bibnamefont{Chiu}, \bibfnamefont{P.~T.}}, \bibinfo
  {author} {\bibfnamefont{B.~W.}\ \bibnamefont{Wessels}}, \bibinfo {author}
  {\bibfnamefont{D.~J.}\ \bibnamefont{Keavney}},\ and\ \bibinfo {author}
  {\bibfnamefont{J.~W.}\ \bibnamefont{Freeland}}}%
  , \bibinfo {year} {2005},\ \bibfield{title}{%
  \enquote{\bibinfo {title} {Local environment of ferromagnetically ordered
  {Mn} in epitaxial {InMnAs}},}\ }%
  \bibfield{journal}{%
  \bibinfo {journal} {Appl. Phys. Lett.}\ }%
  \textbf{\bibinfo {volume} {86}},\ \bibinfo {pages} {072505}%
  \bibAnnoteFile{NoStop}{Chiu:2005_APL}%
\bibitem[{\citenamefont{Chun}\ \emph{et~al.}(2002)\citenamefont{Chun},
  \citenamefont{Potashnik}, \citenamefont{Ku}, \citenamefont{Schiffer},\ and\
  \citenamefont{Samarth}}]{Chun:2002_PRB}%
  \BibitemOpen
  \bibfield{author}{%
  \bibinfo {author} {\bibnamefont{Chun}, \bibfnamefont{S.~H.}}, \bibinfo
  {author} {\bibfnamefont{S.~J.}\ \bibnamefont{Potashnik}}, \bibinfo {author}
  {\bibfnamefont{K.~C.}\ \bibnamefont{Ku}}, \bibinfo {author}
  {\bibfnamefont{P.}~\bibnamefont{Schiffer}},\ and\ \bibinfo {author}
  {\bibfnamefont{N.}~\bibnamefont{Samarth}}}%
  , \bibinfo {year} {2002},\ \bibfield{title}{%
  \enquote{\bibinfo {title} {Spin-polarized tunneling in hybrid
  metal-semiconductor magnetic tunnel junctions},}\ }%
  \bibfield{journal}{%
  \bibinfo {journal} {Phys. Rev. B}\ }%
  \textbf{\bibinfo {volume} {66}},\ \bibinfo {pages} {100408}%
  \bibAnnoteFile{NoStop}{Chun:2002_PRB}%
\bibitem[{\citenamefont{Chung}\ \emph{et~al.}(2010)\citenamefont{Chung},
  \citenamefont{Lee}, \citenamefont{Chung}, \citenamefont{Yoo},
  \citenamefont{Lee}, \citenamefont{Kirby}, \citenamefont{Liu},\ and\
  \citenamefont{Furdyna}}]{Chung:2010_PRB}%
  \BibitemOpen
  \bibfield{author}{%
  \bibinfo {author} {\bibnamefont{Chung}, \bibfnamefont{Sunjae}}, \bibinfo
  {author} {\bibfnamefont{Sanghoon}\ \bibnamefont{Lee}}, \bibinfo {author}
  {\bibfnamefont{J.-H.}\ \bibnamefont{Chung}}, \bibinfo {author}
  {\bibfnamefont{Taehee}\ \bibnamefont{Yoo}}, \bibinfo {author}
  {\bibfnamefont{Hakjoon}\ \bibnamefont{Lee}}, \bibinfo {author}
  {\bibfnamefont{B.}~\bibnamefont{Kirby}}, \bibinfo {author}
  {\bibfnamefont{X.}~\bibnamefont{Liu}},\ and\ \bibinfo {author}
  {\bibfnamefont{J.~K.}\ \bibnamefont{Furdyna}}}%
  , \bibinfo {year} {2010},\ \bibfield{title}{%
  \enquote{\bibinfo {title} {Giant magnetoresistance and long-range
  antiferromagnetic interlayer exchange coupling in {(Ga,Mn)As/GaAs:Be}
  multilayers},}\ }%
  \bibfield{journal}{%
  \bibinfo {journal} {Phys. Rev. B}\ }%
  \textbf{\bibinfo {volume} {82}},\ \bibinfo {pages} {054420}%
  \bibAnnoteFile{NoStop}{Chung:2010_PRB}%
\bibitem[{\citenamefont{Cibert}\ \emph{et~al.}(2008)\citenamefont{Cibert},
  \citenamefont{Besombes}, \citenamefont{Ferrand},\ and\
  \citenamefont{Mariette}}]{Cibert:2008_B}%
  \BibitemOpen
  \bibfield{author}{%
  \bibinfo {author} {\bibnamefont{Cibert}, \bibfnamefont{J.}}, \bibinfo
  {author} {\bibfnamefont{L.}~\bibnamefont{Besombes}}, \bibinfo {author}
  {\bibfnamefont{D.}~\bibnamefont{Ferrand}},\ and\ \bibinfo {author}
  {\bibfnamefont{H.}~\bibnamefont{Mariette}}}%
  , \bibinfo {year} {2008},\ \enquote{\bibinfo {title} {Quantum structures of
  {II-VI} diluted magnetic semiconductors},}\ in\ \emph{\bibinfo {booktitle}
  {Spintronics}},\ \bibinfo {editor} {edited by\ \bibinfo {editor}
  {\bibfnamefont{T.}~\bibnamefont{Dietl}}, \bibinfo {editor}
  {\bibfnamefont{D.~D.}\ \bibnamefont{Awschalom}}, \bibinfo {editor}
  {\bibfnamefont{M.}~\bibnamefont{Kami{\'n}ska}},\ and\ \bibinfo {editor}
  {\bibfnamefont{H.}~\bibnamefont{Ohno}}}\ (\bibinfo {publisher} {Elsevier,
  Amsterdam})\ p.\ \bibinfo {pages} {287}%
  \bibAnnoteFile{NoStop}{Cibert:2008_B}%
\bibitem[{\citenamefont{Ciccarelli}\
  \emph{et~al.}(2012)\citenamefont{Ciccarelli}, \citenamefont{Z\^{a}rbo},
  \citenamefont{Irvine}, \citenamefont{Campion}, \citenamefont{Gallagher},
  \citenamefont{Wunderlich}, \citenamefont{Jungwirth},\ and\
  \citenamefont{Ferguson}}]{Ciccarelli:2012_APL}%
  \BibitemOpen
  \bibfield{author}{%
  \bibinfo {author} {\bibnamefont{Ciccarelli}, \bibfnamefont{C.}}, \bibinfo
  {author} {\bibfnamefont{L.~P.}\ \bibnamefont{Z\^{a}rbo}}, \bibinfo {author}
  {\bibfnamefont{A.~C.}\ \bibnamefont{Irvine}}, \bibinfo {author}
  {\bibfnamefont{R.~P.}\ \bibnamefont{Campion}}, \bibinfo {author}
  {\bibfnamefont{B.~L.}\ \bibnamefont{Gallagher}}, \bibinfo {author}
  {\bibfnamefont{J.}~\bibnamefont{Wunderlich}}, \bibinfo {author}
  {\bibfnamefont{T.}~\bibnamefont{Jungwirth}},\ and\ \bibinfo {author}
  {\bibfnamefont{A.~J.}\ \bibnamefont{Ferguson}}}%
  , \bibinfo {year} {2012},\ \bibfield{title}{%
  \enquote{\bibinfo {title} {Spin gating electrical current},}\ }%
  \bibfield{journal}{%
  \bibinfo {journal} {Appl. Phys. Lett.}\ }%
  \textbf{\bibinfo {volume} {101}},\ \bibinfo {pages} {122411}%
  \bibAnnoteFile{NoStop}{Ciccarelli:2012_APL}%
\bibitem[{\citenamefont{Ciorga}\ \emph{et~al.}(2009)\citenamefont{Ciorga},
  \citenamefont{Einwanger}, \citenamefont{W{\"u}rstbauer},
  \citenamefont{Schuh}, \citenamefont{Wegscheider},\ and\
  \citenamefont{Weiss}}]{Ciorga:2009_PRB}%
  \BibitemOpen
  \bibfield{author}{%
  \bibinfo {author} {\bibnamefont{Ciorga}, \bibfnamefont{M.}}, \bibinfo
  {author} {\bibfnamefont{A.}~\bibnamefont{Einwanger}}, \bibinfo {author}
  {\bibfnamefont{U.}~\bibnamefont{W{\"u}rstbauer}}, \bibinfo {author}
  {\bibfnamefont{D.}~\bibnamefont{Schuh}}, \bibinfo {author}
  {\bibfnamefont{W.}~\bibnamefont{Wegscheider}},\ and\ \bibinfo {author}
  {\bibfnamefont{D.}~\bibnamefont{Weiss}}}%
  , \bibinfo {year} {2009},\ \bibfield{title}{%
  \enquote{\bibinfo {title} {Electrical spin injection and detection in lateral
  all-semiconductor devices},}\ }%
  \bibfield{journal}{%
  \bibinfo {journal} {Phys. Rev. B}\ }%
  \textbf{\bibinfo {volume} {79}},\ \bibinfo {pages} {165321}%
  \bibAnnoteFile{NoStop}{Ciorga:2009_PRB}%
\bibitem[{\citenamefont{Cochrane}\ \emph{et~al.}(1974)\citenamefont{Cochrane},
  \citenamefont{Plischke},\ and\
  \citenamefont{Str\"om-Olsen}}]{Cochrane:1974_PRB}%
  \BibitemOpen
  \bibfield{author}{%
  \bibinfo {author} {\bibnamefont{Cochrane}, \bibfnamefont{R.~W.}}, \bibinfo
  {author} {\bibfnamefont{M.}~\bibnamefont{Plischke}},\ and\ \bibinfo {author}
  {\bibfnamefont{J.~O.}\ \bibnamefont{Str\"om-Olsen}}}%
  , \bibinfo {year} {1974},\ \bibfield{title}{%
  \enquote{\bibinfo {title} {Magnetization studies of
  {(GeTe)$_{1-x}$(MnTe)$_{x}$} pseudobinary alloys},}\ }%
  \bibfield{journal}{%
  \bibinfo {journal} {Phys. Rev. B}\ }%
  \textbf{\bibinfo {volume} {9}},\ \bibinfo {pages} {3013}%
  \bibAnnoteFile{NoStop}{Cochrane:1974_PRB}%
\bibitem[{\citenamefont{Coey}\ \emph{et~al.}(2008)\citenamefont{Coey},
  \citenamefont{Wongsaprom}, \citenamefont{Alaria},\ and\
  \citenamefont{Venkatesan}}]{Coey:2008_JPD}%
  \BibitemOpen
  \bibfield{author}{%
  \bibinfo {author} {\bibnamefont{Coey}, \bibfnamefont{J.~M.~D.}}, \bibinfo
  {author} {\bibfnamefont{Kwanruthai}\ \bibnamefont{Wongsaprom}}, \bibinfo
  {author} {\bibfnamefont{J.}~\bibnamefont{Alaria}},\ and\ \bibinfo {author}
  {\bibfnamefont{M}~\bibnamefont{Venkatesan}}}%
  , \bibinfo {year} {2008},\ \bibfield{title}{%
  \enquote{\bibinfo {title} {Charge-transfer ferromagnetism in oxide
  nanoparticles},}\ }%
  \bibfield{journal}{%
  \bibinfo {journal} {J. Phys. D: Appl. Phys.}\ }%
  \textbf{\bibinfo {volume} {41}},\ \bibinfo {pages} {134012}%
  \bibAnnoteFile{NoStop}{Coey:2008_JPD}%
\bibitem[{\citenamefont{Csontos}\ \emph{et~al.}(2005)\citenamefont{Csontos},
  \citenamefont{{Mih{\'a}ly}}, \citenamefont{{Jank{\'o}}},
  \citenamefont{Wojtowicz}, \citenamefont{Liu},\ and\
  \citenamefont{Furdyna}}]{Csontos:2005_NM}%
  \BibitemOpen
  \bibfield{author}{%
  \bibinfo {author} {\bibnamefont{Csontos}, \bibfnamefont{M.}}, \bibinfo
  {author} {\bibfnamefont{G.}~\bibnamefont{{Mih{\'a}ly}}}, \bibinfo {author}
  {\bibfnamefont{B.}~\bibnamefont{{Jank{\'o}}}}, \bibinfo {author}
  {\bibfnamefont{T.}~\bibnamefont{Wojtowicz}}, \bibinfo {author}
  {\bibfnamefont{X.}~\bibnamefont{Liu}},\ and\ \bibinfo {author}
  {\bibfnamefont{J.~K.}\ \bibnamefont{Furdyna}}}%
  , \bibinfo {year} {2005},\ \bibfield{title}{%
  \enquote{\bibinfo {title} {Pressure-induced ferromagnetism in {(In,Mn)Sb}
  dilute magnetic semiconductor},}\ }%
  \bibfield{journal}{%
  \bibinfo {journal} {Nat. Mater.}\ }%
  \textbf{\bibinfo {volume} {4}},\ \bibinfo {pages} {447}%
  \bibAnnoteFile{NoStop}{Csontos:2005_NM}%
\bibitem[{\citenamefont{Cubukcu}\ \emph{et~al.}(2010)\citenamefont{Cubukcu},
  \citenamefont{von Bardeleben}, \citenamefont{Khazen}, \citenamefont{Cantin},
  \citenamefont{Mauguin}, \citenamefont{Largeau},\ and\
  \citenamefont{{Lema\^itre}}}]{Cubukcu:2010_PRB}%
  \BibitemOpen
  \bibfield{author}{%
  \bibinfo {author} {\bibnamefont{Cubukcu}, \bibfnamefont{M.}}, \bibinfo
  {author} {\bibfnamefont{H.~J.}\ \bibnamefont{von Bardeleben}}, \bibinfo
  {author} {\bibfnamefont{Kh.}\ \bibnamefont{Khazen}}, \bibinfo {author}
  {\bibfnamefont{J.~L.}\ \bibnamefont{Cantin}}, \bibinfo {author}
  {\bibfnamefont{O.}~\bibnamefont{Mauguin}}, \bibinfo {author}
  {\bibfnamefont{L.}~\bibnamefont{Largeau}},\ and\ \bibinfo {author}
  {\bibfnamefont{A.}~\bibnamefont{{Lema\^itre}}}}%
  , \bibinfo {year} {2010},\ \bibfield{title}{%
  \enquote{\bibinfo {title} {Adjustable anisotropy in ferromagnetic
  {(Ga,Mn)(As,P)} layered alloys},}\ }%
  \bibfield{journal}{%
  \bibinfo {journal} {Phys. Rev. B}\ }%
  \textbf{\bibinfo {volume} {81}},\ \bibinfo {pages} {041202}%
  \bibAnnoteFile{NoStop}{Cubukcu:2010_PRB}%
\bibitem[{\citenamefont{Curiale}\ \emph{et~al.}(2012)\citenamefont{Curiale},
  \citenamefont{Lema\^itre}, \citenamefont{Ulysse}, \citenamefont{Faini},\ and\
  \citenamefont{Jeudy}}]{Curiale:2012_PRL}%
  \BibitemOpen
  \bibfield{author}{%
  \bibinfo {author} {\bibnamefont{Curiale}, \bibfnamefont{J.}}, \bibinfo
  {author} {\bibfnamefont{A.}~\bibnamefont{Lema\^itre}}, \bibinfo {author}
  {\bibfnamefont{C.}~\bibnamefont{Ulysse}}, \bibinfo {author}
  {\bibfnamefont{G.}~\bibnamefont{Faini}},\ and\ \bibinfo {author}
  {\bibfnamefont{V.}~\bibnamefont{Jeudy}}}%
  , \bibinfo {year} {2012},\ \bibfield{title}{%
  \enquote{\bibinfo {title} {Spin drift velocity, polarization, and
  current-driven domain-wall motion in {(Ga,Mn)(As,P)}},}\ }%
  \bibfield{journal}{%
  \bibinfo {journal} {Phys. Rev. Lett.}\ }%
  \textbf{\bibinfo {volume} {108}},\ \bibinfo {pages} {076604}%
  \bibAnnoteFile{NoStop}{Curiale:2012_PRL}%
\bibitem[{\citenamefont{Cywi\'nski}\ and\
  \citenamefont{Sham}(2007)}]{Cywinski:2007_PRB}%
  \BibitemOpen
  \bibfield{author}{%
  \bibinfo {author} {\bibnamefont{Cywi\'nski}, \bibfnamefont{\L.}},\ and\
  \bibinfo {author} {\bibfnamefont{L.~J.}\ \bibnamefont{Sham}}}%
  , \bibinfo {year} {2007},\ \bibfield{title}{%
  \enquote{\bibinfo {title} {Ultrafast demagnetization in the sp-d model: a
  theoretical study},}\ }%
  \bibfield{journal}{%
  \bibinfo {journal} {Phys. Rev. B}\ }%
  \textbf{\bibinfo {volume} {76}},\ \bibinfo {pages} {045205}%
  \bibAnnoteFile{NoStop}{Cywinski:2007_PRB}%
\bibitem[{\citenamefont{Da~Silva}\ \emph{et~al.}(2008)\citenamefont{Da~Silva},
  \citenamefont{Dalpian},\ and\ \citenamefont{Wei}}]{Da_Silva:2008_NJP}%
  \BibitemOpen
  \bibfield{author}{%
  \bibinfo {author} {\bibnamefont{Da~Silva}, \bibfnamefont{Juarez L.~F.}},
  \bibinfo {author} {\bibfnamefont{Gustavo~M.}\ \bibnamefont{Dalpian}},\ and\
  \bibinfo {author} {\bibfnamefont{Su-Huai}\ \bibnamefont{Wei}}}%
  , \bibinfo {year} {2008},\ \bibfield{title}{%
  \enquote{\bibinfo {title} {Carrier-induced enhancement and suppression of
  ferromagnetism in {Zn$_{1-x}$Cr$_{x}$Te} and {Ga$_{1-x}$Cr$_{x}$As}: origin
  of the spinodal decomposition},}\ }%
  \bibfield{journal}{%
  \bibinfo {journal} {New J. Phys.}\ }%
  \textbf{\bibinfo {volume} {10}},\ \bibinfo {pages} {113007}%
  \bibAnnoteFile{NoStop}{Da_Silva:2008_NJP}%
\bibitem[{\citenamefont{{d'Acapito}}\
  \emph{et~al.}(2006)\citenamefont{{d'Acapito}}, \citenamefont{Smolentsev},
  \citenamefont{Boscherini}, \citenamefont{Piccin}, \citenamefont{Bais},
  \citenamefont{Rubini}, \citenamefont{Martelli},\ and\
  \citenamefont{Franciosi}}]{DAcapito:2006_PRB}%
  \BibitemOpen
  \bibfield{author}{%
  \bibinfo {author} {\bibnamefont{{d'Acapito}}, \bibfnamefont{F.}}, \bibinfo
  {author} {\bibfnamefont{G.}~\bibnamefont{Smolentsev}}, \bibinfo {author}
  {\bibfnamefont{F.}~\bibnamefont{Boscherini}}, \bibinfo {author}
  {\bibfnamefont{M.}~\bibnamefont{Piccin}}, \bibinfo {author}
  {\bibfnamefont{G.}~\bibnamefont{Bais}}, \bibinfo {author}
  {\bibfnamefont{S.}~\bibnamefont{Rubini}}, \bibinfo {author}
  {\bibfnamefont{F.}~\bibnamefont{Martelli}},\ and\ \bibinfo {author}
  {\bibfnamefont{A.}~\bibnamefont{Franciosi}}}%
  , \bibinfo {year} {2006},\ \bibfield{title}{%
  \enquote{\bibinfo {title} {Site of {Mn} in {Mn $\delta$-doped GaAs}: {X}-ray
  absorption spectroscopy},}\ }%
  \bibfield{journal}{%
  \bibinfo {journal} {Phys. Rev. B}\ }%
  \textbf{\bibinfo {volume} {73}},\ \bibinfo {pages} {035314}%
  \bibAnnoteFile{NoStop}{DAcapito:2006_PRB}%
\bibitem[{\citenamefont{Dagotto}\ \emph{et~al.}(2001)\citenamefont{Dagotto},
  \citenamefont{Hotta},\ and\ \citenamefont{Moreo}}]{Dagotto:2001_PR}%
  \BibitemOpen
  \bibfield{author}{%
  \bibinfo {author} {\bibnamefont{Dagotto}, \bibfnamefont{E.}}, \bibinfo
  {author} {\bibfnamefont{T.}~\bibnamefont{Hotta}},\ and\ \bibinfo {author}
  {\bibfnamefont{A.}~\bibnamefont{Moreo}}}%
  , \bibinfo {year} {2001},\ \bibfield{title}{%
  \enquote{\bibinfo {title} {Colossal magnetoresistant materials: the key role
  of phase separation},}\ }%
  \bibfield{journal}{%
  \bibinfo {journal} {Phys. Rep.}\ }%
  \textbf{\bibinfo {volume} {344}},\ \bibinfo {pages} {1}%
  \bibAnnoteFile{NoStop}{Dagotto:2001_PR}%
\bibitem[{\citenamefont{Deng}\ \emph{et~al.}(2011)\citenamefont{Deng},
  \citenamefont{Jin}, \citenamefont{Liu}, \citenamefont{Wang},
  \citenamefont{Zhu}, \citenamefont{Feng}, \citenamefont{Chen},
  \citenamefont{Yu}, \citenamefont{Arguello}, \citenamefont{Goko},
  \citenamefont{Ning}, \citenamefont{Zhang}, \citenamefont{Wang},
  \citenamefont{Aczel}, \citenamefont{Munsie}, \citenamefont{Williams},
  \citenamefont{Luke}, \citenamefont{Kakeshita}, \citenamefont{Uchida},
  \citenamefont{Higemoto}, \citenamefont{Ito}, \citenamefont{Gu},
  \citenamefont{Maekawa}, \citenamefont{Morris},\ and\
  \citenamefont{Uemura}}]{Deng:2011_NC}%
  \BibitemOpen
  \bibfield{author}{%
  \bibinfo {author} {\bibnamefont{Deng}, \bibfnamefont{Z.}}, \bibinfo {author}
  {\bibfnamefont{C.~Q.}\ \bibnamefont{Jin}}, \bibinfo {author}
  {\bibfnamefont{Q.~Q.}\ \bibnamefont{Liu}}, \bibinfo {author}
  {\bibfnamefont{X.~C.}\ \bibnamefont{Wang}}, \bibinfo {author}
  {\bibfnamefont{J.~L.}\ \bibnamefont{Zhu}}, \bibinfo {author}
  {\bibfnamefont{S.~M.}\ \bibnamefont{Feng}}, \bibinfo {author}
  {\bibfnamefont{L.~C.}\ \bibnamefont{Chen}}, \bibinfo {author}
  {\bibfnamefont{R.~C.}\ \bibnamefont{Yu}}, \bibinfo {author}
  {\bibfnamefont{C.}~\bibnamefont{Arguello}}, \bibinfo {author}
  {\bibfnamefont{T.}~\bibnamefont{Goko}}, \bibinfo {author}
  {\bibfnamefont{Fanlong}\ \bibnamefont{Ning}}, \bibinfo {author}
  {\bibfnamefont{Jinsong}\ \bibnamefont{Zhang}}, \bibinfo {author}
  {\bibfnamefont{Yayu}\ \bibnamefont{Wang}}, \bibinfo {author}
  {\bibfnamefont{A.~A.}\ \bibnamefont{Aczel}}, \bibinfo {author}
  {\bibfnamefont{T.}~\bibnamefont{Munsie}}, \bibinfo {author}
  {\bibfnamefont{T.~J.}\ \bibnamefont{Williams}}, \bibinfo {author}
  {\bibfnamefont{G.~M.}\ \bibnamefont{Luke}}, \bibinfo {author}
  {\bibfnamefont{T.}~\bibnamefont{Kakeshita}}, \bibinfo {author}
  {\bibfnamefont{S.}~\bibnamefont{Uchida}}, \bibinfo {author}
  {\bibfnamefont{W.}~\bibnamefont{Higemoto}}, \bibinfo {author}
  {\bibfnamefont{T.~U.}\ \bibnamefont{Ito}}, \bibinfo {author}
  {\bibfnamefont{Bo}~\bibnamefont{Gu}}, \bibinfo {author}
  {\bibfnamefont{S.}~\bibnamefont{Maekawa}}, \bibinfo {author}
  {\bibfnamefont{G.~D.}\ \bibnamefont{Morris}},\ and\ \bibinfo {author}
  {\bibfnamefont{Y.~J.}\ \bibnamefont{Uemura}}}%
  , \bibinfo {year} {2011},\ \bibfield{title}{%
  \enquote{\bibinfo {title} {{Li(Zn,Mn)As} as a new generation ferromagnet
  based on a {I-II-V} semiconductor},}\ }%
  \bibfield{journal}{%
  \bibinfo {journal} {Nat. Commun.}\ }%
  \textbf{\bibinfo {volume} {2}},\ \bibinfo {pages} {422}%
  \bibAnnoteFile{NoStop}{Deng:2011_NC}%
\bibitem[{\citenamefont{Devillers}\
  \emph{et~al.}(2012)\citenamefont{Devillers}, \citenamefont{Rovezzi},
  \citenamefont{Gonzalez~Szwacki}, \citenamefont{Dobkowska},
  \citenamefont{Stefanowicz}, \citenamefont{Sztenkiel}, \citenamefont{Grois},
  \citenamefont{Suffczy{\'n}ski}, \citenamefont{Navarro-Quezada},
  \citenamefont{Faina}, \citenamefont{Li}, \citenamefont{Glatzel},
  \citenamefont{d'Acapito}, \citenamefont{Jakie{\l}a}, \citenamefont{Sawicki},
  \citenamefont{Majewski}, \citenamefont{Dietl},\ and\
  \citenamefont{Bonanni}}]{Devillers:2012_SR}%
  \BibitemOpen
  \bibfield{author}{%
  \bibinfo {author} {\bibnamefont{Devillers}, \bibfnamefont{T.}}, \bibinfo
  {author} {\bibfnamefont{M.}~\bibnamefont{Rovezzi}}, \bibinfo {author}
  {\bibfnamefont{N.}~\bibnamefont{Gonzalez~Szwacki}}, \bibinfo {author}
  {\bibfnamefont{S.}~\bibnamefont{Dobkowska}}, \bibinfo {author}
  {\bibfnamefont{W.}~\bibnamefont{Stefanowicz}}, \bibinfo {author}
  {\bibfnamefont{D.}~\bibnamefont{Sztenkiel}}, \bibinfo {author}
  {\bibfnamefont{A.}~\bibnamefont{Grois}}, \bibinfo {author}
  {\bibfnamefont{J.}~\bibnamefont{Suffczy{\'n}ski}}, \bibinfo {author}
  {\bibfnamefont{A.}~\bibnamefont{Navarro-Quezada}}, \bibinfo {author}
  {\bibfnamefont{B.}~\bibnamefont{Faina}}, \bibinfo {author}
  {\bibfnamefont{T.}~\bibnamefont{Li}}, \bibinfo {author}
  {\bibfnamefont{P.}~\bibnamefont{Glatzel}}, \bibinfo {author}
  {\bibfnamefont{F.}~\bibnamefont{d'Acapito}}, \bibinfo {author}
  {\bibfnamefont{R.}~\bibnamefont{Jakie{\l}a}}, \bibinfo {author}
  {\bibfnamefont{M.}~\bibnamefont{Sawicki}}, \bibinfo {author}
  {\bibfnamefont{J.~A.}\ \bibnamefont{Majewski}}, \bibinfo {author}
  {\bibfnamefont{T.}~\bibnamefont{Dietl}},\ and\ \bibinfo {author}
  {\bibfnamefont{A.}~\bibnamefont{Bonanni}}}%
  , \bibinfo {year} {2012},\ \bibfield{title}{%
  \enquote{\bibinfo {title} {Manipulating {Mn--Mg$_{k}$} cation complexes to
  control the charge- and spin-state of {Mn} in {GaN}},}\ }%
  \bibfield{journal}{%
  \bibinfo {journal} {Sci. Rep.}\ }%
  \textbf{\bibinfo {volume} {2}},\ \bibinfo {pages} {722}%
  \bibAnnoteFile{NoStop}{Devillers:2012_SR}%
\bibitem[{\citenamefont{Dhar}\ \emph{et~al.}(2003)\citenamefont{Dhar},
  \citenamefont{Brandt}, \citenamefont{Trampert}, \citenamefont{Friedland},
  \citenamefont{Sun},\ and\ \citenamefont{Ploog}}]{Dhar:2003_PRB}%
  \BibitemOpen
  \bibfield{author}{%
  \bibinfo {author} {\bibnamefont{Dhar}, \bibfnamefont{S.}}, \bibinfo {author}
  {\bibfnamefont{O.}~\bibnamefont{Brandt}}, \bibinfo {author}
  {\bibfnamefont{A.}~\bibnamefont{Trampert}}, \bibinfo {author}
  {\bibfnamefont{K.~J.}\ \bibnamefont{Friedland}}, \bibinfo {author}
  {\bibfnamefont{Y.~J.}\ \bibnamefont{Sun}},\ and\ \bibinfo {author}
  {\bibfnamefont{K.~H.}\ \bibnamefont{Ploog}}}%
  , \bibinfo {year} {2003},\ \bibfield{title}{%
  \enquote{\bibinfo {title} {Observation of spin-glass behavior in homogeneous
  {(Ga,Mn)N} layers grown by reactive molecular-beam epitaxy},}\ }%
  \bibfield{journal}{%
  \bibinfo {journal} {Phys. Rev. B}\ }%
  \textbf{\bibinfo {volume} {67}},\ \bibinfo {pages} {165205}%
  \bibAnnoteFile{NoStop}{Dhar:2003_PRB}%
\bibitem[{\citenamefont{{Di Marco}}\ \emph{et~al.}(2013)\citenamefont{{Di
  Marco}}, \citenamefont{Thunstr\"om}, \citenamefont{Katsnelson},
  \citenamefont{Sadowski}, \citenamefont{Karlsson}, \citenamefont{Leb\`egue},
  \citenamefont{Ka{\'n}ski}, ,\ and\
  \citenamefont{Eriksson}}]{DiMarco:2013_NC}%
  \BibitemOpen
  \bibfield{author}{%
  \bibinfo {author} {\bibnamefont{{Di Marco}}, \bibfnamefont{I.}}, \bibinfo
  {author} {\bibfnamefont{P.}~\bibnamefont{Thunstr\"om}}, \bibinfo {author}
  {\bibfnamefont{M.~I.}\ \bibnamefont{Katsnelson}}, \bibinfo {author}
  {\bibfnamefont{J.}~\bibnamefont{Sadowski}}, \bibinfo {author}
  {\bibfnamefont{K.}~\bibnamefont{Karlsson}}, \bibinfo {author}
  {\bibfnamefont{S.}~\bibnamefont{Leb\`egue}}, \bibinfo {author}
  {\bibfnamefont{J.}~\bibnamefont{Ka{\'n}ski}}, ,\ and\ \bibinfo {author}
  {\bibfnamefont{O.}~\bibnamefont{Eriksson}}}%
  , \bibinfo {year} {2013},\ \bibfield{title}{%
  \enquote{\bibinfo {title} {Electron correlations in mn$_x$ga$_{1–x}$as as
  seen by resonant electron spectroscopy and dynamical mean field theory},}\ }%
  \bibfield{journal}{%
  \bibinfo {journal} {Nat. Commun.}\ }%
  \textbf{\bibinfo {volume} {4}},\ \bibinfo {pages} {2645}%
  \bibAnnoteFile{NoStop}{DiMarco:2013_NC}%
\bibitem[{\citenamefont{Dietl}(1981)}]{Dietl:1981_B}%
  \BibitemOpen
  \bibfield{author}{%
  \bibinfo {author} {\bibnamefont{Dietl}, \bibfnamefont{T.}}}%
  , \bibinfo {year} {1981},\ \enquote{\bibinfo {title} {{Semimagnetic
  Semiconductors in High Magnetic Fields}},}\ in\ \emph{\bibinfo {booktitle}
  {{Physics in High Magnetic Fields}}},\ \bibinfo {editor} {edited by\ \bibinfo
  {editor} {\bibfnamefont{S.}~\bibnamefont{Chikazumi}}\ and\ \bibinfo {editor}
  {\bibfnamefont{N.}~\bibnamefont{Miura}}}\ (\bibinfo {publisher} {Springer,
  Berlin})\ p.\ \bibinfo {pages} {344}%
  \bibAnnoteFile{NoStop}{Dietl:1981_B}%
\bibitem[{\citenamefont{Dietl}(1983)}]{Dietl:1983_JMMM}%
  \BibitemOpen
  \bibfield{author}{%
  \bibinfo {author} {\bibnamefont{Dietl}, \bibfnamefont{T.}}}%
  , \bibinfo {year} {1983},\ \bibfield{title}{%
  \enquote{\bibinfo {title} {Optical properties of donor electrons in
  semimagnetic semiconductors},}\ }%
  \bibfield{journal}{%
  \bibinfo {journal} {J. Magn. Magn. Mater.}\ }%
  \textbf{\bibinfo {volume} {38}},\ \bibinfo {pages} {34}%
  \bibAnnoteFile{NoStop}{Dietl:1983_JMMM}%
\bibitem[{\citenamefont{Dietl}(1994)}]{Dietl:1994_B}%
  \BibitemOpen
  \bibfield{author}{%
  \bibinfo {author} {\bibnamefont{Dietl}, \bibfnamefont{T.}}}%
  , \bibinfo {year} {1994},\ \enquote{\bibinfo {title} {{(Diluted) Magnetic
  Semiconductors}},}\ in\ \emph{\bibinfo {booktitle} {Handbook of
  Semiconductors}},\ Vol.~\bibinfo {volume} {3B},\ \bibinfo {editor} {edited
  by\ \bibinfo {editor} {\bibfnamefont{S.}~\bibnamefont{Mahajan}}}\ (\bibinfo
  {publisher} {North Holland, Amsterdam})\ p.\ \bibinfo {pages} {1251}%
  \bibAnnoteFile{NoStop}{Dietl:1994_B}%
\bibitem[{\citenamefont{Dietl}(2002)}]{Dietl:2002_SST}%
  \BibitemOpen
  \bibfield{author}{%
  \bibinfo {author} {\bibnamefont{Dietl}, \bibfnamefont{T.}}}%
  , \bibinfo {year} {2002},\ \bibfield{title}{%
  \enquote{\bibinfo {title} {Ferromagnetic semiconductors},}\ }%
  \bibfield{journal}{%
  \bibinfo {journal} {Semicond. Sci. Technol.}\ }%
  \textbf{\bibinfo {volume} {17}},\ \bibinfo {pages} {377}%
  \bibAnnoteFile{NoStop}{Dietl:2002_SST}%
\bibitem[{\citenamefont{Dietl}(2007)}]{Dietl:2007_JPCM}%
  \BibitemOpen
  \bibfield{author}{%
  \bibinfo {author} {\bibnamefont{Dietl}, \bibfnamefont{T.}}}%
  , \bibinfo {year} {2007},\ \bibfield{title}{%
  \enquote{\bibinfo {title} {Origin of ferromagnetic response in diluted
  magnetic semiconductors and oxides},}\ }%
  \bibfield{journal}{%
  \bibinfo {journal} {J. Phys. Condens. Matter}\ }%
  \textbf{\bibinfo {volume} {19}},\ \bibinfo {pages} {165204}%
  \bibAnnoteFile{NoStop}{Dietl:2007_JPCM}%
\bibitem[{\citenamefont{Dietl}(2008{\natexlab{a}})}]{Dietl:2008_PRB}%
  \BibitemOpen
  \bibfield{author}{%
  \bibinfo {author} {\bibnamefont{Dietl}, \bibfnamefont{T.}}}%
  , \bibinfo {year} {2008}{\natexlab{a}},\ \bibfield{title}{%
  \enquote{\bibinfo {title} {Hole states in wide band-gap diluted magnetic
  semiconductors and oxides},}\ }%
  \bibfield{journal}{%
  \bibinfo {journal} {Phys. Rev. B}\ }%
  \textbf{\bibinfo {volume} {77}},\ \bibinfo {pages} {085208}%
  \bibAnnoteFile{NoStop}{Dietl:2008_PRB}%
\bibitem[{\citenamefont{Dietl}(2008{\natexlab{b}})}]{Dietl:2008_JPSJ}%
  \BibitemOpen
  \bibfield{author}{%
  \bibinfo {author} {\bibnamefont{Dietl}, \bibfnamefont{T.}}}%
  , \bibinfo {year} {2008}{\natexlab{b}},\ \bibfield{title}{%
  \enquote{\bibinfo {title} {Interplay between carrier localization and
  magnetism in diluted magnetic and ferromagnetic semiconductors},}\ }%
  \bibfield{journal}{%
  \bibinfo {journal} {J. Phys. Soc. Jpn.}\ }%
  \textbf{\bibinfo {volume} {77}},\ \bibinfo {pages} {031005}%
  \bibAnnoteFile{NoStop}{Dietl:2008_JPSJ}%
\bibitem[{\citenamefont{Dietl}(2010)}]{Dietl:2010_NM}%
  \BibitemOpen
  \bibfield{author}{%
  \bibinfo {author} {\bibnamefont{Dietl}, \bibfnamefont{T.}}}%
  , \bibinfo {year} {2010},\ \bibfield{title}{%
  \enquote{\bibinfo {title} {A ten-year perspective on dilute magnetic
  semiconductors and oxides},}\ }%
  \bibfield{journal}{%
  \bibinfo {journal} {Nat. Mater.}\ }%
  \textbf{\bibinfo {volume} {9}},\ \bibinfo {pages} {965}%
  \bibAnnoteFile{NoStop}{Dietl:2010_NM}%
\bibitem[{\citenamefont{Dietl}\ \emph{et~al.}(1999)\citenamefont{Dietl},
  \citenamefont{Cibert}, \citenamefont{Ferrand},\ and\ \citenamefont{{Merle
  d'Aubign{\'e}}}}]{Dietl:1999_MSEB}%
  \BibitemOpen
  \bibfield{author}{%
  \bibinfo {author} {\bibnamefont{Dietl}, \bibfnamefont{T.}}, \bibinfo {author}
  {\bibfnamefont{J.}~\bibnamefont{Cibert}}, \bibinfo {author}
  {\bibfnamefont{D.}~\bibnamefont{Ferrand}},\ and\ \bibinfo {author}
  {\bibfnamefont{Y.}~\bibnamefont{{Merle d'Aubign{\'e}}}}}%
  , \bibinfo {year} {1999},\ \bibfield{title}{%
  \enquote{\bibinfo {title} {Carrier-mediated ferromagnetic interactions in
  structures of magnetic semiconductors},}\ }%
  \bibfield{journal}{%
  \bibinfo {journal} {Mater. Sci. Eng. B}\ }%
  \textbf{\bibinfo {volume} {63}},\ \bibinfo {pages} {103}%
  \bibAnnoteFile{NoStop}{Dietl:1999_MSEB}%
\bibitem[{\citenamefont{Dietl}\ \emph{et~al.}(1997)\citenamefont{Dietl},
  \citenamefont{Haury},\ and\ \citenamefont{d'Aubigne}}]{Dietl:1997_PRB}%
  \BibitemOpen
  \bibfield{author}{%
  \bibinfo {author} {\bibnamefont{Dietl}, \bibfnamefont{T.}}, \bibinfo {author}
  {\bibfnamefont{A.}~\bibnamefont{Haury}},\ and\ \bibinfo {author}
  {\bibfnamefont{Y.~Merle}\ \bibnamefont{d'Aubigne}}}%
  , \bibinfo {year} {1997},\ \bibfield{title}{%
  \enquote{\bibinfo {title} {Free carrier-induced ferromagnetism in structures
  of diluted magnetic semiconductors},}\ }%
  \bibfield{journal}{%
  \bibinfo {journal} {Phys. Rev. B}\ }%
  \textbf{\bibinfo {volume} {55}},\ \bibinfo {pages} {R3347}%
  \bibAnnoteFile{NoStop}{Dietl:1997_PRB}%
\bibitem[{\citenamefont{Dietl}\
  \emph{et~al.}(2001{\natexlab{a}})\citenamefont{Dietl},
  \citenamefont{{K{\"o}nig}},\ and\
  \citenamefont{MacDonald}}]{Dietl:2001_PRBb}%
  \BibitemOpen
  \bibfield{author}{%
  \bibinfo {author} {\bibnamefont{Dietl}, \bibfnamefont{T.}}, \bibinfo {author}
  {\bibfnamefont{J.}~\bibnamefont{{K{\"o}nig}}},\ and\ \bibinfo {author}
  {\bibfnamefont{A.~H.}\ \bibnamefont{MacDonald}}}%
  , \bibinfo {year} {2001}{\natexlab{a}},\ \bibfield{title}{%
  \enquote{\bibinfo {title} {Magnetic domains in {III--V} magnetic
  semiconductors},}\ }%
  \bibfield{journal}{%
  \bibinfo {journal} {Phys. Rev. B}\ }%
  \textbf{\bibinfo {volume} {64}},\ \bibinfo {pages} {241201}%
  \bibAnnoteFile{NoStop}{Dietl:2001_PRBb}%
\bibitem[{\citenamefont{Dietl}\ \emph{et~al.}(2002)\citenamefont{Dietl},
  \citenamefont{Matsukura},\ and\ \citenamefont{Ohno}}]{Dietl:2002_PRB}%
  \BibitemOpen
  \bibfield{author}{%
  \bibinfo {author} {\bibnamefont{Dietl}, \bibfnamefont{T.}}, \bibinfo {author}
  {\bibfnamefont{F.}~\bibnamefont{Matsukura}},\ and\ \bibinfo {author}
  {\bibfnamefont{H.}~\bibnamefont{Ohno}}}%
  , \bibinfo {year} {2002},\ \bibfield{title}{%
  \enquote{\bibinfo {title} {Ferromagnetism of magnetic semiconductors:
  {Zhang-Rice} limit},}\ }%
  \bibfield{journal}{%
  \bibinfo {journal} {Phys. Rev. B}\ }%
  \textbf{\bibinfo {volume} {66}},\ \bibinfo {pages} {033203}%
  \bibAnnoteFile{NoStop}{Dietl:2002_PRB}%
\bibitem[{\citenamefont{Dietl}\
  \emph{et~al.}(2001{\natexlab{b}})\citenamefont{Dietl}, \citenamefont{Ohno},\
  and\ \citenamefont{Matsukura}}]{Dietl:2001_PRB}%
  \BibitemOpen
  \bibfield{author}{%
  \bibinfo {author} {\bibnamefont{Dietl}, \bibfnamefont{T.}}, \bibinfo {author}
  {\bibfnamefont{H.}~\bibnamefont{Ohno}},\ and\ \bibinfo {author}
  {\bibfnamefont{F.}~\bibnamefont{Matsukura}}}%
  , \bibinfo {year} {2001}{\natexlab{b}},\ \bibfield{title}{%
  \enquote{\bibinfo {title} {Hole-mediated ferromagnetism in tetrahedrally
  coordinated semiconductors},}\ }%
  \bibfield{journal}{%
  \bibinfo {journal} {Phys. Rev. B}\ }%
  \textbf{\bibinfo {volume} {63}},\ \bibinfo {pages} {195205}%
  \bibAnnoteFile{NoStop}{Dietl:2001_PRB}%
\bibitem[{\citenamefont{Dietl}\ \emph{et~al.}(2000)\citenamefont{Dietl},
  \citenamefont{Ohno}, \citenamefont{Matsukura}, \citenamefont{Cibert},\ and\
  \citenamefont{Ferrand}}]{Dietl:2000_S}%
  \BibitemOpen
  \bibfield{author}{%
  \bibinfo {author} {\bibnamefont{Dietl}, \bibfnamefont{T.}}, \bibinfo {author}
  {\bibfnamefont{H.}~\bibnamefont{Ohno}}, \bibinfo {author}
  {\bibfnamefont{F.}~\bibnamefont{Matsukura}}, \bibinfo {author}
  {\bibfnamefont{J.}~\bibnamefont{Cibert}},\ and\ \bibinfo {author}
  {\bibfnamefont{D.}~\bibnamefont{Ferrand}}}%
  , \bibinfo {year} {2000},\ \bibfield{title}{%
  \enquote{\bibinfo {title} {{Zener} model description of ferromagnetism in
  zinc-blende magnetic semiconductors},}\ }%
  \bibfield{journal}{%
  \bibinfo {journal} {Science}\ }%
  \textbf{\bibinfo {volume} {287}},\ \bibinfo {pages} {1019}%
  \bibAnnoteFile{NoStop}{Dietl:2000_S}%
\bibitem[{\citenamefont{Dietl}\ \emph{et~al.}(1994)\citenamefont{Dietl},
  \citenamefont{{\'S}liwa}, \citenamefont{Bauer},\ and\
  \citenamefont{Pascher}}]{Dietl:1994_PRB}%
  \BibitemOpen
  \bibfield{author}{%
  \bibinfo {author} {\bibnamefont{Dietl}, \bibfnamefont{T.}}, \bibinfo {author}
  {\bibfnamefont{C.}~\bibnamefont{{\'S}liwa}}, \bibinfo {author}
  {\bibfnamefont{G.}~\bibnamefont{Bauer}},\ and\ \bibinfo {author}
  {\bibfnamefont{H.}~\bibnamefont{Pascher}}}%
  , \bibinfo {year} {1994},\ \bibfield{title}{%
  \enquote{\bibinfo {title} {Mechanisms of exchange interactions between
  carriers and {Mn} or {Eu} spins in lead chalcogenides},}\ }%
  \bibfield{journal}{%
  \bibinfo {journal} {Phys. Rev. B}\ }%
  \textbf{\bibinfo {volume} {49}},\ \bibinfo {pages} {2230}%
  \bibAnnoteFile{NoStop}{Dietl:1994_PRB}%
\bibitem[{\citenamefont{Dietl}\ and\
  \citenamefont{Spalek}(1983)}]{Dietl:1983_PRB}%
  \BibitemOpen
  \bibfield{author}{%
  \bibinfo {author} {\bibnamefont{Dietl}, \bibfnamefont{T.}},\ and\ \bibinfo
  {author} {\bibfnamefont{J.}~\bibnamefont{Spalek}}}%
  , \bibinfo {year} {1983},\ \bibfield{title}{%
  \enquote{\bibinfo {title} {Effect of thermodynamic fluctuations of
  magnetization on the bound magnetic polaron in dilute magnetic
  semiconductors},}\ }%
  \bibfield{journal}{%
  \bibinfo {journal} {Phys. Rev. B}\ }%
  \textbf{\bibinfo {volume} {28}},\ \bibinfo {pages} {1548}%
  \bibAnnoteFile{NoStop}{Dietl:1983_PRB}%
\bibitem[{\citenamefont{Dietl}\ and\
  \citenamefont{Sztenkiel}(2011)}]{Dietl:2011_arXiv}%
  \BibitemOpen
  \bibfield{author}{%
  \bibinfo {author} {\bibnamefont{Dietl}, \bibfnamefont{T.}},\ and\ \bibinfo
  {author} {\bibfnamefont{D.}~\bibnamefont{Sztenkiel}}}%
  , \bibinfo {year} {2011},\ \bibfield{title}{%
  \enquote{\bibinfo {title} {Reconciling results of tunnelling experiments on
  {(Ga,Mn)As}},}\ }%
  \bibinfo {journal} {arXiv:1102.3267}%
  \bibAnnoteFile{NoStop}{Dietl:2011_arXiv}%
\bibitem[{\citenamefont{Dourlat}\ \emph{et~al.}(2008)\citenamefont{Dourlat},
  \citenamefont{Jeudy}, \citenamefont{Lema\^{i}tre},\ and\
  \citenamefont{Gourdon}}]{Dourlat:2008_PRB}%
  \BibitemOpen
\bibfield{journal}{%
    }%
  \bibfield{author}{%
  \bibinfo {author} {\bibnamefont{Dourlat}, \bibfnamefont{A.}}, \bibinfo
  {author} {\bibfnamefont{V.}~\bibnamefont{Jeudy}}, \bibinfo {author}
  {\bibfnamefont{A.}~\bibnamefont{Lema\^{i}tre}},\ and\ \bibinfo {author}
  {\bibfnamefont{C.}~\bibnamefont{Gourdon}}}%
  , \bibinfo {year} {2008},\ \bibfield{title}{%
  \enquote{\bibinfo {title} {Field-driven domain-wall dynamics in {(Ga,Mn)As}
  films with perpendicular anisotropy},}\ }%
  \bibfield{journal}{%
  \bibinfo {journal} {Phys. Rev. B}\ }%
  \textbf{\bibinfo {volume} {78}},\ \bibinfo {pages} {161303}%
  \bibAnnoteFile{NoStop}{Dourlat:2008_PRB}%
\bibitem[{\citenamefont{Dreher}\ \emph{et~al.}(2010)\citenamefont{Dreher},
  \citenamefont{Donhauser}, \citenamefont{Daeubler}, \citenamefont{Glunk},
  \citenamefont{Rapp}, \citenamefont{Schoch}, \citenamefont{Sauer},\ and\
  \citenamefont{Limmer}}]{Dreher:2010_PRB}%
  \BibitemOpen
  \bibfield{author}{%
  \bibinfo {author} {\bibnamefont{Dreher}, \bibfnamefont{L.}}, \bibinfo
  {author} {\bibfnamefont{D.}~\bibnamefont{Donhauser}}, \bibinfo {author}
  {\bibfnamefont{J.}~\bibnamefont{Daeubler}}, \bibinfo {author}
  {\bibfnamefont{M.}~\bibnamefont{Glunk}}, \bibinfo {author}
  {\bibfnamefont{C.}~\bibnamefont{Rapp}}, \bibinfo {author}
  {\bibfnamefont{W.}~\bibnamefont{Schoch}}, \bibinfo {author}
  {\bibfnamefont{R.}~\bibnamefont{Sauer}},\ and\ \bibinfo {author}
  {\bibfnamefont{W.}~\bibnamefont{Limmer}}}%
  , \bibinfo {year} {2010},\ \bibfield{title}{%
  \enquote{\bibinfo {title} {Strain, magnetic anisotropy, and anisotropic
  magnetoresistance in {(Ga,Mn)As} on high-index substrates: {Application} to
  {(113)A} -oriented layers},}\ }%
  \bibfield{journal}{%
  \bibinfo {journal} {Phys. Rev. B}\ }%
  \textbf{\bibinfo {volume} {81}},\ \bibinfo {pages} {245202}%
  \bibAnnoteFile{NoStop}{Dreher:2010_PRB}%
\bibitem[{\citenamefont{Dunsiger}\ \emph{et~al.}(2010)\citenamefont{Dunsiger},
  \citenamefont{Carlo}, \citenamefont{Goko}, \citenamefont{Nieuwenhuys},
  \citenamefont{Prokscha}, \citenamefont{Suter}, \citenamefont{Morenzoni},
  \citenamefont{Chiba}, \citenamefont{Nishitani}, \citenamefont{Tanikawa},
  \citenamefont{Matsukura}, \citenamefont{Ohno}, \citenamefont{Ohe},
  \citenamefont{Maekawa}, ,\ and\ \citenamefont{Uemura}}]{Dunsiger:2010_NM}%
  \BibitemOpen
  \bibfield{author}{%
  \bibinfo {author} {\bibnamefont{Dunsiger}, \bibfnamefont{S.~R.}}, \bibinfo
  {author} {\bibfnamefont{J.~P.}\ \bibnamefont{Carlo}}, \bibinfo {author}
  {\bibfnamefont{T.}~\bibnamefont{Goko}}, \bibinfo {author}
  {\bibfnamefont{G.}~\bibnamefont{Nieuwenhuys}}, \bibinfo {author}
  {\bibfnamefont{T.}~\bibnamefont{Prokscha}}, \bibinfo {author}
  {\bibfnamefont{A.}~\bibnamefont{Suter}}, \bibinfo {author}
  {\bibfnamefont{E.}~\bibnamefont{Morenzoni}}, \bibinfo {author}
  {\bibfnamefont{D.}~\bibnamefont{Chiba}}, \bibinfo {author}
  {\bibfnamefont{Y.}~\bibnamefont{Nishitani}}, \bibinfo {author}
  {\bibfnamefont{T.}~\bibnamefont{Tanikawa}}, \bibinfo {author}
  {\bibfnamefont{F.}~\bibnamefont{Matsukura}}, \bibinfo {author}
  {\bibfnamefont{H.}~\bibnamefont{Ohno}}, \bibinfo {author}
  {\bibfnamefont{J.}~\bibnamefont{Ohe}}, \bibinfo {author}
  {\bibfnamefont{S.}~\bibnamefont{Maekawa}}, ,\ and\ \bibinfo {author}
  {\bibfnamefont{Y.~J.}\ \bibnamefont{Uemura}}}%
  , \bibinfo {year} {2010},\ \bibfield{title}{%
  \enquote{\bibinfo {title} {Spatially homogeneous ferromagnetism of {(Ga,
  Mn)As}},}\ }%
  \bibfield{journal}{%
  \bibinfo {journal} {Nat. Mater.}\ }%
  \textbf{\bibinfo {volume} {9}},\ \bibinfo {pages} {299}%
  \bibAnnoteFile{NoStop}{Dunsiger:2010_NM}%
\bibitem[{\citenamefont{Dziawa}\ \emph{et~al.}(2008)\citenamefont{Dziawa},
  \citenamefont{Knoff}, \citenamefont{Domukhovski}, \citenamefont{Domagala},
  \citenamefont{Jakiela}, \citenamefont{Lusakowska}, \citenamefont{Osinniy},
  \citenamefont{Swiatek}, \citenamefont{Taliashvili},\ and\
  \citenamefont{Story}}]{Dziawa:2008_B}%
  \BibitemOpen
  \bibfield{author}{%
  \bibinfo {author} {\bibnamefont{Dziawa}, \bibfnamefont{P.}}, \bibinfo
  {author} {\bibfnamefont{W.}~\bibnamefont{Knoff}}, \bibinfo {author}
  {\bibfnamefont{V.}~\bibnamefont{Domukhovski}}, \bibinfo {author}
  {\bibfnamefont{J.}~\bibnamefont{Domagala}}, \bibinfo {author}
  {\bibfnamefont{R.}~\bibnamefont{Jakiela}}, \bibinfo {author}
  {\bibfnamefont{E.}~\bibnamefont{Lusakowska}}, \bibinfo {author}
  {\bibfnamefont{V.}~\bibnamefont{Osinniy}}, \bibinfo {author}
  {\bibfnamefont{K.}~\bibnamefont{Swiatek}}, \bibinfo {author}
  {\bibfnamefont{B.}~\bibnamefont{Taliashvili}},\ and\ \bibinfo {author}
  {\bibfnamefont{T.}~\bibnamefont{Story}}}%
  , \bibinfo {year} {2008},\ \enquote{\bibinfo {title} {Magnetic and structural
  properties of ferromagnetic {GeMnTe} layers},}\ in\ \emph{\bibinfo
  {booktitle} {Narrow Gap Semiconductors 2007}},\ \bibinfo {series} {Springer
  Proceedings in Physics}, Vol.\ \bibinfo {volume} {119},\ \bibinfo {editor}
  {edited by\ \bibinfo {editor} {\bibfnamefont{B.}~\bibnamefont{Murdin}}\ and\
  \bibinfo {editor} {\bibfnamefont{S.}~\bibnamefont{Clowes}}}\ (\bibinfo
  {publisher} {Springer Netherlands})\ p.~\bibinfo {pages} {11}%
  \bibAnnoteFile{NoStop}{Dziawa:2008_B}%
\bibitem[{\citenamefont{Edmonds}\
  \emph{et~al.}(2004{\natexlab{a}})\citenamefont{Edmonds},
  \citenamefont{Bogus{\l}awski}, \citenamefont{Wang}, \citenamefont{Campion},
  \citenamefont{Farley}, \citenamefont{Gallagher}, \citenamefont{Foxon},
  \citenamefont{Sawicki}, \citenamefont{Dietl}, \citenamefont{Nardelli},\ and\
  \citenamefont{Bernholc}}]{Edmonds:2004_PRL}%
  \BibitemOpen
  \bibfield{author}{%
  \bibinfo {author} {\bibnamefont{Edmonds}, \bibfnamefont{K.~W.}}, \bibinfo
  {author} {\bibfnamefont{P.}~\bibnamefont{Bogus{\l}awski}}, \bibinfo {author}
  {\bibfnamefont{K.~Y.}\ \bibnamefont{Wang}}, \bibinfo {author}
  {\bibfnamefont{R.~P.}\ \bibnamefont{Campion}}, \bibinfo {author}
  {\bibfnamefont{N.~R.~S.}\ \bibnamefont{Farley}}, \bibinfo {author}
  {\bibfnamefont{B.~L.}\ \bibnamefont{Gallagher}}, \bibinfo {author}
  {\bibfnamefont{C.~T.}\ \bibnamefont{Foxon}}, \bibinfo {author}
  {\bibfnamefont{M.}~\bibnamefont{Sawicki}}, \bibinfo {author}
  {\bibfnamefont{T.}~\bibnamefont{Dietl}}, \bibinfo {author}
  {\bibfnamefont{M.~B.}\ \bibnamefont{Nardelli}},\ and\ \bibinfo {author}
  {\bibfnamefont{J.}~\bibnamefont{Bernholc}}}%
  , \bibinfo {year} {2004}{\natexlab{a}},\ \bibfield{title}{%
  \enquote{\bibinfo {title} {Mn interstitial diffusion in {(Ga,Mn)As}},}\ }%
  \bibfield{journal}{%
  \bibinfo {journal} {Phys. Rev. Lett.}\ }%
  \textbf{\bibinfo {volume} {92}},\ \bibinfo {pages} {037201}%
  \bibAnnoteFile{NoStop}{Edmonds:2004_PRL}%
\bibitem[{\citenamefont{Edmonds}\
  \emph{et~al.}(2004{\natexlab{b}})\citenamefont{Edmonds},
  \citenamefont{Farley}, \citenamefont{Campion}, \citenamefont{Foxon},
  \citenamefont{Gallagher}, \citenamefont{Johal}, \citenamefont{{van der
  {Laan}}}, \citenamefont{MacKenzie}, \citenamefont{Chapman},\ and\
  \citenamefont{Arenholz}}]{Edmonds:2004_APL}%
  \BibitemOpen
  \bibfield{author}{%
  \bibinfo {author} {\bibnamefont{Edmonds}, \bibfnamefont{K.~W.}}, \bibinfo
  {author} {\bibfnamefont{N.~R.~S.}\ \bibnamefont{Farley}}, \bibinfo {author}
  {\bibfnamefont{R.~P.}\ \bibnamefont{Campion}}, \bibinfo {author}
  {\bibfnamefont{C.~T.}\ \bibnamefont{Foxon}}, \bibinfo {author}
  {\bibfnamefont{B.~L.}\ \bibnamefont{Gallagher}}, \bibinfo {author}
  {\bibfnamefont{T.~K.}\ \bibnamefont{Johal}}, \bibinfo {author}
  {\bibfnamefont{G.}~\bibnamefont{{van der {Laan}}}}, \bibinfo {author}
  {\bibfnamefont{M.}~\bibnamefont{MacKenzie}}, \bibinfo {author}
  {\bibfnamefont{J.~N.}\ \bibnamefont{Chapman}},\ and\ \bibinfo {author}
  {\bibfnamefont{E.}~\bibnamefont{Arenholz}}}%
  , \bibinfo {year} {2004}{\natexlab{b}},\ \bibfield{title}{%
  \enquote{\bibinfo {title} {Surface effects in {Mn} {L$_{3,2}$} x-ray
  absorption spectra from {(Ga,Mn)As}},}\ }%
  \bibfield{journal}{%
  \bibinfo {journal} {Appl. Phys. Lett.}\ }%
  \textbf{\bibinfo {volume} {84}},\ \bibinfo {pages} {4065}%
  \bibAnnoteFile{NoStop}{Edmonds:2004_APL}%
\bibitem[{\citenamefont{Edmonds}\ \emph{et~al.}(2005)\citenamefont{Edmonds},
  \citenamefont{Farley}, \citenamefont{Johal}, \citenamefont{van~der Laan},
  \citenamefont{Campion}, \citenamefont{Gallagher},\ and\
  \citenamefont{Foxon}}]{Edmonds:2005_PRB}%
  \BibitemOpen
  \bibfield{author}{%
  \bibinfo {author} {\bibnamefont{Edmonds}, \bibfnamefont{K.~W.}}, \bibinfo
  {author} {\bibfnamefont{N.~R.~S.}\ \bibnamefont{Farley}}, \bibinfo {author}
  {\bibfnamefont{T.~K.}\ \bibnamefont{Johal}}, \bibinfo {author}
  {\bibfnamefont{G.}~\bibnamefont{van~der Laan}}, \bibinfo {author}
  {\bibfnamefont{R.~P.}\ \bibnamefont{Campion}}, \bibinfo {author}
  {\bibfnamefont{B.~L.}\ \bibnamefont{Gallagher}},\ and\ \bibinfo {author}
  {\bibfnamefont{C.~T.}\ \bibnamefont{Foxon}}}%
  , \bibinfo {year} {2005},\ \bibfield{title}{%
  \enquote{\bibinfo {title} {Ferromagnetic moment and antiferromagnetic
  coupling in {(Ga,Mn)As} thin films},}\ }%
  \bibfield{journal}{%
  \bibinfo {journal} {Phys. Rev. B}\ }%
  \textbf{\bibinfo {volume} {71}},\ \bibinfo {pages} {064418}%
  \bibAnnoteFile{NoStop}{Edmonds:2005_PRB}%
\bibitem[{\citenamefont{Edmonds}\ \emph{et~al.}(2006)\citenamefont{Edmonds},
  \citenamefont{van~der Laan}, \citenamefont{Freeman}, \citenamefont{Farley},
  \citenamefont{Johal}, \citenamefont{Campion}, \citenamefont{Foxon},
  \citenamefont{Gallagher},\ and\ \citenamefont{Arenholz}}]{Edmonds:2006_PRL}%
  \BibitemOpen
  \bibfield{author}{%
  \bibinfo {author} {\bibnamefont{Edmonds}, \bibfnamefont{K.~W.}}, \bibinfo
  {author} {\bibfnamefont{G.}~\bibnamefont{van~der Laan}}, \bibinfo {author}
  {\bibfnamefont{A.~A.}\ \bibnamefont{Freeman}}, \bibinfo {author}
  {\bibfnamefont{N.~R.~S.}\ \bibnamefont{Farley}}, \bibinfo {author}
  {\bibfnamefont{T.~K.}\ \bibnamefont{Johal}}, \bibinfo {author}
  {\bibfnamefont{R.~P.}\ \bibnamefont{Campion}}, \bibinfo {author}
  {\bibfnamefont{C.~T.}\ \bibnamefont{Foxon}}, \bibinfo {author}
  {\bibfnamefont{B.~L.}\ \bibnamefont{Gallagher}},\ and\ \bibinfo {author}
  {\bibfnamefont{E.}~\bibnamefont{Arenholz}}}%
  , \bibinfo {year} {2006},\ \bibfield{title}{%
  \enquote{\bibinfo {title} {Angle-dependent x-ray magnetic circular dichroism
  from {(Ga,Mn)As}: Anisotropy and identification of hybridized states},}\ }%
  \bibfield{journal}{%
  \bibinfo {journal} {Phys. Rev. Lett.}\ }%
  \textbf{\bibinfo {volume} {96}},\ \bibinfo {pages} {117207}%
  \bibAnnoteFile{NoStop}{Edmonds:2006_PRL}%
\bibitem[{\citenamefont{Edmonds}\ \emph{et~al.}(2002)\citenamefont{Edmonds},
  \citenamefont{Wang}, \citenamefont{Campion}, \citenamefont{Neumann},
  \citenamefont{Farley}, \citenamefont{Gallagher},\ and\
  \citenamefont{Foxon}}]{Edmonds:2002_APL}%
  \BibitemOpen
  \bibfield{author}{%
  \bibinfo {author} {\bibnamefont{Edmonds}, \bibfnamefont{K.~W.}}, \bibinfo
  {author} {\bibfnamefont{K.~Y.}\ \bibnamefont{Wang}}, \bibinfo {author}
  {\bibfnamefont{R.~P.}\ \bibnamefont{Campion}}, \bibinfo {author}
  {\bibfnamefont{A.~C.}\ \bibnamefont{Neumann}}, \bibinfo {author}
  {\bibfnamefont{N.~R.~S.}\ \bibnamefont{Farley}}, \bibinfo {author}
  {\bibfnamefont{B.~L.}\ \bibnamefont{Gallagher}},\ and\ \bibinfo {author}
  {\bibfnamefont{C.~T.}\ \bibnamefont{Foxon}}}%
  , \bibinfo {year} {2002},\ \bibfield{title}{%
  \enquote{\bibinfo {title} {High {Curie} temperature {GaMnAs} obtained by
  resistance-monitored annealing},}\ }%
  \bibfield{journal}{%
  \bibinfo {journal} {Appl. Phys. Lett.}\ }%
  \textbf{\bibinfo {volume} {81}},\ \bibinfo {pages} {4991}%
  \bibAnnoteFile{NoStop}{Edmonds:2002_APL}%
\bibitem[{\citenamefont{Edwards}\ and\
  \citenamefont{Sienko}(1978)}]{Edwards:1978_PRB}%
  \BibitemOpen
  \bibfield{author}{%
  \bibinfo {author} {\bibnamefont{Edwards}, \bibfnamefont{P.~P.}},\ and\
  \bibinfo {author} {\bibfnamefont{M.~J.}\ \bibnamefont{Sienko}}}%
  , \bibinfo {year} {1978},\ \bibfield{title}{%
  \enquote{\bibinfo {title} {Universality aspects of the metal-nonmetal
  transition in condensed media},}\ }%
  \bibfield{journal}{%
  \bibinfo {journal} {Phys. Rev. B}\ }%
  \textbf{\bibinfo {volume} {17}},\ \bibinfo {pages} {2575}%
  \bibAnnoteFile{NoStop}{Edwards:1978_PRB}%
\bibitem[{\citenamefont{Eggenkamp}\
  \emph{et~al.}(1995)\citenamefont{Eggenkamp}, \citenamefont{Swagten},
  \citenamefont{Story}, \citenamefont{Litvinov}, \citenamefont{Sw{\"u}ste},\
  and\ \citenamefont{de~Jonge}}]{Eggenkamp:1995_PRB}%
  \BibitemOpen
  \bibfield{author}{%
  \bibinfo {author} {\bibnamefont{Eggenkamp}, \bibfnamefont{P.~J.~T.}},
  \bibinfo {author} {\bibfnamefont{H.~J.~M.}\ \bibnamefont{Swagten}}, \bibinfo
  {author} {\bibfnamefont{T.}~\bibnamefont{Story}}, \bibinfo {author}
  {\bibfnamefont{V.~I.}\ \bibnamefont{Litvinov}}, \bibinfo {author}
  {\bibfnamefont{C.~H.~W.}\ \bibnamefont{Sw{\"u}ste}},\ and\ \bibinfo {author}
  {\bibfnamefont{W.~J.~M.}\ \bibnamefont{de~Jonge}}}%
  , \bibinfo {year} {1995},\ \bibfield{title}{%
  \enquote{\bibinfo {title} {Calculations of the ferromagnet-to-spin-glass
  transition in diluted magnetic systems with an {RKKY} interaction},}\ }%
  \bibfield{journal}{%
  \bibinfo {journal} {Phys. Rev. B}\ }%
  \textbf{\bibinfo {volume} {51}},\ \bibinfo {pages} {15250}%
  \bibAnnoteFile{NoStop}{Eggenkamp:1995_PRB}%
\bibitem[{\citenamefont{Ehlert}\ \emph{et~al.}(2012)\citenamefont{Ehlert},
  \citenamefont{Song}, \citenamefont{Ciorga}, \citenamefont{Utz},
  \citenamefont{Schuh}, \citenamefont{Bougeard},\ and\
  \citenamefont{Weiss}}]{Ehlert:2012_PRB}%
  \BibitemOpen
  \bibfield{author}{%
  \bibinfo {author} {\bibnamefont{Ehlert}, \bibfnamefont{M.}}, \bibinfo
  {author} {\bibfnamefont{C.}~\bibnamefont{Song}}, \bibinfo {author}
  {\bibfnamefont{M.}~\bibnamefont{Ciorga}}, \bibinfo {author}
  {\bibfnamefont{M.}~\bibnamefont{Utz}}, \bibinfo {author}
  {\bibfnamefont{D.}~\bibnamefont{Schuh}}, \bibinfo {author}
  {\bibfnamefont{D.}~\bibnamefont{Bougeard}},\ and\ \bibinfo {author}
  {\bibfnamefont{D.}~\bibnamefont{Weiss}}}%
  , \bibinfo {year} {2012},\ \bibfield{title}{%
  \enquote{\bibinfo {title} {All-electrical measurements of direct spin {Hall}
  effect in {GaAs} with {Esaki} diode electrodes},}\ }%
  \bibfield{journal}{%
  \bibinfo {journal} {Phys. Rev. B}\ }%
  \textbf{\bibinfo {volume} {86}},\ \bibinfo {pages} {205204}%
  \bibAnnoteFile{NoStop}{Ehlert:2012_PRB}%
\bibitem[{\citenamefont{Eid}\ \emph{et~al.}(2005)\citenamefont{Eid},
  \citenamefont{Sheu}, \citenamefont{Maksimov}, \citenamefont{Stone},
  \citenamefont{Schiffer},\ and\ \citenamefont{Samarth}}]{Eid:2005_APL}%
  \BibitemOpen
  \bibfield{author}{%
  \bibinfo {author} {\bibnamefont{Eid}, \bibfnamefont{K.~F.}}, \bibinfo
  {author} {\bibfnamefont{B.~L.}\ \bibnamefont{Sheu}}, \bibinfo {author}
  {\bibfnamefont{O.}~\bibnamefont{Maksimov}}, \bibinfo {author}
  {\bibfnamefont{M.~B.}\ \bibnamefont{Stone}}, \bibinfo {author}
  {\bibfnamefont{P.}~\bibnamefont{Schiffer}},\ and\ \bibinfo {author}
  {\bibfnamefont{N.}~\bibnamefont{Samarth}}}%
  , \bibinfo {year} {2005},\ \bibfield{title}{%
  \enquote{\bibinfo {title} {Nanoengineered {Curie} temperature in
  laterally-patterned ferromagnetic semiconductor heterostructures},}\ }%
  \bibfield{journal}{%
  \bibinfo {journal} {Appl. Phys. Lett.}\ }%
  \textbf{\bibinfo {volume} {86}},\ \bibinfo {pages} {152505}%
  \bibAnnoteFile{NoStop}{Eid:2005_APL}%
\bibitem[{\citenamefont{Eid}\ \emph{et~al.}(2004)\citenamefont{Eid},
  \citenamefont{Stone}, \citenamefont{Ku}, \citenamefont{Schiffer},\ and\
  \citenamefont{Samarth}}]{Eid:2004_APL}%
  \BibitemOpen
  \bibfield{author}{%
  \bibinfo {author} {\bibnamefont{Eid}, \bibfnamefont{K.~F.}}, \bibinfo
  {author} {\bibfnamefont{M.~B.}\ \bibnamefont{Stone}}, \bibinfo {author}
  {\bibfnamefont{K.~C.}\ \bibnamefont{Ku}}, \bibinfo {author}
  {\bibfnamefont{P.}~\bibnamefont{Schiffer}},\ and\ \bibinfo {author}
  {\bibfnamefont{N.}~\bibnamefont{Samarth}}}%
  , \bibinfo {year} {2004},\ \bibfield{title}{%
  \enquote{\bibinfo {title} {Exchange biasing of the ferromagnetic
  semiconductor {Ga$_{1-x}$Mn$_{x}$As}},}\ }%
  \bibfield{journal}{%
  \bibinfo {journal} {Appl. Phys. Lett.}\ }%
  \textbf{\bibinfo {volume} {85}},\ \bibinfo {pages} {1556}%
  \bibAnnoteFile{NoStop}{Eid:2004_APL}%
\bibitem[{\citenamefont{Elsen}\ \emph{et~al.}(2006)\citenamefont{Elsen},
  \citenamefont{Boulle}, \citenamefont{George}, \citenamefont{Jaffr{\'e}s},
  \citenamefont{Mattana}, \citenamefont{Cros}, \citenamefont{Fert},
  \citenamefont{Lema\^itre}, \citenamefont{Giraud},\ and\
  \citenamefont{Faini}}]{Elsen:2006_PRB}%
  \BibitemOpen
  \bibfield{author}{%
  \bibinfo {author} {\bibnamefont{Elsen}, \bibfnamefont{M.}}, \bibinfo {author}
  {\bibfnamefont{O.}~\bibnamefont{Boulle}}, \bibinfo {author}
  {\bibfnamefont{J.-M.}\ \bibnamefont{George}}, \bibinfo {author}
  {\bibfnamefont{H.}~\bibnamefont{Jaffr{\'e}s}}, \bibinfo {author}
  {\bibfnamefont{R.}~\bibnamefont{Mattana}}, \bibinfo {author}
  {\bibfnamefont{V.}~\bibnamefont{Cros}}, \bibinfo {author}
  {\bibfnamefont{A.}~\bibnamefont{Fert}}, \bibinfo {author}
  {\bibfnamefont{A.}~\bibnamefont{Lema\^itre}}, \bibinfo {author}
  {\bibfnamefont{R.}~\bibnamefont{Giraud}},\ and\ \bibinfo {author}
  {\bibfnamefont{G.}~\bibnamefont{Faini}}}%
  , \bibinfo {year} {2006},\ \bibfield{title}{%
  \enquote{\bibinfo {title} {Spin transfer experiments on
  {(Ga,Mn)As/(In,Ga)As/(Ga,Mn)As} tunnel junctions},}\ }%
  \bibfield{journal}{%
  \bibinfo {journal} {Phys. Rev. B}\ }%
  \textbf{\bibinfo {volume} {73}},\ \bibinfo {pages} {035303}%
  \bibAnnoteFile{NoStop}{Elsen:2006_PRB}%
\bibitem[{\citenamefont{Elsen}\ \emph{et~al.}(2007)\citenamefont{Elsen},
  \citenamefont{Jaffr\`es}, \citenamefont{Mattana}, \citenamefont{Tran},
  \citenamefont{George}, \citenamefont{Miard},\ and\
  \citenamefont{Lema\^{i}tre}}]{Elsen:2007_PRL}%
  \BibitemOpen
  \bibfield{author}{%
  \bibinfo {author} {\bibnamefont{Elsen}, \bibfnamefont{M.}}, \bibinfo {author}
  {\bibfnamefont{H.}~\bibnamefont{Jaffr\`es}}, \bibinfo {author}
  {\bibfnamefont{R.}~\bibnamefont{Mattana}}, \bibinfo {author}
  {\bibfnamefont{M.}~\bibnamefont{Tran}}, \bibinfo {author}
  {\bibfnamefont{J.-M.}\ \bibnamefont{George}}, \bibinfo {author}
  {\bibfnamefont{A.}~\bibnamefont{Miard}},\ and\ \bibinfo {author}
  {\bibfnamefont{A.}~\bibnamefont{Lema\^{i}tre}}}%
  , \bibinfo {year} {2007},\ \bibfield{title}{%
  \enquote{\bibinfo {title} {Exchange-mediated anisotropy of {(Ga,Mn)As}
  valence-band probed by resonant tunneling spectroscopy},}\ }%
  \bibfield{journal}{%
  \bibinfo {journal} {Phys. Rev. Lett.}\ }%
  \textbf{\bibinfo {volume} {99}},\ \bibinfo {pages} {127203}%
  \bibAnnoteFile{NoStop}{Elsen:2007_PRL}%
\bibitem[{\citenamefont{Endo}\
  \emph{et~al.}(2010{\natexlab{a}})\citenamefont{Endo}, \citenamefont{Chiba},
  \citenamefont{Shimotani}, \citenamefont{Matsukura}, \citenamefont{Iwasa},\
  and\ \citenamefont{Ohno}}]{Endo:2010_APLb}%
  \BibitemOpen
  \bibfield{author}{%
  \bibinfo {author} {\bibnamefont{Endo}, \bibfnamefont{M.}}, \bibinfo {author}
  {\bibfnamefont{D.}~\bibnamefont{Chiba}}, \bibinfo {author}
  {\bibfnamefont{H.}~\bibnamefont{Shimotani}}, \bibinfo {author}
  {\bibfnamefont{F.}~\bibnamefont{Matsukura}}, \bibinfo {author}
  {\bibfnamefont{Y.}~\bibnamefont{Iwasa}},\ and\ \bibinfo {author}
  {\bibfnamefont{H.}~\bibnamefont{Ohno}}}%
  , \bibinfo {year} {2010}{\natexlab{a}},\ \bibfield{title}{%
  \enquote{\bibinfo {title} {Electric double layer transistor with a
  {(Ga,Mn)As} channel},}\ }%
  \bibfield{journal}{%
  \bibinfo {journal} {Appl. Phys. Lett.}\ }%
  \textbf{\bibinfo {volume} {96}},\ \bibinfo {pages} {022515}%
  \bibAnnoteFile{NoStop}{Endo:2010_APLb}%
\bibitem[{\citenamefont{Endo}\
  \emph{et~al.}(2010{\natexlab{b}})\citenamefont{Endo},
  \citenamefont{Matsukura},\ and\ \citenamefont{Ohno}}]{Endo:2010_APL}%
  \BibitemOpen
  \bibfield{author}{%
  \bibinfo {author} {\bibnamefont{Endo}, \bibfnamefont{M.}}, \bibinfo {author}
  {\bibfnamefont{F.}~\bibnamefont{Matsukura}},\ and\ \bibinfo {author}
  {\bibfnamefont{H.}~\bibnamefont{Ohno}}}%
  , \bibinfo {year} {2010}{\natexlab{b}},\ \bibfield{title}{%
  \enquote{\bibinfo {title} {Current induced effective magnetic field and
  magnetization reversal in uniaxial anisotropy {(Ga,Mn)As}},}\ }%
  \bibfield{journal}{%
  \bibinfo {journal} {Appl. Phys. Lett.}\ }%
  \textbf{\bibinfo {volume} {97}},\ \bibinfo {pages} {222501}%
  \bibAnnoteFile{NoStop}{Endo:2010_APL}%
\bibitem[{\citenamefont{Fang}\ \emph{et~al.}(2011)\citenamefont{Fang},
  \citenamefont{Kurebayashi}, \citenamefont{Wunderlich},
  \citenamefont{V{\'{y}}born{\'{y}}}, \citenamefont{Z{\^{a}}rbo},
  \citenamefont{Campion}, \citenamefont{Casiraghi}, \citenamefont{Gallagher},
  \citenamefont{Jungwirth},\ and\ \citenamefont{Ferguson}}]{Fang:2011_NN}%
  \BibitemOpen
  \bibfield{author}{%
  \bibinfo {author} {\bibnamefont{Fang}, \bibfnamefont{D.}}, \bibinfo {author}
  {\bibfnamefont{H.}~\bibnamefont{Kurebayashi}}, \bibinfo {author}
  {\bibfnamefont{J.}~\bibnamefont{Wunderlich}}, \bibinfo {author}
  {\bibfnamefont{K.}~\bibnamefont{V{\'{y}}born{\'{y}}}}, \bibinfo {author}
  {\bibfnamefont{L.~P.}\ \bibnamefont{Z{\^{a}}rbo}}, \bibinfo {author}
  {\bibfnamefont{R.~P.}\ \bibnamefont{Campion}}, \bibinfo {author}
  {\bibfnamefont{A.}~\bibnamefont{Casiraghi}}, \bibinfo {author}
  {\bibfnamefont{B.~L.}\ \bibnamefont{Gallagher}}, \bibinfo {author}
  {\bibfnamefont{T.}~\bibnamefont{Jungwirth}},\ and\ \bibinfo {author}
  {\bibfnamefont{A.~J.}\ \bibnamefont{Ferguson}}}%
  , \bibinfo {year} {2011},\ \bibfield{title}{%
  \enquote{\bibinfo {title} {Spin-orbit-driven ferromagnetic resonance},}\ }%
  \bibfield{journal}{%
  \bibinfo {journal} {Nat. Nanotech.}\ }%
  \textbf{\bibinfo {volume} {6}},\ \bibinfo {pages} {413}%
  \bibAnnoteFile{NoStop}{Fang:2011_NN}%
\bibitem[{\citenamefont{Farshchi}\ \emph{et~al.}(2007)\citenamefont{Farshchi},
  \citenamefont{Ashby}, \citenamefont{Hwang}, \citenamefont{Grigoropoulos},
  \citenamefont{Chopdekar}, \citenamefont{Suzuki},\ and\
  \citenamefont{Dubon}}]{Farshchi:2007_PB}%
  \BibitemOpen
  \bibfield{author}{%
  \bibinfo {author} {\bibnamefont{Farshchi}, \bibfnamefont{R.}}, \bibinfo
  {author} {\bibfnamefont{P.~D.}\ \bibnamefont{Ashby}}, \bibinfo {author}
  {\bibfnamefont{D.~J.}\ \bibnamefont{Hwang}}, \bibinfo {author}
  {\bibfnamefont{C.~P.}\ \bibnamefont{Grigoropoulos}}, \bibinfo {author}
  {\bibfnamefont{R.~V.}\ \bibnamefont{Chopdekar}}, \bibinfo {author}
  {\bibfnamefont{Y.}~\bibnamefont{Suzuki}},\ and\ \bibinfo {author}
  {\bibfnamefont{O.~D.}\ \bibnamefont{Dubon}}}%
  , \bibinfo {year} {2007},\ \bibfield{title}{%
  \enquote{\bibinfo {title} {Hydrogen patterning of {Ga$_{1-x}$Mn$_{x}$As} for
  planar spintronics},}\ }%
  \bibfield{journal}{%
  \bibinfo {journal} {Physica B}\ }%
  \textbf{\bibinfo {volume} {401}},\ \bibinfo {pages} {447}%
  \bibAnnoteFile{NoStop}{Farshchi:2007_PB}%
\bibitem[{\citenamefont{Fedorych}\ \emph{et~al.}(2002)\citenamefont{Fedorych},
  \citenamefont{Hankiewicz}, \citenamefont{Wilamowski},\ and\
  \citenamefont{Sadowski}}]{Fedorych:2002_PRB}%
  \BibitemOpen
  \bibfield{author}{%
  \bibinfo {author} {\bibnamefont{Fedorych}, \bibfnamefont{O.~M.}}, \bibinfo
  {author} {\bibfnamefont{E.~M.}\ \bibnamefont{Hankiewicz}}, \bibinfo {author}
  {\bibfnamefont{Z.}~\bibnamefont{Wilamowski}},\ and\ \bibinfo {author}
  {\bibfnamefont{J.}~\bibnamefont{Sadowski}}}%
  , \bibinfo {year} {2002},\ \bibfield{title}{%
  \enquote{\bibinfo {title} {Single ion anisotropy of {Mn}-doped {GaAs}
  measured by electron paramagnetic resonance},}\ }%
  \bibfield{journal}{%
  \bibinfo {journal} {Phys. Rev. B}\ }%
  \textbf{\bibinfo {volume} {66}},\ \bibinfo {pages} {045201}%
  \bibAnnoteFile{NoStop}{Fedorych:2002_PRB}%
\bibitem[{\citenamefont{{Fern{\'a}ndez-Rossier}}\ and\
  \citenamefont{Sham}(2002)}]{Fernandez-Rossier:2002_PRB}%
  \BibitemOpen
  \bibfield{author}{%
  \bibinfo {author} {\bibnamefont{{Fern{\'a}ndez-Rossier}},
  \bibfnamefont{J.}},\ and\ \bibinfo {author} {\bibfnamefont{L.~J.}\
  \bibnamefont{Sham}}}%
  , \bibinfo {year} {2002},\ \bibfield{title}{%
  \enquote{\bibinfo {title} {Spin separation in digital ferromagnetic
  heterostructures},}\ }%
  \bibfield{journal}{%
  \bibinfo {journal} {Phys. Rev. B}\ }%
  \textbf{\bibinfo {volume} {66}},\ \bibinfo {pages} {073312}%
  \bibAnnoteFile{NoStop}{Fernandez-Rossier:2002_PRB}%
\bibitem[{\citenamefont{Ferrand}\ \emph{et~al.}(2000)\citenamefont{Ferrand},
  \citenamefont{Cibert}, \citenamefont{Bourgognon}, \citenamefont{Tatarenko},
  \citenamefont{Wasiela}, \citenamefont{Fishman}, \citenamefont{Bonanni},
  \citenamefont{Sitter}, \citenamefont{Kole{\'s}nik},
  \citenamefont{Jaroszy{\'n}ski}, \citenamefont{Barcz},\ and\
  \citenamefont{Dietl}}]{Ferrand:2000_JCG}%
  \BibitemOpen
  \bibfield{author}{%
  \bibinfo {author} {\bibnamefont{Ferrand}, \bibfnamefont{D.}}, \bibinfo
  {author} {\bibfnamefont{J.}~\bibnamefont{Cibert}}, \bibinfo {author}
  {\bibfnamefont{C.}~\bibnamefont{Bourgognon}}, \bibinfo {author}
  {\bibfnamefont{S.}~\bibnamefont{Tatarenko}}, \bibinfo {author}
  {\bibfnamefont{A.}~\bibnamefont{Wasiela}}, \bibinfo {author}
  {\bibfnamefont{G.}~\bibnamefont{Fishman}}, \bibinfo {author}
  {\bibfnamefont{A.}~\bibnamefont{Bonanni}}, \bibinfo {author}
  {\bibfnamefont{H.}~\bibnamefont{Sitter}}, \bibinfo {author}
  {\bibfnamefont{S.}~\bibnamefont{Kole{\'s}nik}}, \bibinfo {author}
  {\bibfnamefont{J.}~\bibnamefont{Jaroszy{\'n}ski}}, \bibinfo {author}
  {\bibfnamefont{A.}~\bibnamefont{Barcz}},\ and\ \bibinfo {author}
  {\bibfnamefont{T.}~\bibnamefont{Dietl}}}%
  , \bibinfo {year} {2000},\ \bibfield{title}{%
  \enquote{\bibinfo {title} {Carrier-induced ferromagnetic interactions in
  p-doped {Zn$_{1-x}$Mn$_{x}$Te} epilayers},}\ }%
  \bibfield{journal}{%
  \bibinfo {journal} {J. Cryst. Growth}\ }%
  \textbf{\bibinfo {volume} {214}},\ \bibinfo {pages} {387}%
  \bibAnnoteFile{NoStop}{Ferrand:2000_JCG}%
\bibitem[{\citenamefont{Ferrand}\ \emph{et~al.}(2001)\citenamefont{Ferrand},
  \citenamefont{Cibert}, \citenamefont{Wasiela}, \citenamefont{Bourgognon},
  \citenamefont{Tatarenko}, \citenamefont{Fishman}, \citenamefont{Andrearczyk},
  \citenamefont{Jaroszy{\'n}ski}, \citenamefont{Kole{\'s}nik},
  \citenamefont{Dietl}, \citenamefont{Barbara},\ and\
  \citenamefont{Dufeu}}]{Ferrand:2001_PRB}%
  \BibitemOpen
  \bibfield{author}{%
  \bibinfo {author} {\bibnamefont{Ferrand}, \bibfnamefont{D.}}, \bibinfo
  {author} {\bibfnamefont{J.}~\bibnamefont{Cibert}}, \bibinfo {author}
  {\bibfnamefont{A.}~\bibnamefont{Wasiela}}, \bibinfo {author}
  {\bibfnamefont{C.}~\bibnamefont{Bourgognon}}, \bibinfo {author}
  {\bibfnamefont{S.}~\bibnamefont{Tatarenko}}, \bibinfo {author}
  {\bibfnamefont{G.}~\bibnamefont{Fishman}}, \bibinfo {author}
  {\bibfnamefont{T.}~\bibnamefont{Andrearczyk}}, \bibinfo {author}
  {\bibfnamefont{J.}~\bibnamefont{Jaroszy{\'n}ski}}, \bibinfo {author}
  {\bibfnamefont{S.}~\bibnamefont{Kole{\'s}nik}}, \bibinfo {author}
  {\bibfnamefont{T.}~\bibnamefont{Dietl}}, \bibinfo {author}
  {\bibfnamefont{B.}~\bibnamefont{Barbara}},\ and\ \bibinfo {author}
  {\bibfnamefont{D.}~\bibnamefont{Dufeu}}}%
  , \bibinfo {year} {2001},\ \bibfield{title}{%
  \enquote{\bibinfo {title} {Carrier-induced ferromagnetism in
  {p-Zn$_{1-x}$Mn$_{x}$Te}},}\ }%
  \bibfield{journal}{%
  \bibinfo {journal} {Phys. Rev. B}\ }%
  \textbf{\bibinfo {volume} {63}},\ \bibinfo {pages} {085201}%
  \bibAnnoteFile{NoStop}{Ferrand:2001_PRB}%
\bibitem[{\citenamefont{Fiederling}\
  \emph{et~al.}(2003)\citenamefont{Fiederling}, \citenamefont{Grabs},
  \citenamefont{Ossau}, \citenamefont{Schmidt},\ and\
  \citenamefont{Molenkamp}}]{Fiederling:2003_APL}%
  \BibitemOpen
  \bibfield{author}{%
  \bibinfo {author} {\bibnamefont{Fiederling}, \bibfnamefont{R.}}, \bibinfo
  {author} {\bibfnamefont{P.}~\bibnamefont{Grabs}}, \bibinfo {author}
  {\bibfnamefont{W.}~\bibnamefont{Ossau}}, \bibinfo {author}
  {\bibfnamefont{G.}~\bibnamefont{Schmidt}},\ and\ \bibinfo {author}
  {\bibfnamefont{L.~W.}\ \bibnamefont{Molenkamp}}}%
  , \bibinfo {year} {2003},\ \bibfield{title}{%
  \enquote{\bibinfo {title} {Detection of electrical spin injection by
  light-emitting diodes in top- and side-emission configurations},}\ }%
  \bibfield{journal}{%
  \bibinfo {journal} {Appl. Phys. Lett.}\ }%
  \textbf{\bibinfo {volume} {82}},\ \bibinfo {pages} {2160}%
  \bibAnnoteFile{NoStop}{Fiederling:2003_APL}%
\bibitem[{\citenamefont{Figielski}\
  \emph{et~al.}(2007)\citenamefont{Figielski}, \citenamefont{Wosinski},
  \citenamefont{Morawski}, \citenamefont{Makosa}, \citenamefont{Wrobel},\ and\
  \citenamefont{Sadowski}}]{Figielski:2007_APL}%
  \BibitemOpen
  \bibfield{author}{%
  \bibinfo {author} {\bibnamefont{Figielski}, \bibfnamefont{T.}}, \bibinfo
  {author} {\bibfnamefont{T.}~\bibnamefont{Wosinski}}, \bibinfo {author}
  {\bibfnamefont{A.}~\bibnamefont{Morawski}}, \bibinfo {author}
  {\bibfnamefont{A.}~\bibnamefont{Makosa}}, \bibinfo {author}
  {\bibfnamefont{J.}~\bibnamefont{Wrobel}},\ and\ \bibinfo {author}
  {\bibfnamefont{J.}~\bibnamefont{Sadowski}}}%
  , \bibinfo {year} {2007},\ \bibfield{title}{%
  \enquote{\bibinfo {title} {Remnant magnetoresistance in ferromagnetic
  {(Ga,Mn)As} nanostructures},}\ }%
  \bibfield{journal}{%
  \bibinfo {journal} {Appl. Phys. Lett.}\ }%
  \textbf{\bibinfo {volume} {90}},\ \bibinfo {eid} {052108}%
  \bibAnnoteFile{NoStop}{Figielski:2007_APL}%
\bibitem[{\citenamefont{Finkelstein}(1990)}]{Finkelstein:1990_SSR}%
  \BibitemOpen
  \bibfield{author}{%
  \bibinfo {author} {\bibnamefont{Finkelstein}, \bibfnamefont{A.~M.}}}%
  , \bibinfo {year} {1990},\ \bibfield{title}{%
  \enquote{\bibinfo {title} {Electron liquid in disordered conductors},}\ }%
  \bibfield{journal}{%
  \bibinfo {journal} {Soviet Sci. Rev.}\ }%
  \textbf{\bibinfo {volume} {14}},\ \bibinfo {pages} {1}%
  \bibAnnoteFile{NoStop}{Finkelstein:1990_SSR}%
\bibitem[{\citenamefont{Fiorentini}(1995)}]{Fiorentini:1995_PRB}%
  \BibitemOpen
  \bibfield{author}{%
  \bibinfo {author} {\bibnamefont{Fiorentini}, \bibfnamefont{V.}}}%
  , \bibinfo {year} {1995},\ \bibfield{title}{%
  \enquote{\bibinfo {title} {Effective-mass single and double acceptor spectra
  in {GaAs}},}\ }%
  \bibfield{journal}{%
  \bibinfo {journal} {Phys. Rev. B}\ }%
  \textbf{\bibinfo {volume} {51}},\ \bibinfo {pages} {10161}%
  \bibAnnoteFile{NoStop}{Fiorentini:1995_PRB}%
\bibitem[{\citenamefont{Fleury}\ and\
  \citenamefont{Waintal}(2008)}]{Fleury:2008_PRL}%
  \BibitemOpen
  \bibfield{author}{%
  \bibinfo {author} {\bibnamefont{Fleury}, \bibfnamefont{G.}},\ and\ \bibinfo
  {author} {\bibfnamefont{X.}~\bibnamefont{Waintal}}}%
  , \bibinfo {year} {2008},\ \bibfield{title}{%
  \enquote{\bibinfo {title} {Many-body localization study in low-density
  electron gases: Do metals exist in two dimensions?}.}\ }%
  \bibfield{journal}{%
  \bibinfo {journal} {Phys. Rev. Lett.}\ }%
  \textbf{\bibinfo {volume} {101}},\ \bibinfo {pages} {226803}%
  \bibAnnoteFile{NoStop}{Fleury:2008_PRL}%
\bibitem[{\citenamefont{Freeman}\ \emph{et~al.}(2007)\citenamefont{Freeman},
  \citenamefont{Edmonds}, \citenamefont{Farley}, \citenamefont{Novikov},
  \citenamefont{Campion}, \citenamefont{Foxon}, \citenamefont{Gallagher},
  \citenamefont{Sarigiannidou},\ and\ \citenamefont{van~der
  Laan}}]{Freeman:2007_PRB}%
  \BibitemOpen
  \bibfield{author}{%
  \bibinfo {author} {\bibnamefont{Freeman}, \bibfnamefont{A.~A.}}, \bibinfo
  {author} {\bibfnamefont{K.~W.}\ \bibnamefont{Edmonds}}, \bibinfo {author}
  {\bibfnamefont{N.~R.~S.}\ \bibnamefont{Farley}}, \bibinfo {author}
  {\bibfnamefont{S.~V.}\ \bibnamefont{Novikov}}, \bibinfo {author}
  {\bibfnamefont{R.~P.}\ \bibnamefont{Campion}}, \bibinfo {author}
  {\bibfnamefont{C.~T.}\ \bibnamefont{Foxon}}, \bibinfo {author}
  {\bibfnamefont{B.~L.}\ \bibnamefont{Gallagher}}, \bibinfo {author}
  {\bibfnamefont{E.}~\bibnamefont{Sarigiannidou}},\ and\ \bibinfo {author}
  {\bibfnamefont{G.}~\bibnamefont{van~der Laan}}}%
  , \bibinfo {year} {2007},\ \bibfield{title}{%
  \enquote{\bibinfo {title} {Depth dependence of the {Mn} valence and {Mn-Mn}
  coupling in {(Ga,Mn)N}},}\ }%
  \bibfield{journal}{%
  \bibinfo {journal} {Phys. Rev. B}\ }%
  \textbf{\bibinfo {volume} {76}},\ \bibinfo {pages} {081201}%
  \bibAnnoteFile{NoStop}{Freeman:2007_PRB}%
\bibitem[{\citenamefont{Freeman}\ \emph{et~al.}(2008)\citenamefont{Freeman},
  \citenamefont{Edmonds}, \citenamefont{van~der Laan}, \citenamefont{Campion},
  \citenamefont{Rushforth}, \citenamefont{Farley}, \citenamefont{Johal},
  \citenamefont{Foxon}, \citenamefont{Gallagher}, \citenamefont{Rogalev},\ and\
  \citenamefont{Wilhelm}}]{Freeman:2008_PRB}%
  \BibitemOpen
  \bibfield{author}{%
  \bibinfo {author} {\bibnamefont{Freeman}, \bibfnamefont{A.~A.}}, \bibinfo
  {author} {\bibfnamefont{K.~W.}\ \bibnamefont{Edmonds}}, \bibinfo {author}
  {\bibfnamefont{G.}~\bibnamefont{van~der Laan}}, \bibinfo {author}
  {\bibfnamefont{R.~P.}\ \bibnamefont{Campion}}, \bibinfo {author}
  {\bibfnamefont{A.~W.}\ \bibnamefont{Rushforth}}, \bibinfo {author}
  {\bibfnamefont{N.~R.~S.}\ \bibnamefont{Farley}}, \bibinfo {author}
  {\bibfnamefont{T.~K.}\ \bibnamefont{Johal}}, \bibinfo {author}
  {\bibfnamefont{C.~T.}\ \bibnamefont{Foxon}}, \bibinfo {author}
  {\bibfnamefont{B.~L.}\ \bibnamefont{Gallagher}}, \bibinfo {author}
  {\bibfnamefont{A.}~\bibnamefont{Rogalev}},\ and\ \bibinfo {author}
  {\bibfnamefont{F.}~\bibnamefont{Wilhelm}}}%
  , \bibinfo {year} {2008},\ \bibfield{title}{%
  \enquote{\bibinfo {title} {Valence band orbital polarization in {III-V}
  ferromagnetic semiconductors},}\ }%
  \bibfield{journal}{%
  \bibinfo {journal} {Phys. Rev. B}\ }%
  \textbf{\bibinfo {volume} {77}},\ \bibinfo {pages} {073304}%
  \bibAnnoteFile{NoStop}{Freeman:2008_PRB}%
\bibitem[{\citenamefont{Fritzsche}\ and\
  \citenamefont{Cuevas}(1960)}]{Fritzsche:1960_PR}%
  \BibitemOpen
  \bibfield{author}{%
  \bibinfo {author} {\bibnamefont{Fritzsche}, \bibfnamefont{H.}},\ and\
  \bibinfo {author} {\bibfnamefont{M.}~\bibnamefont{Cuevas}}}%
  , \bibinfo {year} {1960},\ \bibfield{title}{%
  \enquote{\bibinfo {title} {Impurity conduction in transmutation-doped
  $p$-type {Germanium}},}\ }%
  \bibfield{journal}{%
  \bibinfo {journal} {Phys. Rev.}\ }%
  \textbf{\bibinfo {volume} {119}},\ \bibinfo {pages} {1238}%
  \bibAnnoteFile{NoStop}{Fritzsche:1960_PR}%
\bibitem[{\citenamefont{{Fr{\"o}hlich}}\ and\
  \citenamefont{Nabarro}(1940)}]{Frohlich:1940_PRSL}%
  \BibitemOpen
  \bibfield{author}{%
  \bibinfo {author} {\bibnamefont{{Fr{\"o}hlich}}, \bibfnamefont{F.}},\ and\
  \bibinfo {author} {\bibfnamefont{F.~R.~N.}\ \bibnamefont{Nabarro}}}%
  , \bibinfo {year} {1940},\ \bibfield{title}{%
  \enquote{\bibinfo {title} {Orientation of nuclear spins in metals},}\ }%
  \bibfield{journal}{%
  \bibinfo {journal} {Proc. R. Soc. London, Ser. A}\ }%
  \textbf{\bibinfo {volume} {175}},\ \bibinfo {pages} {382}%
  \bibAnnoteFile{NoStop}{Frohlich:1940_PRSL}%
\bibitem[{\citenamefont{Fujii}\ \emph{et~al.}(2013)\citenamefont{Fujii},
  \citenamefont{Salles}, \citenamefont{Sperl}, \citenamefont{Ueda},
  \citenamefont{Kobata}, \citenamefont{Kobayashi}, \citenamefont{Yamashita},
  \citenamefont{Torelli}, \citenamefont{Utz}, \citenamefont{Fadley},
  \citenamefont{Gray}, \citenamefont{Braun}, \citenamefont{Ebert},
  \citenamefont{Di~Marco}, \citenamefont{Eriksson}, \citenamefont{Thunstr\"om},
  \citenamefont{Fecher}, \citenamefont{Stryhanyuk}, \citenamefont{Ikenaga},
  \citenamefont{Min\'ar}, \citenamefont{Back}, \citenamefont{van~der Laan},\
  and\ \citenamefont{Panaccione}}]{Fujii:2013_PRL}%
  \BibitemOpen
  \bibfield{author}{%
  \bibinfo {author} {\bibnamefont{Fujii}, \bibfnamefont{J.}}, \bibinfo {author}
  {\bibfnamefont{B.~R.}\ \bibnamefont{Salles}}, \bibinfo {author}
  {\bibfnamefont{M.}~\bibnamefont{Sperl}}, \bibinfo {author}
  {\bibfnamefont{S.}~\bibnamefont{Ueda}}, \bibinfo {author}
  {\bibfnamefont{M.}~\bibnamefont{Kobata}}, \bibinfo {author}
  {\bibfnamefont{K.}~\bibnamefont{Kobayashi}}, \bibinfo {author}
  {\bibfnamefont{Y.}~\bibnamefont{Yamashita}}, \bibinfo {author}
  {\bibfnamefont{P.}~\bibnamefont{Torelli}}, \bibinfo {author}
  {\bibfnamefont{M.}~\bibnamefont{Utz}}, \bibinfo {author}
  {\bibfnamefont{C.~S.}\ \bibnamefont{Fadley}}, \bibinfo {author}
  {\bibfnamefont{A.~X.}\ \bibnamefont{Gray}}, \bibinfo {author}
  {\bibfnamefont{J.}~\bibnamefont{Braun}}, \bibinfo {author}
  {\bibfnamefont{H.}~\bibnamefont{Ebert}}, \bibinfo {author}
  {\bibfnamefont{I.}~\bibnamefont{Di~Marco}}, \bibinfo {author}
  {\bibfnamefont{O.}~\bibnamefont{Eriksson}}, \bibinfo {author}
  {\bibfnamefont{P.}~\bibnamefont{Thunstr\"om}}, \bibinfo {author}
  {\bibfnamefont{G.~H.}\ \bibnamefont{Fecher}}, \bibinfo {author}
  {\bibfnamefont{H.}~\bibnamefont{Stryhanyuk}}, \bibinfo {author}
  {\bibfnamefont{E.}~\bibnamefont{Ikenaga}}, \bibinfo {author}
  {\bibfnamefont{J.}~\bibnamefont{Min\'ar}}, \bibinfo {author}
  {\bibfnamefont{C.~H.}\ \bibnamefont{Back}}, \bibinfo {author}
  {\bibfnamefont{G.}~\bibnamefont{van~der Laan}},\ and\ \bibinfo {author}
  {\bibfnamefont{G.}~\bibnamefont{Panaccione}}}%
  , \bibinfo {year} {2013},\ \bibfield{title}{%
  \enquote{\bibinfo {title} {Identifying the electronic character and role of
  the {Mn} states in the valence band of {(Ga,Mn)As}},}\ }%
  \bibfield{journal}{%
  \bibinfo {journal} {Phys. Rev. Lett.}\ }%
  \textbf{\bibinfo {volume} {111}},\ \bibinfo {pages} {097201}%
  \bibAnnoteFile{NoStop}{Fujii:2013_PRL}%
\bibitem[{\citenamefont{Fukuma}\ \emph{et~al.}(2008)\citenamefont{Fukuma},
  \citenamefont{Asada}, \citenamefont{Yamamoto}, \citenamefont{Odawara},\ and\
  \citenamefont{Koyanagi}}]{Fukuma:2008_APLb}%
  \BibitemOpen
  \bibfield{author}{%
  \bibinfo {author} {\bibnamefont{Fukuma}, \bibfnamefont{Y.}}, \bibinfo
  {author} {\bibfnamefont{H.}~\bibnamefont{Asada}}, \bibinfo {author}
  {\bibfnamefont{J.}~\bibnamefont{Yamamoto}}, \bibinfo {author}
  {\bibfnamefont{F.}~\bibnamefont{Odawara}},\ and\ \bibinfo {author}
  {\bibfnamefont{T.}~\bibnamefont{Koyanagi}}}%
  , \bibinfo {year} {2008},\ \bibfield{title}{%
  \enquote{\bibinfo {title} {Large magnetic circular dichroism of {Co} clusters
  in {Co}-doped {ZnO}},}\ }%
  \bibfield{journal}{%
  \bibinfo {journal} {Appl. Phys. Lett.}\ }%
  \textbf{\bibinfo {volume} {93}},\ \bibinfo {pages} {142510}%
  \bibAnnoteFile{NoStop}{Fukuma:2008_APLb}%
\bibitem[{\citenamefont{Fukumura}\ \emph{et~al.}(2005)\citenamefont{Fukumura},
  \citenamefont{Toyosaki},\ and\ \citenamefont{Yamada}}]{Fukumura:2005_SST}%
  \BibitemOpen
  \bibfield{author}{%
  \bibinfo {author} {\bibnamefont{Fukumura}, \bibfnamefont{T.}}, \bibinfo
  {author} {\bibfnamefont{H.}~\bibnamefont{Toyosaki}},\ and\ \bibinfo {author}
  {\bibfnamefont{Y.}~\bibnamefont{Yamada}}}%
  , \bibinfo {year} {2005},\ \bibfield{title}{%
  \enquote{\bibinfo {title} {Magnetic oxide semiconductors},}\ }%
  \bibfield{journal}{%
  \bibinfo {journal} {Semicond. Sci. Technol.}\ }%
  \textbf{\bibinfo {volume} {20}},\ \bibinfo {pages} {S103}%
  \bibAnnoteFile{NoStop}{Fukumura:2005_SST}%
\bibitem[{\citenamefont{Furdyna}\ and\
  \citenamefont{Kossut}(1988)}]{Furdyna:1988_B}%
  \BibitemOpen
  \bibfield{author}{%
  \bibinfo {author} {\bibnamefont{Furdyna}, \bibfnamefont{J.~K.}},\ and\
  \bibinfo {author} {\bibfnamefont{J.}~\bibnamefont{Kossut}}}%
  , \bibinfo {year} {1988},\ \emph{\bibinfo {title} {(eds.) Diluted Magnetic
  Semiconductors}},\ \bibinfo {series} {{Semiconductors and Semimetals}},
  Vol.~\bibinfo {volume} {25}\ (\bibinfo {publisher} {Academic Press},\
  \bibinfo {address} {New York})%
  \bibAnnoteFile{NoStop}{Furdyna:1988_B}%
\bibitem[{\citenamefont{Gaj}\ \emph{et~al.}(1979)\citenamefont{Gaj},
  \citenamefont{Planel},\ and\ \citenamefont{Fishman}}]{Gaj:1979_SSC}%
  \BibitemOpen
  \bibfield{author}{%
  \bibinfo {author} {\bibnamefont{Gaj}, \bibfnamefont{J.~A.}}, \bibinfo
  {author} {\bibfnamefont{R.}~\bibnamefont{Planel}},\ and\ \bibinfo {author}
  {\bibfnamefont{G.}~\bibnamefont{Fishman}}}%
  , \bibinfo {year} {1979},\ \bibfield{journal}{%
  \bibinfo {journal} {Solid State Commun.}\ }%
  \textbf{\bibinfo {volume} {29}},\ \bibinfo {pages} {435}%
  \bibAnnoteFile{NoStop}{Gaj:1979_SSC}%
\bibitem[{\citenamefont{Garate}\ \emph{et~al.}(2009)\citenamefont{Garate},
  \citenamefont{Gilmore}, \citenamefont{Stiles},\ and\
  \citenamefont{MacDonald}}]{Garate:2009_PRBc}%
  \BibitemOpen
  \bibfield{author}{%
  \bibinfo {author} {\bibnamefont{Garate}, \bibfnamefont{I.}}, \bibinfo
  {author} {\bibfnamefont{K.}~\bibnamefont{Gilmore}}, \bibinfo {author}
  {\bibfnamefont{M.~D.}\ \bibnamefont{Stiles}},\ and\ \bibinfo {author}
  {\bibfnamefont{A.~H.}\ \bibnamefont{MacDonald}}}%
  , \bibinfo {year} {2009},\ \bibfield{title}{%
  \enquote{\bibinfo {title} {Nonadiabatic spin-transfer torque in real
  materials},}\ }%
  \bibfield{journal}{%
  \bibinfo {journal} {Phys. Rev. B}\ }%
  \textbf{\bibinfo {volume} {79}},\ \bibinfo {pages} {104416}%
  \bibAnnoteFile{NoStop}{Garate:2009_PRBc}%
\bibitem[{\citenamefont{Garate}\ and\
  \citenamefont{MacDonald}(2009)}]{Garate:2009_PRBb}%
  \BibitemOpen
  \bibfield{author}{%
  \bibinfo {author} {\bibnamefont{Garate}, \bibfnamefont{I.}},\ and\ \bibinfo
  {author} {\bibfnamefont{Allan}\ \bibnamefont{MacDonald}}}%
  , \bibinfo {year} {2009},\ \bibfield{title}{%
  \enquote{\bibinfo {title} {Gilbert damping in conducting ferromagnets. {II}.
  {Model} tests of the torque-correlation formula},}\ }%
  \bibfield{journal}{%
  \bibinfo {journal} {Phys. Rev. B}\ }%
  \textbf{\bibinfo {volume} {79}},\ \bibinfo {pages} {064404}%
  \bibAnnoteFile{NoStop}{Garate:2009_PRBb}%
\bibitem[{\citenamefont{Gareev}\ \emph{et~al.}(2010)\citenamefont{Gareev},
  \citenamefont{Petukhov}, \citenamefont{Schlapps}, \citenamefont{Sadowski},\
  and\ \citenamefont{Wegscheider}}]{Gareev:2010_APL}%
  \BibitemOpen
  \bibfield{author}{%
  \bibinfo {author} {\bibnamefont{Gareev}, \bibfnamefont{R.~R.}}, \bibinfo
  {author} {\bibfnamefont{A.}~\bibnamefont{Petukhov}}, \bibinfo {author}
  {\bibfnamefont{M.}~\bibnamefont{Schlapps}}, \bibinfo {author}
  {\bibfnamefont{J.}~\bibnamefont{Sadowski}},\ and\ \bibinfo {author}
  {\bibfnamefont{W.}~\bibnamefont{Wegscheider}}}%
  , \bibinfo {year} {2010},\ \bibfield{title}{%
  \enquote{\bibinfo {title} {Giant anisotropic magnetoresistance in insulating
  ultrathin {(Ga,Mn)As}},}\ }%
  \bibfield{journal}{%
  \bibinfo {journal} {Appl. Phys. Lett.}\ }%
  \textbf{\bibinfo {volume} {96}},\ \bibinfo {eid} {052114}%
  \bibAnnoteFile{NoStop}{Gareev:2010_APL}%
\bibitem[{\citenamefont{Ge}\ \emph{et~al.}(2007)\citenamefont{Ge},
  \citenamefont{Lim}, \citenamefont{Shen}, \citenamefont{Zhou},
  \citenamefont{Liu}, \citenamefont{Furdyna},\ and\
  \citenamefont{Dobrowolska}}]{Ge:2007_PRB}%
  \BibitemOpen
  \bibfield{author}{%
  \bibinfo {author} {\bibnamefont{Ge}, \bibfnamefont{Z.}}, \bibinfo {author}
  {\bibfnamefont{W.~L.}\ \bibnamefont{Lim}}, \bibinfo {author}
  {\bibfnamefont{S.}~\bibnamefont{Shen}}, \bibinfo {author}
  {\bibfnamefont{Y.~Y.}\ \bibnamefont{Zhou}}, \bibinfo {author}
  {\bibfnamefont{X.}~\bibnamefont{Liu}}, \bibinfo {author}
  {\bibfnamefont{J.~K.}\ \bibnamefont{Furdyna}},\ and\ \bibinfo {author}
  {\bibfnamefont{M.}~\bibnamefont{Dobrowolska}}}%
  , \bibinfo {year} {2007},\ \bibfield{title}{%
  \enquote{\bibinfo {title} {Magnetization reversal in {(Ga,Mn)As/MnO}
  exchange-biased structures: {Investigation} by planar {Hall} effect},}\ }%
  \bibfield{journal}{%
  \bibinfo {journal} {Phys. Rev. B}\ }%
  \textbf{\bibinfo {volume} {75}},\ \bibinfo {pages} {014407}%
  \bibAnnoteFile{NoStop}{Ge:2007_PRB}%
\bibitem[{\citenamefont{Geresdi}\ \emph{et~al.}(2008)\citenamefont{Geresdi},
  \citenamefont{Halbritter}, \citenamefont{Csontos}, \citenamefont{Csonka},
  \citenamefont{Mih\'aly}, \citenamefont{Wojtowicz}, \citenamefont{Liu},
  \citenamefont{Jank\'o},\ and\ \citenamefont{Furdyna}}]{Geresdi:2008_PRB}%
  \BibitemOpen
  \bibfield{author}{%
  \bibinfo {author} {\bibnamefont{Geresdi}, \bibfnamefont{A.}}, \bibinfo
  {author} {\bibfnamefont{A.}~\bibnamefont{Halbritter}}, \bibinfo {author}
  {\bibfnamefont{M.}~\bibnamefont{Csontos}}, \bibinfo {author}
  {\bibfnamefont{Sz.}\ \bibnamefont{Csonka}}, \bibinfo {author}
  {\bibfnamefont{G.}~\bibnamefont{Mih\'aly}}, \bibinfo {author}
  {\bibfnamefont{T.}~\bibnamefont{Wojtowicz}}, \bibinfo {author}
  {\bibfnamefont{X.}~\bibnamefont{Liu}}, \bibinfo {author}
  {\bibfnamefont{B.}~\bibnamefont{Jank\'o}},\ and\ \bibinfo {author}
  {\bibfnamefont{J.~K.}\ \bibnamefont{Furdyna}}}%
  , \bibinfo {year} {2008},\ \bibfield{title}{%
  \enquote{\bibinfo {title} {Nanoscale spin polarization in the dilute magnetic
  semiconductor {(In,Mn)Sb}},}\ }%
  \bibfield{journal}{%
  \bibinfo {journal} {Phys. Rev. B}\ }%
  \textbf{\bibinfo {volume} {77}},\ \bibinfo {pages} {233304}%
  \bibAnnoteFile{NoStop}{Geresdi:2008_PRB}%
\bibitem[{\citenamefont{Giddings}\ \emph{et~al.}(2008)\citenamefont{Giddings},
  \citenamefont{Jungwirth},\ and\
  \citenamefont{Gallagher}}]{Giddings:2008_PRB}%
  \BibitemOpen
  \bibfield{author}{%
  \bibinfo {author} {\bibnamefont{Giddings}, \bibfnamefont{A.~D.}}, \bibinfo
  {author} {\bibfnamefont{T.}~\bibnamefont{Jungwirth}},\ and\ \bibinfo {author}
  {\bibfnamefont{B.~L.}\ \bibnamefont{Gallagher}}}%
  , \bibinfo {year} {2008},\ \bibfield{title}{%
  \enquote{\bibinfo {title} {{(Ga,Mn)As} based superlattices and the search for
  antiferromagnetic interlayer coupling},}\ }%
  \bibfield{journal}{%
  \bibinfo {journal} {Phys. Rev. B}\ }%
  \textbf{\bibinfo {volume} {78}},\ \bibinfo {pages} {165312}%
  \bibAnnoteFile{NoStop}{Giddings:2008_PRB}%
\bibitem[{\citenamefont{Giddings}\ \emph{et~al.}(2005)\citenamefont{Giddings},
  \citenamefont{Khalid}, \citenamefont{Jungwirth}, \citenamefont{Wunderlich},
  \citenamefont{Yasin}, \citenamefont{Campion}, \citenamefont{Edmonds},
  \citenamefont{Sinova}, \citenamefont{Ito}, \citenamefont{Wang},
  \citenamefont{Williams}, \citenamefont{Gallagher},\ and\
  \citenamefont{Foxon}}]{Giddings:2005_PRL}%
  \BibitemOpen
  \bibfield{author}{%
  \bibinfo {author} {\bibnamefont{Giddings}, \bibfnamefont{A.~D.}}, \bibinfo
  {author} {\bibfnamefont{M.~N.}\ \bibnamefont{Khalid}}, \bibinfo {author}
  {\bibfnamefont{T.}~\bibnamefont{Jungwirth}}, \bibinfo {author}
  {\bibfnamefont{J.}~\bibnamefont{Wunderlich}}, \bibinfo {author}
  {\bibfnamefont{S.}~\bibnamefont{Yasin}}, \bibinfo {author}
  {\bibfnamefont{R.~P.}\ \bibnamefont{Campion}}, \bibinfo {author}
  {\bibfnamefont{K.~W.}\ \bibnamefont{Edmonds}}, \bibinfo {author}
  {\bibfnamefont{J.}~\bibnamefont{Sinova}}, \bibinfo {author}
  {\bibfnamefont{K.}~\bibnamefont{Ito}}, \bibinfo {author}
  {\bibfnamefont{K.~Y.}\ \bibnamefont{Wang}}, \bibinfo {author}
  {\bibfnamefont{D.}~\bibnamefont{Williams}}, \bibinfo {author}
  {\bibfnamefont{B.~L.}\ \bibnamefont{Gallagher}},\ and\ \bibinfo {author}
  {\bibfnamefont{C.~T.}\ \bibnamefont{Foxon}}}%
  , \bibinfo {year} {2005},\ \bibfield{title}{%
  \enquote{\bibinfo {title} {Large tunneling anisotropic magnetoresistance in
  {(Ga,Mn)As} nanoconstrictions},}\ }%
  \bibfield{journal}{%
  \bibinfo {journal} {Phys. Rev. Lett.}\ }%
  \textbf{\bibinfo {volume} {94}},\ \bibinfo {pages} {127202}%
  \bibAnnoteFile{NoStop}{Giddings:2005_PRL}%
\bibitem[{\citenamefont{Glas}\ \emph{et~al.}(2004)\citenamefont{Glas},
  \citenamefont{Patriarche}, \citenamefont{Largeau},\ and\
  \citenamefont{Lemaitre}}]{Glas:2004_PRL}%
  \BibitemOpen
  \bibfield{author}{%
  \bibinfo {author} {\bibnamefont{Glas}, \bibfnamefont{F.}}, \bibinfo {author}
  {\bibfnamefont{G.}~\bibnamefont{Patriarche}}, \bibinfo {author}
  {\bibfnamefont{L.}~\bibnamefont{Largeau}},\ and\ \bibinfo {author}
  {\bibfnamefont{A.}~\bibnamefont{Lemaitre}}}%
  , \bibinfo {year} {2004},\ \bibfield{title}{%
  \enquote{\bibinfo {title} {Determination of the local concentrations of {Mn}
  interstitials and antisite defects in {GaMnAs}},}\ }%
  \bibfield{journal}{%
  \bibinfo {journal} {Phys. Rev. Lett.}\ }%
  \textbf{\bibinfo {volume} {93}},\ \bibinfo {pages} {086107}%
  \bibAnnoteFile{NoStop}{Glas:2004_PRL}%
\bibitem[{\citenamefont{Glunk}\ \emph{et~al.}(2009)\citenamefont{Glunk},
  \citenamefont{Daeubler}, \citenamefont{Dreher}, \citenamefont{Schwaiger},
  \citenamefont{Schoch}, \citenamefont{Sauer}, \citenamefont{Limmer},
  \citenamefont{Brandlmaier}, \citenamefont{Goennenwein},
  \citenamefont{Bihler},\ and\ \citenamefont{Brandt}}]{Glunk:2009_PRB}%
  \BibitemOpen
  \bibfield{author}{%
  \bibinfo {author} {\bibnamefont{Glunk}, \bibfnamefont{M.}}, \bibinfo {author}
  {\bibfnamefont{J.}~\bibnamefont{Daeubler}}, \bibinfo {author}
  {\bibfnamefont{L.}~\bibnamefont{Dreher}}, \bibinfo {author}
  {\bibfnamefont{S.}~\bibnamefont{Schwaiger}}, \bibinfo {author}
  {\bibfnamefont{W.}~\bibnamefont{Schoch}}, \bibinfo {author}
  {\bibfnamefont{R.}~\bibnamefont{Sauer}}, \bibinfo {author}
  {\bibfnamefont{W.}~\bibnamefont{Limmer}}, \bibinfo {author}
  {\bibfnamefont{A.}~\bibnamefont{Brandlmaier}}, \bibinfo {author}
  {\bibfnamefont{S.~T.~B.}\ \bibnamefont{Goennenwein}}, \bibinfo {author}
  {\bibfnamefont{C.}~\bibnamefont{Bihler}},\ and\ \bibinfo {author}
  {\bibfnamefont{M.~S.}\ \bibnamefont{Brandt}}}%
  , \bibinfo {year} {2009},\ \bibfield{title}{%
  \enquote{\bibinfo {title} {Magnetic anisotropy in {(Ga,Mn)As}: {Influence} of
  epitaxial strain and hole concentration},}\ }%
  \bibfield{journal}{%
  \bibinfo {journal} {Phys. Rev. B}\ }%
  \textbf{\bibinfo {volume} {79}},\ \bibinfo {pages} {195206}%
  \bibAnnoteFile{NoStop}{Glunk:2009_PRB}%
\bibitem[{\citenamefont{Godlewski}\
  \emph{et~al.}(2010)\citenamefont{Godlewski}, \citenamefont{Wasiakowski},
  \citenamefont{Ivanov}, \citenamefont{{W\'ojcik-G{\l}odowska}},
  \citenamefont{{{\L}ukasiewicz}}, \citenamefont{Guziewicz},
  \citenamefont{Jakie{\l}a}, \citenamefont{Kopalko},
  \citenamefont{Zakrzewski},\ and\ \citenamefont{Dumont}}]{Godlewski:2010_OM}%
  \BibitemOpen
  \bibfield{author}{%
  \bibinfo {author} {\bibnamefont{Godlewski}, \bibfnamefont{M.}}, \bibinfo
  {author} {\bibfnamefont{A.}~\bibnamefont{Wasiakowski}}, \bibinfo {author}
  {\bibfnamefont{V.~Yu.}\ \bibnamefont{Ivanov}}, \bibinfo {author}
  {\bibfnamefont{A.}~\bibnamefont{{W\'ojcik-G{\l}odowska}}}, \bibinfo {author}
  {\bibfnamefont{M.}~\bibnamefont{{{\L}ukasiewicz}}}, \bibinfo {author}
  {\bibfnamefont{E.}~\bibnamefont{Guziewicz}}, \bibinfo {author}
  {\bibfnamefont{R.}~\bibnamefont{Jakie{\l}a}}, \bibinfo {author}
  {\bibfnamefont{K.}~\bibnamefont{Kopalko}}, \bibinfo {author}
  {\bibfnamefont{A.}~\bibnamefont{Zakrzewski}},\ and\ \bibinfo {author}
  {\bibfnamefont{Y.}~\bibnamefont{Dumont}}}%
  , \bibinfo {year} {2010},\ \bibfield{title}{%
  \enquote{\bibinfo {title} {Puzzling magneto-optical properties of {ZnMnO}
  films},}\ }%
  \bibfield{journal}{%
  \bibinfo {journal} {Opt. Mater.}\ }%
  \textbf{\bibinfo {volume} {32}},\ \bibinfo {pages} {680}%
  \bibAnnoteFile{NoStop}{Godlewski:2010_OM}%
\bibitem[{\citenamefont{Goennenwein}\
  \emph{et~al.}(2008)\citenamefont{Goennenwein}, \citenamefont{Althammer},
  \citenamefont{Bihler}, \citenamefont{Brandlmaier}, \citenamefont{Geprags},
  \citenamefont{Opel}, \citenamefont{Schoch}, \citenamefont{Limmer},
  \citenamefont{Gross},\ and\ \citenamefont{Brandt}}]{Goennenwein:2008_pss}%
  \BibitemOpen
  \bibfield{author}{%
  \bibinfo {author} {\bibnamefont{Goennenwein}, \bibfnamefont{S.~T.~B.}},
  \bibinfo {author} {\bibfnamefont{M.}~\bibnamefont{Althammer}}, \bibinfo
  {author} {\bibfnamefont{C.}~\bibnamefont{Bihler}}, \bibinfo {author}
  {\bibfnamefont{A.}~\bibnamefont{Brandlmaier}}, \bibinfo {author}
  {\bibfnamefont{S.}~\bibnamefont{Geprags}}, \bibinfo {author}
  {\bibfnamefont{M.}~\bibnamefont{Opel}}, \bibinfo {author}
  {\bibfnamefont{W.}~\bibnamefont{Schoch}}, \bibinfo {author}
  {\bibfnamefont{W.}~\bibnamefont{Limmer}}, \bibinfo {author}
  {\bibfnamefont{R.}~\bibnamefont{Gross}},\ and\ \bibinfo {author}
  {\bibfnamefont{M.~S.}\ \bibnamefont{Brandt}}}%
  , \bibinfo {year} {2008},\ \bibfield{title}{%
  \enquote{\bibinfo {title} {Piezo-voltage control of magnetization orientation
  in a ferromagnetic semiconductor},}\ }%
  \bibfield{journal}{%
  \bibinfo {journal} {Phys. Status Solidi (RRL)}\ }%
  \textbf{\bibinfo {volume} {2}},\ \bibinfo {pages} {96}%
  \bibAnnoteFile{NoStop}{Goennenwein:2008_pss}%
\bibitem[{\citenamefont{Goennenwein}\
  \emph{et~al.}(2004)\citenamefont{Goennenwein}, \citenamefont{Wassner},
  \citenamefont{Huebl}, \citenamefont{Brandt}, \citenamefont{Philipp},
  \citenamefont{Opel}, \citenamefont{Gross}, \citenamefont{Koeder},
  \citenamefont{Schoch},\ and\ \citenamefont{Waag}}]{Goennenwein:2004_PRL}%
  \BibitemOpen
  \bibfield{author}{%
  \bibinfo {author} {\bibnamefont{Goennenwein}, \bibfnamefont{S.~T.~B.}},
  \bibinfo {author} {\bibfnamefont{T.~A.}\ \bibnamefont{Wassner}}, \bibinfo
  {author} {\bibfnamefont{H.}~\bibnamefont{Huebl}}, \bibinfo {author}
  {\bibfnamefont{M.~S.}\ \bibnamefont{Brandt}}, \bibinfo {author}
  {\bibfnamefont{J.~B.}\ \bibnamefont{Philipp}}, \bibinfo {author}
  {\bibfnamefont{M.}~\bibnamefont{Opel}}, \bibinfo {author}
  {\bibfnamefont{R.}~\bibnamefont{Gross}}, \bibinfo {author}
  {\bibfnamefont{A.}~\bibnamefont{Koeder}}, \bibinfo {author}
  {\bibfnamefont{W.}~\bibnamefont{Schoch}},\ and\ \bibinfo {author}
  {\bibfnamefont{A.}~\bibnamefont{Waag}}}%
  , \bibinfo {year} {2004},\ \bibfield{title}{%
  \enquote{\bibinfo {title} {Hydrogen control of ferromagnetism in a dilute
  magnetic semiconductor},}\ }%
  \bibfield{journal}{%
  \bibinfo {journal} {Phys. Rev. Lett.}\ }%
  \textbf{\bibinfo {volume} {92}},\ \bibinfo {pages} {227202}%
  \bibAnnoteFile{NoStop}{Goennenwein:2004_PRL}%
\bibitem[{\citenamefont{{Gonzalez Szwacki}}\
  \emph{et~al.}(2011)\citenamefont{{Gonzalez Szwacki}},
  \citenamefont{Majewski},\ and\ \citenamefont{Dietl}}]{Gonzalez:2011_PRB}%
  \BibitemOpen
  \bibfield{author}{%
  \bibinfo {author} {\bibnamefont{{Gonzalez Szwacki}}, \bibfnamefont{N.}},
  \bibinfo {author} {\bibfnamefont{J.~A.}\ \bibnamefont{Majewski}},\ and\
  \bibinfo {author} {\bibfnamefont{T.}~\bibnamefont{Dietl}}}%
  , \bibinfo {year} {2011},\ \bibfield{title}{%
  \enquote{\bibinfo {title} {Aggregation and magnetism of {Cr, Mn, and Fe
  cations in GaN}},}\ }%
  \bibfield{journal}{%
  \bibinfo {journal} {Phys. Rev. B}\ }%
  \textbf{\bibinfo {volume} {83}},\ \bibinfo {pages} {184417}%
  \bibAnnoteFile{NoStop}{Gonzalez:2011_PRB}%
\bibitem[{\citenamefont{Goodenough}(1958)}]{Goodenough:1958_JPCS}%
  \BibitemOpen
  \bibfield{author}{%
  \bibinfo {author} {\bibnamefont{Goodenough}, \bibfnamefont{John~B.}}}%
  , \bibinfo {year} {1958},\ \bibfield{title}{%
  \enquote{\bibinfo {title} {An interpretation of the magnetic properties of
  the perovskite-type mixed crystals {La$_{1-x}$Sr$_{x}$CoO}$_{3-\lambda}$},}\
  }%
  \bibfield{journal}{%
  \bibinfo {journal} {J. Phys. Chem. Solids}\ }%
  \textbf{\bibinfo {volume} {6}},\ \bibinfo {pages} {287}%
  \bibAnnoteFile{NoStop}{Goodenough:1958_JPCS}%
\bibitem[{\citenamefont{Goryca}\ \emph{et~al.}(2009)\citenamefont{Goryca},
  \citenamefont{Kazimierczuk}, \citenamefont{Nawrocki}, \citenamefont{Golnik},
  \citenamefont{Gaj}, \citenamefont{Kossacki}, \citenamefont{Wojnar},\ and\
  \citenamefont{Karczewski}}]{Goryca:2009_PRLb}%
  \BibitemOpen
  \bibfield{author}{%
  \bibinfo {author} {\bibnamefont{Goryca}, \bibfnamefont{M.}}, \bibinfo
  {author} {\bibfnamefont{T.}~\bibnamefont{Kazimierczuk}}, \bibinfo {author}
  {\bibfnamefont{M.}~\bibnamefont{Nawrocki}}, \bibinfo {author}
  {\bibfnamefont{A.}~\bibnamefont{Golnik}}, \bibinfo {author}
  {\bibfnamefont{J.~A.}\ \bibnamefont{Gaj}}, \bibinfo {author}
  {\bibfnamefont{P.}~\bibnamefont{Kossacki}}, \bibinfo {author}
  {\bibfnamefont{P.}~\bibnamefont{Wojnar}},\ and\ \bibinfo {author}
  {\bibfnamefont{G.}~\bibnamefont{Karczewski}}}%
  , \bibinfo {year} {2009},\ \bibfield{title}{%
  \enquote{\bibinfo {title} {Optical manipulation of a single {Mn} spin in a
  {CdTe}-based quantum dot},}\ }%
  \bibfield{journal}{%
  \bibinfo {journal} {Phys. Rev. Lett.}\ }%
  \textbf{\bibinfo {volume} {103}},\ \bibinfo {pages} {087401}%
  \bibAnnoteFile{NoStop}{Goryca:2009_PRLb}%
\bibitem[{\citenamefont{Gosk}\ \emph{et~al.}(2005)\citenamefont{Gosk},
  \citenamefont{Zaj{\c{a}}c}, \citenamefont{Wo{\l}o{\'s}},
  \citenamefont{Kami{\'n}ska}, \citenamefont{Twardowski},
  \citenamefont{Grzegory}, \citenamefont{Bockowski},\ and\
  \citenamefont{Porowski}}]{Gosk:2005_PRB}%
  \BibitemOpen
  \bibfield{author}{%
  \bibinfo {author} {\bibnamefont{Gosk}, \bibfnamefont{J.}}, \bibinfo {author}
  {\bibfnamefont{M.}~\bibnamefont{Zaj{\c{a}}c}}, \bibinfo {author}
  {\bibfnamefont{A.}~\bibnamefont{Wo{\l}o{\'s}}}, \bibinfo {author}
  {\bibfnamefont{M.}~\bibnamefont{Kami{\'n}ska}}, \bibinfo {author}
  {\bibfnamefont{A.}~\bibnamefont{Twardowski}}, \bibinfo {author}
  {\bibfnamefont{I.}~\bibnamefont{Grzegory}}, \bibinfo {author}
  {\bibfnamefont{M.}~\bibnamefont{Bockowski}},\ and\ \bibinfo {author}
  {\bibfnamefont{S.}~\bibnamefont{Porowski}}}%
  , \bibinfo {year} {2005},\ \bibfield{title}{%
  \enquote{\bibinfo {title} {Magnetic anisotropy of bulk {GaN:Mn} single
  crystals codoped with {Mg} acceptors},}\ }%
  \bibfield{journal}{%
  \bibinfo {journal} {Phys. Rev. B}\ }%
  \textbf{\bibinfo {volume} {71}},\ \bibinfo {pages} {094432}%
  \bibAnnoteFile{NoStop}{Gosk:2005_PRB}%
\bibitem[{\citenamefont{Gould}\ \emph{et~al.}(2008)\citenamefont{Gould},
  \citenamefont{Mark}, \citenamefont{Pappert}, \citenamefont{Dengel},
  \citenamefont{Wenisch}, \citenamefont{Campion}, \citenamefont{Rushforth},
  \citenamefont{Chiba}, \citenamefont{Li}, \citenamefont{Liu},
  \citenamefont{Roy}, \citenamefont{Ohno}, \citenamefont{Furdyna},
  \citenamefont{Gallagher}, \citenamefont{Brunner}, \citenamefont{Schmidt},\
  and\ \citenamefont{Molenkamp}}]{Gould:2008_NJP}%
  \BibitemOpen
  \bibfield{author}{%
  \bibinfo {author} {\bibnamefont{Gould}, \bibfnamefont{C.}}, \bibinfo {author}
  {\bibfnamefont{S.}~\bibnamefont{Mark}}, \bibinfo {author}
  {\bibfnamefont{K.}~\bibnamefont{Pappert}}, \bibinfo {author}
  {\bibfnamefont{R.~G.}\ \bibnamefont{Dengel}}, \bibinfo {author}
  {\bibfnamefont{J.}~\bibnamefont{Wenisch}}, \bibinfo {author}
  {\bibfnamefont{R.~P.}\ \bibnamefont{Campion}}, \bibinfo {author}
  {\bibfnamefont{A.~W.}\ \bibnamefont{Rushforth}}, \bibinfo {author}
  {\bibfnamefont{D.}~\bibnamefont{Chiba}}, \bibinfo {author}
  {\bibfnamefont{Z.}~\bibnamefont{Li}}, \bibinfo {author}
  {\bibfnamefont{X.}~\bibnamefont{Liu}}, \bibinfo {author}
  {\bibfnamefont{W.~Van}\ \bibnamefont{Roy}}, \bibinfo {author}
  {\bibfnamefont{H.}~\bibnamefont{Ohno}}, \bibinfo {author}
  {\bibfnamefont{J.~K.}\ \bibnamefont{Furdyna}}, \bibinfo {author}
  {\bibfnamefont{B.}~\bibnamefont{Gallagher}}, \bibinfo {author}
  {\bibfnamefont{K.}~\bibnamefont{Brunner}}, \bibinfo {author}
  {\bibfnamefont{G.}~\bibnamefont{Schmidt}},\ and\ \bibinfo {author}
  {\bibfnamefont{L.~W.}\ \bibnamefont{Molenkamp}}}%
  , \bibinfo {year} {2008},\ \bibfield{title}{%
  \enquote{\bibinfo {title} {An extensive comparison of anisotropies in {MBE}
  grown {(Ga,Mn)As} material},}\ }%
  \bibfield{journal}{%
  \bibinfo {journal} {New J. Phys.}\ }%
  \textbf{\bibinfo {volume} {10}},\ \bibinfo {pages} {055007}%
  \bibAnnoteFile{NoStop}{Gould:2008_NJP}%
\bibitem[{\citenamefont{Gould}\ \emph{et~al.}(2004)\citenamefont{Gould},
  \citenamefont{{R{\"u}ster}}, \citenamefont{Jungwirth}, \citenamefont{Girgis},
  \citenamefont{Schott}, \citenamefont{Giraud}, \citenamefont{Brunner},
  \citenamefont{Schmidt},\ and\ \citenamefont{Molenkamp}}]{Gould:2004_PRL}%
  \BibitemOpen
  \bibfield{author}{%
  \bibinfo {author} {\bibnamefont{Gould}, \bibfnamefont{C.}}, \bibinfo {author}
  {\bibfnamefont{C.}~\bibnamefont{{R{\"u}ster}}}, \bibinfo {author}
  {\bibfnamefont{T.}~\bibnamefont{Jungwirth}}, \bibinfo {author}
  {\bibfnamefont{E.}~\bibnamefont{Girgis}}, \bibinfo {author}
  {\bibfnamefont{G.~M.}\ \bibnamefont{Schott}}, \bibinfo {author}
  {\bibfnamefont{R.}~\bibnamefont{Giraud}}, \bibinfo {author}
  {\bibfnamefont{K.}~\bibnamefont{Brunner}}, \bibinfo {author}
  {\bibfnamefont{G.}~\bibnamefont{Schmidt}},\ and\ \bibinfo {author}
  {\bibfnamefont{L.~W.}\ \bibnamefont{Molenkamp}}}%
  , \bibinfo {year} {2004},\ \bibfield{title}{%
  \enquote{\bibinfo {title} {Tunneling anisotropic magnetoresistance: A
  spin-valve like tunnel magnetoresistance using a single magnetic layer},}\ }%
  \bibfield{journal}{%
  \bibinfo {journal} {Phys. Rev. Lett.}\ }%
  \textbf{\bibinfo {volume} {93}},\ \bibinfo {pages} {117203}%
  \bibAnnoteFile{NoStop}{Gould:2004_PRL}%
\bibitem[{\citenamefont{Gourdon}\ \emph{et~al.}(2007)\citenamefont{Gourdon},
  \citenamefont{Dourlat}, \citenamefont{Jeudy}, \citenamefont{Khazen},
  \citenamefont{von Bardeleben}, \citenamefont{Thevenard},\ and\
  \citenamefont{Lema\^itre}}]{Gourdon:2007_PRB}%
  \BibitemOpen
  \bibfield{author}{%
  \bibinfo {author} {\bibnamefont{Gourdon}, \bibfnamefont{C.}}, \bibinfo
  {author} {\bibfnamefont{A.}~\bibnamefont{Dourlat}}, \bibinfo {author}
  {\bibfnamefont{V.}~\bibnamefont{Jeudy}}, \bibinfo {author}
  {\bibfnamefont{K.}~\bibnamefont{Khazen}}, \bibinfo {author}
  {\bibfnamefont{H.~J.}\ \bibnamefont{von Bardeleben}}, \bibinfo {author}
  {\bibfnamefont{L.}~\bibnamefont{Thevenard}},\ and\ \bibinfo {author}
  {\bibfnamefont{A.}~\bibnamefont{Lema\^itre}}}%
  , \bibinfo {year} {2007},\ \bibfield{title}{%
  \enquote{\bibinfo {title} {Determination of the micromagnetic parameters in
  {(Ga,Mn)As} using domain theory},}\ }%
  \bibfield{journal}{%
  \bibinfo {journal} {Phys. Rev. B}\ }%
  \textbf{\bibinfo {volume} {76}},\ \bibinfo {pages} {241301}%
  \bibAnnoteFile{NoStop}{Gourdon:2007_PRB}%
\bibitem[{\citenamefont{Grace}\ \emph{et~al.}(2009)\citenamefont{Grace},
  \citenamefont{Venkatesan}, \citenamefont{Alaria}, \citenamefont{Coey},
  \citenamefont{Kopnov},\ and\ \citenamefont{Naaman}}]{Grace:2009_AM}%
  \BibitemOpen
  \bibfield{author}{%
  \bibinfo {author} {\bibnamefont{Grace}, \bibfnamefont{P.~J.}}, \bibinfo
  {author} {\bibfnamefont{M.}~\bibnamefont{Venkatesan}}, \bibinfo {author}
  {\bibfnamefont{J.}~\bibnamefont{Alaria}}, \bibinfo {author}
  {\bibfnamefont{J.~M.~D.}\ \bibnamefont{Coey}}, \bibinfo {author}
  {\bibfnamefont{G.}~\bibnamefont{Kopnov}},\ and\ \bibinfo {author}
  {\bibfnamefont{R.}~\bibnamefont{Naaman}}}%
  , \bibinfo {year} {2009},\ \bibfield{title}{%
  \enquote{\bibinfo {title} {The origin of the magnetism of etched
  {Silicon}},}\ }%
  \bibfield{journal}{%
  \bibinfo {journal} {Adv. Mater.}\ }%
  \textbf{\bibinfo {volume} {21}},\ \bibinfo {pages} {71}%
  \bibAnnoteFile{NoStop}{Grace:2009_AM}%
\bibitem[{\citenamefont{Graf}\
  \emph{et~al.}(2003{\natexlab{a}})\citenamefont{Graf}, \citenamefont{Gjukic},
  \citenamefont{Hermann}, \citenamefont{Brandt}, \citenamefont{Stutzmann},\
  and\ \citenamefont{Ambacher}}]{Graf:2003_PRB}%
  \BibitemOpen
  \bibfield{author}{%
  \bibinfo {author} {\bibnamefont{Graf}, \bibfnamefont{T.}}, \bibinfo {author}
  {\bibfnamefont{M.}~\bibnamefont{Gjukic}}, \bibinfo {author}
  {\bibfnamefont{M.}~\bibnamefont{Hermann}}, \bibinfo {author}
  {\bibfnamefont{M.~S.}\ \bibnamefont{Brandt}}, \bibinfo {author}
  {\bibfnamefont{M.}~\bibnamefont{Stutzmann}},\ and\ \bibinfo {author}
  {\bibfnamefont{O.}~\bibnamefont{Ambacher}}}%
  , \bibinfo {year} {2003}{\natexlab{a}},\ \bibfield{title}{%
  \enquote{\bibinfo {title} {Spin resonance investigations of {Mn}$^{2+}$ in
  wurtzite {GaN} and {AlN} films},}\ }%
  \bibfield{journal}{%
  \bibinfo {journal} {Phys. Rev. B}\ }%
  \textbf{\bibinfo {volume} {67}},\ \bibinfo {pages} {165215}%
  \bibAnnoteFile{NoStop}{Graf:2003_PRB}%
\bibitem[{\citenamefont{Graf}\
  \emph{et~al.}(2003{\natexlab{b}})\citenamefont{Graf},
  \citenamefont{Goennenwein},\ and\ \citenamefont{Brandt}}]{Graf:2003_PSSB}%
  \BibitemOpen
  \bibfield{author}{%
  \bibinfo {author} {\bibnamefont{Graf}, \bibfnamefont{T.}}, \bibinfo {author}
  {\bibfnamefont{S.~T.~B.}\ \bibnamefont{Goennenwein}},\ and\ \bibinfo {author}
  {\bibfnamefont{M.~S.}\ \bibnamefont{Brandt}}}%
  , \bibinfo {year} {2003}{\natexlab{b}},\ \bibfield{title}{%
  \enquote{\bibinfo {title} {Prospects for carrier-mediated ferromagnetism in
  {GaN}},}\ }%
  \bibfield{journal}{%
  \bibinfo {journal} {Phys. Status Solidi B}\ }%
  \textbf{\bibinfo {volume} {239}},\ \bibinfo {pages} {277}%
  \bibAnnoteFile{NoStop}{Graf:2003_PSSB}%
\bibitem[{\citenamefont{Granville}\
  \emph{et~al.}(2010)\citenamefont{Granville}, \citenamefont{Ruck},
  \citenamefont{Budde}, \citenamefont{Trodahl},\ and\
  \citenamefont{Williams}}]{Granville:2010_PRB}%
  \BibitemOpen
  \bibfield{author}{%
  \bibinfo {author} {\bibnamefont{Granville}, \bibfnamefont{S.}}, \bibinfo
  {author} {\bibfnamefont{B.~J.}\ \bibnamefont{Ruck}}, \bibinfo {author}
  {\bibfnamefont{F.}~\bibnamefont{Budde}}, \bibinfo {author}
  {\bibfnamefont{H.~J.}\ \bibnamefont{Trodahl}},\ and\ \bibinfo {author}
  {\bibfnamefont{G.~V.~M.}\ \bibnamefont{Williams}}}%
  , \bibinfo {year} {2010},\ \bibfield{title}{%
  \enquote{\bibinfo {title} {Nearest-neighbor mn antiferromagnetic exchange in
  {Ga$_{1-x}$Mn$_{x}$N}},}\ }%
  \bibfield{journal}{%
  \bibinfo {journal} {Phys. Rev. B}\ }%
  \textbf{\bibinfo {volume} {81}},\ \bibinfo {pages} {184425}%
  \bibAnnoteFile{NoStop}{Granville:2010_PRB}%
\bibitem[{\citenamefont{Gray}\ \emph{et~al.}(2012)\citenamefont{Gray},
  \citenamefont{Min\'ar}, \citenamefont{Ueda}, \citenamefont{Stone},
  \citenamefont{Yamashita}, \citenamefont{Fujii}, \citenamefont{Braun},
  \citenamefont{Plucinski}, \citenamefont{Schneider},
  \citenamefont{Panaccione}, \citenamefont{Ebert}, \citenamefont{Dubon},
  \citenamefont{Kobayashi},\ and\ \citenamefont{Fadley}}]{Gray:2012_NM}%
  \BibitemOpen
  \bibfield{author}{%
  \bibinfo {author} {\bibnamefont{Gray}, \bibfnamefont{A.~X.}}, \bibinfo
  {author} {\bibfnamefont{J.}~\bibnamefont{Min\'ar}}, \bibinfo {author}
  {\bibfnamefont{S.}~\bibnamefont{Ueda}}, \bibinfo {author}
  {\bibfnamefont{P.~R.}\ \bibnamefont{Stone}}, \bibinfo {author}
  {\bibfnamefont{Y.}~\bibnamefont{Yamashita}}, \bibinfo {author}
  {\bibfnamefont{J.}~\bibnamefont{Fujii}}, \bibinfo {author}
  {\bibfnamefont{J.}~\bibnamefont{Braun}}, \bibinfo {author}
  {\bibfnamefont{L.}~\bibnamefont{Plucinski}}, \bibinfo {author}
  {\bibfnamefont{C.~M.}\ \bibnamefont{Schneider}}, \bibinfo {author}
  {\bibfnamefont{G.}~\bibnamefont{Panaccione}}, \bibinfo {author}
  {\bibfnamefont{H.}~\bibnamefont{Ebert}}, \bibinfo {author}
  {\bibfnamefont{O.~D.}\ \bibnamefont{Dubon}}, \bibinfo {author}
  {\bibfnamefont{K.}~\bibnamefont{Kobayashi}},\ and\ \bibinfo {author}
  {\bibfnamefont{C.~S.}\ \bibnamefont{Fadley}}}%
  , \bibinfo {year} {2012},\ \bibfield{title}{%
  \enquote{\bibinfo {title} {Bulk electronic structure of the dilute magnetic
  semiconductor {Ga$_{1-x}$Mn$_{x}$As} through hard {X}-ray angle-resolved
  photoemission},}\ }%
  \bibfield{journal}{%
  \bibinfo {journal} {Nat. Mater.}\ }%
  \textbf{\bibinfo {volume} {11}},\ \bibinfo {pages} {957}%
  \bibAnnoteFile{NoStop}{Gray:2012_NM}%
\bibitem[{\citenamefont{Gryglas-Borysiewicz}\
  \emph{et~al.}(2010)\citenamefont{Gryglas-Borysiewicz},
  \citenamefont{Kwiatkowski}, \citenamefont{Baj}, \citenamefont{Wasik},
  \citenamefont{Przybytek},\ and\ \citenamefont{Sadowski}}]{Gryglas:2010_PRB}%
  \BibitemOpen
  \bibfield{author}{%
  \bibinfo {author} {\bibnamefont{Gryglas-Borysiewicz}, \bibfnamefont{M.}},
  \bibinfo {author} {\bibfnamefont{A.}~\bibnamefont{Kwiatkowski}}, \bibinfo
  {author} {\bibfnamefont{M.}~\bibnamefont{Baj}}, \bibinfo {author}
  {\bibfnamefont{D.}~\bibnamefont{Wasik}}, \bibinfo {author}
  {\bibfnamefont{J.}~\bibnamefont{Przybytek}},\ and\ \bibinfo {author}
  {\bibfnamefont{J.}~\bibnamefont{Sadowski}}}%
  , \bibinfo {year} {2010},\ \bibfield{title}{%
  \enquote{\bibinfo {title} {Hydrostatic pressure study of the
  paramagnetic-ferromagnetic phase transition in {(Ga,Mn)As}},}\ }%
  \bibfield{journal}{%
  \bibinfo {journal} {Phys. Rev. B}\ }%
  \textbf{\bibinfo {volume} {82}},\ \bibinfo {pages} {153204}%
  \bibAnnoteFile{NoStop}{Gryglas:2010_PRB}%
\bibitem[{\citenamefont{Haghgoo}\ \emph{et~al.}(2010)\citenamefont{Haghgoo},
  \citenamefont{Cubukcu}, \citenamefont{von Bardeleben},
  \citenamefont{Thevenard}, \citenamefont{{Lema\^itre}},\ and\
  \citenamefont{Gourdon}}]{Haghgoo:2010_PRB}%
  \BibitemOpen
  \bibfield{author}{%
  \bibinfo {author} {\bibnamefont{Haghgoo}, \bibfnamefont{S.}}, \bibinfo
  {author} {\bibfnamefont{M.}~\bibnamefont{Cubukcu}}, \bibinfo {author}
  {\bibfnamefont{H.~J.}\ \bibnamefont{von Bardeleben}}, \bibinfo {author}
  {\bibfnamefont{L.}~\bibnamefont{Thevenard}}, \bibinfo {author}
  {\bibfnamefont{A.}~\bibnamefont{{Lema\^itre}}},\ and\ \bibinfo {author}
  {\bibfnamefont{C.}~\bibnamefont{Gourdon}}}%
  , \bibinfo {year} {2010},\ \bibfield{title}{%
  \enquote{\bibinfo {title} {Exchange constant and domain wall width in
  {(Ga,Mn)(As,P)} films with self-organization of magnetic domains},}\ }%
  \bibfield{journal}{%
  \bibinfo {journal} {Phys. Rev. B}\ }%
  \textbf{\bibinfo {volume} {82}},\ \bibinfo {pages} {041301}%
  \bibAnnoteFile{NoStop}{Haghgoo:2010_PRB}%
\bibitem[{\citenamefont{Hals}\ \emph{et~al.}(2009)\citenamefont{Hals},
  \citenamefont{Nguyen},\ and\ \citenamefont{Brataas}}]{Hals:2009_PRL}%
  \BibitemOpen
  \bibfield{author}{%
  \bibinfo {author} {\bibnamefont{Hals}, \bibfnamefont{K.~M.~D.}}, \bibinfo
  {author} {\bibfnamefont{A.~K.}\ \bibnamefont{Nguyen}},\ and\ \bibinfo
  {author} {\bibfnamefont{A.}~\bibnamefont{Brataas}}}%
  , \bibinfo {year} {2009},\ \bibfield{title}{%
  \enquote{\bibinfo {title} {Intrinsic coupling between current and domain wall
  motion in {(Ga,Mn)As}},}\ }%
  \bibfield{journal}{%
  \bibinfo {journal} {Phys. Rev. Lett.}\ }%
  \textbf{\bibinfo {volume} {102}},\ \bibinfo {pages} {256601}%
  \bibAnnoteFile{NoStop}{Hals:2009_PRL}%
\bibitem[{\citenamefont{Hankiewicz}\
  \emph{et~al.}(2004)\citenamefont{Hankiewicz}, \citenamefont{Jungwirth},
  \citenamefont{Dietl}, \citenamefont{Timm},\ and\
  \citenamefont{Sinova}}]{Hankiewicz:2004_PRB}%
  \BibitemOpen
  \bibfield{author}{%
  \bibinfo {author} {\bibnamefont{Hankiewicz}, \bibfnamefont{E.~M.}}, \bibinfo
  {author} {\bibfnamefont{T.}~\bibnamefont{Jungwirth}}, \bibinfo {author}
  {\bibfnamefont{T.}~\bibnamefont{Dietl}}, \bibinfo {author}
  {\bibfnamefont{C.}~\bibnamefont{Timm}},\ and\ \bibinfo {author}
  {\bibfnamefont{Jairo}\ \bibnamefont{Sinova}}}%
  , \bibinfo {year} {2004},\ \bibfield{title}{%
  \enquote{\bibinfo {title} {Optical properties of metallic {(III,Mn)V}
  ferromagnetic semiconductors in the infrared to visible range},}\ }%
  \bibfield{journal}{%
  \bibinfo {journal} {Phys. Rev. B}\ }%
  \textbf{\bibinfo {volume} {70}},\ \bibinfo {pages} {245211}%
  \bibAnnoteFile{NoStop}{Hankiewicz:2004_PRB}%
\bibitem[{\citenamefont{Hansen}\ \emph{et~al.}(2001)\citenamefont{Hansen},
  \citenamefont{Ferrand}, \citenamefont{Richter}, \citenamefont{Thierley},
  \citenamefont{Hock}, \citenamefont{Schwarz}, \citenamefont{Reuscher},
  \citenamefont{Schmidt}, \citenamefont{Molenkamp},\ and\
  \citenamefont{Waag}}]{Hansen:2001_APL}%
  \BibitemOpen
  \bibfield{author}{%
  \bibinfo {author} {\bibnamefont{Hansen}, \bibfnamefont{L.}}, \bibinfo
  {author} {\bibfnamefont{D.}~\bibnamefont{Ferrand}}, \bibinfo {author}
  {\bibfnamefont{G.}~\bibnamefont{Richter}}, \bibinfo {author}
  {\bibfnamefont{M.}~\bibnamefont{Thierley}}, \bibinfo {author}
  {\bibfnamefont{V.}~\bibnamefont{Hock}}, \bibinfo {author}
  {\bibfnamefont{N.}~\bibnamefont{Schwarz}}, \bibinfo {author}
  {\bibfnamefont{G.}~\bibnamefont{Reuscher}}, \bibinfo {author}
  {\bibfnamefont{G.}~\bibnamefont{Schmidt}}, \bibinfo {author}
  {\bibfnamefont{L.~W.}\ \bibnamefont{Molenkamp}},\ and\ \bibinfo {author}
  {\bibfnamefont{A.}~\bibnamefont{Waag}}}%
  , \bibinfo {year} {2001},\ \bibfield{title}{%
  \enquote{\bibinfo {title} {Epitaxy and magnetotransport properties of the
  diluted magnetic semiconductor {p-Be$_{1-x}$Mn$_{x}$Te}},}\ }%
  \bibfield{journal}{%
  \bibinfo {journal} {Appl. Phys. Lett.}\ }%
  \textbf{\bibinfo {volume} {79}},\ \bibinfo {pages} {3125}%
  \bibAnnoteFile{NoStop}{Hansen:2001_APL}%
\bibitem[{\citenamefont{Hashimoto}\
  \emph{et~al.}(2002)\citenamefont{Hashimoto}, \citenamefont{Hayashi},
  \citenamefont{Katsumoto},\ and\ \citenamefont{Iye}}]{Hashimoto:2002_JCG}%
  \BibitemOpen
  \bibfield{author}{%
  \bibinfo {author} {\bibnamefont{Hashimoto}, \bibfnamefont{Y.}}, \bibinfo
  {author} {\bibfnamefont{T.}~\bibnamefont{Hayashi}}, \bibinfo {author}
  {\bibfnamefont{S.}~\bibnamefont{Katsumoto}},\ and\ \bibinfo {author}
  {\bibfnamefont{Y.}~\bibnamefont{Iye}}}%
  , \bibinfo {year} {2002},\ \bibfield{title}{%
  \enquote{\bibinfo {title} {Effect of low-temperature annealing on the
  crystallinity of {III--V}-based diluted magnetic semiconductors},}\ }%
  \bibfield{journal}{%
  \bibinfo {journal} {J. Cryst. Growth}\ }%
  \textbf{\bibinfo {volume} {237}},\ \bibinfo {pages} {1334}%
  \bibAnnoteFile{NoStop}{Hashimoto:2002_JCG}%
\bibitem[{\citenamefont{Hashimoto}\
  \emph{et~al.}(2008)\citenamefont{Hashimoto}, \citenamefont{Kobayashi},\ and\
  \citenamefont{Munekata}}]{Hashimoto:2008_PRL}%
  \BibitemOpen
  \bibfield{author}{%
  \bibinfo {author} {\bibnamefont{Hashimoto}, \bibfnamefont{Y.}}, \bibinfo
  {author} {\bibfnamefont{S.}~\bibnamefont{Kobayashi}},\ and\ \bibinfo {author}
  {\bibfnamefont{H.}~\bibnamefont{Munekata}}}%
  , \bibinfo {year} {2008},\ \bibfield{title}{%
  \enquote{\bibinfo {title} {Photoinduced precession of magnetization in
  ferromagnetic {(Ga,Mn)As}},}\ }%
  \bibfield{journal}{%
  \bibinfo {journal} {Phys. Rev. Lett.}\ }%
  \textbf{\bibinfo {volume} {100}},\ \bibinfo {pages} {067202}%
  \bibAnnoteFile{NoStop}{Hashimoto:2008_PRL}%
\bibitem[{\citenamefont{Hassan}\ \emph{et~al.}(2011)\citenamefont{Hassan},
  \citenamefont{Springholz}, \citenamefont{Lechner}, \citenamefont{Groiss},
  \citenamefont{Kirchschlager},\ and\ \citenamefont{Bauer}}]{Hassan:2011_JCG}%
  \BibitemOpen
  \bibfield{author}{%
  \bibinfo {author} {\bibnamefont{Hassan}, \bibfnamefont{M.}}, \bibinfo
  {author} {\bibfnamefont{G.}~\bibnamefont{Springholz}}, \bibinfo {author}
  {\bibfnamefont{R.T.}\ \bibnamefont{Lechner}}, \bibinfo {author}
  {\bibfnamefont{H.}~\bibnamefont{Groiss}}, \bibinfo {author}
  {\bibfnamefont{R.}~\bibnamefont{Kirchschlager}},\ and\ \bibinfo {author}
  {\bibfnamefont{G.}~\bibnamefont{Bauer}}}%
  , \bibinfo {year} {2011},\ \bibfield{title}{%
  \enquote{\bibinfo {title} {Molecular beam epitaxy of single phase {GeMnTe}
  with high ferromagnetic transition temperature},}\ }%
  \bibfield{journal}{%
  \bibinfo {journal} {J. Cryst. Growth}\ }%
  \textbf{\bibinfo {volume} {323}},\ \bibinfo {pages} {363}%
  \bibAnnoteFile{NoStop}{Hassan:2011_JCG}%
\bibitem[{\citenamefont{Haury}\ \emph{et~al.}(1997)\citenamefont{Haury},
  \citenamefont{Wasiela}, \citenamefont{Arnoult}, \citenamefont{Cibert},
  \citenamefont{Tatarenko}, \citenamefont{Dietl},\ and\
  \citenamefont{d'Aubigne}}]{Haury:1997_PRL}%
  \BibitemOpen
  \bibfield{author}{%
  \bibinfo {author} {\bibnamefont{Haury}, \bibfnamefont{A.}}, \bibinfo {author}
  {\bibfnamefont{A.}~\bibnamefont{Wasiela}}, \bibinfo {author}
  {\bibfnamefont{A.}~\bibnamefont{Arnoult}}, \bibinfo {author}
  {\bibfnamefont{J.}~\bibnamefont{Cibert}}, \bibinfo {author}
  {\bibfnamefont{S.}~\bibnamefont{Tatarenko}}, \bibinfo {author}
  {\bibfnamefont{T.}~\bibnamefont{Dietl}},\ and\ \bibinfo {author}
  {\bibfnamefont{Y.~Merle}\ \bibnamefont{d'Aubigne}}}%
  , \bibinfo {year} {1997},\ \bibfield{title}{%
  \enquote{\bibinfo {title} {Observation of a ferromagnetic transition induced
  by two-dimensional hole gas in modulation-doped {CdMnTe} quantum wells},}\ }%
  \bibfield{journal}{%
  \bibinfo {journal} {Phys. Rev. Lett.}\ }%
  \textbf{\bibinfo {volume} {79}},\ \bibinfo {pages} {511}%
  \bibAnnoteFile{NoStop}{Haury:1997_PRL}%
\bibitem[{\citenamefont{Hayashi}\ \emph{et~al.}(2001)\citenamefont{Hayashi},
  \citenamefont{Hashimoto}, \citenamefont{Katsumoto},\ and\
  \citenamefont{Iye}}]{Hayashi:2001_APL}%
  \BibitemOpen
  \bibfield{author}{%
  \bibinfo {author} {\bibnamefont{Hayashi}, \bibfnamefont{T.}}, \bibinfo
  {author} {\bibfnamefont{Y.}~\bibnamefont{Hashimoto}}, \bibinfo {author}
  {\bibfnamefont{S.}~\bibnamefont{Katsumoto}},\ and\ \bibinfo {author}
  {\bibfnamefont{Y.}~\bibnamefont{Iye}}}%
  , \bibinfo {year} {2001},\ \bibfield{title}{%
  \enquote{\bibinfo {title} {Effect of low-temperature annealing on transport
  and magnetism of diluted magnetic semiconductor {(Ga, Mn)As}},}\ }%
  \bibfield{journal}{%
  \bibinfo {journal} {Appl. Phys. Lett.}\ }%
  \textbf{\bibinfo {volume} {78}},\ \bibinfo {pages} {1691}%
  \bibAnnoteFile{NoStop}{Hayashi:2001_APL}%
\bibitem[{\citenamefont{Heremans}\ \emph{et~al.}(2012)\citenamefont{Heremans},
  \citenamefont{Wiendlocha},\ and\
  \citenamefont{Chamoire}}]{Heremans:2012_EES}%
  \BibitemOpen
  \bibfield{author}{%
  \bibinfo {author} {\bibnamefont{Heremans}, \bibfnamefont{J.~P.}}, \bibinfo
  {author} {\bibfnamefont{B.}~\bibnamefont{Wiendlocha}},\ and\ \bibinfo
  {author} {\bibfnamefont{A.~M.}\ \bibnamefont{Chamoire}}}%
  , \bibinfo {year} {2012},\ \bibfield{title}{%
  \enquote{\bibinfo {title} {Resonant levels in bulk thermoelectric
  semiconductors},}\ }%
  \bibfield{journal}{%
  \bibinfo {journal} {Energy Environ. Sci.}\ }%
  \textbf{\bibinfo {volume} {5}},\ \bibinfo {pages} {5510}%
  \bibAnnoteFile{NoStop}{Heremans:2012_EES}%
\bibitem[{\citenamefont{{Hol{\'y}}}\
  \emph{et~al.}(2006)\citenamefont{{Hol{\'y}}}, \citenamefont{{Mat\v{e}j}},
  \citenamefont{{Pacherov{\'a}}}, \citenamefont{{Nov{\'a}k}},
  \citenamefont{Cukr}, \citenamefont{{Olejn{\'i}k}},\ and\
  \citenamefont{Jungwirth}}]{Holy:2006_PRB}%
  \BibitemOpen
  \bibfield{author}{%
  \bibinfo {author} {\bibnamefont{{Hol{\'y}}}, \bibfnamefont{V.}}, \bibinfo
  {author} {\bibfnamefont{Z.}~\bibnamefont{{Mat\v{e}j}}}, \bibinfo {author}
  {\bibfnamefont{O.}~\bibnamefont{{Pacherov{\'a}}}}, \bibinfo {author}
  {\bibfnamefont{V.}~\bibnamefont{{Nov{\'a}k}}}, \bibinfo {author}
  {\bibfnamefont{M.}~\bibnamefont{Cukr}}, \bibinfo {author}
  {\bibfnamefont{K.}~\bibnamefont{{Olejn{\'i}k}}},\ and\ \bibinfo {author}
  {\bibfnamefont{T.}~\bibnamefont{Jungwirth}}}%
  , \bibinfo {year} {2006},\ \bibfield{title}{%
  \enquote{\bibinfo {title} {Mn incorporation in as-grown and annealed
  {(Ga,Mn)As} layers studied by x-ray diffraction and standing-wave
  fluorescence},}\ }%
  \bibfield{journal}{%
  \bibinfo {journal} {Phys. Rev. B}\ }%
  \textbf{\bibinfo {volume} {74}},\ \bibinfo {pages} {245205}%
  \bibAnnoteFile{NoStop}{Holy:2006_PRB}%
\bibitem[{\citenamefont{Honolka}\ \emph{et~al.}(2007)\citenamefont{Honolka},
  \citenamefont{Masmanidis}, \citenamefont{Tang}, \citenamefont{Awschalom},\
  and\ \citenamefont{Roukes}}]{Honolka:2007_PRB}%
  \BibitemOpen
  \bibfield{author}{%
  \bibinfo {author} {\bibnamefont{Honolka}, \bibfnamefont{J.}}, \bibinfo
  {author} {\bibfnamefont{S.}~\bibnamefont{Masmanidis}}, \bibinfo {author}
  {\bibfnamefont{H.~X.}\ \bibnamefont{Tang}}, \bibinfo {author}
  {\bibfnamefont{D.~D.}\ \bibnamefont{Awschalom}},\ and\ \bibinfo {author}
  {\bibfnamefont{M.~L.}\ \bibnamefont{Roukes}}}%
  , \bibinfo {year} {2007},\ \bibfield{title}{%
  \enquote{\bibinfo {title} {Magnetotransport properties of strained
  {Ga$_{0.95}$Mn$_{0.05}$As} epilayers close to the metal-insulator transition:
  Description using {Aronov-Altshuler} three-dimensional scaling theory},}\ }%
  \bibfield{journal}{%
  \bibinfo {journal} {Phys. Rev. B}\ }%
  \textbf{\bibinfo {volume} {75}},\ \bibinfo {pages} {245310}%
  \bibAnnoteFile{NoStop}{Honolka:2007_PRB}%
\bibitem[{\citenamefont{Hor}\ \emph{et~al.}(2010)\citenamefont{Hor},
  \citenamefont{Roushan}, \citenamefont{Beidenkopf}, \citenamefont{Seo},
  \citenamefont{Qu}, \citenamefont{Checkelsky}, \citenamefont{Wray},
  \citenamefont{Hsieh}, \citenamefont{Xia}, \citenamefont{Xu},
  \citenamefont{Qian}, \citenamefont{Hasan}, \citenamefont{Ong},
  \citenamefont{Yazdani},\ and\ \citenamefont{Cava}}]{Hor:2010_PRB}%
  \BibitemOpen
  \bibfield{author}{%
  \bibinfo {author} {\bibnamefont{Hor}, \bibfnamefont{Y.~S.}}, \bibinfo
  {author} {\bibfnamefont{P.}~\bibnamefont{Roushan}}, \bibinfo {author}
  {\bibfnamefont{H.}~\bibnamefont{Beidenkopf}}, \bibinfo {author}
  {\bibfnamefont{J.}~\bibnamefont{Seo}}, \bibinfo {author}
  {\bibfnamefont{D.}~\bibnamefont{Qu}}, \bibinfo {author}
  {\bibfnamefont{J.~G.}\ \bibnamefont{Checkelsky}}, \bibinfo {author}
  {\bibfnamefont{L.~A.}\ \bibnamefont{Wray}}, \bibinfo {author}
  {\bibfnamefont{D.}~\bibnamefont{Hsieh}}, \bibinfo {author}
  {\bibfnamefont{Y.}~\bibnamefont{Xia}}, \bibinfo {author}
  {\bibfnamefont{S.-Y.}\ \bibnamefont{Xu}}, \bibinfo {author}
  {\bibfnamefont{D.}~\bibnamefont{Qian}}, \bibinfo {author}
  {\bibfnamefont{M.~Z.}\ \bibnamefont{Hasan}}, \bibinfo {author}
  {\bibfnamefont{N.~P.}\ \bibnamefont{Ong}}, \bibinfo {author}
  {\bibfnamefont{A.}~\bibnamefont{Yazdani}},\ and\ \bibinfo {author}
  {\bibfnamefont{R.~J.}\ \bibnamefont{Cava}}}%
  , \bibinfo {year} {2010},\ \bibfield{title}{%
  \enquote{\bibinfo {title} {Development of ferromagnetism in the doped
  topological insulator {Bi$_{2-x}$Mn$_x$Te$_3$}},}\ }%
  \bibfield{journal}{%
  \bibinfo {journal} {Phys. Rev. B}\ }%
  \textbf{\bibinfo {volume} {81}},\ \bibinfo {pages} {195203}%
  \bibAnnoteFile{NoStop}{Hor:2010_PRB}%
\bibitem[{\citenamefont{Hrabovsky}\
  \emph{et~al.}(2002)\citenamefont{Hrabovsky}, \citenamefont{Vanelle},
  \citenamefont{Fert}, \citenamefont{Yee}, \citenamefont{Redoules},
  \citenamefont{Sadowski}, \citenamefont{Ka{\'n}ski},\ and\
  \citenamefont{Ilver}}]{Hrabovsky:2002_APL}%
  \BibitemOpen
  \bibfield{author}{%
  \bibinfo {author} {\bibnamefont{Hrabovsky}, \bibfnamefont{D.}}, \bibinfo
  {author} {\bibfnamefont{E.}~\bibnamefont{Vanelle}}, \bibinfo {author}
  {\bibfnamefont{A.~R.}\ \bibnamefont{Fert}}, \bibinfo {author}
  {\bibfnamefont{D.~S.}\ \bibnamefont{Yee}}, \bibinfo {author}
  {\bibfnamefont{J.~P.}\ \bibnamefont{Redoules}}, \bibinfo {author}
  {\bibfnamefont{J.}~\bibnamefont{Sadowski}}, \bibinfo {author}
  {\bibfnamefont{J.}~\bibnamefont{Ka{\'n}ski}},\ and\ \bibinfo {author}
  {\bibfnamefont{L.}~\bibnamefont{Ilver}}}%
  , \bibinfo {year} {2002},\ \bibfield{title}{%
  \enquote{\bibinfo {title} {Magnetization reversal in {GaMnAs} layers studied
  by kerr effect},}\ }%
  \bibfield{journal}{%
  \bibinfo {journal} {Appl. Phys. Lett.}\ }%
  \textbf{\bibinfo {volume} {81}},\ \bibinfo {pages} {2806}%
  \bibAnnoteFile{NoStop}{Hrabovsky:2002_APL}%
\bibitem[{\citenamefont{{Hsiao Wen Chang}}\
  \emph{et~al.}(2013)\citenamefont{{Hsiao Wen Chang}}, \citenamefont{Akita},
  \citenamefont{Matsukura},\ and\ \citenamefont{Ohno}}]{Chang:2013_APL}%
  \BibitemOpen
  \bibfield{author}{%
  \bibinfo {author} {\bibnamefont{{Hsiao Wen Chang}}}, \bibinfo {author}
  {\bibfnamefont{S.}~\bibnamefont{Akita}}, \bibinfo {author}
  {\bibfnamefont{F.}~\bibnamefont{Matsukura}},\ and\ \bibinfo {author}
  {\bibfnamefont{H.}~\bibnamefont{Ohno}}}%
  , \bibinfo {year} {2013},\ \bibfield{title}{%
  \enquote{\bibinfo {title} {Hole concentration dependence of the {Curie}
  temperature of {(Ga,Mn)Sb} in a field-effect structure},}\ }%
  \bibfield{journal}{%
  \bibinfo {journal} {Appl. Phys. Lett.}\ }%
  \textbf{\bibinfo {volume} {103}},\ \bibinfo {pages} {142402}%
  \bibAnnoteFile{NoStop}{Chang:2013_APL}%
\bibitem[{\citenamefont{H{\"u}mpfner}\
  \emph{et~al.}(2007)\citenamefont{H{\"u}mpfner}, \citenamefont{Pappert},
  \citenamefont{Wenisch}, \citenamefont{Brunner}, \citenamefont{Gould},
  \citenamefont{Schmidt}, \citenamefont{Molenkamp}, \citenamefont{Sawicki},\
  and\ \citenamefont{Dietl}}]{Humpfner:2007_APL}%
  \BibitemOpen
  \bibfield{author}{%
  \bibinfo {author} {\bibnamefont{H{\"u}mpfner}, \bibfnamefont{S.}}, \bibinfo
  {author} {\bibfnamefont{K.}~\bibnamefont{Pappert}}, \bibinfo {author}
  {\bibfnamefont{J.}~\bibnamefont{Wenisch}}, \bibinfo {author}
  {\bibfnamefont{K.}~\bibnamefont{Brunner}}, \bibinfo {author}
  {\bibfnamefont{C.}~\bibnamefont{Gould}}, \bibinfo {author}
  {\bibfnamefont{G.}~\bibnamefont{Schmidt}}, \bibinfo {author}
  {\bibfnamefont{L.~W.}\ \bibnamefont{Molenkamp}}, \bibinfo {author}
  {\bibfnamefont{M.}~\bibnamefont{Sawicki}},\ and\ \bibinfo {author}
  {\bibfnamefont{T.}~\bibnamefont{Dietl}}}%
  , \bibinfo {year} {2007},\ \bibfield{title}{%
  \enquote{\bibinfo {title} {Lithographic engineering of anisotropies in
  {(Ga,Mn)As}},}\ }%
  \bibfield{journal}{%
  \bibinfo {journal} {Appl. Phys. Lett.}\ }%
  \textbf{\bibinfo {volume} {90}},\ \bibinfo {pages} {102102}%
  \bibAnnoteFile{NoStop}{Humpfner:2007_APL}%
\bibitem[{\citenamefont{Hwang}\ \emph{et~al.}(2005)\citenamefont{Hwang},
  \citenamefont{Ishida}, \citenamefont{Kobayashi}, \citenamefont{Hirata},
  \citenamefont{Takubo}, \citenamefont{Mizokawa}, \citenamefont{Fujimori},
  \citenamefont{Okamoto}, \citenamefont{Mamiya}, \citenamefont{Saito},
  \citenamefont{Muramatsu}, \citenamefont{Ott}, \citenamefont{Tanaka},
  \citenamefont{Kondo},\ and\ \citenamefont{Munekata}}]{Hwang:2005_PRB}%
  \BibitemOpen
  \bibfield{author}{%
  \bibinfo {author} {\bibnamefont{Hwang}, \bibfnamefont{J.~I.}}, \bibinfo
  {author} {\bibfnamefont{Y.}~\bibnamefont{Ishida}}, \bibinfo {author}
  {\bibfnamefont{M.}~\bibnamefont{Kobayashi}}, \bibinfo {author}
  {\bibfnamefont{H.}~\bibnamefont{Hirata}}, \bibinfo {author}
  {\bibfnamefont{K.}~\bibnamefont{Takubo}}, \bibinfo {author}
  {\bibfnamefont{T.}~\bibnamefont{Mizokawa}}, \bibinfo {author}
  {\bibfnamefont{A.}~\bibnamefont{Fujimori}}, \bibinfo {author}
  {\bibfnamefont{J.}~\bibnamefont{Okamoto}}, \bibinfo {author}
  {\bibfnamefont{K.}~\bibnamefont{Mamiya}}, \bibinfo {author}
  {\bibfnamefont{Y.}~\bibnamefont{Saito}}, \bibinfo {author}
  {\bibfnamefont{Y.}~\bibnamefont{Muramatsu}}, \bibinfo {author}
  {\bibfnamefont{H.}~\bibnamefont{Ott}}, \bibinfo {author}
  {\bibfnamefont{A.}~\bibnamefont{Tanaka}}, \bibinfo {author}
  {\bibfnamefont{T.}~\bibnamefont{Kondo}},\ and\ \bibinfo {author}
  {\bibfnamefont{H.}~\bibnamefont{Munekata}}}%
  , \bibinfo {year} {2005},\ \bibfield{title}{%
  \enquote{\bibinfo {title} {High-energy spectroscopic study of the {III-V}
  nitride-based diluted magnetic semiconductor {Ga$_{1-x}$Mn$_{x}$N}},}\ }%
  \bibfield{journal}{%
  \bibinfo {journal} {Phys. Rev. B}\ }%
  \textbf{\bibinfo {volume} {72}},\ \bibinfo {pages} {085216}%
  \bibAnnoteFile{NoStop}{Hwang:2005_PRB}%
\bibitem[{\citenamefont{Ikeda}\ \emph{et~al.}(2008)\citenamefont{Ikeda},
  \citenamefont{Hayakawa}, \citenamefont{Ashizawa}, \citenamefont{Lee},
  \citenamefont{Miura}, \citenamefont{Hasegawa}, \citenamefont{Tsunoda},
  \citenamefont{Matsukura},\ and\ \citenamefont{Ohno}}]{Ikeda:2008_APL}%
  \BibitemOpen
  \bibfield{author}{%
  \bibinfo {author} {\bibnamefont{Ikeda}, \bibfnamefont{S.}}, \bibinfo {author}
  {\bibfnamefont{J.}~\bibnamefont{Hayakawa}}, \bibinfo {author}
  {\bibfnamefont{Y.}~\bibnamefont{Ashizawa}}, \bibinfo {author}
  {\bibfnamefont{Y.~M.}\ \bibnamefont{Lee}}, \bibinfo {author}
  {\bibfnamefont{K.}~\bibnamefont{Miura}}, \bibinfo {author}
  {\bibfnamefont{H.}~\bibnamefont{Hasegawa}}, \bibinfo {author}
  {\bibfnamefont{M.}~\bibnamefont{Tsunoda}}, \bibinfo {author}
  {\bibfnamefont{F.}~\bibnamefont{Matsukura}},\ and\ \bibinfo {author}
  {\bibfnamefont{H.}~\bibnamefont{Ohno}}}%
  , \bibinfo {year} {2008},\ \bibfield{title}{%
  \enquote{\bibinfo {title} {Tunnel magnetoresistance of 604\% at 300 {K} by
  suppression of {Ta} diffusion in {CoFeB/MgO/CoFeB} pseudo-spin-valves
  annealed at high temperature},}\ }%
  \bibfield{journal}{%
  \bibinfo {journal} {Appl. Phys. Lett.}\ }%
  \textbf{\bibinfo {volume} {93}},\ \bibinfo {pages} {082508}%
  \bibAnnoteFile{NoStop}{Ikeda:2008_APL}%
\bibitem[{\citenamefont{Ivchenko}\ and\
  \citenamefont{Pikus}(1995)}]{Ivchenko:1995_B}%
  \BibitemOpen
  \bibfield{author}{%
  \bibinfo {author} {\bibnamefont{Ivchenko}, \bibfnamefont{E.L.}},\ and\
  \bibinfo {author} {\bibfnamefont{G.E.}\ \bibnamefont{Pikus}}}%
  , \bibinfo {year} {1995},\ \emph{\bibinfo {title} {{Superlattices and other
  heterostructures. Symmetry and optical phenomena}}}\ (\bibinfo {publisher}
  {Springer})%
  \bibAnnoteFile{NoStop}{Ivchenko:1995_B}%
\bibitem[{\citenamefont{Jamet}\ \emph{et~al.}(2006)\citenamefont{Jamet},
  \citenamefont{Barski}, \citenamefont{Devillers}, \citenamefont{Poydenot},
  \citenamefont{Dujardin}, \citenamefont{Bayle-Guillmaud},
  \citenamefont{Rotheman}, \citenamefont{Bellet-Amalric}, \citenamefont{Marty},
  \citenamefont{Cibert}, \citenamefont{Mattana},\ and\
  \citenamefont{Tatarenko}}]{Jamet:2006_NM}%
  \BibitemOpen
  \bibfield{author}{%
  \bibinfo {author} {\bibnamefont{Jamet}, \bibfnamefont{M.}}, \bibinfo {author}
  {\bibfnamefont{A.}~\bibnamefont{Barski}}, \bibinfo {author}
  {\bibfnamefont{T.}~\bibnamefont{Devillers}}, \bibinfo {author}
  {\bibfnamefont{V.}~\bibnamefont{Poydenot}}, \bibinfo {author}
  {\bibfnamefont{R.}~\bibnamefont{Dujardin}}, \bibinfo {author}
  {\bibfnamefont{P.}~\bibnamefont{Bayle-Guillmaud}}, \bibinfo {author}
  {\bibfnamefont{J.}~\bibnamefont{Rotheman}}, \bibinfo {author}
  {\bibfnamefont{E.}~\bibnamefont{Bellet-Amalric}}, \bibinfo {author}
  {\bibfnamefont{A.}~\bibnamefont{Marty}}, \bibinfo {author}
  {\bibfnamefont{J.}~\bibnamefont{Cibert}}, \bibinfo {author}
  {\bibfnamefont{R.}~\bibnamefont{Mattana}},\ and\ \bibinfo {author}
  {\bibfnamefont{S.}~\bibnamefont{Tatarenko}}}%
  , \bibinfo {year} {2006},\ \bibfield{title}{%
  \enquote{\bibinfo {title} {High-{Curie}-temperature ferromagnetism in
  self-organized {Ge$_{1-x}$Mn$_x$} nanocolumns},}\ }%
  \bibfield{journal}{%
  \bibinfo {journal} {Nat. Mater.}\ }%
  \textbf{\bibinfo {volume} {5}},\ \bibinfo {pages} {653}%
  \bibAnnoteFile{NoStop}{Jamet:2006_NM}%
\bibitem[{\citenamefont{Jaroszy{\'n}ski}\ and\
  \citenamefont{Dietl}(1985)}]{Jaroszynski:1985_SSC}%
  \BibitemOpen
  \bibfield{author}{%
  \bibinfo {author} {\bibnamefont{Jaroszy{\'n}ski}, \bibfnamefont{J.}},\ and\
  \bibinfo {author} {\bibfnamefont{T.}~\bibnamefont{Dietl}}}%
  , \bibinfo {year} {1985},\ \bibfield{title}{%
  \enquote{\bibinfo {title} {Magnetoresistance studies of
  {Cd$_{1-x}$Mn$_{x}$Te}},}\ }%
  \bibfield{journal}{%
  \bibinfo {journal} {Solid State Commun.}\ }%
  \textbf{\bibinfo {volume} {55}},\ \bibinfo {pages} {491}%
  \bibAnnoteFile{NoStop}{Jaroszynski:1985_SSC}%
\bibitem[{\citenamefont{Jaworski}\ \emph{et~al.}(2010)\citenamefont{Jaworski},
  \citenamefont{Yang}, \citenamefont{Mack}, \citenamefont{Awschalom},
  \citenamefont{Heremans},\ and\ \citenamefont{Myers}}]{Jaworski:2010_NM}%
  \BibitemOpen
  \bibfield{author}{%
  \bibinfo {author} {\bibnamefont{Jaworski}, \bibfnamefont{C.~M.}}, \bibinfo
  {author} {\bibfnamefont{J.}~\bibnamefont{Yang}}, \bibinfo {author}
  {\bibfnamefont{S.}~\bibnamefont{Mack}}, \bibinfo {author}
  {\bibfnamefont{D.~D.}\ \bibnamefont{Awschalom}}, \bibinfo {author}
  {\bibfnamefont{J.~P.}\ \bibnamefont{Heremans}},\ and\ \bibinfo {author}
  {\bibfnamefont{R.~C.}\ \bibnamefont{Myers}}}%
  , \bibinfo {year} {2010},\ \bibfield{title}{%
  \enquote{\bibinfo {title} {Observation of the spin-{Seebeck} effect in a
  ferromagnetic semiconductor},}\ }%
  \bibfield{journal}{%
  \bibinfo {journal} {Nat. Mater.}\ }%
  \textbf{\bibinfo {volume} {9}},\ \bibinfo {pages} {898}%
  \bibAnnoteFile{NoStop}{Jaworski:2010_NM}%
\bibitem[{\citenamefont{Jaworski}\ \emph{et~al.}(2011)\citenamefont{Jaworski},
  \citenamefont{Yang}, \citenamefont{Mack}, \citenamefont{Awschalom},
  \citenamefont{Myers},\ and\ \citenamefont{Heremans}}]{Jaworski:2011_PRL}%
  \BibitemOpen
  \bibfield{author}{%
  \bibinfo {author} {\bibnamefont{Jaworski}, \bibfnamefont{C.~M.}}, \bibinfo
  {author} {\bibfnamefont{J.}~\bibnamefont{Yang}}, \bibinfo {author}
  {\bibfnamefont{S.}~\bibnamefont{Mack}}, \bibinfo {author}
  {\bibfnamefont{D.~D.}\ \bibnamefont{Awschalom}}, \bibinfo {author}
  {\bibfnamefont{R.~C.}\ \bibnamefont{Myers}},\ and\ \bibinfo {author}
  {\bibfnamefont{J.~P.}\ \bibnamefont{Heremans}}}%
  , \bibinfo {year} {2011},\ \bibfield{title}{%
  \enquote{\bibinfo {title} {Spin-{Seebeck} effect: {A} phonon driven spin
  distribution},}\ }%
  \bibfield{journal}{%
  \bibinfo {journal} {Phys. Rev. Lett.}\ }%
  \textbf{\bibinfo {volume} {106}},\ \bibinfo {pages} {186601}%
  \bibAnnoteFile{NoStop}{Jaworski:2011_PRL}%
\bibitem[{\citenamefont{Johnston-Halperin}\
  \emph{et~al.}(2002)\citenamefont{Johnston-Halperin}, \citenamefont{Lofgreen},
  \citenamefont{Kawakami}, \citenamefont{Young}, \citenamefont{Coldren},
  \citenamefont{Gossard},\ and\
  \citenamefont{Awschalom}}]{Johnston-Halperin:2002_PRB}%
  \BibitemOpen
  \bibfield{author}{%
  \bibinfo {author} {\bibnamefont{Johnston-Halperin}, \bibfnamefont{E.}},
  \bibinfo {author} {\bibfnamefont{D.}~\bibnamefont{Lofgreen}}, \bibinfo
  {author} {\bibfnamefont{R.K.}\ \bibnamefont{Kawakami}}, \bibinfo {author}
  {\bibfnamefont{D.K.}\ \bibnamefont{Young}}, \bibinfo {author}
  {\bibfnamefont{L.}~\bibnamefont{Coldren}}, \bibinfo {author}
  {\bibfnamefont{A.C.}\ \bibnamefont{Gossard}},\ and\ \bibinfo {author}
  {\bibfnamefont{D.D.}\ \bibnamefont{Awschalom}}}%
  , \bibinfo {year} {2002},\ \bibfield{title}{%
  \enquote{\bibinfo {title} {Spin-polarized {Zener} tunneling in
  {(Ga,Mn)As}},}\ }%
  \bibfield{journal}{%
  \bibinfo {journal} {Phys. Rev. B}\ }%
  \textbf{\bibinfo {volume} {65}},\ \bibinfo {pages} {041306}%
  \bibAnnoteFile{NoStop}{Johnston-Halperin:2002_PRB}%
\bibitem[{\citenamefont{Jonker}\ \emph{et~al.}(2000)\citenamefont{Jonker},
  \citenamefont{Park}, \citenamefont{Bennett}, \citenamefont{Cheong},
  \citenamefont{Kioseoglou},\ and\ \citenamefont{Petrou}}]{Jonker:2000_PRB}%
  \BibitemOpen
  \bibfield{author}{%
  \bibinfo {author} {\bibnamefont{Jonker}, \bibfnamefont{B.~T.}}, \bibinfo
  {author} {\bibfnamefont{Y.~D.}\ \bibnamefont{Park}}, \bibinfo {author}
  {\bibfnamefont{B.~R.}\ \bibnamefont{Bennett}}, \bibinfo {author}
  {\bibfnamefont{H.~D.}\ \bibnamefont{Cheong}}, \bibinfo {author}
  {\bibfnamefont{G.}~\bibnamefont{Kioseoglou}},\ and\ \bibinfo {author}
  {\bibfnamefont{A.}~\bibnamefont{Petrou}}}%
  , \bibinfo {year} {2000},\ \bibfield{title}{%
  \enquote{\bibinfo {title} {Robust electrical spin injection into a
  semiconductor heterostructure},}\ }%
  \bibfield{journal}{%
  \bibinfo {journal} {Phys. Rev. B}\ }%
  \textbf{\bibinfo {volume} {62}},\ \bibinfo {pages} {8180}%
  \bibAnnoteFile{NoStop}{Jonker:2000_PRB}%
\bibitem[{\citenamefont{Jungwirth}\
  \emph{et~al.}(1999)\citenamefont{Jungwirth}, \citenamefont{Atkinson},
  \citenamefont{Lee},\ and\ \citenamefont{MacDonald}}]{Jungwirth:1999_PRB}%
  \BibitemOpen
  \bibfield{author}{%
  \bibinfo {author} {\bibnamefont{Jungwirth}, \bibfnamefont{T.}}, \bibinfo
  {author} {\bibfnamefont{W.~A.}\ \bibnamefont{Atkinson}}, \bibinfo {author}
  {\bibfnamefont{B.}~\bibnamefont{Lee}},\ and\ \bibinfo {author}
  {\bibfnamefont{A.~H.}\ \bibnamefont{MacDonald}}}%
  , \bibinfo {year} {1999},\ \bibfield{title}{%
  \enquote{\bibinfo {title} {Interlayer coupling in ferromagnetic semiconductor
  superlattices},}\ }%
  \bibfield{journal}{%
  \bibinfo {journal} {Phys. Rev. B}\ }%
  \textbf{\bibinfo {volume} {59}},\ \bibinfo {pages} {9818}%
  \bibAnnoteFile{NoStop}{Jungwirth:1999_PRB}%
\bibitem[{\citenamefont{Jungwirth}\
  \emph{et~al.}(2008)\citenamefont{Jungwirth}, \citenamefont{Gallagher},\ and\
  \citenamefont{Wunderlich}}]{Jungwirth:2008_B}%
  \BibitemOpen
  \bibfield{author}{%
  \bibinfo {author} {\bibnamefont{Jungwirth}, \bibfnamefont{T.}}, \bibinfo
  {author} {\bibfnamefont{B.~L.}\ \bibnamefont{Gallagher}},\ and\ \bibinfo
  {author} {\bibfnamefont{J.}~\bibnamefont{Wunderlich}}}%
  , \bibinfo {year} {2008},\ \enquote{\bibinfo {title} {Transport properties of
  ferromagnetic semiconductors},}\ in\ \emph{\bibinfo {booktitle}
  {Spintronics}},\ \bibinfo {editor} {edited by\ \bibinfo {editor}
  {\bibfnamefont{T.}~\bibnamefont{Dietl}}, \bibinfo {editor}
  {\bibfnamefont{D.~D.}\ \bibnamefont{Awschalom}}, \bibinfo {editor}
  {\bibfnamefont{M.}~\bibnamefont{Kami{\'n}ska}},\ and\ \bibinfo {editor}
  {\bibfnamefont{H.}~\bibnamefont{Ohno}}}\ (\bibinfo {publisher} {Elsevier,
  Amsterdam})\ p.\ \bibinfo {pages} {135}%
  \bibAnnoteFile{NoStop}{Jungwirth:2008_B}%
\bibitem[{\citenamefont{Jungwirth}\
  \emph{et~al.}(2010)\citenamefont{Jungwirth}, \citenamefont{Horodysk{\'a}},
  \citenamefont{Tesa\ifmmode~\check{r}\else \v{r}\fi{}ov\'a},
  \citenamefont{N\ifmmode~\check{e}\else \v{e}\fi{}mec},
  \citenamefont{\ifmmode~\check{S}\else \v{S}\fi{}ubrt},
  \citenamefont{Mal{\'y}}, \citenamefont{Ku\ifmmode~\check{z}\else
  \v{z}\fi{}el}, \citenamefont{Kadlec}, \citenamefont{Ma\ifmmode~\check{s}\else
  \v{s}\fi{}ek}, \citenamefont{N\ifmmode~\check{e}\else \v{e}\fi{}mec},
  \citenamefont{Orlita}, \citenamefont{Nov\'ak}, \citenamefont{Olejn{\'i}k},
  \citenamefont{\ifmmode \check{S}\else \v{S}\fi{}ob\'a\ifmmode~\check{n}\else
  \v{n}\fi{}}, \citenamefont{Va\ifmmode~\check{s}\else \v{s}\fi{}ek},
  \citenamefont{Svoboda},\ and\ \citenamefont{Sinova}}]{Jungwirth:2010_PRL}%
  \BibitemOpen
  \bibfield{author}{%
  \bibinfo {author} {\bibnamefont{Jungwirth}, \bibfnamefont{T.}}, \bibinfo
  {author} {\bibfnamefont{P.}~\bibnamefont{Horodysk{\'a}}}, \bibinfo {author}
  {\bibfnamefont{N.}~\bibnamefont{Tesa\ifmmode~\check{r}\else
  \v{r}\fi{}ov\'a}}, \bibinfo {author}
  {\bibfnamefont{P.}~\bibnamefont{N\ifmmode~\check{e}\else \v{e}\fi{}mec}},
  \bibinfo {author} {\bibfnamefont{J.}~\bibnamefont{\ifmmode~\check{S}\else
  \v{S}\fi{}ubrt}}, \bibinfo {author}
  {\bibfnamefont{P.}~\bibnamefont{Mal{\'y}}}, \bibinfo {author}
  {\bibfnamefont{P.}~\bibnamefont{Ku\ifmmode~\check{z}\else \v{z}\fi{}el}},
  \bibinfo {author} {\bibfnamefont{C.}~\bibnamefont{Kadlec}}, \bibinfo {author}
  {\bibfnamefont{J.}~\bibnamefont{Ma\ifmmode~\check{s}\else \v{s}\fi{}ek}},
  \bibinfo {author} {\bibfnamefont{I.}~\bibnamefont{N\ifmmode~\check{e}\else
  \v{e}\fi{}mec}}, \bibinfo {author} {\bibfnamefont{M.}~\bibnamefont{Orlita}},
  \bibinfo {author} {\bibfnamefont{V.}~\bibnamefont{Nov\'ak}}, \bibinfo
  {author} {\bibfnamefont{K.}~\bibnamefont{Olejn{\'i}k}}, \bibinfo {author}
  {\bibfnamefont{Z.}~\bibnamefont{\ifmmode \check{S}\else
  \v{S}\fi{}ob\'a\ifmmode~\check{n}\else \v{n}\fi{}}}, \bibinfo {author}
  {\bibfnamefont{P.}~\bibnamefont{Va\ifmmode~\check{s}\else \v{s}\fi{}ek}},
  \bibinfo {author} {\bibfnamefont{P.}~\bibnamefont{Svoboda}},\ and\ \bibinfo
  {author} {\bibfnamefont{Jairo}\ \bibnamefont{Sinova}}}%
  , \bibinfo {year} {2010},\ \bibfield{title}{%
  \enquote{\bibinfo {title} {Systematic study of {Mn}-doping trends in optical
  properties of {(Ga,Mn)As}},}\ }%
  \bibfield{journal}{%
  \bibinfo {journal} {Phys. Rev. Lett.}\ }%
  \textbf{\bibinfo {volume} {105}},\ \bibinfo {pages} {227201}%
  \bibAnnoteFile{NoStop}{Jungwirth:2010_PRL}%
\bibitem[{\citenamefont{Jungwirth}\
  \emph{et~al.}(2002)\citenamefont{Jungwirth}, \citenamefont{{K{\"o}nig}},
  \citenamefont{Sinova}, \citenamefont{{Ku\v{c}era}},\ and\
  \citenamefont{MacDonald}}]{Jungwirth:2002_PRB}%
  \BibitemOpen
  \bibfield{author}{%
  \bibinfo {author} {\bibnamefont{Jungwirth}, \bibfnamefont{T.}}, \bibinfo
  {author} {\bibfnamefont{J.}~\bibnamefont{{K{\"o}nig}}}, \bibinfo {author}
  {\bibfnamefont{J.}~\bibnamefont{Sinova}}, \bibinfo {author}
  {\bibfnamefont{J.}~\bibnamefont{{Ku\v{c}era}}},\ and\ \bibinfo {author}
  {\bibfnamefont{A.~H.}\ \bibnamefont{MacDonald}}}%
  , \bibinfo {year} {2002},\ \bibfield{title}{%
  \enquote{\bibinfo {title} {Curie temperature trends in {(III,Mn)V}
  ferromagnetic semiconductors},}\ }%
  \bibfield{journal}{%
  \bibinfo {journal} {Phys. Rev. B}\ }%
  \textbf{\bibinfo {volume} {66}},\ \bibinfo {pages} {012402}%
  \bibAnnoteFile{NoStop}{Jungwirth:2002_PRB}%
\bibitem[{\citenamefont{Jungwirth}\
  \emph{et~al.}(2011)\citenamefont{Jungwirth}, \citenamefont{Nov\'ak},
  \citenamefont{Mart\'i}, \citenamefont{Cukr}, \citenamefont{M\'aca},
  \citenamefont{Shick}, \citenamefont{Ma\ifmmode~\check{s}\else \v{s}\fi{}ek},
  \citenamefont{Horodysk\'a}, \citenamefont{N\ifmmode~\check{e}\else
  \v{e}\fi{}mec}, \citenamefont{Hol\'y}, \citenamefont{Zemek},
  \citenamefont{Ku\ifmmode~\check{z}\else \v{z}\fi{}el},
  \citenamefont{N\ifmmode~\check{e}\else \v{e}\fi{}mec},
  \citenamefont{Gallagher}, \citenamefont{Campion}, \citenamefont{Foxon},\ and\
  \citenamefont{Wunderlich}}]{Jungwirth:2011_PRB}%
  \BibitemOpen
  \bibfield{author}{%
  \bibinfo {author} {\bibnamefont{Jungwirth}, \bibfnamefont{T.}}, \bibinfo
  {author} {\bibfnamefont{V.}~\bibnamefont{Nov\'ak}}, \bibinfo {author}
  {\bibfnamefont{X.}~\bibnamefont{Mart\'i}}, \bibinfo {author}
  {\bibfnamefont{M.}~\bibnamefont{Cukr}}, \bibinfo {author}
  {\bibfnamefont{F.}~\bibnamefont{M\'aca}}, \bibinfo {author}
  {\bibfnamefont{A.~B.}\ \bibnamefont{Shick}}, \bibinfo {author}
  {\bibfnamefont{J.}~\bibnamefont{Ma\ifmmode~\check{s}\else \v{s}\fi{}ek}},
  \bibinfo {author} {\bibfnamefont{P.}~\bibnamefont{Horodysk\'a}}, \bibinfo
  {author} {\bibfnamefont{P.}~\bibnamefont{N\ifmmode~\check{e}\else
  \v{e}\fi{}mec}}, \bibinfo {author} {\bibfnamefont{V.}~\bibnamefont{Hol\'y}},
  \bibinfo {author} {\bibfnamefont{J.}~\bibnamefont{Zemek}}, \bibinfo {author}
  {\bibfnamefont{P.}~\bibnamefont{Ku\ifmmode~\check{z}\else \v{z}\fi{}el}},
  \bibinfo {author} {\bibfnamefont{I.}~\bibnamefont{N\ifmmode~\check{e}\else
  \v{e}\fi{}mec}}, \bibinfo {author} {\bibfnamefont{B.~L.}\
  \bibnamefont{Gallagher}}, \bibinfo {author} {\bibfnamefont{R.~P.}\
  \bibnamefont{Campion}}, \bibinfo {author} {\bibfnamefont{C.~T.}\
  \bibnamefont{Foxon}},\ and\ \bibinfo {author}
  {\bibfnamefont{J.}~\bibnamefont{Wunderlich}}}%
  , \bibinfo {year} {2011},\ \bibfield{title}{%
  \enquote{\bibinfo {title} {Demonstration of molecular beam epitaxy and a
  semiconducting band structure for {I-Mn-V} compounds},}\ }%
  \bibfield{journal}{%
  \bibinfo {journal} {Phys. Rev. B}\ }%
  \textbf{\bibinfo {volume} {83}},\ \bibinfo {pages} {035321}%
  \bibAnnoteFile{NoStop}{Jungwirth:2011_PRB}%
\bibitem[{\citenamefont{Jungwirth}\
  \emph{et~al.}(2007)\citenamefont{Jungwirth}, \citenamefont{Sinova},
  \citenamefont{MacDonald}, \citenamefont{Gallagher}, \citenamefont{Nov{\'a}k},
  \citenamefont{Edmonds}, \citenamefont{Rushforth}, \citenamefont{Campion},
  \citenamefont{Foxon}, \citenamefont{Eaves}, \citenamefont{Olejn{\'i}k},
  \citenamefont{{Ma\v{s}ek}}, \citenamefont{Yang}, \citenamefont{Wunderlich},
  \citenamefont{Gould}, \citenamefont{Molenkamp}, \citenamefont{Dietl},\ and\
  \citenamefont{Ohno}}]{Jungwirth:2007_PRB}%
  \BibitemOpen
  \bibfield{author}{%
  \bibinfo {author} {\bibnamefont{Jungwirth}, \bibfnamefont{T.}}, \bibinfo
  {author} {\bibfnamefont{Jairo}\ \bibnamefont{Sinova}}, \bibinfo {author}
  {\bibfnamefont{A.~H.}\ \bibnamefont{MacDonald}}, \bibinfo {author}
  {\bibfnamefont{B.~L.}\ \bibnamefont{Gallagher}}, \bibinfo {author}
  {\bibfnamefont{V.}~\bibnamefont{Nov{\'a}k}}, \bibinfo {author}
  {\bibfnamefont{K.~W.}\ \bibnamefont{Edmonds}}, \bibinfo {author}
  {\bibfnamefont{A.~W.}\ \bibnamefont{Rushforth}}, \bibinfo {author}
  {\bibfnamefont{R.~P.}\ \bibnamefont{Campion}}, \bibinfo {author}
  {\bibfnamefont{C.~T.}\ \bibnamefont{Foxon}}, \bibinfo {author}
  {\bibfnamefont{L.}~\bibnamefont{Eaves}}, \bibinfo {author}
  {\bibfnamefont{E.}~\bibnamefont{Olejn{\'i}k}}, \bibinfo {author}
  {\bibfnamefont{J.}~\bibnamefont{{Ma\v{s}ek}}}, \bibinfo {author}
  {\bibfnamefont{S.-R.~Eric}\ \bibnamefont{Yang}}, \bibinfo {author}
  {\bibfnamefont{J.}~\bibnamefont{Wunderlich}}, \bibinfo {author}
  {\bibfnamefont{C.}~\bibnamefont{Gould}}, \bibinfo {author}
  {\bibfnamefont{L.~W.}\ \bibnamefont{Molenkamp}}, \bibinfo {author}
  {\bibfnamefont{T.}~\bibnamefont{Dietl}},\ and\ \bibinfo {author}
  {\bibfnamefont{H.}~\bibnamefont{Ohno}}}%
  , \bibinfo {year} {2007},\ \bibfield{title}{%
  \enquote{\bibinfo {title} {Character of states near the {Fermi} level in
  {(Ga,Mn)As}: Impurity to valence band crossover},}\ }%
  \bibfield{journal}{%
  \bibinfo {journal} {Phys. Rev. B}\ }%
  \textbf{\bibinfo {volume} {76}},\ \bibinfo {pages} {125206}%
  \bibAnnoteFile{NoStop}{Jungwirth:2007_PRB}%
\bibitem[{\citenamefont{Jungwirth}\
  \emph{et~al.}(2006{\natexlab{a}})\citenamefont{Jungwirth},
  \citenamefont{Sinova}, \citenamefont{{Ma\v{s}ek}},
  \citenamefont{{Ku\v{c}era}},\ and\
  \citenamefont{MacDonald}}]{Jungwirth:2006_RMP}%
  \BibitemOpen
  \bibfield{author}{%
  \bibinfo {author} {\bibnamefont{Jungwirth}, \bibfnamefont{T.}}, \bibinfo
  {author} {\bibfnamefont{Jairo}\ \bibnamefont{Sinova}}, \bibinfo {author}
  {\bibfnamefont{J.}~\bibnamefont{{Ma\v{s}ek}}}, \bibinfo {author}
  {\bibfnamefont{J.}~\bibnamefont{{Ku\v{c}era}}},\ and\ \bibinfo {author}
  {\bibfnamefont{A.~H.}\ \bibnamefont{MacDonald}}}%
  , \bibinfo {year} {2006}{\natexlab{a}},\ \bibfield{title}{%
  \enquote{\bibinfo {title} {Theory of ferromagnetic {(III,Mn)V}
  semiconductors},}\ }%
  \bibfield{journal}{%
  \bibinfo {journal} {Rev. Mod. Phys.}\ }%
  \textbf{\bibinfo {volume} {78}},\ \bibinfo {pages} {809}%
  \bibAnnoteFile{NoStop}{Jungwirth:2006_RMP}%
\bibitem[{\citenamefont{Jungwirth}\
  \emph{et~al.}(2005)\citenamefont{Jungwirth}, \citenamefont{Wang},
  \citenamefont{{Ma\v{s}ek}}, \citenamefont{Edmonds},
  \citenamefont{{K{\"o}nig}}, \citenamefont{Sinova}, \citenamefont{Polini},
  \citenamefont{Goncharuk}, \citenamefont{MacDonald}, \citenamefont{Sawicki},
  \citenamefont{Campion}, \citenamefont{Zhao}, \citenamefont{Foxon},\ and\
  \citenamefont{Gallagher}}]{Jungwirth:2005_PRBa}%
  \BibitemOpen
  \bibfield{author}{%
  \bibinfo {author} {\bibnamefont{Jungwirth}, \bibfnamefont{T.}}, \bibinfo
  {author} {\bibfnamefont{K.~Y.}\ \bibnamefont{Wang}}, \bibinfo {author}
  {\bibfnamefont{J.}~\bibnamefont{{Ma\v{s}ek}}}, \bibinfo {author}
  {\bibfnamefont{K.~W.}\ \bibnamefont{Edmonds}}, \bibinfo {author}
  {\bibfnamefont{{J{\"u}rgen}}\ \bibnamefont{{K{\"o}nig}}}, \bibinfo {author}
  {\bibfnamefont{Jairo}\ \bibnamefont{Sinova}}, \bibinfo {author}
  {\bibfnamefont{M.}~\bibnamefont{Polini}}, \bibinfo {author}
  {\bibfnamefont{N.~A.}\ \bibnamefont{Goncharuk}}, \bibinfo {author}
  {\bibfnamefont{A.~H.}\ \bibnamefont{MacDonald}}, \bibinfo {author}
  {\bibfnamefont{M.}~\bibnamefont{Sawicki}}, \bibinfo {author}
  {\bibfnamefont{R.~P.}\ \bibnamefont{Campion}}, \bibinfo {author}
  {\bibfnamefont{L.~X.}\ \bibnamefont{Zhao}}, \bibinfo {author}
  {\bibfnamefont{C.~T.}\ \bibnamefont{Foxon}},\ and\ \bibinfo {author}
  {\bibfnamefont{B.~L.}\ \bibnamefont{Gallagher}}}%
  , \bibinfo {year} {2005},\ \bibfield{title}{%
  \enquote{\bibinfo {title} {Prospects for high temperature ferromagnetism in
  {(Ga,Mn)As} semiconductors},}\ }%
  \bibfield{journal}{%
  \bibinfo {journal} {Phys. Rev. B}\ }%
  \textbf{\bibinfo {volume} {72}},\ \bibinfo {pages} {165204}%
  \bibAnnoteFile{NoStop}{Jungwirth:2005_PRBa}%
\bibitem[{\citenamefont{Jungwirth}\
  \emph{et~al.}(2006{\natexlab{b}})\citenamefont{Jungwirth},
  \citenamefont{Wang}, \citenamefont{{Ma\v{s}ek}}, \citenamefont{Edmonds},
  \citenamefont{{K{\"o}nig}}, \citenamefont{Sinova}, \citenamefont{Polini},
  \citenamefont{Goncharuk}, \citenamefont{MacDonald}, \citenamefont{Sawicki},
  \citenamefont{Campion}, \citenamefont{Zhao}, \citenamefont{Foxon},\ and\
  \citenamefont{Gallagher}}]{Jungwirth:2006_PRB}%
  \BibitemOpen
  \bibfield{author}{%
  \bibinfo {author} {\bibnamefont{Jungwirth}, \bibfnamefont{T.}}, \bibinfo
  {author} {\bibfnamefont{K.~Y.}\ \bibnamefont{Wang}}, \bibinfo {author}
  {\bibfnamefont{J.}~\bibnamefont{{Ma\v{s}ek}}}, \bibinfo {author}
  {\bibfnamefont{K.~W.}\ \bibnamefont{Edmonds}}, \bibinfo {author}
  {\bibfnamefont{{J{\"u}rgen}}\ \bibnamefont{{K{\"o}nig}}}, \bibinfo {author}
  {\bibfnamefont{Jairo}\ \bibnamefont{Sinova}}, \bibinfo {author}
  {\bibfnamefont{M.}~\bibnamefont{Polini}}, \bibinfo {author}
  {\bibfnamefont{N.A.}\ \bibnamefont{Goncharuk}}, \bibinfo {author}
  {\bibfnamefont{A.~H.}\ \bibnamefont{MacDonald}}, \bibinfo {author}
  {\bibfnamefont{M.}~\bibnamefont{Sawicki}}, \bibinfo {author}
  {\bibfnamefont{R.~P.}\ \bibnamefont{Campion}}, \bibinfo {author}
  {\bibfnamefont{L.~X.}\ \bibnamefont{Zhao}}, \bibinfo {author}
  {\bibfnamefont{C.~T.}\ \bibnamefont{Foxon}},\ and\ \bibinfo {author}
  {\bibfnamefont{B.~L.}\ \bibnamefont{Gallagher}}}%
  , \bibinfo {year} {2006}{\natexlab{b}},\ \bibfield{title}{%
  \enquote{\bibinfo {title} {Low tempearature magnetization of {(Ga,Mn)As}
  semiconductors},}\ }%
  \bibfield{journal}{%
  \bibinfo {journal} {Phys. Rev. B}\ }%
  \textbf{\bibinfo {volume} {73}},\ \bibinfo {pages} {165205}%
  \bibAnnoteFile{NoStop}{Jungwirth:2006_PRB}%
\bibitem[{\citenamefont{Kacman}(2001)}]{Kacman:2001_SST}%
  \BibitemOpen
  \bibfield{author}{%
  \bibinfo {author} {\bibnamefont{Kacman}, \bibfnamefont{P.}}}%
  , \bibinfo {year} {2001},\ \bibfield{title}{%
  \enquote{\bibinfo {title} {Spin interactions in diluted magnetic
  semiconductors and magnetic semiconductor structures},}\ }%
  \bibfield{journal}{%
  \bibinfo {journal} {Semicond. Sci. Technol.}\ }%
  \textbf{\bibinfo {volume} {16}},\ \bibinfo {pages} {R25}%
  \bibAnnoteFile{NoStop}{Kacman:2001_SST}%
\bibitem[{\citenamefont{Kakazei}\ \emph{et~al.}(2005)\citenamefont{Kakazei},
  \citenamefont{Pogorelov}, \citenamefont{Costa}, \citenamefont{Golub},
  \citenamefont{Sousa}, \citenamefont{Freitas}, \citenamefont{Cardoso},\ and\
  \citenamefont{Wigen}}]{Kakazei:2005_JAP}%
  \BibitemOpen
  \bibfield{author}{%
  \bibinfo {author} {\bibnamefont{Kakazei}, \bibfnamefont{G.~N.}}, \bibinfo
  {author} {\bibfnamefont{Yu.~G.}\ \bibnamefont{Pogorelov}}, \bibinfo {author}
  {\bibfnamefont{M.~D.}\ \bibnamefont{Costa}}, \bibinfo {author}
  {\bibfnamefont{V.~O.}\ \bibnamefont{Golub}}, \bibinfo {author}
  {\bibfnamefont{J.~B.}\ \bibnamefont{Sousa}}, \bibinfo {author}
  {\bibfnamefont{P.~P.}\ \bibnamefont{Freitas}}, \bibinfo {author}
  {\bibfnamefont{S.}~\bibnamefont{Cardoso}},\ and\ \bibinfo {author}
  {\bibfnamefont{P.~E.}\ \bibnamefont{Wigen}}}%
  , \bibinfo {year} {2005},\ \bibfield{title}{%
  \enquote{\bibinfo {title} {Interlayer dipolar interactions in multilayered
  granular films},}\ }%
  \bibfield{journal}{%
  \bibinfo {journal} {J. Appl. Phys.}\ }%
  \textbf{\bibinfo {volume} {97}},\ \bibinfo {pages} {10A723}%
  \bibAnnoteFile{NoStop}{Kakazei:2005_JAP}%
\bibitem[{\citenamefont{Kamara}\ \emph{et~al.}(2012)\citenamefont{Kamara},
  \citenamefont{Terki}, \citenamefont{Dumas}, \citenamefont{Dehbaoui},
  \citenamefont{Sadowski}, \citenamefont{Galera}, \citenamefont{Tran},\ and\
  \citenamefont{Charar}}]{Kamara:2012_JNN}%
  \BibitemOpen
  \bibfield{author}{%
  \bibinfo {author} {\bibnamefont{Kamara}, \bibfnamefont{S.}}, \bibinfo
  {author} {\bibfnamefont{F.}~\bibnamefont{Terki}}, \bibinfo {author}
  {\bibfnamefont{R.}~\bibnamefont{Dumas}}, \bibinfo {author}
  {\bibfnamefont{M.}~\bibnamefont{Dehbaoui}}, \bibinfo {author}
  {\bibfnamefont{J.}~\bibnamefont{Sadowski}}, \bibinfo {author}
  {\bibfnamefont{R.~M.}\ \bibnamefont{Galera}}, \bibinfo {author}
  {\bibfnamefont{Q.~H.}\ \bibnamefont{Tran}},\ and\ \bibinfo {author}
  {\bibfnamefont{S.}~\bibnamefont{Charar}}}%
  , \bibinfo {year} {2012},\ \bibfield{title}{%
  \enquote{\bibinfo {title} {In-plane magnetic anisotropy and temperature
  dependence of switching field in {(Ga,Mn)As} as ferromagnetic
  semiconductors},}\ }%
  \bibfield{journal}{%
  \bibinfo {journal} {J. Nanosci. Nanotechnol.}\ }%
  \textbf{\bibinfo {volume} {12}},\ \bibinfo {pages} {4868}%
  \bibAnnoteFile{NoStop}{Kamara:2012_JNN}%
\bibitem[{\citenamefont{Kanamori}(1959)}]{Kanamori:1959_JPCS}%
  \BibitemOpen
  \bibfield{author}{%
  \bibinfo {author} {\bibnamefont{Kanamori}, \bibfnamefont{J.}}}%
  , \bibinfo {year} {1959},\ \bibfield{title}{%
  \enquote{\bibinfo {title} {Superexchange interaction and symmetry properties
  of electron orbitals},}\ }%
  \bibfield{journal}{%
  \bibinfo {journal} {J. Phys. Chem. Solids}\ }%
  \textbf{\bibinfo {volume} {10}},\ \bibinfo {pages} {87}%
  \bibAnnoteFile{NoStop}{Kanamori:1959_JPCS}%
\bibitem[{\citenamefont{Kapetanakis}\
  \emph{et~al.}(2011)\citenamefont{Kapetanakis}, \citenamefont{Lingos},
  \citenamefont{Piermarocchi}, \citenamefont{Wang},\ and\
  \citenamefont{Perakis}}]{Kapetanakis:2011_APL}%
  \BibitemOpen
  \bibfield{author}{%
  \bibinfo {author} {\bibnamefont{Kapetanakis}, \bibfnamefont{M.~D.}}, \bibinfo
  {author} {\bibfnamefont{P.~C.}\ \bibnamefont{Lingos}}, \bibinfo {author}
  {\bibfnamefont{C.}~\bibnamefont{Piermarocchi}}, \bibinfo {author}
  {\bibfnamefont{J.}~\bibnamefont{Wang}},\ and\ \bibinfo {author}
  {\bibfnamefont{I.~E.}\ \bibnamefont{Perakis}}}%
  , \bibinfo {year} {2011},\ \bibfield{title}{%
  \enquote{\bibinfo {title} {All-optical four-state magnetization reversal in
  {(Ga,Mn)As} ferromagnetic semiconductors},}\ }%
  \bibfield{journal}{%
  \bibinfo {journal} {Appl. Phys. Lett.}\ }%
  \textbf{\bibinfo {volume} {99}},\ \bibinfo {eid} {091111}%
  \bibAnnoteFile{NoStop}{Kapetanakis:2011_APL}%
\bibitem[{\citenamefont{Kapetanakis}\
  \emph{et~al.}(2009)\citenamefont{Kapetanakis}, \citenamefont{Perakis},
  \citenamefont{Wickey}, \citenamefont{Piermarocchi},\ and\
  \citenamefont{Wang}}]{Kapetanakis:2009_PRL}%
  \BibitemOpen
  \bibfield{author}{%
  \bibinfo {author} {\bibnamefont{Kapetanakis}, \bibfnamefont{M.~D.}}, \bibinfo
  {author} {\bibfnamefont{I.~E.}\ \bibnamefont{Perakis}}, \bibinfo {author}
  {\bibfnamefont{K.~J.}\ \bibnamefont{Wickey}}, \bibinfo {author}
  {\bibfnamefont{C.}~\bibnamefont{Piermarocchi}},\ and\ \bibinfo {author}
  {\bibfnamefont{J.}~\bibnamefont{Wang}}}%
  , \bibinfo {year} {2009},\ \bibfield{title}{%
  \enquote{\bibinfo {title} {Femtosecond coherent control of spins in
  {(Ga,Mn)As} ferromagnetic semiconductors using light},}\ }%
  \bibfield{journal}{%
  \bibinfo {journal} {Phys. Rev. Lett.}\ }%
  \textbf{\bibinfo {volume} {103}},\ \bibinfo {pages} {047404}%
  \bibAnnoteFile{NoStop}{Kapetanakis:2009_PRL}%
\bibitem[{\citenamefont{Karczewski}\
  \emph{et~al.}(2003)\citenamefont{Karczewski}, \citenamefont{Sawicki},
  \citenamefont{Ivanov}, \citenamefont{Ruester}, \citenamefont{Grabecki},
  \citenamefont{Matsukura}, \citenamefont{Molenkamp},\ and\
  \citenamefont{Dietl}}]{Karczewski:2003_JSNM}%
  \BibitemOpen
  \bibfield{author}{%
  \bibinfo {author} {\bibnamefont{Karczewski}, \bibfnamefont{G.}}, \bibinfo
  {author} {\bibfnamefont{M.}~\bibnamefont{Sawicki}}, \bibinfo {author}
  {\bibfnamefont{V.}~\bibnamefont{Ivanov}}, \bibinfo {author}
  {\bibfnamefont{C.}~\bibnamefont{Ruester}}, \bibinfo {author}
  {\bibfnamefont{G.}~\bibnamefont{Grabecki}}, \bibinfo {author}
  {\bibfnamefont{F.}~\bibnamefont{Matsukura}}, \bibinfo {author}
  {\bibfnamefont{L.W.}\ \bibnamefont{Molenkamp}},\ and\ \bibinfo {author}
  {\bibfnamefont{T.}~\bibnamefont{Dietl}}}%
  , \bibinfo {year} {2003},\ \bibfield{title}{%
  \enquote{\bibinfo {title} {Ferromagnetism in {(Zn,Cr)Se} layers grown by
  molecular beam epitaxy},}\ }%
  \bibfield{journal}{%
  \bibinfo {journal} {J. Supercond. Nov. Magn.}\ }%
  \textbf{\bibinfo {volume} {16}},\ \bibinfo {pages} {55}%
  \bibAnnoteFile{NoStop}{Karczewski:2003_JSNM}%
\bibitem[{\citenamefont{Katsumoto}\
  \emph{et~al.}(1998)\citenamefont{Katsumoto}, \citenamefont{Oiwa},
  \citenamefont{Iye}, \citenamefont{Ohno}, \citenamefont{Matsukura},
  \citenamefont{Shen},\ and\ \citenamefont{Sugawara}}]{Katsumoto:1998_pssb}%
  \BibitemOpen
  \bibfield{author}{%
  \bibinfo {author} {\bibnamefont{Katsumoto}, \bibfnamefont{S.}}, \bibinfo
  {author} {\bibfnamefont{A.}~\bibnamefont{Oiwa}}, \bibinfo {author}
  {\bibfnamefont{Y.}~\bibnamefont{Iye}}, \bibinfo {author}
  {\bibfnamefont{H.}~\bibnamefont{Ohno}}, \bibinfo {author}
  {\bibfnamefont{F.}~\bibnamefont{Matsukura}}, \bibinfo {author}
  {\bibfnamefont{A.}~\bibnamefont{Shen}},\ and\ \bibinfo {author}
  {\bibfnamefont{Y.}~\bibnamefont{Sugawara}}}%
  , \bibinfo {year} {1998},\ \bibfield{title}{%
  \enquote{\bibinfo {title} {Strongly anisotropic hopping conduction in
  {(Ga,Mn)As/GaAs}},}\ }%
  \bibfield{journal}{%
  \bibinfo {journal} {Phys. Status Solidi B}\ }%
  \textbf{\bibinfo {volume} {205}},\ \bibinfo {pages} {115}%
  \bibAnnoteFile{NoStop}{Katsumoto:1998_pssb}%
\bibitem[{\citenamefont{{K{\c{e}}pa}}\
  \emph{et~al.}(2003)\citenamefont{{K{\c{e}}pa}}, \citenamefont{Khoi},
  \citenamefont{Brown}, \citenamefont{Sawicki}, \citenamefont{Furdyna},
  \citenamefont{{Giebu{\l}towicz}},\ and\
  \citenamefont{Dietl}}]{Kepa:2003_PRL}%
  \BibitemOpen
  \bibfield{author}{%
  \bibinfo {author} {\bibnamefont{{K{\c{e}}pa}}, \bibfnamefont{H.}}, \bibinfo
  {author} {\bibfnamefont{Le~Van}\ \bibnamefont{Khoi}}, \bibinfo {author}
  {\bibfnamefont{C.~M.}\ \bibnamefont{Brown}}, \bibinfo {author}
  {\bibfnamefont{M.}~\bibnamefont{Sawicki}}, \bibinfo {author}
  {\bibfnamefont{J.~K.}\ \bibnamefont{Furdyna}}, \bibinfo {author}
  {\bibfnamefont{T.~M.}\ \bibnamefont{{Giebu{\l}towicz}}},\ and\ \bibinfo
  {author} {\bibfnamefont{T.}~\bibnamefont{Dietl}}}%
  , \bibinfo {year} {2003},\ \bibfield{title}{%
  \enquote{\bibinfo {title} {Probing hole-induced ferromagnetic exchange in
  magnetic semiconductors by inelastic neutron scattering},}\ }%
  \bibfield{journal}{%
  \bibinfo {journal} {Phys. Rev. Lett.}\ }%
  \textbf{\bibinfo {volume} {91}},\ \bibinfo {pages} {087205}%
  \bibAnnoteFile{NoStop}{Kepa:2003_PRL}%
\bibitem[{\citenamefont{K\c{e}pa}\ \emph{et~al.}(2001)\citenamefont{K\c{e}pa},
  \citenamefont{Kutner-Pielaszek}, \citenamefont{Twardowski},
  \citenamefont{Majkrzak}, \citenamefont{Sadowski}, \citenamefont{Story},\ and\
  \citenamefont{Giebultowicz}}]{Kepa:2001_PRB}%
  \BibitemOpen
  \bibfield{author}{%
  \bibinfo {author} {\bibnamefont{K\c{e}pa}, \bibfnamefont{H.}}, \bibinfo
  {author} {\bibfnamefont{J.}~\bibnamefont{Kutner-Pielaszek}}, \bibinfo
  {author} {\bibfnamefont{A.}~\bibnamefont{Twardowski}}, \bibinfo {author}
  {\bibfnamefont{C.~F.}\ \bibnamefont{Majkrzak}}, \bibinfo {author}
  {\bibfnamefont{J.}~\bibnamefont{Sadowski}}, \bibinfo {author}
  {\bibfnamefont{T.}~\bibnamefont{Story}},\ and\ \bibinfo {author}
  {\bibfnamefont{T.~M.}\ \bibnamefont{Giebultowicz}}}%
  , \bibinfo {year} {2001},\ \bibfield{title}{%
  \enquote{\bibinfo {title} {Ferromagnetism of {GaMnAs} studied by polarized
  neutron reflectometry},}\ }%
  \bibfield{journal}{%
  \bibinfo {journal} {Phys. Rev. B}\ }%
  \textbf{\bibinfo {volume} {64}},\ \bibinfo {pages} {121302}%
  \bibAnnoteFile{NoStop}{Kepa:2001_PRB}%
\bibitem[{\citenamefont{Khazen}\ \emph{et~al.}(2008)\citenamefont{Khazen},
  \citenamefont{von Bardeleben}, \citenamefont{Cantin},
  \citenamefont{Thevenard}, \citenamefont{Largeau}, \citenamefont{Mauguin},\
  and\ \citenamefont{Lema\^itre}}]{Khazen:2008_PRB}%
  \BibitemOpen
  \bibfield{author}{%
  \bibinfo {author} {\bibnamefont{Khazen}, \bibfnamefont{Kh.}}, \bibinfo
  {author} {\bibfnamefont{H.~J.}\ \bibnamefont{von Bardeleben}}, \bibinfo
  {author} {\bibfnamefont{J.~L.}\ \bibnamefont{Cantin}}, \bibinfo {author}
  {\bibfnamefont{L.}~\bibnamefont{Thevenard}}, \bibinfo {author}
  {\bibfnamefont{L.}~\bibnamefont{Largeau}}, \bibinfo {author}
  {\bibfnamefont{O.}~\bibnamefont{Mauguin}},\ and\ \bibinfo {author}
  {\bibfnamefont{A.}~\bibnamefont{Lema\^itre}}}%
  , \bibinfo {year} {2008},\ \bibfield{title}{%
  \enquote{\bibinfo {title} {Ferromagnetic resonance of
  {Ga$_{0.93}$Mn$_{0.07}$As} thin films with constant {Mn} and variable
  free-hole concentrations},}\ }%
  \bibfield{journal}{%
  \bibinfo {journal} {Phys. Rev. B}\ }%
  \textbf{\bibinfo {volume} {77}},\ \bibinfo {pages} {165204}%
  \bibAnnoteFile{NoStop}{Khazen:2008_PRB}%
\bibitem[{\citenamefont{Kimel}\ \emph{et~al.}(2004)\citenamefont{Kimel},
  \citenamefont{Astakhov}, \citenamefont{Schott}, \citenamefont{Kirilyuk},
  \citenamefont{Yakovlev}, \citenamefont{Karczewski}, \citenamefont{Ossau},
  \citenamefont{Schmidt}, \citenamefont{Molenkamp},\ and\
  \citenamefont{Rasing}}]{Kimel:2004_PRL}%
  \BibitemOpen
  \bibfield{author}{%
  \bibinfo {author} {\bibnamefont{Kimel}, \bibfnamefont{A.~V.}}, \bibinfo
  {author} {\bibfnamefont{G.~V.}\ \bibnamefont{Astakhov}}, \bibinfo {author}
  {\bibfnamefont{G.~M.}\ \bibnamefont{Schott}}, \bibinfo {author}
  {\bibfnamefont{A.}~\bibnamefont{Kirilyuk}}, \bibinfo {author}
  {\bibfnamefont{D.~R.}\ \bibnamefont{Yakovlev}}, \bibinfo {author}
  {\bibfnamefont{G.}~\bibnamefont{Karczewski}}, \bibinfo {author}
  {\bibfnamefont{W.}~\bibnamefont{Ossau}}, \bibinfo {author}
  {\bibfnamefont{G.}~\bibnamefont{Schmidt}}, \bibinfo {author}
  {\bibfnamefont{L.~W.}\ \bibnamefont{Molenkamp}},\ and\ \bibinfo {author}
  {\bibfnamefont{Th.}\ \bibnamefont{Rasing}}}%
  , \bibinfo {year} {2004},\ \bibfield{title}{%
  \enquote{\bibinfo {title} {Picosecond dynamics of the photoinduced spin
  polarization in epitaxial {(Ga,Mn)As} films},}\ }%
  \bibfield{journal}{%
  \bibinfo {journal} {Phys. Rev. Lett.}\ }%
  \textbf{\bibinfo {volume} {92}},\ \bibinfo {pages} {237203}%
  \bibAnnoteFile{NoStop}{Kimel:2004_PRL}%
\bibitem[{\citenamefont{King}\ \emph{et~al.}(2011)\citenamefont{King},
  \citenamefont{Zemen}, \citenamefont{Olejn\'ik}, \citenamefont{Hor\'ak},
  \citenamefont{Haigh}, \citenamefont{Nov\'ak}, \citenamefont{Irvine},
  \citenamefont{Ku\ifmmode~\check{c}\else \v{c}\fi{}era},
  \citenamefont{Hol\'y}, \citenamefont{Campion}, \citenamefont{Gallagher},\
  and\ \citenamefont{Jungwirth}}]{King:2011_PRB}%
  \BibitemOpen
  \bibfield{author}{%
  \bibinfo {author} {\bibnamefont{King}, \bibfnamefont{C.~S.}}, \bibinfo
  {author} {\bibfnamefont{J.}~\bibnamefont{Zemen}}, \bibinfo {author}
  {\bibfnamefont{K.}~\bibnamefont{Olejn\'ik}}, \bibinfo {author}
  {\bibfnamefont{L.}~\bibnamefont{Hor\'ak}}, \bibinfo {author}
  {\bibfnamefont{J.~A.}\ \bibnamefont{Haigh}}, \bibinfo {author}
  {\bibfnamefont{V.}~\bibnamefont{Nov\'ak}}, \bibinfo {author}
  {\bibfnamefont{A.}~\bibnamefont{Irvine}}, \bibinfo {author}
  {\bibfnamefont{J.}~\bibnamefont{Ku\ifmmode~\check{c}\else \v{c}\fi{}era}},
  \bibinfo {author} {\bibfnamefont{V.}~\bibnamefont{Hol\'y}}, \bibinfo {author}
  {\bibfnamefont{R.~P.}\ \bibnamefont{Campion}}, \bibinfo {author}
  {\bibfnamefont{B.~L.}\ \bibnamefont{Gallagher}},\ and\ \bibinfo {author}
  {\bibfnamefont{T.}~\bibnamefont{Jungwirth}}}%
  , \bibinfo {year} {2011},\ \bibfield{title}{%
  \enquote{\bibinfo {title} {Strain control of magnetic anisotropy in
  {(Ga,Mn)As} microbars},}\ }%
  \bibfield{journal}{%
  \bibinfo {journal} {Phys. Rev. B}\ }%
  \textbf{\bibinfo {volume} {83}},\ \bibinfo {pages} {115312}%
  \bibAnnoteFile{NoStop}{King:2011_PRB}%
\bibitem[{\citenamefont{Kirby}\ \emph{et~al.}(2006)\citenamefont{Kirby},
  \citenamefont{Borchers}, \citenamefont{Rhyne}, \citenamefont{O'Donovan},
  \citenamefont{{te {Velthuis}}}, \citenamefont{Roy},
  \citenamefont{Sanchez-Hanke}, \citenamefont{Wojtowicz}, \citenamefont{Liu},
  \citenamefont{Lim}, \citenamefont{Dobrowolska},\ and\
  \citenamefont{Furdyna}}]{Kirby:2006_PRB}%
  \BibitemOpen
  \bibfield{author}{%
  \bibinfo {author} {\bibnamefont{Kirby}, \bibfnamefont{B.~J.}}, \bibinfo
  {author} {\bibfnamefont{J.~A.}\ \bibnamefont{Borchers}}, \bibinfo {author}
  {\bibfnamefont{J.~J.}\ \bibnamefont{Rhyne}}, \bibinfo {author}
  {\bibfnamefont{K.~V.}\ \bibnamefont{O'Donovan}}, \bibinfo {author}
  {\bibfnamefont{S.~G.~E.}\ \bibnamefont{{te {Velthuis}}}}, \bibinfo {author}
  {\bibfnamefont{S.}~\bibnamefont{Roy}}, \bibinfo {author}
  {\bibfnamefont{Cecilia}\ \bibnamefont{Sanchez-Hanke}}, \bibinfo {author}
  {\bibfnamefont{T.}~\bibnamefont{Wojtowicz}}, \bibinfo {author}
  {\bibfnamefont{X.}~\bibnamefont{Liu}}, \bibinfo {author}
  {\bibfnamefont{W.~L.}\ \bibnamefont{Lim}}, \bibinfo {author}
  {\bibfnamefont{M.}~\bibnamefont{Dobrowolska}},\ and\ \bibinfo {author}
  {\bibfnamefont{J.~K.}\ \bibnamefont{Furdyna}}}%
  , \bibinfo {year} {2006},\ \bibfield{title}{%
  \enquote{\bibinfo {title} {Magnetic and chemical nonuniformity in
  {Ga$_{1-x}$Mn$_{x}$As} as probed by neutron and x-ray reflectometry},}\ }%
  \bibfield{journal}{%
  \bibinfo {journal} {Phys. Rev. B}\ }%
  \textbf{\bibinfo {volume} {74}},\ \bibinfo {pages} {245304}%
  \bibAnnoteFile{NoStop}{Kirby:2006_PRB}%
\bibitem[{\citenamefont{Knoff}\ \emph{et~al.}(2009)\citenamefont{Knoff},
  \citenamefont{Domukhovski}, \citenamefont{Dybko}, \citenamefont{Dziawa},
  \citenamefont{Jakie{\l}a}, \citenamefont{{\L}usakowska},
  \citenamefont{Reszka}, \citenamefont{Swiatek}, \citenamefont{Taliashvili},
  \citenamefont{Story}, \citenamefont{Sza{\l}owski},\ and\
  \citenamefont{Balcerzak}}]{Knoff:2009_APPA}%
  \BibitemOpen
  \bibfield{author}{%
  \bibinfo {author} {\bibnamefont{Knoff}, \bibfnamefont{W.}}, \bibinfo {author}
  {\bibfnamefont{V.}~\bibnamefont{Domukhovski}}, \bibinfo {author}
  {\bibfnamefont{K.}~\bibnamefont{Dybko}}, \bibinfo {author}
  {\bibfnamefont{P.}~\bibnamefont{Dziawa}}, \bibinfo {author}
  {\bibfnamefont{R.}~\bibnamefont{Jakie{\l}a}}, \bibinfo {author}
  {\bibfnamefont{E.}~\bibnamefont{{\L}usakowska}}, \bibinfo {author}
  {\bibfnamefont{A.}~\bibnamefont{Reszka}}, \bibinfo {author}
  {\bibfnamefont{K.}~\bibnamefont{Swiatek}}, \bibinfo {author}
  {\bibfnamefont{B.}~\bibnamefont{Taliashvili}}, \bibinfo {author}
  {\bibfnamefont{T.}~\bibnamefont{Story}}, \bibinfo {author}
  {\bibfnamefont{K.}~\bibnamefont{Sza{\l}owski}},\ and\ \bibinfo {author}
  {\bibfnamefont{T.}~\bibnamefont{Balcerzak}}}%
  , \bibinfo {year} {2009},\ \bibfield{title}{%
  \enquote{\bibinfo {title} {Ferromagnetic transition in {Ge$_{1-x}$Mn$_x$Te}
  layers},}\ }%
  \bibfield{journal}{%
  \bibinfo {journal} {Acta Phys. Pol. A}\ }%
  \textbf{\bibinfo {volume} {116}},\ \bibinfo {pages} {904}%
  \bibAnnoteFile{NoStop}{Knoff:2009_APPA}%
\bibitem[{\citenamefont{Knoff}\ \emph{et~al.}(2011)\citenamefont{Knoff},
  \citenamefont{\'Swi\c{a}tek}, \citenamefont{Andrearczyk},
  \citenamefont{Domukhovski}, \citenamefont{Dziawa}, \citenamefont{Kowalczyk},
  \citenamefont{{\L}usakowska}, \citenamefont{\v{S}iu\v{s}ys},
  \citenamefont{Taliashvili}, \citenamefont{Wr\'obel},\ and\
  \citenamefont{Story}}]{Knoff:2011_PSSB}%
  \BibitemOpen
  \bibfield{author}{%
  \bibinfo {author} {\bibnamefont{Knoff}, \bibfnamefont{W.}}, \bibinfo {author}
  {\bibfnamefont{K.}~\bibnamefont{\'Swi\c{a}tek}}, \bibinfo {author}
  {\bibfnamefont{T.}~\bibnamefont{Andrearczyk}}, \bibinfo {author}
  {\bibfnamefont{V.}~\bibnamefont{Domukhovski}}, \bibinfo {author}
  {\bibfnamefont{P.}~\bibnamefont{Dziawa}}, \bibinfo {author}
  {\bibfnamefont{L.}~\bibnamefont{Kowalczyk}}, \bibinfo {author}
  {\bibfnamefont{E.}~\bibnamefont{{\L}usakowska}}, \bibinfo {author}
  {\bibfnamefont{A.}~\bibnamefont{\v{S}iu\v{s}ys}}, \bibinfo {author}
  {\bibfnamefont{B.}~\bibnamefont{Taliashvili}}, \bibinfo {author}
  {\bibfnamefont{J.}~\bibnamefont{Wr\'obel}},\ and\ \bibinfo {author}
  {\bibfnamefont{T.}~\bibnamefont{Story}}}%
  , \bibinfo {year} {2011},\ \bibfield{title}{%
  \enquote{\bibinfo {title} {Magnetic anisotropy of semiconductor {(Ge,Mn)Te}
  microstructures produced by laser and electron beam induced
  crystallization},}\ }%
  \bibfield{journal}{%
  \bibinfo {journal} {Phys. Staus Solidi B}\ }%
  \textbf{\bibinfo {volume} {248}},\ \bibinfo {pages} {1605}%
  \bibAnnoteFile{NoStop}{Knoff:2011_PSSB}%
\bibitem[{\citenamefont{Kobayashi}\
  \emph{et~al.}(2013)\citenamefont{Kobayashi}, \citenamefont{Muneta},
  \citenamefont{Takeda}, \citenamefont{Harada}, \citenamefont{Fujimori},
  \citenamefont{Krempask\v{y}}, \citenamefont{Schmitt}, \citenamefont{Ohya},
  \citenamefont{Tanaka}, \citenamefont{Oshima},\ and\
  \citenamefont{Strocov}}]{Kobayashi:2013_arXiv}%
  \BibitemOpen
  \bibfield{author}{%
  \bibinfo {author} {\bibnamefont{Kobayashi}, \bibfnamefont{M.}}, \bibinfo
  {author} {\bibfnamefont{I.}~\bibnamefont{Muneta}}, \bibinfo {author}
  {\bibfnamefont{Y.}~\bibnamefont{Takeda}}, \bibinfo {author}
  {\bibfnamefont{Y.}~\bibnamefont{Harada}}, \bibinfo {author}
  {\bibfnamefont{A.}~\bibnamefont{Fujimori}}, \bibinfo {author}
  {\bibfnamefont{J.}~\bibnamefont{Krempask\v{y}}}, \bibinfo {author}
  {\bibfnamefont{T.}~\bibnamefont{Schmitt}}, \bibinfo {author}
  {\bibfnamefont{S.}~\bibnamefont{Ohya}}, \bibinfo {author}
  {\bibfnamefont{M.}~\bibnamefont{Tanaka}}, \bibinfo {author}
  {\bibfnamefont{M.}~\bibnamefont{Oshima}},\ and\ \bibinfo {author}
  {\bibfnamefont{V.~N.}\ \bibnamefont{Strocov}}}%
  , \bibinfo {year} {2013},\ \bibfield{title}{%
  \enquote{\bibinfo {title} {Unveiling the impurity band inducing
  ferromagnetism in magnetic semiconductor {(Ga,Mn)As}},}\ }%
  \bibinfo {journal} {arXiv:1302.0063}%
  \bibAnnoteFile{NoStop}{Kobayashi:2013_arXiv}%
\bibitem[{\citenamefont{Kodzuka}\ \emph{et~al.}(2009)\citenamefont{Kodzuka},
  \citenamefont{Ohkubo}, \citenamefont{Hono}, \citenamefont{Matsukura},\ and\
  \citenamefont{Ohno}}]{Kodzuka:2009_U}%
  \BibitemOpen
\bibfield{journal}{%
    }%
  \bibfield{author}{%
  \bibinfo {author} {\bibnamefont{Kodzuka}, \bibfnamefont{M.}}, \bibinfo
  {author} {\bibfnamefont{T.}~\bibnamefont{Ohkubo}}, \bibinfo {author}
  {\bibfnamefont{K.}~\bibnamefont{Hono}}, \bibinfo {author}
  {\bibfnamefont{F.}~\bibnamefont{Matsukura}},\ and\ \bibinfo {author}
  {\bibfnamefont{H.}~\bibnamefont{Ohno}}}%
  , \bibinfo {year} {2009},\ \bibfield{title}{%
  \enquote{\bibinfo {title} {{3DAP} analysis of {(Ga,Mn)As} diluted magnetic
  semiconductor thin film},}\ }%
  \bibfield{journal}{%
  \bibinfo {journal} {Ultramicroscopy}\ }%
  \textbf{\bibinfo {volume} {109}},\ \bibinfo {pages} {644}%
  \bibAnnoteFile{NoStop}{Kodzuka:2009_U}%
\bibitem[{\citenamefont{Koenraad}\ and\
  \citenamefont{Flatt\'e}(2011)}]{Koenraad:2011_NM}%
  \BibitemOpen
  \bibfield{author}{%
  \bibinfo {author} {\bibnamefont{Koenraad}, \bibfnamefont{P.~M.}},\ and\
  \bibinfo {author} {\bibfnamefont{M.~E.}\ \bibnamefont{Flatt\'e}}}%
  , \bibinfo {year} {2011},\ \bibfield{title}{%
  \enquote{\bibinfo {title} {Single dopants in semiconductors},}\ }%
  \bibfield{journal}{%
  \bibinfo {journal} {Nat. Mater.}\ }%
  \textbf{\bibinfo {volume} {10}},\ \bibinfo {pages} {91}%
  \bibAnnoteFile{NoStop}{Koenraad:2011_NM}%
\bibitem[{\citenamefont{Kohda}\ \emph{et~al.}(2006)\citenamefont{Kohda},
  \citenamefont{Kita}, \citenamefont{Ohno}, \citenamefont{Matsukura},\ and\
  \citenamefont{Ohno}}]{Kohda:2006_APL}%
  \BibitemOpen
  \bibfield{author}{%
  \bibinfo {author} {\bibnamefont{Kohda}, \bibfnamefont{M.}}, \bibinfo {author}
  {\bibfnamefont{T.}~\bibnamefont{Kita}}, \bibinfo {author}
  {\bibfnamefont{Y.}~\bibnamefont{Ohno}}, \bibinfo {author}
  {\bibfnamefont{F.}~\bibnamefont{Matsukura}},\ and\ \bibinfo {author}
  {\bibfnamefont{H.}~\bibnamefont{Ohno}}}%
  , \bibinfo {year} {2006},\ \bibfield{title}{%
  \enquote{\bibinfo {title} {Spectroscopic analysis of ballistic spin injection
  in a three-terminal device based on a {p-(Ga,Mn)As/n$^+$-GaAs} {Esaki}
  diode},}\ }%
  \bibfield{journal}{%
  \bibinfo {journal} {Appl. Phys. Lett.}\ }%
  \textbf{\bibinfo {volume} {89}},\ \bibinfo {pages} {012103}%
  \bibAnnoteFile{NoStop}{Kohda:2006_APL}%
\bibitem[{\citenamefont{Kohda}\ \emph{et~al.}(2001)\citenamefont{Kohda},
  \citenamefont{Ohno}, \citenamefont{Takamura}, \citenamefont{Matsukura},\ and\
  \citenamefont{Ohno}}]{Kohda:2001_JJAP}%
  \BibitemOpen
  \bibfield{author}{%
  \bibinfo {author} {\bibnamefont{Kohda}, \bibfnamefont{M.}}, \bibinfo {author}
  {\bibfnamefont{Y.}~\bibnamefont{Ohno}}, \bibinfo {author}
  {\bibfnamefont{K.}~\bibnamefont{Takamura}}, \bibinfo {author}
  {\bibfnamefont{F.}~\bibnamefont{Matsukura}},\ and\ \bibinfo {author}
  {\bibfnamefont{H.}~\bibnamefont{Ohno}}}%
  , \bibinfo {year} {2001},\ \bibfield{title}{%
  \enquote{\bibinfo {title} {A spin {Esaki} diode},}\ }%
  \bibfield{journal}{%
  \bibinfo {journal} {Jpn. J. Appl. Phys.}\ }%
  \textbf{\bibinfo {volume} {40}},\ \bibinfo {pages} {L1274}%
  \bibAnnoteFile{NoStop}{Kohda:2001_JJAP}%
\bibitem[{\citenamefont{Kojima}\ \emph{et~al.}(2007)\citenamefont{Kojima},
  \citenamefont{H\'eroux}, \citenamefont{Shimano}, \citenamefont{Hashimoto},
  \citenamefont{Katsumoto}, \citenamefont{Iye},\ and\
  \citenamefont{Kuwata-Gonokami}}]{Kojima:2007_PRB}%
  \BibitemOpen
  \bibfield{author}{%
  \bibinfo {author} {\bibnamefont{Kojima}, \bibfnamefont{E.}}, \bibinfo
  {author} {\bibfnamefont{J.~B.}\ \bibnamefont{H\'eroux}}, \bibinfo {author}
  {\bibfnamefont{R.}~\bibnamefont{Shimano}}, \bibinfo {author}
  {\bibfnamefont{Y.}~\bibnamefont{Hashimoto}}, \bibinfo {author}
  {\bibfnamefont{S.}~\bibnamefont{Katsumoto}}, \bibinfo {author}
  {\bibfnamefont{Y.}~\bibnamefont{Iye}},\ and\ \bibinfo {author}
  {\bibfnamefont{M.}~\bibnamefont{Kuwata-Gonokami}}}%
  , \bibinfo {year} {2007},\ \bibfield{title}{%
  \enquote{\bibinfo {title} {Experimental investigation of polaron effects in
  {Ga$_{1-x}$Mn$_{x}$As} by time-resolved and continuous-wave midinfrared
  spectroscopy},}\ }%
  \bibfield{journal}{%
  \bibinfo {journal} {Phys. Rev. B}\ }%
  \textbf{\bibinfo {volume} {76}},\ \bibinfo {pages} {195323}%
  \bibAnnoteFile{NoStop}{Kojima:2007_PRB}%
\bibitem[{\citenamefont{Kojima}\ \emph{et~al.}(2003)\citenamefont{Kojima},
  \citenamefont{Shimano}, \citenamefont{Hashimoto}, \citenamefont{Katsumoto},
  \citenamefont{Iye},\ and\ \citenamefont{Kuwata-Gonokami}}]{Kojima:2003_PRB}%
  \BibitemOpen
  \bibfield{author}{%
  \bibinfo {author} {\bibnamefont{Kojima}, \bibfnamefont{E.}}, \bibinfo
  {author} {\bibfnamefont{R.}~\bibnamefont{Shimano}}, \bibinfo {author}
  {\bibfnamefont{Y.}~\bibnamefont{Hashimoto}}, \bibinfo {author}
  {\bibfnamefont{S.}~\bibnamefont{Katsumoto}}, \bibinfo {author}
  {\bibfnamefont{Y.}~\bibnamefont{Iye}},\ and\ \bibinfo {author}
  {\bibfnamefont{M.}~\bibnamefont{Kuwata-Gonokami}}}%
  , \bibinfo {year} {2003},\ \bibfield{title}{%
  \enquote{\bibinfo {title} {Observation of the spin-charge thermal isolation
  of ferromagnetic {Ga$_{0.94}$Mn$_{0.06}$As} by time-resolved magneto-optical
  measurements},}\ }%
  \bibfield{journal}{%
  \bibinfo {journal} {Phys. Rev. B}\ }%
  \textbf{\bibinfo {volume} {68}},\ \bibinfo {pages} {193203}%
  \bibAnnoteFile{NoStop}{Kojima:2003_PRB}%
\bibitem[{\citenamefont{Kong}\ \emph{et~al.}(2005)\citenamefont{Kong},
  \citenamefont{Trampert}, \citenamefont{Guo}, \citenamefont{Daweritz},\ and\
  \citenamefont{Ploog}}]{Kong:2005_JAP}%
  \BibitemOpen
  \bibfield{author}{%
  \bibinfo {author} {\bibnamefont{Kong}, \bibfnamefont{X.}}, \bibinfo {author}
  {\bibfnamefont{A.}~\bibnamefont{Trampert}}, \bibinfo {author}
  {\bibfnamefont{X.~X.}\ \bibnamefont{Guo}}, \bibinfo {author}
  {\bibfnamefont{L.}~\bibnamefont{Daweritz}},\ and\ \bibinfo {author}
  {\bibfnamefont{K.~H.}\ \bibnamefont{Ploog}}}%
  , \bibinfo {year} {2005},\ \bibfield{title}{%
  \enquote{\bibinfo {title} {Anisotropic distribution of stacking faults in
  {(Ga,Mn)As} digital ferromagnetic heterostructures grown by low-temperature
  molecular-beam epitaxy},}\ }%
  \bibfield{journal}{%
  \bibinfo {journal} {J. Appl. Phys.}\ }%
  \textbf{\bibinfo {volume} {97}},\ \bibinfo {pages} {036105}%
  \bibAnnoteFile{NoStop}{Kong:2005_JAP}%
\bibitem[{\citenamefont{{K{\"o}nig}}\
  \emph{et~al.}(2001)\citenamefont{{K{\"o}nig}}, \citenamefont{Jungwirth},\
  and\ \citenamefont{MacDonald}}]{Konig:2001_PRB}%
  \BibitemOpen
  \bibfield{author}{%
  \bibinfo {author} {\bibnamefont{{K{\"o}nig}}, \bibfnamefont{J.}}, \bibinfo
  {author} {\bibfnamefont{T.}~\bibnamefont{Jungwirth}},\ and\ \bibinfo {author}
  {\bibfnamefont{A.~H.}\ \bibnamefont{MacDonald}}}%
  , \bibinfo {year} {2001},\ \bibfield{title}{%
  \enquote{\bibinfo {title} {Theory of magnetic properties and spin-wave
  dispersion for ferromagnetic {(Ga,Mn)As}},}\ }%
  \bibfield{journal}{%
  \bibinfo {journal} {Phys. Rev. B}\ }%
  \textbf{\bibinfo {volume} {64}},\ \bibinfo {pages} {184423}%
  \bibAnnoteFile{NoStop}{Konig:2001_PRB}%
\bibitem[{\citenamefont{Kopeck\'y}\
  \emph{et~al.}(2011)\citenamefont{Kopeck\'y}, \citenamefont{Kub},
  \citenamefont{M\'aca}, \citenamefont{Ma\ifmmode~\check{s}\else \v{s}\fi{}ek},
  \citenamefont{Pacherov\'a}, \citenamefont{Rushforth},
  \citenamefont{Gallagher}, \citenamefont{Campion}, \citenamefont{Nov\'ak},\
  and\ \citenamefont{Jungwirth}}]{Kopecky:2011_PRB}%
  \BibitemOpen
  \bibfield{author}{%
  \bibinfo {author} {\bibnamefont{Kopeck\'y}, \bibfnamefont{M.}}, \bibinfo
  {author} {\bibfnamefont{J.}~\bibnamefont{Kub}}, \bibinfo {author}
  {\bibfnamefont{F.}~\bibnamefont{M\'aca}}, \bibinfo {author}
  {\bibfnamefont{J.}~\bibnamefont{Ma\ifmmode~\check{s}\else \v{s}\fi{}ek}},
  \bibinfo {author} {\bibfnamefont{O.}~\bibnamefont{Pacherov\'a}}, \bibinfo
  {author} {\bibfnamefont{A.~W.}\ \bibnamefont{Rushforth}}, \bibinfo {author}
  {\bibfnamefont{B.~L.}\ \bibnamefont{Gallagher}}, \bibinfo {author}
  {\bibfnamefont{R.~P.}\ \bibnamefont{Campion}}, \bibinfo {author}
  {\bibfnamefont{V.}~\bibnamefont{Nov\'ak}},\ and\ \bibinfo {author}
  {\bibfnamefont{T.}~\bibnamefont{Jungwirth}}}%
  , \bibinfo {year} {2011},\ \bibfield{title}{%
  \enquote{\bibinfo {title} {Detection of stacking faults breaking the
  [110]/[1$\overline{1}$0] symmetry in ferromagnetic semiconductors {(Ga,Mn)As}
  and {(Ga,Mn)(As,P)}},}\ }%
  \bibfield{journal}{%
  \bibinfo {journal} {Phys. Rev. B}\ }%
  \textbf{\bibinfo {volume} {83}},\ \bibinfo {pages} {235324}%
  \bibAnnoteFile{NoStop}{Kopecky:2011_PRB}%
\bibitem[{\citenamefont{Koshihara}\
  \emph{et~al.}(1997)\citenamefont{Koshihara}, \citenamefont{Oiwa},
  \citenamefont{Hirasawa}, \citenamefont{Katsumoto}, \citenamefont{Iye},
  \citenamefont{Urano}, \citenamefont{Takagi},\ and\
  \citenamefont{Munekata}}]{Koshihara:1997_PRB}%
  \BibitemOpen
  \bibfield{author}{%
  \bibinfo {author} {\bibnamefont{Koshihara}, \bibfnamefont{S.}}, \bibinfo
  {author} {\bibfnamefont{A.}~\bibnamefont{Oiwa}}, \bibinfo {author}
  {\bibfnamefont{M.}~\bibnamefont{Hirasawa}}, \bibinfo {author}
  {\bibfnamefont{S.}~\bibnamefont{Katsumoto}}, \bibinfo {author}
  {\bibfnamefont{Y.}~\bibnamefont{Iye}}, \bibinfo {author}
  {\bibfnamefont{C.}~\bibnamefont{Urano}}, \bibinfo {author}
  {\bibfnamefont{H.}~\bibnamefont{Takagi}},\ and\ \bibinfo {author}
  {\bibfnamefont{H.}~\bibnamefont{Munekata}}}%
  , \bibinfo {year} {1997},\ \bibfield{title}{%
  \enquote{\bibinfo {title} {Ferromagnetic order induced by photogenerated
  carriers in magnetic {III-V} semiconductor heterostructures of
  {(In,Mn)As/GaSb}},}\ }%
  \bibfield{journal}{%
  \bibinfo {journal} {Phys. Rev. Lett.}\ }%
  \textbf{\bibinfo {volume} {78}},\ \bibinfo {pages} {4617}%
  \bibAnnoteFile{NoStop}{Koshihara:1997_PRB}%
\bibitem[{\citenamefont{Kossacki}\
  \emph{et~al.}(2004{\natexlab{a}})\citenamefont{Kossacki},
  \citenamefont{Boukari}, \citenamefont{Bertolini}, \citenamefont{Ferrand},
  \citenamefont{Cibert}, \citenamefont{Tatarenko}, \citenamefont{Gaj},
  \citenamefont{Deveaud}, \citenamefont{Ciulin},\ and\
  \citenamefont{Potemski}}]{Kossacki:2004_PRB}%
  \BibitemOpen
  \bibfield{author}{%
  \bibinfo {author} {\bibnamefont{Kossacki}, \bibfnamefont{P.}}, \bibinfo
  {author} {\bibfnamefont{H.}~\bibnamefont{Boukari}}, \bibinfo {author}
  {\bibfnamefont{M.}~\bibnamefont{Bertolini}}, \bibinfo {author}
  {\bibfnamefont{D.}~\bibnamefont{Ferrand}}, \bibinfo {author}
  {\bibfnamefont{J.}~\bibnamefont{Cibert}}, \bibinfo {author}
  {\bibfnamefont{S.}~\bibnamefont{Tatarenko}}, \bibinfo {author}
  {\bibfnamefont{J.~A.}\ \bibnamefont{Gaj}}, \bibinfo {author}
  {\bibfnamefont{B.}~\bibnamefont{Deveaud}}, \bibinfo {author}
  {\bibfnamefont{V.}~\bibnamefont{Ciulin}},\ and\ \bibinfo {author}
  {\bibfnamefont{M.}~\bibnamefont{Potemski}}}%
  , \bibinfo {year} {2004}{\natexlab{a}},\ \bibfield{title}{%
  \enquote{\bibinfo {title} {Photoluminescence of $p$ -doped quantum wells with
  strong spin splitting},}\ }%
  \bibfield{journal}{%
  \bibinfo {journal} {Phys. Rev. B}\ }%
  \textbf{\bibinfo {volume} {70}},\ \bibinfo {pages} {195337}%
  \bibAnnoteFile{NoStop}{Kossacki:2004_PRB}%
\bibitem[{\citenamefont{Kossacki}\
  \emph{et~al.}(2004{\natexlab{b}})\citenamefont{Kossacki},
  \citenamefont{Pacuski}, \citenamefont{{Ma{\'s}lana}}, \citenamefont{Gaj},
  \citenamefont{Bertolini}, \citenamefont{Ferrand}, \citenamefont{Bleuse},
  \citenamefont{Tatarenko},\ and\ \citenamefont{Cibert}}]{Kossacki:2004_PE}%
  \BibitemOpen
  \bibfield{author}{%
  \bibinfo {author} {\bibnamefont{Kossacki}, \bibfnamefont{P.}}, \bibinfo
  {author} {\bibfnamefont{W.}~\bibnamefont{Pacuski}}, \bibinfo {author}
  {\bibfnamefont{W.}~\bibnamefont{{Ma{\'s}lana}}}, \bibinfo {author}
  {\bibfnamefont{J.~A.}\ \bibnamefont{Gaj}}, \bibinfo {author}
  {\bibfnamefont{M.}~\bibnamefont{Bertolini}}, \bibinfo {author}
  {\bibfnamefont{D.}~\bibnamefont{Ferrand}}, \bibinfo {author}
  {\bibfnamefont{J.}~\bibnamefont{Bleuse}}, \bibinfo {author}
  {\bibfnamefont{S.}~\bibnamefont{Tatarenko}},\ and\ \bibinfo {author}
  {\bibfnamefont{J.}~\bibnamefont{Cibert}}}%
  , \bibinfo {year} {2004}{\natexlab{b}},\ \bibfield{title}{%
  \enquote{\bibinfo {title} {Strain engineering of carrier-induced magnetic
  ordering in {(Cd,Mn)Te} quantum wells},}\ }%
  \bibfield{journal}{%
  \bibinfo {journal} {Physica E}\ }%
  \textbf{\bibinfo {volume} {21}},\ \bibinfo {pages} {943}%
  \bibAnnoteFile{NoStop}{Kossacki:2004_PE}%
\bibitem[{\citenamefont{Kreissl}\ \emph{et~al.}(1996)\citenamefont{Kreissl},
  \citenamefont{Ulrici}, \citenamefont{El-Metoui}, \citenamefont{Vasson},
  \citenamefont{Vasson},\ and\ \citenamefont{Gavaix}}]{Kreissl:1996_PRB}%
  \BibitemOpen
  \bibfield{author}{%
  \bibinfo {author} {\bibnamefont{Kreissl}, \bibfnamefont{J.}}, \bibinfo
  {author} {\bibfnamefont{W.}~\bibnamefont{Ulrici}}, \bibinfo {author}
  {\bibfnamefont{M.}~\bibnamefont{El-Metoui}}, \bibinfo {author}
  {\bibfnamefont{A.~M.}\ \bibnamefont{Vasson}}, \bibinfo {author}
  {\bibfnamefont{A.}~\bibnamefont{Vasson}},\ and\ \bibinfo {author}
  {\bibfnamefont{A.}~\bibnamefont{Gavaix}}}%
  , \bibinfo {year} {1996},\ \bibfield{title}{%
  \enquote{\bibinfo {title} {Neutral manganese acceptor in {GaP}: An
  electron-paramagnetic-resonance study},}\ }%
  \bibfield{journal}{%
  \bibinfo {journal} {Phys. Rev. B}\ }%
  \textbf{\bibinfo {volume} {54}},\ \bibinfo {pages} {10508}%
  \bibAnnoteFile{NoStop}{Kreissl:1996_PRB}%
\bibitem[{\citenamefont{Kronast}\ \emph{et~al.}(2006)\citenamefont{Kronast},
  \citenamefont{Ovsyannikov}, \citenamefont{Vollmer}, \citenamefont{D{\"u}rr},
  \citenamefont{Eberhardt}, \citenamefont{Imperia}, \citenamefont{Schmitz},
  \citenamefont{Schott}, \citenamefont{Ruester}, \citenamefont{Gould},
  \citenamefont{Schmidt}, \citenamefont{Brunner}, \citenamefont{Sawicki},\ and\
  \citenamefont{Molenkamp}}]{Kronast:2006_PRB}%
  \BibitemOpen
  \bibfield{author}{%
  \bibinfo {author} {\bibnamefont{Kronast}, \bibfnamefont{F.}}, \bibinfo
  {author} {\bibfnamefont{R.}~\bibnamefont{Ovsyannikov}}, \bibinfo {author}
  {\bibfnamefont{A.}~\bibnamefont{Vollmer}}, \bibinfo {author}
  {\bibfnamefont{H.~A.}\ \bibnamefont{D{\"u}rr}}, \bibinfo {author}
  {\bibfnamefont{W.}~\bibnamefont{Eberhardt}}, \bibinfo {author}
  {\bibfnamefont{P.}~\bibnamefont{Imperia}}, \bibinfo {author}
  {\bibfnamefont{D.}~\bibnamefont{Schmitz}}, \bibinfo {author}
  {\bibfnamefont{G.~M.}\ \bibnamefont{Schott}}, \bibinfo {author}
  {\bibfnamefont{C.}~\bibnamefont{Ruester}}, \bibinfo {author}
  {\bibfnamefont{C.}~\bibnamefont{Gould}}, \bibinfo {author}
  {\bibfnamefont{G.}~\bibnamefont{Schmidt}}, \bibinfo {author}
  {\bibfnamefont{K.}~\bibnamefont{Brunner}}, \bibinfo {author}
  {\bibfnamefont{M.}~\bibnamefont{Sawicki}},\ and\ \bibinfo {author}
  {\bibfnamefont{L.~W.}\ \bibnamefont{Molenkamp}}}%
  , \bibinfo {year} {2006},\ \bibfield{title}{%
  \enquote{\bibinfo {title} {Mn $3d$ electronic configurations in
  {Ga$_{1-x}$Mn$_{x}$As} ferromagnetic semiconductors and their influence on
  magnetic ordering},}\ }%
  \bibfield{journal}{%
  \bibinfo {journal} {Phys. Rev. B}\ }%
  \textbf{\bibinfo {volume} {74}},\ \bibinfo {pages} {235213}%
  \bibAnnoteFile{NoStop}{Kronast:2006_PRB}%
\bibitem[{\citenamefont{Ku}\ \emph{et~al.}(2003)\citenamefont{Ku},
  \citenamefont{Potashnik}, \citenamefont{Wang}, \citenamefont{Seong},
  \citenamefont{Johnston-Halperin}, \citenamefont{Meyers}, \citenamefont{Chun},
  \citenamefont{Mascarenhas}, \citenamefont{Gossard}, \citenamefont{Awschalom},
  \citenamefont{Schiffer},\ and\ \citenamefont{Samarth}}]{Ku:2003_APL}%
  \BibitemOpen
  \bibfield{author}{%
  \bibinfo {author} {\bibnamefont{Ku}, \bibfnamefont{K.~C.}}, \bibinfo {author}
  {\bibfnamefont{S.~J.}\ \bibnamefont{Potashnik}}, \bibinfo {author}
  {\bibfnamefont{R.~F.}\ \bibnamefont{Wang}}, \bibinfo {author}
  {\bibfnamefont{M.~J.}\ \bibnamefont{Seong}}, \bibinfo {author}
  {\bibfnamefont{E.}~\bibnamefont{Johnston-Halperin}}, \bibinfo {author}
  {\bibfnamefont{R.~C.}\ \bibnamefont{Meyers}}, \bibinfo {author}
  {\bibfnamefont{S.~H.}\ \bibnamefont{Chun}}, \bibinfo {author}
  {\bibfnamefont{A.}~\bibnamefont{Mascarenhas}}, \bibinfo {author}
  {\bibfnamefont{A.~C.}\ \bibnamefont{Gossard}}, \bibinfo {author}
  {\bibfnamefont{D.~D.}\ \bibnamefont{Awschalom}}, \bibinfo {author}
  {\bibfnamefont{P.}~\bibnamefont{Schiffer}},\ and\ \bibinfo {author}
  {\bibfnamefont{N.}~\bibnamefont{Samarth}}}%
  , \bibinfo {year} {2003},\ \bibfield{title}{%
  \enquote{\bibinfo {title} {Highly enhanced {Curie} temperatures in low
  temperature annealed {(Ga,Mn)As} epilayers},}\ }%
  \bibfield{journal}{%
  \bibinfo {journal} {Appl. Phys. Lett.}\ }%
  \textbf{\bibinfo {volume} {82}},\ \bibinfo {pages} {2302}%
  \bibAnnoteFile{NoStop}{Ku:2003_APL}%
\bibitem[{\citenamefont{Kudelski}\ \emph{et~al.}(2007)\citenamefont{Kudelski},
  \citenamefont{Lema\^{i}tre}, \citenamefont{Miard}, \citenamefont{Voisin},
  \citenamefont{Graham}, \citenamefont{Warburton},\ and\
  \citenamefont{Krebs}}]{Kudelski:2007_PRL}%
  \BibitemOpen
  \bibfield{author}{%
  \bibinfo {author} {\bibnamefont{Kudelski}, \bibfnamefont{A.}}, \bibinfo
  {author} {\bibfnamefont{A.}~\bibnamefont{Lema\^{i}tre}}, \bibinfo {author}
  {\bibfnamefont{A.}~\bibnamefont{Miard}}, \bibinfo {author}
  {\bibfnamefont{P.}~\bibnamefont{Voisin}}, \bibinfo {author}
  {\bibfnamefont{T.~C.~M.}\ \bibnamefont{Graham}}, \bibinfo {author}
  {\bibfnamefont{R.~J.}\ \bibnamefont{Warburton}},\ and\ \bibinfo {author}
  {\bibfnamefont{O.}~\bibnamefont{Krebs}}}%
  , \bibinfo {year} {2007},\ \bibfield{title}{%
  \enquote{\bibinfo {title} {Optically probing the fine structure of a single
  {Mn} atom in an {InAs} quantum dot},}\ }%
  \bibfield{journal}{%
  \bibinfo {journal} {Phys. Rev. Lett.}\ }%
  \textbf{\bibinfo {volume} {99}},\ \bibinfo {pages} {247209}%
  \bibAnnoteFile{NoStop}{Kudelski:2007_PRL}%
\bibitem[{\citenamefont{Kudrnovsk\'y}\
  \emph{et~al.}(2004)\citenamefont{Kudrnovsk\'y}, \citenamefont{Turek},
  \citenamefont{Drchal}, \citenamefont{M\'aca}, \citenamefont{Weinberger},\
  and\ \citenamefont{Bruno}}]{Kudrnovsky:2004_PRB}%
  \BibitemOpen
  \bibfield{author}{%
  \bibinfo {author} {\bibnamefont{Kudrnovsk\'y}, \bibfnamefont{J.}}, \bibinfo
  {author} {\bibfnamefont{I.}~\bibnamefont{Turek}}, \bibinfo {author}
  {\bibfnamefont{V.}~\bibnamefont{Drchal}}, \bibinfo {author}
  {\bibfnamefont{F.}~\bibnamefont{M\'aca}}, \bibinfo {author}
  {\bibfnamefont{P.}~\bibnamefont{Weinberger}},\ and\ \bibinfo {author}
  {\bibfnamefont{P.}~\bibnamefont{Bruno}}}%
  , \bibinfo {year} {2004},\ \bibfield{title}{%
  \enquote{\bibinfo {title} {Exchange interactions in {III--V} and group-{IV}
  diluted magnetic semiconductors},}\ }%
  \bibfield{journal}{%
  \bibinfo {journal} {Phys. Rev. B}\ }%
  \textbf{\bibinfo {volume} {69}},\ \bibinfo {pages} {115208}%
  \bibAnnoteFile{NoStop}{Kudrnovsky:2004_PRB}%
\bibitem[{\citenamefont{Kunert}\ \emph{et~al.}(2012)\citenamefont{Kunert},
  \citenamefont{Dobkowska}, \citenamefont{Li}, \citenamefont{Reuther},
  \citenamefont{Kruse}, \citenamefont{Figge}, \citenamefont{Jakie{\l}a},
  \citenamefont{Bonanni}, \citenamefont{Grenzer}, \citenamefont{Stefanowicz},
  \citenamefont{von Borany}, \citenamefont{Sawicki}, \citenamefont{Dietl},\
  and\ \citenamefont{Hommel}}]{Kunert:2012_APL}%
  \BibitemOpen
  \bibfield{author}{%
  \bibinfo {author} {\bibnamefont{Kunert}, \bibfnamefont{G.}}, \bibinfo
  {author} {\bibfnamefont{S.}~\bibnamefont{Dobkowska}}, \bibinfo {author}
  {\bibfnamefont{Tian}\ \bibnamefont{Li}}, \bibinfo {author}
  {\bibfnamefont{H.}~\bibnamefont{Reuther}}, \bibinfo {author}
  {\bibfnamefont{C.}~\bibnamefont{Kruse}}, \bibinfo {author}
  {\bibfnamefont{S.}~\bibnamefont{Figge}}, \bibinfo {author}
  {\bibfnamefont{R.}~\bibnamefont{Jakie{\l}a}}, \bibinfo {author}
  {\bibfnamefont{A.}~\bibnamefont{Bonanni}}, \bibinfo {author}
  {\bibfnamefont{J.}~\bibnamefont{Grenzer}}, \bibinfo {author}
  {\bibfnamefont{W.}~\bibnamefont{Stefanowicz}}, \bibinfo {author}
  {\bibfnamefont{J.}~\bibnamefont{von Borany}}, \bibinfo {author}
  {\bibfnamefont{M.}~\bibnamefont{Sawicki}}, \bibinfo {author}
  {\bibfnamefont{T.}~\bibnamefont{Dietl}},\ and\ \bibinfo {author}
  {\bibfnamefont{D.}~\bibnamefont{Hommel}}}%
  , \bibinfo {year} {2012},\ \bibfield{title}{%
  \enquote{\bibinfo {title} {{Ga$_{1 - x}$Mn$_{x}$N} epitaxial films with high
  magnetization},}\ }%
  \bibfield{journal}{%
  \bibinfo {journal} {Appl. Phys. Lett.}\ }%
  \textbf{\bibinfo {volume} {101}},\ \bibinfo {pages} {022413}%
  \bibAnnoteFile{NoStop}{Kunert:2012_APL}%
\bibitem[{\citenamefont{Kuroda}\ \emph{et~al.}(2007)\citenamefont{Kuroda},
  \citenamefont{Nishizawa}, \citenamefont{Takita}, \citenamefont{Mitome},
  \citenamefont{Bando}, \citenamefont{Osuch},\ and\
  \citenamefont{Dietl}}]{Kuroda:2007_NM}%
  \BibitemOpen
  \bibfield{author}{%
  \bibinfo {author} {\bibnamefont{Kuroda}, \bibfnamefont{S.}}, \bibinfo
  {author} {\bibfnamefont{N.}~\bibnamefont{Nishizawa}}, \bibinfo {author}
  {\bibfnamefont{K.}~\bibnamefont{Takita}}, \bibinfo {author}
  {\bibfnamefont{M.}~\bibnamefont{Mitome}}, \bibinfo {author}
  {\bibfnamefont{Y.}~\bibnamefont{Bando}}, \bibinfo {author}
  {\bibfnamefont{K.}~\bibnamefont{Osuch}},\ and\ \bibinfo {author}
  {\bibfnamefont{T.}~\bibnamefont{Dietl}}}%
  , \bibinfo {year} {2007},\ \bibfield{title}{%
  \enquote{\bibinfo {title} {Origin and control of high temperature
  ferromagnetism in semiconductors},}\ }%
  \bibfield{journal}{%
  \bibinfo {journal} {Nat. Mater.}\ }%
  \textbf{\bibinfo {volume} {6}},\ \bibinfo {pages} {440}%
  \bibAnnoteFile{NoStop}{Kuroda:2007_NM}%
\bibitem[{\citenamefont{Larson}\ \emph{et~al.}(1988)\citenamefont{Larson},
  \citenamefont{Hass}, \citenamefont{Ehrenreich},\ and\
  \citenamefont{Carlsson}}]{Larson:1988_PRB}%
  \BibitemOpen
  \bibfield{author}{%
  \bibinfo {author} {\bibnamefont{Larson}, \bibfnamefont{B.~E.}}, \bibinfo
  {author} {\bibfnamefont{K.~C.}\ \bibnamefont{Hass}}, \bibinfo {author}
  {\bibfnamefont{H.}~\bibnamefont{Ehrenreich}},\ and\ \bibinfo {author}
  {\bibfnamefont{A.~E.}\ \bibnamefont{Carlsson}}}%
  , \bibinfo {year} {1988},\ \bibfield{title}{%
  \enquote{\bibinfo {title} {Theory of exchange interactions and chemical
  trends in diluted magnetic semiconductors},}\ }%
  \bibfield{journal}{%
  \bibinfo {journal} {Phys. Rev. B}\ }%
  \textbf{\bibinfo {volume} {37}},\ \bibinfo {pages} {4137}%
  \bibAnnoteFile{NoStop}{Larson:1988_PRB}%
\bibitem[{\citenamefont{Lechner}\ \emph{et~al.}(2010)\citenamefont{Lechner},
  \citenamefont{Springholz}, \citenamefont{Hassan}, \citenamefont{Groiss},
  \citenamefont{Kirchschlager}, \citenamefont{Stangl}, \citenamefont{Hrauda},\
  and\ \citenamefont{Bauer}}]{Lechner:2010_APL}%
  \BibitemOpen
  \bibfield{author}{%
  \bibinfo {author} {\bibnamefont{Lechner}, \bibfnamefont{R.~T.}}, \bibinfo
  {author} {\bibfnamefont{G.}~\bibnamefont{Springholz}}, \bibinfo {author}
  {\bibfnamefont{M.}~\bibnamefont{Hassan}}, \bibinfo {author}
  {\bibfnamefont{H.}~\bibnamefont{Groiss}}, \bibinfo {author}
  {\bibfnamefont{R.}~\bibnamefont{Kirchschlager}}, \bibinfo {author}
  {\bibfnamefont{J.}~\bibnamefont{Stangl}}, \bibinfo {author}
  {\bibfnamefont{N.}~\bibnamefont{Hrauda}},\ and\ \bibinfo {author}
  {\bibfnamefont{G.}~\bibnamefont{Bauer}}}%
  , \bibinfo {year} {2010},\ \bibfield{title}{%
  \enquote{\bibinfo {title} {Phase separation and exchange biasing in the
  ferromagnetic {IV-VI} semiconductor {Ge$_{1 - x}$Mn$_x$Te}},}\ }%
  \bibfield{journal}{%
  \bibinfo {journal} {Appl. Phys. Lett.}\ }%
  \textbf{\bibinfo {volume} {97}},\ \bibinfo {eid} {023101}%
  \bibAnnoteFile{NoStop}{Lechner:2010_APL}%
\bibitem[{\citenamefont{Lee}\ and\
  \citenamefont{Ramakrishnan}(1985)}]{Lee:1985_RMP}%
  \BibitemOpen
  \bibfield{author}{%
  \bibinfo {author} {\bibnamefont{Lee}, \bibfnamefont{Patrick~A.}},\ and\
  \bibinfo {author} {\bibfnamefont{T.~V.}\ \bibnamefont{Ramakrishnan}}}%
  , \bibinfo {year} {1985},\ \bibfield{title}{%
  \enquote{\bibinfo {title} {Disordered electronic systems},}\ }%
  \bibfield{journal}{%
  \bibinfo {journal} {Rev. Mod. Phys.}\ }%
  \textbf{\bibinfo {volume} {57}},\ \bibinfo {pages} {287}%
  \bibAnnoteFile{NoStop}{Lee:1985_RMP}%
\bibitem[{\citenamefont{Ley}\ \emph{et~al.}(1987)\citenamefont{Ley},
  \citenamefont{Taniguchi}, \citenamefont{Ghijsen}, \citenamefont{Johnson},\
  and\ \citenamefont{Fujimori}}]{Ley:1987_PRB}%
  \BibitemOpen
  \bibfield{author}{%
  \bibinfo {author} {\bibnamefont{Ley}, \bibfnamefont{L.}}, \bibinfo {author}
  {\bibfnamefont{M.}~\bibnamefont{Taniguchi}}, \bibinfo {author}
  {\bibfnamefont{J.}~\bibnamefont{Ghijsen}}, \bibinfo {author}
  {\bibfnamefont{R.~L.}\ \bibnamefont{Johnson}},\ and\ \bibinfo {author}
  {\bibfnamefont{A.}~\bibnamefont{Fujimori}}}%
  , \bibinfo {year} {1987},\ \bibfield{title}{%
  \enquote{\bibinfo {title} {Manganese-derived partial density of states in
  {Cd$_{1-x}$Mn$_x$Te}},}\ }%
  \bibfield{journal}{%
  \bibinfo {journal} {Phys. Rev. B}\ }%
  \textbf{\bibinfo {volume} {35}},\ \bibinfo {pages} {2839}%
  \bibAnnoteFile{NoStop}{Ley:1987_PRB}%
\bibitem[{\citenamefont{Likovich}\ \emph{et~al.}(2009)\citenamefont{Likovich},
  \citenamefont{Russell}, \citenamefont{Yi}, \citenamefont{Narayanamurti},
  \citenamefont{Ku}, \citenamefont{Zhu},\ and\
  \citenamefont{Samarth}}]{Likovich:2009_PRB}%
  \BibitemOpen
  \bibfield{author}{%
  \bibinfo {author} {\bibnamefont{Likovich}, \bibfnamefont{E.}}, \bibinfo
  {author} {\bibfnamefont{K.}~\bibnamefont{Russell}}, \bibinfo {author}
  {\bibfnamefont{Wei}\ \bibnamefont{Yi}}, \bibinfo {author}
  {\bibfnamefont{V.}~\bibnamefont{Narayanamurti}}, \bibinfo {author}
  {\bibfnamefont{Keh-Chiang}\ \bibnamefont{Ku}}, \bibinfo {author}
  {\bibfnamefont{Meng}\ \bibnamefont{Zhu}},\ and\ \bibinfo {author}
  {\bibfnamefont{N.}~\bibnamefont{Samarth}}}%
  , \bibinfo {year} {2009},\ \bibfield{title}{%
  \enquote{\bibinfo {title} {Magnetoresistance in an asymmetric
  {Ga$_{1-x}$Mn$_x$As} resonant tunneling diode},}\ }%
  \bibfield{journal}{%
  \bibinfo {journal} {Phys. Rev. B}\ }%
  \textbf{\bibinfo {volume} {80}},\ \bibinfo {pages} {201307}%
  \bibAnnoteFile{NoStop}{Likovich:2009_PRB}%
\bibitem[{\citenamefont{Lim}\ \emph{et~al.}(2011)\citenamefont{Lim},
  \citenamefont{Bi}, \citenamefont{Hui},\ and\
  \citenamefont{Teo}}]{Lim:2011_JAP}%
  \BibitemOpen
  \bibfield{author}{%
  \bibinfo {author} {\bibnamefont{Lim}, \bibfnamefont{S.~T.}}, \bibinfo
  {author} {\bibfnamefont{J.~F.}\ \bibnamefont{Bi}}, \bibinfo {author}
  {\bibfnamefont{L.}~\bibnamefont{Hui}},\ and\ \bibinfo {author}
  {\bibfnamefont{K.~L.}\ \bibnamefont{Teo}}}%
  , \bibinfo {year} {2011},\ \bibfield{title}{%
  \enquote{\bibinfo {title} {Exchange interaction and {Curie} temperature in
  {Ge$_{1-x}$Mn$_{x}$Te} ferromagnetic semiconductors},}\ }%
  \bibfield{journal}{%
  \bibinfo {journal} {J. Appl. Phys.}\ }%
  \textbf{\bibinfo {volume} {110}},\ \bibinfo {pages} {023905}%
  \bibAnnoteFile{NoStop}{Lim:2011_JAP}%
\bibitem[{\citenamefont{Lim}\ \emph{et~al.}(2012)\citenamefont{Lim},
  \citenamefont{Hui}, \citenamefont{Bi}, \citenamefont{Liew},\ and\
  \citenamefont{Teo}}]{Lim:2012_JAP}%
  \BibitemOpen
  \bibfield{author}{%
  \bibinfo {author} {\bibnamefont{Lim}, \bibfnamefont{S.~T.}}, \bibinfo
  {author} {\bibfnamefont{Lu}~\bibnamefont{Hui}}, \bibinfo {author}
  {\bibfnamefont{J.~F.}\ \bibnamefont{Bi}}, \bibinfo {author}
  {\bibfnamefont{T.}~\bibnamefont{Liew}},\ and\ \bibinfo {author}
  {\bibfnamefont{K.~L.}\ \bibnamefont{Teo}}}%
  , \bibinfo {year} {2012},\ \bibfield{title}{%
  \enquote{\bibinfo {title} {Exchange bias effect of {Ge$_{1 - x}$Mn$_x$Te}
  with antiferromagnetic mnte and mno materials},}\ }%
  \bibfield{journal}{%
  \bibinfo {journal} {J. Appl. Phys.}\ }%
  \textbf{\bibinfo {volume} {111}},\ \bibinfo {eid} {07C308}%
  \bibAnnoteFile{NoStop}{Lim:2012_JAP}%
\bibitem[{\citenamefont{Limmer}\ \emph{et~al.}(2006)\citenamefont{Limmer},
  \citenamefont{Glunk}, \citenamefont{Daeubler}, \citenamefont{Hummel},
  \citenamefont{Schoch}, \citenamefont{Sauer}, \citenamefont{Bihler},
  \citenamefont{Huebl}, \citenamefont{Brandt},\ and\
  \citenamefont{Goennenwein}}]{Limmer:2006_PRB}%
  \BibitemOpen
  \bibfield{author}{%
  \bibinfo {author} {\bibnamefont{Limmer}, \bibfnamefont{W.}}, \bibinfo
  {author} {\bibfnamefont{M.}~\bibnamefont{Glunk}}, \bibinfo {author}
  {\bibfnamefont{J.}~\bibnamefont{Daeubler}}, \bibinfo {author}
  {\bibfnamefont{T.}~\bibnamefont{Hummel}}, \bibinfo {author}
  {\bibfnamefont{W.}~\bibnamefont{Schoch}}, \bibinfo {author}
  {\bibfnamefont{R.}~\bibnamefont{Sauer}}, \bibinfo {author}
  {\bibfnamefont{C.}~\bibnamefont{Bihler}}, \bibinfo {author}
  {\bibfnamefont{H.}~\bibnamefont{Huebl}}, \bibinfo {author}
  {\bibfnamefont{M.~S.}\ \bibnamefont{Brandt}},\ and\ \bibinfo {author}
  {\bibfnamefont{S.~T.~B.}\ \bibnamefont{Goennenwein}}}%
  , \bibinfo {year} {2006},\ \bibfield{title}{%
  \enquote{\bibinfo {title} {Angle-dependent magnetotransport in cubic and
  tetragonal ferromagnets: Application to (001)- and (113){A}-oriented
  {(Ga,Mn)As}},}\ }%
  \bibfield{journal}{%
  \bibinfo {journal} {Phys. Rev. B}\ }%
  \textbf{\bibinfo {volume} {74}},\ \bibinfo {pages} {205205}%
  \bibAnnoteFile{NoStop}{Limmer:2006_PRB}%
\bibitem[{\citenamefont{Linnarsson}\
  \emph{et~al.}(1997)\citenamefont{Linnarsson}, \citenamefont{{Janz{\'e}n}},
  \citenamefont{Monemar}, \citenamefont{Kleverman},\ and\
  \citenamefont{Thilderkvist}}]{Linnarsson:1997_PRB}%
  \BibitemOpen
  \bibfield{author}{%
  \bibinfo {author} {\bibnamefont{Linnarsson}, \bibfnamefont{M.}}, \bibinfo
  {author} {\bibfnamefont{E.}~\bibnamefont{{Janz{\'e}n}}}, \bibinfo {author}
  {\bibfnamefont{B.}~\bibnamefont{Monemar}}, \bibinfo {author}
  {\bibfnamefont{M.}~\bibnamefont{Kleverman}},\ and\ \bibinfo {author}
  {\bibfnamefont{A.}~\bibnamefont{Thilderkvist}}}%
  , \bibinfo {year} {1997},\ \bibfield{title}{%
  \enquote{\bibinfo {title} {Electronic structure of the {GaAs:Mn$_{Ga}$}
  center},}\ }%
  \bibfield{journal}{%
  \bibinfo {journal} {Phys. Rev. B}\ }%
  \textbf{\bibinfo {volume} {55}},\ \bibinfo {pages} {6938}%
  \bibAnnoteFile{NoStop}{Linnarsson:1997_PRB}%
\bibitem[{\citenamefont{Lipi\'{n}ska}\
  \emph{et~al.}(2009)\citenamefont{Lipi\'{n}ska}, \citenamefont{Simserides},
  \citenamefont{Trohidou}, \citenamefont{Goryca}, \citenamefont{Kossacki},
  \citenamefont{Majhofer},\ and\ \citenamefont{Dietl}}]{Lipinska:2009_PRB}%
  \BibitemOpen
  \bibfield{author}{%
  \bibinfo {author} {\bibnamefont{Lipi\'{n}ska}, \bibfnamefont{A.}}, \bibinfo
  {author} {\bibfnamefont{C.}~\bibnamefont{Simserides}}, \bibinfo {author}
  {\bibfnamefont{K.~N.}\ \bibnamefont{Trohidou}}, \bibinfo {author}
  {\bibfnamefont{M.}~\bibnamefont{Goryca}}, \bibinfo {author}
  {\bibfnamefont{P.}~\bibnamefont{Kossacki}}, \bibinfo {author}
  {\bibfnamefont{A.}~\bibnamefont{Majhofer}},\ and\ \bibinfo {author}
  {\bibfnamefont{T.}~\bibnamefont{Dietl}}}%
  , \bibinfo {year} {2009},\ \bibfield{title}{%
  \enquote{\bibinfo {title} {Ferromagnetic properties of p-{(Cd,Mn)Te} quantum
  wells: {Interpretation} of magneto-optical measurements by {Monte} {Carlo}
  simulations},}\ }%
  \bibfield{journal}{%
  \bibinfo {journal} {Phys. Rev. B}\ }%
  \textbf{\bibinfo {volume} {79}},\ \bibinfo {pages} {235322}%
  \bibAnnoteFile{NoStop}{Lipinska:2009_PRB}%
\bibitem[{\citenamefont{Liu}\ \emph{et~al.}(2007)\citenamefont{Liu},
  \citenamefont{Bihlmayer}, \citenamefont{Bl\"ugel},\ and\
  \citenamefont{Chang}}]{Liu:2007_PRB}%
  \BibitemOpen
  \bibfield{author}{%
  \bibinfo {author} {\bibnamefont{Liu}, \bibfnamefont{M.}}, \bibinfo {author}
  {\bibfnamefont{G.}~\bibnamefont{Bihlmayer}}, \bibinfo {author}
  {\bibfnamefont{S.}~\bibnamefont{Bl\"ugel}},\ and\ \bibinfo {author}
  {\bibfnamefont{C.}~\bibnamefont{Chang}}}%
  , \bibinfo {year} {2007},\ \bibfield{title}{%
  \enquote{\bibinfo {title} {Intrinsic spin-{Hall} accumulation in honeycomb
  lattices: Band structure effects},}\ }%
  \bibfield{journal}{%
  \bibinfo {journal} {Phys. Rev. B}\ }%
  \textbf{\bibinfo {volume} {76}},\ \bibinfo {pages} {121301}%
  \bibAnnoteFile{NoStop}{Liu:2007_PRB}%
\bibitem[{\citenamefont{Liu}\ \emph{et~al.}(2009)\citenamefont{Liu},
  \citenamefont{Liu}, \citenamefont{Xu}, \citenamefont{Qi},\ and\
  \citenamefont{Zhang}}]{Liu:2009_PRL}%
  \BibitemOpen
  \bibfield{author}{%
  \bibinfo {author} {\bibnamefont{Liu}, \bibfnamefont{{Qin}}}, \bibinfo
  {author} {\bibfnamefont{{Chao-Xing}}\ \bibnamefont{Liu}}, \bibinfo {author}
  {\bibfnamefont{{Cenke}}\ \bibnamefont{Xu}}, \bibinfo {author}
  {\bibfnamefont{{Xiao-Liang}}\ \bibnamefont{Qi}},\ and\ \bibinfo {author}
  {\bibfnamefont{{Shou-Cheng}}\ \bibnamefont{Zhang}}}%
  , \bibinfo {year} {2009},\ \bibfield{title}{%
  \enquote{\bibinfo {title} {Magnetic impurities on the surface of a
  topological insulator},}\ }%
  \bibfield{journal}{%
  \bibinfo {journal} {Phys. Rev. Lett.}\ }%
  \textbf{\bibinfo {volume} {102}},\ \bibinfo {pages} {156603}%
  \bibAnnoteFile{NoStop}{Liu:2009_PRL}%
\bibitem[{\citenamefont{Liu}\ \emph{et~al.}(2005)\citenamefont{Liu},
  \citenamefont{Lim}, \citenamefont{Ge}, \citenamefont{Shen},
  \citenamefont{Dobrowolska}, \citenamefont{Furdyna}, \citenamefont{Wojtowicz},
  \citenamefont{Yu},\ and\ \citenamefont{Walukiewicz}}]{Liu:2005_APL}%
  \BibitemOpen
  \bibfield{author}{%
  \bibinfo {author} {\bibnamefont{Liu}, \bibfnamefont{X.}}, \bibinfo {author}
  {\bibfnamefont{W.~L.}\ \bibnamefont{Lim}}, \bibinfo {author}
  {\bibfnamefont{Z.}~\bibnamefont{Ge}}, \bibinfo {author}
  {\bibfnamefont{S.}~\bibnamefont{Shen}}, \bibinfo {author}
  {\bibfnamefont{M.}~\bibnamefont{Dobrowolska}}, \bibinfo {author}
  {\bibfnamefont{J.~K.}\ \bibnamefont{Furdyna}}, \bibinfo {author}
  {\bibfnamefont{T.}~\bibnamefont{Wojtowicz}}, \bibinfo {author}
  {\bibfnamefont{K.~M.}\ \bibnamefont{Yu}},\ and\ \bibinfo {author}
  {\bibfnamefont{W.}~\bibnamefont{Walukiewicz}}}%
  , \bibinfo {year} {2005},\ \bibfield{title}{%
  \enquote{\bibinfo {title} {Strain-engineered ferromagnetic {In$_{1 -
  x}$Mn$_{x}$As} films with in-plane easy axis},}\ }%
  \bibfield{journal}{%
  \bibinfo {journal} {Appl. Phys. Lett.}\ }%
  \textbf{\bibinfo {volume} {86}},\ \bibinfo {pages} {112512}%
  \bibAnnoteFile{NoStop}{Liu:2005_APL}%
\bibitem[{\citenamefont{Liu}\ \emph{et~al.}(2004)\citenamefont{Liu},
  \citenamefont{Lim}, \citenamefont{Titova}, \citenamefont{Wojtowicz},
  \citenamefont{Kutrowski}, \citenamefont{Yee}, \citenamefont{Dobrowolska},
  \citenamefont{Furdyna}, \citenamefont{Potashnik}, \citenamefont{Stone},
  \citenamefont{Schiffer}, \citenamefont{Vurgaftman},\ and\
  \citenamefont{Meyer}}]{Liu:2004_PE}%
  \BibitemOpen
  \bibfield{author}{%
  \bibinfo {author} {\bibnamefont{Liu}, \bibfnamefont{X.}}, \bibinfo {author}
  {\bibfnamefont{W.~L.}\ \bibnamefont{Lim}}, \bibinfo {author}
  {\bibfnamefont{L.~V.}\ \bibnamefont{Titova}}, \bibinfo {author}
  {\bibfnamefont{T.}~\bibnamefont{Wojtowicz}}, \bibinfo {author}
  {\bibfnamefont{M.}~\bibnamefont{Kutrowski}}, \bibinfo {author}
  {\bibfnamefont{K.~J.}\ \bibnamefont{Yee}}, \bibinfo {author}
  {\bibfnamefont{M.}~\bibnamefont{Dobrowolska}}, \bibinfo {author}
  {\bibfnamefont{J.~K.}\ \bibnamefont{Furdyna}}, \bibinfo {author}
  {\bibfnamefont{S.~J.}\ \bibnamefont{Potashnik}}, \bibinfo {author}
  {\bibfnamefont{M.~B.}\ \bibnamefont{Stone}}, \bibinfo {author}
  {\bibfnamefont{P.}~\bibnamefont{Schiffer}}, \bibinfo {author}
  {\bibfnamefont{I.}~\bibnamefont{Vurgaftman}},\ and\ \bibinfo {author}
  {\bibfnamefont{J.~R.}\ \bibnamefont{Meyer}}}%
  , \bibinfo {year} {2004},\ \bibfield{title}{%
  \enquote{\bibinfo {title} {External control of the direction of magnetization
  in ferromagnetic {InMnAs/GaSb} heterostructures},}\ }%
  \bibfield{journal}{%
  \bibinfo {journal} {Physica E}\ }%
  \textbf{\bibinfo {volume} {E 20}},\ \bibinfo {pages} {370}%
  \bibAnnoteFile{NoStop}{Liu:2004_PE}%
\bibitem[{\citenamefont{Liu}\ and\
  \citenamefont{Furdyna}(2006)}]{Liu:2006_JPCM}%
  \BibitemOpen
  \bibfield{author}{%
  \bibinfo {author} {\bibnamefont{Liu}, \bibfnamefont{Xinyu}},\ and\ \bibinfo
  {author} {\bibfnamefont{Jacek~K.}\ \bibnamefont{Furdyna}}}%
  , \bibinfo {year} {2006},\ \bibfield{title}{%
  \enquote{\bibinfo {title} {Ferromagnetic resonance in {Ga$_{1-x}$Mn$_{x}$As}
  dilute magnetic semiconductors},}\ }%
  \bibfield{journal}{%
  \bibinfo {journal} {J. Phys. Condens. Matter}\ }%
  \textbf{\bibinfo {volume} {18}},\ \bibinfo {pages} {R245}%
  \bibAnnoteFile{NoStop}{Liu:2006_JPCM}%
\bibitem[{\citenamefont{{\L}ukasiewicz}\
  \emph{et~al.}(2012)\citenamefont{{\L}ukasiewicz},
  \citenamefont{W{\'o}jcik-G{\l}odowska}, \citenamefont{Guziewicz},
  \citenamefont{Wolska}, \citenamefont{Klepka},
  \citenamefont{D{\l}u{\.z}ewski}, \citenamefont{Jakie{\l}a},
  \citenamefont{{\L}usakowska}, \citenamefont{Kopalko},
  \citenamefont{Paszkowicz}, \citenamefont{Wachnicki},
  \citenamefont{Witkowski}, \citenamefont{Lisowski}, \citenamefont{Krawczyk},
  \citenamefont{Sobczak}, \citenamefont{Jab{\l}o\'nski},\ and\
  \citenamefont{Godlewski}}]{Lukasiewicz:2012_SST}%
  \BibitemOpen
  \bibfield{author}{%
  \bibinfo {author} {\bibnamefont{{\L}ukasiewicz}, \bibfnamefont{M.~I.}},
  \bibinfo {author} {\bibfnamefont{A.}~\bibnamefont{W{\'o}jcik-G{\l}odowska}},
  \bibinfo {author} {\bibfnamefont{E.}~\bibnamefont{Guziewicz}}, \bibinfo
  {author} {\bibfnamefont{A.}~\bibnamefont{Wolska}}, \bibinfo {author}
  {\bibfnamefont{M.~T.}\ \bibnamefont{Klepka}}, \bibinfo {author}
  {\bibfnamefont{P.}~\bibnamefont{D{\l}u{\.z}ewski}}, \bibinfo {author}
  {\bibfnamefont{R.}~\bibnamefont{Jakie{\l}a}}, \bibinfo {author}
  {\bibfnamefont{E.}~\bibnamefont{{\L}usakowska}}, \bibinfo {author}
  {\bibfnamefont{K.}~\bibnamefont{Kopalko}}, \bibinfo {author}
  {\bibfnamefont{W.}~\bibnamefont{Paszkowicz}}, \bibinfo {author}
  {\bibfnamefont{{\L}.}~\bibnamefont{Wachnicki}}, \bibinfo {author}
  {\bibfnamefont{B.~S.}\ \bibnamefont{Witkowski}}, \bibinfo {author}
  {\bibfnamefont{W.}~\bibnamefont{Lisowski}}, \bibinfo {author}
  {\bibfnamefont{M.}~\bibnamefont{Krawczyk}}, \bibinfo {author}
  {\bibfnamefont{J.~W.}\ \bibnamefont{Sobczak}}, \bibinfo {author}
  {\bibfnamefont{A.}~\bibnamefont{Jab{\l}o\'nski}},\ and\ \bibinfo {author}
  {\bibfnamefont{M.}~\bibnamefont{Godlewski}}}%
  , \bibinfo {year} {2012},\ \bibfield{title}{%
  \enquote{\bibinfo {title} {{ZnO, ZnMnO and ZnCoO} films grown by atomic layer
  deposition},}\ }%
  \bibfield{journal}{%
  \bibinfo {journal} {Semicon. Sci. Technol.}\ }%
  \textbf{\bibinfo {volume} {27}},\ \bibinfo {pages} {074009}%
  \bibAnnoteFile{NoStop}{Lukasiewicz:2012_SST}%
\bibitem[{\citenamefont{Luo}\ \emph{et~al.}(2010)\citenamefont{Luo},
  \citenamefont{Zheng}, \citenamefont{Shen}, \citenamefont{Zhang},
  \citenamefont{Zhu}, \citenamefont{Zhu}, \citenamefont{Liu},
  \citenamefont{Li}, \citenamefont{Ji},\ and\
  \citenamefont{Zhao}}]{Luo:2010_SCh}%
  \BibitemOpen
  \bibfield{author}{%
  \bibinfo {author} {\bibnamefont{Luo}, \bibfnamefont{Jing}}, \bibinfo {author}
  {\bibfnamefont{Hou-Zhi}\ \bibnamefont{Zheng}}, \bibinfo {author}
  {\bibfnamefont{Chao}\ \bibnamefont{Shen}}, \bibinfo {author}
  {\bibfnamefont{Hao}\ \bibnamefont{Zhang}}, \bibinfo {author}
  {\bibfnamefont{Ke}~\bibnamefont{Zhu}}, \bibinfo {author} {\bibfnamefont{Hui}\
  \bibnamefont{Zhu}}, \bibinfo {author} {\bibfnamefont{Jian}\
  \bibnamefont{Liu}}, \bibinfo {author} {\bibfnamefont{Gui-Rong}\
  \bibnamefont{Li}}, \bibinfo {author} {\bibfnamefont{Yang}\
  \bibnamefont{Ji}},\ and\ \bibinfo {author} {\bibfnamefont{Jian-Hua}\
  \bibnamefont{Zhao}}}%
  , \bibinfo {year} {2010},\ \bibfield{title}{%
  \enquote{\bibinfo {title} {Ultrafast photo-induced turning of magnetization
  and its relaxation dynamics in {GaMnAs}},}\ }%
  \bibfield{journal}{%
  \bibinfo {journal} {Sci. China Phys. Mechanis, Astron.}\ }%
  \textbf{\bibinfo {volume} {53}},\ \bibinfo {pages} {779}%
  \bibAnnoteFile{NoStop}{Luo:2010_SCh}%
\bibitem[{\citenamefont{Luttinger}\ and\
  \citenamefont{Kohn}(1955)}]{Kohn:1955_PR}%
  \BibitemOpen
  \bibfield{author}{%
  \bibinfo {author} {\bibnamefont{Luttinger}, \bibfnamefont{J.~M}},\ and\
  \bibinfo {author} {\bibfnamefont{W.}~\bibnamefont{Kohn}}}%
  , \bibinfo {year} {1955},\ \bibfield{title}{%
  \enquote{\bibinfo {title} {Motion of electrons and holes in perturbed
  periodic fields},}\ }%
  \bibfield{journal}{%
  \bibinfo {journal} {Phys. Rev.}\ }%
  \textbf{\bibinfo {volume} {97}},\ \bibinfo {pages} {869}%
  \bibAnnoteFile{NoStop}{Kohn:1955_PR}%
\bibitem[{\citenamefont{Maccherozzi}\
  \emph{et~al.}(2008)\citenamefont{Maccherozzi}, \citenamefont{Sperl},
  \citenamefont{Panaccione}, \citenamefont{Minar}, \citenamefont{Polesya},
  \citenamefont{Ebert}, \citenamefont{W{\"u}rstbauer},
  \citenamefont{Hochstrasser}, \citenamefont{Rossi},
  \citenamefont{Woltersdorf}, \citenamefont{Wegscheider},\ and\
  \citenamefont{Back}}]{Maccherozzi:2008_PRL}%
  \BibitemOpen
  \bibfield{author}{%
  \bibinfo {author} {\bibnamefont{Maccherozzi}, \bibfnamefont{F.}}, \bibinfo
  {author} {\bibfnamefont{M.}~\bibnamefont{Sperl}}, \bibinfo {author}
  {\bibfnamefont{G.}~\bibnamefont{Panaccione}}, \bibinfo {author}
  {\bibfnamefont{J.}~\bibnamefont{Minar}}, \bibinfo {author}
  {\bibfnamefont{S.}~\bibnamefont{Polesya}}, \bibinfo {author}
  {\bibfnamefont{H.}~\bibnamefont{Ebert}}, \bibinfo {author}
  {\bibfnamefont{U.}~\bibnamefont{W{\"u}rstbauer}}, \bibinfo {author}
  {\bibfnamefont{M.}~\bibnamefont{Hochstrasser}}, \bibinfo {author}
  {\bibfnamefont{G.}~\bibnamefont{Rossi}}, \bibinfo {author}
  {\bibfnamefont{G.}~\bibnamefont{Woltersdorf}}, \bibinfo {author}
  {\bibfnamefont{W.}~\bibnamefont{Wegscheider}},\ and\ \bibinfo {author}
  {\bibfnamefont{C.~H.}\ \bibnamefont{Back}}}%
  , \bibinfo {year} {2008},\ \bibfield{title}{%
  \enquote{\bibinfo {title} {Evidence for a magnetic proximity effect up to
  room temperature at {Fe/(Ga,Mn)As} interfaces},}\ }%
  \bibfield{journal}{%
  \bibinfo {journal} {Phys. Rev. Lett.}\ }%
  \textbf{\bibinfo {volume} {101}},\ \bibinfo {pages} {267201}%
  \bibAnnoteFile{NoStop}{Maccherozzi:2008_PRL}%
\bibitem[{\citenamefont{Mack}\ \emph{et~al.}(2008)\citenamefont{Mack},
  \citenamefont{Myers}, \citenamefont{Heron}, \citenamefont{Gossard},\ and\
  \citenamefont{Awschalom}}]{Mack:2008_APL}%
  \BibitemOpen
  \bibfield{author}{%
  \bibinfo {author} {\bibnamefont{Mack}, \bibfnamefont{S.}}, \bibinfo {author}
  {\bibfnamefont{R.~C.}\ \bibnamefont{Myers}}, \bibinfo {author}
  {\bibfnamefont{J.~T.}\ \bibnamefont{Heron}}, \bibinfo {author}
  {\bibfnamefont{A.~C.}\ \bibnamefont{Gossard}},\ and\ \bibinfo {author}
  {\bibfnamefont{D.~D.}\ \bibnamefont{Awschalom}}}%
  , \bibinfo {year} {2008},\ \bibfield{title}{%
  \enquote{\bibinfo {title} {Stoichiometric growth of high {Curie} temperature
  heavily alloyed {GaMnAs}},}\ }%
  \bibfield{journal}{%
  \bibinfo {journal} {Appl. Phys. Lett.}\ }%
  \textbf{\bibinfo {volume} {92}},\ \bibinfo {pages} {192502}%
  \bibAnnoteFile{NoStop}{Mack:2008_APL}%
\bibitem[{\citenamefont{Mahadevan}\ and\
  \citenamefont{Zunger}(2004)}]{Mahadevan:2004_APL}%
  \BibitemOpen
  \bibfield{author}{%
  \bibinfo {author} {\bibnamefont{Mahadevan}, \bibfnamefont{P.}},\ and\
  \bibinfo {author} {\bibfnamefont{A.}~\bibnamefont{Zunger}}}%
  , \bibinfo {year} {2004},\ \bibfield{title}{%
  \enquote{\bibinfo {title} {Trends in ferromagnetism, hole localization, and
  acceptor level depth for {Mn} substitution in {GaN}, {GaP}, {GaAs},
  {GaSb}},}\ }%
  \bibfield{journal}{%
  \bibinfo {journal} {Appl. Phys. Lett.}\ }%
  \textbf{\bibinfo {volume} {85}},\ \bibinfo {pages} {2860}%
  \bibAnnoteFile{NoStop}{Mahadevan:2004_APL}%
\bibitem[{\citenamefont{Mankovsky}\
  \emph{et~al.}(2011)\citenamefont{Mankovsky}, \citenamefont{Polesya},
  \citenamefont{Bornemann}, \citenamefont{Min{\'{a}}r},
  \citenamefont{Hoffmann}, \citenamefont{Back},\ and\
  \citenamefont{Ebert}}]{Mankovsky:2011_PRB}%
  \BibitemOpen
  \bibfield{author}{%
  \bibinfo {author} {\bibnamefont{Mankovsky}, \bibfnamefont{S.}}, \bibinfo
  {author} {\bibfnamefont{S.}~\bibnamefont{Polesya}}, \bibinfo {author}
  {\bibfnamefont{S.}~\bibnamefont{Bornemann}}, \bibinfo {author}
  {\bibfnamefont{J.}~\bibnamefont{Min{\'{a}}r}}, \bibinfo {author}
  {\bibfnamefont{F.}~\bibnamefont{Hoffmann}}, \bibinfo {author}
  {\bibfnamefont{C.~H.}\ \bibnamefont{Back}},\ and\ \bibinfo {author}
  {\bibfnamefont{H.}~\bibnamefont{Ebert}}}%
  , \bibinfo {year} {2011},\ \bibfield{title}{%
  \enquote{\bibinfo {title} {Spin-orbit coupling effect in {(Ga,Mn)As} films:
  {Anisotropic} exchange interactions and magnetocrystalline anisotropy},}\ }%
  \bibfield{journal}{%
  \bibinfo {journal} {Phys. Rev. B}\ }%
  \textbf{\bibinfo {volume} {84}},\ \bibinfo {pages} {201201}%
  \bibAnnoteFile{NoStop}{Mankovsky:2011_PRB}%
\bibitem[{\citenamefont{Mark}\ \emph{et~al.}(2011)\citenamefont{Mark},
  \citenamefont{D\"urrenfeld}, \citenamefont{Pappert}, \citenamefont{Ebel},
  \citenamefont{Brunner}, \citenamefont{Gould},\ and\
  \citenamefont{Molenkamp}}]{Mark:2011_PRL}%
  \BibitemOpen
  \bibfield{author}{%
  \bibinfo {author} {\bibnamefont{Mark}, \bibfnamefont{S.}}, \bibinfo {author}
  {\bibfnamefont{P.}~\bibnamefont{D\"urrenfeld}}, \bibinfo {author}
  {\bibfnamefont{K.}~\bibnamefont{Pappert}}, \bibinfo {author}
  {\bibfnamefont{L.}~\bibnamefont{Ebel}}, \bibinfo {author}
  {\bibfnamefont{K.}~\bibnamefont{Brunner}}, \bibinfo {author}
  {\bibfnamefont{C.}~\bibnamefont{Gould}},\ and\ \bibinfo {author}
  {\bibfnamefont{L.~W.}\ \bibnamefont{Molenkamp}}}%
  , \bibinfo {year} {2011},\ \bibfield{title}{%
  \enquote{\bibinfo {title} {Fully electrical read-write device out of a
  ferromagnetic semiconductor},}\ }%
  \bibfield{journal}{%
  \bibinfo {journal} {Phys. Rev. Lett.}\ }%
  \textbf{\bibinfo {volume} {106}},\ \bibinfo {pages} {057204}%
  \bibAnnoteFile{NoStop}{Mark:2011_PRL}%
\bibitem[{\citenamefont{Masmanidis}\
  \emph{et~al.}(2005)\citenamefont{Masmanidis}, \citenamefont{Tang},
  \citenamefont{Myers}, \citenamefont{Li}, \citenamefont{De~Greve},
  \citenamefont{Vermeulen}, \citenamefont{{Van Roy}},\ and\
  \citenamefont{Roukes}}]{Masmanidis:2005_PRL}%
  \BibitemOpen
  \bibfield{author}{%
  \bibinfo {author} {\bibnamefont{Masmanidis}, \bibfnamefont{S.~C.}}, \bibinfo
  {author} {\bibfnamefont{H.~X.}\ \bibnamefont{Tang}}, \bibinfo {author}
  {\bibfnamefont{E.~B.}\ \bibnamefont{Myers}}, \bibinfo {author}
  {\bibfnamefont{M.}~\bibnamefont{Li}}, \bibinfo {author}
  {\bibfnamefont{K.}~\bibnamefont{De~Greve}}, \bibinfo {author}
  {\bibfnamefont{G.}~\bibnamefont{Vermeulen}}, \bibinfo {author}
  {\bibfnamefont{W.}~\bibnamefont{{Van Roy}}},\ and\ \bibinfo {author}
  {\bibfnamefont{M.~L.}\ \bibnamefont{Roukes}}}%
  , \bibinfo {year} {2005},\ \bibfield{title}{%
  \enquote{\bibinfo {title} {Nanomechanical measurement of magnetostriction and
  magnetic anisotropy in {(Ga,Mn)As}},}\ }%
  \bibfield{journal}{%
  \bibinfo {journal} {Phys. Rev. Lett.}\ }%
  \textbf{\bibinfo {volume} {95}},\ \bibinfo {pages} {187206}%
  \bibAnnoteFile{NoStop}{Masmanidis:2005_PRL}%
\bibitem[{\citenamefont{Matsuda}\ \emph{et~al.}(2011)\citenamefont{Matsuda},
  \citenamefont{Khodaparast}, \citenamefont{Shen}, \citenamefont{Takeyama},
  \citenamefont{Liu}, \citenamefont{Furdyna},\ and\
  \citenamefont{Wessels}}]{Matsuda:2011_JPCS}%
  \BibitemOpen
  \bibfield{author}{%
  \bibinfo {author} {\bibnamefont{Matsuda}, \bibfnamefont{Y.~H.}}, \bibinfo
  {author} {\bibfnamefont{G.~A.}\ \bibnamefont{Khodaparast}}, \bibinfo {author}
  {\bibfnamefont{R.}~\bibnamefont{Shen}}, \bibinfo {author}
  {\bibfnamefont{S.}~\bibnamefont{Takeyama}}, \bibinfo {author}
  {\bibfnamefont{X.}~\bibnamefont{Liu}}, \bibinfo {author}
  {\bibfnamefont{J.}~\bibnamefont{Furdyna}},\ and\ \bibinfo {author}
  {\bibfnamefont{B.~W.}\ \bibnamefont{Wessels}}}%
  , \bibinfo {year} {2011},\ \bibfield{title}{%
  \enquote{\bibinfo {title} {Cyclotron resonance in {InMnAs} and {InMnSb}
  ferromagnetic films},}\ }%
  \bibfield{journal}{%
  \bibinfo {journal} {J. Phys. Conf. Ser.}\ }%
  \textbf{\bibinfo {volume} {334}},\ \bibinfo {pages} {012056}%
  \bibAnnoteFile{NoStop}{Matsuda:2011_JPCS}%
\bibitem[{\citenamefont{Matsukura}\
  \emph{et~al.}(2002)\citenamefont{Matsukura}, \citenamefont{Ohno},\ and\
  \citenamefont{Dietl}}]{Matsukura:2002_B}%
  \BibitemOpen
  \bibfield{author}{%
  \bibinfo {author} {\bibnamefont{Matsukura}, \bibfnamefont{F.}}, \bibinfo
  {author} {\bibfnamefont{H.}~\bibnamefont{Ohno}},\ and\ \bibinfo {author}
  {\bibfnamefont{T.}~\bibnamefont{Dietl}}}%
  , \bibinfo {year} {2002},\ \enquote{\bibinfo {title} {{III-V} ferromagnetic
  semiconductors},}\ in\ \emph{\bibinfo {booktitle} {{Handbook of Magnetic
  Materials}}},\ Vol.~\bibinfo {volume} {14},\ \bibinfo {editor} {edited by\
  \bibinfo {editor} {\bibfnamefont{K.~H.~J.}\ \bibnamefont{Buschow}}}\
  (\bibinfo {publisher} {Elsevier})\ p.~\bibinfo {pages} {1}%
  \bibAnnoteFile{NoStop}{Matsukura:2002_B}%
\bibitem[{\citenamefont{Matsukura}\
  \emph{et~al.}(1998)\citenamefont{Matsukura}, \citenamefont{Ohno},
  \citenamefont{Shen},\ and\ \citenamefont{Sugawara}}]{Matsukura:1998_PRB}%
  \BibitemOpen
  \bibfield{author}{%
  \bibinfo {author} {\bibnamefont{Matsukura}, \bibfnamefont{F.}}, \bibinfo
  {author} {\bibfnamefont{H.}~\bibnamefont{Ohno}}, \bibinfo {author}
  {\bibfnamefont{A.}~\bibnamefont{Shen}},\ and\ \bibinfo {author}
  {\bibfnamefont{Y.}~\bibnamefont{Sugawara}}}%
  , \bibinfo {year} {1998},\ \bibfield{title}{%
  \enquote{\bibinfo {title} {Transport properties and origin of ferromagnetism
  in {(Ga,Mn)As}},}\ }%
  \bibfield{journal}{%
  \bibinfo {journal} {Phys. Rev. B}\ }%
  \textbf{\bibinfo {volume} {57}},\ \bibinfo {pages} {R2037}%
  \bibAnnoteFile{NoStop}{Matsukura:1998_PRB}%
\bibitem[{\citenamefont{Matsukura}\
  \emph{et~al.}(2004)\citenamefont{Matsukura}, \citenamefont{Sawicki},
  \citenamefont{Dietl}, \citenamefont{Chiba},\ and\
  \citenamefont{Ohno}}]{Matsukura:2004_PE}%
  \BibitemOpen
  \bibfield{author}{%
  \bibinfo {author} {\bibnamefont{Matsukura}, \bibfnamefont{F.}}, \bibinfo
  {author} {\bibfnamefont{M.}~\bibnamefont{Sawicki}}, \bibinfo {author}
  {\bibfnamefont{T.}~\bibnamefont{Dietl}}, \bibinfo {author}
  {\bibfnamefont{D.}~\bibnamefont{Chiba}},\ and\ \bibinfo {author}
  {\bibfnamefont{H.}~\bibnamefont{Ohno}}}%
  , \bibinfo {year} {2004},\ \bibfield{title}{%
  \enquote{\bibinfo {title} {Magnetotransport properties of metallic
  {(Ga,Mn)As} films with compressive and tensile strain},}\ }%
  \bibfield{journal}{%
  \bibinfo {journal} {Physica E}\ }%
  \textbf{\bibinfo {volume} {E 21}},\ \bibinfo {pages} {1032}%
  \bibAnnoteFile{NoStop}{Matsukura:2004_PE}%
\bibitem[{\citenamefont{Mattana}\ \emph{et~al.}(2003)\citenamefont{Mattana},
  \citenamefont{George}, \citenamefont{{Jaffr\`{e}s}}, \citenamefont{Dau},
  \citenamefont{Fert}, \citenamefont{{L{\'e}pine}}, \citenamefont{Guivarc'h},\
  and\ \citenamefont{{J{\'e}z{\'e}quel}}}]{Mattana:2003_PRL}%
  \BibitemOpen
  \bibfield{author}{%
  \bibinfo {author} {\bibnamefont{Mattana}, \bibfnamefont{R.}}, \bibinfo
  {author} {\bibfnamefont{J.~M.}\ \bibnamefont{George}}, \bibinfo {author}
  {\bibfnamefont{H.}~\bibnamefont{{Jaffr\`{e}s}}}, \bibinfo {author}
  {\bibfnamefont{F.~Nguyen~Van}\ \bibnamefont{Dau}}, \bibinfo {author}
  {\bibfnamefont{A.}~\bibnamefont{Fert}}, \bibinfo {author}
  {\bibfnamefont{B.}~\bibnamefont{{L{\'e}pine}}}, \bibinfo {author}
  {\bibfnamefont{A.}~\bibnamefont{Guivarc'h}},\ and\ \bibinfo {author}
  {\bibfnamefont{G.}~\bibnamefont{{J{\'e}z{\'e}quel}}}}%
  , \bibinfo {year} {2003},\ \bibfield{title}{%
  \enquote{\bibinfo {title} {Electrical detection of spin accumulation in a
  p-type {GaAs} quantum well},}\ }%
  \bibfield{journal}{%
  \bibinfo {journal} {Phys. Rev. Lett.}\ }%
  \textbf{\bibinfo {volume} {90}},\ \bibinfo {pages} {166601}%
  \bibAnnoteFile{NoStop}{Mattana:2003_PRL}%
\bibitem[{\citenamefont{Matthias}\ \emph{et~al.}(2002)\citenamefont{Matthias},
  \citenamefont{Gonzalo},\ and\ \citenamefont{Elbio}}]{Mayr:2002_PRB}%
  \BibitemOpen
  \bibfield{author}{%
  \bibinfo {author} {\bibnamefont{Matthias}, \bibfnamefont{M.}}, \bibinfo
  {author} {\bibfnamefont{A.}~\bibnamefont{Gonzalo}},\ and\ \bibinfo {author}
  {\bibfnamefont{D.}~\bibnamefont{Elbio}}}%
  , \bibinfo {year} {2002},\ \bibfield{title}{%
  \enquote{\bibinfo {title} {Global versus local ferromagnetism in a model for
  diluted magnetic semiconductors studied with {Monte Carlo} techniques},}\ }%
  \bibfield{journal}{%
  \bibinfo {journal} {Phys. Rev. B}\ }%
  \textbf{\bibinfo {volume} {65}},\ \bibinfo {pages} {241202}%
  \bibAnnoteFile{NoStop}{Mayr:2002_PRB}%
\bibitem[{\citenamefont{{Ma\v{s}ek}}\
  \emph{et~al.}(2003)\citenamefont{{Ma\v{s}ek}},
  \citenamefont{{Kudrnovsk{\'y}}},\ and\
  \citenamefont{{M{\'a}ca}}}]{Masek:2003_PRB}%
  \BibitemOpen
  \bibfield{author}{%
  \bibinfo {author} {\bibnamefont{{Ma\v{s}ek}}, \bibfnamefont{J.}}, \bibinfo
  {author} {\bibfnamefont{J.}~\bibnamefont{{Kudrnovsk{\'y}}}},\ and\ \bibinfo
  {author} {\bibfnamefont{F.}~\bibnamefont{{M{\'a}ca}}}}%
  , \bibinfo {year} {2003},\ \bibfield{title}{%
  \enquote{\bibinfo {title} {Lattice constant in diluted magnetic
  semiconductors {(Ga,Mn)As}},}\ }%
  \bibfield{journal}{%
  \bibinfo {journal} {Phys. Rev. B}\ }%
  \textbf{\bibinfo {volume} {67}},\ \bibinfo {pages} {153203}%
  \bibAnnoteFile{NoStop}{Masek:2003_PRB}%
\bibitem[{\citenamefont{{Ma\v{s}ek}}\ and\
  \citenamefont{{M{\'a}ca}}(2001)}]{Masek:2001_APP}%
  \BibitemOpen
  \bibfield{author}{%
  \bibinfo {author} {\bibnamefont{{Ma\v{s}ek}}, \bibfnamefont{J.}},\ and\
  \bibinfo {author} {\bibfnamefont{F.}~\bibnamefont{{M{\'a}ca}}}}%
  , \bibinfo {year} {2001},\ \bibfield{title}{%
  \enquote{\bibinfo {title} {Self-compensating incorporation of {Mn} in
  {Ga$_{1-x}$Mn$_{x}$As}},}\ }%
  \bibfield{journal}{%
  \bibinfo {journal} {Acta Phys. Pol. A}\ }%
  \textbf{\bibinfo {volume} {100}},\ \bibinfo {pages} {319}%
  \bibAnnoteFile{NoStop}{Masek:2001_APP}%
\bibitem[{\citenamefont{Mayer}\ \emph{et~al.}(2010)\citenamefont{Mayer},
  \citenamefont{Stone}, \citenamefont{Miller}, \citenamefont{Smith},
  \citenamefont{Dubon}, \citenamefont{Haller}, \citenamefont{Yu},
  \citenamefont{Walukiewicz}, \citenamefont{Liu},\ and\
  \citenamefont{Furdyna}}]{Mayer:2010_PRB}%
  \BibitemOpen
  \bibfield{author}{%
  \bibinfo {author} {\bibnamefont{Mayer}, \bibfnamefont{M.~A.}}, \bibinfo
  {author} {\bibfnamefont{P.~R.}\ \bibnamefont{Stone}}, \bibinfo {author}
  {\bibfnamefont{N.}~\bibnamefont{Miller}}, \bibinfo {author}
  {\bibfnamefont{H.~M.}\ \bibnamefont{Smith}}, \bibinfo {author}
  {\bibfnamefont{O.~D.}\ \bibnamefont{Dubon}}, \bibinfo {author}
  {\bibfnamefont{E.~E.}\ \bibnamefont{Haller}}, \bibinfo {author}
  {\bibfnamefont{K.~M.}\ \bibnamefont{Yu}}, \bibinfo {author}
  {\bibfnamefont{W.}~\bibnamefont{Walukiewicz}}, \bibinfo {author}
  {\bibfnamefont{X.}~\bibnamefont{Liu}},\ and\ \bibinfo {author}
  {\bibfnamefont{J.~K.}\ \bibnamefont{Furdyna}}}%
  , \bibinfo {year} {2010},\ \bibfield{title}{%
  \enquote{\bibinfo {title} {Electronic structure of {Ga$_{1-x}$Mn$_{x}$As}
  analyzed according to hole-concentration-dependent measurements},}\ }%
  \bibfield{journal}{%
  \bibinfo {journal} {Phys. Rev. B}\ }%
  \textbf{\bibinfo {volume} {81}},\ \bibinfo {pages} {045205}%
  \bibAnnoteFile{NoStop}{Mayer:2010_PRB}%
\bibitem[{\citenamefont{Mih\'aly}\ \emph{et~al.}(2008)\citenamefont{Mih\'aly},
  \citenamefont{Csontos}, \citenamefont{Bord\'acs},
  \citenamefont{K\'ezsm\'arki}, \citenamefont{Wojtowicz}, \citenamefont{Liu},
  \citenamefont{Jank\'o},\ and\ \citenamefont{Furdyna}}]{Mihaly:2008_PRL}%
  \BibitemOpen
  \bibfield{author}{%
  \bibinfo {author} {\bibnamefont{Mih\'aly}, \bibfnamefont{G.}}, \bibinfo
  {author} {\bibfnamefont{M.}~\bibnamefont{Csontos}}, \bibinfo {author}
  {\bibfnamefont{S.}~\bibnamefont{Bord\'acs}}, \bibinfo {author}
  {\bibfnamefont{I.}~\bibnamefont{K\'ezsm\'arki}}, \bibinfo {author}
  {\bibfnamefont{T.}~\bibnamefont{Wojtowicz}}, \bibinfo {author}
  {\bibfnamefont{X.}~\bibnamefont{Liu}}, \bibinfo {author}
  {\bibfnamefont{B.}~\bibnamefont{Jank\'o}},\ and\ \bibinfo {author}
  {\bibfnamefont{J.~K.}\ \bibnamefont{Furdyna}}}%
  , \bibinfo {year} {2008},\ \bibfield{title}{%
  \enquote{\bibinfo {title} {Anomalous {Hall} effect in the {(In,Mn)Sb} dilute
  magnetic semiconductor},}\ }%
  \bibfield{journal}{%
  \bibinfo {journal} {Phys. Rev. Lett.}\ }%
  \textbf{\bibinfo {volume} {100}},\ \bibinfo {pages} {107201}%
  \bibAnnoteFile{NoStop}{Mihaly:2008_PRL}%
\bibitem[{\citenamefont{Mitra}\ \emph{et~al.}(2010)\citenamefont{Mitra},
  \citenamefont{Kumar},\ and\ \citenamefont{Samarth}}]{Mitra:2010_PRB}%
  \BibitemOpen
  \bibfield{author}{%
  \bibinfo {author} {\bibnamefont{Mitra}, \bibfnamefont{P.}}, \bibinfo {author}
  {\bibfnamefont{N.}~\bibnamefont{Kumar}},\ and\ \bibinfo {author}
  {\bibfnamefont{N.}~\bibnamefont{Samarth}}}%
  , \bibinfo {year} {2010},\ \bibfield{title}{%
  \enquote{\bibinfo {title} {Localization and the anomalous {Hall} effect in a
  dirty metallic ferromagnet},}\ }%
  \bibfield{journal}{%
  \bibinfo {journal} {Phys. Rev. B}\ }%
  \textbf{\bibinfo {volume} {82}},\ \bibinfo {pages} {035205}%
  \bibAnnoteFile{NoStop}{Mitra:2010_PRB}%
\bibitem[{\citenamefont{Mizokawa}\ and\
  \citenamefont{Fujimori}(1993)}]{Mizokawa:1993_PRB}%
  \BibitemOpen
  \bibfield{author}{%
  \bibinfo {author} {\bibnamefont{Mizokawa}, \bibfnamefont{T.}},\ and\ \bibinfo
  {author} {\bibfnamefont{A.}~\bibnamefont{Fujimori}}}%
  , \bibinfo {year} {1993},\ \bibfield{title}{%
  \enquote{\bibinfo {title} {Configuration-interaction description of
  transition-metal impurities in {II-VI} semiconductors},}\ }%
  \bibfield{journal}{%
  \bibinfo {journal} {Phys. Rev. B}\ }%
  \textbf{\bibinfo {volume} {48}},\ \bibinfo {pages} {14150}%
  \bibAnnoteFile{NoStop}{Mizokawa:1993_PRB}%
\bibitem[{\citenamefont{Mizokawa}\ \emph{et~al.}(2002)\citenamefont{Mizokawa},
  \citenamefont{Nambu}, \citenamefont{Fujimori}, \citenamefont{Fukumura},\ and\
  \citenamefont{Kawasaki}}]{Mizokawa:2002_PRB}%
  \BibitemOpen
  \bibfield{author}{%
  \bibinfo {author} {\bibnamefont{Mizokawa}, \bibfnamefont{T.}}, \bibinfo
  {author} {\bibfnamefont{T.}~\bibnamefont{Nambu}}, \bibinfo {author}
  {\bibfnamefont{A.}~\bibnamefont{Fujimori}}, \bibinfo {author}
  {\bibfnamefont{T.}~\bibnamefont{Fukumura}},\ and\ \bibinfo {author}
  {\bibfnamefont{M.}~\bibnamefont{Kawasaki}}}%
  , \bibinfo {year} {2002},\ \bibfield{title}{%
  \enquote{\bibinfo {title} {Electronic structure of the oxide-diluted magnetic
  semiconductor {Zn$_{1-x}$Mn$_{x}$O}},}\ }%
  \bibfield{journal}{%
  \bibinfo {journal} {Phys. Rev. B}\ }%
  \textbf{\bibinfo {volume} {65}},\ \bibinfo {pages} {085209}%
  \bibAnnoteFile{NoStop}{Mizokawa:2002_PRB}%
\bibitem[{\citenamefont{Munekata}\ \emph{et~al.}(1989)\citenamefont{Munekata},
  \citenamefont{Ohno}, \citenamefont{von Molnar}, \citenamefont{Segm\"uller},
  \citenamefont{Chang},\ and\ \citenamefont{Esaki}}]{Munekata:1989_PRL}%
  \BibitemOpen
  \bibfield{author}{%
  \bibinfo {author} {\bibnamefont{Munekata}, \bibfnamefont{H.}}, \bibinfo
  {author} {\bibfnamefont{H.}~\bibnamefont{Ohno}}, \bibinfo {author}
  {\bibfnamefont{S.}~\bibnamefont{von Molnar}}, \bibinfo {author}
  {\bibfnamefont{A.}~\bibnamefont{Segm\"uller}}, \bibinfo {author}
  {\bibfnamefont{L.~L.}\ \bibnamefont{Chang}},\ and\ \bibinfo {author}
  {\bibfnamefont{L.}~\bibnamefont{Esaki}}}%
  , \bibinfo {year} {1989},\ \bibfield{title}{%
  \enquote{\bibinfo {title} {Diluted magnetic {III--V} semiconductors},}\ }%
  \bibfield{journal}{%
  \bibinfo {journal} {Phys. Rev. Lett.}\ }%
  \textbf{\bibinfo {volume} {63}},\ \bibinfo {pages} {1849}%
  \bibAnnoteFile{NoStop}{Munekata:1989_PRL}%
\bibitem[{\citenamefont{Munekata}\ \emph{et~al.}(1993)\citenamefont{Munekata},
  \citenamefont{Zaslavsky}, \citenamefont{Fumagalli},\ and\
  \citenamefont{Gambino}}]{Munekata:1993_APL}%
  \BibitemOpen
  \bibfield{author}{%
  \bibinfo {author} {\bibnamefont{Munekata}, \bibfnamefont{H.}}, \bibinfo
  {author} {\bibfnamefont{A.}~\bibnamefont{Zaslavsky}}, \bibinfo {author}
  {\bibfnamefont{P.}~\bibnamefont{Fumagalli}},\ and\ \bibinfo {author}
  {\bibfnamefont{R.~J.}\ \bibnamefont{Gambino}}}%
  , \bibinfo {year} {1993},\ \bibfield{title}{%
  \enquote{\bibinfo {title} {Preparation of {(In,Mn)As/(Ga,Al)Sb} magnetic
  semiconductor heterostructures and their ferromagnetic characteristics},}\ }%
  \bibfield{journal}{%
  \bibinfo {journal} {Appl. Phys. Lett.}\ }%
  \textbf{\bibinfo {volume} {63}},\ \bibinfo {pages} {2929}%
  \bibAnnoteFile{NoStop}{Munekata:1993_APL}%
\bibitem[{\citenamefont{Muneta}\ \emph{et~al.}(2012)\citenamefont{Muneta},
  \citenamefont{Ohya},\ and\ \citenamefont{Tanaka}}]{Muneta:2012_APL}%
  \BibitemOpen
  \bibfield{author}{%
  \bibinfo {author} {\bibnamefont{Muneta}, \bibfnamefont{I.}}, \bibinfo
  {author} {\bibfnamefont{S.}~\bibnamefont{Ohya}},\ and\ \bibinfo {author}
  {\bibfnamefont{M.}~\bibnamefont{Tanaka}}}%
  , \bibinfo {year} {2012},\ \bibfield{title}{%
  \enquote{\bibinfo {title} {Spin-dependent tunneling transport in a
  ferromagnetic {GaMnAs} and un-doped {GaAs} double-quantum-well
  heterostructure},}\ }%
  \bibfield{journal}{%
  \bibinfo {journal} {Appl. Phys. Lett.}\ }%
  \textbf{\bibinfo {volume} {100}},\ \bibinfo {pages} {162409}%
  \bibAnnoteFile{NoStop}{Muneta:2012_APL}%
\bibitem[{\citenamefont{Myers}\ \emph{et~al.}(2006)\citenamefont{Myers},
  \citenamefont{Sheu}, \citenamefont{Jackson}, \citenamefont{Gossard},
  \citenamefont{Schiffer}, \citenamefont{Samarth},\ and\
  \citenamefont{Awschalom}}]{Myers:2006_PRB}%
  \BibitemOpen
  \bibfield{author}{%
  \bibinfo {author} {\bibnamefont{Myers}, \bibfnamefont{R.~C.}}, \bibinfo
  {author} {\bibfnamefont{B.~L.}\ \bibnamefont{Sheu}}, \bibinfo {author}
  {\bibfnamefont{A.~W.}\ \bibnamefont{Jackson}}, \bibinfo {author}
  {\bibfnamefont{A.~C.}\ \bibnamefont{Gossard}}, \bibinfo {author}
  {\bibfnamefont{P.}~\bibnamefont{Schiffer}}, \bibinfo {author}
  {\bibfnamefont{N.}~\bibnamefont{Samarth}},\ and\ \bibinfo {author}
  {\bibfnamefont{D.~D.}\ \bibnamefont{Awschalom}}}%
  , \bibinfo {year} {2006},\ \bibfield{title}{%
  \enquote{\bibinfo {title} {Antisite effect on hole-mediated ferromagnetism in
  {(Ga,Mn)As}},}\ }%
  \bibfield{journal}{%
  \bibinfo {journal} {Phys. Rev. B}\ }%
  \textbf{\bibinfo {volume} {74}},\ \bibinfo {pages} {155203}%
  \bibAnnoteFile{NoStop}{Myers:2006_PRB}%
\bibitem[{\citenamefont{Nagaev}(1993)}]{Nagaev:1993_B}%
  \BibitemOpen
  \bibfield{author}{%
  \bibinfo {author} {\bibnamefont{Nagaev}, \bibfnamefont{E.~L.}}}%
  , \bibinfo {year} {1993},\ \emph{\bibinfo {title} {{Physics of Magnetic
  Semiconductors}}}\ (\bibinfo {publisher} {Mir, Moscow})%
  \bibAnnoteFile{NoStop}{Nagaev:1993_B}%
\bibitem[{\citenamefont{Nagaosa}\ \emph{et~al.}(2010)\citenamefont{Nagaosa},
  \citenamefont{Sinova}, \citenamefont{Onoda}, \citenamefont{MacDonald},\ and\
  \citenamefont{Ong}}]{Nagaosa:2010_RMP}%
  \BibitemOpen
  \bibfield{author}{%
  \bibinfo {author} {\bibnamefont{Nagaosa}, \bibfnamefont{N.}}, \bibinfo
  {author} {\bibfnamefont{J.}~\bibnamefont{Sinova}}, \bibinfo {author}
  {\bibfnamefont{S.}~\bibnamefont{Onoda}}, \bibinfo {author}
  {\bibfnamefont{A.~H.}\ \bibnamefont{MacDonald}},\ and\ \bibinfo {author}
  {\bibfnamefont{N.~P.}\ \bibnamefont{Ong}}}%
  , \bibinfo {year} {2010},\ \bibfield{title}{%
  \enquote{\bibinfo {title} {Anomalous {Hall} effect},}\ }%
  \bibfield{journal}{%
  \bibinfo {journal} {Rev. Mod. Phys.}\ }%
  \textbf{\bibinfo {volume} {82}},\ \bibinfo {pages} {1539}%
  \bibAnnoteFile{NoStop}{Nagaosa:2010_RMP}%
\bibitem[{\citenamefont{Navarro-Quezada}\
  \emph{et~al.}(2011)\citenamefont{Navarro-Quezada},
  \citenamefont{Gonzalez~Szwacki}, \citenamefont{Stefanowicz},
  \citenamefont{Li}, \citenamefont{Grois}, \citenamefont{Devillers},
  \citenamefont{Rovezzi}, \citenamefont{Jakie{\l}a}, \citenamefont{Faina},
  \citenamefont{Majewski}, \citenamefont{Sawicki}, \citenamefont{Dietl},\ and\
  \citenamefont{Bonanni}}]{Navarro:2011_PRB}%
  \BibitemOpen
  \bibfield{author}{%
  \bibinfo {author} {\bibnamefont{Navarro-Quezada}, \bibfnamefont{A.}},
  \bibinfo {author} {\bibfnamefont{N.}~\bibnamefont{Gonzalez~Szwacki}},
  \bibinfo {author} {\bibfnamefont{W.}~\bibnamefont{Stefanowicz}}, \bibinfo
  {author} {\bibfnamefont{Tian}\ \bibnamefont{Li}}, \bibinfo {author}
  {\bibfnamefont{A.}~\bibnamefont{Grois}}, \bibinfo {author}
  {\bibfnamefont{T.}~\bibnamefont{Devillers}}, \bibinfo {author}
  {\bibfnamefont{M.}~\bibnamefont{Rovezzi}}, \bibinfo {author}
  {\bibfnamefont{R.}~\bibnamefont{Jakie{\l}a}}, \bibinfo {author}
  {\bibfnamefont{B.}~\bibnamefont{Faina}}, \bibinfo {author}
  {\bibfnamefont{J.~A.}\ \bibnamefont{Majewski}}, \bibinfo {author}
  {\bibfnamefont{M.}~\bibnamefont{Sawicki}}, \bibinfo {author}
  {\bibfnamefont{T.}~\bibnamefont{Dietl}},\ and\ \bibinfo {author}
  {\bibfnamefont{A.}~\bibnamefont{Bonanni}}}%
  , \bibinfo {year} {2011},\ \bibfield{title}{%
  \enquote{\bibinfo {title} {{Fe-Mg} interplay and the effect of deposition
  mode in {(Ga,Fe)N} doped with {Mg}},}\ }%
  \bibfield{journal}{%
  \bibinfo {journal} {Phys. Rev. B}\ }%
  \textbf{\bibinfo {volume} {84}},\ \bibinfo {pages} {155321}%
  \bibAnnoteFile{NoStop}{Navarro:2011_PRB}%
\bibitem[{\citenamefont{Naydenova}\
  \emph{et~al.}(2011)\citenamefont{Naydenova}, \citenamefont{D\"urrenfeld},
  \citenamefont{Tavakoli}, \citenamefont{P\'egard}, \citenamefont{Ebel},
  \citenamefont{Pappert}, \citenamefont{Brunner}, \citenamefont{Gould},\ and\
  \citenamefont{Molenkamp}}]{Naydenova:2011_PRL}%
  \BibitemOpen
  \bibfield{author}{%
  \bibinfo {author} {\bibnamefont{Naydenova}, \bibfnamefont{Ts.}}, \bibinfo
  {author} {\bibfnamefont{P.}~\bibnamefont{D\"urrenfeld}}, \bibinfo {author}
  {\bibfnamefont{K.}~\bibnamefont{Tavakoli}}, \bibinfo {author}
  {\bibfnamefont{N.}~\bibnamefont{P\'egard}}, \bibinfo {author}
  {\bibfnamefont{L.}~\bibnamefont{Ebel}}, \bibinfo {author}
  {\bibfnamefont{K.}~\bibnamefont{Pappert}}, \bibinfo {author}
  {\bibfnamefont{K.}~\bibnamefont{Brunner}}, \bibinfo {author}
  {\bibfnamefont{C.}~\bibnamefont{Gould}},\ and\ \bibinfo {author}
  {\bibfnamefont{L.~W.}\ \bibnamefont{Molenkamp}}}%
  , \bibinfo {year} {2011},\ \bibfield{title}{%
  \enquote{\bibinfo {title} {Diffusion thermopower of {(Ga,Mn)As/GaAs} tunnel
  junctions},}\ }%
  \bibfield{journal}{%
  \bibinfo {journal} {Phys. Rev. Lett.}\ }%
  \textbf{\bibinfo {volume} {107}},\ \bibinfo {pages} {197201}%
  \bibAnnoteFile{NoStop}{Naydenova:2011_PRL}%
\bibitem[{\citenamefont{Nazmul}\ \emph{et~al.}(2004)\citenamefont{Nazmul},
  \citenamefont{Kobayashi}, \citenamefont{Sugahara},\ and\
  \citenamefont{Tanaka}}]{Nazmul:2004_JJAP}%
  \BibitemOpen
  \bibfield{author}{%
  \bibinfo {author} {\bibnamefont{Nazmul}, \bibfnamefont{A.~M.}}, \bibinfo
  {author} {\bibfnamefont{S.}~\bibnamefont{Kobayashi}}, \bibinfo {author}
  {\bibfnamefont{S.}~\bibnamefont{Sugahara}},\ and\ \bibinfo {author}
  {\bibfnamefont{M.}~\bibnamefont{Tanaka}}}%
  , \bibinfo {year} {2004},\ \bibfield{title}{%
  \enquote{\bibinfo {title} {Electrical and optical control of ferromagnetism
  in {III-V} semiconductor heterostructures at high temperature (similar to 100
  {K})},}\ }%
  \bibfield{journal}{%
  \bibinfo {journal} {Jpn. J. Appl. Phys.}\ }%
  \textbf{\bibinfo {volume} {43}},\ \bibinfo {pages} {L233}%
  \bibAnnoteFile{NoStop}{Nazmul:2004_JJAP}%
\bibitem[{\citenamefont{Neumaier}\ \emph{et~al.}(2008)\citenamefont{Neumaier},
  \citenamefont{Schlapps}, \citenamefont{W{\"u}rstbauer},
  \citenamefont{Sadowski}, \citenamefont{Reinwald},
  \citenamefont{Wegscheider},\ and\ \citenamefont{Weiss}}]{Neumaier:2008_PRB}%
  \BibitemOpen
  \bibfield{author}{%
  \bibinfo {author} {\bibnamefont{Neumaier}, \bibfnamefont{D.}}, \bibinfo
  {author} {\bibfnamefont{M.}~\bibnamefont{Schlapps}}, \bibinfo {author}
  {\bibfnamefont{U.}~\bibnamefont{W{\"u}rstbauer}}, \bibinfo {author}
  {\bibfnamefont{J.}~\bibnamefont{Sadowski}}, \bibinfo {author}
  {\bibfnamefont{M.}~\bibnamefont{Reinwald}}, \bibinfo {author}
  {\bibfnamefont{W.}~\bibnamefont{Wegscheider}},\ and\ \bibinfo {author}
  {\bibfnamefont{D.}~\bibnamefont{Weiss}}}%
  , \bibinfo {year} {2008},\ \bibfield{title}{%
  \enquote{\bibinfo {title} {Electron-electron interaction in one- and
  two-dimensional ferromagnetic {(Ga,Mn)As}},}\ }%
  \bibfield{journal}{%
  \bibinfo {journal} {Phys. Rev. B}\ }%
  \textbf{\bibinfo {volume} {77}},\ \bibinfo {pages} {041306(R)}%
  \bibAnnoteFile{NoStop}{Neumaier:2008_PRB}%
\bibitem[{\citenamefont{Neumaier}\ \emph{et~al.}(2009)\citenamefont{Neumaier},
  \citenamefont{Turek}, \citenamefont{W{\"u}rstbauer}, \citenamefont{Vogl},
  \citenamefont{Utz}, \citenamefont{Wegscheider},\ and\
  \citenamefont{Weiss}}]{Neumaier:2009_PRL}%
  \BibitemOpen
  \bibfield{author}{%
  \bibinfo {author} {\bibnamefont{Neumaier}, \bibfnamefont{D.}}, \bibinfo
  {author} {\bibfnamefont{M.}~\bibnamefont{Turek}}, \bibinfo {author}
  {\bibfnamefont{U.}~\bibnamefont{W{\"u}rstbauer}}, \bibinfo {author}
  {\bibfnamefont{A.}~\bibnamefont{Vogl}}, \bibinfo {author}
  {\bibfnamefont{M.}~\bibnamefont{Utz}}, \bibinfo {author}
  {\bibfnamefont{W.}~\bibnamefont{Wegscheider}},\ and\ \bibinfo {author}
  {\bibfnamefont{D.}~\bibnamefont{Weiss}}}%
  , \bibinfo {year} {2009},\ \bibfield{title}{%
  \enquote{\bibinfo {title} {All-electrical measurement of the density of
  states in {(Ga,Mn)As}},}\ }%
  \bibfield{journal}{%
  \bibinfo {journal} {Phys. Rev. Lett.}\ }%
  \textbf{\bibinfo {volume} {103}},\ \bibinfo {pages} {087203}%
  \bibAnnoteFile{NoStop}{Neumaier:2009_PRL}%
\bibitem[{\citenamefont{Nguyen}\ \emph{et~al.}(2006)\citenamefont{Nguyen},
  \citenamefont{Shchelushkin},\ and\ \citenamefont{Brataas}}]{Nguyen:2006_PRL}%
  \BibitemOpen
  \bibfield{author}{%
  \bibinfo {author} {\bibnamefont{Nguyen}, \bibfnamefont{{Anh Kiet}}}, \bibinfo
  {author} {\bibfnamefont{R.~V.}\ \bibnamefont{Shchelushkin}},\ and\ \bibinfo
  {author} {\bibfnamefont{A.}~\bibnamefont{Brataas}}}%
  , \bibinfo {year} {2006},\ \bibfield{title}{%
  \enquote{\bibinfo {title} {Intrinsic domain-wall resistance in ferromagnetic
  semiconductors},}\ }%
  \bibfield{journal}{%
  \bibinfo {journal} {Phys. Rev. Lett.}\ }%
  \textbf{\bibinfo {volume} {97}},\ \bibinfo {pages} {136603}%
  \bibAnnoteFile{NoStop}{Nguyen:2006_PRL}%
\bibitem[{\citenamefont{Niazi}\ \emph{et~al.}(2013)\citenamefont{Niazi},
  \citenamefont{Cormier}, \citenamefont{Lucot}, \citenamefont{Largeau},
  \citenamefont{Jeudy}, \citenamefont{Cibert},\ and\
  \citenamefont{Lema\^{i}tre}}]{Niazi:2013_APL}%
  \BibitemOpen
  \bibfield{author}{%
  \bibinfo {author} {\bibnamefont{Niazi}, \bibfnamefont{T.}}, \bibinfo {author}
  {\bibfnamefont{M.}~\bibnamefont{Cormier}}, \bibinfo {author}
  {\bibfnamefont{D.}~\bibnamefont{Lucot}}, \bibinfo {author}
  {\bibfnamefont{L.}~\bibnamefont{Largeau}}, \bibinfo {author}
  {\bibfnamefont{V.}~\bibnamefont{Jeudy}}, \bibinfo {author}
  {\bibfnamefont{J.}~\bibnamefont{Cibert}},\ and\ \bibinfo {author}
  {\bibfnamefont{A.}~\bibnamefont{Lema\^{i}tre}}}%
  , \bibinfo {year} {2013},\ \bibfield{title}{%
  \enquote{\bibinfo {title} {Electric-field control of the magnetic anisotropy
  in an ultrathin {(Ga,Mn)As/(Ga,Mn)(As,P)} bilayer},}\ }%
  \bibfield{journal}{%
  \bibinfo {journal} {Appl. Phys. Lett.}\ }%
  \textbf{\bibinfo {volume} {102}},\ \bibinfo {pages} {122403}%
  \bibAnnoteFile{NoStop}{Niazi:2013_APL}%
\bibitem[{\citenamefont{Nishitani}\
  \emph{et~al.}(2010)\citenamefont{Nishitani}, \citenamefont{Chiba},
  \citenamefont{Endo}, \citenamefont{Sawicki}, \citenamefont{Matsukura},
  \citenamefont{Dietl},\ and\ \citenamefont{Ohno}}]{Nishitani:2010_PRB}%
  \BibitemOpen
  \bibfield{author}{%
  \bibinfo {author} {\bibnamefont{Nishitani}, \bibfnamefont{Y.}}, \bibinfo
  {author} {\bibfnamefont{D.}~\bibnamefont{Chiba}}, \bibinfo {author}
  {\bibfnamefont{M.}~\bibnamefont{Endo}}, \bibinfo {author}
  {\bibfnamefont{M.}~\bibnamefont{Sawicki}}, \bibinfo {author}
  {\bibfnamefont{F.}~\bibnamefont{Matsukura}}, \bibinfo {author}
  {\bibfnamefont{T.}~\bibnamefont{Dietl}},\ and\ \bibinfo {author}
  {\bibfnamefont{H.}~\bibnamefont{Ohno}}}%
  , \bibinfo {year} {2010},\ \bibfield{title}{%
  \enquote{\bibinfo {title} {Curie temperature versus hole concentration in
  field-effect structures of {Ga$_{1-x}$Mn$_x$As}},}\ }%
  \bibfield{journal}{%
  \bibinfo {journal} {Phys. Rev. B}\ }%
  \textbf{\bibinfo {volume} {81}},\ \bibinfo {pages} {045208}%
  \bibAnnoteFile{NoStop}{Nishitani:2010_PRB}%
\bibitem[{\citenamefont{Nov{\'a}k}\
  \emph{et~al.}(2008)\citenamefont{Nov{\'a}k}, \citenamefont{Olejn\'{\i}k},
  \citenamefont{Wunderlich}, \citenamefont{Cukr}, \citenamefont{andA.
  W.~Rushforth}, \citenamefont{Edmonds}, \citenamefont{Campion},
  \citenamefont{Gallagher}, \citenamefont{Sinova},\ and\
  \citenamefont{Jungwirth}}]{Novak:2008_PRL}%
  \BibitemOpen
  \bibfield{author}{%
  \bibinfo {author} {\bibnamefont{Nov{\'a}k}, \bibfnamefont{V.}}, \bibinfo
  {author} {\bibfnamefont{K.}~\bibnamefont{Olejn\'{\i}k}}, \bibinfo {author}
  {\bibfnamefont{J.}~\bibnamefont{Wunderlich}}, \bibinfo {author}
  {\bibfnamefont{M.}~\bibnamefont{Cukr}}, \bibinfo {author}
  {\bibfnamefont{K.~V{\'y}born{\'y}}\ \bibnamefont{andA. W.~Rushforth}},
  \bibinfo {author} {\bibfnamefont{K.~W.}\ \bibnamefont{Edmonds}}, \bibinfo
  {author} {\bibfnamefont{R.~P.}\ \bibnamefont{Campion}}, \bibinfo {author}
  {\bibfnamefont{B.~L.}\ \bibnamefont{Gallagher}}, \bibinfo {author}
  {\bibfnamefont{Jairo}\ \bibnamefont{Sinova}},\ and\ \bibinfo {author}
  {\bibfnamefont{T.}~\bibnamefont{Jungwirth}}}%
  , \bibinfo {year} {2008},\ \bibfield{title}{%
  \enquote{\bibinfo {title} {Curie point singularity in the temperature
  derivative of resistivity in {(Ga,Mn)As}},}\ }%
  \bibfield{journal}{%
  \bibinfo {journal} {Phys. Rev. Lett.}\ }%
  \textbf{\bibinfo {volume} {101}},\ \bibinfo {pages} {077201}%
  \bibAnnoteFile{NoStop}{Novak:2008_PRL}%
\bibitem[{\citenamefont{N\v{e}mec}\
  \emph{et~al.}(2013)\citenamefont{N\v{e}mec}, \citenamefont{Nov\'ak},
  \citenamefont{Tesa\v{r}ov\'a}, \citenamefont{Rozkotov\'a},
  \citenamefont{Reichlov\'a}, \citenamefont{Butkovi\v{c}ov\'a},
  \citenamefont{Troj\'anek}, \citenamefont{Olejn\'ik}, \citenamefont{Mal\'y},
  \citenamefont{Campion}, \citenamefont{Gallagher}, \citenamefont{Sinova},\
  and\ \citenamefont{Jungwirth}}]{Nemec:2013_NC}%
  \BibitemOpen
  \bibfield{author}{%
  \bibinfo {author} {\bibnamefont{N\v{e}mec}, \bibfnamefont{P.}}, \bibinfo
  {author} {\bibfnamefont{V.}~\bibnamefont{Nov\'ak}}, \bibinfo {author}
  {\bibfnamefont{N.}~\bibnamefont{Tesa\v{r}ov\'a}}, \bibinfo {author}
  {\bibfnamefont{E.}~\bibnamefont{Rozkotov\'a}}, \bibinfo {author}
  {\bibfnamefont{H.}~\bibnamefont{Reichlov\'a}}, \bibinfo {author}
  {\bibfnamefont{D.}~\bibnamefont{Butkovi\v{c}ov\'a}}, \bibinfo {author}
  {\bibfnamefont{F.}~\bibnamefont{Troj\'anek}}, \bibinfo {author}
  {\bibfnamefont{K.}~\bibnamefont{Olejn\'ik}}, \bibinfo {author}
  {\bibfnamefont{P.}~\bibnamefont{Mal\'y}}, \bibinfo {author}
  {\bibfnamefont{R.~P.}\ \bibnamefont{Campion}}, \bibinfo {author}
  {\bibfnamefont{B.~L.}\ \bibnamefont{Gallagher}}, \bibinfo {author}
  {\bibfnamefont{Jairo}\ \bibnamefont{Sinova}},\ and\ \bibinfo {author}
  {\bibfnamefont{T.}~\bibnamefont{Jungwirth}}}%
  , \bibinfo {year} {2013},\ \bibfield{title}{%
  \enquote{\bibinfo {title} {Establishing micromagnetic parameters of
  ferromagnetic semiconductor {(Ga,Mn)As}},}\ }%
  \bibfield{journal}{%
  \bibinfo {journal} {Nat. Commun.}\ }%
  \textbf{\bibinfo {volume} {4}},\ \bibinfo {pages} {1422}%
  \bibAnnoteFile{NoStop}{Nemec:2013_NC}%
\bibitem[{\citenamefont{N\v{e}mec}\
  \emph{et~al.}(2012)\citenamefont{N\v{e}mec}, \citenamefont{Rozkotov\'a},
  \citenamefont{end F.~Troj\'anek}, \citenamefont{{De Ranieri}},
  \citenamefont{Olejn\'{i}k}, \citenamefont{Zemen}, \citenamefont{Nov\'ak},
  \citenamefont{Cukr}, \citenamefont{Mal\'{y}},\ and\
  \citenamefont{Jungwirth}}]{Nemec:2012_NP}%
  \BibitemOpen
  \bibfield{author}{%
  \bibinfo {author} {\bibnamefont{N\v{e}mec}, \bibfnamefont{P.}}, \bibinfo
  {author} {\bibfnamefont{E.}~\bibnamefont{Rozkotov\'a}}, \bibinfo {author}
  {\bibfnamefont{N.~Tesa\v{r}ov\'a}\ \bibnamefont{end F.~Troj\'anek}}, \bibinfo
  {author} {\bibfnamefont{E.}~\bibnamefont{{De Ranieri}}}, \bibinfo {author}
  {\bibfnamefont{K.}~\bibnamefont{Olejn\'{i}k}}, \bibinfo {author}
  {\bibfnamefont{J.}~\bibnamefont{Zemen}}, \bibinfo {author}
  {\bibfnamefont{V.}~\bibnamefont{Nov\'ak}}, \bibinfo {author}
  {\bibfnamefont{M.}~\bibnamefont{Cukr}}, \bibinfo {author}
  {\bibfnamefont{P.}~\bibnamefont{Mal\'{y}}},\ and\ \bibinfo {author}
  {\bibfnamefont{T.}~\bibnamefont{Jungwirth}}}%
  , \bibinfo {year} {2012},\ \bibfield{title}{%
  \enquote{\bibinfo {title} {Experimental observation of the optical spin
  transfer torque},}\ }%
  \bibfield{journal}{%
  \bibinfo {journal} {Nat. Phys.}\ }%
  \textbf{\bibinfo {volume} {8}},\ \bibinfo {pages} {411}%
  \bibAnnoteFile{NoStop}{Nemec:2012_NP}%
\bibitem[{\citenamefont{Oestreich}(1999)}]{Oestreich:1999_N}%
  \BibitemOpen
  \bibfield{author}{%
  \bibinfo {author} {\bibnamefont{Oestreich}, \bibfnamefont{M.}}}%
  , \bibinfo {year} {1999},\ \bibfield{title}{%
  \enquote{\bibinfo {title} {Injecting spin into electronics},}\ }%
  \bibfield{journal}{%
  \bibinfo {journal} {Nature}\ }%
  \textbf{\bibinfo {volume} {402}},\ \bibinfo {pages} {735}%
  \bibAnnoteFile{NoStop}{Oestreich:1999_N}%
\bibitem[{\citenamefont{Ohno}(1998)}]{Ohno:1998_S}%
  \BibitemOpen
  \bibfield{author}{%
  \bibinfo {author} {\bibnamefont{Ohno}, \bibfnamefont{H.}}}%
  , \bibinfo {year} {1998},\ \bibfield{title}{%
  \enquote{\bibinfo {title} {Making nonmagnetic semiconductors
  ferromagnetic},}\ }%
  \bibfield{journal}{%
  \bibinfo {journal} {Science}\ }%
  \textbf{\bibinfo {volume} {281}},\ \bibinfo {pages} {951}%
  \bibAnnoteFile{NoStop}{Ohno:1998_S}%
\bibitem[{\citenamefont{Ohno}(2010)}]{Ohno:2010_NM}%
  \BibitemOpen
  \bibfield{author}{%
  \bibinfo {author} {\bibnamefont{Ohno}, \bibfnamefont{H.}}}%
  , \bibinfo {year} {2010},\ \bibfield{title}{%
  \enquote{\bibinfo {title} {A window on the future of spintronics},}\ }%
  \bibfield{journal}{%
  \bibinfo {journal} {Nat. Mater.}\ }%
  \textbf{\bibinfo {volume} {9}},\ \bibinfo {pages} {952}%
  \bibAnnoteFile{NoStop}{Ohno:2010_NM}%
\bibitem[{\citenamefont{Ohno}(2013)}]{Ohno:2013_JAP}%
  \BibitemOpen
  \bibfield{author}{%
  \bibinfo {author} {\bibnamefont{Ohno}, \bibfnamefont{H.}}}%
  , \bibinfo {year} {2013},\ \bibfield{title}{%
  \enquote{\bibinfo {title} {Bridging semiconductor and magnetism},}\ }%
  \bibfield{journal}{%
  \bibinfo {journal} {J. Appl. Phys.}\ }%
  \textbf{\bibinfo {volume} {113}},\ \bibinfo {pages} {136509}%
  \bibAnnoteFile{NoStop}{Ohno:2013_JAP}%
\bibitem[{\citenamefont{Ohno}\ \emph{et~al.}(1998)\citenamefont{Ohno},
  \citenamefont{Akiba}, \citenamefont{Matsukura}, \citenamefont{Shen},
  \citenamefont{Ohtani},\ and\ \citenamefont{Ohno}}]{Ohno:1998_APL}%
  \BibitemOpen
  \bibfield{author}{%
  \bibinfo {author} {\bibnamefont{Ohno}, \bibfnamefont{H.}}, \bibinfo {author}
  {\bibfnamefont{N.}~\bibnamefont{Akiba}}, \bibinfo {author}
  {\bibfnamefont{F.}~\bibnamefont{Matsukura}}, \bibinfo {author}
  {\bibfnamefont{A.}~\bibnamefont{Shen}}, \bibinfo {author}
  {\bibfnamefont{K.}~\bibnamefont{Ohtani}},\ and\ \bibinfo {author}
  {\bibfnamefont{Y.}~\bibnamefont{Ohno}}}%
  , \bibinfo {year} {1998},\ \bibfield{title}{%
  \enquote{\bibinfo {title} {Spontaneous splitting of ferromagnetic {(Ga,Mn)As}
  valence band observed by resonant tunneling spectroscopy},}\ }%
  \bibfield{journal}{%
  \bibinfo {journal} {Appl. Phys. Lett.}\ }%
  \textbf{\bibinfo {volume} {73}},\ \bibinfo {pages} {363}%
  \bibAnnoteFile{NoStop}{Ohno:1998_APL}%
\bibitem[{\citenamefont{Ohno}\ \emph{et~al.}(2000)\citenamefont{Ohno},
  \citenamefont{Chiba}, \citenamefont{Matsukura}, \citenamefont{Omiya},
  \citenamefont{Abe}, \citenamefont{Dietl}, \citenamefont{Ohno},\ and\
  \citenamefont{Ohtani}}]{Ohno:2000_N}%
  \BibitemOpen
  \bibfield{author}{%
  \bibinfo {author} {\bibnamefont{Ohno}, \bibfnamefont{H.}}, \bibinfo {author}
  {\bibfnamefont{D.}~\bibnamefont{Chiba}}, \bibinfo {author}
  {\bibfnamefont{F.}~\bibnamefont{Matsukura}}, \bibinfo {author}
  {\bibfnamefont{T.}~\bibnamefont{Omiya}}, \bibinfo {author}
  {\bibfnamefont{E.}~\bibnamefont{Abe}}, \bibinfo {author}
  {\bibfnamefont{T.}~\bibnamefont{Dietl}}, \bibinfo {author}
  {\bibfnamefont{Y.}~\bibnamefont{Ohno}},\ and\ \bibinfo {author}
  {\bibfnamefont{K.}~\bibnamefont{Ohtani}}}%
  , \bibinfo {year} {2000},\ \bibfield{title}{%
  \enquote{\bibinfo {title} {Electric-field control of ferromagnetism},}\ }%
  \bibfield{journal}{%
  \bibinfo {journal} {Nature}\ }%
  \textbf{\bibinfo {volume} {408}},\ \bibinfo {pages} {944}%
  \bibAnnoteFile{NoStop}{Ohno:2000_N}%
\bibitem[{\citenamefont{Ohno}\ \emph{et~al.}(1992)\citenamefont{Ohno},
  \citenamefont{Munekata}, \citenamefont{Penney}, \citenamefont{von
  {Moln{\'a}r}},\ and\ \citenamefont{Chang}}]{Ohno:1992_PRL}%
  \BibitemOpen
  \bibfield{author}{%
  \bibinfo {author} {\bibnamefont{Ohno}, \bibfnamefont{H.}}, \bibinfo {author}
  {\bibfnamefont{H.}~\bibnamefont{Munekata}}, \bibinfo {author}
  {\bibfnamefont{T.}~\bibnamefont{Penney}}, \bibinfo {author}
  {\bibfnamefont{S.}~\bibnamefont{von {Moln{\'a}r}}},\ and\ \bibinfo {author}
  {\bibfnamefont{L.~L.}\ \bibnamefont{Chang}}}%
  , \bibinfo {year} {1992},\ \bibfield{title}{%
  \enquote{\bibinfo {title} {Magnetotransport properties of p-type {(In,Mn)As}
  diluted magnetic {III-V} semiconductors},}\ }%
  \bibfield{journal}{%
  \bibinfo {journal} {Phys. Rev. Lett.}\ }%
  \textbf{\bibinfo {volume} {68}},\ \bibinfo {pages} {2664}%
  \bibAnnoteFile{NoStop}{Ohno:1992_PRL}%
\bibitem[{\citenamefont{Ohno}\ \emph{et~al.}(1996)\citenamefont{Ohno},
  \citenamefont{Shen}, \citenamefont{Matsukura}, \citenamefont{Oiwa},
  \citenamefont{Endo}, \citenamefont{Katsumoto},\ and\
  \citenamefont{Iye}}]{Ohno:1996_APL}%
  \BibitemOpen
  \bibfield{author}{%
  \bibinfo {author} {\bibnamefont{Ohno}, \bibfnamefont{H.}}, \bibinfo {author}
  {\bibfnamefont{A.}~\bibnamefont{Shen}}, \bibinfo {author}
  {\bibfnamefont{F.}~\bibnamefont{Matsukura}}, \bibinfo {author}
  {\bibfnamefont{A.}~\bibnamefont{Oiwa}}, \bibinfo {author}
  {\bibfnamefont{A.}~\bibnamefont{Endo}}, \bibinfo {author}
  {\bibfnamefont{S.}~\bibnamefont{Katsumoto}},\ and\ \bibinfo {author}
  {\bibfnamefont{Y.}~\bibnamefont{Iye}}}%
  , \bibinfo {year} {1996},\ \bibfield{title}{%
  \enquote{\bibinfo {title} {{(Ga,Mn)As}: A new diluted magnetic semiconductor
  based on {GaAs}},}\ }%
  \bibfield{journal}{%
  \bibinfo {journal} {Appl. Phys. Lett.}\ }%
  \textbf{\bibinfo {volume} {69}},\ \bibinfo {pages} {363}%
  \bibAnnoteFile{NoStop}{Ohno:1996_APL}%
\bibitem[{\citenamefont{Ohno}\ \emph{et~al.}(2002)\citenamefont{Ohno},
  \citenamefont{Arata}, \citenamefont{Matsukura},\ and\
  \citenamefont{Ohno}}]{Ohno:2002_PE}%
  \BibitemOpen
  \bibfield{author}{%
  \bibinfo {author} {\bibnamefont{Ohno}, \bibfnamefont{Y.}}, \bibinfo {author}
  {\bibfnamefont{I.}~\bibnamefont{Arata}}, \bibinfo {author}
  {\bibfnamefont{F.}~\bibnamefont{Matsukura}},\ and\ \bibinfo {author}
  {\bibfnamefont{H.}~\bibnamefont{Ohno}}}%
  , \bibinfo {year} {2002},\ \bibfield{title}{%
  \enquote{\bibinfo {title} {Valence band barrier at {(Ga,Mn)As/GaAs}
  interfaces},}\ }%
  \bibfield{journal}{%
  \bibinfo {journal} {Physica E}\ }%
  \textbf{\bibinfo {volume} {13}},\ \bibinfo {pages} {521}%
  \bibAnnoteFile{NoStop}{Ohno:2002_PE}%
\bibitem[{\citenamefont{Ohno}\ \emph{et~al.}(1999)\citenamefont{Ohno},
  \citenamefont{Young}, \citenamefont{Beschoten}, \citenamefont{Matsukura},
  \citenamefont{Ohno},\ and\ \citenamefont{Awschalom}}]{Ohno:1999_N}%
  \BibitemOpen
  \bibfield{author}{%
  \bibinfo {author} {\bibnamefont{Ohno}, \bibfnamefont{Y.}}, \bibinfo {author}
  {\bibfnamefont{D.~K.}\ \bibnamefont{Young}}, \bibinfo {author}
  {\bibfnamefont{B.}~\bibnamefont{Beschoten}}, \bibinfo {author}
  {\bibfnamefont{F.}~\bibnamefont{Matsukura}}, \bibinfo {author}
  {\bibfnamefont{H.}~\bibnamefont{Ohno}},\ and\ \bibinfo {author}
  {\bibfnamefont{D.~D.}\ \bibnamefont{Awschalom}}}%
  , \bibinfo {year} {1999},\ \bibfield{title}{%
  \enquote{\bibinfo {title} {Electrical spin injection in a ferromagnetic
  semiconductor heterostructure},}\ }%
  \bibfield{journal}{%
  \bibinfo {journal} {Nature}\ }%
  \textbf{\bibinfo {volume} {402}},\ \bibinfo {pages} {790}%
  \bibAnnoteFile{NoStop}{Ohno:1999_N}%
\bibitem[{\citenamefont{Ohya}\
  \emph{et~al.}(2007{\natexlab{a}})\citenamefont{Ohya}, \citenamefont{Hai},
  \citenamefont{Mizuno},\ and\ \citenamefont{Tanaka}}]{Ohya:2007_PRB}%
  \BibitemOpen
  \bibfield{author}{%
  \bibinfo {author} {\bibnamefont{Ohya}, \bibfnamefont{S.}}, \bibinfo {author}
  {\bibfnamefont{P.~N.}\ \bibnamefont{Hai}}, \bibinfo {author}
  {\bibfnamefont{Y.}~\bibnamefont{Mizuno}},\ and\ \bibinfo {author}
  {\bibfnamefont{M.}~\bibnamefont{Tanaka}}}%
  , \bibinfo {year} {2007}{\natexlab{a}},\ \bibfield{title}{%
  \enquote{\bibinfo {title} {Quantum-size effect and tunneling
  magnetoresistance in ferromagnetic-semiconductor quantum heterostructures},}\
  }%
  \bibfield{journal}{%
  \bibinfo {journal} {Phys. Rev. B}\ }%
  \textbf{\bibinfo {volume} {75}},\ \bibinfo {pages} {155328}%
  \bibAnnoteFile{NoStop}{Ohya:2007_PRB}%
\bibitem[{\citenamefont{Ohya}\ \emph{et~al.}(2005)\citenamefont{Ohya},
  \citenamefont{Hai},\ and\ \citenamefont{Tanaka}}]{Ohya:2005_APL}%
  \BibitemOpen
  \bibfield{author}{%
  \bibinfo {author} {\bibnamefont{Ohya}, \bibfnamefont{S.}}, \bibinfo {author}
  {\bibfnamefont{P.~N.}\ \bibnamefont{Hai}},\ and\ \bibinfo {author}
  {\bibfnamefont{M.}~\bibnamefont{Tanaka}}}%
  , \bibinfo {year} {2005},\ \bibfield{title}{%
  \enquote{\bibinfo {title} {Tunneling magnetoresistance in
  {GaMnAs/AlAs/InGaAs/AlAs/GaMnAs} double-barrier magnetic tunnel junctions},}\
  }%
  \bibfield{journal}{%
  \bibinfo {journal} {Appl. Phys. Lett.}\ }%
  \textbf{\bibinfo {volume} {87}},\ \bibinfo {pages} {012105}%
  \bibAnnoteFile{NoStop}{Ohya:2005_APL}%
\bibitem[{\citenamefont{Ohya}\ \emph{et~al.}(2009)\citenamefont{Ohya},
  \citenamefont{Muneta}, \citenamefont{Hai},\ and\
  \citenamefont{Tanaka}}]{Ohya:2009_APL}%
  \BibitemOpen
  \bibfield{author}{%
  \bibinfo {author} {\bibnamefont{Ohya}, \bibfnamefont{S.}}, \bibinfo {author}
  {\bibfnamefont{I.}~\bibnamefont{Muneta}}, \bibinfo {author}
  {\bibfnamefont{P.~N.}\ \bibnamefont{Hai}},\ and\ \bibinfo {author}
  {\bibfnamefont{M.}~\bibnamefont{Tanaka}}}%
  , \bibinfo {year} {2009},\ \bibfield{title}{%
  \enquote{\bibinfo {title} {{GaMnAs}-based magnetic tunnel junctions with an
  {AlMnAs} barrier},}\ }%
  \bibfield{journal}{%
  \bibinfo {journal} {Appl. Phys. Lett.}\ }%
  \textbf{\bibinfo {volume} {95}},\ \bibinfo {pages} {242503}%
  \bibAnnoteFile{NoStop}{Ohya:2009_APL}%
\bibitem[{\citenamefont{Ohya}\
  \emph{et~al.}(2010{\natexlab{a}})\citenamefont{Ohya}, \citenamefont{Muneta},
  \citenamefont{Hai},\ and\ \citenamefont{Tanaka}}]{Ohya:2010_PRL}%
  \BibitemOpen
  \bibfield{author}{%
  \bibinfo {author} {\bibnamefont{Ohya}, \bibfnamefont{S.}}, \bibinfo {author}
  {\bibfnamefont{I.}~\bibnamefont{Muneta}}, \bibinfo {author}
  {\bibfnamefont{P.~N.}\ \bibnamefont{Hai}},\ and\ \bibinfo {author}
  {\bibfnamefont{Masaaki}\ \bibnamefont{Tanaka}}}%
  , \bibinfo {year} {2010}{\natexlab{a}},\ \bibfield{title}{%
  \enquote{\bibinfo {title} {Valence-band structure of the ferromagnetic
  semiconductor {GaMnAs} studied by spin-dependent resonant tunneling
  spectroscopy},}\ }%
  \bibfield{journal}{%
  \bibinfo {journal} {Phys. Rev. Lett.}\ }%
  \textbf{\bibinfo {volume} {104}},\ \bibinfo {pages} {167204}%
  \bibAnnoteFile{NoStop}{Ohya:2010_PRL}%
\bibitem[{\citenamefont{Ohya}\
  \emph{et~al.}(2010{\natexlab{b}})\citenamefont{Ohya}, \citenamefont{Muneta},\
  and\ \citenamefont{Tanaka}}]{Ohya:2010_APL}%
  \BibitemOpen
  \bibfield{author}{%
  \bibinfo {author} {\bibnamefont{Ohya}, \bibfnamefont{S.}}, \bibinfo {author}
  {\bibfnamefont{I.}~\bibnamefont{Muneta}},\ and\ \bibinfo {author}
  {\bibfnamefont{M.}~\bibnamefont{Tanaka}}}%
  , \bibinfo {year} {2010}{\natexlab{b}},\ \bibfield{title}{%
  \enquote{\bibinfo {title} {Quantum-level control in a {III--V}-based
  ferromagnetic-semiconductor heterostructure with a {GaMnAs} quantum well and
  double barriers},}\ }%
  \bibfield{journal}{%
  \bibinfo {journal} {Appl. Phys. Lett.}\ }%
  \textbf{\bibinfo {volume} {96}},\ \bibinfo {pages} {052505}%
  \bibAnnoteFile{NoStop}{Ohya:2010_APL}%
\bibitem[{\citenamefont{Ohya}\ \emph{et~al.}(2012)\citenamefont{Ohya},
  \citenamefont{Muneta}, \citenamefont{Xin}, \citenamefont{Takata},\ and\
  \citenamefont{Tanaka}}]{Ohya:2012_PRB}%
  \BibitemOpen
  \bibfield{author}{%
  \bibinfo {author} {\bibnamefont{Ohya}, \bibfnamefont{S.}}, \bibinfo {author}
  {\bibfnamefont{I.}~\bibnamefont{Muneta}}, \bibinfo {author}
  {\bibfnamefont{Y.}~\bibnamefont{Xin}}, \bibinfo {author}
  {\bibfnamefont{K.}~\bibnamefont{Takata}},\ and\ \bibinfo {author}
  {\bibfnamefont{M.}~\bibnamefont{Tanaka}}}%
  , \bibinfo {year} {2012},\ \bibfield{title}{%
  \enquote{\bibinfo {title} {Valence-band structure of ferromagnetic
  semiconductor {(In,Ga,Mn)As}},}\ }%
  \bibfield{journal}{%
  \bibinfo {journal} {Phys. Rev. B}\ }%
  \textbf{\bibinfo {volume} {86}},\ \bibinfo {pages} {094418}%
  \bibAnnoteFile{NoStop}{Ohya:2012_PRB}%
\bibitem[{\citenamefont{Ohya}\
  \emph{et~al.}(2007{\natexlab{b}})\citenamefont{Ohya}, \citenamefont{Ohno},\
  and\ \citenamefont{Tanaka}}]{Ohya:2007_APL}%
  \BibitemOpen
  \bibfield{author}{%
  \bibinfo {author} {\bibnamefont{Ohya}, \bibfnamefont{S.}}, \bibinfo {author}
  {\bibfnamefont{K.}~\bibnamefont{Ohno}},\ and\ \bibinfo {author}
  {\bibfnamefont{M.}~\bibnamefont{Tanaka}}}%
  , \bibinfo {year} {2007}{\natexlab{b}},\ \bibfield{title}{%
  \enquote{\bibinfo {title} {Magneto-optical and magnetotransport properties of
  heavily {Mn}-doped {GaMnAs}},}\ }%
  \bibfield{journal}{%
  \bibinfo {journal} {Appl. Phys. Lett.}\ }%
  \textbf{\bibinfo {volume} {90}},\ \bibinfo {pages} {112503}%
  \bibAnnoteFile{NoStop}{Ohya:2007_APL}%
\bibitem[{\citenamefont{Ohya}\
  \emph{et~al.}(2011{\natexlab{a}})\citenamefont{Ohya}, \citenamefont{Takata},
  \citenamefont{Muneta}, \citenamefont{Hai},\ and\
  \citenamefont{Tanaka}}]{Ohya:2011_arXiv}%
  \BibitemOpen
  \bibfield{author}{%
  \bibinfo {author} {\bibnamefont{Ohya}, \bibfnamefont{S.}}, \bibinfo {author}
  {\bibfnamefont{K.}~\bibnamefont{Takata}}, \bibinfo {author}
  {\bibfnamefont{I.}~\bibnamefont{Muneta}}, \bibinfo {author}
  {\bibfnamefont{P.~N.}\ \bibnamefont{Hai}},\ and\ \bibinfo {author}
  {\bibfnamefont{M.}~\bibnamefont{Tanaka}}}%
  , \bibinfo {year} {2011}{\natexlab{a}},\ \bibfield{title}{%
  \enquote{\bibinfo {title} {Comment on '{Reconciling} results of tunnelling
  experiments on {(Ga,Mn)As}' {arXiv:1102.3267} by {Dietl} and {Sztenkiel}},}\
  }%
  \bibinfo {journal} {arXiv:1102.4459}%
  \bibAnnoteFile{NoStop}{Ohya:2011_arXiv}%
\bibitem[{\citenamefont{Ohya}\
  \emph{et~al.}(2011{\natexlab{b}})\citenamefont{Ohya}, \citenamefont{Takata},
  \citenamefont{Muneta}, \citenamefont{Hai},\ and\
  \citenamefont{Tanaka}}]{Ohya:2011_NP}%
  \BibitemOpen
\bibfield{journal}{%
    }%
  \bibfield{author}{%
  \bibinfo {author} {\bibnamefont{Ohya}, \bibfnamefont{S.}}, \bibinfo {author}
  {\bibfnamefont{K.}~\bibnamefont{Takata}}, \bibinfo {author}
  {\bibfnamefont{I.}~\bibnamefont{Muneta}}, \bibinfo {author}
  {\bibfnamefont{P.~N.}\ \bibnamefont{Hai}},\ and\ \bibinfo {author}
  {\bibfnamefont{M.}~\bibnamefont{Tanaka}}}%
  , \bibinfo {year} {2011}{\natexlab{b}},\ \bibfield{title}{%
  \enquote{\bibinfo {title} {Nearly non-magnetic valence band of the
  ferromagnetic semiconductor {GaMnAs}},}\ }%
  \bibfield{journal}{%
  \bibinfo {journal} {Nat. Phys.}\ }%
  \textbf{\bibinfo {volume} {7}},\ \bibinfo {pages} {342}%
  \bibAnnoteFile{NoStop}{Ohya:2011_NP}%
\bibitem[{\citenamefont{Oiwa}\ \emph{et~al.}(1998)\citenamefont{Oiwa},
  \citenamefont{Katsumoto}, \citenamefont{Endo}, \citenamefont{Hirasawa},
  \citenamefont{Iye}, \citenamefont{Matsukura}, \citenamefont{Shen},\ and\
  \citenamefont{Ohno}}]{Oiwa:1998_PB}%
  \BibitemOpen
  \bibfield{author}{%
  \bibinfo {author} {\bibnamefont{Oiwa}, \bibfnamefont{A.}}, \bibinfo {author}
  {\bibfnamefont{S.}~\bibnamefont{Katsumoto}}, \bibinfo {author}
  {\bibfnamefont{A.}~\bibnamefont{Endo}}, \bibinfo {author}
  {\bibfnamefont{M.}~\bibnamefont{Hirasawa}}, \bibinfo {author}
  {\bibfnamefont{Y.}~\bibnamefont{Iye}}, \bibinfo {author}
  {\bibfnamefont{F.}~\bibnamefont{Matsukura}}, \bibinfo {author}
  {\bibfnamefont{A.}~\bibnamefont{Shen}},\ and\ \bibinfo {author}
  {\bibfnamefont{Y.~Sugawara~H.}\ \bibnamefont{Ohno}}}%
  , \bibinfo {year} {1998},\ \bibfield{title}{%
  \enquote{\bibinfo {title} {Low-temperature conduction and giant negative
  magnetoresistance in {III-V}-based diluted magnetic semiconductor:
  {(Ga,Mn)As/GaAs}},}\ }%
  \bibfield{journal}{%
  \bibinfo {journal} {Physica B}\ }%
  \textbf{\bibinfo {volume} {249}},\ \bibinfo {pages} {775}%
  \bibAnnoteFile{NoStop}{Oiwa:1998_PB}%
\bibitem[{\citenamefont{Oiwa}\ \emph{et~al.}(2002)\citenamefont{Oiwa},
  \citenamefont{Mitsumori}, \citenamefont{Moriya},
  \citenamefont{S{\l}upi{\'n}ski},\ and\
  \citenamefont{Munekata}}]{Oiwa:2002_PRL}%
  \BibitemOpen
  \bibfield{author}{%
  \bibinfo {author} {\bibnamefont{Oiwa}, \bibfnamefont{A.}}, \bibinfo {author}
  {\bibfnamefont{Y.}~\bibnamefont{Mitsumori}}, \bibinfo {author}
  {\bibfnamefont{R.}~\bibnamefont{Moriya}}, \bibinfo {author}
  {\bibfnamefont{T.}~\bibnamefont{S{\l}upi{\'n}ski}},\ and\ \bibinfo {author}
  {\bibfnamefont{H.}~\bibnamefont{Munekata}}}%
  , \bibinfo {year} {2002},\ \bibfield{title}{%
  \enquote{\bibinfo {title} {Effect of optical spin injection on
  ferromagnetically coupled {Mn} spins in the {III-V} magnetic alloy
  semiconductor {(Ga, Mn)As}},}\ }%
  \bibfield{journal}{%
  \bibinfo {journal} {Phys. Rev. Lett.}\ }%
  \textbf{\bibinfo {volume} {88}},\ \bibinfo {pages} {137202}%
  \bibAnnoteFile{NoStop}{Oiwa:2002_PRL}%
\bibitem[{\citenamefont{Oiwa}\ \emph{et~al.}(2001)\citenamefont{Oiwa},
  \citenamefont{{S{\l}upi{\'n}ski}},\ and\
  \citenamefont{Munekata}}]{Oiwa:2001_APL}%
  \BibitemOpen
  \bibfield{author}{%
  \bibinfo {author} {\bibnamefont{Oiwa}, \bibfnamefont{A.}}, \bibinfo {author}
  {\bibfnamefont{T.}~\bibnamefont{{S{\l}upi{\'n}ski}}},\ and\ \bibinfo {author}
  {\bibfnamefont{H.}~\bibnamefont{Munekata}}}%
  , \bibinfo {year} {2001},\ \bibfield{title}{%
  \enquote{\bibinfo {title} {Control of magnetization reversal process by light
  illumination in ferromagnetic semiconductor heterosturucture
  p-{(In,Mn)As/GaSb}},}\ }%
  \bibfield{journal}{%
  \bibinfo {journal} {Appl. Phys. Lett.}\ }%
  \textbf{\bibinfo {volume} {78}},\ \bibinfo {pages} {518}%
  \bibAnnoteFile{NoStop}{Oiwa:2001_APL}%
\bibitem[{\citenamefont{Okabayashi}\
  \emph{et~al.}(1999)\citenamefont{Okabayashi}, \citenamefont{Kimura},
  \citenamefont{Mizokawa}, \citenamefont{Fujimori}, \citenamefont{Hayashi},\
  and\ \citenamefont{Tanaka}}]{Okabayashi:1999_PRB}%
  \BibitemOpen
  \bibfield{author}{%
  \bibinfo {author} {\bibnamefont{Okabayashi}, \bibfnamefont{J.}}, \bibinfo
  {author} {\bibfnamefont{A.}~\bibnamefont{Kimura}}, \bibinfo {author}
  {\bibfnamefont{T.}~\bibnamefont{Mizokawa}}, \bibinfo {author}
  {\bibfnamefont{A.}~\bibnamefont{Fujimori}}, \bibinfo {author}
  {\bibfnamefont{T.}~\bibnamefont{Hayashi}},\ and\ \bibinfo {author}
  {\bibfnamefont{M.}~\bibnamefont{Tanaka}}}%
  , \bibinfo {year} {1999},\ \bibfield{title}{%
  \enquote{\bibinfo {title} {Mn 3d partial density of states in
  {Ga$_{1-x}$Mn$_{x}$As} studied by resonant photoemission spectroscopy},}\ }%
  \bibfield{journal}{%
  \bibinfo {journal} {Phys. Rev. B}\ }%
  \textbf{\bibinfo {volume} {59}},\ \bibinfo {pages} {R2486}%
  \bibAnnoteFile{NoStop}{Okabayashi:1999_PRB}%
\bibitem[{\citenamefont{Okabayashi}\
  \emph{et~al.}(1998)\citenamefont{Okabayashi}, \citenamefont{Kimura},
  \citenamefont{Rader}, \citenamefont{Mizokawa}, \citenamefont{Fujimori},
  \citenamefont{Hayashi},\ and\ \citenamefont{Tanaka}}]{Okabayashi:1998_PRB}%
  \BibitemOpen
  \bibfield{author}{%
  \bibinfo {author} {\bibnamefont{Okabayashi}, \bibfnamefont{J.}}, \bibinfo
  {author} {\bibfnamefont{A.}~\bibnamefont{Kimura}}, \bibinfo {author}
  {\bibfnamefont{O.}~\bibnamefont{Rader}}, \bibinfo {author}
  {\bibfnamefont{T.}~\bibnamefont{Mizokawa}}, \bibinfo {author}
  {\bibfnamefont{A.}~\bibnamefont{Fujimori}}, \bibinfo {author}
  {\bibfnamefont{T.}~\bibnamefont{Hayashi}},\ and\ \bibinfo {author}
  {\bibfnamefont{M.}~\bibnamefont{Tanaka}}}%
  , \bibinfo {year} {1998},\ \bibfield{title}{%
  \enquote{\bibinfo {title} {Core-level photoemission study of
  {Ga$_{1-x}$Mn$_{x}$As}},}\ }%
  \bibfield{journal}{%
  \bibinfo {journal} {Phys. Rev. B}\ }%
  \textbf{\bibinfo {volume} {58}},\ \bibinfo {pages} {R4211}%
  \bibAnnoteFile{NoStop}{Okabayashi:1998_PRB}%
\bibitem[{\citenamefont{Okabayashi}\
  \emph{et~al.}(2001)\citenamefont{Okabayashi}, \citenamefont{Kimura},
  \citenamefont{Rader}, \citenamefont{Mizokawa}, \citenamefont{Fujimori},
  \citenamefont{Hayashi},\ and\ \citenamefont{Tanaka}}]{Okabayashi:2001_PRB}%
  \BibitemOpen
  \bibfield{author}{%
  \bibinfo {author} {\bibnamefont{Okabayashi}, \bibfnamefont{J.}}, \bibinfo
  {author} {\bibfnamefont{A.}~\bibnamefont{Kimura}}, \bibinfo {author}
  {\bibfnamefont{O.}~\bibnamefont{Rader}}, \bibinfo {author}
  {\bibfnamefont{T.}~\bibnamefont{Mizokawa}}, \bibinfo {author}
  {\bibfnamefont{A.}~\bibnamefont{Fujimori}}, \bibinfo {author}
  {\bibfnamefont{T.}~\bibnamefont{Hayashi}},\ and\ \bibinfo {author}
  {\bibfnamefont{M.}~\bibnamefont{Tanaka}}}%
  , \bibinfo {year} {2001},\ \bibfield{title}{%
  \enquote{\bibinfo {title} {Angle-resolved photoemission study of
  {Ga$_{1-x}$Mn$_{x}$As}},}\ }%
  \bibfield{journal}{%
  \bibinfo {journal} {Phys. Rev. B}\ }%
  \textbf{\bibinfo {volume} {64}},\ \bibinfo {pages} {125304}%
  \bibAnnoteFile{NoStop}{Okabayashi:2001_PRB}%
\bibitem[{\citenamefont{Okabayashi}\
  \emph{et~al.}(2002)\citenamefont{Okabayashi}, \citenamefont{Mizokawa},
  \citenamefont{Sarma}, \citenamefont{Fujimori},
  \citenamefont{S{\l}upi{\'n}ski}, \citenamefont{Oiwa},\ and\
  \citenamefont{Munekata}}]{Okabayashi:2002_PRB}%
  \BibitemOpen
  \bibfield{author}{%
  \bibinfo {author} {\bibnamefont{Okabayashi}, \bibfnamefont{J.}}, \bibinfo
  {author} {\bibfnamefont{T.}~\bibnamefont{Mizokawa}}, \bibinfo {author}
  {\bibfnamefont{D.~D.}\ \bibnamefont{Sarma}}, \bibinfo {author}
  {\bibfnamefont{A.}~\bibnamefont{Fujimori}}, \bibinfo {author}
  {\bibfnamefont{T.}~\bibnamefont{S{\l}upi{\'n}ski}}, \bibinfo {author}
  {\bibfnamefont{A.}~\bibnamefont{Oiwa}},\ and\ \bibinfo {author}
  {\bibfnamefont{H.}~\bibnamefont{Munekata}}}%
  , \bibinfo {year} {2002},\ \bibfield{title}{%
  \enquote{\bibinfo {title} {Electronic structure of {In$_{1-x}$Mn$_x$As}
  studied by photoemission spectroscopy: Comparison with
  {Ga$_{1-x}$Mn$_x$As}},}\ }%
  \bibfield{journal}{%
  \bibinfo {journal} {Phys. Rev. B}\ }%
  \textbf{\bibinfo {volume} {65}},\ \bibinfo {pages} {161203}%
  \bibAnnoteFile{NoStop}{Okabayashi:2002_PRB}%
\bibitem[{\citenamefont{Okabayashi}\
  \emph{et~al.}(2004)\citenamefont{Okabayashi}, \citenamefont{Ono},
  \citenamefont{Mizuguchi}, \citenamefont{Oshima}, \citenamefont{Gupta},
  \citenamefont{Sarma}, \citenamefont{Mizokawa}, \citenamefont{Fujimori},
  \citenamefont{Yuri}, \citenamefont{Chen}, \citenamefont{Fukumura},
  \citenamefont{Kawasaki},\ and\ \citenamefont{Koinuma}}]{Okabayashi:2004_JAP}%
  \BibitemOpen
  \bibfield{author}{%
  \bibinfo {author} {\bibnamefont{Okabayashi}, \bibfnamefont{J.}}, \bibinfo
  {author} {\bibfnamefont{K.}~\bibnamefont{Ono}}, \bibinfo {author}
  {\bibfnamefont{M.}~\bibnamefont{Mizuguchi}}, \bibinfo {author}
  {\bibfnamefont{M.}~\bibnamefont{Oshima}}, \bibinfo {author}
  {\bibfnamefont{Subhra~Sen}\ \bibnamefont{Gupta}}, \bibinfo {author}
  {\bibfnamefont{D.~D.}\ \bibnamefont{Sarma}}, \bibinfo {author}
  {\bibfnamefont{T.}~\bibnamefont{Mizokawa}}, \bibinfo {author}
  {\bibfnamefont{A.}~\bibnamefont{Fujimori}}, \bibinfo {author}
  {\bibfnamefont{M.}~\bibnamefont{Yuri}}, \bibinfo {author}
  {\bibfnamefont{C.~T.}\ \bibnamefont{Chen}}, \bibinfo {author}
  {\bibfnamefont{T.}~\bibnamefont{Fukumura}}, \bibinfo {author}
  {\bibfnamefont{M.}~\bibnamefont{Kawasaki}},\ and\ \bibinfo {author}
  {\bibfnamefont{H.}~\bibnamefont{Koinuma}}}%
  , \bibinfo {year} {2004},\ \bibfield{title}{%
  \enquote{\bibinfo {title} {X-ray absorption spectroscopy of transition-metal
  doped diluted magnetic semiconductors {Zn$_{1-x}$M$_x$O}},}\ }%
  \bibfield{journal}{%
  \bibinfo {journal} {J. Appl. Phys.}\ }%
  \textbf{\bibinfo {volume} {95}},\ \bibinfo {pages} {3573}%
  \bibAnnoteFile{NoStop}{Okabayashi:2004_JAP}%
\bibitem[{\citenamefont{{Olejn{\'i}k}}\
  \emph{et~al.}(2008)\citenamefont{{Olejn{\'i}k}}, \citenamefont{Owen},
  \citenamefont{{Nov{\'a}k}}, \citenamefont{{Ma\v{s}ek}},
  \citenamefont{Irvine}, \citenamefont{Wunderlich},\ and\
  \citenamefont{Jungwirth}}]{Olejnik:2008_PRB}%
  \BibitemOpen
  \bibfield{author}{%
  \bibinfo {author} {\bibnamefont{{Olejn{\'i}k}}, \bibfnamefont{K.}}, \bibinfo
  {author} {\bibfnamefont{M.~H.~S.}\ \bibnamefont{Owen}}, \bibinfo {author}
  {\bibfnamefont{V.}~\bibnamefont{{Nov{\'a}k}}}, \bibinfo {author}
  {\bibfnamefont{J.}~\bibnamefont{{Ma\v{s}ek}}}, \bibinfo {author}
  {\bibfnamefont{A.~C.}\ \bibnamefont{Irvine}}, \bibinfo {author}
  {\bibfnamefont{J.}~\bibnamefont{Wunderlich}},\ and\ \bibinfo {author}
  {\bibfnamefont{T.}~\bibnamefont{Jungwirth}}}%
  , \bibinfo {year} {2008},\ \bibfield{title}{%
  \enquote{\bibinfo {title} {Enhanced annealing, high {Curie} temperature and
  low-voltage gating in {(Ga,Mn)As}: A surface oxide control study},}\ }%
  \bibfield{journal}{%
  \bibinfo {journal} {Phys. Rev. B}\ }%
  \textbf{\bibinfo {volume} {78}},\ \bibinfo {pages} {054403}%
  \bibAnnoteFile{NoStop}{Olejnik:2008_PRB}%
\bibitem[{\citenamefont{Olejnik}\ \emph{et~al.}(2010)\citenamefont{Olejnik},
  \citenamefont{Wadley}, \citenamefont{Haigh}, \citenamefont{Edmonds},
  \citenamefont{Campion}, \citenamefont{Rushforth}, \citenamefont{Gallagher},
  \citenamefont{Foxon}, \citenamefont{Jungwirth}, \citenamefont{Wunderlich},
  \citenamefont{Dhesi}, \citenamefont{Cavill}, \citenamefont{van~der Laan},\
  and\ \citenamefont{Arenholz}}]{Olejnik:2010_PRB}%
  \BibitemOpen
  \bibfield{author}{%
  \bibinfo {author} {\bibnamefont{Olejnik}, \bibfnamefont{K.}}, \bibinfo
  {author} {\bibfnamefont{P.}~\bibnamefont{Wadley}}, \bibinfo {author}
  {\bibfnamefont{J.~A.}\ \bibnamefont{Haigh}}, \bibinfo {author}
  {\bibfnamefont{K.~W.}\ \bibnamefont{Edmonds}}, \bibinfo {author}
  {\bibfnamefont{R.~P.}\ \bibnamefont{Campion}}, \bibinfo {author}
  {\bibfnamefont{A.~W.}\ \bibnamefont{Rushforth}}, \bibinfo {author}
  {\bibfnamefont{B.~L.}\ \bibnamefont{Gallagher}}, \bibinfo {author}
  {\bibfnamefont{C.~T.}\ \bibnamefont{Foxon}}, \bibinfo {author}
  {\bibfnamefont{T.}~\bibnamefont{Jungwirth}}, \bibinfo {author}
  {\bibfnamefont{J.}~\bibnamefont{Wunderlich}}, \bibinfo {author}
  {\bibfnamefont{S.~S.}\ \bibnamefont{Dhesi}}, \bibinfo {author}
  {\bibfnamefont{S.~A.}\ \bibnamefont{Cavill}}, \bibinfo {author}
  {\bibfnamefont{G.}~\bibnamefont{van~der Laan}},\ and\ \bibinfo {author}
  {\bibfnamefont{E.}~\bibnamefont{Arenholz}}}%
  , \bibinfo {year} {2010},\ \bibfield{title}{%
  \enquote{\bibinfo {title} {Exchange bias in a ferromagnetic semiconductor
  induced by a ferromagnetic metal: {Fe/(Ga,Mn)As} bilayer films studied by
  {XMCD} measurements and {SQUID} magnetometry},}\ }%
  \bibfield{journal}{%
  \bibinfo {journal} {Phys. Rev. B}\ }%
  \textbf{\bibinfo {volume} {81}},\ \bibinfo {pages} {104402}%
  \bibAnnoteFile{NoStop}{Olejnik:2010_PRB}%
\bibitem[{\citenamefont{Omiya}\ \emph{et~al.}(2001)\citenamefont{Omiya},
  \citenamefont{Matsukura}, \citenamefont{Shen}, \citenamefont{Ohno},\ and\
  \citenamefont{Ohno}}]{Omiya:2001_PE}%
  \BibitemOpen
  \bibfield{author}{%
  \bibinfo {author} {\bibnamefont{Omiya}, \bibfnamefont{T.}}, \bibinfo {author}
  {\bibfnamefont{F.}~\bibnamefont{Matsukura}}, \bibinfo {author}
  {\bibfnamefont{A.}~\bibnamefont{Shen}}, \bibinfo {author}
  {\bibfnamefont{Y.}~\bibnamefont{Ohno}},\ and\ \bibinfo {author}
  {\bibfnamefont{H.}~\bibnamefont{Ohno}}}%
  , \bibinfo {year} {2001},\ \bibfield{title}{%
  \enquote{\bibinfo {title} {Magnetotransport properties of {(Ga,Mn)As} grown
  on {GaAs (411)A} substrates},}\ }%
  \bibfield{journal}{%
  \bibinfo {journal} {Physica E}\ }%
  \textbf{\bibinfo {volume} {10}},\ \bibinfo {pages} {206}%
  \bibAnnoteFile{NoStop}{Omiya:2001_PE}%
\bibitem[{\citenamefont{Oszwa{\l}dowski}\
  \emph{et~al.}(2006)\citenamefont{Oszwa{\l}dowski}, \citenamefont{Majewski},\
  and\ \citenamefont{Dietl}}]{Oszwaldowski:2006_PRB}%
  \BibitemOpen
  \bibfield{author}{%
  \bibinfo {author} {\bibnamefont{Oszwa{\l}dowski}, \bibfnamefont{R.}},
  \bibinfo {author} {\bibfnamefont{J.~A.}\ \bibnamefont{Majewski}},\ and\
  \bibinfo {author} {\bibfnamefont{T.}~\bibnamefont{Dietl}}}%
  , \bibinfo {year} {2006},\ \bibfield{title}{%
  \enquote{\bibinfo {title} {Influence of band structure effects on domain-wall
  resistance in diluted ferromagnetic semiconductors},}\ }%
  \bibfield{journal}{%
  \bibinfo {journal} {Phys. Rev. B}\ }%
  \textbf{\bibinfo {volume} {74}},\ \bibinfo {pages} {153310}%
  \bibAnnoteFile{NoStop}{Oszwaldowski:2006_PRB}%
\bibitem[{\citenamefont{Ouardi}\ \emph{et~al.}(2013)\citenamefont{Ouardi},
  \citenamefont{Fecher}, \citenamefont{Felser},\ and\
  \citenamefont{K\"ubler}}]{Ouardi:2013_PRL}%
  \BibitemOpen
  \bibfield{author}{%
  \bibinfo {author} {\bibnamefont{Ouardi}, \bibfnamefont{S.}}, \bibinfo
  {author} {\bibfnamefont{G.~H.}\ \bibnamefont{Fecher}}, \bibinfo {author}
  {\bibfnamefont{C.}~\bibnamefont{Felser}},\ and\ \bibinfo {author}
  {\bibfnamefont{J.}~\bibnamefont{K\"ubler}}}%
  , \bibinfo {year} {2013},\ \bibfield{title}{%
  \enquote{\bibinfo {title} {Realization of spin gapless semiconductors: the
  {Heusler} compound {Mn$_2$CoAl}},}\ }%
  \bibfield{journal}{%
  \bibinfo {journal} {Phys. Rev. Lett.}\ }%
  \textbf{\bibinfo {volume} {110}},\ \bibinfo {pages} {100401}%
  \bibAnnoteFile{NoStop}{Ouardi:2013_PRL}%
\bibitem[{\citenamefont{Overby}\ \emph{et~al.}(2008)\citenamefont{Overby},
  \citenamefont{Chernyshov}, \citenamefont{Rokhinson}, \citenamefont{Liu},\
  and\ \citenamefont{Furdyna}}]{Overby:2008_APL}%
  \BibitemOpen
  \bibfield{author}{%
  \bibinfo {author} {\bibnamefont{Overby}, \bibfnamefont{M.}}, \bibinfo
  {author} {\bibfnamefont{A.}~\bibnamefont{Chernyshov}}, \bibinfo {author}
  {\bibfnamefont{L.~P.}\ \bibnamefont{Rokhinson}}, \bibinfo {author}
  {\bibfnamefont{X.}~\bibnamefont{Liu}},\ and\ \bibinfo {author}
  {\bibfnamefont{J.~K.}\ \bibnamefont{Furdyna}}}%
  , \bibinfo {year} {2008},\ \bibfield{title}{%
  \enquote{\bibinfo {title} {{GaMnAs}-based hybrid multiferroic memory
  device},}\ }%
  \bibfield{journal}{%
  \bibinfo {journal} {Appl. Phys. Lett.}\ }%
  \textbf{\bibinfo {volume} {92}},\ \bibinfo {pages} {192501}%
  \bibAnnoteFile{NoStop}{Overby:2008_APL}%
\bibitem[{\citenamefont{Paalanen}\ and\
  \citenamefont{Bhatt}(1991)}]{Paalanen:1991_PB}%
  \BibitemOpen
  \bibfield{author}{%
  \bibinfo {author} {\bibnamefont{Paalanen}, \bibfnamefont{M.~A.}},\ and\
  \bibinfo {author} {\bibfnamefont{R.~N.}\ \bibnamefont{Bhatt}}}%
  , \bibinfo {year} {1991},\ \bibfield{title}{%
  \enquote{\bibinfo {title} {Transport and thermodynamic properties across the
  metal-insulator transition},}\ }%
  \bibfield{journal}{%
  \bibinfo {journal} {Physica B}\ }%
  \textbf{\bibinfo {volume} {169}},\ \bibinfo {pages} {223}%
  \bibAnnoteFile{NoStop}{Paalanen:1991_PB}%
\bibitem[{\citenamefont{Pacuski}\ \emph{et~al.}(2007)\citenamefont{Pacuski},
  \citenamefont{Ferrand}, \citenamefont{Cibert}, \citenamefont{Gaj},
  \citenamefont{Golnik}, \citenamefont{Kossacki}, \citenamefont{Marcet},
  \citenamefont{Sarigiannidou},\ and\
  \citenamefont{Mariette}}]{Pacuski:2007_PRB}%
  \BibitemOpen
  \bibfield{author}{%
  \bibinfo {author} {\bibnamefont{Pacuski}, \bibfnamefont{W.}}, \bibinfo
  {author} {\bibfnamefont{D.}~\bibnamefont{Ferrand}}, \bibinfo {author}
  {\bibfnamefont{J.}~\bibnamefont{Cibert}}, \bibinfo {author}
  {\bibfnamefont{J.~A.}\ \bibnamefont{Gaj}}, \bibinfo {author}
  {\bibfnamefont{A.}~\bibnamefont{Golnik}}, \bibinfo {author}
  {\bibfnamefont{P.}~\bibnamefont{Kossacki}}, \bibinfo {author}
  {\bibfnamefont{S.}~\bibnamefont{Marcet}}, \bibinfo {author}
  {\bibfnamefont{E.}~\bibnamefont{Sarigiannidou}},\ and\ \bibinfo {author}
  {\bibfnamefont{H.}~\bibnamefont{Mariette}}}%
  , \bibinfo {year} {2007},\ \bibfield{title}{%
  \enquote{\bibinfo {title} {Excitonic giant {Zeeman} effect in
  {GaN:Mn}$^{3+}$},}\ }%
  \bibfield{journal}{%
  \bibinfo {journal} {Phys. Rev. B}\ }%
  \textbf{\bibinfo {volume} {76}},\ \bibinfo {pages} {165304}%
  \bibAnnoteFile{NoStop}{Pacuski:2007_PRB}%
\bibitem[{\citenamefont{Pacuski}\ \emph{et~al.}(2011)\citenamefont{Pacuski},
  \citenamefont{Suffczy\'{n}ski}, \citenamefont{Osewski},
  \citenamefont{Kossacki}, \citenamefont{Golnik}, \citenamefont{Gaj},
  \citenamefont{Deparis}, \citenamefont{Morhain}, \citenamefont{Chikoidze},
  \citenamefont{Dumont}, \citenamefont{Ferrand}, \citenamefont{Cibert},\ and\
  \citenamefont{Dietl}}]{Pacuski:2011_PRB}%
  \BibitemOpen
  \bibfield{author}{%
  \bibinfo {author} {\bibnamefont{Pacuski}, \bibfnamefont{W.}}, \bibinfo
  {author} {\bibfnamefont{J.}~\bibnamefont{Suffczy\'{n}ski}}, \bibinfo {author}
  {\bibfnamefont{P.}~\bibnamefont{Osewski}}, \bibinfo {author}
  {\bibfnamefont{P.}~\bibnamefont{Kossacki}}, \bibinfo {author}
  {\bibfnamefont{A.}~\bibnamefont{Golnik}}, \bibinfo {author}
  {\bibfnamefont{J.~A.}\ \bibnamefont{Gaj}}, \bibinfo {author}
  {\bibfnamefont{C.}~\bibnamefont{Deparis}}, \bibinfo {author}
  {\bibfnamefont{C.}~\bibnamefont{Morhain}}, \bibinfo {author}
  {\bibfnamefont{E.}~\bibnamefont{Chikoidze}}, \bibinfo {author}
  {\bibfnamefont{Y.}~\bibnamefont{Dumont}}, \bibinfo {author}
  {\bibfnamefont{D.}~\bibnamefont{Ferrand}}, \bibinfo {author}
  {\bibfnamefont{J.}~\bibnamefont{Cibert}},\ and\ \bibinfo {author}
  {\bibfnamefont{T.}~\bibnamefont{Dietl}}}%
  , \bibinfo {year} {2011},\ \bibfield{title}{%
  \enquote{\bibinfo {title} {Influence of $s$,$p$-$d$ and $s$-$p$ exchange
  couplings on exciton splitting in {Zn$_{1-x}$Mn$_x$O}},}\ }%
  \bibfield{journal}{%
  \bibinfo {journal} {Phys. Rev. B}\ }%
  \textbf{\bibinfo {volume} {84}},\ \bibinfo {pages} {035214}%
  \bibAnnoteFile{NoStop}{Pacuski:2011_PRB}%
\bibitem[{\citenamefont{Paj{\c{a}}czkowska}(1978)}]{Pajaczkowska:1978_PCGC}%
  \BibitemOpen
  \bibfield{author}{%
  \bibinfo {author} {\bibnamefont{Paj{\c{a}}czkowska}, \bibfnamefont{A.}}}%
  , \bibinfo {year} {1978},\ \bibfield{title}{%
  \enquote{\bibinfo {title} {Physicochemical properties and crystal growth of
  {A$^{II}$B$^{VI}$ - MnB$^{VI}$} systems},}\ }%
  \bibfield{journal}{%
  \bibinfo {journal} {Prog. Crystal Growth Charact.}\ }%
  \textbf{\bibinfo {volume} {1}},\ \bibinfo {pages} {289}%
  \bibAnnoteFile{NoStop}{Pajaczkowska:1978_PCGC}%
\bibitem[{\citenamefont{Panguluri}\
  \emph{et~al.}(2005)\citenamefont{Panguluri}, \citenamefont{Ku},
  \citenamefont{Wojtowicz}, \citenamefont{Liu}, \citenamefont{Furdyna},
  \citenamefont{Lyanda-Geller}, \citenamefont{Samarth},\ and\
  \citenamefont{Nadgorny}}]{Panguluri:2005_PRB}%
  \BibitemOpen
  \bibfield{author}{%
  \bibinfo {author} {\bibnamefont{Panguluri}, \bibfnamefont{R.~P.}}, \bibinfo
  {author} {\bibfnamefont{K.~C.}\ \bibnamefont{Ku}}, \bibinfo {author}
  {\bibfnamefont{T.}~\bibnamefont{Wojtowicz}}, \bibinfo {author}
  {\bibfnamefont{X.}~\bibnamefont{Liu}}, \bibinfo {author}
  {\bibfnamefont{J.~K.}\ \bibnamefont{Furdyna}}, \bibinfo {author}
  {\bibfnamefont{Y.~B.}\ \bibnamefont{Lyanda-Geller}}, \bibinfo {author}
  {\bibfnamefont{N.}~\bibnamefont{Samarth}},\ and\ \bibinfo {author}
  {\bibfnamefont{B.}~\bibnamefont{Nadgorny}}}%
  , \bibinfo {year} {2005},\ \bibfield{title}{%
  \enquote{\bibinfo {title} {Andreev reflection and pair-breaking effects at
  the superconductor/magnetic semiconductor interface},}\ }%
  \bibfield{journal}{%
  \bibinfo {journal} {Phys. Rev. B}\ }%
  \textbf{\bibinfo {volume} {72}},\ \bibinfo {pages} {054510}%
  \bibAnnoteFile{NoStop}{Panguluri:2005_PRB}%
\bibitem[{\citenamefont{Panguluri}\
  \emph{et~al.}(2004)\citenamefont{Panguluri}, \citenamefont{Nadgorny},
  \citenamefont{Wojtowicz}, \citenamefont{Lim}, \citenamefont{Liu},\ and\
  \citenamefont{Furdyna}}]{Panguluri:2004_APL}%
  \BibitemOpen
  \bibfield{author}{%
  \bibinfo {author} {\bibnamefont{Panguluri}, \bibfnamefont{R.~P.}}, \bibinfo
  {author} {\bibfnamefont{B.}~\bibnamefont{Nadgorny}}, \bibinfo {author}
  {\bibfnamefont{T.}~\bibnamefont{Wojtowicz}}, \bibinfo {author}
  {\bibfnamefont{W.~L.}\ \bibnamefont{Lim}}, \bibinfo {author}
  {\bibfnamefont{X.}~\bibnamefont{Liu}},\ and\ \bibinfo {author}
  {\bibfnamefont{J.~K.}\ \bibnamefont{Furdyna}}}%
  , \bibinfo {year} {2004},\ \bibfield{title}{%
  \enquote{\bibinfo {title} {Measurement of spin polarization by andreev
  reflection in ferromagnetic {In$_{1-x}$Mn$_{x}$Sb} epilayers},}\ }%
  \bibfield{journal}{%
  \bibinfo {journal} {Appl. Phys. Lett.}\ }%
  \textbf{\bibinfo {volume} {84}},\ \bibinfo {pages} {4947}%
  \bibAnnoteFile{NoStop}{Panguluri:2004_APL}%
\bibitem[{\citenamefont{Pappert}\ \emph{et~al.}(2007)\citenamefont{Pappert},
  \citenamefont{H\"umpfner}, \citenamefont{Gould}, \citenamefont{Wenish},
  \citenamefont{Brunner}, \citenamefont{Schmidt},\ and\
  \citenamefont{Molekamp}}]{Pappert:2007_NP}%
  \BibitemOpen
  \bibfield{author}{%
  \bibinfo {author} {\bibnamefont{Pappert}, \bibfnamefont{K.}}, \bibinfo
  {author} {\bibfnamefont{S.}~\bibnamefont{H\"umpfner}}, \bibinfo {author}
  {\bibfnamefont{C.}~\bibnamefont{Gould}}, \bibinfo {author}
  {\bibfnamefont{J.}~\bibnamefont{Wenish}}, \bibinfo {author}
  {\bibfnamefont{K.}~\bibnamefont{Brunner}}, \bibinfo {author}
  {\bibfnamefont{G.}~\bibnamefont{Schmidt}},\ and\ \bibinfo {author}
  {\bibfnamefont{L.~M.}\ \bibnamefont{Molekamp}}}%
  , \bibinfo {year} {2007},\ \bibfield{title}{%
  \enquote{\bibinfo {title} {A non-volatile-memory device on the basis of
  engineered anisotropies in {(Ga,Mn)As}},}\ }%
  \bibfield{journal}{%
  \bibinfo {journal} {Nat. Phys.}\ }%
  \textbf{\bibinfo {volume} {3}},\ \bibinfo {pages} {573}%
  \bibAnnoteFile{NoStop}{Pappert:2007_NP}%
\bibitem[{\citenamefont{Pappert}\ \emph{et~al.}(2006)\citenamefont{Pappert},
  \citenamefont{Schmidt}, \citenamefont{{H{\"u}mpfner}},
  \citenamefont{{R{\"u}ster}}, \citenamefont{Schott}, \citenamefont{Brunner},
  \citenamefont{Gould}, \citenamefont{Schmidt},\ and\
  \citenamefont{Molenkamp}}]{Pappert:2006_PRL}%
  \BibitemOpen
  \bibfield{author}{%
  \bibinfo {author} {\bibnamefont{Pappert}, \bibfnamefont{K.}}, \bibinfo
  {author} {\bibfnamefont{M.~J.}\ \bibnamefont{Schmidt}}, \bibinfo {author}
  {\bibfnamefont{S.}~\bibnamefont{{H{\"u}mpfner}}}, \bibinfo {author}
  {\bibfnamefont{C.}~\bibnamefont{{R{\"u}ster}}}, \bibinfo {author}
  {\bibfnamefont{G.~M.}\ \bibnamefont{Schott}}, \bibinfo {author}
  {\bibfnamefont{K.}~\bibnamefont{Brunner}}, \bibinfo {author}
  {\bibfnamefont{C.}~\bibnamefont{Gould}}, \bibinfo {author}
  {\bibfnamefont{G.}~\bibnamefont{Schmidt}},\ and\ \bibinfo {author}
  {\bibfnamefont{L.~W.}\ \bibnamefont{Molenkamp}}}%
  , \bibinfo {year} {2006},\ \bibfield{title}{%
  \enquote{\bibinfo {title} {Magnetization-switched metal-insulator transition
  in a {(Ga,Mn)As} tunnel device},}\ }%
  \bibfield{journal}{%
  \bibinfo {journal} {Phys. Rev. Lett.}\ }%
  \textbf{\bibinfo {volume} {97}},\ \bibinfo {pages} {186402}%
  \bibAnnoteFile{NoStop}{Pappert:2006_PRL}%
\bibitem[{\citenamefont{Park}\ \emph{et~al.}(2002)\citenamefont{Park},
  \citenamefont{Hanbicki}, \citenamefont{Erwin}, \citenamefont{Hellberg},
  \citenamefont{Sullivan}, \citenamefont{Mattson}, \citenamefont{Ambrose},
  \citenamefont{Wilson}, \citenamefont{Spanos},\ and\
  \citenamefont{Jonker}}]{Park:2002_S}%
  \BibitemOpen
  \bibfield{author}{%
  \bibinfo {author} {\bibnamefont{Park}, \bibfnamefont{Y.~D.}}, \bibinfo
  {author} {\bibfnamefont{A.~T.}\ \bibnamefont{Hanbicki}}, \bibinfo {author}
  {\bibfnamefont{S.~C.}\ \bibnamefont{Erwin}}, \bibinfo {author}
  {\bibfnamefont{C.~S.}\ \bibnamefont{Hellberg}}, \bibinfo {author}
  {\bibfnamefont{J.~M.}\ \bibnamefont{Sullivan}}, \bibinfo {author}
  {\bibfnamefont{J.~E.}\ \bibnamefont{Mattson}}, \bibinfo {author}
  {\bibfnamefont{T.~F.}\ \bibnamefont{Ambrose}}, \bibinfo {author}
  {\bibfnamefont{A.}~\bibnamefont{Wilson}}, \bibinfo {author}
  {\bibfnamefont{G.}~\bibnamefont{Spanos}},\ and\ \bibinfo {author}
  {\bibfnamefont{B.~T.}\ \bibnamefont{Jonker}}}%
  , \bibinfo {year} {2002},\ \bibfield{title}{%
  \enquote{\bibinfo {title} {A group-{IV} ferromagnetic semiconductor:
  {Mn$_{x}$Ge$_{1-x}$}},}\ }%
  \bibfield{journal}{%
  \bibinfo {journal} {Science}\ }%
  \textbf{\bibinfo {volume} {295}},\ \bibinfo {pages} {651}%
  \bibAnnoteFile{NoStop}{Park:2002_S}%
\bibitem[{\citenamefont{Parmenter}(1973)}]{Parmenter:1973_PRB}%
  \BibitemOpen
  \bibfield{author}{%
  \bibinfo {author} {\bibnamefont{Parmenter}, \bibfnamefont{R.~H.}}}%
  , \bibinfo {year} {1973},\ \bibfield{title}{%
  \enquote{\bibinfo {title} {Effect of orbital degeneracy on the {Anderson}
  model of a localized moment in a metal},}\ }%
  \bibfield{journal}{%
  \bibinfo {journal} {Phys. Rev. B}\ }%
  \textbf{\bibinfo {volume} {8}},\ \bibinfo {pages} {1273}%
  \bibAnnoteFile{NoStop}{Parmenter:1973_PRB}%
\bibitem[{\citenamefont{Pashitskii}\ and\
  \citenamefont{Ryabchenko}(1979)}]{Pashitskii:1979_SPSS}%
  \BibitemOpen
  \bibfield{author}{%
  \bibinfo {author} {\bibnamefont{Pashitskii}, \bibfnamefont{E.~A.}},\ and\
  \bibinfo {author} {\bibfnamefont{S.~M.}\ \bibnamefont{Ryabchenko}}}%
  , \bibinfo {year} {1979},\ \bibfield{title}{%
  \enquote{\bibinfo {title} {Magnetic ordering in semiconductors with magnetic
  impurities},}\ }%
  \bibfield{journal}{%
  \bibinfo {journal} {Sov. Phys. Solid State}\ }%
  \textbf{\bibinfo {volume} {21}},\ \bibinfo {pages} {322}%
  \bibAnnoteFile{NoStop}{Pashitskii:1979_SPSS}%
\bibitem[{\citenamefont{Pearton}\ \emph{et~al.}(2003)\citenamefont{Pearton},
  \citenamefont{Abernathy}, \citenamefont{Overberg}, \citenamefont{Thaler},
  \citenamefont{Norton}, \citenamefont{Theodoropoulou}, \citenamefont{Hebard},
  \citenamefont{Park}, \citenamefont{Ren}, \citenamefont{Kim},\ and\
  \citenamefont{Boatner}}]{Pearton:2003_JAP}%
  \BibitemOpen
  \bibfield{author}{%
  \bibinfo {author} {\bibnamefont{Pearton}, \bibfnamefont{S.~J.}}, \bibinfo
  {author} {\bibfnamefont{C.~R.}\ \bibnamefont{Abernathy}}, \bibinfo {author}
  {\bibfnamefont{M.~E.}\ \bibnamefont{Overberg}}, \bibinfo {author}
  {\bibfnamefont{G.~T.}\ \bibnamefont{Thaler}}, \bibinfo {author}
  {\bibfnamefont{D.~P.}\ \bibnamefont{Norton}}, \bibinfo {author}
  {\bibfnamefont{N.}~\bibnamefont{Theodoropoulou}}, \bibinfo {author}
  {\bibfnamefont{A.~F.}\ \bibnamefont{Hebard}}, \bibinfo {author}
  {\bibfnamefont{Y.~D.}\ \bibnamefont{Park}}, \bibinfo {author}
  {\bibfnamefont{F.}~\bibnamefont{Ren}}, \bibinfo {author}
  {\bibfnamefont{J.}~\bibnamefont{Kim}},\ and\ \bibinfo {author}
  {\bibfnamefont{L.~A.}\ \bibnamefont{Boatner}}}%
  , \bibinfo {year} {2003},\ \bibfield{title}{%
  \enquote{\bibinfo {title} {Wide band gap ferromagnetic semiconductors and
  oxides},}\ }%
  \bibfield{journal}{%
  \bibinfo {journal} {J. Appl. Phys.}\ }%
  \textbf{\bibinfo {volume} {93}},\ \bibinfo {pages} {1}%
  \bibAnnoteFile{NoStop}{Pearton:2003_JAP}%
\bibitem[{\citenamefont{Petukhov}\ \emph{et~al.}(2002)\citenamefont{Petukhov},
  \citenamefont{Chantis},\ and\ \citenamefont{Demchenko}}]{Petukhov:2002_PRL}%
  \BibitemOpen
  \bibfield{author}{%
  \bibinfo {author} {\bibnamefont{Petukhov}, \bibfnamefont{A.~G.}}, \bibinfo
  {author} {\bibfnamefont{A.~N.}\ \bibnamefont{Chantis}},\ and\ \bibinfo
  {author} {\bibfnamefont{D.~O.}\ \bibnamefont{Demchenko}}}%
  , \bibinfo {year} {2002},\ \bibfield{title}{%
  \enquote{\bibinfo {title} {Resonant enhancement of tunneling
  magnetoresistance in double-barrier magnetic heterostructures},}\ }%
  \bibfield{journal}{%
  \bibinfo {journal} {Phys. Rev. Lett.}\ }%
  \textbf{\bibinfo {volume} {89}},\ \bibinfo {pages} {107205}%
  \bibAnnoteFile{NoStop}{Petukhov:2002_PRL}%
\bibitem[{\citenamefont{Petukhov}\ \emph{et~al.}(2003)\citenamefont{Petukhov},
  \citenamefont{Demchenko},\ and\ \citenamefont{Chantis}}]{Petukhov:2003_PRB}%
  \BibitemOpen
  \bibfield{author}{%
  \bibinfo {author} {\bibnamefont{Petukhov}, \bibfnamefont{A.~G.}}, \bibinfo
  {author} {\bibfnamefont{D.~O.}\ \bibnamefont{Demchenko}},\ and\ \bibinfo
  {author} {\bibfnamefont{A.~N.}\ \bibnamefont{Chantis}}}%
  , \bibinfo {year} {2003},\ \bibfield{title}{%
  \enquote{\bibinfo {title} {Electron spin polarization in resonant interband
  tunneling devices},}\ }%
  \bibfield{journal}{%
  \bibinfo {journal} {Phys. Rev. B}\ }%
  \textbf{\bibinfo {volume} {68}},\ \bibinfo {pages} {125332}%
  \bibAnnoteFile{NoStop}{Petukhov:2003_PRB}%
\bibitem[{\citenamefont{Piano}\ \emph{et~al.}(2011)\citenamefont{Piano},
  \citenamefont{Grein}, \citenamefont{Mellor}, \citenamefont{V\'yborn\'y},
  \citenamefont{Campion}, \citenamefont{Wang}, \citenamefont{Eschrig},\ and\
  \citenamefont{Gallagher}}]{Piano:2011_PRB}%
  \BibitemOpen
  \bibfield{author}{%
  \bibinfo {author} {\bibnamefont{Piano}, \bibfnamefont{S.}}, \bibinfo {author}
  {\bibfnamefont{R.}~\bibnamefont{Grein}}, \bibinfo {author}
  {\bibfnamefont{C.~J.}\ \bibnamefont{Mellor}}, \bibinfo {author}
  {\bibfnamefont{K.}~\bibnamefont{V\'yborn\'y}}, \bibinfo {author}
  {\bibfnamefont{R.}~\bibnamefont{Campion}}, \bibinfo {author}
  {\bibfnamefont{M.}~\bibnamefont{Wang}}, \bibinfo {author}
  {\bibfnamefont{M.}~\bibnamefont{Eschrig}},\ and\ \bibinfo {author}
  {\bibfnamefont{B.~L.}\ \bibnamefont{Gallagher}}}%
  , \bibinfo {year} {2011},\ \bibfield{title}{%
  \enquote{\bibinfo {title} {Spin polarization of {(Ga,Mn)As} measured by
  {Andreev} spectroscopy: {The} role of spin-active scattering},}\ }%
  \bibfield{journal}{%
  \bibinfo {journal} {Phys. Rev. B}\ }%
  \textbf{\bibinfo {volume} {83}},\ \bibinfo {pages} {081305}%
  \bibAnnoteFile{NoStop}{Piano:2011_PRB}%
\bibitem[{\citenamefont{Popescu}\ \emph{et~al.}(2006)\citenamefont{Popescu},
  \citenamefont{Yildirim}, \citenamefont{Alvarez}, \citenamefont{Moreo},\ and\
  \citenamefont{Dagotto}}]{Popescu:2006_PRB}%
  \BibitemOpen
  \bibfield{author}{%
  \bibinfo {author} {\bibnamefont{Popescu}, \bibfnamefont{F.}}, \bibinfo
  {author} {\bibfnamefont{Y.}~\bibnamefont{Yildirim}}, \bibinfo {author}
  {\bibfnamefont{G.}~\bibnamefont{Alvarez}}, \bibinfo {author}
  {\bibfnamefont{A.}~\bibnamefont{Moreo}},\ and\ \bibinfo {author}
  {\bibfnamefont{E.}~\bibnamefont{Dagotto}}}%
  , \bibinfo {year} {2006},\ \bibfield{title}{%
  \enquote{\bibinfo {title} {Critical temperatures of the two-band model for
  diluted magnetic semiconductors},}\ }%
  \bibfield{journal}{%
  \bibinfo {journal} {Phys. Rev. B}\ }%
  \textbf{\bibinfo {volume} {73}},\ \bibinfo {pages} {075206}%
  \bibAnnoteFile{NoStop}{Popescu:2006_PRB}%
\bibitem[{\citenamefont{Potashnik}\
  \emph{et~al.}(2001)\citenamefont{Potashnik}, \citenamefont{Ku},
  \citenamefont{Chun}, \citenamefont{Berry}, \citenamefont{Samarth},\ and\
  \citenamefont{Schiffer}}]{Potashnik:2001_APL}%
  \BibitemOpen
  \bibfield{author}{%
  \bibinfo {author} {\bibnamefont{Potashnik}, \bibfnamefont{S.~J.}}, \bibinfo
  {author} {\bibfnamefont{K.~C.}\ \bibnamefont{Ku}}, \bibinfo {author}
  {\bibfnamefont{S.~H.}\ \bibnamefont{Chun}}, \bibinfo {author}
  {\bibfnamefont{J.~J.}\ \bibnamefont{Berry}}, \bibinfo {author}
  {\bibfnamefont{N.}~\bibnamefont{Samarth}},\ and\ \bibinfo {author}
  {\bibfnamefont{P.}~\bibnamefont{Schiffer}}}%
  , \bibinfo {year} {2001},\ \bibfield{title}{%
  \enquote{\bibinfo {title} {Effects of annealing time on defect-controlled
  ferromagnetism in {Ga$_{1-x}$Mn$_{x}$As}},}\ }%
  \bibfield{journal}{%
  \bibinfo {journal} {Appl. Phys. Lett.}\ }%
  \textbf{\bibinfo {volume} {79}},\ \bibinfo {pages} {1495}%
  \bibAnnoteFile{NoStop}{Potashnik:2001_APL}%
\bibitem[{\citenamefont{Potashnik}\
  \emph{et~al.}(2002)\citenamefont{Potashnik}, \citenamefont{Ku},
  \citenamefont{Mahendiran}, \citenamefont{Chun}, \citenamefont{Wang},
  \citenamefont{Samarth},\ and\ \citenamefont{Schiffer}}]{Potashnik:2002_PRB}%
  \BibitemOpen
  \bibfield{author}{%
  \bibinfo {author} {\bibnamefont{Potashnik}, \bibfnamefont{S.~J.}}, \bibinfo
  {author} {\bibfnamefont{K.~C.}\ \bibnamefont{Ku}}, \bibinfo {author}
  {\bibfnamefont{R.}~\bibnamefont{Mahendiran}}, \bibinfo {author}
  {\bibfnamefont{S.~H.}\ \bibnamefont{Chun}}, \bibinfo {author}
  {\bibfnamefont{R.~F.}\ \bibnamefont{Wang}}, \bibinfo {author}
  {\bibfnamefont{N.}~\bibnamefont{Samarth}},\ and\ \bibinfo {author}
  {\bibfnamefont{P.}~\bibnamefont{Schiffer}}}%
  , \bibinfo {year} {2002},\ \bibfield{title}{%
  \enquote{\bibinfo {title} {Saturated ferromagnetism and magnetization deficit
  in in optimally annealed {(Ga,Mn)As} epilayers},}\ }%
  \bibfield{journal}{%
  \bibinfo {journal} {Phys. Rev. B}\ }%
  \textbf{\bibinfo {volume} {66}},\ \bibinfo {pages} {012408}%
  \bibAnnoteFile{NoStop}{Potashnik:2002_PRB}%
\bibitem[{\citenamefont{Proselkov}\
  \emph{et~al.}(2012)\citenamefont{Proselkov}, \citenamefont{Sztenkiel},
  \citenamefont{Stefanowicz}, \citenamefont{Aleszkiewicz},
  \citenamefont{Sadowski}, \citenamefont{Dietl},\ and\
  \citenamefont{Sawicki}}]{Proselkov:2012_APL}%
  \BibitemOpen
  \bibfield{author}{%
  \bibinfo {author} {\bibnamefont{Proselkov}, \bibfnamefont{O.}}, \bibinfo
  {author} {\bibfnamefont{D.}~\bibnamefont{Sztenkiel}}, \bibinfo {author}
  {\bibfnamefont{W.}~\bibnamefont{Stefanowicz}}, \bibinfo {author}
  {\bibfnamefont{M.}~\bibnamefont{Aleszkiewicz}}, \bibinfo {author}
  {\bibfnamefont{J.}~\bibnamefont{Sadowski}}, \bibinfo {author}
  {\bibfnamefont{T.}~\bibnamefont{Dietl}},\ and\ \bibinfo {author}
  {\bibfnamefont{M.}~\bibnamefont{Sawicki}}}%
  , \bibinfo {year} {2012},\ \bibfield{title}{%
  \enquote{\bibinfo {title} {Thickness dependent magnetic properties of
  {(Ga,Mn)As} ultrathin films},}\ }%
  \bibfield{journal}{%
  \bibinfo {journal} {Appl. Phys. Lett.}\ }%
  \textbf{\bibinfo {volume} {100}},\ \bibinfo {pages} {262405}%
  \bibAnnoteFile{NoStop}{Proselkov:2012_APL}%
\bibitem[{\citenamefont{Pu}\ \emph{et~al.}(2008)\citenamefont{Pu},
  \citenamefont{Chiba}, \citenamefont{Matsukura}, \citenamefont{Ohno},\ and\
  \citenamefont{Shi}}]{Pu:2008_PRL}%
  \BibitemOpen
  \bibfield{author}{%
  \bibinfo {author} {\bibnamefont{Pu}, \bibfnamefont{Yong}}, \bibinfo {author}
  {\bibfnamefont{Daichi}\ \bibnamefont{Chiba}}, \bibinfo {author}
  {\bibfnamefont{Fumihiro}\ \bibnamefont{Matsukura}}, \bibinfo {author}
  {\bibfnamefont{Hideo}\ \bibnamefont{Ohno}},\ and\ \bibinfo {author}
  {\bibfnamefont{Jing}\ \bibnamefont{Shi}}}%
  , \bibinfo {year} {2008},\ \bibfield{title}{%
  \enquote{\bibinfo {title} {Mott relation for anomalous {Hall} and {Nernst}
  effects in {Ga$_{1-x}$Mn$_x$As} ferromagnetic semiconductors},}\ }%
  \bibfield{journal}{%
  \bibinfo {journal} {Phys. Rev. Lett.}\ }%
  \textbf{\bibinfo {volume} {101}},\ \bibinfo {pages} {117208}%
  \bibAnnoteFile{NoStop}{Pu:2008_PRL}%
\bibitem[{\citenamefont{Pu}\ \emph{et~al.}(2006)\citenamefont{Pu},
  \citenamefont{Johnston-Halperin}, \citenamefont{Awschalom},\ and\
  \citenamefont{Shi}}]{Pu:2006_PRL}%
  \BibitemOpen
  \bibfield{author}{%
  \bibinfo {author} {\bibnamefont{Pu}, \bibfnamefont{Yong}}, \bibinfo {author}
  {\bibfnamefont{E.}~\bibnamefont{Johnston-Halperin}}, \bibinfo {author}
  {\bibfnamefont{D.~D.}\ \bibnamefont{Awschalom}},\ and\ \bibinfo {author}
  {\bibfnamefont{Jing}\ \bibnamefont{Shi}}}%
  , \bibinfo {year} {2006},\ \bibfield{title}{%
  \enquote{\bibinfo {title} {Anisotropic thermopower and planar nernst effect
  in {Ga$_{1-x}$Mn$_x$As} ferromagnetic semiconductors},}\ }%
  \bibfield{journal}{%
  \bibinfo {journal} {Phys. Rev. Lett.}\ }%
  \textbf{\bibinfo {volume} {97}},\ \bibinfo {pages} {036601}%
  \bibAnnoteFile{NoStop}{Pu:2006_PRL}%
\bibitem[{\citenamefont{Qazzaz}\ \emph{et~al.}(1995)\citenamefont{Qazzaz},
  \citenamefont{Yang}, \citenamefont{Kin}, \citenamefont{Montes},
  \citenamefont{Luo},\ and\ \citenamefont{Furdyna}}]{Qazzaz:1995_SSC}%
  \BibitemOpen
  \bibfield{author}{%
  \bibinfo {author} {\bibnamefont{Qazzaz}, \bibfnamefont{M.}}, \bibinfo
  {author} {\bibfnamefont{G.}~\bibnamefont{Yang}}, \bibinfo {author}
  {\bibfnamefont{S.H.}\ \bibnamefont{Kin}}, \bibinfo {author}
  {\bibfnamefont{L.}~\bibnamefont{Montes}}, \bibinfo {author}
  {\bibfnamefont{H.}~\bibnamefont{Luo}},\ and\ \bibinfo {author}
  {\bibfnamefont{J.K.}\ \bibnamefont{Furdyna}}}%
  , \bibinfo {year} {1995},\ \bibfield{title}{%
  \enquote{\bibinfo {title} {Electron paramagnetic resonance of {Mn$^{2+}$} in
  strained-layer semiconductor superlattices},}\ }%
  \bibfield{journal}{%
  \bibinfo {journal} {Solid State Commun.}\ }%
  \textbf{\bibinfo {volume} {96}},\ \bibinfo {pages} {405}%
  \bibAnnoteFile{NoStop}{Qazzaz:1995_SSC}%
\bibitem[{\citenamefont{Qi}\ \emph{et~al.}(2009)\citenamefont{Qi},
  \citenamefont{Sun}, \citenamefont{Weinert},\ and\
  \citenamefont{Li}}]{Qi:2009_PRB}%
  \BibitemOpen
  \bibfield{author}{%
  \bibinfo {author} {\bibnamefont{Qi}, \bibfnamefont{Y.}}, \bibinfo {author}
  {\bibfnamefont{G.~F.}\ \bibnamefont{Sun}}, \bibinfo {author}
  {\bibfnamefont{M.}~\bibnamefont{Weinert}},\ and\ \bibinfo {author}
  {\bibfnamefont{L.}~\bibnamefont{Li}}}%
  , \bibinfo {year} {2009},\ \bibfield{title}{%
  \enquote{\bibinfo {title} {Electronic structures of {Mn}-induced phases on
  {GaN(0001)}},}\ }%
  \bibfield{journal}{%
  \bibinfo {journal} {Phys. Rev. B}\ }%
  \textbf{\bibinfo {volume} {80}},\ \bibinfo {pages} {235323}%
  \bibAnnoteFile{NoStop}{Qi:2009_PRB}%
\bibitem[{\citenamefont{Rader}\ \emph{et~al.}(2004)\citenamefont{Rader},
  \citenamefont{Pampuch}, \citenamefont{Shikin}, \citenamefont{Gudat},
  \citenamefont{Okabayashi}, \citenamefont{Mizokawa}, \citenamefont{Fujimori},
  \citenamefont{Hayashi}, \citenamefont{Tanaka}, \citenamefont{Tanaka},\ and\
  \citenamefont{Kimura}}]{Rader:2004_PRB}%
  \BibitemOpen
  \bibfield{author}{%
  \bibinfo {author} {\bibnamefont{Rader}, \bibfnamefont{O.}}, \bibinfo {author}
  {\bibfnamefont{C.}~\bibnamefont{Pampuch}}, \bibinfo {author}
  {\bibfnamefont{A.~M.}\ \bibnamefont{Shikin}}, \bibinfo {author}
  {\bibfnamefont{W.}~\bibnamefont{Gudat}}, \bibinfo {author}
  {\bibfnamefont{J.}~\bibnamefont{Okabayashi}}, \bibinfo {author}
  {\bibfnamefont{T.}~\bibnamefont{Mizokawa}}, \bibinfo {author}
  {\bibfnamefont{A.}~\bibnamefont{Fujimori}}, \bibinfo {author}
  {\bibfnamefont{T.}~\bibnamefont{Hayashi}}, \bibinfo {author}
  {\bibfnamefont{M.}~\bibnamefont{Tanaka}}, \bibinfo {author}
  {\bibfnamefont{A.}~\bibnamefont{Tanaka}},\ and\ \bibinfo {author}
  {\bibfnamefont{A.}~\bibnamefont{Kimura}}}%
  , \bibinfo {year} {2004},\ \bibfield{title}{%
  \enquote{\bibinfo {title} {Resonant photoemission of {Ga$_{1-x}$Mn$_{x}$As}
  at the {Mn L} edge},}\ }%
  \bibfield{journal}{%
  \bibinfo {journal} {Phys. Rev. B}\ }%
  \textbf{\bibinfo {volume} {69}},\ \bibinfo {pages} {075202}%
  \bibAnnoteFile{NoStop}{Rader:2004_PRB}%
\bibitem[{\citenamefont{Rader}\ \emph{et~al.}(2009)\citenamefont{Rader},
  \citenamefont{Valencia}, \citenamefont{Gudat}, \citenamefont{Edmonds},
  \citenamefont{Campion}, \citenamefont{Gallagher}, \citenamefont{Foxon},
  \citenamefont{Emtsev},\ and\ \citenamefont{Seyller}}]{Rader:2009_PSSB}%
  \BibitemOpen
  \bibfield{author}{%
  \bibinfo {author} {\bibnamefont{Rader}, \bibfnamefont{O.}}, \bibinfo {author}
  {\bibfnamefont{S.}~\bibnamefont{Valencia}}, \bibinfo {author}
  {\bibfnamefont{W.}~\bibnamefont{Gudat}}, \bibinfo {author}
  {\bibfnamefont{K.~W.}\ \bibnamefont{Edmonds}}, \bibinfo {author}
  {\bibfnamefont{R.~P.}\ \bibnamefont{Campion}}, \bibinfo {author}
  {\bibfnamefont{B.~L.}\ \bibnamefont{Gallagher}}, \bibinfo {author}
  {\bibfnamefont{C.~T.}\ \bibnamefont{Foxon}}, \bibinfo {author}
  {\bibfnamefont{K.~V.}\ \bibnamefont{Emtsev}},\ and\ \bibinfo {author}
  {\bibfnamefont{Th.}\ \bibnamefont{Seyller}}}%
  , \bibinfo {year} {2009},\ \bibfield{title}{%
  \enquote{\bibinfo {title} {Photoemission of {Ga$_{1-x}$Mn$_x$As} with high
  {Curie} temperature and transformation into {MnAs} of zincblende
  structure},}\ }%
  \bibfield{journal}{%
  \bibinfo {journal} {Phys. Status Solidi B}\ }%
  \textbf{\bibinfo {volume} {246}},\ \bibinfo {pages} {1435}%
  \bibAnnoteFile{NoStop}{Rader:2009_PSSB}%
\bibitem[{\citenamefont{Rammal}\ and\
  \citenamefont{Souletie}(1982)}]{Rammal:1982_B}%
  \BibitemOpen
  \bibfield{author}{%
  \bibinfo {author} {\bibnamefont{Rammal}, \bibfnamefont{R.}},\ and\ \bibinfo
  {author} {\bibfnamefont{J.}~\bibnamefont{Souletie}}}%
  , \bibinfo {year} {1982},\ \emph{\bibinfo {title} {{Magnetism of Metals and
  Alloys}}}\ (\bibinfo {publisher} {North-Holland})%
  \bibAnnoteFile{NoStop}{Rammal:1982_B}%
\bibitem[{\citenamefont{Richardella}\
  \emph{et~al.}(2010)\citenamefont{Richardella}, \citenamefont{Roushan},
  \citenamefont{Mack}, \citenamefont{Zhou}, \citenamefont{Huse},
  \citenamefont{Awschalom},\ and\ \citenamefont{Yazdani}}]{Richardella:2010_S}%
  \BibitemOpen
  \bibfield{author}{%
  \bibinfo {author} {\bibnamefont{Richardella}, \bibfnamefont{A.}}, \bibinfo
  {author} {\bibfnamefont{P.}~\bibnamefont{Roushan}}, \bibinfo {author}
  {\bibfnamefont{S.}~\bibnamefont{Mack}}, \bibinfo {author}
  {\bibfnamefont{B.}~\bibnamefont{Zhou}}, \bibinfo {author}
  {\bibfnamefont{D.~A.}\ \bibnamefont{Huse}}, \bibinfo {author}
  {\bibfnamefont{D.~D.}\ \bibnamefont{Awschalom}},\ and\ \bibinfo {author}
  {\bibfnamefont{A.}~\bibnamefont{Yazdani}}}%
  , \bibinfo {year} {2010},\ \bibfield{title}{%
  \enquote{\bibinfo {title} {Visualizing critical correlations near the
  metal-insulator transition in {Ga$_{1-x}$Mn$_{x}$As}},}\ }%
  \bibfield{journal}{%
  \bibinfo {journal} {Science}\ }%
  \textbf{\bibinfo {volume} {327}},\ \bibinfo {pages} {665}%
  \bibAnnoteFile{NoStop}{Richardella:2010_S}%
\bibitem[{\citenamefont{Riester}\ \emph{et~al.}(2009)\citenamefont{Riester},
  \citenamefont{Stolichnov}, \citenamefont{Trodahl}, \citenamefont{Setter},
  \citenamefont{Rushforth}, \citenamefont{Edmonds}, \citenamefont{Campion},
  \citenamefont{Foxon}, \citenamefont{Gallagher},\ and\
  \citenamefont{Jungwirth}}]{Riester:2009_APL}%
  \BibitemOpen
  \bibfield{author}{%
  \bibinfo {author} {\bibnamefont{Riester}, \bibfnamefont{S.~W.~E.}}, \bibinfo
  {author} {\bibfnamefont{I.}~\bibnamefont{Stolichnov}}, \bibinfo {author}
  {\bibfnamefont{H.~J.}\ \bibnamefont{Trodahl}}, \bibinfo {author}
  {\bibfnamefont{N.}~\bibnamefont{Setter}}, \bibinfo {author}
  {\bibfnamefont{A.~W.}\ \bibnamefont{Rushforth}}, \bibinfo {author}
  {\bibfnamefont{K.~W.}\ \bibnamefont{Edmonds}}, \bibinfo {author}
  {\bibfnamefont{R.~P.}\ \bibnamefont{Campion}}, \bibinfo {author}
  {\bibfnamefont{C.~T.}\ \bibnamefont{Foxon}}, \bibinfo {author}
  {\bibfnamefont{B.~L.}\ \bibnamefont{Gallagher}},\ and\ \bibinfo {author}
  {\bibfnamefont{T.}~\bibnamefont{Jungwirth}}}%
  , \bibinfo {year} {2009},\ \bibfield{title}{%
  \enquote{\bibinfo {title} {Toward a low-voltage multiferroic transistor:
  {Magnetic} {(Ga,Mn)As} under ferroelectric control},}\ }%
  \bibfield{journal}{%
  \bibinfo {journal} {Appl. Phys. Lett.}\ }%
  \textbf{\bibinfo {volume} {94}},\ \bibinfo {pages} {063504}%
  \bibAnnoteFile{NoStop}{Riester:2009_APL}%
\bibitem[{\citenamefont{Roberts}\ \emph{et~al.}(2007)\citenamefont{Roberts},
  \citenamefont{Crampin},\ and\ \citenamefont{Bending}}]{Roberts:2007_PRB}%
  \BibitemOpen
  \bibfield{author}{%
  \bibinfo {author} {\bibnamefont{Roberts}, \bibfnamefont{H.~G.}}, \bibinfo
  {author} {\bibfnamefont{S.}~\bibnamefont{Crampin}},\ and\ \bibinfo {author}
  {\bibfnamefont{S.~J.}\ \bibnamefont{Bending}}}%
  , \bibinfo {year} {2007},\ \bibfield{title}{%
  \enquote{\bibinfo {title} {Extrinsic anisotropic magnetoresistance
  contribution to measured domain wall resistances of in-plane magnetized
  {(Ga,Mn)As}},}\ }%
  \bibfield{journal}{%
  \bibinfo {journal} {Phys. Rev. B}\ }%
  \textbf{\bibinfo {volume} {76}},\ \bibinfo {pages} {035323}%
  \bibAnnoteFile{NoStop}{Roberts:2007_PRB}%
\bibitem[{\citenamefont{Rudolph}\ \emph{et~al.}(2009)\citenamefont{Rudolph},
  \citenamefont{Soda}, \citenamefont{Kiessling}, \citenamefont{Wojtowicz},
  \citenamefont{Schuh}, \citenamefont{Wegscheider}, \citenamefont{Zweck},
  \citenamefont{Back}, ,\ and\ \citenamefont{Reiger}}]{Rudolph:2009_NL}%
  \BibitemOpen
  \bibfield{author}{%
  \bibinfo {author} {\bibnamefont{Rudolph}, \bibfnamefont{A.}}, \bibinfo
  {author} {\bibfnamefont{M.}~\bibnamefont{Soda}}, \bibinfo {author}
  {\bibfnamefont{M.}~\bibnamefont{Kiessling}}, \bibinfo {author}
  {\bibfnamefont{T.}~\bibnamefont{Wojtowicz}}, \bibinfo {author}
  {\bibfnamefont{D.}~\bibnamefont{Schuh}}, \bibinfo {author}
  {\bibfnamefont{W.}~\bibnamefont{Wegscheider}}, \bibinfo {author}
  {\bibfnamefont{J.}~\bibnamefont{Zweck}}, \bibinfo {author}
  {\bibfnamefont{Ch.}\ \bibnamefont{Back}}, ,\ and\ \bibinfo {author}
  {\bibfnamefont{E.}~\bibnamefont{Reiger}}}%
  , \bibinfo {year} {2009},\ \bibfield{title}{%
  \enquote{\bibinfo {title} {Ferromagnetic {GaAs/GaMnAs} core-shell nanowires
  grown by molecular beam epitaxy},}\ }%
  \bibfield{journal}{%
  \bibinfo {journal} {Nano Lett.}\ }%
  \textbf{\bibinfo {volume} {9}},\ \bibinfo {pages} {3860}%
  \bibAnnoteFile{NoStop}{Rudolph:2009_NL}%
\bibitem[{\citenamefont{Rushforth}\
  \emph{et~al.}(2008)\citenamefont{Rushforth}, \citenamefont{Farley},
  \citenamefont{Campion}, \citenamefont{Edmonds}, \citenamefont{Staddon},
  \citenamefont{Foxon}, \citenamefont{Gallagher},\ and\
  \citenamefont{Yu}}]{Rushforth:2008_PRB}%
  \BibitemOpen
  \bibfield{author}{%
  \bibinfo {author} {\bibnamefont{Rushforth}, \bibfnamefont{A.~W.}}, \bibinfo
  {author} {\bibfnamefont{N.~R.~S.}\ \bibnamefont{Farley}}, \bibinfo {author}
  {\bibfnamefont{R.~P.}\ \bibnamefont{Campion}}, \bibinfo {author}
  {\bibfnamefont{K.~W.}\ \bibnamefont{Edmonds}}, \bibinfo {author}
  {\bibfnamefont{C.~R.}\ \bibnamefont{Staddon}}, \bibinfo {author}
  {\bibfnamefont{C.~T.}\ \bibnamefont{Foxon}}, \bibinfo {author}
  {\bibfnamefont{B.~L.}\ \bibnamefont{Gallagher}},\ and\ \bibinfo {author}
  {\bibfnamefont{K.~M.}\ \bibnamefont{Yu}}}%
  , \bibinfo {year} {2008},\ \bibfield{title}{%
  \enquote{\bibinfo {title} {Compositional dependence of ferromagnetism in
  {(Al,Ga,Mn)As} magnetic semiconductors},}\ }%
  \bibfield{journal}{%
  \bibinfo {journal} {Phys. Rev. B}\ }%
  \textbf{\bibinfo {volume} {78}},\ \bibinfo {pages} {085209}%
  \bibAnnoteFile{NoStop}{Rushforth:2008_PRB}%
\bibitem[{\citenamefont{R{\"u}ster}\
  \emph{et~al.}(2005)\citenamefont{R{\"u}ster}, \citenamefont{Gould},
  \citenamefont{Jungwirth}, \citenamefont{Sinova}, \citenamefont{Schott},
  \citenamefont{Giraud}, \citenamefont{Brunner}, \citenamefont{Schmidt},\ and\
  \citenamefont{Molenkamp}}]{Ruster:2005_PRL}%
  \BibitemOpen
  \bibfield{author}{%
  \bibinfo {author} {\bibnamefont{R{\"u}ster}, \bibfnamefont{C.}}, \bibinfo
  {author} {\bibfnamefont{C.}~\bibnamefont{Gould}}, \bibinfo {author}
  {\bibfnamefont{T.}~\bibnamefont{Jungwirth}}, \bibinfo {author}
  {\bibfnamefont{J.}~\bibnamefont{Sinova}}, \bibinfo {author}
  {\bibfnamefont{G.~M.}\ \bibnamefont{Schott}}, \bibinfo {author}
  {\bibfnamefont{R.}~\bibnamefont{Giraud}}, \bibinfo {author}
  {\bibfnamefont{K.}~\bibnamefont{Brunner}}, \bibinfo {author}
  {\bibfnamefont{G.}~\bibnamefont{Schmidt}},\ and\ \bibinfo {author}
  {\bibfnamefont{L.~W.}\ \bibnamefont{Molenkamp}}}%
  , \bibinfo {year} {2005},\ \bibfield{title}{%
  \enquote{\bibinfo {title} {Very large tunneling anisotropic magnetoresistance
  of a {(Ga,Mn)As/GaAs/(Ga,Mn)As} stack},}\ }%
  \bibfield{journal}{%
  \bibinfo {journal} {Phys. Rev. Lett.}\ }%
  \textbf{\bibinfo {volume} {94}},\ \bibinfo {pages} {027203}%
  \bibAnnoteFile{NoStop}{Ruster:2005_PRL}%
\bibitem[{\citenamefont{Sadowski}\ \emph{et~al.}(2007)\citenamefont{Sadowski},
  \citenamefont{D{\l}u{\.z}ewski}, \citenamefont{Kret}, \citenamefont{Janik},
  \citenamefont{{\L}usakowska}, \citenamefont{Ka{\'n}ski},
  \citenamefont{Presz}, \citenamefont{Terki}, \citenamefont{Charar},\ and\
  \citenamefont{Tang}}]{Sadowski:2007_NL}%
  \BibitemOpen
  \bibfield{author}{%
  \bibinfo {author} {\bibnamefont{Sadowski}, \bibfnamefont{J.}}, \bibinfo
  {author} {\bibfnamefont{P.}~\bibnamefont{D{\l}u{\.z}ewski}}, \bibinfo
  {author} {\bibfnamefont{S.}~\bibnamefont{Kret}}, \bibinfo {author}
  {\bibfnamefont{E.}~\bibnamefont{Janik}}, \bibinfo {author}
  {\bibfnamefont{E.}~\bibnamefont{{\L}usakowska}}, \bibinfo {author}
  {\bibfnamefont{J.}~\bibnamefont{Ka{\'n}ski}}, \bibinfo {author}
  {\bibfnamefont{A.}~\bibnamefont{Presz}}, \bibinfo {author}
  {\bibfnamefont{F.}~\bibnamefont{Terki}}, \bibinfo {author}
  {\bibfnamefont{S.}~\bibnamefont{Charar}},\ and\ \bibinfo {author}
  {\bibfnamefont{D.}~\bibnamefont{Tang}}}%
  , \bibinfo {year} {2007},\ \bibfield{title}{%
  \enquote{\bibinfo {title} {{GaAs:Mn} nanowires grown by molecular beam
  epitaxy of {(Ga,Mn)As} at {MnAs} segregation conditions},}\ }%
  \bibfield{journal}{%
  \bibinfo {journal} {Nano Lett.}\ }%
  \textbf{\bibinfo {volume} {7}},\ \bibinfo {pages} {2724}%
  \bibAnnoteFile{NoStop}{Sadowski:2007_NL}%
\bibitem[{\citenamefont{Sadowski}\ and\
  \citenamefont{Domagala}(2004)}]{Sadowski:2004_PRB}%
  \BibitemOpen
  \bibfield{author}{%
  \bibinfo {author} {\bibnamefont{Sadowski}, \bibfnamefont{J.}},\ and\ \bibinfo
  {author} {\bibfnamefont{J.~Z.}\ \bibnamefont{Domagala}}}%
  , \bibinfo {year} {2004},\ \bibfield{title}{%
  \enquote{\bibinfo {title} {Influence of defects on lattice constant of
  {GaMnAs}},}\ }%
  \bibfield{journal}{%
  \bibinfo {journal} {Phys. Rev. B}\ }%
  \textbf{\bibinfo {volume} {69}},\ \bibinfo {pages} {075206}%
  \bibAnnoteFile{NoStop}{Sadowski:2004_PRB}%
\bibitem[{\citenamefont{Sadowski}\ \emph{et~al.}(2002)\citenamefont{Sadowski},
  \citenamefont{Mathieu}, \citenamefont{Svedlindh}, \citenamefont{Karlsteen},
  \citenamefont{Kanski}, \citenamefont{Fu}, \citenamefont{Domagala},
  \citenamefont{Szuszkiewicz}, \citenamefont{Hennion}, \citenamefont{Maude},
  \citenamefont{Airey},\ and\ \citenamefont{Hill}}]{Sadowski:2002_TSF}%
  \BibitemOpen
  \bibfield{author}{%
  \bibinfo {author} {\bibnamefont{Sadowski}, \bibfnamefont{J.}}, \bibinfo
  {author} {\bibfnamefont{R.}~\bibnamefont{Mathieu}}, \bibinfo {author}
  {\bibfnamefont{P.}~\bibnamefont{Svedlindh}}, \bibinfo {author}
  {\bibfnamefont{M.}~\bibnamefont{Karlsteen}}, \bibinfo {author}
  {\bibfnamefont{J.}~\bibnamefont{Kanski}}, \bibinfo {author}
  {\bibfnamefont{Y.}~\bibnamefont{Fu}}, \bibinfo {author}
  {\bibfnamefont{J.~T.}\ \bibnamefont{Domagala}}, \bibinfo {author}
  {\bibfnamefont{W.}~\bibnamefont{Szuszkiewicz}}, \bibinfo {author}
  {\bibfnamefont{B.}~\bibnamefont{Hennion}}, \bibinfo {author}
  {\bibfnamefont{D.~K.}\ \bibnamefont{Maude}}, \bibinfo {author}
  {\bibfnamefont{R.}~\bibnamefont{Airey}},\ and\ \bibinfo {author}
  {\bibfnamefont{G.}~\bibnamefont{Hill}}}%
  , \bibinfo {year} {2002},\ \bibfield{title}{%
  \enquote{\bibinfo {title} {Ferromagnetic {GaMnAs/GaAs} superlattices ---
  {MBE} growth and magnetic properties},}\ }%
  \bibfield{journal}{%
  \bibinfo {journal} {Thin Solid Films}\ }%
  \textbf{\bibinfo {volume} {412}},\ \bibinfo {pages} {122}%
  \bibAnnoteFile{NoStop}{Sadowski:2002_TSF}%
\bibitem[{\citenamefont{Saito}\ \emph{et~al.}(2003)\citenamefont{Saito},
  \citenamefont{H.}, \citenamefont{Zayets}, \citenamefont{V.},
  \citenamefont{Yamagata}, \citenamefont{S.},\ and\
  \citenamefont{A}}]{Saito:2003_PRL}%
  \BibitemOpen
  \bibfield{author}{%
  \bibinfo {author} {\bibnamefont{Saito}}, \bibinfo {author}
  {\bibnamefont{H.}}, \bibinfo {author} {\bibnamefont{Zayets}}, \bibinfo
  {author} {\bibnamefont{V.}}, \bibinfo {author} {\bibnamefont{Yamagata}},
  \bibinfo {author} {\bibnamefont{S.}},\ and\ \bibinfo {author}
  {\bibfnamefont{Ando}\ \bibnamefont{A}}}%
  , \bibinfo {year} {2003},\ \bibfield{title}{%
  \enquote{\bibinfo {title} {Room-temperature ferromagnetism in a {II-VI}
  diluted magnetic semiconductor {Zn$_{1-x}$Cr$_{x}$Te}},}\ }%
  \bibfield{journal}{%
  \bibinfo {journal} {Phys. Rev. Lett.}\ }%
  \textbf{\bibinfo {volume} {90}},\ \bibinfo {pages} {207202}%
  \bibAnnoteFile{NoStop}{Saito:2003_PRL}%
\bibitem[{\citenamefont{Saito}\ \emph{et~al.}(2005)\citenamefont{Saito},
  \citenamefont{Yuasa},\ and\ \citenamefont{Ando}}]{Saito:2005_PRL}%
  \BibitemOpen
  \bibfield{author}{%
  \bibinfo {author} {\bibnamefont{Saito}, \bibfnamefont{H.}}, \bibinfo {author}
  {\bibfnamefont{S.}~\bibnamefont{Yuasa}},\ and\ \bibinfo {author}
  {\bibfnamefont{K.}~\bibnamefont{Ando}}}%
  , \bibinfo {year} {2005},\ \bibfield{title}{%
  \enquote{\bibinfo {title} {Origin of the tunnel anisotropic magnetoresistance
  in {Ga$_{1-x}$Mn$_{x}$As/ZnSe/Ga$_{1-x}$Mn$_{x}$As} magnetic tunnel junctions
  of {II-VI/III-V} heterostructures},}\ }%
  \bibfield{journal}{%
  \bibinfo {journal} {Phys. Rev. Lett.}\ }%
  \textbf{\bibinfo {volume} {95}},\ \bibinfo {pages} {086604}%
  \bibAnnoteFile{NoStop}{Saito:2005_PRL}%
\bibitem[{\citenamefont{Samarth}(2012)}]{Samarth:2012_NM}%
  \BibitemOpen
  \bibfield{author}{%
  \bibinfo {author} {\bibnamefont{Samarth}, \bibfnamefont{N.}}}%
  , \bibinfo {year} {2012},\ \bibfield{title}{%
  \enquote{\bibinfo {title} {Battle of the bands},}\ }%
  \bibfield{journal}{%
  \bibinfo {journal} {Nat. Mater.}\ }%
  \textbf{\bibinfo {volume} {11}},\ \bibinfo {pages} {360}%
  \bibAnnoteFile{NoStop}{Samarth:2012_NM}%
\bibitem[{\citenamefont{Sankowski}\ and\
  \citenamefont{Kacman}(2005)}]{Sankowski:2005_PRB}%
  \BibitemOpen
  \bibfield{author}{%
  \bibinfo {author} {\bibnamefont{Sankowski}, \bibfnamefont{P.}},\ and\
  \bibinfo {author} {\bibfnamefont{P.}~\bibnamefont{Kacman}}}%
  , \bibinfo {year} {2005},\ \bibfield{title}{%
  \enquote{\bibinfo {title} {Interlayer exchange coupling in {(Ga,Mn)As}-based
  superlattices},}\ }%
  \bibfield{journal}{%
  \bibinfo {journal} {Phys. Rev. B}\ }%
  \textbf{\bibinfo {volume} {71}},\ \bibinfo {pages} {201303}%
  \bibAnnoteFile{NoStop}{Sankowski:2005_PRB}%
\bibitem[{\citenamefont{Sankowski}\
  \emph{et~al.}(2006)\citenamefont{Sankowski}, \citenamefont{Kacman},
  \citenamefont{Majewski},\ and\ \citenamefont{Dietl}}]{Sankowski:2006_PE}%
  \BibitemOpen
  \bibfield{author}{%
  \bibinfo {author} {\bibnamefont{Sankowski}, \bibfnamefont{P.}}, \bibinfo
  {author} {\bibfnamefont{P.}~\bibnamefont{Kacman}}, \bibinfo {author}
  {\bibfnamefont{J.}~\bibnamefont{Majewski}},\ and\ \bibinfo {author}
  {\bibfnamefont{T.}~\bibnamefont{Dietl}}}%
  , \bibinfo {year} {2006},\ \bibfield{title}{%
  \enquote{\bibinfo {title} {Tight-binding model of spin-polarized tunnelling
  in {(Ga,Mn)As}-based structures},}\ }%
  \bibfield{journal}{%
  \bibinfo {journal} {Physica E}\ }%
  \textbf{\bibinfo {volume} {32}},\ \bibinfo {pages} {375}%
  \bibAnnoteFile{NoStop}{Sankowski:2006_PE}%
\bibitem[{\citenamefont{Sankowski}\
  \emph{et~al.}(2007)\citenamefont{Sankowski}, \citenamefont{Kacman},
  \citenamefont{Majewski},\ and\ \citenamefont{Dietl}}]{Sankowski:2007_PRB}%
  \BibitemOpen
  \bibfield{author}{%
  \bibinfo {author} {\bibnamefont{Sankowski}, \bibfnamefont{P.}}, \bibinfo
  {author} {\bibfnamefont{P.}~\bibnamefont{Kacman}}, \bibinfo {author}
  {\bibfnamefont{J.~A.}\ \bibnamefont{Majewski}},\ and\ \bibinfo {author}
  {\bibfnamefont{T.}~\bibnamefont{Dietl}}}%
  , \bibinfo {year} {2007},\ \bibfield{title}{%
  \enquote{\bibinfo {title} {Spin-dependent tunneling in modulated structures
  of {(Ga,Mn)As}},}\ }%
  \bibfield{journal}{%
  \bibinfo {journal} {Phys. Rev. B}\ }%
  \textbf{\bibinfo {volume} {75}},\ \bibinfo {pages} {045306}%
  \bibAnnoteFile{NoStop}{Sankowski:2007_PRB}%
\bibitem[{\citenamefont{Sarigiannidou}\
  \emph{et~al.}(2006)\citenamefont{Sarigiannidou}, \citenamefont{Wilhelm},
  \citenamefont{Monroy}, \citenamefont{Galera}, \citenamefont{Bellet-Amalric},
  \citenamefont{Rogalev}, \citenamefont{Goulon}, \citenamefont{Cibert},\ and\
  \citenamefont{Mariette}}]{Sarigiannidou:2006_PRB}%
  \BibitemOpen
  \bibfield{author}{%
  \bibinfo {author} {\bibnamefont{Sarigiannidou}, \bibfnamefont{E.}}, \bibinfo
  {author} {\bibfnamefont{F.}~\bibnamefont{Wilhelm}}, \bibinfo {author}
  {\bibfnamefont{E.}~\bibnamefont{Monroy}}, \bibinfo {author}
  {\bibfnamefont{R.~M.}\ \bibnamefont{Galera}}, \bibinfo {author}
  {\bibfnamefont{E.}~\bibnamefont{Bellet-Amalric}}, \bibinfo {author}
  {\bibfnamefont{A.}~\bibnamefont{Rogalev}}, \bibinfo {author}
  {\bibfnamefont{J.}~\bibnamefont{Goulon}}, \bibinfo {author}
  {\bibfnamefont{J.}~\bibnamefont{Cibert}},\ and\ \bibinfo {author}
  {\bibfnamefont{H.}~\bibnamefont{Mariette}}}%
  , \bibinfo {year} {2006},\ \bibfield{title}{%
  \enquote{\bibinfo {title} {Intrinsic ferromagnetism in wurtzite {(Ga,Mn)N}
  semiconductor},}\ }%
  \bibfield{journal}{%
  \bibinfo {journal} {Phys. Rev. B}\ }%
  \textbf{\bibinfo {volume} {74}},\ \bibinfo {pages} {041306(R)}%
  \bibAnnoteFile{NoStop}{Sarigiannidou:2006_PRB}%
\bibitem[{\citenamefont{Sato}\ \emph{et~al.}(2010)\citenamefont{Sato},
  \citenamefont{Bergqvist}, \citenamefont{Kudrnovsk\'y},
  \citenamefont{Dederichs}, \citenamefont{Eriksson}, \citenamefont{Turek},
  \citenamefont{Sanyal}, \citenamefont{Bouzerar},
  \citenamefont{Katayama-Yoshida}, \citenamefont{Dinh},
  \citenamefont{Fukushima}, \citenamefont{Kizaki},\ and\
  \citenamefont{Zeller}}]{Sato:2010_RMP}%
  \BibitemOpen
  \bibfield{author}{%
  \bibinfo {author} {\bibnamefont{Sato}, \bibfnamefont{K.}}, \bibinfo {author}
  {\bibfnamefont{L.}~\bibnamefont{Bergqvist}}, \bibinfo {author}
  {\bibfnamefont{J.}~\bibnamefont{Kudrnovsk\'y}}, \bibinfo {author}
  {\bibfnamefont{P.~H.}\ \bibnamefont{Dederichs}}, \bibinfo {author}
  {\bibfnamefont{O.}~\bibnamefont{Eriksson}}, \bibinfo {author}
  {\bibfnamefont{I.}~\bibnamefont{Turek}}, \bibinfo {author}
  {\bibfnamefont{B.}~\bibnamefont{Sanyal}}, \bibinfo {author}
  {\bibfnamefont{G.}~\bibnamefont{Bouzerar}}, \bibinfo {author}
  {\bibfnamefont{H.}~\bibnamefont{Katayama-Yoshida}}, \bibinfo {author}
  {\bibfnamefont{V.~A.}\ \bibnamefont{Dinh}}, \bibinfo {author}
  {\bibfnamefont{T.}~\bibnamefont{Fukushima}}, \bibinfo {author}
  {\bibfnamefont{H.}~\bibnamefont{Kizaki}},\ and\ \bibinfo {author}
  {\bibfnamefont{R.}~\bibnamefont{Zeller}}}%
  , \bibinfo {year} {2010},\ \bibfield{title}{%
  \enquote{\bibinfo {title} {First-principles theory of dilute magnetic
  semiconductors},}\ }%
  \bibfield{journal}{%
  \bibinfo {journal} {Rev. Mod. Phys.}\ }%
  \textbf{\bibinfo {volume} {82}},\ \bibinfo {pages} {1633}%
  \bibAnnoteFile{NoStop}{Sato:2010_RMP}%
\bibitem[{\citenamefont{Sato}\ \emph{et~al.}(2005)\citenamefont{Sato},
  \citenamefont{Katayama-Yoshida},\ and\
  \citenamefont{Dederichs}}]{Sato:2005_JJAP}%
  \BibitemOpen
  \bibfield{author}{%
  \bibinfo {author} {\bibnamefont{Sato}, \bibfnamefont{K.}}, \bibinfo {author}
  {\bibfnamefont{H.}~\bibnamefont{Katayama-Yoshida}},\ and\ \bibinfo {author}
  {\bibfnamefont{P.~H.}\ \bibnamefont{Dederichs}}}%
  , \bibinfo {year} {2005},\ \bibfield{title}{%
  \enquote{\bibinfo {title} {High {Curie} temperature and nano-scale spinodal
  decomposition phase in dilute magnetic semiconductors},}\ }%
  \bibfield{journal}{%
  \bibinfo {journal} {Japan. J. Appl. Phys.}\ }%
  \textbf{\bibinfo {volume} {44}},\ \bibinfo {pages} {L948}%
  \bibAnnoteFile{NoStop}{Sato:2005_JJAP}%
\bibitem[{\citenamefont{Satoh}\ \emph{et~al.}(2001)\citenamefont{Satoh},
  \citenamefont{Okazawa}, \citenamefont{Nagashima},\ and\
  \citenamefont{Yoshino}}]{Satoh:2001_PE}%
  \BibitemOpen
  \bibfield{author}{%
  \bibinfo {author} {\bibnamefont{Satoh}, \bibfnamefont{Y.}}, \bibinfo {author}
  {\bibfnamefont{D.}~\bibnamefont{Okazawa}}, \bibinfo {author}
  {\bibfnamefont{A.}~\bibnamefont{Nagashima}},\ and\ \bibinfo {author}
  {\bibfnamefont{J.}~\bibnamefont{Yoshino}}}%
  , \bibinfo {year} {2001},\ \bibfield{title}{%
  \enquote{\bibinfo {title} {Carrier concentration dependence of electronic and
  magnetic properties of {Sn}-doped {(GaMn)As}},}\ }%
  \bibfield{journal}{%
  \bibinfo {journal} {Physica E}\ }%
  \textbf{\bibinfo {volume} {10}},\ \bibinfo {pages} {196}%
  \bibAnnoteFile{NoStop}{Satoh:2001_PE}%
\bibitem[{\citenamefont{Sawicki}(2006)}]{Sawicki:2006_JMMM}%
  \BibitemOpen
  \bibfield{author}{%
  \bibinfo {author} {\bibnamefont{Sawicki}, \bibfnamefont{M.}}}%
  , \bibinfo {year} {2006},\ \bibfield{title}{%
  \enquote{\bibinfo {title} {Magnetic properties of {(Ga,Mn)As}},}\ }%
  \bibfield{journal}{%
  \bibinfo {journal} {J. Mag. Magn. Mater.}\ }%
  \textbf{\bibinfo {volume} {300}},\ \bibinfo {pages} {1}%
  \bibAnnoteFile{NoStop}{Sawicki:2006_JMMM}%
\bibitem[{\citenamefont{Sawicki}\ \emph{et~al.}(2010)\citenamefont{Sawicki},
  \citenamefont{Chiba}, \citenamefont{Korbecka}, \citenamefont{Nishitani},
  \citenamefont{Majewski}, \citenamefont{Matsukura}, \citenamefont{Dietl},\
  and\ \citenamefont{Ohno}}]{Sawicki:2010_NP}%
  \BibitemOpen
  \bibfield{author}{%
  \bibinfo {author} {\bibnamefont{Sawicki}, \bibfnamefont{M.}}, \bibinfo
  {author} {\bibfnamefont{D.}~\bibnamefont{Chiba}}, \bibinfo {author}
  {\bibfnamefont{A.}~\bibnamefont{Korbecka}}, \bibinfo {author}
  {\bibfnamefont{Yu}~\bibnamefont{Nishitani}}, \bibinfo {author}
  {\bibfnamefont{J.~A.}\ \bibnamefont{Majewski}}, \bibinfo {author}
  {\bibfnamefont{F.}~\bibnamefont{Matsukura}}, \bibinfo {author}
  {\bibfnamefont{T.}~\bibnamefont{Dietl}},\ and\ \bibinfo {author}
  {\bibfnamefont{H.}~\bibnamefont{Ohno}}}%
  , \bibinfo {year} {2010},\ \bibfield{title}{%
  \enquote{\bibinfo {title} {Experimental probing of the interplay between
  ferromagnetism and localization in {(Ga, Mn)As}},}\ }%
  \bibfield{journal}{%
  \bibinfo {journal} {Nat. Phys.}\ }%
  \textbf{\bibinfo {volume} {6}},\ \bibinfo {pages} {22}%
  \bibAnnoteFile{NoStop}{Sawicki:2010_NP}%
\bibitem[{\citenamefont{Sawicki}\ \emph{et~al.}(2012)\citenamefont{Sawicki},
  \citenamefont{Devillers}, \citenamefont{Ga{\l}\c{e}ski},
  \citenamefont{Simserides}, \citenamefont{Dobkowska}, \citenamefont{Faina},
  \citenamefont{Grois}, \citenamefont{Navarro-Quezada},
  \citenamefont{Trohidou}, \citenamefont{Majewski}, \citenamefont{Dietl},\ and\
  \citenamefont{Bonanni}}]{Sawicki:2012_PRB}%
  \BibitemOpen
  \bibfield{author}{%
  \bibinfo {author} {\bibnamefont{Sawicki}, \bibfnamefont{M.}}, \bibinfo
  {author} {\bibfnamefont{T.}~\bibnamefont{Devillers}}, \bibinfo {author}
  {\bibfnamefont{S.}~\bibnamefont{Ga{\l}\c{e}ski}}, \bibinfo {author}
  {\bibfnamefont{C.}~\bibnamefont{Simserides}}, \bibinfo {author}
  {\bibfnamefont{S.}~\bibnamefont{Dobkowska}}, \bibinfo {author}
  {\bibfnamefont{B.}~\bibnamefont{Faina}}, \bibinfo {author}
  {\bibfnamefont{A.}~\bibnamefont{Grois}}, \bibinfo {author}
  {\bibfnamefont{A.}~\bibnamefont{Navarro-Quezada}}, \bibinfo {author}
  {\bibfnamefont{K.~N.}\ \bibnamefont{Trohidou}}, \bibinfo {author}
  {\bibfnamefont{J.~A.}\ \bibnamefont{Majewski}}, \bibinfo {author}
  {\bibfnamefont{T.}~\bibnamefont{Dietl}},\ and\ \bibinfo {author}
  {\bibfnamefont{A.}~\bibnamefont{Bonanni}}}%
  , \bibinfo {year} {2012},\ \bibfield{title}{%
  \enquote{\bibinfo {title} {Origin of low-temperature magnetic ordering in
  {Ga$_{1-x}$Mn$_{x}$N}},}\ }%
  \bibfield{journal}{%
  \bibinfo {journal} {Phys. Rev. B}\ }%
  \textbf{\bibinfo {volume} {85}},\ \bibinfo {pages} {205204}%
  \bibAnnoteFile{NoStop}{Sawicki:2012_PRB}%
\bibitem[{\citenamefont{Sawicki}\ \emph{et~al.}(2002)\citenamefont{Sawicki},
  \citenamefont{Khoi}, \citenamefont{Hansen}, \citenamefont{Ferrand},
  \citenamefont{Molenkamp}, \citenamefont{Waag},\ and\
  \citenamefont{Dietl}}]{Sawicki:2002_PSSB}%
  \BibitemOpen
  \bibfield{author}{%
  \bibinfo {author} {\bibnamefont{Sawicki}, \bibfnamefont{M.}}, \bibinfo
  {author} {\bibfnamefont{Le~Van}\ \bibnamefont{Khoi}}, \bibinfo {author}
  {\bibfnamefont{L.}~\bibnamefont{Hansen}}, \bibinfo {author}
  {\bibfnamefont{D.}~\bibnamefont{Ferrand}}, \bibinfo {author}
  {\bibfnamefont{L.~W.}\ \bibnamefont{Molenkamp}}, \bibinfo {author}
  {\bibfnamefont{A.}~\bibnamefont{Waag}},\ and\ \bibinfo {author}
  {\bibfnamefont{T.}~\bibnamefont{Dietl}}}%
  , \bibinfo {year} {2002},\ \bibfield{title}{%
  \enquote{\bibinfo {title} {Magnetic characterisation of highly doped {MBE}
  grown {Be$_{1-x}$Mn$_{x}$Te} and bulk {Zn$_{1-x}$Mn$_{x}$Te}},}\ }%
  \bibfield{journal}{%
  \bibinfo {journal} {Phys. Status Solidi B}\ }%
  \textbf{\bibinfo {volume} {229}},\ \bibinfo {pages} {717}%
  \bibAnnoteFile{NoStop}{Sawicki:2002_PSSB}%
\bibitem[{\citenamefont{Sawicki}\ \emph{et~al.}(2004)\citenamefont{Sawicki},
  \citenamefont{Matsukura}, \citenamefont{Idziaszek}, \citenamefont{Dietl},
  \citenamefont{Schott}, \citenamefont{R{\"u}ster}, \citenamefont{Gould},
  \citenamefont{Karczewski}, \citenamefont{Schmidt},\ and\
  \citenamefont{Molenkamp}}]{Sawicki:2004_PRB}%
  \BibitemOpen
  \bibfield{author}{%
  \bibinfo {author} {\bibnamefont{Sawicki}, \bibfnamefont{M.}}, \bibinfo
  {author} {\bibfnamefont{F.}~\bibnamefont{Matsukura}}, \bibinfo {author}
  {\bibfnamefont{A.}~\bibnamefont{Idziaszek}}, \bibinfo {author}
  {\bibfnamefont{T.}~\bibnamefont{Dietl}}, \bibinfo {author}
  {\bibfnamefont{G.~M.}\ \bibnamefont{Schott}}, \bibinfo {author}
  {\bibfnamefont{C.}~\bibnamefont{R{\"u}ster}}, \bibinfo {author}
  {\bibfnamefont{C.}~\bibnamefont{Gould}}, \bibinfo {author}
  {\bibfnamefont{G.}~\bibnamefont{Karczewski}}, \bibinfo {author}
  {\bibfnamefont{G.}~\bibnamefont{Schmidt}},\ and\ \bibinfo {author}
  {\bibfnamefont{L.~W.}\ \bibnamefont{Molenkamp}}}%
  , \bibinfo {year} {2004},\ \bibfield{title}{%
  \enquote{\bibinfo {title} {Temperature dependent magnetic anisotropy in
  {(Ga,Mn)As} layers},}\ }%
  \bibfield{journal}{%
  \bibinfo {journal} {Phys. Rev. B}\ }%
  \textbf{\bibinfo {volume} {70}},\ \bibinfo {pages} {245325}%
  \bibAnnoteFile{NoStop}{Sawicki:2004_PRB}%
\bibitem[{\citenamefont{Sawicki}\ \emph{et~al.}(2011)\citenamefont{Sawicki},
  \citenamefont{Stefanowicz},\ and\ \citenamefont{Ney}}]{Sawicki:2011_SST}%
  \BibitemOpen
  \bibfield{author}{%
  \bibinfo {author} {\bibnamefont{Sawicki}, \bibfnamefont{M.}}, \bibinfo
  {author} {\bibfnamefont{W.}~\bibnamefont{Stefanowicz}},\ and\ \bibinfo
  {author} {\bibfnamefont{A.}~\bibnamefont{Ney}}}%
  , \bibinfo {year} {2011},\ \bibfield{title}{%
  \enquote{\bibinfo {title} {Sensitive {SQUID} magnetometry for studying
  nanomagnetism},}\ }%
  \bibfield{journal}{%
  \bibinfo {journal} {Semicon. Sci. Technol.}\ }%
  \textbf{\bibinfo {volume} {26}},\ \bibinfo {pages} {064006}%
  \bibAnnoteFile{NoStop}{Sawicki:2011_SST}%
\bibitem[{\citenamefont{Sawicki}\ \emph{et~al.}(2005)\citenamefont{Sawicki},
  \citenamefont{Wang}, \citenamefont{Edmonds}, \citenamefont{Campion},
  \citenamefont{Staddon}, \citenamefont{Farley}, \citenamefont{Foxon},
  \citenamefont{Papis}, \citenamefont{Kami{\'n}ska}, \citenamefont{Piotrowska},
  \citenamefont{Dietl},\ and\ \citenamefont{Gallagher}}]{Sawicki:2005_PRB}%
  \BibitemOpen
  \bibfield{author}{%
  \bibinfo {author} {\bibnamefont{Sawicki}, \bibfnamefont{M.}}, \bibinfo
  {author} {\bibfnamefont{K-Y.}\ \bibnamefont{Wang}}, \bibinfo {author}
  {\bibfnamefont{K.~W.}\ \bibnamefont{Edmonds}}, \bibinfo {author}
  {\bibfnamefont{R.~P.}\ \bibnamefont{Campion}}, \bibinfo {author}
  {\bibfnamefont{C.~R.}\ \bibnamefont{Staddon}}, \bibinfo {author}
  {\bibfnamefont{N.~R.~S.}\ \bibnamefont{Farley}}, \bibinfo {author}
  {\bibfnamefont{C.~T.}\ \bibnamefont{Foxon}}, \bibinfo {author}
  {\bibfnamefont{E.}~\bibnamefont{Papis}}, \bibinfo {author}
  {\bibfnamefont{E.}~\bibnamefont{Kami{\'n}ska}}, \bibinfo {author}
  {\bibfnamefont{A.}~\bibnamefont{Piotrowska}}, \bibinfo {author}
  {\bibfnamefont{T.}~\bibnamefont{Dietl}},\ and\ \bibinfo {author}
  {\bibfnamefont{B.~L.}\ \bibnamefont{Gallagher}}}%
  , \bibinfo {year} {2005},\ \bibfield{title}{%
  \enquote{\bibinfo {title} {In-plane uniaxial anisotropy rotations in
  {(Ga,Mn)As} thin films},}\ }%
  \bibfield{journal}{%
  \bibinfo {journal} {Phys. Rev. B}\ }%
  \textbf{\bibinfo {volume} {71}},\ \bibinfo {pages} {121302(R)}%
  \bibAnnoteFile{NoStop}{Sawicki:2005_PRB}%
\bibitem[{\citenamefont{Scarpulla}\
  \emph{et~al.}(2005)\citenamefont{Scarpulla}, \citenamefont{Cardozo},
  \citenamefont{Oo}, \citenamefont{McCluskey}, \citenamefont{Yu},\ and\
  \citenamefont{Dubon}}]{Scarpulla:2005_PRL}%
  \BibitemOpen
  \bibfield{author}{%
  \bibinfo {author} {\bibnamefont{Scarpulla}, \bibfnamefont{M.~A.}}, \bibinfo
  {author} {\bibfnamefont{B.~L.}\ \bibnamefont{Cardozo}}, \bibinfo {author}
  {\bibfnamefont{W.~M.~Hlaing}\ \bibnamefont{Oo}}, \bibinfo {author}
  {\bibfnamefont{M.~D.}\ \bibnamefont{McCluskey}}, \bibinfo {author}
  {\bibfnamefont{K.~M.}\ \bibnamefont{Yu}},\ and\ \bibinfo {author}
  {\bibfnamefont{O.~D.}\ \bibnamefont{Dubon}}}%
  , \bibinfo {year} {2005},\ \bibfield{title}{%
  \enquote{\bibinfo {title} {Ferromagnetism in {Ga$_{1-x}$Mn$_{x}$P:} evidence
  for inter-{Mn} exchange mediated by localized holes within a detached
  impurity band},}\ }%
  \bibfield{journal}{%
  \bibinfo {journal} {Phys. Rev. Lett.}\ }%
  \textbf{\bibinfo {volume} {95}},\ \bibinfo {pages} {207204}%
  \bibAnnoteFile{NoStop}{Scarpulla:2005_PRL}%
\bibitem[{\citenamefont{Scarpulla}\
  \emph{et~al.}(2008)\citenamefont{Scarpulla}, \citenamefont{Stone},
  \citenamefont{Sharp}, \citenamefont{Haller}, \citenamefont{Dubon},
  \citenamefont{Beeman},\ and\ \citenamefont{Yu}}]{Scarpulla:2008_JAP}%
  \BibitemOpen
  \bibfield{author}{%
  \bibinfo {author} {\bibnamefont{Scarpulla}, \bibfnamefont{M.~A.}}, \bibinfo
  {author} {\bibfnamefont{P.~R.}\ \bibnamefont{Stone}}, \bibinfo {author}
  {\bibfnamefont{I.~D.}\ \bibnamefont{Sharp}}, \bibinfo {author}
  {\bibfnamefont{E.~E.}\ \bibnamefont{Haller}}, \bibinfo {author}
  {\bibfnamefont{O.~D.}\ \bibnamefont{Dubon}}, \bibinfo {author}
  {\bibfnamefont{J.~W.}\ \bibnamefont{Beeman}},\ and\ \bibinfo {author}
  {\bibfnamefont{K.~M.}\ \bibnamefont{Yu}}}%
  , \bibinfo {year} {2008},\ \bibfield{title}{%
  \enquote{\bibinfo {title} {Nonmagnetic compensation in ferromagnetic
  {Ga$_{1-x}$Mn$_{x}$As} and {Ga$_{1-x}$Mn$_{x}$P} synthesized by ion
  implantation and pulsed-laser melting},}\ }%
  \bibfield{journal}{%
  \bibinfo {journal} {J. Appl. Phys.}\ }%
  \textbf{\bibinfo {volume} {103}},\ \bibinfo {pages} {123906}%
  \bibAnnoteFile{NoStop}{Scarpulla:2008_JAP}%
\bibitem[{\citenamefont{Schallenberg}\ and\
  \citenamefont{Munekata}(2006)}]{Schallenberg:2006_APL}%
  \BibitemOpen
  \bibfield{author}{%
  \bibinfo {author} {\bibnamefont{Schallenberg}, \bibfnamefont{T.}},\ and\
  \bibinfo {author} {\bibfnamefont{H.}~\bibnamefont{Munekata}}}%
  , \bibinfo {year} {2006},\ \bibfield{title}{%
  \enquote{\bibinfo {title} {Preparation of ferromagnetic {(In,Mn)As} with a
  high {Curie} temperature of 90 {K}},}\ }%
  \bibfield{journal}{%
  \bibinfo {journal} {Appl. Phys. Lett.}\ }%
  \textbf{\bibinfo {volume} {89}},\ \bibinfo {pages} {042507}%
  \bibAnnoteFile{NoStop}{Schallenberg:2006_APL}%
\bibitem[{\citenamefont{Scherbakov}\
  \emph{et~al.}(2010)\citenamefont{Scherbakov}, \citenamefont{Salasyuk},
  \citenamefont{Akimov}, \citenamefont{Liu}, \citenamefont{Bombeck},
  \citenamefont{Br\"uggemann}, \citenamefont{Yakovlev}, \citenamefont{Sapega},
  \citenamefont{Furdyna},\ and\ \citenamefont{Bayer}}]{Scherbakov:2010_PRL}%
  \BibitemOpen
  \bibfield{author}{%
  \bibinfo {author} {\bibnamefont{Scherbakov}, \bibfnamefont{A.~V.}}, \bibinfo
  {author} {\bibfnamefont{A.~S.}\ \bibnamefont{Salasyuk}}, \bibinfo {author}
  {\bibfnamefont{A.~V.}\ \bibnamefont{Akimov}}, \bibinfo {author}
  {\bibfnamefont{X.}~\bibnamefont{Liu}}, \bibinfo {author}
  {\bibfnamefont{M.}~\bibnamefont{Bombeck}}, \bibinfo {author}
  {\bibfnamefont{C.}~\bibnamefont{Br\"uggemann}}, \bibinfo {author}
  {\bibfnamefont{D.~R.}\ \bibnamefont{Yakovlev}}, \bibinfo {author}
  {\bibfnamefont{V.~F.}\ \bibnamefont{Sapega}}, \bibinfo {author}
  {\bibfnamefont{J.~K.}\ \bibnamefont{Furdyna}},\ and\ \bibinfo {author}
  {\bibfnamefont{M.}~\bibnamefont{Bayer}}}%
  , \bibinfo {year} {2010},\ \bibfield{title}{%
  \enquote{\bibinfo {title} {Coherent magnetization precession in ferromagnetic
  {(Ga,Mn)As} induced by picosecond acoustic pulses},}\ }%
  \bibfield{journal}{%
  \bibinfo {journal} {Phys. Rev. Lett.}\ }%
  \textbf{\bibinfo {volume} {105}},\ \bibinfo {pages} {117204}%
  \bibAnnoteFile{NoStop}{Scherbakov:2010_PRL}%
\bibitem[{\citenamefont{van Schilfgaarde}\ and\
  \citenamefont{Mryasov}(2001)}]{Schilfgaarde:2001_PRB}%
  \BibitemOpen
  \bibfield{author}{%
  \bibinfo {author} {\bibnamefont{van Schilfgaarde}, \bibfnamefont{Mark}},\
  and\ \bibinfo {author} {\bibfnamefont{O.~N.}\ \bibnamefont{Mryasov}}}%
  , \bibinfo {year} {2001},\ \bibfield{title}{%
  \enquote{\bibinfo {title} {Anomalous exchange interactions in {III-V} dilute
  magnetic semiconductors},}\ }%
  \bibfield{journal}{%
  \bibinfo {journal} {Phys. Rev. B}\ }%
  \textbf{\bibinfo {volume} {63}},\ \bibinfo {pages} {233205}%
  \bibAnnoteFile{NoStop}{Schilfgaarde:2001_PRB}%
\bibitem[{\citenamefont{Schlapps}\ \emph{et~al.}(2009)\citenamefont{Schlapps},
  \citenamefont{Lermer}, \citenamefont{Geissler}, \citenamefont{Neumaier},
  \citenamefont{Sadowski}, \citenamefont{Schuh}, \citenamefont{Wegscheider},\
  and\ \citenamefont{Weiss}}]{Schlapps:2009_PRB}%
  \BibitemOpen
  \bibfield{author}{%
  \bibinfo {author} {\bibnamefont{Schlapps}, \bibfnamefont{M.}}, \bibinfo
  {author} {\bibfnamefont{T.}~\bibnamefont{Lermer}}, \bibinfo {author}
  {\bibfnamefont{S.}~\bibnamefont{Geissler}}, \bibinfo {author}
  {\bibfnamefont{D.}~\bibnamefont{Neumaier}}, \bibinfo {author}
  {\bibfnamefont{J.}~\bibnamefont{Sadowski}}, \bibinfo {author}
  {\bibfnamefont{D.}~\bibnamefont{Schuh}}, \bibinfo {author}
  {\bibfnamefont{W.}~\bibnamefont{Wegscheider}},\ and\ \bibinfo {author}
  {\bibfnamefont{D.}~\bibnamefont{Weiss}}}%
  , \bibinfo {year} {2009},\ \bibfield{title}{%
  \enquote{\bibinfo {title} {Transport through {(Ga,Mn)As} nanoislands:
  {Coulomb} blockade and temperature dependence of the conductance},}\ }%
  \bibfield{journal}{%
  \bibinfo {journal} {Phys. Rev. B}\ }%
  \textbf{\bibinfo {volume} {80}},\ \bibinfo {pages} {125330}%
  \bibAnnoteFile{NoStop}{Schlapps:2009_PRB}%
\bibitem[{\citenamefont{Schmid}\ \emph{et~al.}(2008)\citenamefont{Schmid},
  \citenamefont{M{\"u}ller}, \citenamefont{Sing}, \citenamefont{Claessen},
  \citenamefont{Wenisch}, \citenamefont{Gould}, \citenamefont{Brunner},
  \citenamefont{Molenkamp},\ and\ \citenamefont{Drube}}]{Schmid:2008_PRB}%
  \BibitemOpen
  \bibfield{author}{%
  \bibinfo {author} {\bibnamefont{Schmid}, \bibfnamefont{B.}}, \bibinfo
  {author} {\bibfnamefont{A.}~\bibnamefont{M{\"u}ller}}, \bibinfo {author}
  {\bibfnamefont{M.}~\bibnamefont{Sing}}, \bibinfo {author}
  {\bibfnamefont{R.}~\bibnamefont{Claessen}}, \bibinfo {author}
  {\bibfnamefont{J.}~\bibnamefont{Wenisch}}, \bibinfo {author}
  {\bibfnamefont{C.}~\bibnamefont{Gould}}, \bibinfo {author}
  {\bibfnamefont{K.}~\bibnamefont{Brunner}}, \bibinfo {author}
  {\bibfnamefont{L.}~\bibnamefont{Molenkamp}},\ and\ \bibinfo {author}
  {\bibfnamefont{W.}~\bibnamefont{Drube}}}%
  , \bibinfo {year} {2008},\ \bibfield{title}{%
  \enquote{\bibinfo {title} {Surface segregation of interstitial manganese in
  {Ga$_{1-x}$Mn$_{x}$As} studied by hard x-ray photoemission spectroscopy},}\
  }%
  \bibfield{journal}{%
  \bibinfo {journal} {Phys. Rev. B}\ }%
  \textbf{\bibinfo {volume} {78}},\ \bibinfo {pages} {075319}%
  \bibAnnoteFile{NoStop}{Schmid:2008_PRB}%
\bibitem[{\citenamefont{Schneider}\
  \emph{et~al.}(1987)\citenamefont{Schneider}, \citenamefont{Kaufmann},
  \citenamefont{Wilkening}, \citenamefont{Baeumler},\ and\
  \citenamefont{{K{\"o}hl}}}]{Schneider:1987_PRL}%
  \BibitemOpen
  \bibfield{author}{%
  \bibinfo {author} {\bibnamefont{Schneider}, \bibfnamefont{J.}}, \bibinfo
  {author} {\bibfnamefont{U.}~\bibnamefont{Kaufmann}}, \bibinfo {author}
  {\bibfnamefont{W.}~\bibnamefont{Wilkening}}, \bibinfo {author}
  {\bibfnamefont{M.}~\bibnamefont{Baeumler}},\ and\ \bibinfo {author}
  {\bibfnamefont{F.}~\bibnamefont{{K{\"o}hl}}}}%
  , \bibinfo {year} {1987},\ \bibfield{title}{%
  \enquote{\bibinfo {title} {Electronic structure of the neutral manganese
  acceptor in gallium arsenide},}\ }%
  \bibfield{journal}{%
  \bibinfo {journal} {Phys. Rev. Lett.}\ }%
  \textbf{\bibinfo {volume} {59}},\ \bibinfo {pages} {240}%
  \bibAnnoteFile{NoStop}{Schneider:1987_PRL}%
\bibitem[{\citenamefont{Seong}\ \emph{et~al.}(2002)\citenamefont{Seong},
  \citenamefont{Chun}, \citenamefont{Cheong}, \citenamefont{Samarth},\ and\
  \citenamefont{Mascarenhas}}]{Seong:2002_PRB}%
  \BibitemOpen
  \bibfield{author}{%
  \bibinfo {author} {\bibnamefont{Seong}, \bibfnamefont{M.~J.}}, \bibinfo
  {author} {\bibfnamefont{S.~H.}\ \bibnamefont{Chun}}, \bibinfo {author}
  {\bibfnamefont{H.~M.}\ \bibnamefont{Cheong}}, \bibinfo {author}
  {\bibfnamefont{N.}~\bibnamefont{Samarth}},\ and\ \bibinfo {author}
  {\bibfnamefont{A.}~\bibnamefont{Mascarenhas}}}%
  , \bibinfo {year} {2002},\ \bibfield{title}{%
  \enquote{\bibinfo {title} {Spectroscopic determination of hole density in the
  ferromagnetic semiconductor {Ga$_{1-x}$Mn$_{x}$As}},}\ }%
  \bibfield{journal}{%
  \bibinfo {journal} {Phys. Rev. B}\ }%
  \textbf{\bibinfo {volume} {66}},\ \bibinfo {pages} {033202}%
  \bibAnnoteFile{NoStop}{Seong:2002_PRB}%
\bibitem[{\citenamefont{Serrate}\ \emph{et~al.}(2007)\citenamefont{Serrate},
  \citenamefont{Teresa},\ and\ \citenamefont{Ibarra}}]{Serrate:2007_JPCM}%
  \BibitemOpen
  \bibfield{author}{%
  \bibinfo {author} {\bibnamefont{Serrate}, \bibfnamefont{D.}}, \bibinfo
  {author} {\bibfnamefont{J.~M.~De}\ \bibnamefont{Teresa}},\ and\ \bibinfo
  {author} {\bibfnamefont{M~R}\ \bibnamefont{Ibarra}}}%
  , \bibinfo {year} {2007},\ \bibfield{title}{%
  \enquote{\bibinfo {title} {Double perovskites with ferromagnetism above room
  temperature},}\ }%
  \bibfield{journal}{%
  \bibinfo {journal} {J. Phys. Cond. Matter}\ }%
  \textbf{\bibinfo {volume} {19}},\ \bibinfo {pages} {023201}%
  \bibAnnoteFile{NoStop}{Serrate:2007_JPCM}%
\bibitem[{\citenamefont{Shapira}\ and\
  \citenamefont{Bindilatti}(2002)}]{Shapira:2002_JAP}%
  \BibitemOpen
  \bibfield{author}{%
  \bibinfo {author} {\bibnamefont{Shapira}, \bibfnamefont{Y.}},\ and\ \bibinfo
  {author} {\bibfnamefont{V.}~\bibnamefont{Bindilatti}}}%
  , \bibinfo {year} {2002},\ \bibfield{title}{%
  \enquote{\bibinfo {title} {Magnetization-step studies of antiferromagnetic
  clusters and single ions: {Exchange}, anisotropy, and statistics},}\ }%
  \bibfield{journal}{%
  \bibinfo {journal} {J. Appl. Phys.}\ }%
  \textbf{\bibinfo {volume} {92}},\ \bibinfo {pages} {4155}%
  \bibAnnoteFile{NoStop}{Shapira:2002_JAP}%
\bibitem[{\citenamefont{Sheu}\ \emph{et~al.}(2007)\citenamefont{Sheu},
  \citenamefont{Myers}, \citenamefont{Tang}, \citenamefont{Samarth},
  \citenamefont{Awschalom}, \citenamefont{Schiffer},\ and\
  \citenamefont{Flatt{\'e}}}]{Sheu:2007_PRL}%
  \BibitemOpen
  \bibfield{author}{%
  \bibinfo {author} {\bibnamefont{Sheu}, \bibfnamefont{B.~L.}}, \bibinfo
  {author} {\bibfnamefont{R.~C.}\ \bibnamefont{Myers}}, \bibinfo {author}
  {\bibfnamefont{J.-M.}\ \bibnamefont{Tang}}, \bibinfo {author}
  {\bibfnamefont{N.}~\bibnamefont{Samarth}}, \bibinfo {author}
  {\bibfnamefont{D.~D.}\ \bibnamefont{Awschalom}}, \bibinfo {author}
  {\bibfnamefont{P.}~\bibnamefont{Schiffer}},\ and\ \bibinfo {author}
  {\bibfnamefont{M.~E.}\ \bibnamefont{Flatt{\'e}}}}%
  , \bibinfo {year} {2007},\ \bibfield{title}{%
  \enquote{\bibinfo {title} {Onset of ferromagnetism in low-doped
  {Ga$_{1-x}$Mn$_x$As}},}\ }%
  \bibfield{journal}{%
  \bibinfo {journal} {Phys. Rev. Lett.}\ }%
  \textbf{\bibinfo {volume} {99}},\ \bibinfo {pages} {227205}%
  \bibAnnoteFile{NoStop}{Sheu:2007_PRL}%
\bibitem[{\citenamefont{Sheu}\ \emph{et~al.}(2013)\citenamefont{Sheu},
  \citenamefont{Huang}, \citenamefont{Lee}, \citenamefont{Lee},
  \citenamefont{Yeh}, \citenamefont{Chen},\ and\
  \citenamefont{Lai}}]{Sheu:2013_APL}%
  \BibitemOpen
  \bibfield{author}{%
  \bibinfo {author} {\bibnamefont{Sheu}, \bibfnamefont{Jinn-Kong}}, \bibinfo
  {author} {\bibfnamefont{Feng-Wen}\ \bibnamefont{Huang}}, \bibinfo {author}
  {\bibfnamefont{Chia-Hui}\ \bibnamefont{Lee}}, \bibinfo {author}
  {\bibfnamefont{Ming-Lun}\ \bibnamefont{Lee}}, \bibinfo {author}
  {\bibfnamefont{Yu-Hsiang}\ \bibnamefont{Yeh}}, \bibinfo {author}
  {\bibfnamefont{Po-Cheng}\ \bibnamefont{Chen}},\ and\ \bibinfo {author}
  {\bibfnamefont{Wei-Chih}\ \bibnamefont{Lai}}}%
  , \bibinfo {year} {2013},\ \bibfield{title}{%
  \enquote{\bibinfo {title} {Improved conversion efficiency of {GaN}-based
  solar cells with {Mn}-doped absorption layer},}\ }%
  \bibfield{journal}{%
  \bibinfo {journal} {Appl. Phys. Lett.}\ }%
  \textbf{\bibinfo {volume} {103}},\ \bibinfo {pages} {063906}%
  \bibAnnoteFile{NoStop}{Sheu:2013_APL}%
\bibitem[{\citenamefont{Shono}\ \emph{et~al.}(2000)\citenamefont{Shono},
  \citenamefont{Hasegawa}, \citenamefont{Fukumura}, \citenamefont{Matsukura},\
  and\ \citenamefont{Ohno}}]{Shono:2000_APL}%
  \BibitemOpen
  \bibfield{author}{%
  \bibinfo {author} {\bibnamefont{Shono}, \bibfnamefont{T.}}, \bibinfo {author}
  {\bibfnamefont{T.}~\bibnamefont{Hasegawa}}, \bibinfo {author}
  {\bibfnamefont{T.}~\bibnamefont{Fukumura}}, \bibinfo {author}
  {\bibfnamefont{F.}~\bibnamefont{Matsukura}},\ and\ \bibinfo {author}
  {\bibfnamefont{H.}~\bibnamefont{Ohno}}}%
  , \bibinfo {year} {2000},\ \bibfield{title}{%
  \enquote{\bibinfo {title} {Observation of magnetic domain structure in a
  ferromagnetic semiconductor {(Ga, Mn)As} with a scanning {Hall} probe
  microscope},}\ }%
  \bibfield{journal}{%
  \bibinfo {journal} {Appl. Phys. Lett.}\ }%
  \textbf{\bibinfo {volume} {77}},\ \bibinfo {pages} {1363}%
  \bibAnnoteFile{NoStop}{Shono:2000_APL}%
\bibitem[{\citenamefont{Sinova}\ \emph{et~al.}(2002)\citenamefont{Sinova},
  \citenamefont{Jungwirth}, \citenamefont{Yang}, \citenamefont{{Ku\v{c}era}},\
  and\ \citenamefont{MacDonald}}]{Sinova:2002_PRB}%
  \BibitemOpen
  \bibfield{author}{%
  \bibinfo {author} {\bibnamefont{Sinova}, \bibfnamefont{J.}}, \bibinfo
  {author} {\bibfnamefont{T.}~\bibnamefont{Jungwirth}}, \bibinfo {author}
  {\bibfnamefont{S.-R.~Eric}\ \bibnamefont{Yang}}, \bibinfo {author}
  {\bibfnamefont{J.}~\bibnamefont{{Ku\v{c}era}}},\ and\ \bibinfo {author}
  {\bibfnamefont{A.~H.}\ \bibnamefont{MacDonald}}}%
  , \bibinfo {year} {2002},\ \bibfield{title}{%
  \enquote{\bibinfo {title} {Infrared conductivity of metallic {(III,Mn)V}
  ferromagnets},}\ }%
  \bibfield{journal}{%
  \bibinfo {journal} {Phys. Rev. B}\ }%
  \textbf{\bibinfo {volume} {66}},\ \bibinfo {pages} {041202}%
  \bibAnnoteFile{NoStop}{Sinova:2002_PRB}%
\bibitem[{\citenamefont{Sinova}\ \emph{et~al.}(2004)\citenamefont{Sinova},
  \citenamefont{Jungwirth}, \citenamefont{Liu}, \citenamefont{Sasaki},
  \citenamefont{Furdyna}, \citenamefont{Atkinson},\ and\
  \citenamefont{MacDonald}}]{Sinova:2004_PRB}%
  \BibitemOpen
  \bibfield{author}{%
  \bibinfo {author} {\bibnamefont{Sinova}, \bibfnamefont{Jairo}}, \bibinfo
  {author} {\bibfnamefont{T.}~\bibnamefont{Jungwirth}}, \bibinfo {author}
  {\bibfnamefont{X.}~\bibnamefont{Liu}}, \bibinfo {author}
  {\bibfnamefont{Y.}~\bibnamefont{Sasaki}}, \bibinfo {author}
  {\bibfnamefont{J.~K.}\ \bibnamefont{Furdyna}}, \bibinfo {author}
  {\bibfnamefont{W.~A.}\ \bibnamefont{Atkinson}},\ and\ \bibinfo {author}
  {\bibfnamefont{A.~H.}\ \bibnamefont{MacDonald}}}%
  , \bibinfo {year} {2004},\ \bibfield{title}{%
  \enquote{\bibinfo {title} {Magnetization relaxation in {(Ga,Mn)As}
  ferromagnetic semiconductors},}\ }%
  \bibfield{journal}{%
  \bibinfo {journal} {Phys. Rev. B}\ }%
  \textbf{\bibinfo {volume} {69}},\ \bibinfo {pages} {085209}%
  \bibAnnoteFile{NoStop}{Sinova:2004_PRB}%
\bibitem[{\citenamefont{{\'S}liwa}\ and\
  \citenamefont{Dietl}(2006)}]{Sliwa:2006_PRB}%
  \BibitemOpen
  \bibfield{author}{%
  \bibinfo {author} {\bibnamefont{{\'S}liwa}, \bibfnamefont{C.}},\ and\
  \bibinfo {author} {\bibfnamefont{T.}~\bibnamefont{Dietl}}}%
  , \bibinfo {year} {2006},\ \bibfield{title}{%
  \enquote{\bibinfo {title} {Magnitude and crystalline anisotropy of hole
  magnetization in {(Ga,Mn)As}},}\ }%
  \bibfield{journal}{%
  \bibinfo {journal} {Phys. Rev. B}\ }%
  \textbf{\bibinfo {volume} {74}},\ \bibinfo {pages} {245215}%
  \bibAnnoteFile{NoStop}{Sliwa:2006_PRB}%
\bibitem[{\citenamefont{{\'S}liwa}\ and\
  \citenamefont{Dietl}(2011)}]{Sliwa:2011_PRB}%
  \BibitemOpen
  \bibfield{author}{%
  \bibinfo {author} {\bibnamefont{{\'S}liwa}, \bibfnamefont{C.}},\ and\
  \bibinfo {author} {\bibfnamefont{T.}~\bibnamefont{Dietl}}}%
  , \bibinfo {year} {2011},\ \bibfield{title}{%
  \enquote{\bibinfo {title} {Thermodynamic and thermoelectric properties of
  {(Ga,Mn)As} and related compounds},}\ }%
  \bibfield{journal}{%
  \bibinfo {journal} {Phys. Rev. B}\ }%
  \textbf{\bibinfo {volume} {83}},\ \bibinfo {pages} {245210}%
  \bibAnnoteFile{NoStop}{Sliwa:2011_PRB}%
\bibitem[{\citenamefont{{\'S}liwa}\ and\
  \citenamefont{Dietl}(2013)}]{Sliwa:2013_arXiv}%
  \BibitemOpen
  \bibfield{author}{%
  \bibinfo {author} {\bibnamefont{{\'S}liwa}, \bibfnamefont{C.}},\ and\
  \bibinfo {author} {\bibfnamefont{T.}~\bibnamefont{Dietl}}}%
  , \bibinfo {year} {2013},\ \bibfield{title}{%
  \enquote{\bibinfo {title} {Orbital magnetization in dilute ferromagnetic
  semiconductors},}\ }%
  \bibinfo {journal} {arXiv:1312.xxxx}%
  \bibAnnoteFile{NoStop}{Sliwa:2013_arXiv}%
\bibitem[{\citenamefont{Song}\ \emph{et~al.}(2011)\citenamefont{Song},
  \citenamefont{Sperl}, \citenamefont{Utz}, \citenamefont{Ciorga},
  \citenamefont{Woltersdorf}, \citenamefont{Schuh}, \citenamefont{Bougeard},
  \citenamefont{Back},\ and\ \citenamefont{Weiss}}]{Song:2011_PRL}%
  \BibitemOpen
\bibfield{journal}{%
    }%
  \bibfield{author}{%
  \bibinfo {author} {\bibnamefont{Song}, \bibfnamefont{C.}}, \bibinfo {author}
  {\bibfnamefont{M.}~\bibnamefont{Sperl}}, \bibinfo {author}
  {\bibfnamefont{M.}~\bibnamefont{Utz}}, \bibinfo {author}
  {\bibfnamefont{M.}~\bibnamefont{Ciorga}}, \bibinfo {author}
  {\bibfnamefont{G.}~\bibnamefont{Woltersdorf}}, \bibinfo {author}
  {\bibfnamefont{D.}~\bibnamefont{Schuh}}, \bibinfo {author}
  {\bibfnamefont{D.}~\bibnamefont{Bougeard}}, \bibinfo {author}
  {\bibfnamefont{C.~H.}\ \bibnamefont{Back}},\ and\ \bibinfo {author}
  {\bibfnamefont{D.}~\bibnamefont{Weiss}}}%
  , \bibinfo {year} {2011},\ \bibfield{title}{%
  \enquote{\bibinfo {title} {Proximity induced enhancement of the {Curie}
  temperature in hybrid spin injection devices},}\ }%
  \bibfield{journal}{%
  \bibinfo {journal} {Phys. Rev. Lett.}\ }%
  \textbf{\bibinfo {volume} {107}},\ \bibinfo {pages} {056601}%
  \bibAnnoteFile{NoStop}{Song:2011_PRL}%
\bibitem[{\citenamefont{Sonoda}\ \emph{et~al.}(2006)\citenamefont{Sonoda},
  \citenamefont{Tanaka}, \citenamefont{Ikeno}, \citenamefont{Yamamoto},
  \citenamefont{Oba}, \citenamefont{Araki}, \citenamefont{Yamamoto},
  \citenamefont{Suga}, \citenamefont{Nanishi}, \citenamefont{Akasaka},
  \citenamefont{Kindo},\ and\ \citenamefont{Hori}}]{Sonoda:2006_JPCM}%
  \BibitemOpen
  \bibfield{author}{%
  \bibinfo {author} {\bibnamefont{Sonoda}, \bibfnamefont{S.}}, \bibinfo
  {author} {\bibfnamefont{I.}~\bibnamefont{Tanaka}}, \bibinfo {author}
  {\bibfnamefont{H.}~\bibnamefont{Ikeno}}, \bibinfo {author}
  {\bibfnamefont{T.}~\bibnamefont{Yamamoto}}, \bibinfo {author}
  {\bibfnamefont{F.}~\bibnamefont{Oba}}, \bibinfo {author}
  {\bibfnamefont{T.}~\bibnamefont{Araki}}, \bibinfo {author}
  {\bibfnamefont{Y.}~\bibnamefont{Yamamoto}}, \bibinfo {author}
  {\bibfnamefont{K.}~\bibnamefont{Suga}}, \bibinfo {author}
  {\bibfnamefont{Y.}~\bibnamefont{Nanishi}}, \bibinfo {author}
  {\bibfnamefont{Y.}~\bibnamefont{Akasaka}}, \bibinfo {author}
  {\bibfnamefont{K.}~\bibnamefont{Kindo}},\ and\ \bibinfo {author}
  {\bibfnamefont{H.}~\bibnamefont{Hori}}}%
  , \bibinfo {year} {2006},\ \bibfield{title}{%
  \enquote{\bibinfo {title} {Coexistence of {Mn$^{2+}$} and {Mn$^{3+}$} in
  ferromagnetic {GaMnN}},}\ }%
  \bibfield{journal}{%
  \bibinfo {journal} {J. Phys. Condens. Matter}\ }%
  \textbf{\bibinfo {volume} {18}},\ \bibinfo {pages} {4615}%
  \bibAnnoteFile{NoStop}{Sonoda:2006_JPCM}%
\bibitem[{\citenamefont{{S\o{}rensen}}\
  \emph{et~al.}(2003)\citenamefont{{S\o{}rensen}}, \citenamefont{Lindelof},
  \citenamefont{Sadowski}, \citenamefont{Mathieu},\ and\
  \citenamefont{Svedlindh}}]{Sorensen:2003_APL}%
  \BibitemOpen
  \bibfield{author}{%
  \bibinfo {author} {\bibnamefont{{S\o{}rensen}}, \bibfnamefont{B.}}, \bibinfo
  {author} {\bibfnamefont{P.~E.}\ \bibnamefont{Lindelof}}, \bibinfo {author}
  {\bibfnamefont{J.}~\bibnamefont{Sadowski}}, \bibinfo {author}
  {\bibfnamefont{R.}~\bibnamefont{Mathieu}},\ and\ \bibinfo {author}
  {\bibfnamefont{P.}~\bibnamefont{Svedlindh}}}%
  , \bibinfo {year} {2003},\ \bibfield{title}{%
  \enquote{\bibinfo {title} {Effect of annealing on carrier density and {Curie}
  temperature in epitaxial {(Ga,Mn)As} thin films},}\ }%
  \bibfield{journal}{%
  \bibinfo {journal} {Appl. Phys. Lett.}\ }%
  \textbf{\bibinfo {volume} {82}},\ \bibinfo {pages} {2287}%
  \bibAnnoteFile{NoStop}{Sorensen:2003_APL}%
\bibitem[{\citenamefont{{Spa{\l}ek}}\
  \emph{et~al.}(1986)\citenamefont{{Spa{\l}ek}}, \citenamefont{Lewicki},
  \citenamefont{Tarnawski}, \citenamefont{Furdyna},
  \citenamefont{Ga{\l}{\c{a}}zka},\ and\
  \citenamefont{Obuszko}}]{Spalek:1986_PRB}%
  \BibitemOpen
  \bibfield{author}{%
  \bibinfo {author} {\bibnamefont{{Spa{\l}ek}}, \bibfnamefont{J.}}, \bibinfo
  {author} {\bibfnamefont{A.}~\bibnamefont{Lewicki}}, \bibinfo {author}
  {\bibfnamefont{Z.}~\bibnamefont{Tarnawski}}, \bibinfo {author}
  {\bibfnamefont{J.~K.}\ \bibnamefont{Furdyna}}, \bibinfo {author}
  {\bibfnamefont{R.~R.}\ \bibnamefont{Ga{\l}{\c{a}}zka}},\ and\ \bibinfo
  {author} {\bibfnamefont{Z.}~\bibnamefont{Obuszko}}}%
  , \bibinfo {year} {1986},\ \bibfield{title}{%
  \enquote{\bibinfo {title} {Magnetic susceptibility of semimagnetic
  semiconductors: The high-temperature regime and the role of superexchange},}\
  }%
  \bibfield{journal}{%
  \bibinfo {journal} {Phys. Rev. B}\ }%
  \textbf{\bibinfo {volume} {33}},\ \bibinfo {pages} {3407}%
  \bibAnnoteFile{NoStop}{Spalek:1986_PRB}%
\bibitem[{\citenamefont{Stefanowicz}\
  \emph{et~al.}(2013)\citenamefont{Stefanowicz}, \citenamefont{Kunert},
  \citenamefont{Simserides}, \citenamefont{Majewski},
  \citenamefont{Stefanowicz}, \citenamefont{Kruse}, \citenamefont{Figge},
  \citenamefont{Li}, \citenamefont{Jakie{\l}a}, \citenamefont{Trohidou},
  \citenamefont{Bonanni}, \citenamefont{Hommel}, \citenamefont{Sawicki},\ and\
  \citenamefont{Dietl}}]{Stefanowicz:2013_PRB}%
  \BibitemOpen
  \bibfield{author}{%
  \bibinfo {author} {\bibnamefont{Stefanowicz}, \bibfnamefont{S.}}, \bibinfo
  {author} {\bibfnamefont{G.}~\bibnamefont{Kunert}}, \bibinfo {author}
  {\bibfnamefont{C.}~\bibnamefont{Simserides}}, \bibinfo {author}
  {\bibfnamefont{J.~A.}\ \bibnamefont{Majewski}}, \bibinfo {author}
  {\bibfnamefont{W.}~\bibnamefont{Stefanowicz}}, \bibinfo {author}
  {\bibfnamefont{C.}~\bibnamefont{Kruse}}, \bibinfo {author}
  {\bibfnamefont{S.}~\bibnamefont{Figge}}, \bibinfo {author}
  {\bibfnamefont{{Tian}}\ \bibnamefont{Li}}, \bibinfo {author}
  {\bibfnamefont{R.}~\bibnamefont{Jakie{\l}a}}, \bibinfo {author}
  {\bibfnamefont{K.~N.}\ \bibnamefont{Trohidou}}, \bibinfo {author}
  {\bibfnamefont{A.}~\bibnamefont{Bonanni}}, \bibinfo {author}
  {\bibfnamefont{D.}~\bibnamefont{Hommel}}, \bibinfo {author}
  {\bibfnamefont{M.}~\bibnamefont{Sawicki}},\ and\ \bibinfo {author}
  {\bibfnamefont{T.}~\bibnamefont{Dietl}}}%
  , \bibinfo {year} {2013},\ \bibfield{title}{%
  \enquote{\bibinfo {title} {Phase diagram and critical behavior of a random
  ferromagnet {Ga$_{1-x}$Mn$_x$N}},}\ }%
  \bibfield{journal}{%
  \bibinfo {journal} {Phys. Rev. B}\ }%
  \textbf{\bibinfo {volume} {88}},\ \bibinfo {pages} {081201(R)}%
  \bibAnnoteFile{NoStop}{Stefanowicz:2013_PRB}%
\bibitem[{\citenamefont{Stefanowicz}\
  \emph{et~al.}(2010{\natexlab{a}})\citenamefont{Stefanowicz},
  \citenamefont{C.}, \citenamefont{{\'S}liwa}, \citenamefont{Aleshkievych},
  \citenamefont{Dietl}, \citenamefont{D{\"o}ppe},
  \citenamefont{W{\"u}rstbauer}, \citenamefont{Wegscheider},
  \citenamefont{Weiss},\ and\ \citenamefont{Sawicki}}]{Stefanowicz:2010_PRBb}%
  \BibitemOpen
  \bibfield{author}{%
  \bibinfo {author} {\bibnamefont{Stefanowicz}, \bibfnamefont{W.}}, \bibinfo
  {author} {\bibnamefont{C.}}, \bibinfo {author} {\bibnamefont{{\'S}liwa}},
  \bibinfo {author} {\bibfnamefont{P.}~\bibnamefont{Aleshkievych}}, \bibinfo
  {author} {\bibfnamefont{T.}~\bibnamefont{Dietl}}, \bibinfo {author}
  {\bibfnamefont{M.}~\bibnamefont{D{\"o}ppe}}, \bibinfo {author}
  {\bibfnamefont{U.}~\bibnamefont{W{\"u}rstbauer}}, \bibinfo {author}
  {\bibfnamefont{W.}~\bibnamefont{Wegscheider}}, \bibinfo {author}
  {\bibfnamefont{D.}~\bibnamefont{Weiss}},\ and\ \bibinfo {author}
  {\bibfnamefont{M.}~\bibnamefont{Sawicki}}}%
  , \bibinfo {year} {2010}{\natexlab{a}},\ \bibfield{title}{%
  \enquote{\bibinfo {title} {Magnetic anisotropy of epitaxial {(Ga,Mn)As} on
  {(113)A} {GaAs}},}\ }%
  \bibfield{journal}{%
  \bibinfo {journal} {Phys. Rev. B}\ }%
  \textbf{\bibinfo {volume} {81}},\ \bibinfo {pages} {155203}%
  \bibAnnoteFile{NoStop}{Stefanowicz:2010_PRBb}%
\bibitem[{\citenamefont{Stefanowicz}\
  \emph{et~al.}(2010{\natexlab{b}})\citenamefont{Stefanowicz},
  \citenamefont{Sztenkiel}, \citenamefont{Faina}, \citenamefont{Grois},
  \citenamefont{Rovezzi}, \citenamefont{Devillers},
  \citenamefont{Navarro-Quezada}, \citenamefont{Li}, \citenamefont{Jakie{\l}a},
  \citenamefont{Sawicki}, \citenamefont{Dietl},\ and\
  \citenamefont{Bonanni}}]{Stefanowicz:2010_PRBa}%
  \BibitemOpen
  \bibfield{author}{%
  \bibinfo {author} {\bibnamefont{Stefanowicz}, \bibfnamefont{W.}}, \bibinfo
  {author} {\bibfnamefont{D.}~\bibnamefont{Sztenkiel}}, \bibinfo {author}
  {\bibfnamefont{B.}~\bibnamefont{Faina}}, \bibinfo {author}
  {\bibfnamefont{A.}~\bibnamefont{Grois}}, \bibinfo {author}
  {\bibfnamefont{M.}~\bibnamefont{Rovezzi}}, \bibinfo {author}
  {\bibfnamefont{T.}~\bibnamefont{Devillers}}, \bibinfo {author}
  {\bibfnamefont{Andrea}\ \bibnamefont{Navarro-Quezada}}, \bibinfo {author}
  {\bibfnamefont{{Tian}}\ \bibnamefont{Li}}, \bibinfo {author}
  {\bibfnamefont{Rafa{\l}}\ \bibnamefont{Jakie{\l}a}}, \bibinfo {author}
  {\bibfnamefont{Maciej}\ \bibnamefont{Sawicki}}, \bibinfo {author}
  {\bibfnamefont{T.}~\bibnamefont{Dietl}},\ and\ \bibinfo {author}
  {\bibfnamefont{A.}~\bibnamefont{Bonanni}}}%
  , \bibinfo {year} {2010}{\natexlab{b}},\ \bibfield{title}{%
  \enquote{\bibinfo {title} {Structural and paramagnetic properties of dilute
  {Ga$_{1-x}$Mn$_{x}$N}},}\ }%
  \bibfield{journal}{%
  \bibinfo {journal} {Phys. Rev. B}\ }%
  \textbf{\bibinfo {volume} {81}},\ \bibinfo {pages} {235210}%
  \bibAnnoteFile{NoStop}{Stefanowicz:2010_PRBa}%
\bibitem[{\citenamefont{Stolichnov}\
  \emph{et~al.}(2011)\citenamefont{Stolichnov}, \citenamefont{Riester},
  \citenamefont{Mikheev}, \citenamefont{Setter}, \citenamefont{Rushforth},
  \citenamefont{Edmonds}, \citenamefont{Campion}, \citenamefont{Foxon},
  \citenamefont{Gallagher}, \citenamefont{Jungwirth},\ and\
  \citenamefont{Trodahl}}]{Stolichnov:2011_PRB}%
  \BibitemOpen
  \bibfield{author}{%
  \bibinfo {author} {\bibnamefont{Stolichnov}, \bibfnamefont{I.}}, \bibinfo
  {author} {\bibfnamefont{S.~W.~E.}\ \bibnamefont{Riester}}, \bibinfo {author}
  {\bibfnamefont{E.}~\bibnamefont{Mikheev}}, \bibinfo {author}
  {\bibfnamefont{N.}~\bibnamefont{Setter}}, \bibinfo {author}
  {\bibfnamefont{A.~W.}\ \bibnamefont{Rushforth}}, \bibinfo {author}
  {\bibfnamefont{K.~W.}\ \bibnamefont{Edmonds}}, \bibinfo {author}
  {\bibfnamefont{R.~P.}\ \bibnamefont{Campion}}, \bibinfo {author}
  {\bibfnamefont{C.~T.}\ \bibnamefont{Foxon}}, \bibinfo {author}
  {\bibfnamefont{B.~L.}\ \bibnamefont{Gallagher}}, \bibinfo {author}
  {\bibfnamefont{T.}~\bibnamefont{Jungwirth}},\ and\ \bibinfo {author}
  {\bibfnamefont{H.~J.}\ \bibnamefont{Trodahl}}}%
  , \bibinfo {year} {2011},\ \bibfield{title}{%
  \enquote{\bibinfo {title} {Enhanced {Curie} temperature and nonvolatile
  switching of ferromagnetism in ultrathin {(Ga,Mn)As} channels},}\ }%
  \bibfield{journal}{%
  \bibinfo {journal} {Phys. Rev. B}\ }%
  \textbf{\bibinfo {volume} {83}},\ \bibinfo {pages} {115203}%
  \bibAnnoteFile{NoStop}{Stolichnov:2011_PRB}%
\bibitem[{\citenamefont{Stolichnov}\
  \emph{et~al.}(2008)\citenamefont{Stolichnov}, \citenamefont{Riester},
  \citenamefont{Trodahl}, \citenamefont{Setter}, \citenamefont{Rushforth},
  \citenamefont{Edmonds}, \citenamefont{Campion}, \citenamefont{Foxon},
  \citenamefont{Gallagher},\ and\
  \citenamefont{Jungwirth}}]{Stolichnov:2008_NM}%
  \BibitemOpen
  \bibfield{author}{%
  \bibinfo {author} {\bibnamefont{Stolichnov}, \bibfnamefont{I.}}, \bibinfo
  {author} {\bibfnamefont{S.~W.~E.}\ \bibnamefont{Riester}}, \bibinfo {author}
  {\bibfnamefont{H.~J.}\ \bibnamefont{Trodahl}}, \bibinfo {author}
  {\bibfnamefont{N.}~\bibnamefont{Setter}}, \bibinfo {author}
  {\bibfnamefont{A.~W.}\ \bibnamefont{Rushforth}}, \bibinfo {author}
  {\bibfnamefont{K.~W.}\ \bibnamefont{Edmonds}}, \bibinfo {author}
  {\bibfnamefont{R.~P.}\ \bibnamefont{Campion}}, \bibinfo {author}
  {\bibfnamefont{C.~T.}\ \bibnamefont{Foxon}}, \bibinfo {author}
  {\bibfnamefont{B.~L.}\ \bibnamefont{Gallagher}},\ and\ \bibinfo {author}
  {\bibfnamefont{T.}~\bibnamefont{Jungwirth}}}%
  , \bibinfo {year} {2008},\ \bibfield{title}{%
  \enquote{\bibinfo {title} {Non-volatile ferroelectric control of
  ferromagnetism in {(Ga, Mn)As}},}\ }%
  \bibfield{journal}{%
  \bibinfo {journal} {Nat. Mater.}\ }%
  \textbf{\bibinfo {volume} {7}},\ \bibinfo {pages} {464}%
  \bibAnnoteFile{NoStop}{Stolichnov:2008_NM}%
\bibitem[{\citenamefont{Stone}\ \emph{et~al.}(2008)\citenamefont{Stone},
  \citenamefont{Alberi}, \citenamefont{Tardif}, \citenamefont{Beeman},
  \citenamefont{Yu}, \citenamefont{Walukiewicz},\ and\
  \citenamefont{Dubon}}]{Stone:2008_PRL}%
  \BibitemOpen
  \bibfield{author}{%
  \bibinfo {author} {\bibnamefont{Stone}, \bibfnamefont{P.~R.}}, \bibinfo
  {author} {\bibfnamefont{K.}~\bibnamefont{Alberi}}, \bibinfo {author}
  {\bibfnamefont{S.~K.~Z.}\ \bibnamefont{Tardif}}, \bibinfo {author}
  {\bibfnamefont{J.~W.}\ \bibnamefont{Beeman}}, \bibinfo {author}
  {\bibfnamefont{K.~M.}\ \bibnamefont{Yu}}, \bibinfo {author}
  {\bibfnamefont{W.}~\bibnamefont{Walukiewicz}},\ and\ \bibinfo {author}
  {\bibfnamefont{O.~D.}\ \bibnamefont{Dubon}}}%
  , \bibinfo {year} {2008},\ \bibfield{title}{%
  \enquote{\bibinfo {title} {Metal-insulator transition by isovalent anion
  substitution in {Ga$_{1-x}$Mn$_{x}$As}: {Implications} to ferromagnetism},}\
  }%
  \bibfield{journal}{%
  \bibinfo {journal} {Phys. Rev. Lett.}\ }%
  \textbf{\bibinfo {volume} {101}},\ \bibinfo {pages} {087203}%
  \bibAnnoteFile{NoStop}{Stone:2008_PRL}%
\bibitem[{\citenamefont{Stone}\ \emph{et~al.}(2006)\citenamefont{Stone},
  \citenamefont{Scarpulla}, \citenamefont{Farshchi}, \citenamefont{Sharp},
  \citenamefont{Haller}, \citenamefont{Dubon}, \citenamefont{Yu},
  \citenamefont{Beeman}, \citenamefont{Arenholz}, \citenamefont{Denlinger},\
  and\ \citenamefont{Ohldag}}]{Stone:2006_APL}%
  \BibitemOpen
  \bibfield{author}{%
  \bibinfo {author} {\bibnamefont{Stone}, \bibfnamefont{P.~R.}}, \bibinfo
  {author} {\bibfnamefont{M.~A.}\ \bibnamefont{Scarpulla}}, \bibinfo {author}
  {\bibfnamefont{R.}~\bibnamefont{Farshchi}}, \bibinfo {author}
  {\bibfnamefont{I.~D.}\ \bibnamefont{Sharp}}, \bibinfo {author}
  {\bibfnamefont{E.~E.}\ \bibnamefont{Haller}}, \bibinfo {author}
  {\bibfnamefont{O.~D.}\ \bibnamefont{Dubon}}, \bibinfo {author}
  {\bibfnamefont{K.~M.}\ \bibnamefont{Yu}}, \bibinfo {author}
  {\bibfnamefont{J.~W.}\ \bibnamefont{Beeman}}, \bibinfo {author}
  {\bibfnamefont{E.}~\bibnamefont{Arenholz}}, \bibinfo {author}
  {\bibfnamefont{J.~D.}\ \bibnamefont{Denlinger}},\ and\ \bibinfo {author}
  {\bibfnamefont{H.}~\bibnamefont{Ohldag}}}%
  , \bibinfo {year} {2006},\ \bibfield{title}{%
  \enquote{\bibinfo {title} {{Mn L$_{3,2}$} x-ray absorption and magnetic
  circular dichroism in ferromagnetic {Ga$_{1 - x}$Mn$_x$P}},}\ }%
  \bibfield{journal}{%
  \bibinfo {journal} {Appl. Phys. Lett.}\ }%
  \textbf{\bibinfo {volume} {89}},\ \bibinfo {pages} {012504}%
  \bibAnnoteFile{NoStop}{Stone:2006_APL}%
\bibitem[{\citenamefont{Story}\ \emph{et~al.}(1986)\citenamefont{Story},
  \citenamefont{Ga{\l}{\c{a}}zka}, \citenamefont{Frankel},\ and\
  \citenamefont{Wolff}}]{Story:1986_PRL}%
  \BibitemOpen
  \bibfield{author}{%
  \bibinfo {author} {\bibnamefont{Story}, \bibfnamefont{T.}}, \bibinfo {author}
  {\bibfnamefont{R.~R.}\ \bibnamefont{Ga{\l}{\c{a}}zka}}, \bibinfo {author}
  {\bibfnamefont{R.~B.}\ \bibnamefont{Frankel}},\ and\ \bibinfo {author}
  {\bibfnamefont{P.~A.}\ \bibnamefont{Wolff}}}%
  , \bibinfo {year} {1986},\ \bibfield{title}{%
  \enquote{\bibinfo {title} {Carrier-concentration--induced ferromagnetism in
  {PbSnMnTe}},}\ }%
  \bibfield{journal}{%
  \bibinfo {journal} {Phys. Rev. Lett.}\ }%
  \textbf{\bibinfo {volume} {56}},\ \bibinfo {pages} {777}%
  \bibAnnoteFile{NoStop}{Story:1986_PRL}%
\bibitem[{\citenamefont{Strandberg}\
  \emph{et~al.}(2010)\citenamefont{Strandberg}, \citenamefont{Canali},\ and\
  \citenamefont{MacDonald}}]{Strandberg:2010_PRB}%
  \BibitemOpen
  \bibfield{author}{%
  \bibinfo {author} {\bibnamefont{Strandberg}, \bibfnamefont{T.~O.}}, \bibinfo
  {author} {\bibfnamefont{C.~M.}\ \bibnamefont{Canali}},\ and\ \bibinfo
  {author} {\bibfnamefont{A.~H.}\ \bibnamefont{MacDonald}}}%
  , \bibinfo {year} {2010},\ \bibfield{title}{%
  \enquote{\bibinfo {title} {Magnetic interactions of substitutional {Mn} pairs
  in {GaAs}},}\ }%
  \bibfield{journal}{%
  \bibinfo {journal} {Phys. Rev. B}\ }%
  \textbf{\bibinfo {volume} {81}},\ \bibinfo {pages} {054401}%
  \bibAnnoteFile{NoStop}{Strandberg:2010_PRB}%
\bibitem[{\citenamefont{Stroppa}\ and\
  \citenamefont{Kresse}(2009)}]{Stroppa:2009_PRB}%
  \BibitemOpen
  \bibfield{author}{%
  \bibinfo {author} {\bibnamefont{Stroppa}, \bibfnamefont{A.}},\ and\ \bibinfo
  {author} {\bibfnamefont{G.}~\bibnamefont{Kresse}}}%
  , \bibinfo {year} {2009},\ \bibfield{title}{%
  \enquote{\bibinfo {title} {Unraveling the {Jahn-Teller} effect in {Mn}-doped
  {GaN} using the {Heyd-Scuseria-Ernzerhof} hybrid functional},}\ }%
  \bibfield{journal}{%
  \bibinfo {journal} {Phys. Rev. B}\ }%
  \textbf{\bibinfo {volume} {79}},\ \bibinfo {pages} {201201(R)}%
  \bibAnnoteFile{NoStop}{Stroppa:2009_PRB}%
\bibitem[{\citenamefont{Suffczy{\'n}ski}\
  \emph{et~al.}(2011)\citenamefont{Suffczy{\'n}ski}, \citenamefont{Grois},
  \citenamefont{Pacuski}, \citenamefont{Golnik}, \citenamefont{Gaj},
  \citenamefont{Navarro-Quezada}, \citenamefont{Faina},
  \citenamefont{Devillers},\ and\
  \citenamefont{Bonanni}}]{Suffczynski:2011_PRB}%
  \BibitemOpen
  \bibfield{author}{%
  \bibinfo {author} {\bibnamefont{Suffczy{\'n}ski}, \bibfnamefont{J.}},
  \bibinfo {author} {\bibfnamefont{A.}~\bibnamefont{Grois}}, \bibinfo {author}
  {\bibfnamefont{W.}~\bibnamefont{Pacuski}}, \bibinfo {author}
  {\bibfnamefont{A.}~\bibnamefont{Golnik}}, \bibinfo {author}
  {\bibfnamefont{J.~A.}\ \bibnamefont{Gaj}}, \bibinfo {author}
  {\bibfnamefont{A.}~\bibnamefont{Navarro-Quezada}}, \bibinfo {author}
  {\bibfnamefont{B.}~\bibnamefont{Faina}}, \bibinfo {author}
  {\bibfnamefont{T.}~\bibnamefont{Devillers}},\ and\ \bibinfo {author}
  {\bibfnamefont{A.}~\bibnamefont{Bonanni}}}%
  , \bibinfo {year} {2011},\ \bibfield{title}{%
  \enquote{\bibinfo {title} {Effects of $s$,$p$-$d$ and $s$-$p$ exchange
  interactions probed by exciton magnetospectroscopy in {(Ga,Mn)N}},}\ }%
  \bibfield{journal}{%
  \bibinfo {journal} {Phys. Rev. B}\ }%
  \textbf{\bibinfo {volume} {83}},\ \bibinfo {pages} {094421}%
  \bibAnnoteFile{NoStop}{Suffczynski:2011_PRB}%
\bibitem[{\citenamefont{Sugawara}\ \emph{et~al.}(2008)\citenamefont{Sugawara},
  \citenamefont{Kasai}, \citenamefont{Tonomura}, \citenamefont{Brown},
  \citenamefont{Campion}, \citenamefont{Edmonds}, \citenamefont{Gallagher},
  \citenamefont{Zemen},\ and\ \citenamefont{Jungwirth}}]{Sugawara:2008_PRL}%
  \BibitemOpen
  \bibfield{author}{%
  \bibinfo {author} {\bibnamefont{Sugawara}, \bibfnamefont{A.}}, \bibinfo
  {author} {\bibfnamefont{H.}~\bibnamefont{Kasai}}, \bibinfo {author}
  {\bibfnamefont{A.}~\bibnamefont{Tonomura}}, \bibinfo {author}
  {\bibfnamefont{P.~D.}\ \bibnamefont{Brown}}, \bibinfo {author}
  {\bibfnamefont{R.~P.}\ \bibnamefont{Campion}}, \bibinfo {author}
  {\bibfnamefont{K.~W.}\ \bibnamefont{Edmonds}}, \bibinfo {author}
  {\bibfnamefont{B.~L.}\ \bibnamefont{Gallagher}}, \bibinfo {author}
  {\bibfnamefont{J.}~\bibnamefont{Zemen}},\ and\ \bibinfo {author}
  {\bibfnamefont{T.}~\bibnamefont{Jungwirth}}}%
  , \bibinfo {year} {2008},\ \bibfield{title}{%
  \enquote{\bibinfo {title} {Domain walls in the {(Ga,Mn)As} diluted magnetic
  semiconductor},}\ }%
  \bibfield{journal}{%
  \bibinfo {journal} {Phys. Rev. Lett.}\ }%
  \textbf{\bibinfo {volume} {100}},\ \bibinfo {pages} {047202}%
  \bibAnnoteFile{NoStop}{Sugawara:2008_PRL}%
\bibitem[{\citenamefont{Szczytko}\ \emph{et~al.}(2001)\citenamefont{Szczytko},
  \citenamefont{Bardyszewski},\ and\
  \citenamefont{Twardowski}}]{Szczytko:2001_PRB}%
  \BibitemOpen
  \bibfield{author}{%
  \bibinfo {author} {\bibnamefont{Szczytko}, \bibfnamefont{J.}}, \bibinfo
  {author} {\bibfnamefont{W.}~\bibnamefont{Bardyszewski}},\ and\ \bibinfo
  {author} {\bibfnamefont{A.}~\bibnamefont{Twardowski}}}%
  , \bibinfo {year} {2001},\ \bibfield{title}{%
  \enquote{\bibinfo {title} {Optical absorption in random media: Application to
  {Ga$_{1-x}$Mn$_{x}$As} epilayers},}\ }%
  \bibfield{journal}{%
  \bibinfo {journal} {Phys. Rev. B}\ }%
  \textbf{\bibinfo {volume} {64}},\ \bibinfo {pages} {075306}%
  \bibAnnoteFile{NoStop}{Szczytko:2001_PRB}%
\bibitem[{\citenamefont{Szczytko}\ \emph{et~al.}(1999)\citenamefont{Szczytko},
  \citenamefont{Mac}, \citenamefont{Twardowski}, \citenamefont{Matsukura},\
  and\ \citenamefont{Ohno}}]{Szczytko:1999_PRB}%
  \BibitemOpen
  \bibfield{author}{%
  \bibinfo {author} {\bibnamefont{Szczytko}, \bibfnamefont{J.}}, \bibinfo
  {author} {\bibfnamefont{W.}~\bibnamefont{Mac}}, \bibinfo {author}
  {\bibfnamefont{A.}~\bibnamefont{Twardowski}}, \bibinfo {author}
  {\bibfnamefont{F.}~\bibnamefont{Matsukura}},\ and\ \bibinfo {author}
  {\bibfnamefont{H.}~\bibnamefont{Ohno}}}%
  , \bibinfo {year} {1999},\ \bibfield{title}{%
  \enquote{\bibinfo {title} {Antiferromagnetic $p-d$ exchange in ferromagnetic
  {Ga$_{1-x}$Mn$_{x}$As} epilayers},}\ }%
  \bibfield{journal}{%
  \bibinfo {journal} {Phys. Rev. B}\ }%
  \textbf{\bibinfo {volume} {59}},\ \bibinfo {pages} {12935}%
  \bibAnnoteFile{NoStop}{Szczytko:1999_PRB}%
\bibitem[{\citenamefont{Takamura}\ \emph{et~al.}(2002)\citenamefont{Takamura},
  \citenamefont{Matsukura}, \citenamefont{Chiba},\ and\
  \citenamefont{Ohno}}]{Takamura:2002_APL}%
  \BibitemOpen
  \bibfield{author}{%
  \bibinfo {author} {\bibnamefont{Takamura}, \bibfnamefont{K.}}, \bibinfo
  {author} {\bibfnamefont{F.}~\bibnamefont{Matsukura}}, \bibinfo {author}
  {\bibfnamefont{D.}~\bibnamefont{Chiba}},\ and\ \bibinfo {author}
  {\bibfnamefont{H.}~\bibnamefont{Ohno}}}%
  , \bibinfo {year} {2002},\ \bibfield{title}{%
  \enquote{\bibinfo {title} {Magnetic properties of {(Al,Ga,Mn)As}},}\ }%
  \bibfield{journal}{%
  \bibinfo {journal} {Appl. Phys. Lett.}\ }%
  \textbf{\bibinfo {volume} {81}},\ \bibinfo {pages} {2590}%
  \bibAnnoteFile{NoStop}{Takamura:2002_APL}%
\bibitem[{\citenamefont{Takeda}\ \emph{et~al.}(2008)\citenamefont{Takeda},
  \citenamefont{Kobayashi}, \citenamefont{Okane}, \citenamefont{Ohkochi},
  \citenamefont{Okamoto}, \citenamefont{Saitoh}, \citenamefont{Kobayashi},
  \citenamefont{Yamagami}, \citenamefont{Fujimori}, \citenamefont{Tanaka},
  \citenamefont{Okabayashi}, \citenamefont{Oshima}, \citenamefont{Ohya},
  \citenamefont{Hai},\ and\ \citenamefont{Tanaka}}]{Takeda:2008_PRL}%
  \BibitemOpen
  \bibfield{author}{%
  \bibinfo {author} {\bibnamefont{Takeda}, \bibfnamefont{Y.}}, \bibinfo
  {author} {\bibfnamefont{M.}~\bibnamefont{Kobayashi}}, \bibinfo {author}
  {\bibfnamefont{T.}~\bibnamefont{Okane}}, \bibinfo {author}
  {\bibfnamefont{T.}~\bibnamefont{Ohkochi}}, \bibinfo {author}
  {\bibfnamefont{J.}~\bibnamefont{Okamoto}}, \bibinfo {author}
  {\bibfnamefont{Y.}~\bibnamefont{Saitoh}}, \bibinfo {author}
  {\bibfnamefont{K.}~\bibnamefont{Kobayashi}}, \bibinfo {author}
  {\bibfnamefont{H.}~\bibnamefont{Yamagami}}, \bibinfo {author}
  {\bibfnamefont{A.}~\bibnamefont{Fujimori}}, \bibinfo {author}
  {\bibfnamefont{A.}~\bibnamefont{Tanaka}}, \bibinfo {author}
  {\bibfnamefont{J.}~\bibnamefont{Okabayashi}}, \bibinfo {author}
  {\bibfnamefont{M.}~\bibnamefont{Oshima}}, \bibinfo {author}
  {\bibfnamefont{S.}~\bibnamefont{Ohya}}, \bibinfo {author}
  {\bibfnamefont{P.~N.}\ \bibnamefont{Hai}},\ and\ \bibinfo {author}
  {\bibfnamefont{M.}~\bibnamefont{Tanaka}}}%
  , \bibinfo {year} {2008},\ \bibfield{title}{%
  \enquote{\bibinfo {title} {Nature of magnetic coupling between {Mn} ions in
  as-grown {Ga$_{1-x}$Mn$_{x}$As} studied by {X}-ray magnetic circular
  dichroism},}\ }%
  \bibfield{journal}{%
  \bibinfo {journal} {Phys. Rev. Lett.}\ }%
  \textbf{\bibinfo {volume} {100}},\ \bibinfo {pages} {247202}%
  \bibAnnoteFile{NoStop}{Takeda:2008_PRL}%
\bibitem[{\citenamefont{Tanaka}\ and\
  \citenamefont{Higo}(2001)}]{Tanaka:2001_PRL}%
  \BibitemOpen
  \bibfield{author}{%
  \bibinfo {author} {\bibnamefont{Tanaka}, \bibfnamefont{M.}},\ and\ \bibinfo
  {author} {\bibfnamefont{Y.}~\bibnamefont{Higo}}}%
  , \bibinfo {year} {2001},\ \bibfield{title}{%
  \enquote{\bibinfo {title} {Large tunneling magnetoresistance in
  {GaMnAs/AlAs/GaMnAs} ferromagnetic semiconductor tunnel junctions},}\ }%
  \bibfield{journal}{%
  \bibinfo {journal} {Phys. Rev. Lett.}\ }%
  \textbf{\bibinfo {volume} {87}},\ \bibinfo {pages} {026602}%
  \bibAnnoteFile{NoStop}{Tanaka:2001_PRL}%
\bibitem[{\citenamefont{Tang}\ \emph{et~al.}(2003)\citenamefont{Tang},
  \citenamefont{Kawakami}, \citenamefont{Awschalom},\ and\
  \citenamefont{Roukes}}]{Tang:2003_PRL}%
  \BibitemOpen
  \bibfield{author}{%
  \bibinfo {author} {\bibnamefont{Tang}, \bibfnamefont{H.~X.}}, \bibinfo
  {author} {\bibfnamefont{R.~K.}\ \bibnamefont{Kawakami}}, \bibinfo {author}
  {\bibfnamefont{D.~D.}\ \bibnamefont{Awschalom}},\ and\ \bibinfo {author}
  {\bibfnamefont{M.~L.}\ \bibnamefont{Roukes}}}%
  , \bibinfo {year} {2003},\ \bibfield{title}{%
  \enquote{\bibinfo {title} {Giant planar hall effect in epitaxial {(Ga,Mn)As}
  devices},}\ }%
  \bibfield{journal}{%
  \bibinfo {journal} {Phys. Rev. Lett.}\ }%
  \textbf{\bibinfo {volume} {90}},\ \bibinfo {pages} {107201}%
  \bibAnnoteFile{NoStop}{Tang:2003_PRL}%
\bibitem[{\citenamefont{Tang}\ \emph{et~al.}(2006)\citenamefont{Tang},
  \citenamefont{Kawakami}, \citenamefont{Awschalom},\ and\
  \citenamefont{Roukes}}]{Tang:2006_PRB}%
  \BibitemOpen
  \bibfield{author}{%
  \bibinfo {author} {\bibnamefont{Tang}, \bibfnamefont{H.~X.}}, \bibinfo
  {author} {\bibfnamefont{R.~K.}\ \bibnamefont{Kawakami}}, \bibinfo {author}
  {\bibfnamefont{D.~D.}\ \bibnamefont{Awschalom}},\ and\ \bibinfo {author}
  {\bibfnamefont{M.~L.}\ \bibnamefont{Roukes}}}%
  , \bibinfo {year} {2006},\ \bibfield{title}{%
  \enquote{\bibinfo {title} {Propagation dynamics of individual domain walls in
  {Ga$_{1 - x}$Mn$_x$As} microdevices},}\ }%
  \bibfield{journal}{%
  \bibinfo {journal} {Phys. Rev. B}\ }%
  \textbf{\bibinfo {volume} {74}},\ \bibinfo {pages} {041310}%
  \bibAnnoteFile{NoStop}{Tang:2006_PRB}%
\bibitem[{\citenamefont{Tang}\ \emph{et~al.}(2004)\citenamefont{Tang},
  \citenamefont{Masmanidis}, \citenamefont{Kawakami},
  \citenamefont{Awschalom},\ and\ \citenamefont{Roukes}}]{Tang:2004_N}%
  \BibitemOpen
  \bibfield{author}{%
  \bibinfo {author} {\bibnamefont{Tang}, \bibfnamefont{H.~X.}}, \bibinfo
  {author} {\bibfnamefont{S.}~\bibnamefont{Masmanidis}}, \bibinfo {author}
  {\bibfnamefont{R.~K.}\ \bibnamefont{Kawakami}}, \bibinfo {author}
  {\bibfnamefont{D.~D.}\ \bibnamefont{Awschalom}},\ and\ \bibinfo {author}
  {\bibfnamefont{M.~L.}\ \bibnamefont{Roukes}}}%
  , \bibinfo {year} {2004},\ \bibfield{title}{%
  \enquote{\bibinfo {title} {Negative intrinsic resistivity of an individual
  domain wall in epitaxial {(Ga,Mn)As} microdevices},}\ }%
  \bibfield{journal}{%
  \bibinfo {journal} {Nature}\ }%
  \textbf{\bibinfo {volume} {431}},\ \bibinfo {pages} {52}%
  \bibAnnoteFile{NoStop}{Tang:2004_N}%
\bibitem[{\citenamefont{Tatara}\ and\
  \citenamefont{Kohno}(2004)}]{Tatara:2004_PRL}%
  \BibitemOpen
  \bibfield{author}{%
  \bibinfo {author} {\bibnamefont{Tatara}, \bibfnamefont{G.}},\ and\ \bibinfo
  {author} {\bibfnamefont{H.}~\bibnamefont{Kohno}}}%
  , \bibinfo {year} {2004},\ \bibfield{title}{%
  \enquote{\bibinfo {title} {Theory of current-driven domain wall motion: Spin
  transfer versus momentum transfer},}\ }%
  \bibfield{journal}{%
  \bibinfo {journal} {Phys. Rev. Lett.}\ }%
  \textbf{\bibinfo {volume} {92}},\ \bibinfo {pages} {086601}%
  \bibAnnoteFile{NoStop}{Tatara:2004_PRL}%
\bibitem[{\citenamefont{Terletska}\ and\
  \citenamefont{Dobrosavljevi\'c}(2011)}]{Terletska:2011_PRL}%
  \BibitemOpen
  \bibfield{author}{%
  \bibinfo {author} {\bibnamefont{Terletska}, \bibfnamefont{H.}},\ and\
  \bibinfo {author} {\bibfnamefont{V.}~\bibnamefont{Dobrosavljevi\'c}}}%
  , \bibinfo {year} {2011},\ \bibfield{title}{%
  \enquote{\bibinfo {title} {Fingerprints of intrinsic phase separation:
  Magnetically doped two-dimensional electron gas},}\ }%
  \bibfield{journal}{%
  \bibinfo {journal} {Phys. Rev. Lett.}\ }%
  \textbf{\bibinfo {volume} {106}},\ \bibinfo {pages} {186402}%
  \bibAnnoteFile{NoStop}{Terletska:2011_PRL}%
\bibitem[{\citenamefont{Thevenard}\
  \emph{et~al.}(2011)\citenamefont{Thevenard}, \citenamefont{Gourdon},
  \citenamefont{Haghgoo}, \citenamefont{Adam}, \citenamefont{von Bardeleben},
  \citenamefont{Lema\^{i}tre}, \citenamefont{Schoch},\ and\
  \citenamefont{Thiaville}}]{Thevenard:2011_PRB}%
  \BibitemOpen
  \bibfield{author}{%
  \bibinfo {author} {\bibnamefont{Thevenard}, \bibfnamefont{L.}}, \bibinfo
  {author} {\bibfnamefont{C.}~\bibnamefont{Gourdon}}, \bibinfo {author}
  {\bibfnamefont{S.}~\bibnamefont{Haghgoo}}, \bibinfo {author}
  {\bibfnamefont{J-P.}\ \bibnamefont{Adam}}, \bibinfo {author}
  {\bibfnamefont{H.~J.}\ \bibnamefont{von Bardeleben}}, \bibinfo {author}
  {\bibfnamefont{A.}~\bibnamefont{Lema\^{i}tre}}, \bibinfo {author}
  {\bibfnamefont{W.}~\bibnamefont{Schoch}},\ and\ \bibinfo {author}
  {\bibfnamefont{A.}~\bibnamefont{Thiaville}}}%
  , \bibinfo {year} {2011},\ \bibfield{title}{%
  \enquote{\bibinfo {title} {Domain wall propagation in ferromagnetic
  semiconductors: {Beyond} the one-dimensional model},}\ }%
  \bibfield{journal}{%
  \bibinfo {journal} {Phys. Rev. B}\ }%
  \textbf{\bibinfo {volume} {83}},\ \bibinfo {pages} {245211}%
  \bibAnnoteFile{NoStop}{Thevenard:2011_PRB}%
\bibitem[{\citenamefont{Thevenard}\
  \emph{et~al.}(2007)\citenamefont{Thevenard}, \citenamefont{Largeau},
  \citenamefont{Mauguin}, \citenamefont{{Lema\^{i}tre}},
  \citenamefont{Khazen},\ and\ \citenamefont{von
  Bardeleben}}]{Thevenard:2007_PRB}%
  \BibitemOpen
  \bibfield{author}{%
  \bibinfo {author} {\bibnamefont{Thevenard}, \bibfnamefont{L.}}, \bibinfo
  {author} {\bibfnamefont{L.}~\bibnamefont{Largeau}}, \bibinfo {author}
  {\bibfnamefont{O.}~\bibnamefont{Mauguin}}, \bibinfo {author}
  {\bibfnamefont{A.}~\bibnamefont{{Lema\^{i}tre}}}, \bibinfo {author}
  {\bibfnamefont{K.}~\bibnamefont{Khazen}},\ and\ \bibinfo {author}
  {\bibfnamefont{H.~J.}\ \bibnamefont{von Bardeleben}}}%
  , \bibinfo {year} {2007},\ \bibfield{title}{%
  \enquote{\bibinfo {title} {Evolution of the magnetic anisotropy with carrier
  density in hydrogenated {Ga$_{1-x}$Mn$_{x}$As}},}\ }%
  \bibfield{journal}{%
  \bibinfo {journal} {Phys. Rev. B}\ }%
  \textbf{\bibinfo {volume} {75}},\ \bibinfo {pages} {195218}%
  \bibAnnoteFile{NoStop}{Thevenard:2007_PRB}%
\bibitem[{\citenamefont{Thevenard}\
  \emph{et~al.}(2005)\citenamefont{Thevenard}, \citenamefont{Largeau},
  \citenamefont{Mauguin}, \citenamefont{{Lema\^{i}tre}},\ and\
  \citenamefont{Theys}}]{Thevenard:2005_APL}%
  \BibitemOpen
  \bibfield{author}{%
  \bibinfo {author} {\bibnamefont{Thevenard}, \bibfnamefont{L.}}, \bibinfo
  {author} {\bibfnamefont{L.}~\bibnamefont{Largeau}}, \bibinfo {author}
  {\bibfnamefont{O.}~\bibnamefont{Mauguin}}, \bibinfo {author}
  {\bibfnamefont{A.}~\bibnamefont{{Lema\^{i}tre}}},\ and\ \bibinfo {author}
  {\bibfnamefont{B.}~\bibnamefont{Theys}}}%
  , \bibinfo {year} {2005},\ \bibfield{title}{%
  \enquote{\bibinfo {title} {Tuning the ferromagnetic properties of
  hydrogenated {GaMnAs}},}\ }%
  \bibfield{journal}{%
  \bibinfo {journal} {Appl. Phys. Lett.}\ }%
  \textbf{\bibinfo {volume} {87}},\ \bibinfo {pages} {182506}%
  \bibAnnoteFile{NoStop}{Thevenard:2005_APL}%
\bibitem[{\citenamefont{Thevenard}\
  \emph{et~al.}(2006)\citenamefont{Thevenard}, \citenamefont{Largeau},
  \citenamefont{Mauguin}, \citenamefont{Patriarche},
  \citenamefont{{Lema\^{i}tre}}, \citenamefont{Vernier},\ and\
  \citenamefont{{Ferr{\'e}}}}]{Thevenard:2006_PRB}%
  \BibitemOpen
  \bibfield{author}{%
  \bibinfo {author} {\bibnamefont{Thevenard}, \bibfnamefont{L.}}, \bibinfo
  {author} {\bibfnamefont{L.}~\bibnamefont{Largeau}}, \bibinfo {author}
  {\bibfnamefont{O.}~\bibnamefont{Mauguin}}, \bibinfo {author}
  {\bibfnamefont{G.}~\bibnamefont{Patriarche}}, \bibinfo {author}
  {\bibfnamefont{A.}~\bibnamefont{{Lema\^{i}tre}}}, \bibinfo {author}
  {\bibfnamefont{N.}~\bibnamefont{Vernier}},\ and\ \bibinfo {author}
  {\bibfnamefont{J.}~\bibnamefont{{Ferr{\'e}}}}}%
  , \bibinfo {year} {2006},\ \bibfield{title}{%
  \enquote{\bibinfo {title} {Magnetic properties and domain structure of
  {(Ga,Mn)As} films with perpendicular anisotropy},}\ }%
  \bibfield{journal}{%
  \bibinfo {journal} {Phys. Rev. B}\ }%
  \textbf{\bibinfo {volume} {73}},\ \bibinfo {pages} {195331}%
  \bibAnnoteFile{NoStop}{Thevenard:2006_PRB}%
\bibitem[{\citenamefont{Thomas}\ \emph{et~al.}(2007)\citenamefont{Thomas},
  \citenamefont{Makarovsky}, \citenamefont{Patane}, \citenamefont{Eaves},
  \citenamefont{Campion}, \citenamefont{Edmonds}, \citenamefont{Foxon},\ and\
  \citenamefont{Gallagher}}]{Thomas:2007_APL}%
  \BibitemOpen
  \bibfield{author}{%
  \bibinfo {author} {\bibnamefont{Thomas}, \bibfnamefont{O.}}, \bibinfo
  {author} {\bibfnamefont{O.}~\bibnamefont{Makarovsky}}, \bibinfo {author}
  {\bibfnamefont{A.}~\bibnamefont{Patane}}, \bibinfo {author}
  {\bibfnamefont{L.}~\bibnamefont{Eaves}}, \bibinfo {author}
  {\bibfnamefont{R.~P.}\ \bibnamefont{Campion}}, \bibinfo {author}
  {\bibfnamefont{K.~W.}\ \bibnamefont{Edmonds}}, \bibinfo {author}
  {\bibfnamefont{C.~T.}\ \bibnamefont{Foxon}},\ and\ \bibinfo {author}
  {\bibfnamefont{B.~L.}\ \bibnamefont{Gallagher}}}%
  , \bibinfo {year} {2007},\ \bibfield{title}{%
  \enquote{\bibinfo {title} {Measuring the hole chemical potential in
  ferromagnetic {Ga$_{1-x}$Mn$_{x}$As/GaAs} heterostructures by photoexcited
  resonant tunneling},}\ }%
  \bibfield{journal}{%
  \bibinfo {journal} {Appl. Phys. Lett.}\ }%
  \textbf{\bibinfo {volume} {90}},\ \bibinfo {pages} {082106}%
  \bibAnnoteFile{NoStop}{Thomas:2007_APL}%
\bibitem[{\citenamefont{Timm}(2006)}]{Timm:2005_PRL}%
  \BibitemOpen
  \bibfield{author}{%
  \bibinfo {author} {\bibnamefont{Timm}, \bibfnamefont{C.}}}%
  , \bibinfo {year} {2006},\ \bibfield{title}{%
  \enquote{\bibinfo {title} {Charge and magnetization inhomogeneities in
  diluted magnetic semiconductors},}\ }%
  \bibfield{journal}{%
  \bibinfo {journal} {Phys. Rev. Lett.}\ }%
  \textbf{\bibinfo {volume} {96}},\ \bibinfo {pages} {117201}%
  \bibAnnoteFile{NoStop}{Timm:2005_PRL}%
\bibitem[{\citenamefont{Timm}\ and\
  \citenamefont{MacDonald}(2005)}]{Timm:2005_PRB}%
  \BibitemOpen
  \bibfield{author}{%
  \bibinfo {author} {\bibnamefont{Timm}, \bibfnamefont{C.}},\ and\ \bibinfo
  {author} {\bibfnamefont{A.~H.}\ \bibnamefont{MacDonald}}}%
  , \bibinfo {year} {2005},\ \bibfield{title}{%
  \enquote{\bibinfo {title} {Influence of non-local exchange on {RKKY}
  interactions in {III-V} diluted magnetic semiconductors},}\ }%
  \bibfield{journal}{%
  \bibinfo {journal} {Phys. Rev. B}\ }%
  \textbf{\bibinfo {volume} {71}},\ \bibinfo {pages} {155206}%
  \bibAnnoteFile{NoStop}{Timm:2005_PRB}%
\bibitem[{\citenamefont{Tran}\ \emph{et~al.}(2009)\citenamefont{Tran},
  \citenamefont{Peiro}, \citenamefont{Jaffres}, \citenamefont{George},
  \citenamefont{Mauguin}, \citenamefont{Largeau},\ and\
  \citenamefont{Lemaitre}}]{Tran:2009_APL}%
  \BibitemOpen
  \bibfield{author}{%
  \bibinfo {author} {\bibnamefont{Tran}, \bibfnamefont{M.}}, \bibinfo {author}
  {\bibfnamefont{J.}~\bibnamefont{Peiro}}, \bibinfo {author}
  {\bibfnamefont{H.}~\bibnamefont{Jaffres}}, \bibinfo {author}
  {\bibfnamefont{J.-M.}\ \bibnamefont{George}}, \bibinfo {author}
  {\bibfnamefont{O.}~\bibnamefont{Mauguin}}, \bibinfo {author}
  {\bibfnamefont{L.}~\bibnamefont{Largeau}},\ and\ \bibinfo {author}
  {\bibfnamefont{A.}~\bibnamefont{Lemaitre}}}%
  , \bibinfo {year} {2009},\ \bibfield{title}{%
  \enquote{\bibinfo {title} {Magnetization-controlled conductance in
  {(Ga,Mn)As}-based resonant tunneling devices},}\ }%
  \bibfield{journal}{%
  \bibinfo {journal} {Appl. Phys. Lett.}\ }%
  \textbf{\bibinfo {volume} {95}},\ \bibinfo {pages} {172101}%
  \bibAnnoteFile{NoStop}{Tran:2009_APL}%
\bibitem[{\citenamefont{Tserkovnyak}\
  \emph{et~al.}(2004)\citenamefont{Tserkovnyak}, \citenamefont{Fiete},\ and\
  \citenamefont{Halperin}}]{Tserkovnyak:2004_APL}%
  \BibitemOpen
  \bibfield{author}{%
  \bibinfo {author} {\bibnamefont{Tserkovnyak}, \bibfnamefont{Y.}}, \bibinfo
  {author} {\bibfnamefont{G.~A.}\ \bibnamefont{Fiete}},\ and\ \bibinfo {author}
  {\bibfnamefont{B.~I.}\ \bibnamefont{Halperin}}}%
  , \bibinfo {year} {2004},\ \bibfield{title}{%
  \enquote{\bibinfo {title} {Mean-field magnetization relaxation in conducting
  ferromagnets},}\ }%
  \bibfield{journal}{%
  \bibinfo {journal} {Appl. Phys. Lett.}\ }%
  \textbf{\bibinfo {volume} {84}},\ \bibinfo {pages} {5234}%
  \bibAnnoteFile{NoStop}{Tserkovnyak:2004_APL}%
\bibitem[{\citenamefont{Turek}\ \emph{et~al.}(2009)\citenamefont{Turek},
  \citenamefont{Siewert},\ and\ \citenamefont{Fabian}}]{Turek:2009_PRB}%
  \BibitemOpen
  \bibfield{author}{%
  \bibinfo {author} {\bibnamefont{Turek}, \bibfnamefont{M.}}, \bibinfo {author}
  {\bibfnamefont{J.}~\bibnamefont{Siewert}},\ and\ \bibinfo {author}
  {\bibfnamefont{J.}~\bibnamefont{Fabian}}}%
  , \bibinfo {year} {2009},\ \bibfield{title}{%
  \enquote{\bibinfo {title} {Magnetic circular dichroism in
  {Ga$_{x}$Mn$_{1-x}$As}: {Theoretica}l evidence for and against an impurity
  band},}\ }%
  \bibfield{journal}{%
  \bibinfo {journal} {Phys. Rev. B}\ }%
  \textbf{\bibinfo {volume} {80}},\ \bibinfo {pages} {161201}%
  \bibAnnoteFile{NoStop}{Turek:2009_PRB}%
\bibitem[{\citenamefont{Twardowski}\
  \emph{et~al.}(1984{\natexlab{a}})\citenamefont{Twardowski}, \citenamefont{von
  Ortenberg}, \citenamefont{Demianiuk},\ and\
  \citenamefont{Pauthenet}}]{Twardowski:1984_SSCa}%
  \BibitemOpen
  \bibfield{author}{%
  \bibinfo {author} {\bibnamefont{Twardowski}, \bibfnamefont{A.}}, \bibinfo
  {author} {\bibfnamefont{M.}~\bibnamefont{von Ortenberg}}, \bibinfo {author}
  {\bibfnamefont{M.}~\bibnamefont{Demianiuk}},\ and\ \bibinfo {author}
  {\bibfnamefont{R.}~\bibnamefont{Pauthenet}}}%
  , \bibinfo {year} {1984}{\natexlab{a}},\ \bibfield{title}{%
  \enquote{\bibinfo {title} {Magnetization and exchange constants in
  {Zn$_{1-x}$Mn$_x$Se}},}\ }%
  \bibfield{journal}{%
  \bibinfo {journal} {Solid State Commun.}\ }%
  \textbf{\bibinfo {volume} {51}},\ \bibinfo {pages} {849}%
  \bibAnnoteFile{NoStop}{Twardowski:1984_SSCa}%
\bibitem[{\citenamefont{Twardowski}\
  \emph{et~al.}(1987)\citenamefont{Twardowski}, \citenamefont{Swagten},
  \citenamefont{de~Jonge},\ and\
  \citenamefont{Demianiuk}}]{Twardowski:1987_PRB}%
  \BibitemOpen
  \bibfield{author}{%
  \bibinfo {author} {\bibnamefont{Twardowski}, \bibfnamefont{A.}}, \bibinfo
  {author} {\bibfnamefont{H.~J.~M.}\ \bibnamefont{Swagten}}, \bibinfo {author}
  {\bibfnamefont{W.~J.~M.}\ \bibnamefont{de~Jonge}},\ and\ \bibinfo {author}
  {\bibfnamefont{M.}~\bibnamefont{Demianiuk}}}%
  , \bibinfo {year} {1987},\ \bibfield{title}{%
  \enquote{\bibinfo {title} {Magnetic behavior of the diluted magnetic
  semiconductor {Zn$_{1-x}$Mn$_{x}$Se}},}\ }%
  \bibfield{journal}{%
  \bibinfo {journal} {Phys. Rev. B}\ }%
  \textbf{\bibinfo {volume} {36}},\ \bibinfo {pages} {7013}%
  \bibAnnoteFile{NoStop}{Twardowski:1987_PRB}%
\bibitem[{\citenamefont{Twardowski}\
  \emph{et~al.}(1984{\natexlab{b}})\citenamefont{Twardowski},
  \citenamefont{Swiderski}, \citenamefont{von Ortenberg},\ and\
  \citenamefont{Pauthenet}}]{Twardowski:1984_SSCb}%
  \BibitemOpen
  \bibfield{author}{%
  \bibinfo {author} {\bibnamefont{Twardowski}, \bibfnamefont{A.}}, \bibinfo
  {author} {\bibfnamefont{P.}~\bibnamefont{Swiderski}}, \bibinfo {author}
  {\bibfnamefont{M.}~\bibnamefont{von Ortenberg}},\ and\ \bibinfo {author}
  {\bibfnamefont{R.}~\bibnamefont{Pauthenet}}}%
  , \bibinfo {year} {1984}{\natexlab{b}},\ \bibfield{title}{%
  \enquote{\bibinfo {title} {Magnetoabsorption and magnetization of
  {Zn$_{1-x}$Mn$_x$Te} mixed crystals},}\ }%
  \bibfield{journal}{%
  \bibinfo {journal} {Solid State Commun.}\ }%
  \textbf{\bibinfo {volume} {50}},\ \bibinfo {pages} {509}%
  \bibAnnoteFile{NoStop}{Twardowski:1984_SSCb}%
\bibitem[{\citenamefont{{Van {Dorpe}}}\ \emph{et~al.}(2005)\citenamefont{{Van
  {Dorpe}}}, \citenamefont{{Van {Roy}}}, \citenamefont{{De {Boeck}}},
  \citenamefont{Borghs}, \citenamefont{Sankowski}, \citenamefont{Kacman},
  \citenamefont{Majewski},\ and\ \citenamefont{Dietl}}]{Dorpe:2005_PRB}%
  \BibitemOpen
  \bibfield{author}{%
  \bibinfo {author} {\bibnamefont{{Van {Dorpe}}}, \bibfnamefont{P.}}, \bibinfo
  {author} {\bibfnamefont{W.}~\bibnamefont{{Van {Roy}}}}, \bibinfo {author}
  {\bibfnamefont{J.}~\bibnamefont{{De {Boeck}}}}, \bibinfo {author}
  {\bibfnamefont{G.}~\bibnamefont{Borghs}}, \bibinfo {author}
  {\bibfnamefont{P.}~\bibnamefont{Sankowski}}, \bibinfo {author}
  {\bibfnamefont{P.}~\bibnamefont{Kacman}}, \bibinfo {author}
  {\bibfnamefont{J.~A.}\ \bibnamefont{Majewski}},\ and\ \bibinfo {author}
  {\bibfnamefont{T.}~\bibnamefont{Dietl}}}%
  , \bibinfo {year} {2005},\ \bibfield{title}{%
  \enquote{\bibinfo {title} {Voltage controlled spin injection in a
  {(Ga,Mn)As/(Al,Ga)As} {Zener} diode},}\ }%
  \bibfield{journal}{%
  \bibinfo {journal} {Phys. Rev. B}\ }%
  \textbf{\bibinfo {volume} {72}},\ \bibinfo {pages} {205322}%
  \bibAnnoteFile{NoStop}{Dorpe:2005_PRB}%
\bibitem[{\citenamefont{{Van Dorpe}}\ \emph{et~al.}(2004)\citenamefont{{Van
  Dorpe}}, \citenamefont{{Van Roy}}, \citenamefont{Motsnyi},
  \citenamefont{Sawicki}, \citenamefont{Borghs},\ and\ \citenamefont{{De
  Boeck}}}]{Dorpe:2004_APL}%
  \BibitemOpen
  \bibfield{author}{%
  \bibinfo {author} {\bibnamefont{{Van Dorpe}}, \bibfnamefont{P.}}, \bibinfo
  {author} {\bibfnamefont{W.}~\bibnamefont{{Van Roy}}}, \bibinfo {author}
  {\bibfnamefont{V.F.}\ \bibnamefont{Motsnyi}}, \bibinfo {author}
  {\bibfnamefont{M.}~\bibnamefont{Sawicki}}, \bibinfo {author}
  {\bibfnamefont{G.}~\bibnamefont{Borghs}},\ and\ \bibinfo {author}
  {\bibfnamefont{J.}~\bibnamefont{{De Boeck}}}}%
  , \bibinfo {year} {2004},\ \bibfield{title}{%
  \enquote{\bibinfo {title} {Very high spin polarization in {GaAs} by injection
  from a {(Ga,Mn)As} {Zener} diode},}\ }%
  \bibfield{journal}{%
  \bibinfo {journal} {Appl. Phys. Lett.}\ }%
  \textbf{\bibinfo {volume} {84}},\ \bibinfo {pages} {3495}%
  \bibAnnoteFile{NoStop}{Dorpe:2004_APL}%
\bibitem[{\citenamefont{Van~Esch}\ \emph{et~al.}(1997)\citenamefont{Van~Esch},
  \citenamefont{Van~Bockstal}, \citenamefont{De~Boeck},
  \citenamefont{Verbanck}, \citenamefont{van Steenbergen},
  \citenamefont{Wellmann}, \citenamefont{Grietens}, \citenamefont{Bogaerts},
  \citenamefont{Herlach},\ and\ \citenamefont{Borghs}}]{Esch:1997_PRB}%
  \BibitemOpen
  \bibfield{author}{%
  \bibinfo {author} {\bibnamefont{Van~Esch}, \bibfnamefont{A.}}, \bibinfo
  {author} {\bibfnamefont{L.}~\bibnamefont{Van~Bockstal}}, \bibinfo {author}
  {\bibfnamefont{J.}~\bibnamefont{De~Boeck}}, \bibinfo {author}
  {\bibfnamefont{G.}~\bibnamefont{Verbanck}}, \bibinfo {author}
  {\bibfnamefont{A.~S.}\ \bibnamefont{van Steenbergen}}, \bibinfo {author}
  {\bibfnamefont{P.~J.}\ \bibnamefont{Wellmann}}, \bibinfo {author}
  {\bibfnamefont{B.}~\bibnamefont{Grietens}}, \bibinfo {author}
  {\bibfnamefont{R.}~\bibnamefont{Bogaerts}}, \bibinfo {author}
  {\bibfnamefont{F.}~\bibnamefont{Herlach}},\ and\ \bibinfo {author}
  {\bibfnamefont{G.}~\bibnamefont{Borghs}}}%
  , \bibinfo {year} {1997},\ \bibfield{title}{%
  \enquote{\bibinfo {title} {Interplay between the magnetic and transport
  properties in the {III--V} diluted magnetic semiconductor
  {Ga$_{1-x}$Mn$_{x}$As}},}\ }%
  \bibfield{journal}{%
  \bibinfo {journal} {Phys. Rev. B}\ }%
  \textbf{\bibinfo {volume} {56}},\ \bibinfo {pages} {13103}%
  \bibAnnoteFile{NoStop}{Esch:1997_PRB}%
\bibitem[{\citenamefont{Vonsovsky}(1946)}]{Vonsovsky:1946_ZETF}%
  \BibitemOpen
  \bibfield{author}{%
  \bibinfo {author} {\bibnamefont{Vonsovsky}, \bibfnamefont{S.~V.}}}%
  , \bibinfo {year} {1946},\ \bibfield{title}{%
  \enquote{\bibinfo {title} {On the exchange interaction of {\em s} and {\em d}
  electrons in ferromagnets},}\ }%
  \bibfield{journal}{%
  \bibinfo {journal} {Zh. Eksp. Teor. Phys.}\ }%
  \textbf{\bibinfo {volume} {16}},\ \bibinfo {pages} {981}%
  \bibAnnoteFile{NoStop}{Vonsovsky:1946_ZETF}%
\bibitem[{\citenamefont{Vurgaftman}\ and\
  \citenamefont{Meyer}(2001)}]{Vurgaftman:2001_PRB}%
  \BibitemOpen
  \bibfield{author}{%
  \bibinfo {author} {\bibnamefont{Vurgaftman}, \bibfnamefont{I.}},\ and\
  \bibinfo {author} {\bibfnamefont{J.~R.}\ \bibnamefont{Meyer}}}%
  , \bibinfo {year} {2001},\ \bibfield{title}{%
  \enquote{\bibinfo {title} {Curie-temperature enhancement in ferromagnetic
  semiconductor superlattices},}\ }%
  \bibfield{journal}{%
  \bibinfo {journal} {Phys. Rev. B}\ }%
  \textbf{\bibinfo {volume} {64}},\ \bibinfo {pages} {245207}%
  \bibAnnoteFile{NoStop}{Vurgaftman:2001_PRB}%
\bibitem[{\citenamefont{Wadley}\ \emph{et~al.}(2010)\citenamefont{Wadley},
  \citenamefont{Freeman}, \citenamefont{Edmonds}, \citenamefont{{van der
  Laan}}, \citenamefont{Chauhan}, \citenamefont{Campion},
  \citenamefont{Rushforth}, \citenamefont{Gallagher}, \citenamefont{Foxon},
  \citenamefont{Wilhelm}, \citenamefont{Smekhova},\ and\
  \citenamefont{Rogalev}}]{Wadley:2010_PRB}%
  \BibitemOpen
  \bibfield{author}{%
  \bibinfo {author} {\bibnamefont{Wadley}, \bibfnamefont{P.}}, \bibinfo
  {author} {\bibfnamefont{A.~A.}\ \bibnamefont{Freeman}}, \bibinfo {author}
  {\bibfnamefont{K.~W.}\ \bibnamefont{Edmonds}}, \bibinfo {author}
  {\bibfnamefont{G.}~\bibnamefont{{van der Laan}}}, \bibinfo {author}
  {\bibfnamefont{J.~S.}\ \bibnamefont{Chauhan}}, \bibinfo {author}
  {\bibfnamefont{R.~P.}\ \bibnamefont{Campion}}, \bibinfo {author}
  {\bibfnamefont{A.~W.}\ \bibnamefont{Rushforth}}, \bibinfo {author}
  {\bibfnamefont{B.~L.}\ \bibnamefont{Gallagher}}, \bibinfo {author}
  {\bibfnamefont{C.~T.}\ \bibnamefont{Foxon}}, \bibinfo {author}
  {\bibfnamefont{F.}~\bibnamefont{Wilhelm}}, \bibinfo {author}
  {\bibfnamefont{A.~G.}\ \bibnamefont{Smekhova}},\ and\ \bibinfo {author}
  {\bibfnamefont{A.}~\bibnamefont{Rogalev}}}%
  , \bibinfo {year} {2010},\ \bibfield{title}{%
  \enquote{\bibinfo {title} {Element-resolved orbital polarization in
  {(III,Mn)As} ferromagnetic semiconductors from $k$-edge x-ray magnetic
  circular dichroism},}\ }%
  \bibfield{journal}{%
  \bibinfo {journal} {Phys. Rev. B}\ }%
  \textbf{\bibinfo {volume} {81}},\ \bibinfo {pages} {235208}%
  \bibAnnoteFile{NoStop}{Wadley:2010_PRB}%
\bibitem[{\citenamefont{Wang}\
  \emph{et~al.}(2007{\natexlab{a}})\citenamefont{Wang}, \citenamefont{Ren},
  \citenamefont{Liu}, \citenamefont{Furdyna}, \citenamefont{Grimsditch},\ and\
  \citenamefont{Merlin}}]{Wang:2007_PRB}%
  \BibitemOpen
  \bibfield{author}{%
  \bibinfo {author} {\bibnamefont{Wang}, \bibfnamefont{D.~M.}}, \bibinfo
  {author} {\bibfnamefont{Y.~H.}\ \bibnamefont{Ren}}, \bibinfo {author}
  {\bibfnamefont{X.}~\bibnamefont{Liu}}, \bibinfo {author}
  {\bibfnamefont{J.~K.}\ \bibnamefont{Furdyna}}, \bibinfo {author}
  {\bibfnamefont{M.}~\bibnamefont{Grimsditch}},\ and\ \bibinfo {author}
  {\bibfnamefont{R.}~\bibnamefont{Merlin}}}%
  , \bibinfo {year} {2007}{\natexlab{a}},\ \bibfield{title}{%
  \enquote{\bibinfo {title} {Light-induced magnetic precession in {(Ga,Mn)As}
  slabs: {Hybrid} standing-wave {Damon-Eshbach modes}},}\ }%
  \bibfield{journal}{%
  \bibinfo {journal} {Phys. Rev. B}\ }%
  \textbf{\bibinfo {volume} {75}},\ \bibinfo {pages} {233308}%
  \bibAnnoteFile{NoStop}{Wang:2007_PRB}%
\bibitem[{\citenamefont{Wang}\ \emph{et~al.}(2009)\citenamefont{Wang},
  \citenamefont{Cotoros}, \citenamefont{Chemla}, \citenamefont{Liu},
  \citenamefont{Furdyna}, \citenamefont{Chovan},\ and\
  \citenamefont{Perakis}}]{Wang:2009_APL}%
  \BibitemOpen
  \bibfield{author}{%
  \bibinfo {author} {\bibnamefont{Wang}, \bibfnamefont{J.}}, \bibinfo {author}
  {\bibfnamefont{I.}~\bibnamefont{Cotoros}}, \bibinfo {author}
  {\bibfnamefont{D.~S.}\ \bibnamefont{Chemla}}, \bibinfo {author}
  {\bibfnamefont{X.}~\bibnamefont{Liu}}, \bibinfo {author}
  {\bibfnamefont{J.~K.}\ \bibnamefont{Furdyna}}, \bibinfo {author}
  {\bibfnamefont{J.}~\bibnamefont{Chovan}},\ and\ \bibinfo {author}
  {\bibfnamefont{I.~E.}\ \bibnamefont{Perakis}}}%
  , \bibinfo {year} {2009},\ \bibfield{title}{%
  \enquote{\bibinfo {title} {Memory effects in photoinduced femtosecond
  magnetization rotation in ferromagnetic {GaMnAs}},}\ }%
  \bibfield{journal}{%
  \bibinfo {journal} {Appl. Phys. Lett.}\ }%
  \textbf{\bibinfo {volume} {94}},\ \bibinfo {pages} {021101}%
  \bibAnnoteFile{NoStop}{Wang:2009_APL}%
\bibitem[{\citenamefont{Wang}\
  \emph{et~al.}(2007{\natexlab{b}})\citenamefont{Wang}, \citenamefont{Cotoros},
  \citenamefont{Dani}, \citenamefont{Liu}, \citenamefont{Furdyna},\ and\
  \citenamefont{Chemla}}]{Wang:2007_PRL}%
  \BibitemOpen
  \bibfield{author}{%
  \bibinfo {author} {\bibnamefont{Wang}, \bibfnamefont{J.}}, \bibinfo {author}
  {\bibfnamefont{I.}~\bibnamefont{Cotoros}}, \bibinfo {author}
  {\bibfnamefont{K.~M.}\ \bibnamefont{Dani}}, \bibinfo {author}
  {\bibfnamefont{X.}~\bibnamefont{Liu}}, \bibinfo {author}
  {\bibfnamefont{J.~K.}\ \bibnamefont{Furdyna}},\ and\ \bibinfo {author}
  {\bibfnamefont{D.~S.}\ \bibnamefont{Chemla}}}%
  , \bibinfo {year} {2007}{\natexlab{b}},\ \bibfield{title}{%
  \enquote{\bibinfo {title} {Ultrafast enhancement of ferromagnetism via
  photoexcited holes in {GaMnAs}},}\ }%
  \bibfield{journal}{%
  \bibinfo {journal} {Phys. Rev. Lett.}\ }%
  \textbf{\bibinfo {volume} {98}},\ \bibinfo {pages} {217401}%
  \bibAnnoteFile{NoStop}{Wang:2007_PRL}%
\bibitem[{\citenamefont{Wang}\
  \emph{et~al.}(2008{\natexlab{a}})\citenamefont{Wang},
  \citenamefont{Cywi\'{n}ski}, \citenamefont{Sun}, \citenamefont{Kono},
  \citenamefont{Munekata},\ and\ \citenamefont{Sham}}]{Wang:2008_PRB}%
  \BibitemOpen
  \bibfield{author}{%
  \bibinfo {author} {\bibnamefont{Wang}, \bibfnamefont{J.}}, \bibinfo {author}
  {\bibfnamefont{\L.}\ \bibnamefont{Cywi\'{n}ski}}, \bibinfo {author}
  {\bibfnamefont{C.}~\bibnamefont{Sun}}, \bibinfo {author}
  {\bibfnamefont{J.}~\bibnamefont{Kono}}, \bibinfo {author}
  {\bibfnamefont{H.}~\bibnamefont{Munekata}},\ and\ \bibinfo {author}
  {\bibfnamefont{L.~J.}\ \bibnamefont{Sham}}}%
  , \bibinfo {year} {2008}{\natexlab{a}},\ \bibfield{title}{%
  \enquote{\bibinfo {title} {Femtosecond demagnetization and hot-hole
  relaxation in ferromagnetic {Ga$_{1-x}$Mn$_{x}$As}},}\ }%
  \bibfield{journal}{%
  \bibinfo {journal} {Phys. Rev. B}\ }%
  \textbf{\bibinfo {volume} {77}},\ \bibinfo {pages} {235308}%
  \bibAnnoteFile{NoStop}{Wang:2008_PRB}%
\bibitem[{\citenamefont{Wang}\
  \emph{et~al.}(2005{\natexlab{a}})\citenamefont{Wang}, \citenamefont{Sun},
  \citenamefont{Kono}, \citenamefont{Oiwa}, \citenamefont{Munekata},
  \citenamefont{Cywinski},\ and\ \citenamefont{Sham}}]{Wang:2005_PRLb}%
  \BibitemOpen
  \bibfield{author}{%
  \bibinfo {author} {\bibnamefont{Wang}, \bibfnamefont{J.}}, \bibinfo {author}
  {\bibfnamefont{C.}~\bibnamefont{Sun}}, \bibinfo {author}
  {\bibfnamefont{J.}~\bibnamefont{Kono}}, \bibinfo {author}
  {\bibfnamefont{A.}~\bibnamefont{Oiwa}}, \bibinfo {author}
  {\bibfnamefont{H.}~\bibnamefont{Munekata}}, \bibinfo {author}
  {\bibfnamefont{L.}~\bibnamefont{Cywinski}},\ and\ \bibinfo {author}
  {\bibfnamefont{L.~J.}\ \bibnamefont{Sham}}}%
  , \bibinfo {year} {2005}{\natexlab{a}},\ \bibfield{title}{%
  \enquote{\bibinfo {title} {Ultrafast quenching of ferromagnetism in {InMnAs}
  induced by intense laser irradiation},}\ }%
  \bibfield{journal}{%
  \bibinfo {journal} {Phys. Rev. Lett.}\ }%
  \textbf{\bibinfo {volume} {95}},\ \bibinfo {pages} {167401}%
  \bibAnnoteFile{NoStop}{Wang:2005_PRLb}%
\bibitem[{\citenamefont{Wang}\
  \emph{et~al.}(2010{\natexlab{a}})\citenamefont{Wang}, \citenamefont{Edmonds},
  \citenamefont{Irvine}, \citenamefont{Tatara}, \citenamefont{Ranieri},
  \citenamefont{Wunderlich}, \citenamefont{Olejnik}, \citenamefont{Rushforth},
  \citenamefont{Campion}, \citenamefont{Williams}, \citenamefont{Foxon},\ and\
  \citenamefont{Gallagher}}]{Wang:2010_APL}%
  \BibitemOpen
  \bibfield{author}{%
  \bibinfo {author} {\bibnamefont{Wang}, \bibfnamefont{K.~Y.}}, \bibinfo
  {author} {\bibfnamefont{K.~W.}\ \bibnamefont{Edmonds}}, \bibinfo {author}
  {\bibfnamefont{A.~C.}\ \bibnamefont{Irvine}}, \bibinfo {author}
  {\bibfnamefont{G.}~\bibnamefont{Tatara}}, \bibinfo {author}
  {\bibfnamefont{E.~De}\ \bibnamefont{Ranieri}}, \bibinfo {author}
  {\bibfnamefont{J.}~\bibnamefont{Wunderlich}}, \bibinfo {author}
  {\bibfnamefont{K.}~\bibnamefont{Olejnik}}, \bibinfo {author}
  {\bibfnamefont{A.~W.}\ \bibnamefont{Rushforth}}, \bibinfo {author}
  {\bibfnamefont{R.~P.}\ \bibnamefont{Campion}}, \bibinfo {author}
  {\bibfnamefont{D.~A.}\ \bibnamefont{Williams}}, \bibinfo {author}
  {\bibfnamefont{C.~T.}\ \bibnamefont{Foxon}},\ and\ \bibinfo {author}
  {\bibfnamefont{B.~L.}\ \bibnamefont{Gallagher}}}%
  , \bibinfo {year} {2010}{\natexlab{a}},\ \bibfield{title}{%
  \enquote{\bibinfo {title} {Current-driven domain wall motion across a wide
  temperature range in a {(Ga,Mn)(As,P)} device},}\ }%
  \bibfield{journal}{%
  \bibinfo {journal} {Appl. Phys. Lett.}\ }%
  \textbf{\bibinfo {volume} {97}},\ \bibinfo {pages} {262102}%
  \bibAnnoteFile{NoStop}{Wang:2010_APL}%
\bibitem[{\citenamefont{Wang}\
  \emph{et~al.}(2010{\natexlab{b}})\citenamefont{Wang}, \citenamefont{Edmonds},
  \citenamefont{Irvine}, \citenamefont{Wunderlich}, \citenamefont{Olejnik},
  \citenamefont{Rushforth}, \citenamefont{Campion}, \citenamefont{Williams},
  \citenamefont{Foxon},\ and\ \citenamefont{Gallagher}}]{Wang:2010_JMMM}%
  \BibitemOpen
  \bibfield{author}{%
  \bibinfo {author} {\bibnamefont{Wang}, \bibfnamefont{K.~Y.}}, \bibinfo
  {author} {\bibfnamefont{K.~W.}\ \bibnamefont{Edmonds}}, \bibinfo {author}
  {\bibfnamefont{A.~C.}\ \bibnamefont{Irvine}}, \bibinfo {author}
  {\bibfnamefont{J.}~\bibnamefont{Wunderlich}}, \bibinfo {author}
  {\bibfnamefont{K.}~\bibnamefont{Olejnik}}, \bibinfo {author}
  {\bibfnamefont{A.~W.}\ \bibnamefont{Rushforth}}, \bibinfo {author}
  {\bibfnamefont{R.~P.}\ \bibnamefont{Campion}}, \bibinfo {author}
  {\bibfnamefont{D.~A.}\ \bibnamefont{Williams}}, \bibinfo {author}
  {\bibfnamefont{C.~T.}\ \bibnamefont{Foxon}},\ and\ \bibinfo {author}
  {\bibfnamefont{B.~L.}\ \bibnamefont{Gallagher}}}%
  , \bibinfo {year} {2010}{\natexlab{b}},\ \bibfield{title}{%
  \enquote{\bibinfo {title} {Domain wall resistance in perpendicular
  {(Ga,Mn)As}: dependence on pinning},}\ }%
  \bibfield{journal}{%
  \bibinfo {journal} {J. Magn. Magn. Mater.}\ }%
  \textbf{\bibinfo {volume} {322}},\ \bibinfo {pages} {3481}%
  \bibAnnoteFile{NoStop}{Wang:2010_JMMM}%
\bibitem[{\citenamefont{Wang}\
  \emph{et~al.}(2005{\natexlab{b}})\citenamefont{Wang}, \citenamefont{Edmonds},
  \citenamefont{Zhao}, \citenamefont{Sawicki}, \citenamefont{Campion},
  \citenamefont{Gallagher},\ and\ \citenamefont{Foxon}}]{Wang:2005_PRBb}%
  \BibitemOpen
  \bibfield{author}{%
  \bibinfo {author} {\bibnamefont{Wang}, \bibfnamefont{K.~Y.}}, \bibinfo
  {author} {\bibfnamefont{K.~W.}\ \bibnamefont{Edmonds}}, \bibinfo {author}
  {\bibfnamefont{L.~X.}\ \bibnamefont{Zhao}}, \bibinfo {author}
  {\bibfnamefont{M.}~\bibnamefont{Sawicki}}, \bibinfo {author}
  {\bibfnamefont{R.~P.}\ \bibnamefont{Campion}}, \bibinfo {author}
  {\bibfnamefont{B.~L.}\ \bibnamefont{Gallagher}},\ and\ \bibinfo {author}
  {\bibfnamefont{C.~T.}\ \bibnamefont{Foxon}}}%
  , \bibinfo {year} {2005}{\natexlab{b}},\ \bibfield{title}{%
  \enquote{\bibinfo {title} {{GaMnAs} grown on (311) {GaAs} substrates:
  modified {Mn} incorporation and new magnetic anisotropies},}\ }%
  \bibfield{journal}{%
  \bibinfo {journal} {Phys. Rev. B}\ }%
  \textbf{\bibinfo {volume} {72}},\ \bibinfo {pages} {115207}%
  \bibAnnoteFile{NoStop}{Wang:2005_PRBb}%
\bibitem[{\citenamefont{Wang}\
  \emph{et~al.}(2005{\natexlab{c}})\citenamefont{Wang}, \citenamefont{Sawicki},
  \citenamefont{Edmonds}, \citenamefont{Campion}, \citenamefont{Maat},
  \citenamefont{Foxon}, \citenamefont{Gallagher},\ and\
  \citenamefont{Dietl}}]{Wang:2005_PRL}%
  \BibitemOpen
  \bibfield{author}{%
  \bibinfo {author} {\bibnamefont{Wang}, \bibfnamefont{K.-Y.}}, \bibinfo
  {author} {\bibfnamefont{M.}~\bibnamefont{Sawicki}}, \bibinfo {author}
  {\bibfnamefont{K.~W.}\ \bibnamefont{Edmonds}}, \bibinfo {author}
  {\bibfnamefont{R.~P.}\ \bibnamefont{Campion}}, \bibinfo {author}
  {\bibfnamefont{S.}~\bibnamefont{Maat}}, \bibinfo {author}
  {\bibfnamefont{C.~T.}\ \bibnamefont{Foxon}}, \bibinfo {author}
  {\bibfnamefont{B.~L.}\ \bibnamefont{Gallagher}},\ and\ \bibinfo {author}
  {\bibfnamefont{T.}~\bibnamefont{Dietl}}}%
  , \bibinfo {year} {2005}{\natexlab{c}},\ \bibfield{title}{%
  \enquote{\bibinfo {title} {Spin reorientation transition in single-domain
  {(Ga,Mn)As}},}\ }%
  \bibfield{journal}{%
  \bibinfo {journal} {Phys. Rev. Lett.}\ }%
  \textbf{\bibinfo {volume} {95}},\ \bibinfo {pages} {217204}%
  \bibAnnoteFile{NoStop}{Wang:2005_PRL}%
\bibitem[{\citenamefont{Wang}\ \emph{et~al.}(2006)\citenamefont{Wang},
  \citenamefont{Sawicki}, \citenamefont{Edmonds}, \citenamefont{Campion},
  \citenamefont{Rushforth}, \citenamefont{Freeman}, \citenamefont{Foxon},
  \citenamefont{Gallagher},\ and\ \citenamefont{Dietl}}]{Wang:2006_APL}%
  \BibitemOpen
  \bibfield{author}{%
  \bibinfo {author} {\bibnamefont{Wang}, \bibfnamefont{K.-Y.}}, \bibinfo
  {author} {\bibfnamefont{M.}~\bibnamefont{Sawicki}}, \bibinfo {author}
  {\bibfnamefont{K.W.}\ \bibnamefont{Edmonds}}, \bibinfo {author}
  {\bibfnamefont{R.P.}\ \bibnamefont{Campion}}, \bibinfo {author}
  {\bibfnamefont{A.W.}\ \bibnamefont{Rushforth}}, \bibinfo {author}
  {\bibfnamefont{A.A.}\ \bibnamefont{Freeman}}, \bibinfo {author}
  {\bibfnamefont{C.T.}\ \bibnamefont{Foxon}}, \bibinfo {author}
  {\bibfnamefont{B.L.}\ \bibnamefont{Gallagher}},\ and\ \bibinfo {author}
  {\bibfnamefont{T.}~\bibnamefont{Dietl}}}%
  , \bibinfo {year} {2006},\ \bibfield{title}{%
  \enquote{\bibinfo {title} {Control of coercivities in {(Ga,Mn)As} thin films
  by small concentrations of {MnAs} nanoclusters},}\ }%
  \bibfield{journal}{%
  \bibinfo {journal} {Appl. Phys. Lett.}\ }%
  \textbf{\bibinfo {volume} {88}},\ \bibinfo {pages} {022510}%
  \bibAnnoteFile{NoStop}{Wang:2006_APL}%
\bibitem[{\citenamefont{Wang}\ \emph{et~al.}(2004)\citenamefont{Wang},
  \citenamefont{Edmonds}, \citenamefont{Campion}, \citenamefont{Gallagher},
  \citenamefont{Farley}, \citenamefont{Foxon}, \citenamefont{Sawicki},
  \citenamefont{Bogus{\l}awski},\ and\ \citenamefont{Dietl}}]{Wang:2004_JAP}%
  \BibitemOpen
  \bibfield{author}{%
  \bibinfo {author} {\bibnamefont{Wang}, \bibfnamefont{K.Y.}}, \bibinfo
  {author} {\bibfnamefont{K.W.}\ \bibnamefont{Edmonds}}, \bibinfo {author}
  {\bibfnamefont{R.P.}\ \bibnamefont{Campion}}, \bibinfo {author}
  {\bibfnamefont{B.L.}\ \bibnamefont{Gallagher}}, \bibinfo {author}
  {\bibfnamefont{N.R.S.}\ \bibnamefont{Farley}}, \bibinfo {author}
  {\bibfnamefont{C.T.}\ \bibnamefont{Foxon}}, \bibinfo {author}
  {\bibfnamefont{M.}~\bibnamefont{Sawicki}}, \bibinfo {author}
  {\bibfnamefont{P.}~\bibnamefont{Bogus{\l}awski}},\ and\ \bibinfo {author}
  {\bibfnamefont{T.}~\bibnamefont{Dietl}}}%
  , \bibinfo {year} {2004},\ \bibfield{title}{%
  \enquote{\bibinfo {title} {Influence of the {Mn} interstistitials on the
  magnetic and transport properties of {(Ga,Mn)As}},}\ }%
  \bibfield{journal}{%
  \bibinfo {journal} {J. Appl. Phys.}\ }%
  \textbf{\bibinfo {volume} {95}},\ \bibinfo {pages} {6512}%
  \bibAnnoteFile{NoStop}{Wang:2004_JAP}%
\bibitem[{\citenamefont{Wang}\
  \emph{et~al.}(2008{\natexlab{b}})\citenamefont{Wang}, \citenamefont{Campion},
  \citenamefont{Rushforth}, \citenamefont{Edmonds}, \citenamefont{Foxon},\ and\
  \citenamefont{Gallagher}}]{Wang:2008_APL}%
  \BibitemOpen
  \bibfield{author}{%
  \bibinfo {author} {\bibnamefont{Wang}, \bibfnamefont{M.}}, \bibinfo {author}
  {\bibfnamefont{R.~P.}\ \bibnamefont{Campion}}, \bibinfo {author}
  {\bibfnamefont{A.~W.}\ \bibnamefont{Rushforth}}, \bibinfo {author}
  {\bibfnamefont{K.~W.}\ \bibnamefont{Edmonds}}, \bibinfo {author}
  {\bibfnamefont{C.~T.}\ \bibnamefont{Foxon}},\ and\ \bibinfo {author}
  {\bibfnamefont{B.~L.}\ \bibnamefont{Gallagher}}}%
  , \bibinfo {year} {2008}{\natexlab{b}},\ \bibfield{title}{%
  \enquote{\bibinfo {title} {Achieving high {Curie} temperature in
  {(Ga,Mn)As}},}\ }%
  \bibfield{journal}{%
  \bibinfo {journal} {Appl. Phys. Lett.}\ }%
  \textbf{\bibinfo {volume} {93}},\ \bibinfo {pages} {132103}%
  \bibAnnoteFile{NoStop}{Wang:2008_APL}%
\bibitem[{\citenamefont{Wang}\ \emph{et~al.}(2013)\citenamefont{Wang},
  \citenamefont{Edmonds}, \citenamefont{Gallagher}, \citenamefont{Rushforth},
  \citenamefont{Makarovsky}, \citenamefont{Patan\'e}, \citenamefont{Campion},
  \citenamefont{Foxon}, \citenamefont{Novak},\ and\
  \citenamefont{Jungwirth}}]{Wang:2013_PRB}%
  \BibitemOpen
  \bibfield{author}{%
  \bibinfo {author} {\bibnamefont{Wang}, \bibfnamefont{M.}}, \bibinfo {author}
  {\bibfnamefont{K.~W.}\ \bibnamefont{Edmonds}}, \bibinfo {author}
  {\bibfnamefont{B.~L.}\ \bibnamefont{Gallagher}}, \bibinfo {author}
  {\bibfnamefont{A.~W.}\ \bibnamefont{Rushforth}}, \bibinfo {author}
  {\bibfnamefont{O.}~\bibnamefont{Makarovsky}}, \bibinfo {author}
  {\bibfnamefont{A.}~\bibnamefont{Patan\'e}}, \bibinfo {author}
  {\bibfnamefont{R.~P.}\ \bibnamefont{Campion}}, \bibinfo {author}
  {\bibfnamefont{C.~T.}\ \bibnamefont{Foxon}}, \bibinfo {author}
  {\bibfnamefont{V.}~\bibnamefont{Novak}},\ and\ \bibinfo {author}
  {\bibfnamefont{T.}~\bibnamefont{Jungwirth}}}%
  , \bibinfo {year} {2013},\ \bibfield{title}{%
  \enquote{\bibinfo {title} {High {Curie} temperatures at low compensation in
  the ferromagnetic semiconductor {(Ga,Mn)As}},}\ }%
  \bibfield{journal}{%
  \bibinfo {journal} {Phys. Rev. B}\ }%
  \textbf{\bibinfo {volume} {87}},\ \bibinfo {pages} {121301}%
  \bibAnnoteFile{NoStop}{Wang:2013_PRB}%
\bibitem[{\citenamefont{Wang}\
  \emph{et~al.}(2008{\natexlab{c}})\citenamefont{Wang}, \citenamefont{Deng},
  \citenamefont{Lu}, \citenamefont{Chen}, \citenamefont{Ji},\ and\
  \citenamefont{Zhao}}]{Wang:2008_PE}%
  \BibitemOpen
  \bibfield{author}{%
  \bibinfo {author} {\bibnamefont{Wang}, \bibfnamefont{W.~Z.}}, \bibinfo
  {author} {\bibfnamefont{J.~J.}\ \bibnamefont{Deng}}, \bibinfo {author}
  {\bibfnamefont{J.}~\bibnamefont{Lu}}, \bibinfo {author}
  {\bibfnamefont{L.}~\bibnamefont{Chen}}, \bibinfo {author}
  {\bibfnamefont{Y.}~\bibnamefont{Ji}},\ and\ \bibinfo {author}
  {\bibfnamefont{J.~H.}\ \bibnamefont{Zhao}}}%
  , \bibinfo {year} {2008}{\natexlab{c}},\ \bibfield{title}{%
  \enquote{\bibinfo {title} {Influence of {Si} doping on magnetic properties of
  {(Ga,Mn)As}},}\ }%
  \bibfield{journal}{%
  \bibinfo {journal} {Physica E}\ }%
  \textbf{\bibinfo {volume} {41}},\ \bibinfo {pages} {84}%
  \bibAnnoteFile{NoStop}{Wang:2008_PE}%
\bibitem[{\citenamefont{Wei}\ and\ \citenamefont{Zunger}(1987)}]{Wei:1987_PRL}%
  \BibitemOpen
  \bibfield{author}{%
  \bibinfo {author} {\bibnamefont{Wei}, \bibfnamefont{Su-Huai}},\ and\ \bibinfo
  {author} {\bibfnamefont{Alex}\ \bibnamefont{Zunger}}}%
  , \bibinfo {year} {1987},\ \bibfield{title}{%
  \enquote{\bibinfo {title} {Role of \textit{d} orbitals in valence-band
  offsets of common-anion semiconductors},}\ }%
  \bibfield{journal}{%
  \bibinfo {journal} {Phys. Rev. Lett.}\ }%
  \textbf{\bibinfo {volume} {59}},\ \bibinfo {pages} {144}%
  \bibAnnoteFile{NoStop}{Wei:1987_PRL}%
\bibitem[{\citenamefont{Welp}\ \emph{et~al.}(2003)\citenamefont{Welp},
  \citenamefont{Vlasko-Vlasov}, \citenamefont{Liu}, \citenamefont{Furdyna},\
  and\ \citenamefont{Wojtowicz}}]{Welp:2003_PRL}%
  \BibitemOpen
  \bibfield{author}{%
  \bibinfo {author} {\bibnamefont{Welp}, \bibfnamefont{U.}}, \bibinfo {author}
  {\bibfnamefont{V.~K.}\ \bibnamefont{Vlasko-Vlasov}}, \bibinfo {author}
  {\bibfnamefont{X.}~\bibnamefont{Liu}}, \bibinfo {author}
  {\bibfnamefont{J.~K.}\ \bibnamefont{Furdyna}},\ and\ \bibinfo {author}
  {\bibfnamefont{T.}~\bibnamefont{Wojtowicz}}}%
  , \bibinfo {year} {2003},\ \bibfield{title}{%
  \enquote{\bibinfo {title} {Magnetic domain structure and magnetic anisotropy
  in {Ga$_{1-x}$Mn$_{x}$As}},}\ }%
  \bibfield{journal}{%
  \bibinfo {journal} {Phys. Rev. Lett.}\ }%
  \textbf{\bibinfo {volume} {90}},\ \bibinfo {pages} {167206}%
  \bibAnnoteFile{NoStop}{Welp:2003_PRL}%
\bibitem[{\citenamefont{Welp}\ \emph{et~al.}(2004)\citenamefont{Welp},
  \citenamefont{Vlasko-Vlasov}, \citenamefont{Menzel}, \citenamefont{You},
  \citenamefont{Liu}, \citenamefont{Furdyna},\ and\
  \citenamefont{Wojtowicz}}]{Welp:2004_APL}%
  \BibitemOpen
  \bibfield{author}{%
  \bibinfo {author} {\bibnamefont{Welp}, \bibfnamefont{U.}}, \bibinfo {author}
  {\bibfnamefont{V.~K.}\ \bibnamefont{Vlasko-Vlasov}}, \bibinfo {author}
  {\bibfnamefont{A.}~\bibnamefont{Menzel}}, \bibinfo {author}
  {\bibfnamefont{H.~D.}\ \bibnamefont{You}}, \bibinfo {author}
  {\bibfnamefont{X.}~\bibnamefont{Liu}}, \bibinfo {author}
  {\bibfnamefont{J.~K.}\ \bibnamefont{Furdyna}},\ and\ \bibinfo {author}
  {\bibfnamefont{T.}~\bibnamefont{Wojtowicz}}}%
  , \bibinfo {year} {2004},\ \bibfield{title}{%
  \enquote{\bibinfo {title} {Uniaxial in-plane magnetic anisotropy of
  {$\mathrm{Ga}_{1-x}\mathrm{Mn}_{x}\mathrm{As}$}},}\ }%
  \bibfield{journal}{%
  \bibinfo {journal} {Appl. Phys. Lett.}\ }%
  \textbf{\bibinfo {volume} {85}},\ \bibinfo {pages} {260}%
  \bibAnnoteFile{NoStop}{Welp:2004_APL}%
\bibitem[{\citenamefont{Wenisch}\ \emph{et~al.}(2007)\citenamefont{Wenisch},
  \citenamefont{Gould}, \citenamefont{Ebel}, \citenamefont{Storz},
  \citenamefont{Pappert}, \citenamefont{Schmidt}, \citenamefont{Kumpf},
  \citenamefont{Schmidt}, \citenamefont{Brunner},\ and\
  \citenamefont{Molenkamp}}]{Wenisch:2007_PRL}%
  \BibitemOpen
  \bibfield{author}{%
  \bibinfo {author} {\bibnamefont{Wenisch}, \bibfnamefont{J.}}, \bibinfo
  {author} {\bibfnamefont{C.}~\bibnamefont{Gould}}, \bibinfo {author}
  {\bibfnamefont{L.}~\bibnamefont{Ebel}}, \bibinfo {author}
  {\bibfnamefont{J.}~\bibnamefont{Storz}}, \bibinfo {author}
  {\bibfnamefont{K.}~\bibnamefont{Pappert}}, \bibinfo {author}
  {\bibfnamefont{M.~J.}\ \bibnamefont{Schmidt}}, \bibinfo {author}
  {\bibfnamefont{C.}~\bibnamefont{Kumpf}}, \bibinfo {author}
  {\bibfnamefont{G.}~\bibnamefont{Schmidt}}, \bibinfo {author}
  {\bibfnamefont{K.}~\bibnamefont{Brunner}},\ and\ \bibinfo {author}
  {\bibfnamefont{L.~W.}\ \bibnamefont{Molenkamp}}}%
  , \bibinfo {year} {2007},\ \bibfield{title}{%
  \enquote{\bibinfo {title} {Control of magnetic anisotropy in {(Ga,Mn)As} by
  lithography-induced strain relaxation},}\ }%
  \bibfield{journal}{%
  \bibinfo {journal} {Phys. Rev. Lett.}\ }%
  \textbf{\bibinfo {volume} {99}},\ \bibinfo {pages} {077201}%
  \bibAnnoteFile{NoStop}{Wenisch:2007_PRL}%
\bibitem[{\citenamefont{Werpachowska}\ and\
  \citenamefont{Dietl}(2010{\natexlab{a}})}]{Werpachowska:2010_PRBb}%
  \BibitemOpen
  \bibfield{author}{%
  \bibinfo {author} {\bibnamefont{Werpachowska}, \bibfnamefont{A.}},\ and\
  \bibinfo {author} {\bibfnamefont{T.}~\bibnamefont{Dietl}}}%
  , \bibinfo {year} {2010}{\natexlab{a}},\ \bibfield{title}{%
  \enquote{\bibinfo {title} {Effect of inversion asymmetry on the intrinsic
  anomalous {Hall} effect in ferromagnetic {(Ga, Mn)As}},}\ }%
  \bibfield{journal}{%
  \bibinfo {journal} {Phys. Rev. B}\ }%
  \textbf{\bibinfo {volume} {81}},\ \bibinfo {pages} {155205}%
  \bibAnnoteFile{NoStop}{Werpachowska:2010_PRBb}%
\bibitem[{\citenamefont{Werpachowska}\ and\
  \citenamefont{Dietl}(2010{\natexlab{b}})}]{Werpachowska:2010_PRBa}%
  \BibitemOpen
  \bibfield{author}{%
  \bibinfo {author} {\bibnamefont{Werpachowska}, \bibfnamefont{A.}},\ and\
  \bibinfo {author} {\bibfnamefont{T.}~\bibnamefont{Dietl}}}%
  , \bibinfo {year} {2010}{\natexlab{b}},\ \bibfield{title}{%
  \enquote{\bibinfo {title} {Theory of spin waves in ferromagnetic
  {(Ga,Mn)As}},}\ }%
  \bibfield{journal}{%
  \bibinfo {journal} {Phys. Rev. B}\ }%
  \textbf{\bibinfo {volume} {82}},\ \bibinfo {pages} {085204}%
  \bibAnnoteFile{NoStop}{Werpachowska:2010_PRBa}%
\bibitem[{\citenamefont{White}\ \emph{et~al.}(2008)\citenamefont{White},
  \citenamefont{Ochsenbein},\ and\ \citenamefont{Gamelin}}]{White:2008_ChM}%
  \BibitemOpen
  \bibfield{author}{%
  \bibinfo {author} {\bibnamefont{White}, \bibfnamefont{M.~A.}}, \bibinfo
  {author} {\bibfnamefont{S.~T.}\ \bibnamefont{Ochsenbein}},\ and\ \bibinfo
  {author} {\bibfnamefont{D.~R.}\ \bibnamefont{Gamelin}}}%
  , \bibinfo {year} {2008},\ \bibfield{title}{%
  \enquote{\bibinfo {title} {Colloidal nanocrystals of wurtzite
  {Zn$_{1-x}$Co$_{x}$O} ${(0 \leq x \leq 1)}$: Models of spinodal decomposition
  in an oxide diluted magnetic semiconductor},}\ }%
  \bibfield{journal}{%
  \bibinfo {journal} {Chem. Mater.}\ }%
  \textbf{\bibinfo {volume} {20}},\ \bibinfo {pages} {7107}%
  \bibAnnoteFile{NoStop}{White:2008_ChM}%
\bibitem[{\citenamefont{Wilamowski}\
  \emph{et~al.}(2007)\citenamefont{Wilamowski}, \citenamefont{Malissa},
  \citenamefont{Sch\"affler},\ and\
  \citenamefont{Jantsch}}]{Wilamowski:2007_PRL}%
  \BibitemOpen
  \bibfield{author}{%
  \bibinfo {author} {\bibnamefont{Wilamowski}, \bibfnamefont{Z.}}, \bibinfo
  {author} {\bibfnamefont{H.}~\bibnamefont{Malissa}}, \bibinfo {author}
  {\bibfnamefont{F.}~\bibnamefont{Sch\"affler}},\ and\ \bibinfo {author}
  {\bibfnamefont{W.}~\bibnamefont{Jantsch}}}%
  , \bibinfo {year} {2007},\ \bibfield{title}{%
  \enquote{\bibinfo {title} {$g$-factor tuning and manipulation of spins by an
  electric current},}\ }%
  \bibfield{journal}{%
  \bibinfo {journal} {Phys. Rev. Lett.}\ }%
  \textbf{\bibinfo {volume} {98}},\ \bibinfo {pages} {187203}%
  \bibAnnoteFile{NoStop}{Wilamowski:2007_PRL}%
\bibitem[{\citenamefont{Wilson}\ \emph{et~al.}(2010)\citenamefont{Wilson},
  \citenamefont{Zhu}, \citenamefont{Myers}, \citenamefont{Awschalom},
  \citenamefont{Schiffer},\ and\ \citenamefont{Samarth}}]{Wilson:2010_PRB}%
  \BibitemOpen
  \bibfield{author}{%
  \bibinfo {author} {\bibnamefont{Wilson}, \bibfnamefont{M.~J.}}, \bibinfo
  {author} {\bibfnamefont{M.}~\bibnamefont{Zhu}}, \bibinfo {author}
  {\bibfnamefont{R.~C.}\ \bibnamefont{Myers}}, \bibinfo {author}
  {\bibfnamefont{D.~D.}\ \bibnamefont{Awschalom}}, \bibinfo {author}
  {\bibfnamefont{P.}~\bibnamefont{Schiffer}},\ and\ \bibinfo {author}
  {\bibfnamefont{N.}~\bibnamefont{Samarth}}}%
  , \bibinfo {year} {2010},\ \bibfield{title}{%
  \enquote{\bibinfo {title} {Interlayer and interfacial exchange coupling in
  ferromagnetic metal/semiconductor heterostructures},}\ }%
  \bibfield{journal}{%
  \bibinfo {journal} {Phys. Rev. B}\ }%
  \textbf{\bibinfo {volume} {81}},\ \bibinfo {pages} {045319}%
  \bibAnnoteFile{NoStop}{Wilson:2010_PRB}%
\bibitem[{\citenamefont{Winkler}\ \emph{et~al.}(2011)\citenamefont{Winkler},
  \citenamefont{Stone}, \citenamefont{Li}, \citenamefont{Yu},
  \citenamefont{Bonanni},\ and\ \citenamefont{Dubon}}]{Winkler:2011_APL}%
  \BibitemOpen
  \bibfield{author}{%
  \bibinfo {author} {\bibnamefont{Winkler}, \bibfnamefont{T.~E.}}, \bibinfo
  {author} {\bibfnamefont{P.~R.}\ \bibnamefont{Stone}}, \bibinfo {author}
  {\bibfnamefont{Tian}\ \bibnamefont{Li}}, \bibinfo {author}
  {\bibfnamefont{K.~M.}\ \bibnamefont{Yu}}, \bibinfo {author}
  {\bibfnamefont{A.}~\bibnamefont{Bonanni}},\ and\ \bibinfo {author}
  {\bibfnamefont{O.~D.}\ \bibnamefont{Dubon}}}%
  , \bibinfo {year} {2011},\ \bibfield{title}{%
  \enquote{\bibinfo {title} {Compensation-dependence of magnetic and electrical
  properties in {Ga$_{1 - x}$Mn$_{x}$P}},}\ }%
  \bibfield{journal}{%
  \bibinfo {journal} {Appl. Phys. Lett.}\ }%
  \textbf{\bibinfo {volume} {98}},\ \bibinfo {pages} {012103}%
  \bibAnnoteFile{NoStop}{Winkler:2011_APL}%
\bibitem[{\citenamefont{Wojnar}\ \emph{et~al.}(2012)\citenamefont{Wojnar},
  \citenamefont{Janik}, \citenamefont{Baczewski}, \citenamefont{Kret},
  \citenamefont{Dynowska}, \citenamefont{Wojciechowski},
  \citenamefont{Suffczy\'nski}, \citenamefont{Papierska},
  \citenamefont{Kossacki}, \citenamefont{Karczewski}, \citenamefont{Kossut},\
  and\ \citenamefont{Wojtowicz}}]{Wojnar:2012_NL}%
  \BibitemOpen
  \bibfield{author}{%
  \bibinfo {author} {\bibnamefont{Wojnar}, \bibfnamefont{P.}}, \bibinfo
  {author} {\bibfnamefont{E.}~\bibnamefont{Janik}}, \bibinfo {author}
  {\bibfnamefont{L.~T.}\ \bibnamefont{Baczewski}}, \bibinfo {author}
  {\bibfnamefont{S.}~\bibnamefont{Kret}}, \bibinfo {author}
  {\bibfnamefont{E.}~\bibnamefont{Dynowska}}, \bibinfo {author}
  {\bibfnamefont{T.}~\bibnamefont{Wojciechowski}}, \bibinfo {author}
  {\bibfnamefont{J.}~\bibnamefont{Suffczy\'nski}}, \bibinfo {author}
  {\bibfnamefont{J.}~\bibnamefont{Papierska}}, \bibinfo {author}
  {\bibfnamefont{P.}~\bibnamefont{Kossacki}}, \bibinfo {author}
  {\bibfnamefont{G.}~\bibnamefont{Karczewski}}, \bibinfo {author}
  {\bibfnamefont{J.}~\bibnamefont{Kossut}},\ and\ \bibinfo {author}
  {\bibfnamefont{T.}~\bibnamefont{Wojtowicz}}}%
  , \bibinfo {year} {2012},\ \bibfield{title}{%
  \enquote{\bibinfo {title} {Giant spin splitting in optically active
  {ZnMnTe/ZnMgTe} core/shell nanowires},}\ }%
  \bibfield{journal}{%
  \bibinfo {journal} {Nano Lett.}\ }%
  \textbf{\bibinfo {volume} {12}},\ \bibinfo {pages} {3404}%
  \bibAnnoteFile{NoStop}{Wojnar:2012_NL}%
\bibitem[{\citenamefont{Wojtowicz}\
  \emph{et~al.}(2003{\natexlab{a}})\citenamefont{Wojtowicz},
  \citenamefont{{Cywi\'{n}ski}}, \citenamefont{Lim}, \citenamefont{Liu},
  \citenamefont{Dobrowolska}, \citenamefont{Furdyna}, \citenamefont{Yu},
  \citenamefont{Walukiewicz}, \citenamefont{Kim}, \citenamefont{Cheon},
  \citenamefont{Chen}, \citenamefont{Wang},\ and\
  \citenamefont{Luo}}]{Wojtowicz:2003_APL}%
  \BibitemOpen
  \bibfield{author}{%
  \bibinfo {author} {\bibnamefont{Wojtowicz}, \bibfnamefont{T.}}, \bibinfo
  {author} {\bibfnamefont{G.}~\bibnamefont{{Cywi\'{n}ski}}}, \bibinfo {author}
  {\bibfnamefont{W.~L.}\ \bibnamefont{Lim}}, \bibinfo {author}
  {\bibfnamefont{X.}~\bibnamefont{Liu}}, \bibinfo {author}
  {\bibfnamefont{M.}~\bibnamefont{Dobrowolska}}, \bibinfo {author}
  {\bibfnamefont{J.~K.}\ \bibnamefont{Furdyna}}, \bibinfo {author}
  {\bibfnamefont{K.~M.}\ \bibnamefont{Yu}}, \bibinfo {author}
  {\bibfnamefont{W.}~\bibnamefont{Walukiewicz}}, \bibinfo {author}
  {\bibfnamefont{G.~B.}\ \bibnamefont{Kim}}, \bibinfo {author}
  {\bibfnamefont{M.}~\bibnamefont{Cheon}}, \bibinfo {author}
  {\bibfnamefont{X.}~\bibnamefont{Chen}}, \bibinfo {author}
  {\bibfnamefont{S.~M.}\ \bibnamefont{Wang}},\ and\ \bibinfo {author}
  {\bibfnamefont{H.}~\bibnamefont{Luo}}}%
  , \bibinfo {year} {2003}{\natexlab{a}},\ \bibfield{title}{%
  \enquote{\bibinfo {title} {{In$_{1-x}$Mn$_{x}$Sb} --- a new narrow gap
  ferromagnetic semiconductor},}\ }%
  \bibfield{journal}{%
  \bibinfo {journal} {Appl. Phys. Lett.}\ }%
  \textbf{\bibinfo {volume} {82}},\ \bibinfo {pages} {4310}%
  \bibAnnoteFile{NoStop}{Wojtowicz:2003_APL}%
\bibitem[{\citenamefont{Wojtowicz}\
  \emph{et~al.}(2003{\natexlab{b}})\citenamefont{Wojtowicz},
  \citenamefont{Lim}, \citenamefont{Liu}, \citenamefont{Dobrowolska},
  \citenamefont{Furdyna}, \citenamefont{Yu}, \citenamefont{Walukiewicz},
  \citenamefont{Vurgaftman},\ and\ \citenamefont{Meyer}}]{Wojtowicz:2003_APLb}%
  \BibitemOpen
  \bibfield{author}{%
  \bibinfo {author} {\bibnamefont{Wojtowicz}, \bibfnamefont{T.}}, \bibinfo
  {author} {\bibfnamefont{W.~L.}\ \bibnamefont{Lim}}, \bibinfo {author}
  {\bibfnamefont{X.}~\bibnamefont{Liu}}, \bibinfo {author}
  {\bibfnamefont{M.}~\bibnamefont{Dobrowolska}}, \bibinfo {author}
  {\bibfnamefont{J.~K.}\ \bibnamefont{Furdyna}}, \bibinfo {author}
  {\bibfnamefont{K.~M.}\ \bibnamefont{Yu}}, \bibinfo {author}
  {\bibfnamefont{W.}~\bibnamefont{Walukiewicz}}, \bibinfo {author}
  {\bibfnamefont{I.}~\bibnamefont{Vurgaftman}},\ and\ \bibinfo {author}
  {\bibfnamefont{J.~R.}\ \bibnamefont{Meyer}}}%
  , \bibinfo {year} {2003}{\natexlab{b}},\ \bibfield{title}{%
  \enquote{\bibinfo {title} {Enhancement of {Curie} temperature in
  {Ga$_{1-x}$Mn$_{x}$As/Ga$_{1-y}$Al$_{y}$As} ferromagnetic heterostructures by
  be modulation doping},}\ }%
  \bibfield{journal}{%
  \bibinfo {journal} {Appl. Phys. Lett.}\ }%
  \textbf{\bibinfo {volume} {83}},\ \bibinfo {pages} {4220}%
  \bibAnnoteFile{NoStop}{Wojtowicz:2003_APLb}%
\bibitem[{\citenamefont{Wo{\l}o{\'s}}\ and\
  \citenamefont{Kami{\'n}ska}(2008)}]{Wolos:2008_B}%
  \BibitemOpen
  \bibfield{author}{%
  \bibinfo {author} {\bibnamefont{Wo{\l}o{\'s}}, \bibfnamefont{A.}},\ and\
  \bibinfo {author} {\bibfnamefont{M.}~\bibnamefont{Kami{\'n}ska}}}%
  , \bibinfo {year} {2008},\ \enquote{\bibinfo {title} {Magnetic impurities in
  semiconductors},}\ in\ \emph{\bibinfo {booktitle} {Spintronics}},\ \bibinfo
  {editor} {edited by\ \bibinfo {editor}
  {\bibfnamefont{T.}~\bibnamefont{Dietl}}, \bibinfo {editor}
  {\bibfnamefont{D.~D.}\ \bibnamefont{Awschalom}}, \bibinfo {editor}
  {\bibfnamefont{M.}~\bibnamefont{Kami{\'n}ska}},\ and\ \bibinfo {editor}
  {\bibfnamefont{H.}~\bibnamefont{Ohno}}}\ (\bibinfo {publisher} {Elsevier,
  Amsterdam})\ p.\ \bibinfo {pages} {325}%
  \bibAnnoteFile{NoStop}{Wolos:2008_B}%
\bibitem[{\citenamefont{Wo{\l}o{\'s}}\
  \emph{et~al.}(2009)\citenamefont{Wo{\l}o{\'s}}, \citenamefont{Piersa},
  \citenamefont{Strzelecka}, \citenamefont{Korona}, \citenamefont{Hruban},\
  and\ \citenamefont{Kami{\'n}ska}}]{Wolos:2009_PSSC}%
  \BibitemOpen
  \bibfield{author}{%
  \bibinfo {author} {\bibnamefont{Wo{\l}o{\'s}}, \bibfnamefont{A.}}, \bibinfo
  {author} {\bibfnamefont{M.}~\bibnamefont{Piersa}}, \bibinfo {author}
  {\bibfnamefont{G.}~\bibnamefont{Strzelecka}}, \bibinfo {author}
  {\bibfnamefont{K.~P.}\ \bibnamefont{Korona}}, \bibinfo {author}
  {\bibfnamefont{A.}~\bibnamefont{Hruban}},\ and\ \bibinfo {author}
  {\bibfnamefont{M.}~\bibnamefont{Kami{\'n}ska}}}%
  , \bibinfo {year} {2009},\ \bibfield{title}{%
  \enquote{\bibinfo {title} {Mn configuration in {III--V} semiconductors and
  its influence on electric transport and semiconductor magnetism},}\ }%
  \bibfield{journal}{%
  \bibinfo {journal} {Phys. Status Solidi C}\ }%
  \textbf{\bibinfo {volume} {6}},\ \bibinfo {pages} {2769}%
  \bibAnnoteFile{NoStop}{Wolos:2009_PSSC}%
\bibitem[{\citenamefont{Wu}\ \emph{et~al.}(2005)\citenamefont{Wu},
  \citenamefont{Keavney}, \citenamefont{Wu}, \citenamefont{Johnston-Halperin},
  \citenamefont{Awschalom}, ,\ and\ \citenamefont{Shi}}]{Wu:2005_PRB}%
  \BibitemOpen
  \bibfield{author}{%
  \bibinfo {author} {\bibnamefont{Wu}, \bibfnamefont{D.}}, \bibinfo {author}
  {\bibfnamefont{D.~J.}\ \bibnamefont{Keavney}}, \bibinfo {author}
  {\bibfnamefont{{Ruqian}}\ \bibnamefont{Wu}}, \bibinfo {author}
  {\bibfnamefont{E.}~\bibnamefont{Johnston-Halperin}}, \bibinfo {author}
  {\bibfnamefont{D.~D.}\ \bibnamefont{Awschalom}}, ,\ and\ \bibinfo {author}
  {\bibfnamefont{{Jing}}\ \bibnamefont{Shi}}}%
  , \bibinfo {year} {2005},\ \bibfield{title}{%
  \enquote{\bibinfo {title} {Concentration-independent local ferromagnetic {Mn}
  configuration in {Ga$_{1-x}$Mn$_x$As}},}\ }%
  \bibfield{journal}{%
  \bibinfo {journal} {Phys. Rev. B}\ }%
  \textbf{\bibinfo {volume} {71}},\ \bibinfo {pages} {153310}%
  \bibAnnoteFile{NoStop}{Wu:2005_PRB}%
\bibitem[{\citenamefont{Wunderlich}\
  \emph{et~al.}(2006)\citenamefont{Wunderlich}, \citenamefont{Jungwirth},
  \citenamefont{Kaestner}, \citenamefont{Irvine}, \citenamefont{Shick},
  \citenamefont{Stone}, \citenamefont{Wang}, \citenamefont{Rana},
  \citenamefont{Giddings}, \citenamefont{Foxon}, \citenamefont{Campion},
  \citenamefont{Williams},\ and\
  \citenamefont{Gallagher}}]{Wunderlich:2006_PRL}%
  \BibitemOpen
  \bibfield{author}{%
  \bibinfo {author} {\bibnamefont{Wunderlich}, \bibfnamefont{J.}}, \bibinfo
  {author} {\bibfnamefont{T.}~\bibnamefont{Jungwirth}}, \bibinfo {author}
  {\bibfnamefont{B.}~\bibnamefont{Kaestner}}, \bibinfo {author}
  {\bibfnamefont{A.~C.}\ \bibnamefont{Irvine}}, \bibinfo {author}
  {\bibfnamefont{A.~B.}\ \bibnamefont{Shick}}, \bibinfo {author}
  {\bibfnamefont{N.}~\bibnamefont{Stone}}, \bibinfo {author}
  {\bibfnamefont{K.-Y.}\ \bibnamefont{Wang}}, \bibinfo {author}
  {\bibfnamefont{U.}~\bibnamefont{Rana}}, \bibinfo {author}
  {\bibfnamefont{A.~D.}\ \bibnamefont{Giddings}}, \bibinfo {author}
  {\bibfnamefont{C.~T.}\ \bibnamefont{Foxon}}, \bibinfo {author}
  {\bibfnamefont{R.~P.}\ \bibnamefont{Campion}}, \bibinfo {author}
  {\bibfnamefont{D.~A.}\ \bibnamefont{Williams}},\ and\ \bibinfo {author}
  {\bibfnamefont{B.~L.}\ \bibnamefont{Gallagher}}}%
  , \bibinfo {year} {2006},\ \bibfield{title}{%
  \enquote{\bibinfo {title} {Coulomb blockade anisotropic magnetoresistance
  effect in a {(Ga,Mn)As} single-electron transistor},}\ }%
  \bibfield{journal}{%
  \bibinfo {journal} {Phys. Rev. Lett.}\ }%
  \textbf{\bibinfo {volume} {97}},\ \bibinfo {pages} {077201}%
  \bibAnnoteFile{NoStop}{Wunderlich:2006_PRL}%
\bibitem[{\citenamefont{Xiang}\ and\
  \citenamefont{Samarth}(2007)}]{Xiang:2007_PRB}%
  \BibitemOpen
  \bibfield{author}{%
  \bibinfo {author} {\bibnamefont{Xiang}, \bibfnamefont{G.}},\ and\ \bibinfo
  {author} {\bibfnamefont{N.}~\bibnamefont{Samarth}}}%
  , \bibinfo {year} {2007},\ \bibfield{title}{%
  \enquote{\bibinfo {title} {Theoretical analysis of the influence of magnetic
  domain walls on longitudinal and transverse magnetoresistance in tensile
  strained {(Ga,Mn)As} epilayers},}\ }%
  \bibfield{journal}{%
  \bibinfo {journal} {Phys. Rev. B}\ }%
  \textbf{\bibinfo {volume} {76}},\ \bibinfo {pages} {054440}%
  \bibAnnoteFile{NoStop}{Xiang:2007_PRB}%
\bibitem[{\citenamefont{Xiu}\ \emph{et~al.}(2010)\citenamefont{Xiu},
  \citenamefont{Wang}, \citenamefont{Kim}, \citenamefont{Hong},
  \citenamefont{Tang}, \citenamefont{Jacob}, \citenamefont{Zou},\ and\
  \citenamefont{Wang}}]{Xiu:2010_NM}%
  \BibitemOpen
  \bibfield{author}{%
  \bibinfo {author} {\bibnamefont{Xiu}, \bibfnamefont{F.}}, \bibinfo {author}
  {\bibfnamefont{Y.}~\bibnamefont{Wang}}, \bibinfo {author}
  {\bibfnamefont{J.}~\bibnamefont{Kim}}, \bibinfo {author}
  {\bibfnamefont{A.}~\bibnamefont{Hong}}, \bibinfo {author}
  {\bibfnamefont{J.}~\bibnamefont{Tang}}, \bibinfo {author}
  {\bibfnamefont{A.~P.}\ \bibnamefont{Jacob}}, \bibinfo {author}
  {\bibfnamefont{J.}~\bibnamefont{Zou}},\ and\ \bibinfo {author}
  {\bibfnamefont{K.~L.}\ \bibnamefont{Wang}}}%
  , \bibinfo {year} {2010},\ \bibfield{title}{%
  \enquote{\bibinfo {title} {Electric-field-controlled ferromagnetism in
  high-{Curie}-temperature {Mn$_{0.05}$Ge$_{0.95}$} quantum dots},}\ }%
  \bibfield{journal}{%
  \bibinfo {journal} {Nat. Mater.}\ }%
  \textbf{\bibinfo {volume} {9}},\ \bibinfo {pages} {337}%
  \bibAnnoteFile{NoStop}{Xiu:2010_NM}%
\bibitem[{\citenamefont{Yamanouchi}\
  \emph{et~al.}(2006)\citenamefont{Yamanouchi}, \citenamefont{Chiba},
  \citenamefont{Matsukura}, \citenamefont{Dietl},\ and\
  \citenamefont{Ohno}}]{Yamanouchi:2006_PRL}%
  \BibitemOpen
  \bibfield{author}{%
  \bibinfo {author} {\bibnamefont{Yamanouchi}, \bibfnamefont{M.}}, \bibinfo
  {author} {\bibfnamefont{D.}~\bibnamefont{Chiba}}, \bibinfo {author}
  {\bibfnamefont{F.}~\bibnamefont{Matsukura}}, \bibinfo {author}
  {\bibfnamefont{T.}~\bibnamefont{Dietl}},\ and\ \bibinfo {author}
  {\bibfnamefont{H.}~\bibnamefont{Ohno}}}%
  , \bibinfo {year} {2006},\ \bibfield{title}{%
  \enquote{\bibinfo {title} {Velocity of domain-wall motion induced by
  electrical current in a ferromagnetic semiconductor {(Ga,Mn)As}},}\ }%
  \bibfield{journal}{%
  \bibinfo {journal} {Phys. Rev. Lett.}\ }%
  \textbf{\bibinfo {volume} {96}},\ \bibinfo {pages} {096601}%
  \bibAnnoteFile{NoStop}{Yamanouchi:2006_PRL}%
\bibitem[{\citenamefont{Yamanouchi}\
  \emph{et~al.}(2004)\citenamefont{Yamanouchi}, \citenamefont{Chiba},
  \citenamefont{Matsukura},\ and\ \citenamefont{Ohno}}]{Yamanouchi:2004_N}%
  \BibitemOpen
  \bibfield{author}{%
  \bibinfo {author} {\bibnamefont{Yamanouchi}, \bibfnamefont{M.}}, \bibinfo
  {author} {\bibfnamefont{D.}~\bibnamefont{Chiba}}, \bibinfo {author}
  {\bibfnamefont{F.}~\bibnamefont{Matsukura}},\ and\ \bibinfo {author}
  {\bibfnamefont{H.}~\bibnamefont{Ohno}}}%
  , \bibinfo {year} {2004},\ \bibfield{title}{%
  \enquote{\bibinfo {title} {Current-induced domain-wall switching in a
  ferromagnetic semiconductor structure},}\ }%
  \bibfield{journal}{%
  \bibinfo {journal} {Nature}\ }%
  \textbf{\bibinfo {volume} {428}},\ \bibinfo {pages} {539}%
  \bibAnnoteFile{NoStop}{Yamanouchi:2004_N}%
\bibitem[{\citenamefont{Yamanouchi}\
  \emph{et~al.}(2007)\citenamefont{Yamanouchi}, \citenamefont{Ieda},
  \citenamefont{Matsukura}, \citenamefont{Barnes}, \citenamefont{Maekawa},\
  and\ \citenamefont{Ohno}}]{Yamanouchi:2007_S}%
  \BibitemOpen
  \bibfield{author}{%
  \bibinfo {author} {\bibnamefont{Yamanouchi}, \bibfnamefont{M.}}, \bibinfo
  {author} {\bibfnamefont{J.}~\bibnamefont{Ieda}}, \bibinfo {author}
  {\bibfnamefont{F.}~\bibnamefont{Matsukura}}, \bibinfo {author}
  {\bibfnamefont{S.~E.}\ \bibnamefont{Barnes}}, \bibinfo {author}
  {\bibfnamefont{S.}~\bibnamefont{Maekawa}},\ and\ \bibinfo {author}
  {\bibfnamefont{H.}~\bibnamefont{Ohno}}}%
  , \bibinfo {year} {2007},\ \bibfield{title}{%
  \enquote{\bibinfo {title} {Universality classes for domain wall motion in the
  ferromagnetic semiconductor {(Ga,Mn)As}},}\ }%
  \bibfield{journal}{%
  \bibinfo {journal} {Science}\ }%
  \textbf{\bibinfo {volume} {317}},\ \bibinfo {pages} {1726}%
  \bibAnnoteFile{NoStop}{Yamanouchi:2007_S}%
\bibitem[{\citenamefont{Yang}\ \emph{et~al.}(2009)\citenamefont{Yang},
  \citenamefont{Zhu}, \citenamefont{Wang}, \citenamefont{Fang},
  \citenamefont{Yu}, \citenamefont{Yang}, \citenamefont{Zhang},
  \citenamefont{Qin}, \citenamefont{Yu},\ and\
  \citenamefont{Wang}}]{Yang:2009_APL}%
  \BibitemOpen
  \bibfield{author}{%
  \bibinfo {author} {\bibnamefont{Yang}, \bibfnamefont{X.~L.}}, \bibinfo
  {author} {\bibfnamefont{W.~X.}\ \bibnamefont{Zhu}}, \bibinfo {author}
  {\bibfnamefont{C.~D.}\ \bibnamefont{Wang}}, \bibinfo {author}
  {\bibfnamefont{H.}~\bibnamefont{Fang}}, \bibinfo {author}
  {\bibfnamefont{T.~J.}\ \bibnamefont{Yu}}, \bibinfo {author}
  {\bibfnamefont{Z.~J.}\ \bibnamefont{Yang}}, \bibinfo {author}
  {\bibfnamefont{G.~Y.}\ \bibnamefont{Zhang}}, \bibinfo {author}
  {\bibfnamefont{X.~B.}\ \bibnamefont{Qin}}, \bibinfo {author}
  {\bibfnamefont{R.~S.}\ \bibnamefont{Yu}},\ and\ \bibinfo {author}
  {\bibfnamefont{B.~Y.}\ \bibnamefont{Wang}}}%
  , \bibinfo {year} {2009},\ \bibfield{title}{%
  \enquote{\bibinfo {title} {Positron annihilation in {(Ga,Mn)N}: {A} study of
  vacancy-type defects},}\ }%
  \bibfield{journal}{%
  \bibinfo {journal} {Appl. Phys. Lett.}\ }%
  \textbf{\bibinfo {volume} {94}},\ \bibinfo {pages} {151907}%
  \bibAnnoteFile{NoStop}{Yang:2009_APL}%
\bibitem[{\citenamefont{Yang}\ \emph{et~al.}(2013)\citenamefont{Yang},
  \citenamefont{Li}, \citenamefont{Shen}, \citenamefont{Si},
  \citenamefont{Sun}, \citenamefont{Tao}, \citenamefont{Cao},
  \citenamefont{Xu},\ and\ \citenamefont{Zhang}}]{Yang:2013_APL}%
  \BibitemOpen
  \bibfield{author}{%
  \bibinfo {author} {\bibnamefont{Yang}, \bibfnamefont{Xiaojun}}, \bibinfo
  {author} {\bibfnamefont{Yuke}\ \bibnamefont{Li}}, \bibinfo {author}
  {\bibfnamefont{Chenyi}\ \bibnamefont{Shen}}, \bibinfo {author}
  {\bibfnamefont{Bingqi}\ \bibnamefont{Si}}, \bibinfo {author}
  {\bibfnamefont{Yunlei}\ \bibnamefont{Sun}}, \bibinfo {author}
  {\bibfnamefont{Qian}\ \bibnamefont{Tao}}, \bibinfo {author}
  {\bibfnamefont{Guanghan}\ \bibnamefont{Cao}}, \bibinfo {author}
  {\bibfnamefont{Zhuan}\ \bibnamefont{Xu}},\ and\ \bibinfo {author}
  {\bibfnamefont{Fuchun}\ \bibnamefont{Zhang}}}%
  , \bibinfo {year} {2013},\ \bibfield{title}{%
  \enquote{\bibinfo {title} {{Sr} and {Mn} co-doped {LaCuSO}: A wide band gap
  oxide diluted magnetic semiconductor with {$T_C$} around {200~K}},}\ }%
  \bibfield{journal}{%
  \bibinfo {journal} {Appl. Phys. Lett.}\ }%
  \textbf{\bibinfo {volume} {103}},\ \bibinfo {eid} {022410}%
  \bibAnnoteFile{NoStop}{Yang:2013_APL}%
\bibitem[{\citenamefont{Yastrubchak}\
  \emph{et~al.}(2011)\citenamefont{Yastrubchak}, \citenamefont{\.{Z}uk},
  \citenamefont{Krzy\.{z}anowska}, \citenamefont{Domagala},
  \citenamefont{Andrearczyk}, \citenamefont{Sadowski},\ and\
  \citenamefont{Wosinski}}]{Yastrubchak:2011_PRB}%
  \BibitemOpen
  \bibfield{author}{%
  \bibinfo {author} {\bibnamefont{Yastrubchak}, \bibfnamefont{O.}}, \bibinfo
  {author} {\bibfnamefont{J.}~\bibnamefont{\.{Z}uk}}, \bibinfo {author}
  {\bibfnamefont{H.}~\bibnamefont{Krzy\.{z}anowska}}, \bibinfo {author}
  {\bibfnamefont{J.~Z.}\ \bibnamefont{Domagala}}, \bibinfo {author}
  {\bibfnamefont{T.}~\bibnamefont{Andrearczyk}}, \bibinfo {author}
  {\bibfnamefont{J.}~\bibnamefont{Sadowski}},\ and\ \bibinfo {author}
  {\bibfnamefont{T.}~\bibnamefont{Wosinski}}}%
  , \bibinfo {year} {2011},\ \bibfield{title}{%
  \enquote{\bibinfo {title} {Photoreflectance study of the fundamental optical
  properties of {(Ga,Mn)As} epitaxial films},}\ }%
  \bibfield{journal}{%
  \bibinfo {journal} {Phys. Rev. B}\ }%
  \textbf{\bibinfo {volume} {83}},\ \bibinfo {pages} {245201}%
  \bibAnnoteFile{NoStop}{Yastrubchak:2011_PRB}%
\bibitem[{\citenamefont{Ye}\ and\ \citenamefont{Freeman}(2006)}]{Ye:2006_PRB}%
  \BibitemOpen
  \bibfield{author}{%
  \bibinfo {author} {\bibnamefont{Ye}, \bibfnamefont{L.-H.}},\ and\ \bibinfo
  {author} {\bibfnamefont{A.~J.}\ \bibnamefont{Freeman}}}%
  , \bibinfo {year} {2006},\ \bibfield{title}{%
  \enquote{\bibinfo {title} {Defect compensation, clustering, and magnetism in
  {Cr-doped} anatase},}\ }%
  \bibfield{journal}{%
  \bibinfo {journal} {Phys. Rev. B}\ }%
  \textbf{\bibinfo {volume} {73}},\ \bibinfo {pages} {081304(R)}%
  \bibAnnoteFile{NoStop}{Ye:2006_PRB}%
\bibitem[{\citenamefont{Yildirim}\ \emph{et~al.}(2007)\citenamefont{Yildirim},
  \citenamefont{Alvarez}, \citenamefont{Moreo},\ and\
  \citenamefont{Dagotto}}]{Yildirim:2007_PRL}%
  \BibitemOpen
  \bibfield{author}{%
  \bibinfo {author} {\bibnamefont{Yildirim}, \bibfnamefont{Y.}}, \bibinfo
  {author} {\bibfnamefont{G.}~\bibnamefont{Alvarez}}, \bibinfo {author}
  {\bibfnamefont{A.}~\bibnamefont{Moreo}},\ and\ \bibinfo {author}
  {\bibfnamefont{E.}~\bibnamefont{Dagotto}}}%
  , \bibinfo {year} {2007},\ \bibfield{title}{%
  \enquote{\bibinfo {title} {Large-scale monte~carlo study of a realistic
  lattice model for {Ga$_{1-x}$Mn$_{x}$As}},}\ }%
  \bibfield{journal}{%
  \bibinfo {journal} {Phys. Rev. Lett.}\ }%
  \textbf{\bibinfo {volume} {99}},\ \bibinfo {pages} {057207}%
  \bibAnnoteFile{NoStop}{Yildirim:2007_PRL}%
\bibitem[{\citenamefont{Young}\ \emph{et~al.}(2002)\citenamefont{Young},
  \citenamefont{Johnston-Halperin}, \citenamefont{Awschalom},
  \citenamefont{Ohno},\ and\ \citenamefont{Ohno}}]{Young:2002_APL}%
  \BibitemOpen
  \bibfield{author}{%
  \bibinfo {author} {\bibnamefont{Young}, \bibfnamefont{D.~K.}}, \bibinfo
  {author} {\bibfnamefont{E.}~\bibnamefont{Johnston-Halperin}}, \bibinfo
  {author} {\bibfnamefont{D.~D.}\ \bibnamefont{Awschalom}}, \bibinfo {author}
  {\bibfnamefont{Y.}~\bibnamefont{Ohno}},\ and\ \bibinfo {author}
  {\bibfnamefont{H.}~\bibnamefont{Ohno}}}%
  , \bibinfo {year} {2002},\ \bibfield{title}{%
  \enquote{\bibinfo {title} {Anisotropic electrical spin injection in
  ferromagnetic semiconductor heterostructures},}\ }%
  \bibfield{journal}{%
  \bibinfo {journal} {Appl. Phys. Lett.}\ }%
  \textbf{\bibinfo {volume} {80}},\ \bibinfo {pages} {1598}%
  \bibAnnoteFile{NoStop}{Young:2002_APL}%
\bibitem[{\citenamefont{Yu}\ \emph{et~al.}(2005)\citenamefont{Yu},
  \citenamefont{Walukiewicz}, \citenamefont{Wojtowicz},
  \citenamefont{Denlinger}, \citenamefont{Scarpulla}, \citenamefont{Liu},\ and\
  \citenamefont{Furdyna}}]{Yu:2005_APL}%
  \BibitemOpen
  \bibfield{author}{%
  \bibinfo {author} {\bibnamefont{Yu}, \bibfnamefont{K.~M.}}, \bibinfo {author}
  {\bibfnamefont{W.}~\bibnamefont{Walukiewicz}}, \bibinfo {author}
  {\bibfnamefont{T.}~\bibnamefont{Wojtowicz}}, \bibinfo {author}
  {\bibfnamefont{J.}~\bibnamefont{Denlinger}}, \bibinfo {author}
  {\bibfnamefont{M.~A.}\ \bibnamefont{Scarpulla}}, \bibinfo {author}
  {\bibfnamefont{X.}~\bibnamefont{Liu}},\ and\ \bibinfo {author}
  {\bibfnamefont{J.~K.}\ \bibnamefont{Furdyna}}}%
  , \bibinfo {year} {2005},\ \bibfield{title}{%
  \enquote{\bibinfo {title} {Effect of film thickness on the incorporation of
  {Mn} interstitials in {Ga$_{1-x}$Mn$_{x}$As}},}\ }%
  \bibfield{journal}{%
  \bibinfo {journal} {Appl. Phys. Lett.}\ }%
  \textbf{\bibinfo {volume} {86}},\ \bibinfo {pages} {042102}%
  \bibAnnoteFile{NoStop}{Yu:2005_APL}%
\bibitem[{\citenamefont{Yu}\ \emph{et~al.}(2002)\citenamefont{Yu},
  \citenamefont{Walukiewicz}, \citenamefont{Wojtowicz},
  \citenamefont{Kuryliszyn}, \citenamefont{Liu}, \citenamefont{Sasaki},\ and\
  \citenamefont{Furdyna}}]{Yu:2002_PRB}%
  \BibitemOpen
  \bibfield{author}{%
  \bibinfo {author} {\bibnamefont{Yu}, \bibfnamefont{K.~M.}}, \bibinfo {author}
  {\bibfnamefont{W.}~\bibnamefont{Walukiewicz}}, \bibinfo {author}
  {\bibfnamefont{T.}~\bibnamefont{Wojtowicz}}, \bibinfo {author}
  {\bibfnamefont{I.}~\bibnamefont{Kuryliszyn}}, \bibinfo {author}
  {\bibfnamefont{X.}~\bibnamefont{Liu}}, \bibinfo {author}
  {\bibfnamefont{Y.}~\bibnamefont{Sasaki}},\ and\ \bibinfo {author}
  {\bibfnamefont{J.~K.}\ \bibnamefont{Furdyna}}}%
  , \bibinfo {year} {2002},\ \bibfield{title}{%
  \enquote{\bibinfo {title} {Effect of the location of {Mn} sites in
  ferromagnetic {Ga$_{1-x}$Mn$_{x}$As} on its {Curie} temperature},}\ }%
  \bibfield{journal}{%
  \bibinfo {journal} {Phys. Rev. B}\ }%
  \textbf{\bibinfo {volume} {65}},\ \bibinfo {pages} {201303}%
  \bibAnnoteFile{NoStop}{Yu:2002_PRB}%
\bibitem[{\citenamefont{Yu}\ \emph{et~al.}(2004)\citenamefont{Yu},
  \citenamefont{Walukiewicz}, \citenamefont{Wojtowicz}, \citenamefont{Lim},
  \citenamefont{Liu}, \citenamefont{Dobrowolska},\ and\
  \citenamefont{Furdyna}}]{Yu:2004_APL}%
  \BibitemOpen
  \bibfield{author}{%
  \bibinfo {author} {\bibnamefont{Yu}, \bibfnamefont{K.~M.}}, \bibinfo {author}
  {\bibfnamefont{W.}~\bibnamefont{Walukiewicz}}, \bibinfo {author}
  {\bibfnamefont{T.}~\bibnamefont{Wojtowicz}}, \bibinfo {author}
  {\bibfnamefont{W.~L.}\ \bibnamefont{Lim}}, \bibinfo {author}
  {\bibfnamefont{X.}~\bibnamefont{Liu}}, \bibinfo {author}
  {\bibfnamefont{M.}~\bibnamefont{Dobrowolska}},\ and\ \bibinfo {author}
  {\bibfnamefont{J.~K.}\ \bibnamefont{Furdyna}}}%
  , \bibinfo {year} {2004},\ \bibfield{title}{%
  \enquote{\bibinfo {title} {Direct evidence of the {Fermi}-energy-dependent
  formation of {Mn} interstitials in modulation-doped
  {Ga$_{1-y}$Al$_{y}$As/Ga$_{1-x}$Mn$_{x}$As/Ga$_{1-y}$Al$_{y}$As}
  heterostructures},}\ }%
  \bibfield{journal}{%
  \bibinfo {journal} {Appl. Phys. Lett.}\ }%
  \textbf{\bibinfo {volume} {84}},\ \bibinfo {pages} {4325}%
  \bibAnnoteFile{NoStop}{Yu:2004_APL}%
\bibitem[{\citenamefont{Yu}\ \emph{et~al.}(2008)\citenamefont{Yu},
  \citenamefont{Wojtowicz}, \citenamefont{Walukiewicz}, \citenamefont{Liu},\
  and\ \citenamefont{Furdyna}}]{Furdyna:2008_B}%
  \BibitemOpen
  \bibfield{author}{%
  \bibinfo {author} {\bibnamefont{Yu}, \bibfnamefont{K.~M.}}, \bibinfo {author}
  {\bibfnamefont{T.}~\bibnamefont{Wojtowicz}}, \bibinfo {author}
  {\bibfnamefont{W.}~\bibnamefont{Walukiewicz}}, \bibinfo {author}
  {\bibfnamefont{X.}~\bibnamefont{Liu}},\ and\ \bibinfo {author}
  {\bibfnamefont{J.~K.}\ \bibnamefont{Furdyna}}}%
  , \bibinfo {year} {2008},\ \enquote{\bibinfo {title} {Fermi level effects on
  {Mn} incorporation in {III-Mn-V} ferromagnetic semiconductors},}\ in\
  \emph{\bibinfo {booktitle} {Spintronics}},\ \bibinfo {editor} {edited by\
  \bibinfo {editor} {\bibfnamefont{T.}~\bibnamefont{Dietl}}, \bibinfo {editor}
  {\bibfnamefont{D.~D.}\ \bibnamefont{Awschalom}}, \bibinfo {editor}
  {\bibfnamefont{M.}~\bibnamefont{Kami{\'n}ska}},\ and\ \bibinfo {editor}
  {\bibfnamefont{H.}~\bibnamefont{Ohno}}}\ (\bibinfo {publisher} {Elsevier,
  Amsterdam})\ p.~\bibinfo {pages} {89}%
  \bibAnnoteFile{NoStop}{Furdyna:2008_B}%
\bibitem[{\citenamefont{Yu}\ \emph{et~al.}(2010)\citenamefont{Yu},
  \citenamefont{Zhang}, \citenamefont{Zhang}, \citenamefont{Zhang},
  \citenamefont{Dai},\ and\ \citenamefont{Fang}}]{Yu:2010_S}%
  \BibitemOpen
  \bibfield{author}{%
  \bibinfo {author} {\bibnamefont{Yu}, \bibfnamefont{R.}}, \bibinfo {author}
  {\bibfnamefont{W.}~\bibnamefont{Zhang}}, \bibinfo {author}
  {\bibfnamefont{H.}~\bibnamefont{Zhang}}, \bibinfo {author}
  {\bibfnamefont{S.}~\bibnamefont{Zhang}}, \bibinfo {author}
  {\bibfnamefont{X.}~\bibnamefont{Dai}},\ and\ \bibinfo {author}
  {\bibfnamefont{Z.}~\bibnamefont{Fang}}}%
  , \bibinfo {year} {2010},\ \bibfield{title}{%
  \enquote{\bibinfo {title} {Quantized anomalous {Hall} effect in magnetic
  topological insulators},}\ }%
  \bibfield{journal}{%
  \bibinfo {journal} {Science}\ }%
  \textbf{\bibinfo {volume} {329}}~(\bibinfo {number} {5987}),\ \bibinfo
  {pages} {61}%
  \bibAnnoteFile{NoStop}{Yu:2010_S}%
\bibitem[{\citenamefont{Yuldashev}\
  \emph{et~al.}(2010)\citenamefont{Yuldashev}, \citenamefont{Igamberdiev},
  \citenamefont{Lee}, \citenamefont{Kwon}, \citenamefont{Kim},
  \citenamefont{Im}, \citenamefont{Shashkov},\ and\
  \citenamefont{Kang}}]{Yuldashev:2010_APE}%
  \BibitemOpen
  \bibfield{author}{%
  \bibinfo {author} {\bibnamefont{Yuldashev}, \bibfnamefont{S.}}, \bibinfo
  {author} {\bibfnamefont{K.}~\bibnamefont{Igamberdiev}}, \bibinfo {author}
  {\bibfnamefont{S.}~\bibnamefont{Lee}}, \bibinfo {author}
  {\bibfnamefont{Y.}~\bibnamefont{Kwon}}, \bibinfo {author}
  {\bibfnamefont{Y.}~\bibnamefont{Kim}}, \bibinfo {author}
  {\bibfnamefont{H.}~\bibnamefont{Im}}, \bibinfo {author}
  {\bibfnamefont{A.}~\bibnamefont{Shashkov}},\ and\ \bibinfo {author}
  {\bibfnamefont{T.~W.}\ \bibnamefont{Kang}}}%
  , \bibinfo {year} {2010},\ \bibfield{title}{%
  \enquote{\bibinfo {title} {Specific heat study of {GaMnAs}},}\ }%
  \bibfield{journal}{%
  \bibinfo {journal} {Appl. Phys. Express}\ }%
  \textbf{\bibinfo {volume} {3}},\ \bibinfo {pages} {073005}%
  \bibAnnoteFile{NoStop}{Yuldashev:2010_APE}%
\bibitem[{\citenamefont{Zaj{\c{a}}c}\
  \emph{et~al.}(2001)\citenamefont{Zaj{\c{a}}c}, \citenamefont{Gosk},
  \citenamefont{Kami{\'n}ska}, \citenamefont{Twardowski},
  \citenamefont{Szyszko},\ and\ \citenamefont{Podsiad{\l}o}}]{Zajac:2001_APL}%
  \BibitemOpen
  \bibfield{author}{%
  \bibinfo {author} {\bibnamefont{Zaj{\c{a}}c}, \bibfnamefont{M.}}, \bibinfo
  {author} {\bibfnamefont{J.}~\bibnamefont{Gosk}}, \bibinfo {author}
  {\bibfnamefont{M.}~\bibnamefont{Kami{\'n}ska}}, \bibinfo {author}
  {\bibfnamefont{A.}~\bibnamefont{Twardowski}}, \bibinfo {author}
  {\bibfnamefont{T.}~\bibnamefont{Szyszko}},\ and\ \bibinfo {author}
  {\bibfnamefont{S.}~\bibnamefont{Podsiad{\l}o}}}%
  , \bibinfo {year} {2001},\ \bibfield{title}{%
  \enquote{\bibinfo {title} {Paramagnetism and antiferromagnetic d--d coupling
  in {GaMnN} magnetic semiconductor},}\ }%
  \bibfield{journal}{%
  \bibinfo {journal} {Appl. Phys. Lett.}\ }%
  \textbf{\bibinfo {volume} {79}},\ \bibinfo {pages} {2432}%
  \bibAnnoteFile{NoStop}{Zajac:2001_APL}%
\bibitem[{\citenamefont{Zemen}\ \emph{et~al.}(2009)\citenamefont{Zemen},
  \citenamefont{Ku\v{c}era}, \citenamefont{Olejn\'ik},\ and\
  \citenamefont{Jungwirth}}]{Zemen:2009_PRB}%
  \BibitemOpen
  \bibfield{author}{%
  \bibinfo {author} {\bibnamefont{Zemen}, \bibfnamefont{J.}}, \bibinfo {author}
  {\bibfnamefont{J.}~\bibnamefont{Ku\v{c}era}}, \bibinfo {author}
  {\bibfnamefont{K.}~\bibnamefont{Olejn\'ik}},\ and\ \bibinfo {author}
  {\bibfnamefont{T.}~\bibnamefont{Jungwirth}}}%
  , \bibinfo {year} {2009},\ \bibfield{title}{%
  \enquote{\bibinfo {title} {Magnetocrystalline anisotropies in {(Ga,Mn)As}:
  {Systematic} theoretical study and comparison with experiment},}\ }%
  \bibfield{journal}{%
  \bibinfo {journal} {Phys. Rev. B}\ }%
  \textbf{\bibinfo {volume} {80}},\ \bibinfo {pages} {155203}%
  \bibAnnoteFile{NoStop}{Zemen:2009_PRB}%
\bibitem[{\citenamefont{Zener}(1951)}]{Zener:1951_PRb}%
  \BibitemOpen
  \bibfield{author}{%
  \bibinfo {author} {\bibnamefont{Zener}, \bibfnamefont{C.}}}%
  , \bibinfo {year} {1951},\ \bibfield{title}{%
  \enquote{\bibinfo {title} {Interaction between the d-shells in the transition
  metals. {II}. ferromagnetic compounds of manganese with perovskite
  structure},}\ }%
  \bibfield{journal}{%
  \bibinfo {journal} {Phys. Rev.}\ }%
  \textbf{\bibinfo {volume} {82}},\ \bibinfo {pages} {403}%
  \bibAnnoteFile{NoStop}{Zener:1951_PRb}%
\bibitem[{\citenamefont{Zhang}\ and\ \citenamefont{Li}(2004)}]{Zhang:2004_PRL}%
  \BibitemOpen
  \bibfield{author}{%
  \bibinfo {author} {\bibnamefont{Zhang}, \bibfnamefont{S.}},\ and\ \bibinfo
  {author} {\bibfnamefont{Z.}~\bibnamefont{Li}}}%
  , \bibinfo {year} {2004},\ \bibfield{title}{%
  \enquote{\bibinfo {title} {Roles of nonequilibrium conduction electrons on
  the magnetization dynamics of ferromagnets},}\ }%
  \bibfield{journal}{%
  \bibinfo {journal} {Phys. Rev. Lett.}\ }%
  \textbf{\bibinfo {volume} {93}},\ \bibinfo {pages} {127204}%
  \bibAnnoteFile{NoStop}{Zhang:2004_PRL}%
\bibitem[{\citenamefont{Zhao}\ \emph{et~al.}(2013)\citenamefont{Zhao},
  \citenamefont{Deng}, \citenamefont{Wang}, \citenamefont{Han},
  \citenamefont{Zhu}, \citenamefont{Li}, \citenamefont{Liu}, \citenamefont{Yu},
  \citenamefont{Goko}, \citenamefont{Frandsen}, \citenamefont{Liu},
  \citenamefont{Ning}, \citenamefont{Uemura}, \citenamefont{Dabkowska},
  \citenamefont{Luke}, \citenamefont{Luetkens}, \citenamefont{Morenzoni},
  \citenamefont{Dunsiger}, \citenamefont{Senyshyn}, \citenamefont{B\"oni},\
  and\ \citenamefont{Jin}}]{Zhao:2013_NC}%
  \BibitemOpen
  \bibfield{author}{%
  \bibinfo {author} {\bibnamefont{Zhao}, \bibfnamefont{K.}}, \bibinfo {author}
  {\bibfnamefont{Z.}~\bibnamefont{Deng}}, \bibinfo {author}
  {\bibfnamefont{X.C.}\ \bibnamefont{Wang}}, \bibinfo {author}
  {\bibfnamefont{W.}~\bibnamefont{Han}}, \bibinfo {author}
  {\bibfnamefont{J.L.}\ \bibnamefont{Zhu}}, \bibinfo {author}
  {\bibfnamefont{X.}~\bibnamefont{Li}}, \bibinfo {author} {\bibfnamefont{Q.Q.}\
  \bibnamefont{Liu}}, \bibinfo {author} {\bibfnamefont{R.C.}\
  \bibnamefont{Yu}}, \bibinfo {author} {\bibfnamefont{T.}~\bibnamefont{Goko}},
  \bibinfo {author} {\bibfnamefont{B.}~\bibnamefont{Frandsen}}, \bibinfo
  {author} {\bibfnamefont{{Lian}}\ \bibnamefont{Liu}}, \bibinfo {author}
  {\bibfnamefont{{Fanlong}}\ \bibnamefont{Ning}}, \bibinfo {author}
  {\bibfnamefont{Y.J.}\ \bibnamefont{Uemura}}, \bibinfo {author}
  {\bibfnamefont{H.}~\bibnamefont{Dabkowska}}, \bibinfo {author}
  {\bibfnamefont{G.M.}\ \bibnamefont{Luke}}, \bibinfo {author}
  {\bibfnamefont{H.}~\bibnamefont{Luetkens}}, \bibinfo {author}
  {\bibfnamefont{E.}~\bibnamefont{Morenzoni}}, \bibinfo {author}
  {\bibfnamefont{S.R.}\ \bibnamefont{Dunsiger}}, \bibinfo {author}
  {\bibfnamefont{A.}~\bibnamefont{Senyshyn}}, \bibinfo {author}
  {\bibfnamefont{P.}~\bibnamefont{B\"oni}},\ and\ \bibinfo {author}
  {\bibfnamefont{C.Q.}\ \bibnamefont{Jin}}}%
  , \bibinfo {year} {2013},\ \bibfield{title}{%
  \enquote{\bibinfo {title} {New diluted ferromagnetic semiconductor with
  {Curie} temperature up to {180~K} and isostructural to the �122�
  iron-based superconductors},}\ }%
  \bibfield{journal}{%
  \bibinfo {journal} {Nat. Commun.}\ }%
  \textbf{\bibinfo {volume} {4}},\ \bibinfo {pages} {1442}%
  \bibAnnoteFile{NoStop}{Zhao:2013_NC}%
\bibitem[{\citenamefont{Zhou}\ \emph{et~al.}(2012)\citenamefont{Zhou},
  \citenamefont{Wang}, \citenamefont{Jiang}, \citenamefont{Weschke},\ and\
  \citenamefont{Helm}}]{Zhou:2012_APEX}%
  \BibitemOpen
  \bibfield{author}{%
  \bibinfo {author} {\bibnamefont{Zhou}, \bibfnamefont{S.}}, \bibinfo {author}
  {\bibfnamefont{Y.}~\bibnamefont{Wang}}, \bibinfo {author}
  {\bibfnamefont{Z.}~\bibnamefont{Jiang}}, \bibinfo {author}
  {\bibfnamefont{E.}~\bibnamefont{Weschke}},\ and\ \bibinfo {author}
  {\bibfnamefont{M.}~\bibnamefont{Helm}}}%
  , \bibinfo {year} {2012},\ \bibfield{title}{%
  \enquote{\bibinfo {title} {Ferromagnetic {InMnAs} on {InAs} prepared by ion
  implantation and pulsed laser annealing},}\ }%
  \bibfield{journal}{%
  \bibinfo {journal} {Appl. Phys. Expr.}\ }%
  \textbf{\bibinfo {volume} {5}},\ \bibinfo {pages} {093007}%
  \bibAnnoteFile{NoStop}{Zhou:2012_APEX}%
\bibitem[{\citenamefont{Zhu}\ \emph{et~al.}(2007)\citenamefont{Zhu},
  \citenamefont{Wilson}, \citenamefont{Sheu}, \citenamefont{Mitra},
  \citenamefont{Schiffer},\ and\ \citenamefont{Samarth}}]{Zhu:2007_APL}%
  \BibitemOpen
  \bibfield{author}{%
  \bibinfo {author} {\bibnamefont{Zhu}, \bibfnamefont{M.}}, \bibinfo {author}
  {\bibfnamefont{M.~J.}\ \bibnamefont{Wilson}}, \bibinfo {author}
  {\bibfnamefont{B.~L.}\ \bibnamefont{Sheu}}, \bibinfo {author}
  {\bibfnamefont{P.}~\bibnamefont{Mitra}}, \bibinfo {author}
  {\bibfnamefont{P.}~\bibnamefont{Schiffer}},\ and\ \bibinfo {author}
  {\bibfnamefont{N.}~\bibnamefont{Samarth}}}%
  , \bibinfo {year} {2007},\ \bibfield{title}{%
  \enquote{\bibinfo {title} {Spin valve effect in self-exchange biased
  ferromagnetic metal/semiconductor bilayers},}\ }%
  \bibfield{journal}{%
  \bibinfo {journal} {Appl. Phys. Lett.}\ }%
  \textbf{\bibinfo {volume} {91}},\ \bibinfo {pages} {192503}%
  \bibAnnoteFile{NoStop}{Zhu:2007_APL}%
\bibitem[{\citenamefont{Zunger}(1986)}]{Zunger:1986_B}%
  \BibitemOpen
  \bibfield{author}{%
  \bibinfo {author} {\bibnamefont{Zunger}, \bibfnamefont{A.}}}%
  , \bibinfo {year} {1986},\ \enquote{\bibinfo {title} {Electronic {Structure}
  of 3d {Transition-Atom} {Impurities} in {Semiconductors}},}\ in\
  \emph{\bibinfo {booktitle} {Solid State Physics}},\ Vol.~\bibinfo {volume}
  {39},\ \bibinfo {editor} {edited by\ \bibinfo {editor}
  {\bibfnamefont{F.}~\bibnamefont{Seitz}}\ and\ \bibinfo {editor}
  {\bibfnamefont{D.}~\bibnamefont{Turnbull}}}\ (\bibinfo {publisher} {Academic
  Press, New York})\ p.\ \bibinfo {pages} {275}%
  \bibAnnoteFile{NoStop}{Zunger:1986_B}%
\bibitem[{\citenamefont{Zunger}\ \emph{et~al.}(2010)\citenamefont{Zunger},
  \citenamefont{Lany},\ and\ \citenamefont{Raebiger}}]{Zunger:2010_P}%
  \BibitemOpen
  \bibfield{author}{%
  \bibinfo {author} {\bibnamefont{Zunger}, \bibfnamefont{A.}}, \bibinfo
  {author} {\bibfnamefont{S.}~\bibnamefont{Lany}},\ and\ \bibinfo {author}
  {\bibfnamefont{H.}~\bibnamefont{Raebiger}}}%
  , \bibinfo {year} {2010},\ \bibfield{title}{%
  \enquote{\bibinfo {title} {The quest for dilute ferromagnetism in
  semiconductors: {Guides} and misguides by theory},}\ }%
  \bibfield{journal}{%
  \bibinfo {journal} {Physics}\ }%
  \textbf{\bibinfo {volume} {3}},\ \bibinfo {pages} {53}%
  \bibAnnoteFile{NoStop}{Zunger:2010_P}%
\bibitem[{\citenamefont{{\v{Z}uti{\'c}}}\
  \emph{et~al.}(2004)\citenamefont{{\v{Z}uti{\'c}}}, \citenamefont{Fabian},\
  and\ \citenamefont{{Das {Sarma}}}}]{Zutic:2004_RMP}%
  \BibitemOpen
  \bibfield{author}{%
  \bibinfo {author} {\bibnamefont{{\v{Z}uti{\'c}}}, \bibfnamefont{I.}},
  \bibinfo {author} {\bibfnamefont{J.}~\bibnamefont{Fabian}},\ and\ \bibinfo
  {author} {\bibfnamefont{S.}~\bibnamefont{{Das {Sarma}}}}}%
  , \bibinfo {year} {2004},\ \bibfield{title}{%
  \enquote{\bibinfo {title} {Spintronics: Fundamentals and applications},}\ }%
  \bibfield{journal}{%
  \bibinfo {journal} {Rev. Mod. Phys.}\ }%
  \textbf{\bibinfo {volume} {76}},\ \bibinfo {pages} {323}%
  \bibAnnoteFile{NoStop}{Zutic:2004_RMP}%
\end{thebibliography}




\end{document}